**AIN SHAMS UNIVERSITY**
**FACULTY OF ENGINEERING**

Electronics and Communication Engineering Department

# Antenna Designs for Recent Millimeter and THz Applications

## A Thesis

**Submitted in Partial Fulfillment of the Requirements**

**For the Degree of Doctor of Philosophy in Electrical Engineering**

**(Electronics and Communication Engineering)**

Submitted By

**Eng. Kamel Salah Kamel Sultan**

Supervised By

**Prof. Dr. Esmat Abdel-Fattah Abdallah**

**Prof. Dr. Hadia Mohamed Saied El Hennawy**

**Prof. Haythem Hussien Abdullah**

**Assoc. Prof. Mohamed Ali Basha**

**AIN SHAMS UNIVERSITY**
**FACULTY OF ENGINEERING**
Electronics and Communication Engineering Department

**Antenna Designs for Recent Millimeter and THz Applications**

By

**Kamel Salah Kamel Sultan**

Doctor of Philosophy in Electrical Engineering
(Electronics and Communications Engineering)
Faculty of Engineering, Ain Shams University, 2021

# Examiners' Committee

Approved by:

| Name and Affiliation | Signature |
|---|---|
| **1- Prof. Mostafa Elsaid Mostafa**<br>Electronics and Electrical Communications Department,<br>Faculty of Engineering, Cairo University. | ………………..<br>(Examiner) |
| **2- Prof. Amr Mohamed Ezt Safwat**<br>Electronics and Communications Department,<br>Faculty of Engineering, Ain Shams University. | ………………..<br>(Examiner) |
| **3- Prof. Esmat Abdel-Fattah Abdallah**<br>Microstrip Department,<br>Electronics Research Institute. | ………………<br>(Supervisor) |
| **1- Prof. Hadia Mohamed Saied El Hennawy**<br>Electronics and Communications Department,<br>Faculty of Engineering, Ain Shams University. | …………………<br>(Supervisor) |

Date: 13/12/2021

# STATEMENT

This dissertation is submitted as a partial fulfillment of the degree of Doctor of Philosophy in Electrical Engineering (Electronics and Communications Engineering) Faculty of Engineering, Ain Shams University.

The work included in this thesis was carried out by the author at the Electronics and Communications Department, Faculty of Engineering, Ain Shams University, Cairo, Egypt.

No part of this thesis was submitted for a degree or a qualification at any other university or institution.

**Student Name**: Kamel Salah Kamel Sultan
**Signature**:……………………………………
**Date**:

# ABSTRACT

## Antenna Designs for Recent Millimeter and THz Applications

## by

**KAMEL SALAH KAMEL SULTAN**

**DOCTOR OF PHILOSOPHY IN ELECTRICAL ENGINEERING THESIS**

**AIN SHAMS UNIVERSITY**

Over the previous few years, the millimetre wave frequency range and sub-THz range have been received a lot of attention as they include unused frequency spectrum resources that are appropriate for providing a lot of applications such as automotive radars, short communication, medical imaging, security, and 5G communications to provide end-user access to Multi-Gbit/s services. On the other hand and because of the restrictions on the communication systems in these ranges such as power, path losses, and attenuation, the present RF antennas can't be used for millimeter and THz applications. The needed antennas should have low-profile, high-gain, acceptable technology, high efficiency and low cost to compensate these ranges restrictions.

Although there are some antenna designs for millimeter and sub THz ranges through the last decay, the antenna designs in these ranges still under investigation and studies to complete the vision of the proposed designs for the aforementioned applications. Thus, they cannot be carried by the optimum response and efficiency for these applications. This thesis aims to introduce a comprehensive study for the main applications that occupy these ranges of frequencies. In addition to introduce efficient antennas for each application with acceptable technology for it.

As the automotive radar sensors play an important role in driver safety and assist him in the actions, the antennas that are used in this application is a key component in the sensor because they need to provide high gain to increase the radar range and to provide wideband to increase the radar resolution. The virtual antenna array (VAA) concept is introduced to provide low profile radar antenna array with high gain and serve the long range radar (LRR) and medium range radar (MRR). The analysis of VAA are introduced and verified. The antenna is fabricated and measured. Furthermore, hybrid linear antenna arrays with two different configurations are introduced to achieve a suitable range and HPBW for LRR. The frequencies that can be used for automotive radar sensors are 24 GHz and 76 GHz; those two bands are covered in this thesis.

The second contribution in this thesis is introducing an antenna for one of the future communication systems (5G). The proposed antenna has a dual-polarization to overcome the high losses at 28 GHz (the best-recommended band for 5G). Furthermore, the multiple-input multiple-output (MIMO) antenna for 5G is introduced with a complete study of the MIMO parameters. This antenna is based on characteristic mode analysis to study the antenna's performance. The metasurface is combined with the slot antenna to enhance its performance and increases the gain. The detailed illustrations of the dual-polarized antenna for handheld 5G systems with the comprehensive study of the interaction of an antenna with the human body and vice versa are considered. The antenna is fabricated and measured.

The third contribution in this dissertation mainly focused on the antennas designed for short communications and multi-Giga-bit data rate applications. Two different endfire on-chip antennas (OCA) using CMOS technology are introduced. These antennas are the Yagi-uda antenna and tapered slot Vivaldi antenna. These antennas succeeded in achieving high performance compared to the previous published OCAs because of the integration of different techniques to increase the radiation characteristics of these antennas. Furthermore, a MIMO on-chip antenna is introduced to overcome the high losses and high attenuation at 60 GHz. Three different configurations from two elements of MIMO are presented in addition to one configuration from four elements of MIMO based on the diversity technique to increase the isolation between the elements is also introduced. In terms of the on-chip antenna, it is observed that the introduced MIMO antenna overcomes high CMOS losses.

The last contribution in this thesis is an antenna array that is based on the dielectric waveguide and silicon on glass technology. The mode analysis, dielectric rod design, a transition between the metallic waveguide and dielectric waveguide, and disc dielectric antenna design are introduced to the antenna array design. The antenna meets the high gain and low profile structure requirements for the sub-THz applications. In addition, the CPW feeding network is compatible with the other components in the THz devices.

The introduced antennas with different techniques and different technologies in this thesis positively contribute to the millimeter and the sub-THz applications. They are expected to enhance the performance of the antennas for automotive radars, 5G handheld devices, multi-giga-bit communications devices, short-range networks, and biomedical imaging.

**Thesis supervisors:**

- Prof. Dr. Esmat Abdel-Fattah Abdallah
  Electronics Research Institute,
  Giza, EGYPT.
- Prof. Dr. Hadia Mohammed Said El-Hennawy
  Ain Shams University,
  Cairo, EGYPT.
- Prof. Dr. Haythm Hussien Abdullah
  Electronics Research Institute,
  Giza, EGYPT.

- Assoc. Prof. Mohamed Basha
  Waterloo University,
  Canada.

# PUBLICATIONS

**A. Journals**

1. K. S. Sultan, H. H. Abdullah, E. A. Abdallah, and H. El-Hennawy "MOM/GA-based virtual array for radar systems" MDPI, Sensors (Basel, Switzerland), vol. 20, no. 3, pp.1-16, 2020.
2. K. S. Sultan, H. H. Abdullah, E. A. Abdallah, and H. S. El-Hennawy "Metasurface based dual polarized MIMO antenna for 5G smartphones using CMA " *IEEE Access*, vol. 8, pp. 37250-37264, 2020.
3. K. S. Sultan, E. A. Abdallah, and H. El-Hennawy "A MIMO on-chip Quasi-Yagi-Uda antenna for Multi-gigabits Communications" Wiley: Engineering Reports, pp. 1-14, March 2020. doi.org/10.1002/eng2.12133.

**B. Conferences**

1. K. S. Sultan, H. H. Abdullah, E. A. Abdallah, M.A. Basha and H. El-Hennawy "A 60-GHz CMOS quasi-Yagi antenna with enhancement of radiation properties" 12th European Conference on Antennas and Propagation (EuCAP 2018), pp. 1-3, April 2018.
2. K. S. Sultan, H. H. Abdullah, E. A. Abdallah, M.A. Basha and H. El-Hennawy "A 60-GHz gain enhanced Vivaldi antenna on-chip" 2018 IEEE International Symposium on Antennas and Propagation, pp. 1821-1822, 8-13 July 2018.
3. K. S. Sultan, H. H. Abdullah, E. A. Abdallah, M.A. Basha and H. El-Hennawy "Dielectric resonator antenna with AMC for long range automotive radar applications at 77 GHz" 2018 IEEE International Symposium on Antennas and Propagation, 8-13 July 2018..
4. K.S. Sultan and M. A. Basha "High gain disc resonator antenna array with CPW coupled for THz applications" 2018 IEEE International Symposium on Antennas and Propagation, pp. 1821-1822, 8-13 July 2018.
5. K. S. Sultan and M. A. Basha "High gain CPW coupled disc resonator antenna for THz applications", IEEE Antennas and Propagation Symposium (APS), Fajardo, Puerto Rico, pp. 263-264, June 26 - July 1, 2016.

# TABLE OF CONTENTS

# LIST OF FIGURES

**PAGE**

# LIST OF TABLES

**PAGE**

# LIST OF ABBREVIATIONS

| | |
|---|---|
| 4G | Fourth Generation |
| 5G | Fifth Generation |
| AA | Atmospheric Absorption |
| Ae | Effective Area |
| AMC | Artificial Magnetic Conductor |
| AOC | Antenna-On-Chip |
| AR | Automotive Radar |
| ARS | Automotive Radar Sensor |
| BW | Bandwidth |
| CCL | Channel Capacity Loss |
| CMA | Characteristics Mode Analysis |
| CMOS | Complementary Metal-Oxide-Semiconductor |
| CPS | Coplanar Slot |
| CPW | Coplanar Waveguide |
| CST | Computer Simulation Technology |
| CST-MS | CST Microwave Studio |
| CW | Continuous-Wave |
| DG | Diversity Gain |
| DR | Dielectric Resonator |
| DRA | Dielectric Resonator Antenna |
| DRW | Dielectric Rode Waveguide |
| DWG | Dielectric Waveguide |
| EBG | Electromagnetic Band Gap |
| ECC | Envelope Correlation Coefficient |
| ETSI | European-Telecommunications-Standards-Institute |
| FCC | Federal-Communications-Commission |
| FMCW | Frequency-Modulated Continuous-Wave |
| FSL | Free Space Loss |
| GA | Genetic Algorithm |
| HDMI | High-Definition-Multimedia-Interface |
| HP | Horizontal Polarization |
| HPBW | Half Power Beam Width |
| IR | Infrared |
| ITU | International Telecommunication Union |
| LAA | Linear Antenna Arrays |
| LRR | Long-Range Radar |
| MGps | Multi-Gigabit-Per-Second |
| MIMO | Multiple Input Multiple Output |
| MLS | Multilayer Solver |
| mm-Wave | Millimeter Wave |
| Mom | Method of Moment |
| MRR | Medium-Range Radar |
| MS | Modal Significance |
| MTS | Metasurface |
| MWG | Metallic Wave Guide |
| NB | Narrowband |
| OCA | On-Chip Antenna |
| ODS | Outdoor Systems. |
| PAA | Planar Antenna Array |
| PET | Positron Emission Tomography |
| PIFA | Printed Inverted F Antenna |
| PR | Pulsed Radar |
| QYA | Quasi Yagi Antenna |
| RA | Resonator Antenna |
| RCS | Radar Cross Section |
| SAR | Specific Absorption Rate |
| SIMO | Single-Input-Multiple-Output |
| SNR | Signal to Noise Ratio |
| SOC | System on Chip |
| SRC | Short Range Communications |
| SRR | Short-Range Radar |
| S-THz | Sub-Terahertz |
| TCM | Theory of Characteristic Mode |
| TSVA | Tapered Slot Vivaldi Antenna |
| ULA | Uniform Linear Array |
| UWB | Ultra-Wide Band |
| VAA | Virtual Antenna Array |
| VP | Vertical Polarization |
| WLAN | Wireless Local Area Network |
| WPAN | Wireless Personal Area Network |
| SAR | Specific Absorption Rate |

# Chapter One: INTRODUCTION

## 1.1 Introduction

New technology and wireless communication systems have developed over the last two decades. This development requires a wider bandwidth, higher data rate, and more compact devices [1]. In order to achieve the desired requirements, future wireless communication systems are likely to work in the millimeter and sub-terahertz (THz) range [2, 3]. The Millimeter-wave (mmW) and Sub-Terahertz (S-THz) frequency ranges offer a number of unique advantages over other spectra for a large number of emerging applications. These applications include high-resolution images, ultra-high-speed short-distance communication systems, bio-medical, pharmaceutical, security, sensing, radar detectors and spectroscopy. This indicates that wireless devices are required to support different technologies and operate in different frequency bands. Remarkable progress has been made toward the development of high-performance technology platforms for the practical realizations of these applications over recent years. A low-cost and low-loss integrated circuit and system technology platform is essential for general purpose applications. The millimeter band ranges from 30 GHz to 0.3 THz and the terahertz waves offer bands in the range from 0.3 THz to 10 THz. The design of antennas in these ranges is considered a challenging task since mastering the fabrication process and the measurement setup are still under investigation worldwide [4-8].

One of the most promising applications in these ranges is 5G mobile communications. The 5G denotes the next major phase of mobile telecommunication standards beyond the 4G standards. The 5G technology will change the way the higher bandwidth users access their phones. The second application in these ranges is the on-chip systems for indoor applications because it provides high speed for a short distance which can be used for video streaming, broadcasting and networking. In contrast, the on-chip antennas suffer from low gain, low efficiency and complicated technology [2, 6, 9-12].

Another application is an automotive radar detector for short, medium and long-range radar. Currently, there are several manufacturers worldwide of the automotive radar system at 24 GHz and 77/94 GHz that offers a maximum range of 250 m for detection depending on the type of radar. This range of the radar is not sufficient for the train to detect any moving or static objects at the railways' cross-sections ahead of time to stop the train safely to avoid collisions at the intersections. So, another critical factor to mention in the radar system is the resolution required to detect the object in a very large range with a very small angle of detection [5, 13-17].

## 1.2 Objectives

The main objective of our thesis is to design and implement antennas capable of achieving the required specifications of popular applications in the millimeter and THz bands. This thesis aims to introduce antennas for 5G, automotive radars, short-range communication and sub-THz applications; all the designed antennas are verified by different methodology and analyses.

## 1.3 Features of mmW and S-THz

The millimeter ranges occupy the frequency from 30 GHz to 300GHz. The design of antennas in these ranges is a challenging task since mastering the fabrication process in addition to the measurement setup are still under investigation worldwide. One of the main restrictions on millimeter-wave communication is atmospheric absorption. The signal attenuation is very high due to the rain, fog, and any moisture in the air that reduce the transmission distances. The international telecommunication union (ITU) introduced a model for the signal attenuation due to the atmospheric gas, and this model is applicable for frequencies from 1 GHz to 1000 GHz and is used for polarized and non-polarized fields (ITU recommendation number (ITU-R P 676-10)) [18]. The atmospheric attenuation can be calculated from the following formula:

$$\gamma = 0.182 f N''(f). \tag{1.1}$$

where $\gamma$ is the atmospheric attenuation, and quantity N"() is the imaginary part of the complex atmospheric refractivity. From Eq. (1.1) and details of quantity (N) in [18], the atmospheric attenuation versus frequency can be introduced, as shown in Figure 1.1. We note that the signal attenuated significantly at frequency [13, 18-20]. First of all, it is clear that in the microwave range (up to 30 GHz), the atmospheric attenuation is rationally low at a few tenths of dB/km. In contrast, the atmospheric attenuation has a large peak around 60 GHz, limiting this band's communication before falling to about 0.3 dB/km around 80 GHz. Following that, the raising in attenuation becomes the widest behaviour. From the graph, we can deduce that the 28/38 GHz is suitable for 5G applications and the 30GHz bandwidth from 70GHz to 100GHz has low attenuation for automotive radar.

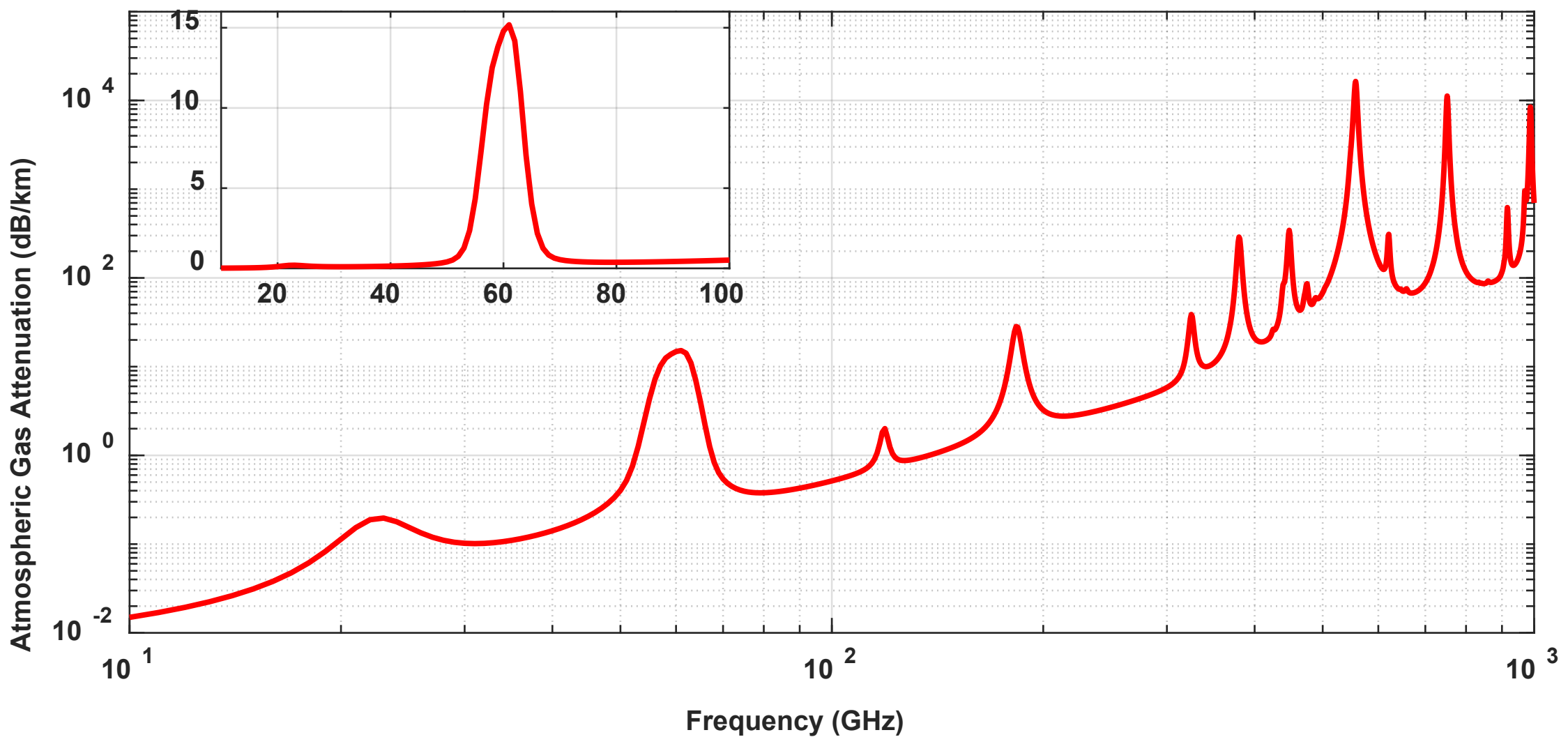

Figure 1.1 Atmospheric attenuation versus frequency.

The rain, fog, and atmospheric gas can be considered as one of the main limiting factors for radar systems when operating above 5 GHz according to the relations of path losses introduced in [18]. Figure 1.2 shows the amount of path losses at different ranges against the frequency range. We notice that path loss is directly proportional to the distance and the frequency.

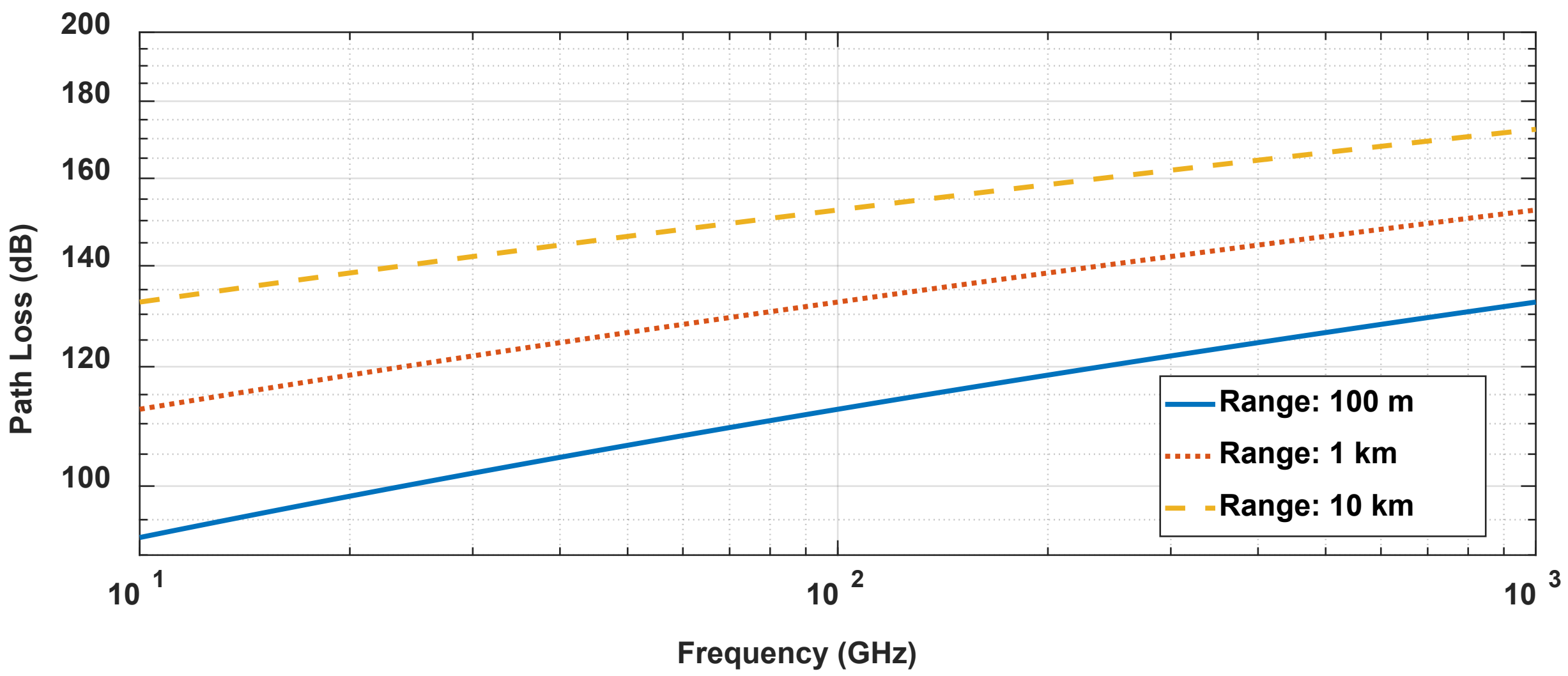

Figure 1.2 Path loss for different ranges (R=100 m, R=1 km, R=10 km)

## 1.4 Applications of mmW and S-THz

Many studies are introduced in the literature to solve the problems of technology used in mmW range and sub-terahertz range because of the mmW and S-THz technology is still under investigation [7, 20-23]. Whereas there are some recommendations for each application through these bands. This thesis focuses on different applications in these bands, as shown in Figure 1.3.

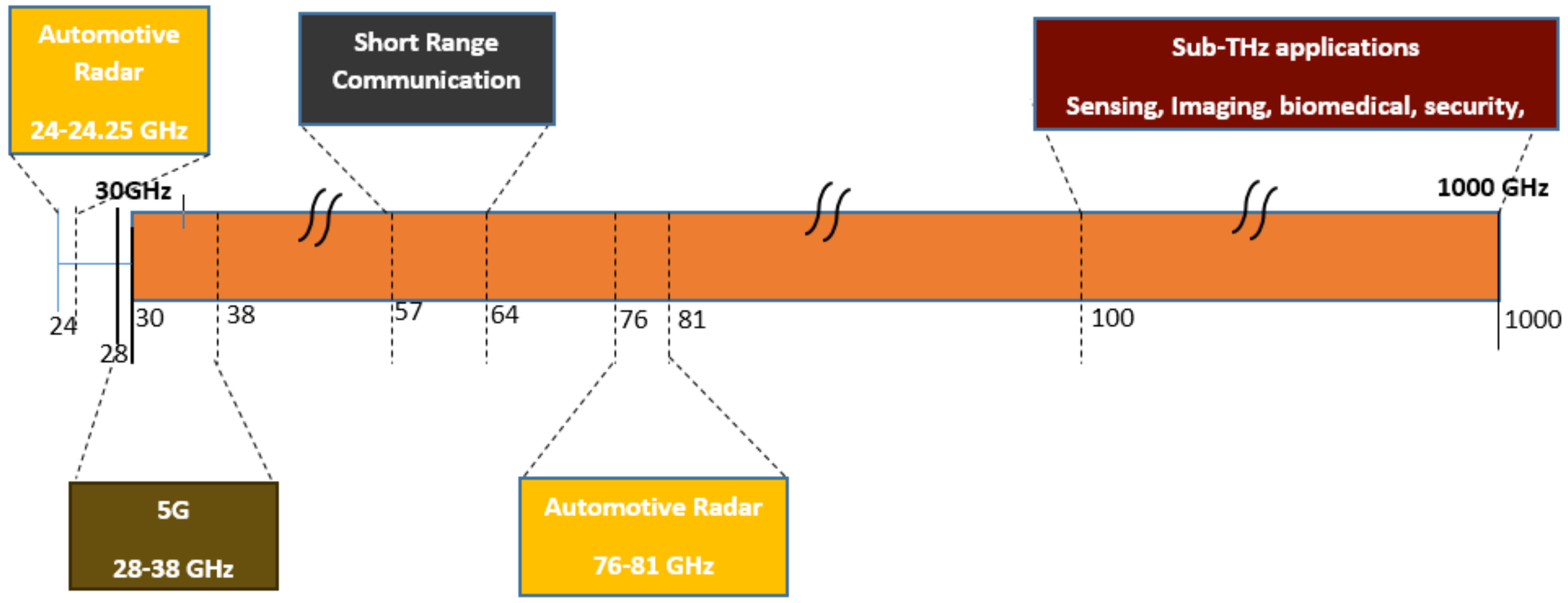

Figure 1.3 Proposed applications in the thesis (spectrum of Millimeter and sub-THz applications)

## 1.4.1 Automotive Radar Applications

Nowadays, there has been growing interest in automotive radar systems to measure the variable of the surrounding environment, such as distance to target and velocity of a target to avoid the collision. Therefore, radar sensors have drawn more attention because of their ability to make driving more comfortable, stable and safer. The automotive radars are classified according to the operating range into long-range radar (LLR) (10- 250 or 300 m), medium-range radar (MRR) (1-100 m), and short-range radar (SRR) (0.15-30 m), where both LLR and MRR are required to detect the forward obstacles [5, 13, 19, 24]. High gain and more efficient antennas are practical components in the automotive radar system.

## 1.4.2 5G Applications

The International Telecommunication Union (ITU) has created several groups to achieve all 5G standards before 2020. The ITU releases the applicable frequencies for the new mobile generation (5G) between 24 GHz to 86 GHz. Even though the range of 5G is still under review, there are several candidate bands [25]. The range from 28 GHz to 38 GHz is highly recommended. In order to design an effective antenna for 5G mobile phone, several fundamental challenges need to be considered. One of these challenges is the free space loss (FSL) and atmospheric absorption (AA) that have high values due to the higher frequency of millimeter ranges[1]. Also, FSL and AA allow for the reuse of the spectrum due to the limit of interference amount between adjacent cells. Although relatively lower losses and ease of technology can be achieved at lower microwave frequencies, these frequencies suffer from a lower data rate, high latency, and vice versa for millimeter.

Nevertheless, most of the antennas in this side are limited to the linearly polarized antennas, while in the real case, the mobile terminal will encounter different sorts of movements in Euler areas in addition to the antenna operating in the MM-wave bands. Therefore, the miss-polarization among the transmitter and the receiver antenna is one of the main significant loss factors in this communication system. The circular polarization antenna loses half of the power in the transmitter or receives linear polarization. Therefore, for full utilization of power in 5G systems, the antenna of dual-polarization candidates to solve the problems of power losses and increase the bit error rate of the communication systems. So, the antenna with different polarization (polarization diversity) plays an essential key to solving the mentioned problems and improving channel capacity. On the other hand, some advanced antenna techniques were reported to solve the problem of high FSL in MM-wave [9, 26, 27]. The researchers still introduce different studies to obtain the optimum antenna specifications that can be used in this range.

### 1.4.3 Short Range Communications

Nowadays, low-frequency bands are very crowded, and with the rapid growth of communication technologies, high-speed short-range wireless communications require wideband, higher data rates, and compact size. In order to achieve the aforementioned requirements, the Millimeter-wave band at 60 GHz has towed more and more attention because it offers unlicensed bandwidth (from 57 GHz to 64 GHz) for several applications such as video streaming, wireless gaming, short-distance communication and wireless personal area network (WPAN) [28, 29]. So, the complementary metal-oxide-semiconductor (CMOS) technology is considered a good solution for cost and circuit integration issues at this frequency. However, the CMOS substrate is inherited losses due to its high permittivity ($\varepsilon_r$=11.9) and low resistivity ($\sigma$=10S/m). Additionally, CMOS antennas at 60 GHz require more enhancements of antenna efficiency and antenna gain [28, 30-32].

### 1.4.4 S-THz Applications

The terahertz waves band ranges from 0.3 THz to 10 THz; the terahertz frequency range offers new specifications over another spectrum for many applications, such as high-resolution imagers, ultra-high-speed, short-distance communication systems, biomedical, pharmaceutical, security, sensing, and spectroscopy.

This indicates that wireless devices are required to support different technologies and operate in different frequency bands [22]. Several studies have been performed to produce an antenna structure able to satisfy the demands for THz applications. A key factor for THz applications is a technology platform for better performance of this band.

# 1.5 Original Contribution

Four major applications in this thesis are illustrated in the previous section. The works were undertaken in this thesis aim to introduce antenna designs compatible with these four independent applications one by one to solve the problems of each application. This thesis possesses four original contributions, which are listed as follows:

1. Virtual antenna array (VAA) to enhance the angular resolution of the radar with a minimum number of antenna elements compared with the conventional planar antenna array (PAA) proposed for the automotive radar sensors. This VAA can be used to increase the radar range and decrease the number of antennas in the antenna array. The proposed VAA is firstly simulated and evaluated on a simple structure. Furthermore, the concept of a hybrid antenna configuration is introduced as another solution to increase the range of the automotive radar. The detailed illustrations of the VAA designs, fabrication, and measurements are exhibited in Chapter 3 (Antenna Design for Automotive Radar), [33, 34].

2. Novel dual linear polarized metasurface antenna based on the characteristic mode analysis (CMA) is proposed for the 5G applications. The dual-polarization is used to solve the problems of isolation and channel capacity for MIMO smartphone designs. Furthermore, the theory of characteristic mode (TCM) or characteristic mode analysis (CMA) is introduced as an accurate analysis for a few unit cells of the metasurface. The detailed illustrations of the dual-polarized antenna for handheld 5G systems with the comprehensive study of the interaction of an antenna with the human body and vice versa are exhibited in Chapter 4 (5G mobile applications), [35].

3. Novel two end-fire antennas based on the hybrid technique to increase the radiation characteristics are proposed of the on-chip systems for short communications at 60 GHz. The proposed two antennas solve the problems of low efficiency and low gain in traditional on chip antennas. Accurate results are introduced by comparison with different simulation tools. The first one, a 60 GHz Yagi-Uda antenna and the second one is a 60 GHz Vivaldi on-chip antenna. The two antennas are presented on standard 0.18 μm CMOS technology. The detailed illustrations of the two antenna designs are presented in Chapter 5 (Short Communications), [36-38].

4. A novel disk resonator antenna (DRA) fed by coplanar waveguide (CPW) technique with compact size and high gain using silicon on glass (SOG) technology platform is proposed. The CPW feed is patterned on the backside of the Si wafer before the bonding process from the Pyrex side. In addition, the dielectric waveguide (DWG) is matched with the disc dielectric antenna using CPW feed. The DRA covers the band from 325 GHz to 600 GHz and can work in broadside and end-fire radiation. This antenna has high efficiency and low cost. The detailed illustrations of the DRA and the antenna array designs are introduced in Chapter 6 (Sub-THz applications), [39, 40].

## 1.6 Software Packages Used

In this thesis, we used different software packages based on different techniques. The first package is CST Microwave Studio (CST-MS) that introduce a wide variety of solvers for different problems. We used four solvers in this thesis namely: time domain solver, Eigen-mode solver, and multilayer solver (MLS) or characteristic mode analysis (CMA). The CST-MS that is based on time-domain solver is used to simulate the VAA and the hybrid antenna array in chapter three. In addition to simulating two dual-polarized 5G antenna in chapter four, two end-fire on-chip antenna in chapter 5 and S-THz antennas in chapter six. On-other hand, the CST-MS that is based on eigenmode solver is used to analyze the operating modes for the DRA in chapter six. Also, the CST-MS based on CMA is used in chapter five to analyze the characteristic mode of metasurface that is used for 5G antenna. A second package is COMSOL software based on multi-physics analyze is used in chapter six to analyze the modes of DRA. The third software package is HFSS (High Frequency Structure Simulation) based on the finite element method is used to ensure the results of on-chip antennas.

## 1.7 Thesis Organization

The work undertaken in this thesis is organized as follows:

1. Chapter 1 presents a brief overview of the proposed research background, in addition to present the motivations, objectives and the original contributions for this thesis.
2. Chapter 2 focuses on a survey about the different applications in the millimeter and sub-THz frequency range. The complete literature survey about the automotive radar sensors at 24 GHz and 77 GHz is introduced. A survey about the 5G antennas and the new guidelines for the new generation of mobile communications are presented. Furthermore, a literature review based on the technology that can be used for short communications at 60 GHz and the on-chip antenna is presented. Finally, the different types of antennas and technology that can be used in sub-THz are given.

3. Chapter 3 presents the concept of VAA which apply on the antenna array for the automotive radar. The usage of VAA concept aims to reduce the number of antenna elements used in the array and decrease the number of channels required in the transceiver system of automotive radar. The VAA and unequal power divider are introduced to achieve flat-shoulder shape (FSS) radiation pattern for covering the long range radar and the short range radar. The number of the VAA elements are defined according to the integration between the method of moment (MOM) and the genetic algorithm (GA) to achieve the optimum number of elements. Furthermore, the hybrid antenna array is introduced with 16 elements to achieve the requirement of the Long range radar (LRR) at 77 GHz. The element of this antenna array consists of a hybrid radiator and dielectric resonator. The hybrid radiator is a circular patch that is fed by aperture method and the dielectric resonator is a ring that is fed by the circular patch to operate at 77 GHz with high gain. The electromagnetic band gap (EBG) structure is implemented on the top layer to widen the proposed band, gives high gain, reduces surface waves and gives good isolation between antenna array elements. Details of these designs, fabrications, and measurements are introduced in this chapter.
4. Chapter 4 introduces a novel dual-polarized MIMO antenna for 5G mobile handset. Also, it presents a complete study for metasurface (MTS) that has a low profile, lightweight, easy integration and low loss in contrast to metamaterial. So, the CMA is used to analyze the proposed antenna with MTS. The isolation coefficients, envelope correlation coefficient (ECC), channel capacity loss (CCL) of the MIMO are also calculated. Furthermore, this chapter presents a comprehensive study on the performance of compact antenna design. Return loss, radiation patterns, specific absorption rate (SAR), and efficiency of this antenna are computed in free space, in the presence of handset as well as in the presence of head and hand. The peak SAR in the head is compared with SAR limits in the safety standards and so the maximum radiated power of each antenna is determined.

5. Chapter 5 introduces two different end-fire antenna configurations to enhance the radiation characteristics of on-chip antennas that is used in short communications at 60 GHz. A hybrid technique that depends on reducing the backward radiation and reduce the surface waves is used to solve the problems of low radiation for on-chip antennas. Also, comprehensive study of the on-chip antenna is introduced.

6. Chapter 6 presents one element, two elements and four elements disc resonator antenna (DRA) with compact size and low profile based on the silicon on glass technology platform. The proposed antenna consists of a silicon straight section waveguide segment connected in series with disc resonator which acts as radiating element. The CPW power divider with compact size is used to the disc resonator. The proposed antenna in this chapter is introduce to cover S- THz band from 325 GHz to 600 GHz. Furthermore, the end-fire and broadside antennas are introduced. The antenna has more compact size when compared to other published antennas.

7. Chapter 7 gives the final conclusion of the presented works as well as suggestions for promising future works.

# Chapter Two:
# ANTENNAS FOR MILLIMETER AND SUB-THZ APPLICATIONS

## 2.1 Introduction

This chapter introduces a literature review on the antennas that are used for the different applications through the mmW and S-THz ranges. We focus here on the antennas for automotive radars, 5G applications, short-range communications and S-THz applications.

## 2.2 Automotive Radar Sensors (ARS)

### 2.2.1 ARS Bands

Nowadays, there has been growing interest in the automotive radar system to measure the variables of the surrounding environment, such as the distance to target and velocity of a target to avoid the collision. So, radar sensors have drawn more attention because of their ability to make driving more comfortable, stable and safer.

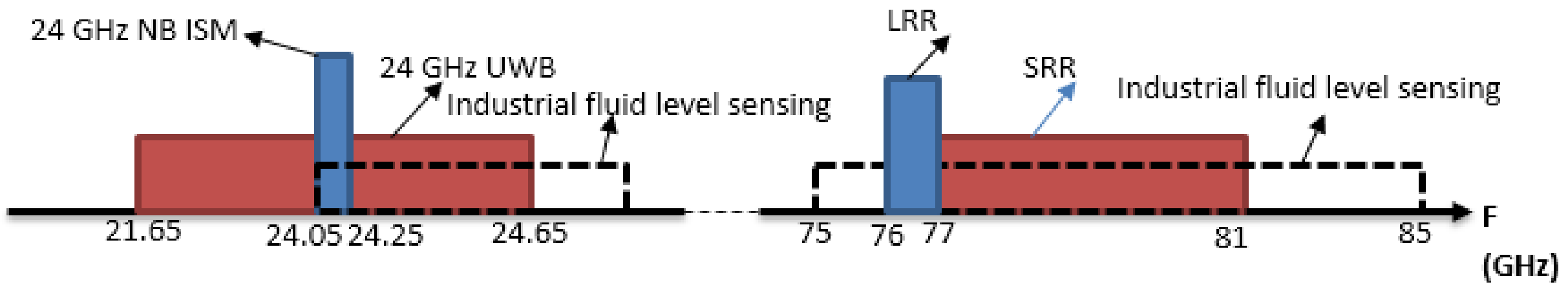

Figure 2.1 Automotive radar frequency bands.

A 24 GHz and 77 GHz are two frequencies that are predominantly used for automotive sensors, as shown in Figure 2.1. In 1999, the automotive radar systems appeared in the international market that was designed to operate as short-range radar (SRR) at 24 GHz and 76 GHz for long-range radar (LRR), and from 77 GHz to 81 GHz for medium and short-range sensors. So, the radar can be classified according to its range to LRR, MRR and SRR. Since then, many different radar systems have been developed; the first radar that operates in 77 GHz was introduced by Daimler S class; after this year, other companies such as Jaguar, Nissan and BMW followed [13].

The 77 GHz band has various benefits that forced the designers to shift the radar applications toward this band. One of these reasons is revolved around 24 GHz radar disadvantages as:

- It includes an ISM band (Industrial, Scientific and Medical) with 200 MHz starting from 24.05 GHz and ending to 24.25 GHz, called the narrowband (NB). Furthermore, it includes 5 GHz to be an ultra-wideband (UWB). For the SRR, the NB and UWB have been used in legacy automotive sensors for the 24-GHz band.
- It doesn't support the long-range radar.
- According to the European-Telecommunications-Standards-Institute (ETSI) and the Federal-Communications-Commission (FCC), the 24-GHz UWB band will not be used from January 1, 2022, and this is called "sunset date" [41, 42].

The other reasons include the benefits of 77 GHz radar compared with the 24 GHz radar, such as shown in Figure 2.2. The 77 GHz radar has advantages such as wideband and small size. On the other hand, it has considerable effort in its design and implementation. Furthermore, the radar cross-section of Pedestrians are small and usually tend to change their directions. So, the radar needs to have a high resolution to track the direction changes and avoid collisions with the pedestrians. The 77 GHz radars include two sub-bands:$76 - 77GHz$ and $77 - 81GHz$ (also called 79 GHz band). The automotive radar sensors at 77-GHz are classified according to the radar distance to three types, as shown in Table 2. 1 and Table 2. 2 [5].

Table 2. 1 Comparison between two bands of SRR

| **Type** | **SRR (24 GHz)** | **SRR (77 GHz)** |
|---|---|---|
| **Operating Band** | 24 GHz – 24.25 GHz Temporary Band | 77 GHz – 81 GHz Permanent Band |
| **Range Resolution** | 75 cm | 4 cm |
| **Impedance BW** | 17 % | 5% |
| **Beam Width** | Wide | More focused |
| **Sensor Size** | 3X | X |
| **Angular Resolution** | 3X | X |

Table 2. 2 Automotive radar classification [5]

| **Type** | **LRR** | **MRR** | **SRR** |
|---|---|---|---|
| **ERIP** | 55 dBm | -9 dBm/MHz | -9 dBm/MHz |
| **Frequency Band** | 76-77 GHz | 77 -81 GHz | 77-81 GHz |
| **Bandwidth** | 600 MHz | 600 MHz | 4 GHz |

| **Main Aspect** | Detection Range | Detection Range | Range accuracy |
|---|---|---|---|
| **Range** | 10 -250 m | 1-100 m | 0.15-30 m |
| **Range Resolution** | 0.5 m | 0.5 m | 0.1 m |
| **Range Accuracy** | 0.1 m | 0.1 m | 0.02 m |
| **Velocity Resolution** | 0.6 m/s | 0.6 m/s | 0.6 m/s |
| **Angular Accuracy** | $0.1^0$ | $0.5^0$ | $0.5^0$ |
| **HPBW in Azimuth** | $\pm 15^0$ | $\pm 40^0$ | $\pm 80^0$ |
| **HPBW in Elevation** | $\pm 5^0$ | $\pm 5^0$ | $\pm 10^0$ |
| **Dimensions** | $74 \times 77 \times 58$ mm | $50 \times 50 \times 50$ mm | $50 \times 50 \times 20$ mm |

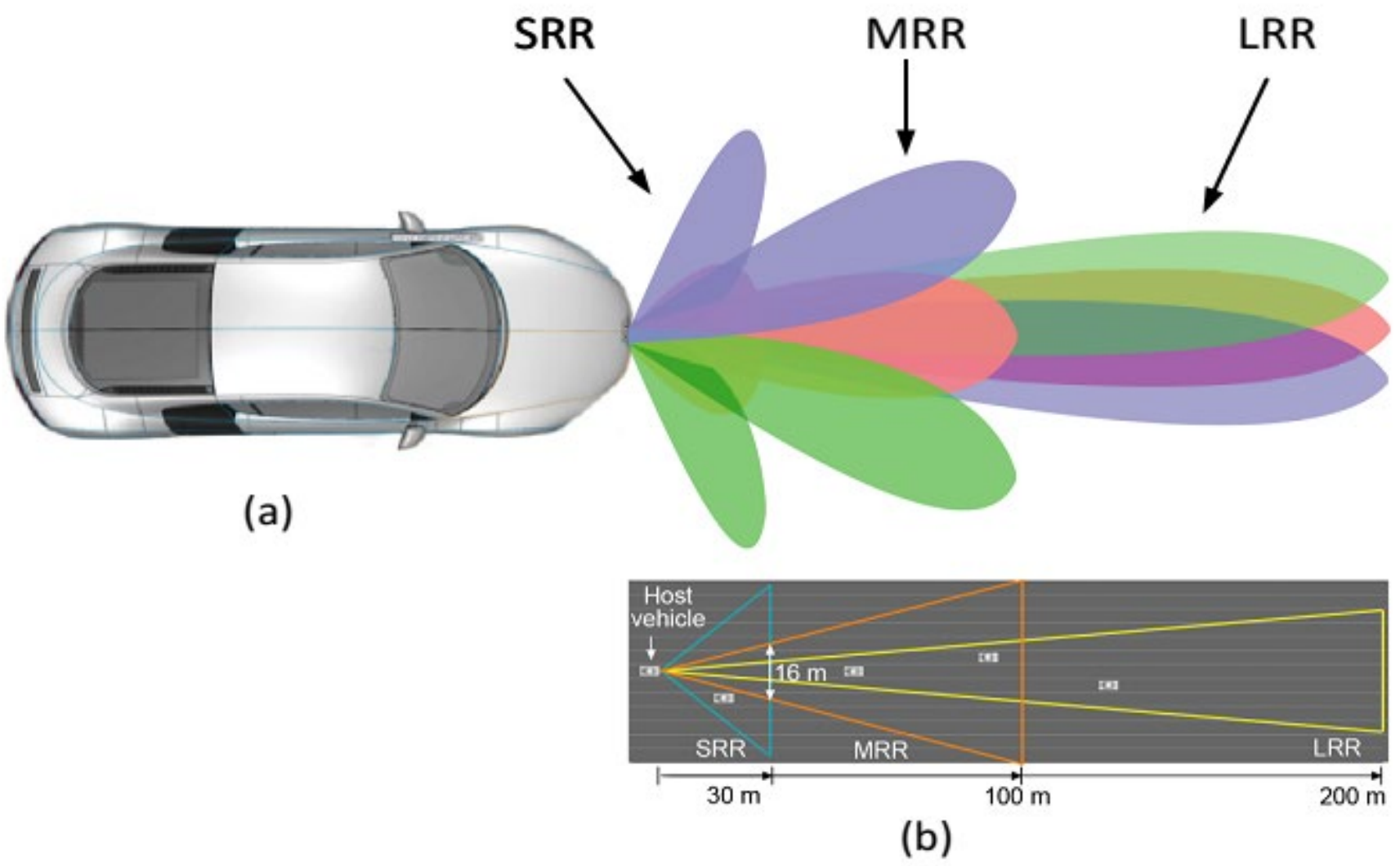

Figure 2.2 (a) The beam coverage of the radar modes and (b) the distance covered in meter [5].

The angular resolution of the radar means the distinction between the two targets. It depends on the two main parameters: wavelength and aperture size of the antenna. The angular accuracy of the radar means the accuracy of angle measurement based on the wavelength, aperture size, and signal to noise ratio. The angular resolution and angular accuracy can be calculated from equation (2.1) and equation (2.2) using the Rayleigh criterion [5].

$$\Delta\varphi = 1.22\frac{\lambda}{d} \tag{2.1}$$

$$\delta\varphi = \frac{\Delta\varphi}{\sqrt{2\,SNR}} \tag{2.2}$$

Where d antenna aperture size, $\lambda$ wavelength, and SNR signal to noise ratio. Figure 2.3 shows the angular resolution and angular accuracy with the frequency variation, when d=30 mm, and SNR=10 dB. We noted that $\Delta\varphi$ =(29.13$^0$, 9.07$^0$) and $\delta\varphi$ =(6.51$^0$, 2.03$^0$) at 24 GHz and 77 GHz, respectively. In conclusion, the radar at 24 GHz needs an antenna with three times larger than that used at 77 GHz to achieve the same angular resolution.

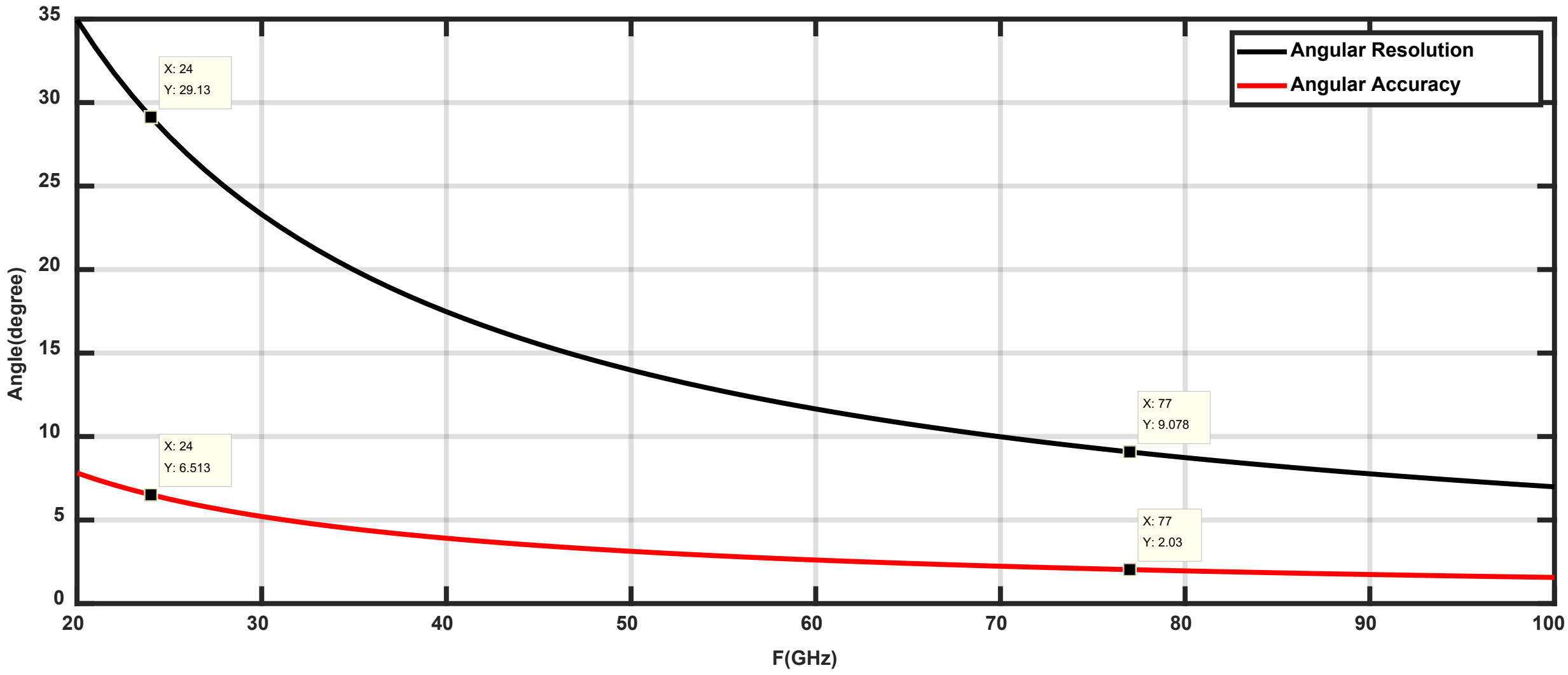

Figure 2.3 Angular resolution and angular accuracy versus frequency,
d=30 cm and SNR=10 dB.

The automotive radar sensors (ARS) are used to eliminate the possibility of collisions occurring or risky situations. So, it is used to alert the driver, control the vehicle to prevent an accident, rearview traffic crossing alert, or blind spot detection. More than one radar sensor is used to detect the obstacles and the relative speed of the target. To avoid the collision and reduce the risk, the appropriate action should be taken by the processing unit; this action depends on the reflected signal from the target.

The functions of any automotive radar system should include the following objectives:

- Detect the obstacles surrounding the vehicle
- The relative position of the target to the vehicle
- The relative speed of the target to the vehicle

Then by decision-maker unit can take one or more actions from the following:

- Alert the driver about the dangerous situation.
- Prevent collision by the control of the vehicle in risk situations
- Adaptive Cruise Control (ACC)
- Assist the driver for car parking

So, the automotive radar can be described as another driver with you, as shown in Figure 2.4.

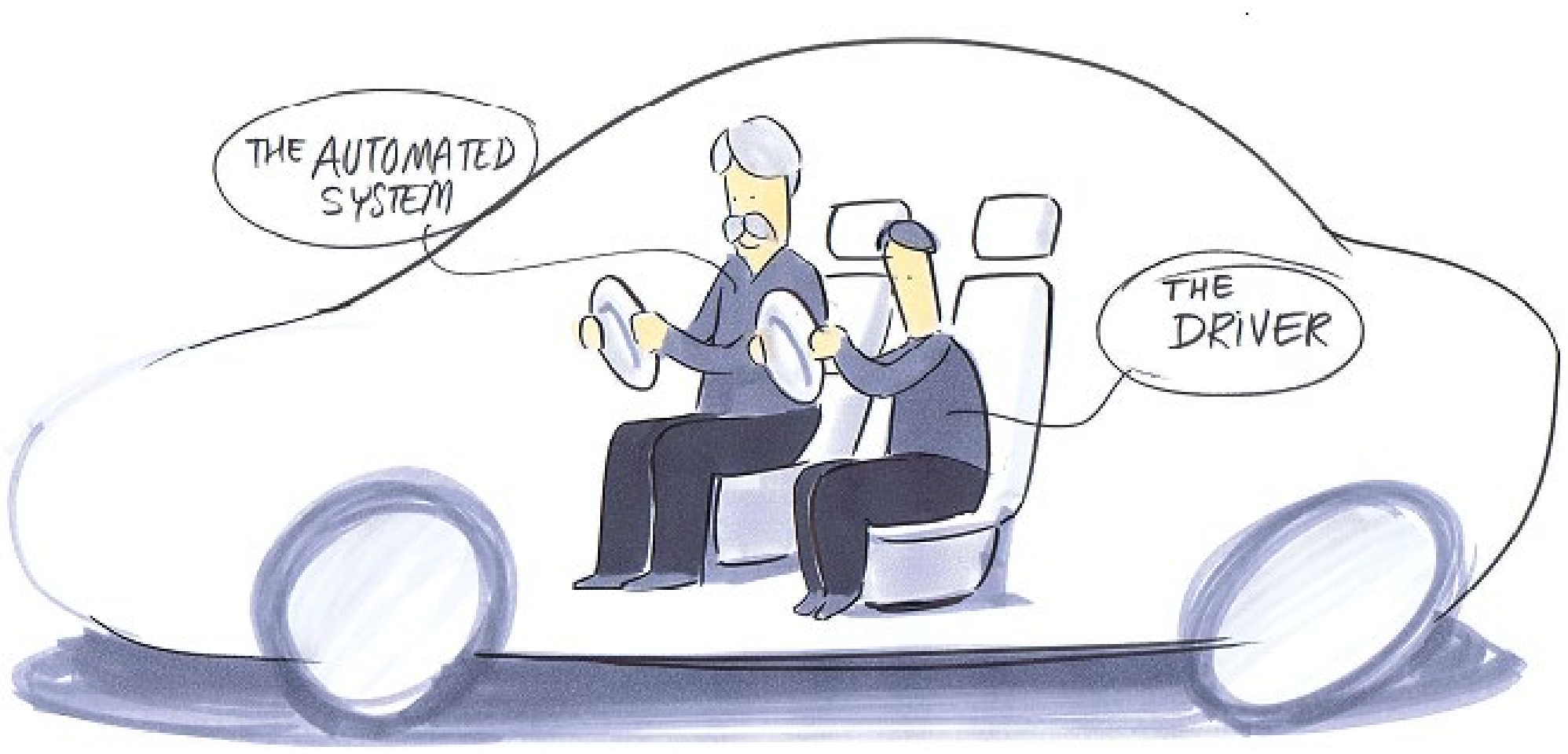

Figure 2.4 Description of automotive radar [43].

There are six main parameters that affect the performance of the automotive radar as follows:

- Detection-Range
- Speed-Detection-Range
- Range-Precision
- Velocity-Precision
- Angular-Resolution
- Angular-Width-of-View

### 2.2.2 Commercial Sensors

In 2008, the Denso Company introduced the third generation of its long-range radar sensor, including digital beamforming for the first time in its radar. The digital beamforming is achieved by one transmit antenna and five switched received antennas. The antenna itself is an array formed of slotted waveguides [44].

In 2009, researchers from BOSCH Company introduced a new long-range transceiver chip as shown in Figure 2.5, to operate in the band 76-77 GHz with a radar range of up to 250m, range accuracy of 0.1 m and its relative speed of -75 to 60 m/s with speed accuracy 0.12 m/s [5, 45]. It has four patches antennas with a dielectric lens to give high gain for long-range radar and to create four slightly offset beams. All four antennas are receiving antennas, with the middle two antennas also simultaneously transmitting, as shown in Figure 2.5. In 2015, BOSCH introduced the fourth generation of the automotive sensor to enhance the radar performance with lens antenna [24].

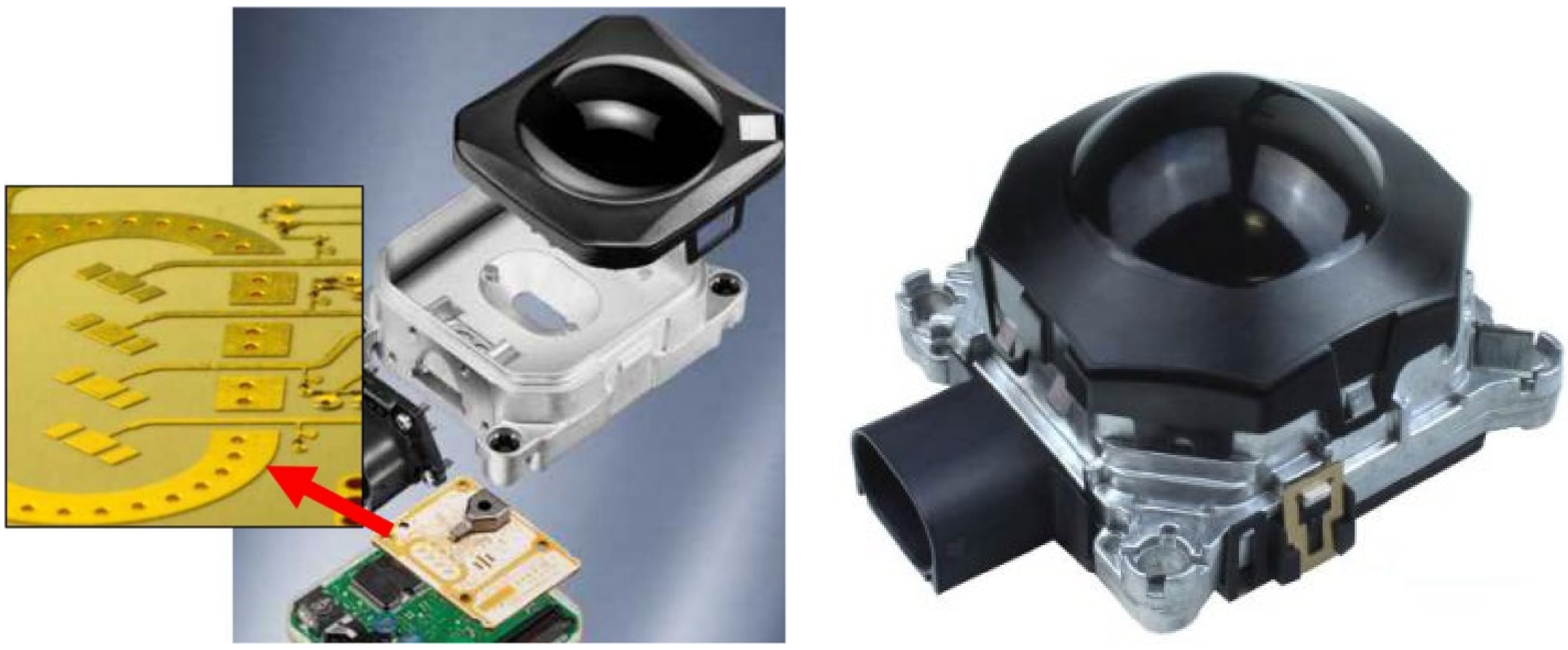

(a)LRR3[5], (b)LRR4[24]

Figure 2.5 BOSCH LRR3 [5] and LRR4 structure [24].

On the other hand, Conti Company introduced ARS 300 as another radar in 2009 that operates for the long and medium-range by using different patterning of the spindle [24]. The ARS system consists of the reflect-array antenna that provides auto alignment by grooved rotating drum, as shown in Figure 2.6. Table 2. 3shows the characteristics and performance parameters for different commercial automotive radar sensors taken from datasheets.

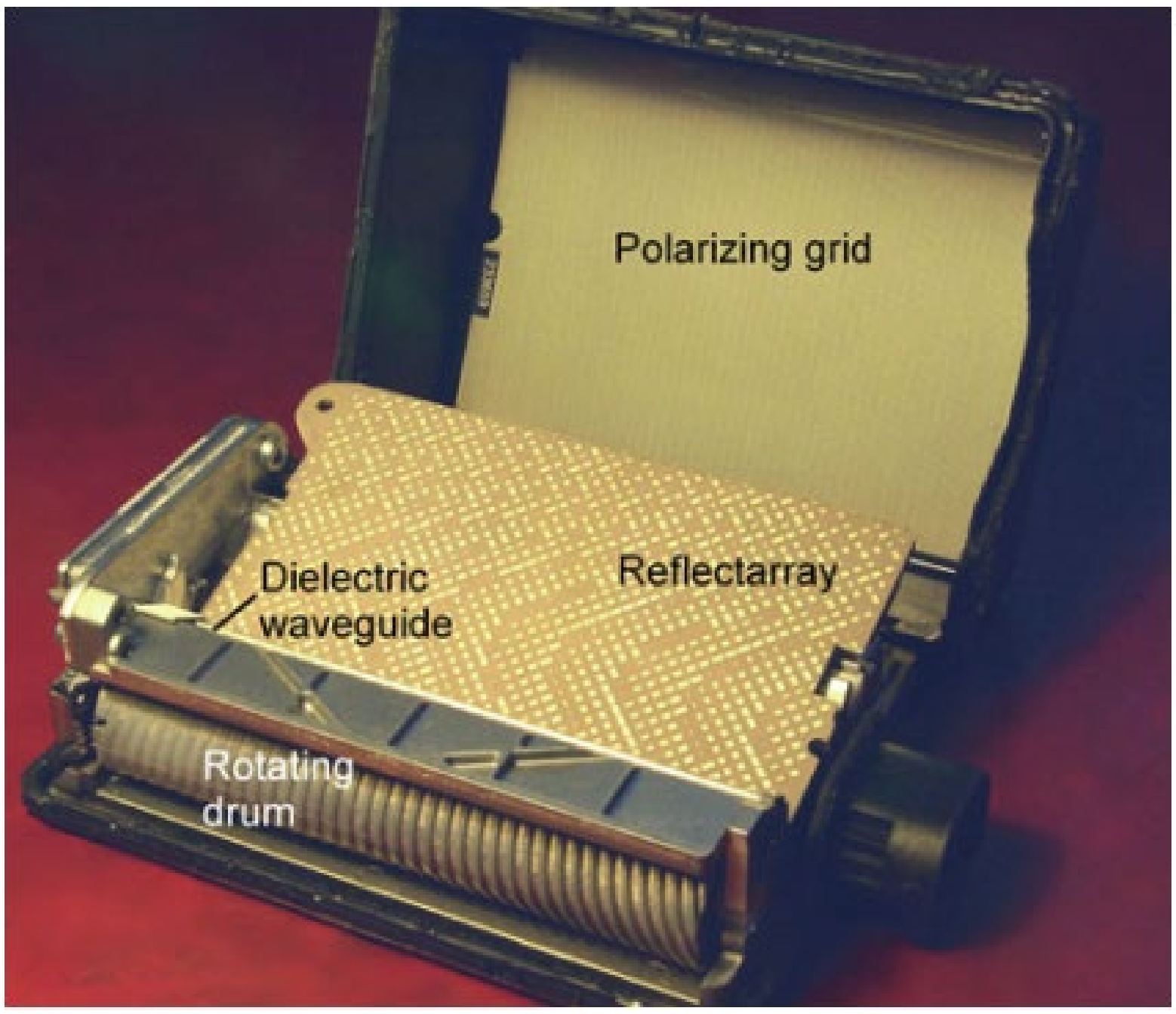

Figure 2.6 ARS 300 sensor with opened cover [24]

Table 2. 3 Commercial radar sensors

| Sensor | F(GHz) | Dimensions (mm) | Range (m) | BW | Azimuth Angle (degree) | Multi-range | Accuracy |
|---|---|---|---|---|---|---|---|
| **Bosch LRR3** | 77 | 74 x 70 x 78 | 250 | 1 GHz | 30 | Single | 0.1m, 0.12 m/s |
| **Delphi ESR** | 77 | 173 x 90 x 49 | 174 | - | 20 | Multi | 1.8m, 0.12m/s |
| **Cont. ARS30x** | 77 | 120 x 90 x 49 | 250 | 1 GHz | 17 | Multi | 1.5% , 0.14 m/s |
| **Denso DNMWR004** | 77 | 78 x 77 x 38 | 150 | 1 GHz | 20 | Single | 1.8, 0.12m/s |
| **SMS UMRR type 40** | 24 | 212 x 154 x 40 | 250 | 250 MHz | 36 | Single | 2.5% , 0.28 |
| **TRW AC100** | 24 | 460 x 460 x 50 | 150 | 100 MHz | 16 | Single | --- |

## 2.2.3 Types of Radars

### 2.2.3.1 Types of Radars According to Configuration

The radar system can be classified according to the transmitter and receiver position relative to the target into two types: Monostatic and Bistatic radar. In the monostatic configuration, a single antenna is used for transmitting the signal and receiving it, whereas the transmitter and receiver are collected in the same device as a transceiver. In this thesis, the monostatic radar configuration will be used (see Figure 2.7(a)). In other words, when the antennas of the transmitter and the receiver are closed to each other and at the same location, the radar is called monostatic. In the configuration of bistatic, two antennas are used, one for the transmitter and the other for the receiver with the displaced distance between them. The antennas are physically separated, as shown in Figure 2.7(b). If Bistatic radar has more than one receiver antenna, then it is known as multi-static radar [19, 24, 46].

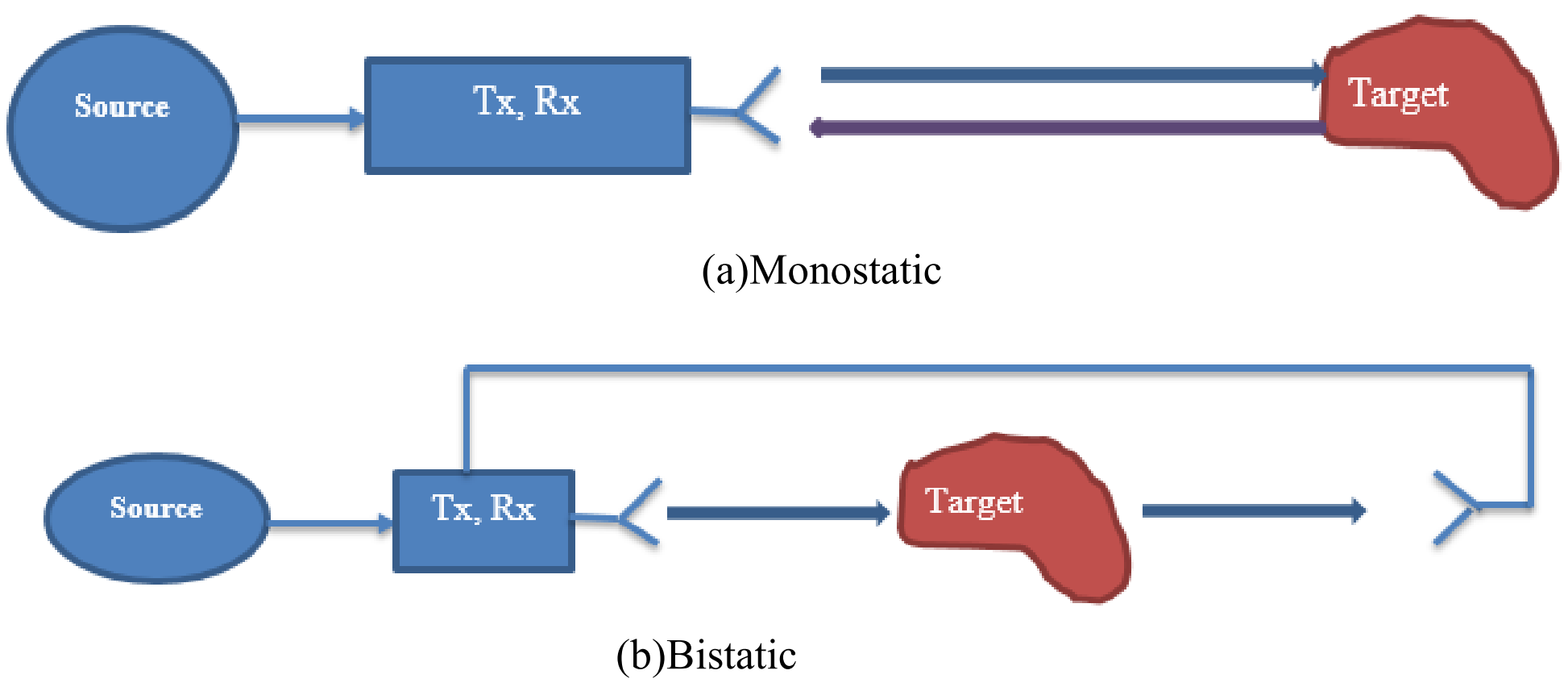

Figure 2.7 Radar Configuration

For one way radar, the received power at the target can be calculated from equation (2.3):

$$P_r = \frac{P_t G_t G_r \lambda^2}{(4\pi R)^2} \tag{2.3}$$

The antenna gain can be expressed as a function of an effective area $A_e$ as presented in equation (2.4)

$$G = \frac{4\pi A_e}{\lambda^2} \tag{2.4}$$

The target reflects a portion of power in a reverse way, in the direction of the radar. This portion of power depends on the Radar Cross Section (RCS) of the target. The RCS describes the target characteristics such as its size and dimension as seen by the radar. For the radar target, the amount of reflected power by the target is equal to the re-radiated power of the antenna with an effective area equal to the RCS of the target. Therefore, the receiving antenna's effective area (Ae) is replaced by the RCS (σ). So, the reflected power from the target can be expressed as:

$$P_r = \frac{P_t G_t \lambda^2}{(4\pi R)^2} \frac{(4\pi\sigma)}{\lambda^2} = \frac{P_t G_t (4\pi\sigma)}{(4\pi R)^2} \tag{2.5}$$

The reflected power back to the radar receiver is:

$$P_r = \frac{P_t G_t (4\pi\sigma)}{(4\pi R)^2} \frac{G_r \lambda^2}{(4\pi R)^2} = \frac{P_t G_t G_r \sigma \lambda^2}{(4\pi)^3 R^4} \tag{2.6}$$

$$P_r = (P_t G_t G_r) \left(\frac{\lambda}{4\pi R}\right)^2 \left(\frac{4\pi\sigma}{\lambda^2}\right) \left(\frac{\lambda}{4\pi R}\right)^2 \tag{2.7}$$

So, the free space loss (FSL) of the monostatic radar $FSL = \left(\frac{\lambda}{4\pi R}\right)^4$.

Where

- $P_r$: Received power in watts.
- $P_t$: Peak transmitted power in watts.
- $G_t$: Transmitter Gain.
- $G_r$: Receiver Gain .
- λ: Wavelength (m).
- σ: RCS of the target ($m^2$) .
- R: Range between radar and target (m).

### 2.2.3.2 Types of Radar According to Operation

**1. Pulsed Radar**

The pulsed radar depends on measuring of the time delay between transmission and reception pulse where the radar transmits a number of pulses then calculate the delay time and the change in pulse width. Due to the delaying of the reflected pulse and the change of pulse width, we can calculate the distance between the sensor and the target in addition to the speed of the target relative to the speed of the vehicle.

Typically, the pulsed radars have a blind speed and ambiguous range issues. In addition, transmitting a narrow pulse in the time domain means that a large amount of power must be transmitted in a short period of time. In order to avoid this issue, spread spectrum techniques may be used.

In pulsed radar, the generated radio signal at a constant frequency $f_0$ passes through the pulse shaping device that converts it to a train of pulses. Suppose the propagation speed c of the electromagnetic wave in the medium is known and the round trip delay is t. In that case, we can calculate the distance between the target and the radar from the simple following equation:

$$R = \frac{ct}{2} \tag{2.8}$$

In case of the motion of target, the relative velocity can be determined from the Doppler shift of received signal frequency$f_r$ as shown in equation (2.9). The Doppler shift is the difference between the transmitted and received frequency$f_d = f_r - f_0$.

$$v_r = \frac{cf_d}{2f_0} \tag{2.9}$$

The maximum range for pulsed radar depends on the pulse repetition rate (PRR) of the transmitted pulses $T_p$ as shown in Figure 2.8(a) and can be determined from equation (2.10). In other words, it is defined as the maximum range of pulse that can send from the transmitter before the next pulse is emitted. Figure 2.8(b) illustrates that the system can receive echo pulses after sending the other pulse. In this case, the range to the target is calculated by false information to be $\Delta t_3$ instead of the real range that equal $T_p + \Delta t_3$. The range accuracy of the radar depends on the operating bandwidth as demonstrated in equation (2.11) [43].

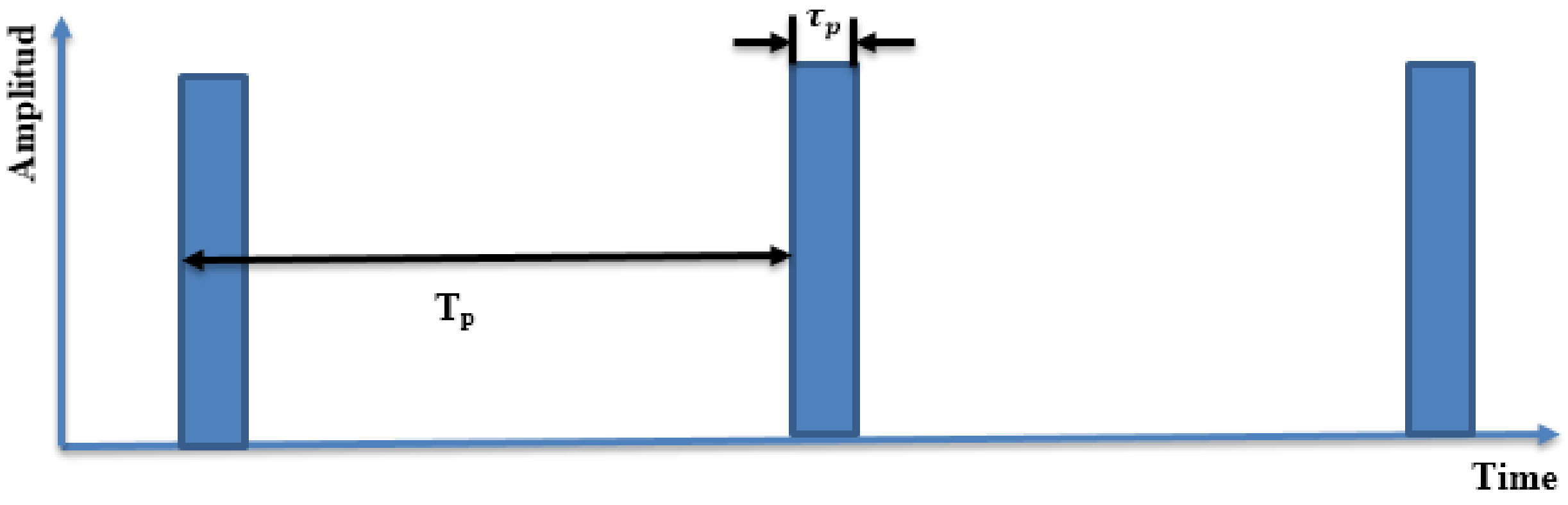

(a)

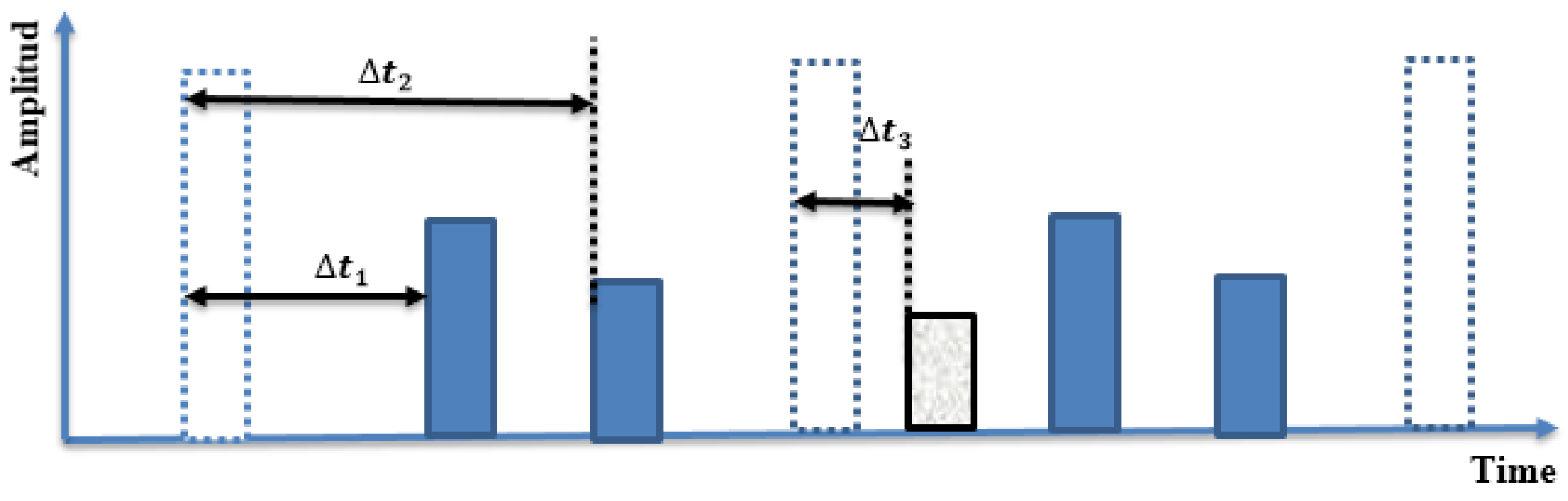

(b)

Figure 2.8 Pulsed radar signal (a) Transmitted signals, (b) Received signals

$$R_{max} = \frac{cT_p}{2} \quad (2.10)$$

$$\Delta R = \frac{c\tau_p}{2} \approx \frac{c}{2B} \quad (2.11)$$

Where B is the transmitted bandwidth.

### 2. Continuous-Wave Radar (CW)

In the CW radars, the speed of the target can be estimated by calculating the Doppler frequency which is the difference between the frequency of the transmitted signal and the frequency of the received signal. These systems are incapable of detecting the target range, and they cannot distinguish between objects moving toward or away from the transmitter [43]. Table 2. 4 shows the comparison between the advantages of CW and pulsed radar.

Table 2. 4 Comparison between CW and pulsed radar

| Parameter | CW Radar Advantages | Pulsed Radar Advantages |
|---|---|---|
| **Hardware complexity** | Low | High |
| **Average Power** | High | Low |
| **Short range target detection** | Good | Poor |
| **Moving target discrimination** | Good | Poor |
| **Target range determination** | Poor | Good |
| **High transmitter receiver isolation** | Poor | Good |

### 3. Frequency-Modulated Continuous-Wave (FMCW) Radar

FMCW radar is the same as CW radar; in contrast, the FMCW radar can change its operating frequency during the measurements that mean the transmission signal is modulated in frequency (frequency modulation). The amplitude of the transmitted signal is proportional to the instantaneous frequency. The generated signal is then transmitted and by measuring the round-trip delay and the frequency difference, the detector can estimate the velocity and the distance of the moving object. A sawtooth function is considered the simplest signal and most often used. The transmitted and received signals of FMCW are shown in Figure 2.9(a); this figure shows the variation of signal frequency versus time. The reflected signal is received with a delayed time $\Delta t$. In case of the target has the relative speed to the transmitter, the received signal shift by$f_d$. The frequency difference between the transmitted signal and the reflected signal is defined as the intermediate frequency (IF) from the mixer output. Figure 2.9(b) illustrates the absolute difference between the transmitted and received signals. The radar range, the relative velocity, and range resolution can be calculated based on $\mathrm{f}_1$and $\mathrm{f}_2$ [47].

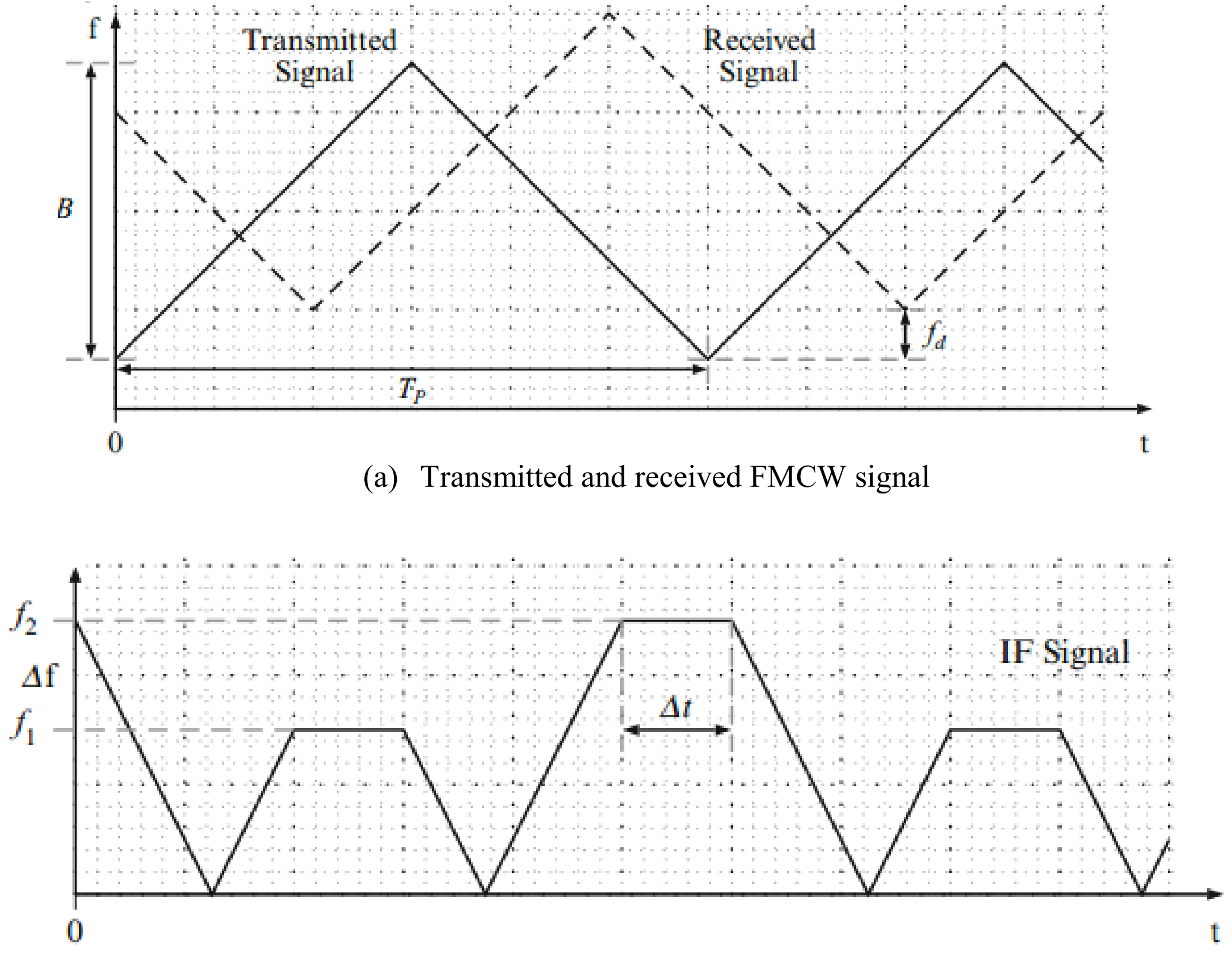

(a) Transmitted and received FMCW signal

(b)IF signal at mixer output

Figure 2.9 Ranging with FMCW radar [47].

$$R = \frac{cT_p}{2B}\frac{f_1 + f_2}{2} \tag{2.12}$$

$$v_r = \frac{c}{2f_c}\frac{f_1 - f_2}{2} \tag{2.13}$$

$$\Delta R = \frac{c}{2B} \tag{2.14}$$

Where $\mathrm{f_c}$ is the center frequency between $\mathrm{f_l}$ and $\mathrm{f_2}$.

### 2.2.4 Antenna Design for ARS

High gain and more efficient antennas in the automotive radar system are effective components. Consequently, the extensive literature review reports several kinds of antennas for automotive radar sensors. Low profile, low cost, compact size, and ease of integration are the main keys to solve the problems of automotive radar antennas design. This section focuses on the state of the art of radar antennas, especially 77GHz automotive radar antennas. In the automotive sensors, the direction of arrival (DOA) should be improved [48, 49].

#### 2.2.4.1 Horn Antenna

In the past years, the horn and parabolic antennas were often used as a first selection for the automotive radar system, as shown in Figure 2.10. Figure 2.10 shows the early radars for the automotive application introduced in the 1970s [47, 50]. The used antennas were huge and bulky to fit the proposed automotive radar. Nevertheless, these radars were remarkable steps to introduce the current professional radars. Due to the high gain of horn antennas become one of the most well-known antennas used in the radar systems to provide narrow HPBW and high power handling, but the gain of horn antennas is directly related to their size. So, the horn antenna with high gain is generally massive.

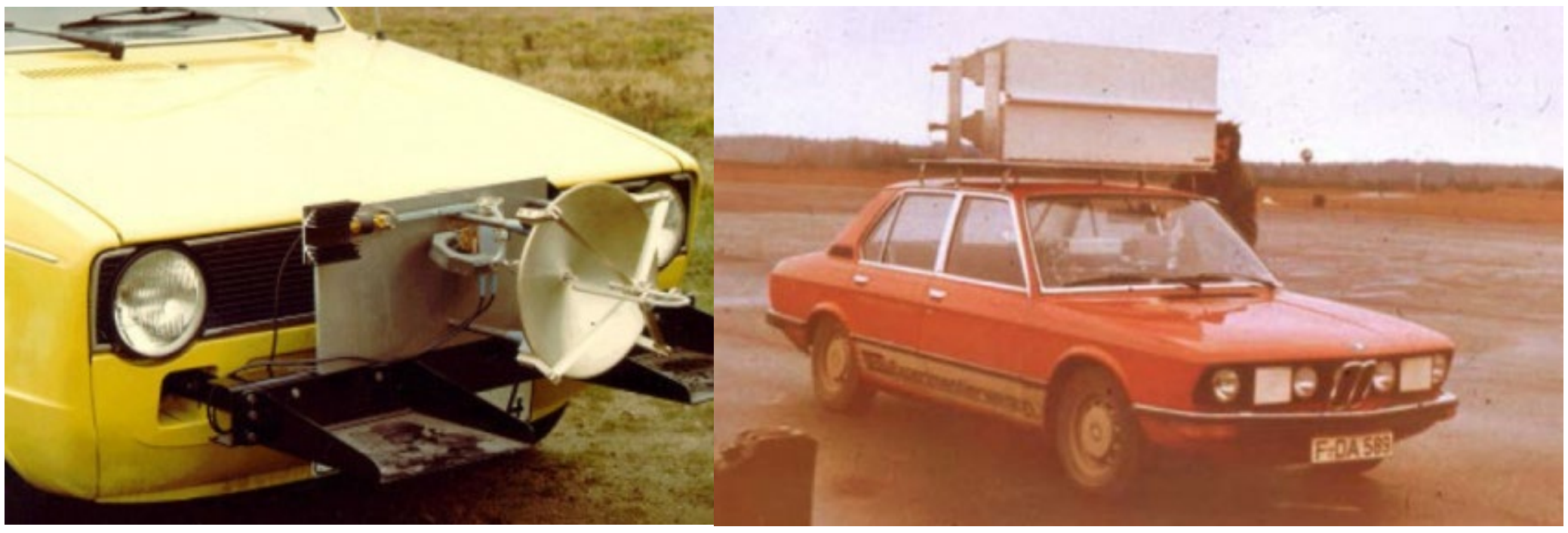

(a) (b)

Figure 2.10 Early automotive radar system (a) parabolic antenna[50], (b) horn antenna[47].

#### 2.2.4.2 Lens Antenna

The lens antenna is one of the famous antennas that has been the first choice for commercial, automotive radar applications to provide high directive beam. The lens antennas with beam switching capability have higher opportunities to the traditional phased antenna array due to excessive metallic and phase shifter losses, specifically at a higher frequency band (77 GHz)[51]. Recently, most commercial sensors used either a rotationally symmetric lens fed by a small array of patches or a large 2-D patch antenna array to overcome the problems of the feeding network of the antenna array. In 2005, Colome et al. [52] introduced cylinder lens fed by microstrip patches with dual feed to increase the isolation between transmitter and receiver for Bi-Static radar at 24 GHz. This antenna achieves a gain of 15.3dB and HPBW $21.3^0$, $37.8^0$ for E-plane and H-plane, respectively. Based on the work in[52], Weing et. al [53] introduced the same idea by placing the uniform series antenna array along the focus line of the lens. Figure 2.11 depicts the configuration of the lens antenna. In the elevation (YZ) plane, the lens is fed by a column of a series microstrip patch antenna. But in the azimuth (XZ) plane, the receiving antenna consists of a lens added to the uniform antenna array that is placed on the focal line of this lens. By 2017, Saleem et al. [4, 51]introduced an integrated lens antenna that consists of 6 layers of cylindrical Luneburg, where the Luneburg lens is the lens that has varying permittivity in the radial direction. Seventeen sources feed the lens; these sources are planar log-periodic dipoles. In the last few days, the fourth generation of commercial sensors introduced by BOSCH uses a lens antenna with a short focal length[54].

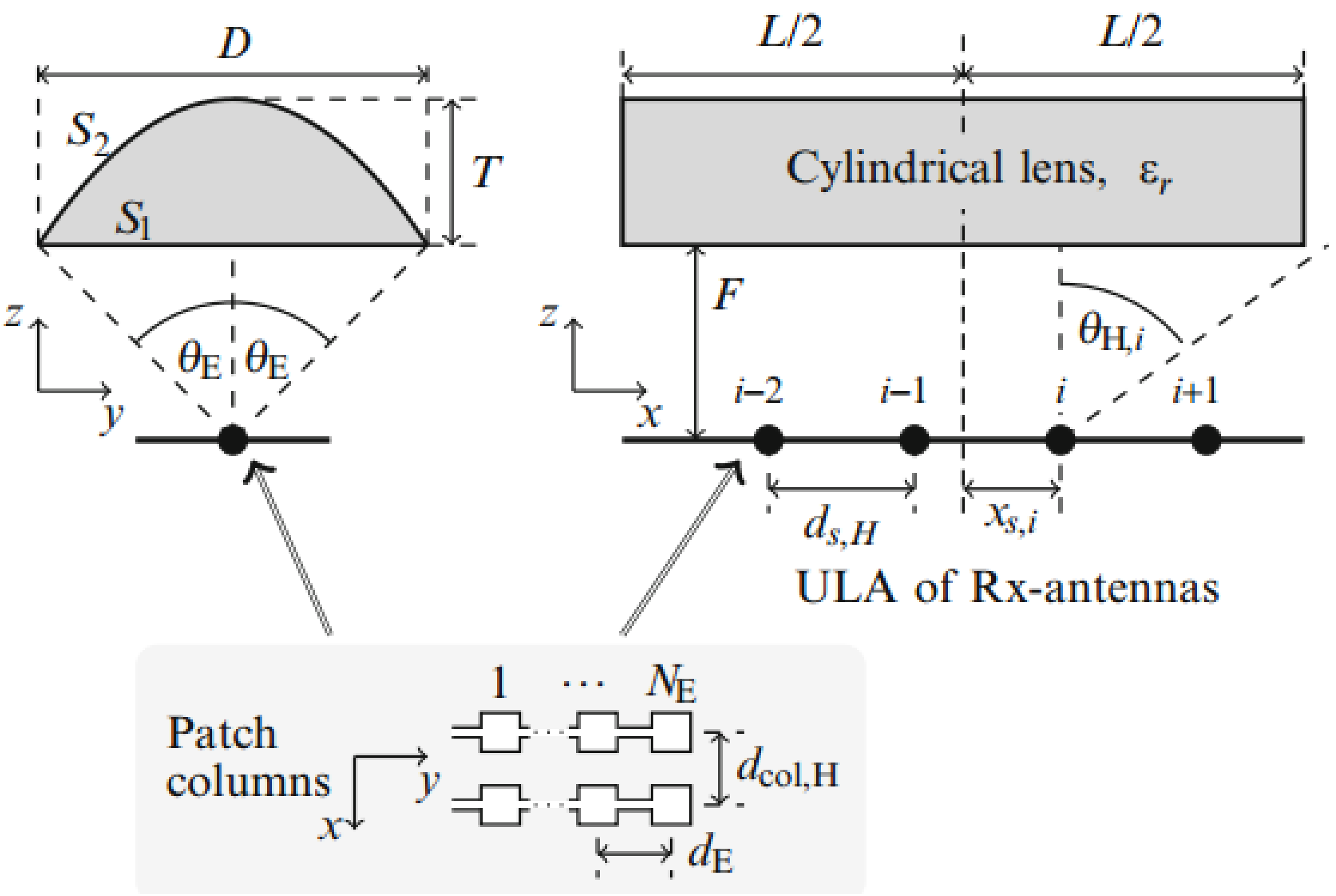

Figure 2.11 Design of cylinder lens antenna in the elevation plane (left) and azimuth plane (right)[53].

### 2.2.4.3 Planar Antennas

Several types of radar sensors use planar antennas due to their simple structure, low profile, ease of integration and low manufacturing cost. The most common types of planar antennas that are used in the automotive radar are series-fed microstrip patch arrays, grid antennas, slotted substrate integrated waveguide, comb line, etc. The antenna array consists of patches that are combined in series and/or parallel arrangements to provide high directivity. Figure 2.12 depicts two different antenna arrays operating at 77 GHz[55, 56]. The first type is a four-column; each column consists of 10 rectangular series patches with an overall size of 20.43× 7.83 mm$^2$. This antenna was printed on Rogers substrate with a thickness of 0.127 mm and a dielectric constant of 3. The second type is printed on the same substrate of the first type and consists of four columns, and each column consists of 12 leaf patches. One additional patch at the end of the feed line is printed for matching. Each six-leaf patch is arranged on one side of the microstrip feed line, and they are inclined by 45$^0$ from the microstrip feed line. This antenna has an overall size of 16.72 × 10.75 mm$^2$. The two antenna types have similar gain 19.8 dBi and angular width 11.6$^0$, 19$^0$ in azimuth-plane, elevation plane, respectively. In contrast, the first type has a smaller size than the second type. Lizuka et al. [57] developed the series antenna array, as shown in Figure 2.13. Vasanelli et al. in [58] introduce a low radar cross-section antenna array by using an artificial magnetic conductor (AMC) set around the series antenna to reduce the reflection in the direction of the car fascia, as depicted in Figure 2.14. The AMC has the capability to eliminate the reflected wave from the antenna.

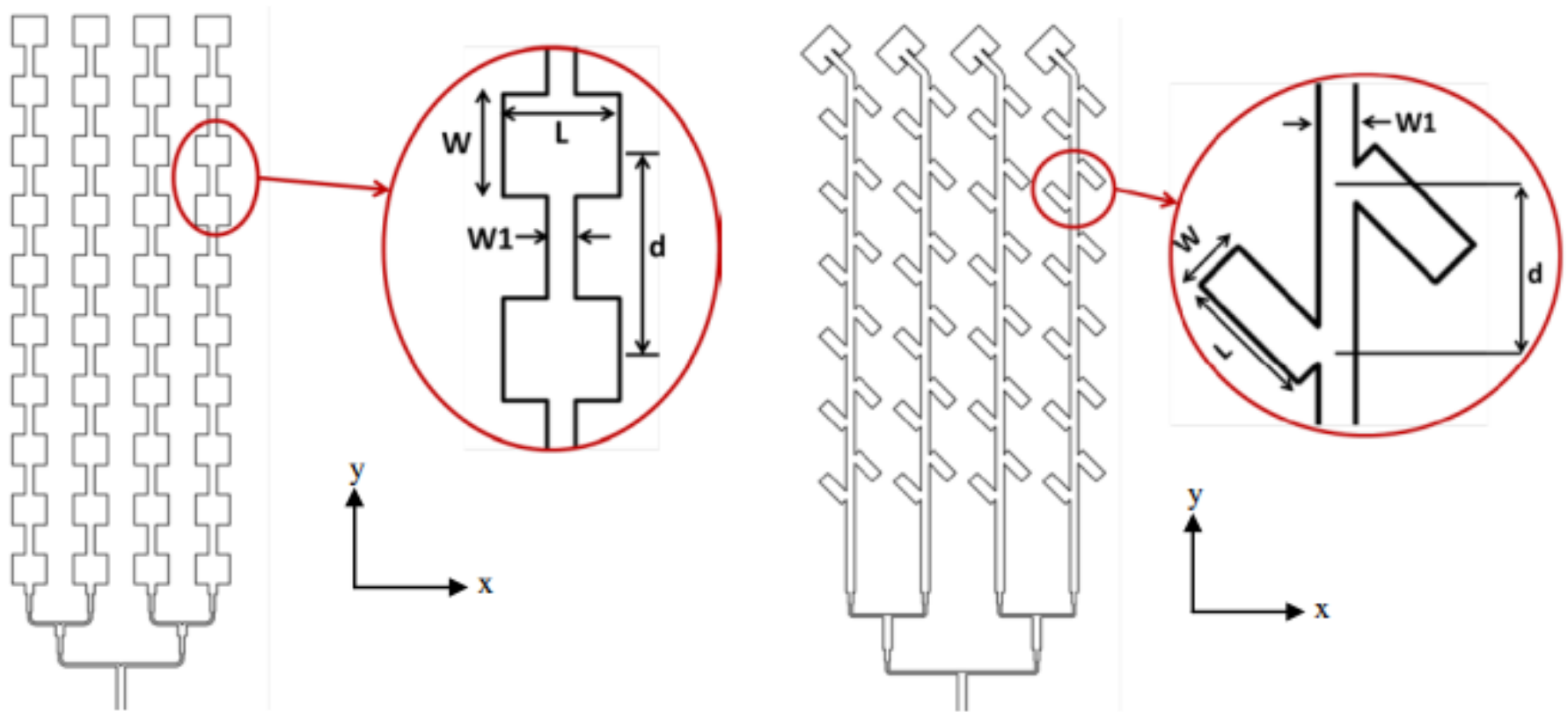

(a)First type (b) Second type

Figure 2.12 Series antenna array[55, 56].

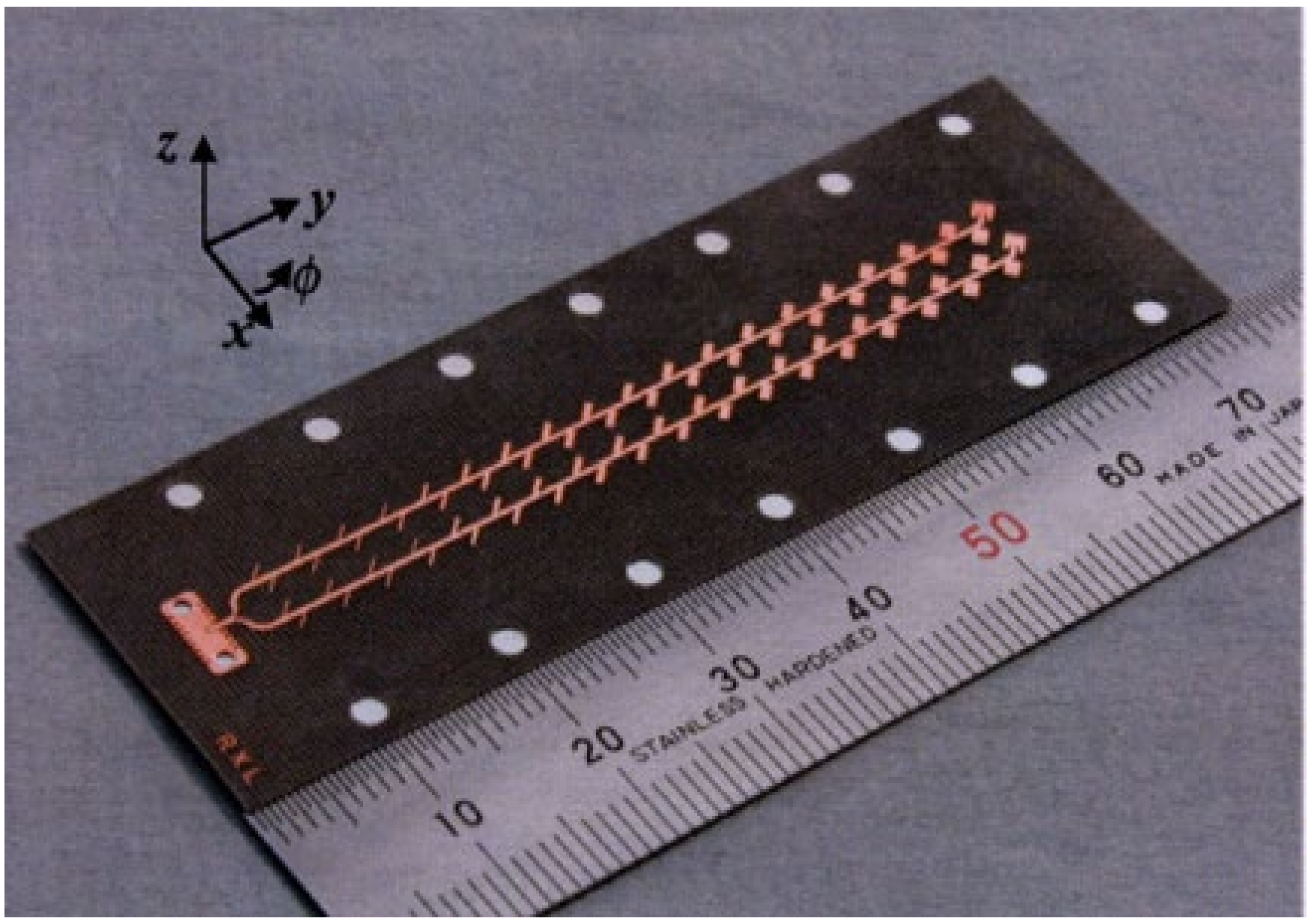

Figure 2.13 Photograph of developed series antenna[57].

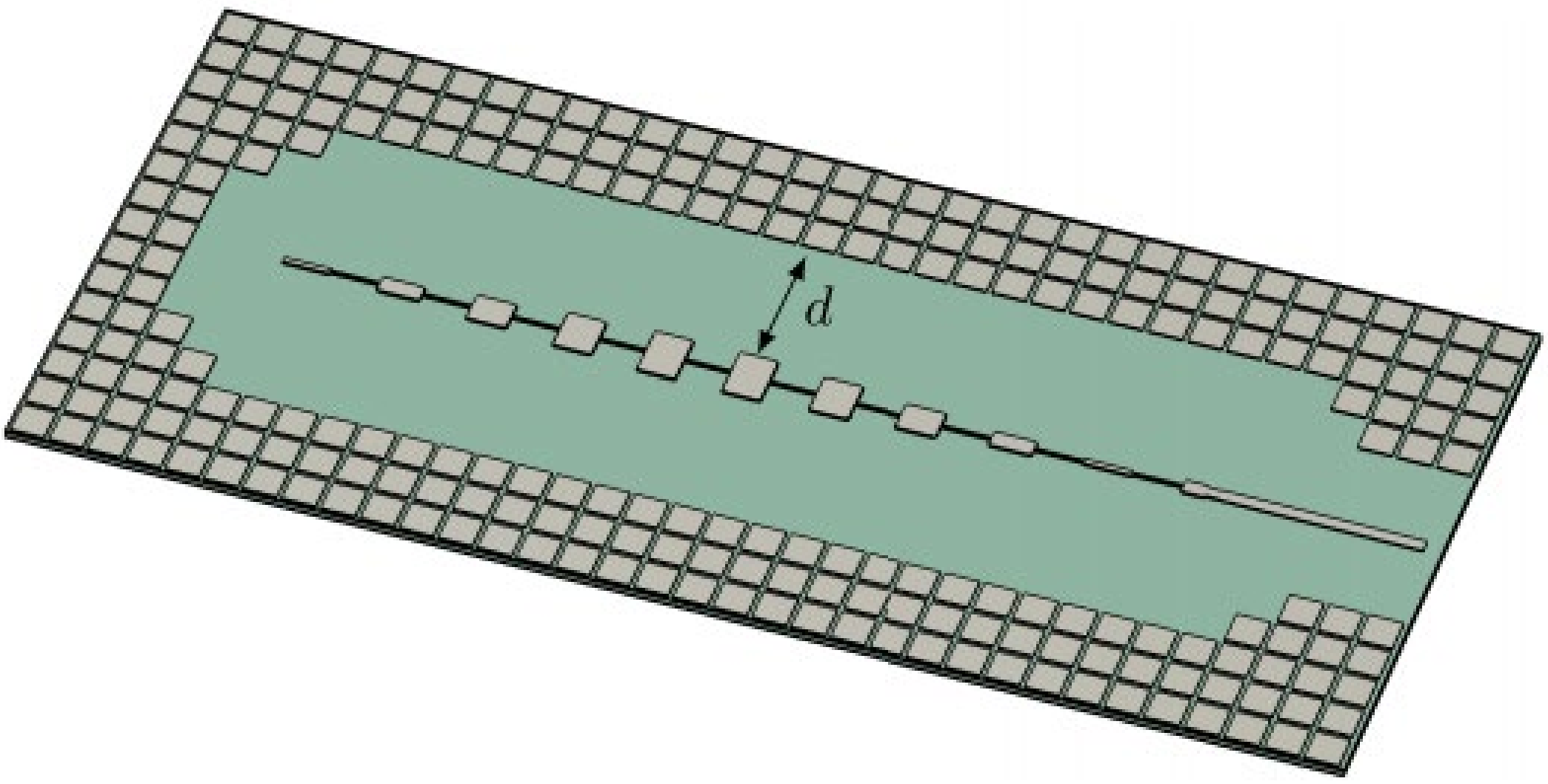

Figure 2.14 Series microstrip antenna array integrated with AMC[58].

## 2.3 Antennas for 5G Mobile Handset

The spectrum of the mobile communications systems below 3 GHz bands has been occupied through the last decade. This spectrum range suffers from high shortages and cannot maintain up with the fast growth in the communications systems rate for the near future. A necessary solution for 5G wireless communication is the use of mmW bands to enhance communication quality [1, 9, 59-66]. There are two candidate spectrums for 5G applications, lower spectrum and higher spectrum bands. The lower candidate bands are 3.3-4.2 GHz, 4.4-5 GHz and 6-7 GHz. Otherwise, the higher candidate bands are (24.25-27.5 GHz, 26.5-27.5 GHz, 26.5-29.5 GHz, 27.5-28.28 GHz, and 37-39 GHz) and some frequencies above 60.0 GHz are 53.3–66.5 GHz, 55.4–66.6 GHz, 56.6–64.8 GHz, 57.0–64.0 GHz and 57.0–65.0 GHz [67].

So, an antenna with wideband, stable radiation features and high gain is desired to overcome high propagation loss within mmW bands. In recent years, some papers have been published to introduce antennas for 5G terminal applications such as phased array [68], switchable antennas[59], dual circular-polarized antennas and dual linear-polarized antennas [69].

## 2.3.1 Beam Steerable Antenna

The phased array antennas have significant challenges to implement inside the mobile handset because of size limitations [70]; the phased array is integrated with a phase shifter and digital beamforming to provide the same functions as a steerable antenna. Furthermore, many studies are introduced to solve this issue which include using patch [64], slot [71] and dipole antenna arrays [72]. These methods mainly use one-dimensional linear arrays, with a fan-shaped beam pattern, on separate substrates, positioned in the cellular handset to achieve a broad beam coverage along with high gain within the restricted mobile size [73-80].

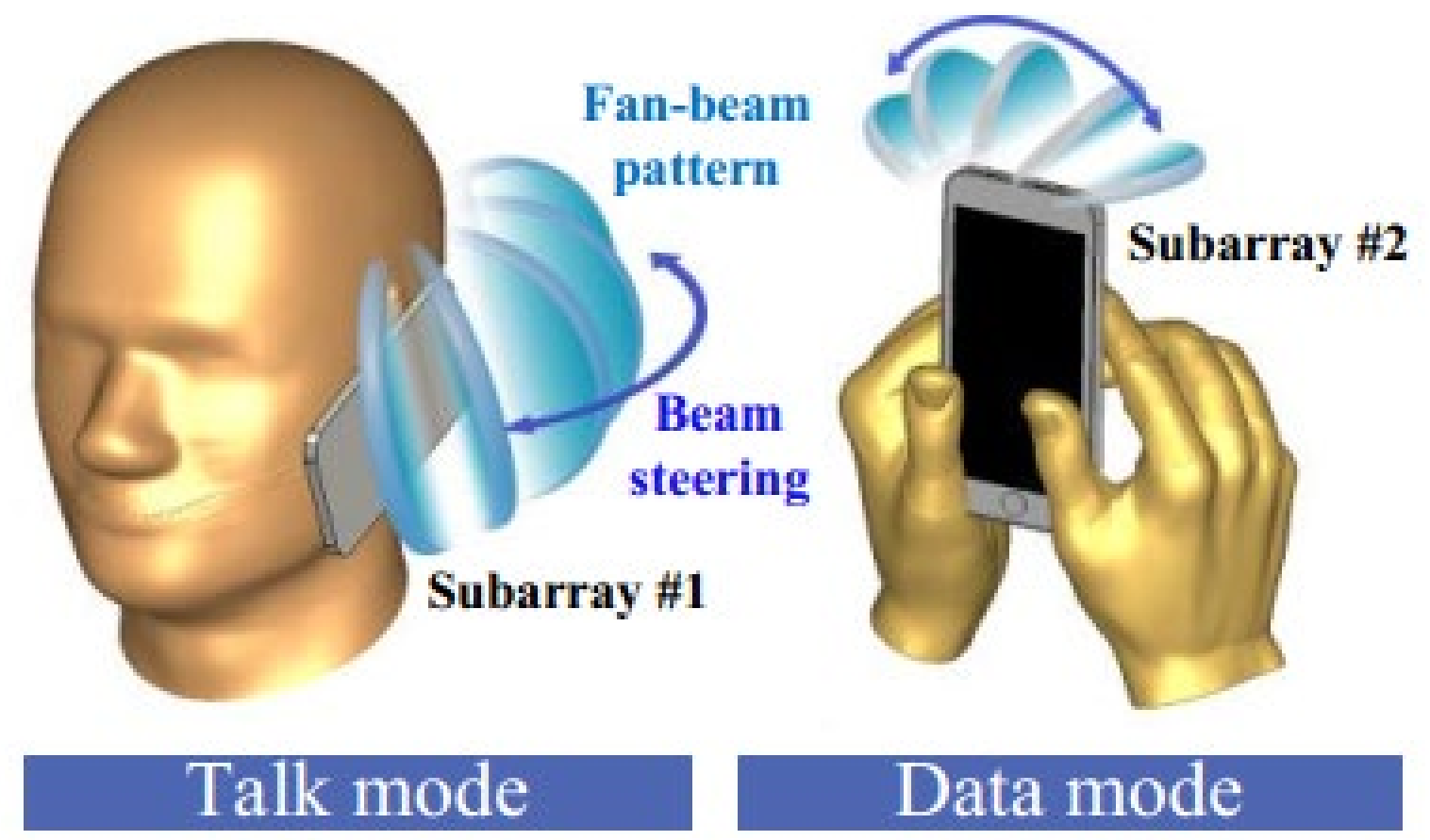

Figure 2.15 Beam steering idea for a talk mode and browsing mode [9]

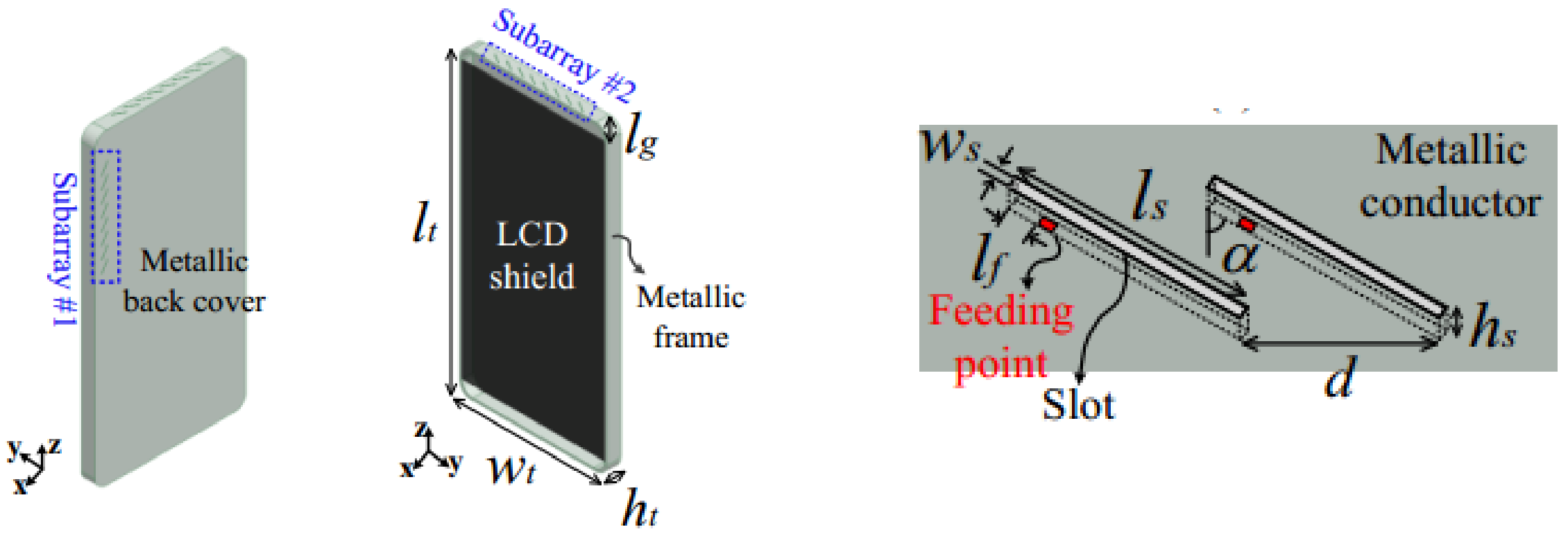

(a) (b)

Figure 2.16 Steerable antenna array with full board (b) feeding details [9]

Lt=67.1, wt=17.28, ht=7.1, hs=1, α=600, Ls=5.43, d=5.45, ws=0.26, Lf=0.5, Lg=1(units (mm))

Recently, Bang et al. [9] introduced a dual-mode scenario of the proposed antenna arrays for the talking and data modes as shown in Figure 2.15 and Figure 2.16. These two modes are introduced for beam-steering to provide high gain and wide coverage. The suggested antenna by the author consists of two subarrays, each array with eight rotated slot antenna elements. The antenna is printed on the top of the upper frame and portion of the handset's back cover. According to the operating mode, the subarray is selected. The first subarray is positioned on the handset's back cover to reduce the effect of the antenna on the user's head and is operated when the handset is in talking mode. In contrast, the second antenna is placed on the front frame of the handset to operate in the browsing mode or data mode because the browsing mode needs a radiation pattern like the hemispherical. Also, Zangh et al. [11] provided an antenna array consisting of two passive parasitic elements and one active element. Two switches are utilized in this design to control the steering beam, as shown in Figure 2.17. Two short circuit microstrip transmission lines with different lengths are connected with the switches. Two printed antenna array is printed on the sidewall of the mobile chassis to provide an $180^0$ coverage angle. However, this antenna provides a good coverage angle with each state of switches but suffers from high complexity, 3D structure, and high loss in switches.

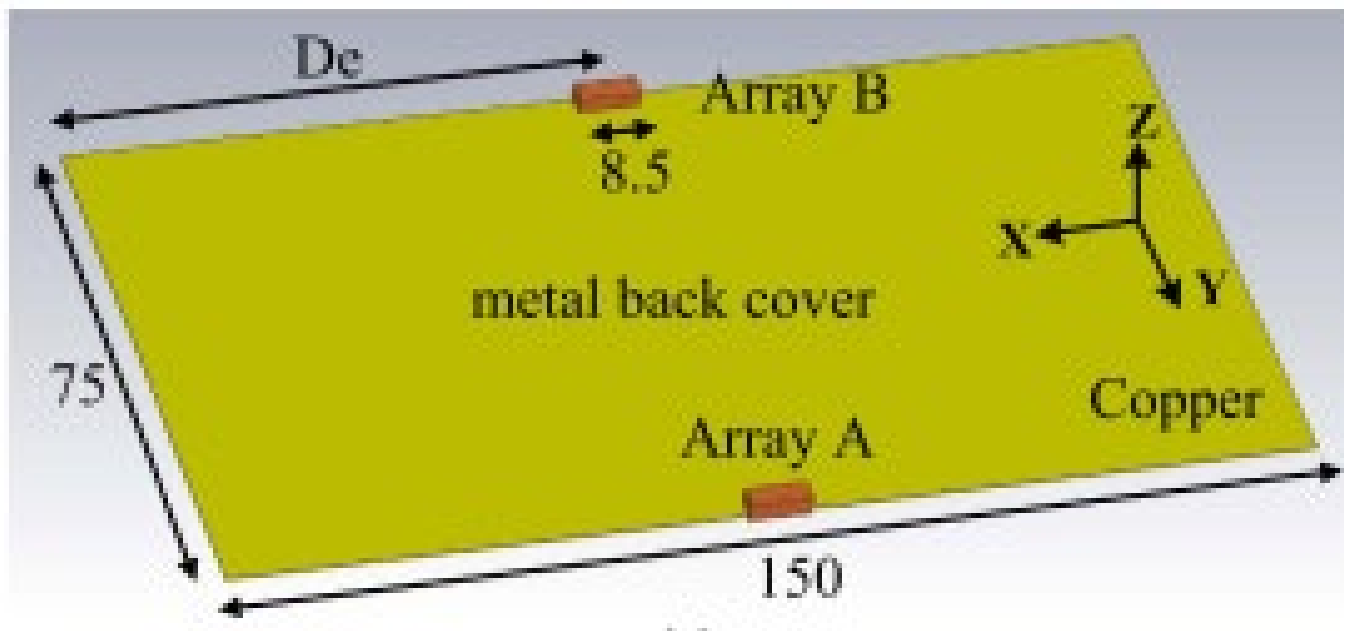

(a)

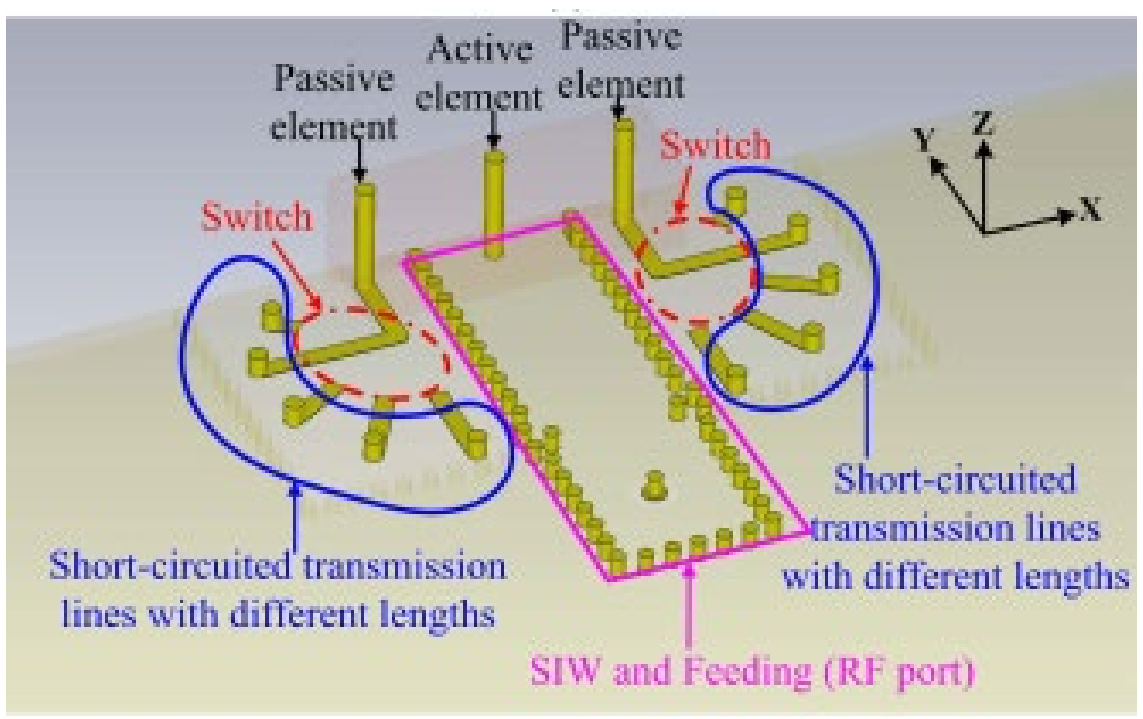

(b)

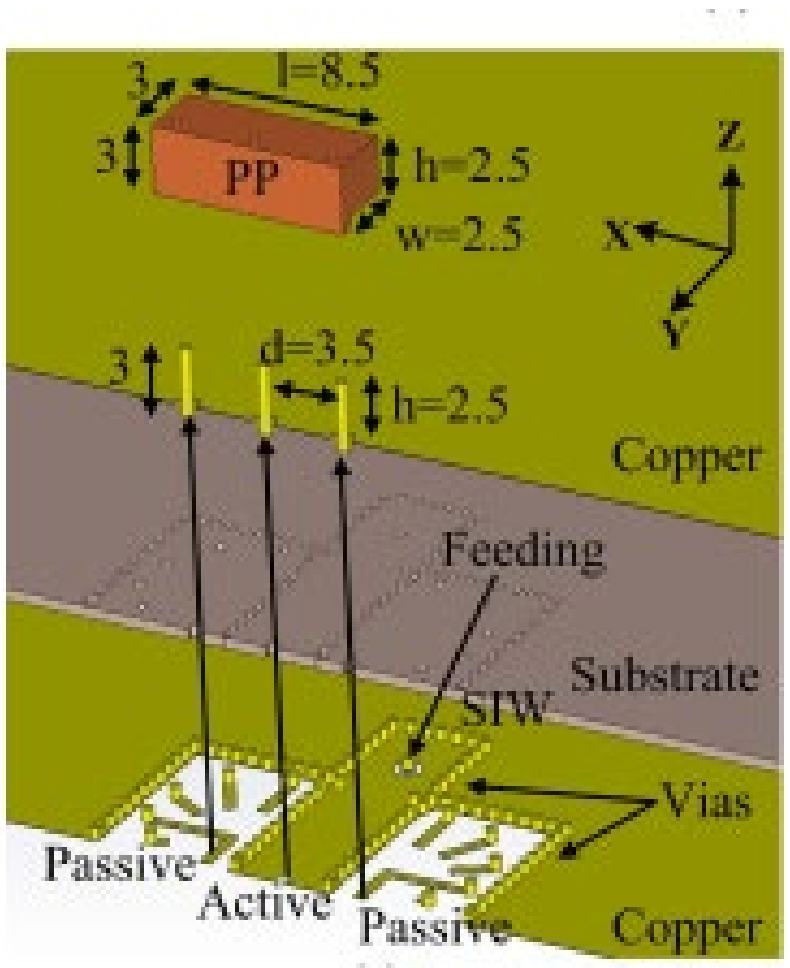

(c)

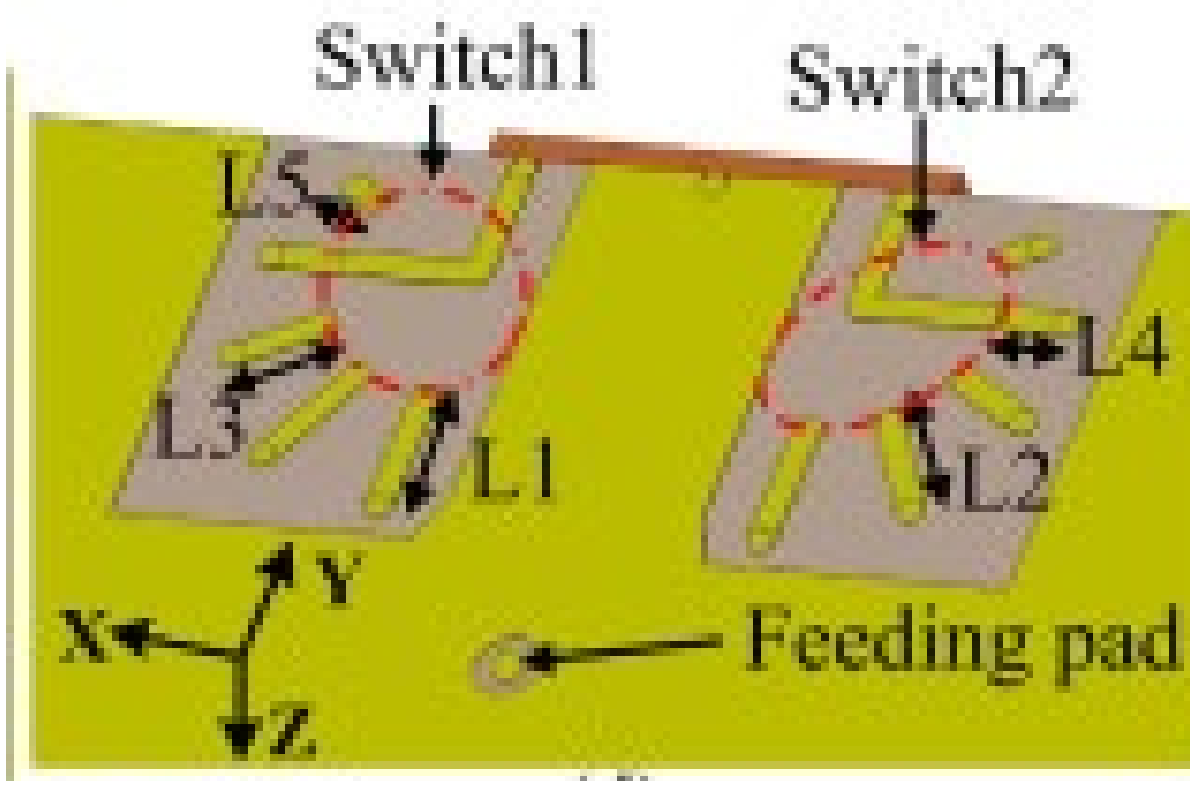

(d)

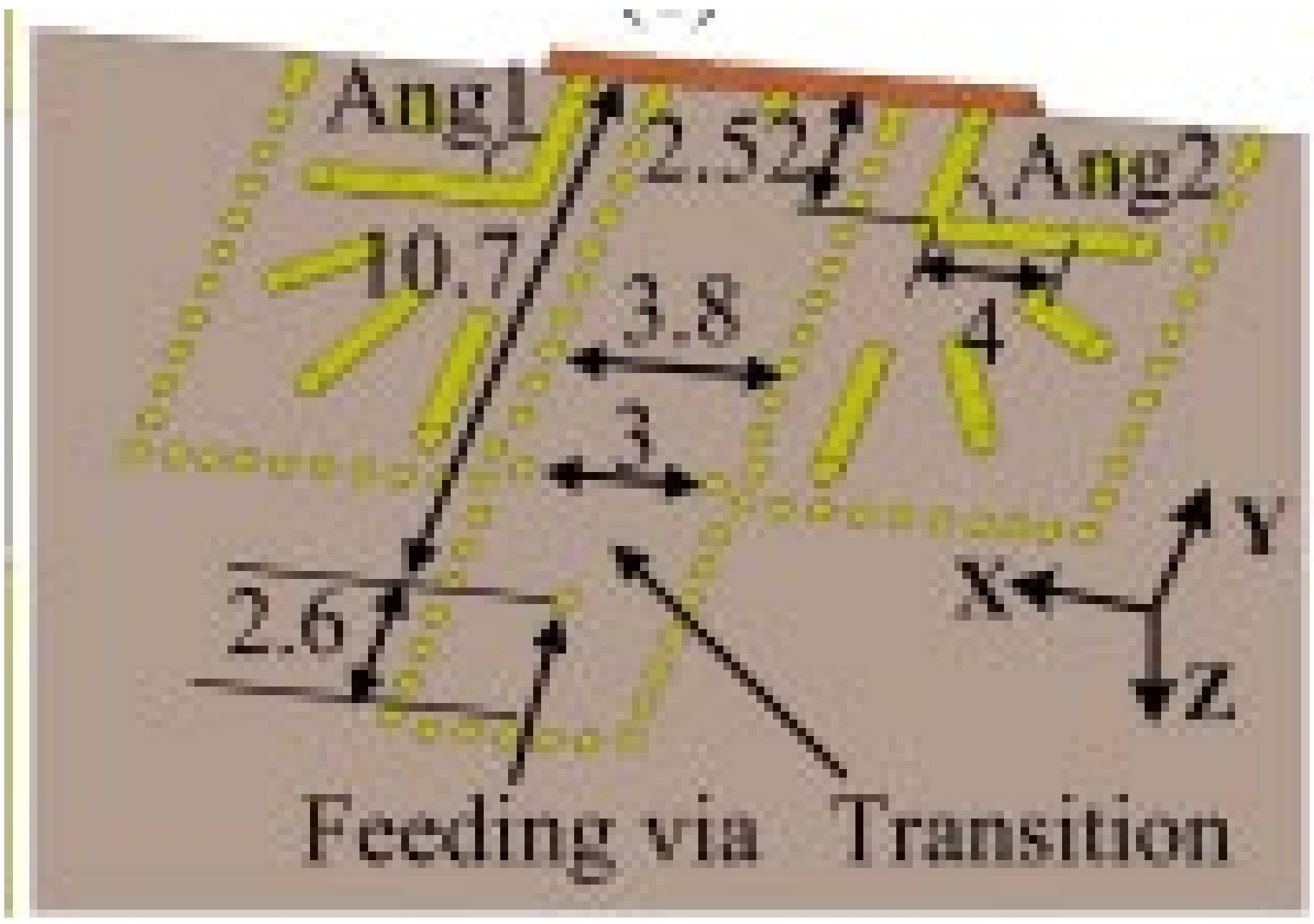

(e)

Figure 2.17 Geometry of antenna array (a) 3-D view, (b) detailed view, (c) exploded view, (d) back view with surface copper (e) back view without surface copper (units (mm)) [11]

## 2.3.2 Switchable Phased Array

The phased array covers only $360^0$, so the switchable phased array is introduced to cover all angles using more than one antenna. Figure 2.18 shows the different radiation pattern shapes that can be used for 4G and 5G mobile phones. According to the shape of the radiation pattern and to cover all angles three antennas will be needed at least for a 5G portable handset. Ojaroudiparchin et al. [64] introduced three printed antenna arrays on the mobile substrate's three different sides, as shown in Figure 2.19. Each sub-array consists of eight rectangular patch antennas with a half-wavelength distance between the patches elements and fed by coaxial probes. This antenna operates from 21 GHz to 22 GHz. To obtain the desired direction for the beam, the feed switches between the three sub-arrays. To select between the three sub-arrays, a microwave switch kit connects the feeding source to a power divider that connects to the phase shifter kit before each array.

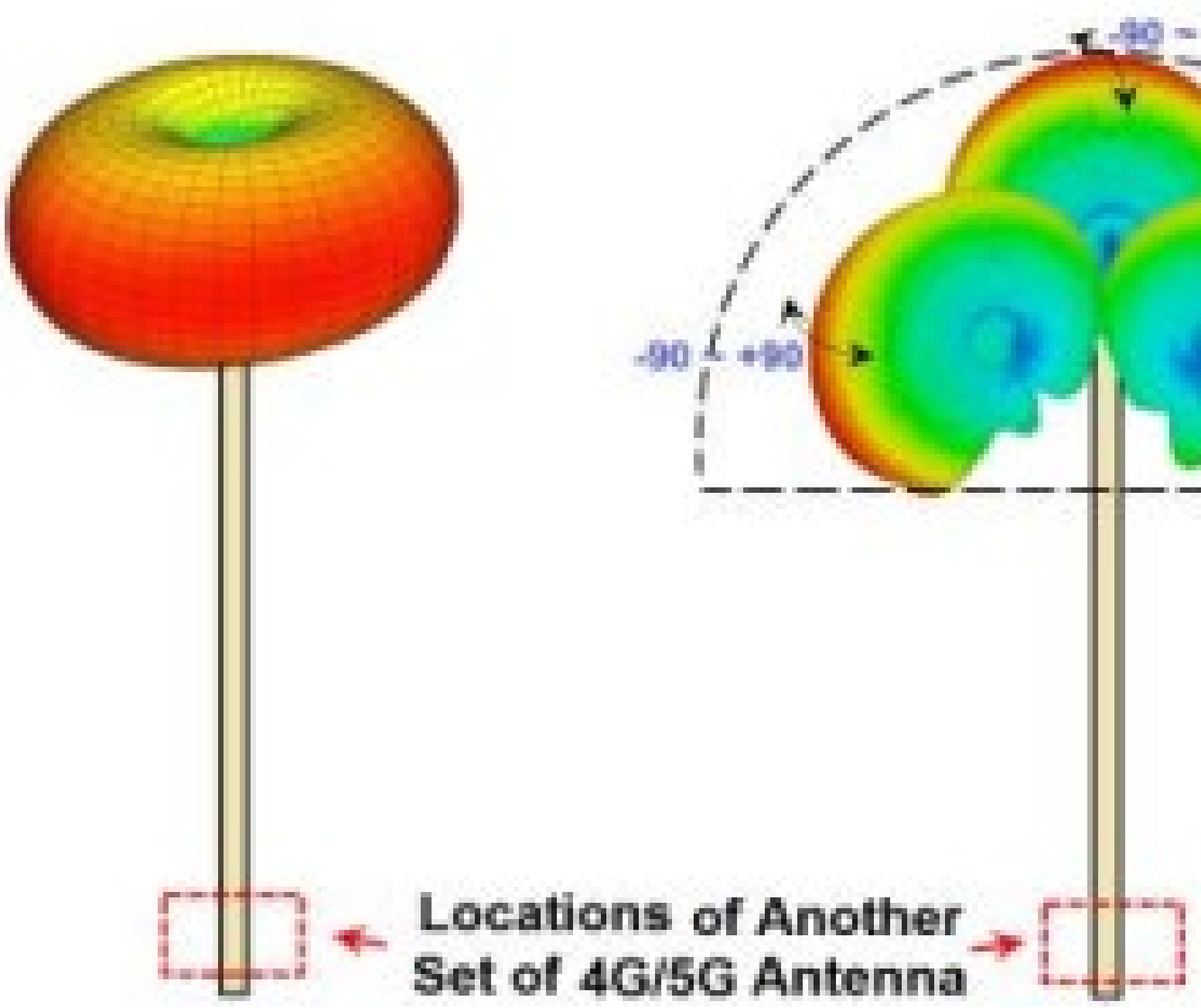

(a) (b)

Figure 2.18 Proposed radiation pattern of mobile phone (a)4G and (b)5G [64]

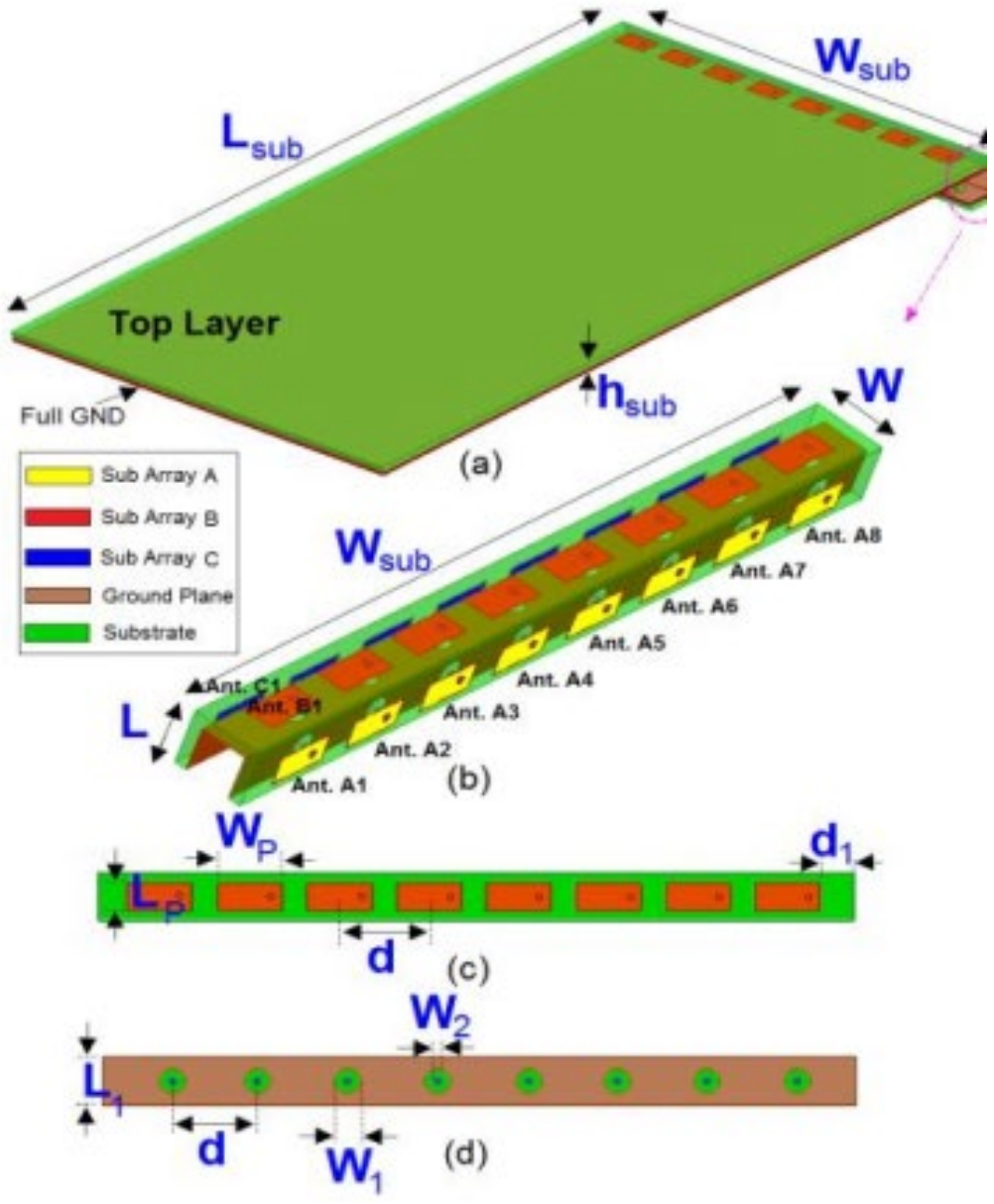

Figure 2.19 Switchable phased arrays (a) side view with full PCB, (b) 3 antenna array, (c) top layer view of one array, (d) bottom layer view of one array [64]. $W_{sub}$=55, $L_{sub}$=110, $h_{sub}$=0.787, $W_p$=4.32, $L_p$=2, d=6.5, $d_1$=4.5, W=4.574, L=3.787, $W_1$=1.72, $L_1$=3, $W_2$=0.5 (units (mm)).

## 2.3.3 Dual Polarized Antenna

Due to mastering of the dual-polarized antennas to introduce a solution in enhancing the isolation and channel capacity, this makes these antennas a good candidate for MIMO smartphone designs [60, 65, 69, 81-86]. In [87] Yang Li et al., introduced a hybrid eight-ports orthogonal dual-polarized antenna for 5G smartphones; this antenna consists of 4 L-shaped monopole slot elements and 4 C-shaped coupled fed elements. The 4 L-shaped elements are printed at the corners and the 4 C-shaped elements are printed at the middle on a thick 1mm FR-4 substrate. This design achieves 12.5 dB, and 15 dB for the isolation and the cross-polarization, respectively. Over the past months, Zaho et. al [88] presented a 5G/WLAN dual-polarized antenna based on the integration between inverted cone monopole antenna and cross bow-tie antenna for VP and HP, respectively. A 90◦ phase difference feeding network feeds the cross bow-tie antenna, so, the separated power divider and phase shifter are introduced to be used as a feeding network. In [89], Huang et al. introduced a dual-polarized antenna that consists of a main radiator, an annulus, and a reflector. The main radiator consists of two pairs of differentially-driven feedlines to transmit the energy to the coplanar patch. This structure achieves 26 dB and 35dB for the isolation and the cross-polarization, respectively. Eight-ports dual-polarized antenna array is reported in[90], the proposed antenna array is composed of four square loops, and each loop is excited by two orthogonal fed coupled feeding strips. Recently, Parchin et al. [65] introduced eight-port MIMO antennas using four square ring slot antennas, as shown in Figure 2.20. Each square ring slot is fed by two microstrip lines to achieve dual-polarization. The antennas are positioned at the four corners of the PCB to provide full coverage with dual-polarization. Two rings are printed with each antenna and operate as parasitic elements to provide isolation between the two ports of the antenna.

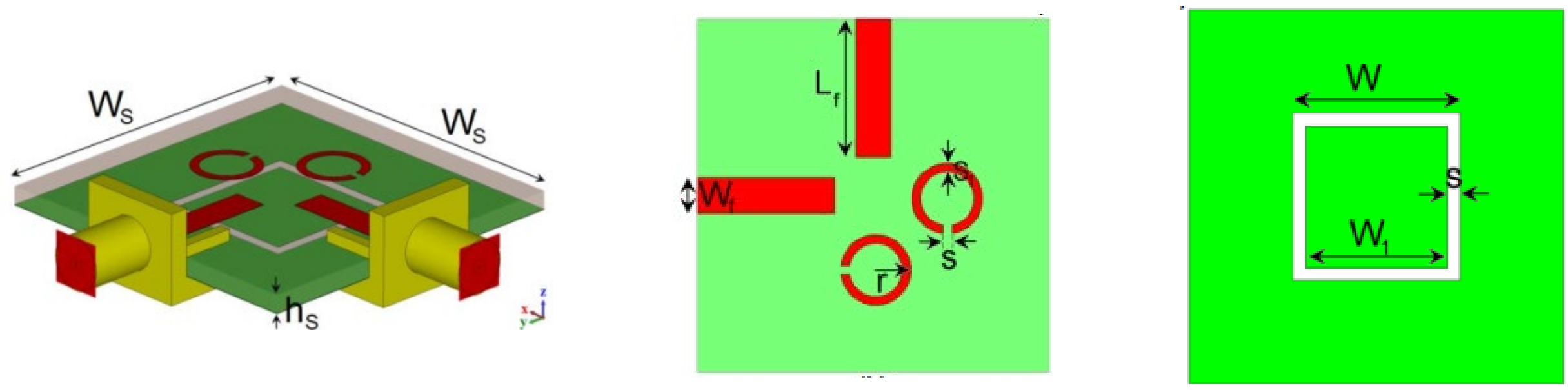

(a)3-D view (b)Front view (c)Back view

Figure 2.20 Dual-polarized square ring slot antenna [65]

# 2.4 Short Range Communications (SRC)

Currently, the applied wireless systems such as Bluetooth and WLANs that operate at lower frequency bands (2.45, 5.2, 5.8 GHz) below 6 GHz suffer from low data rates. Therefore, these technologies became unable to provide similar information rates like the wired standard rates such as gigabit Ethernet and high-definition-multimedia-interface (HDMI). Therefore, with the advent of the unlicensed 60 GHz band, researchers make the best use of the 7 GHz band from 57 GHz to 64 GHz [91-99].

The SRC becomes one of the dominant communication facilities during the last two decay because of its features such as Multi-Gigabit-per-second (MGbps) rate, high video, high streaming, and networking. The availability of broadband at 60 GHz in addition to the high attenuation at this band open the attention doors to the researchers to use this frequency for SRC such as point to point communications and point to multi-points communications. Antenna-on-chip (AoC) integration with other circuits will ensure low-cost SoC because of the removal of costs associated with external antennas [100-102].

## 2.4.1 CMOS Technology

The AoC that is printed on CMOS technology at 60 GHz which use silicon substrate, has low radiation efficiency and low gain. Because of the high permittivity of the silicon substrate, most of the power is confined within the silicon substrate. So, the low resistive silicon substrates are preferred to improve the radiation features of the AoC. In this thesis, we use 180 nm (0.18 $\mu$m) CMOS technology which consists of a base substrate from silicon with thickness of 200$\mu m$ and thin silicon oxide layer (Sio2) with thickness of 10.34$\mu m$ as shown in Figure 2.21. The $Sio_2$ layer consists of six metal layers, the thickness of layers M1 to M5 is 0.53 $\mu m$ and the thickness of M6 is 2.34$\mu m$.

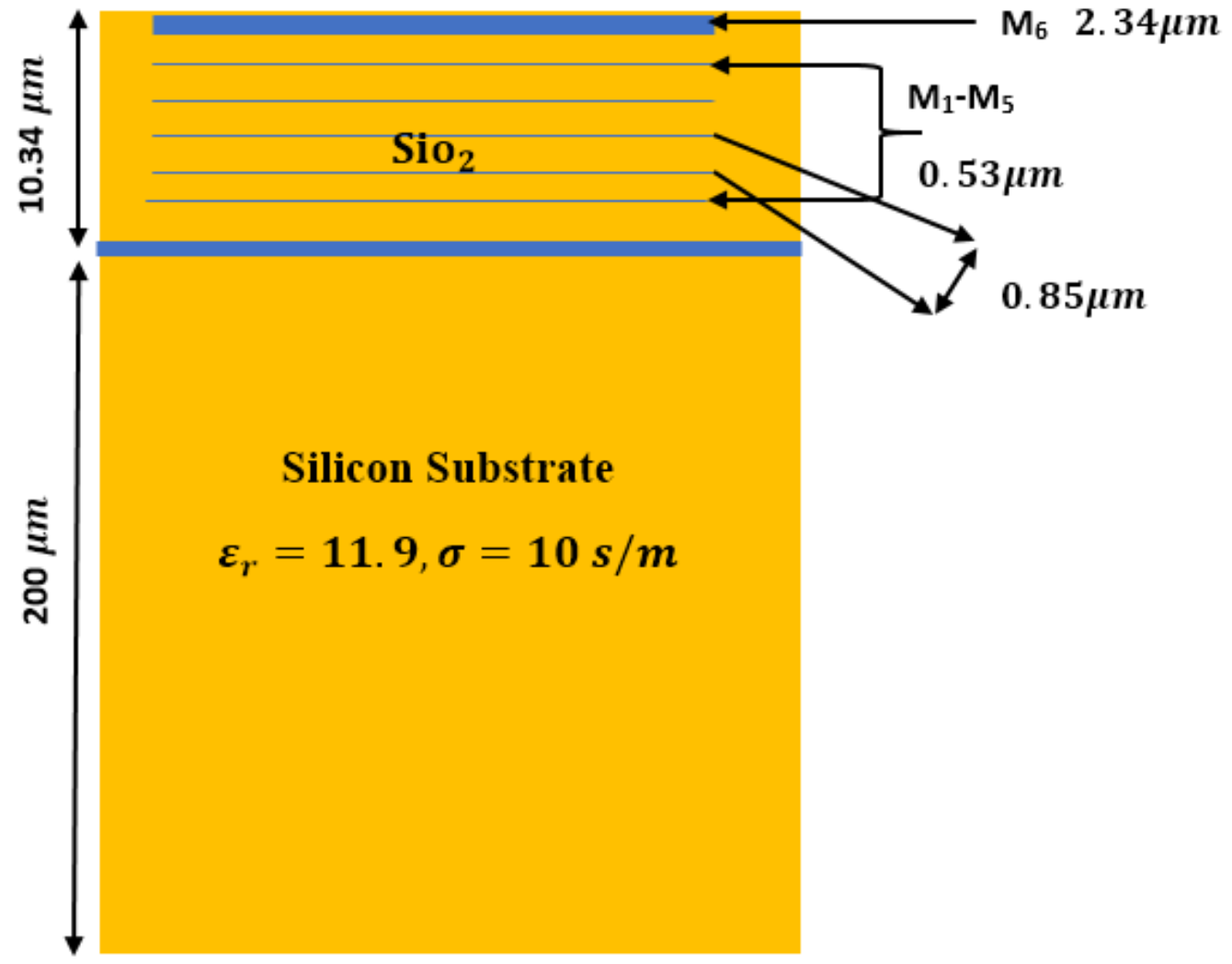

Figure 2.21 Structure of 0.18 µm CMOS

## 2.4.2 AoC Designs

Several studies have been introduced in the literature about the CMOS technology to meet the requirements of the system on-chip such as dipole [103-105], monopole[106, 107], triangular [31, 108-110], Yagi-Uda [29, 96, 98], Vivaldi [111, 112], bow-tie [32], and printed inverted F antenna (PIFA) [113]. Figure 2.22 illustrates different configurations of on-chip antennas that are introduced in the literature.

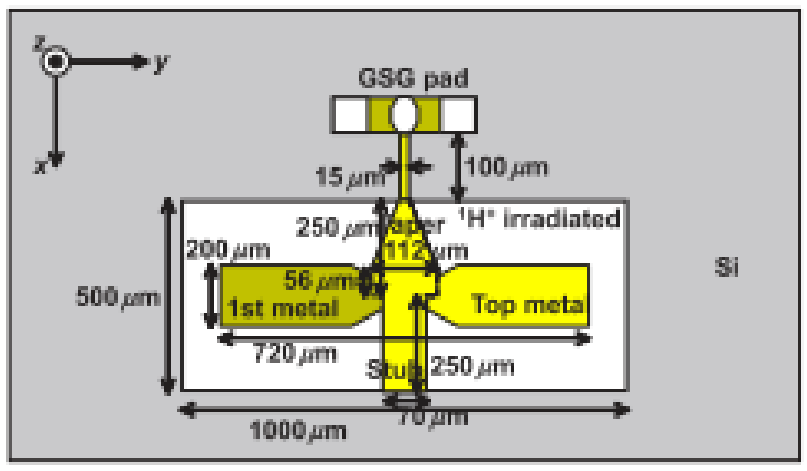

Dipole [103]

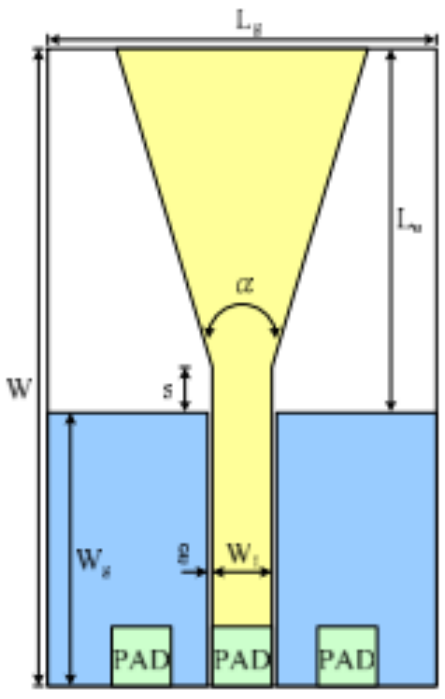

Triangular [110]

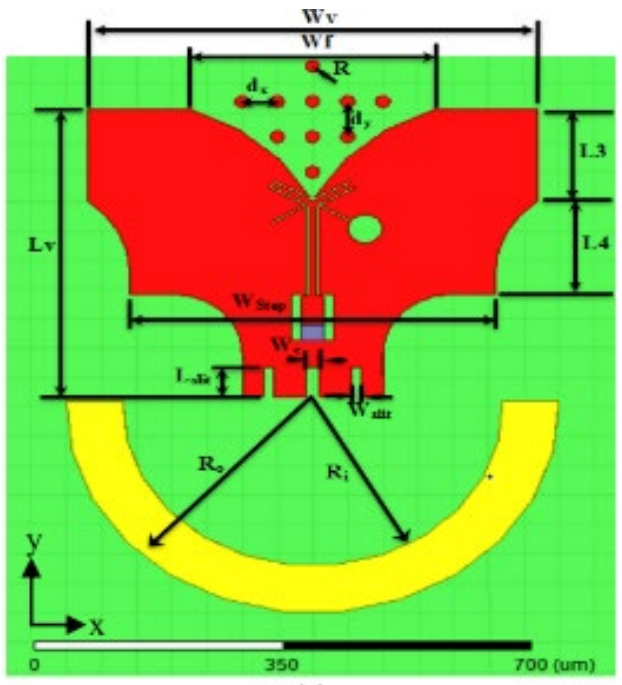

Vivaldi [111]

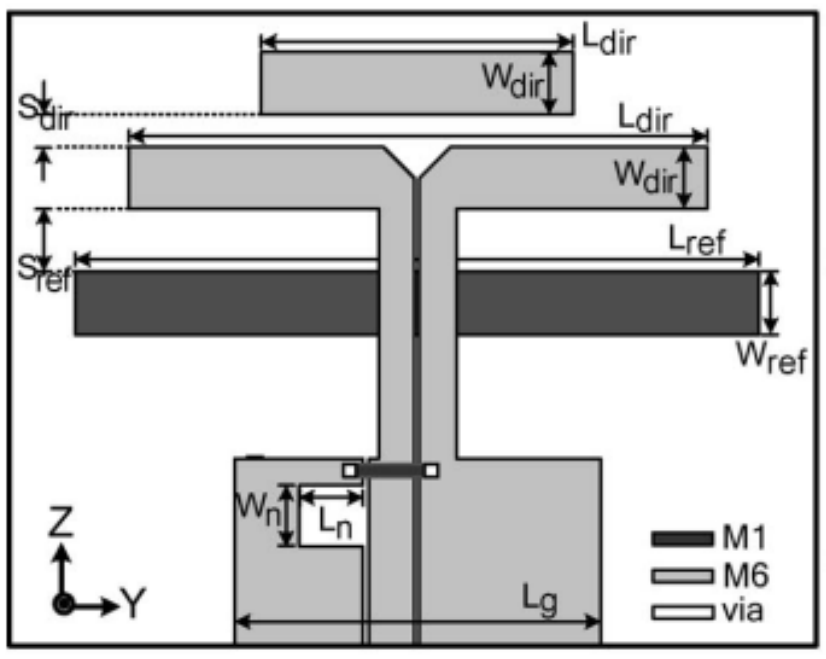

Yagi-Uda [96]

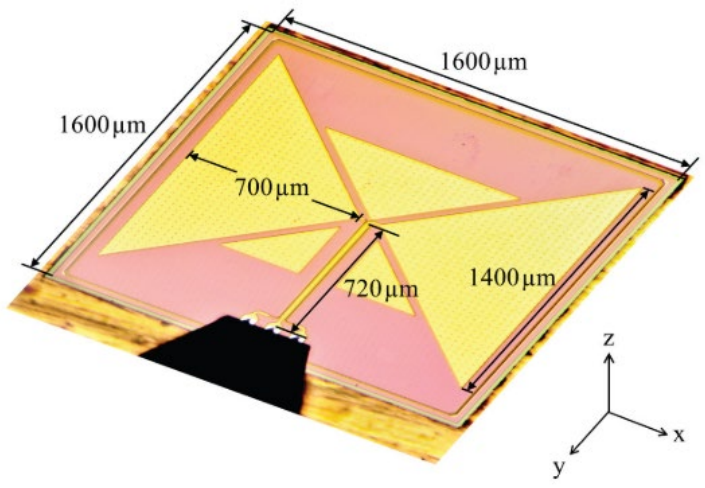

Bowe-Tie [32]

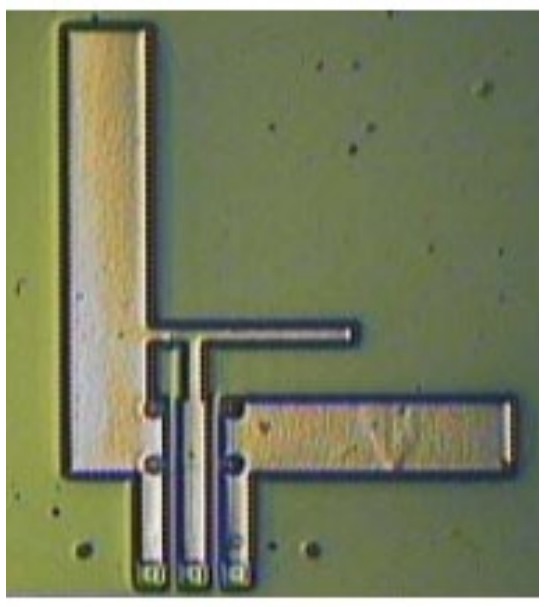

PIFA [113]

Figure 2.22 Different configurations of on-chip antenna designs.

# 2.5 Sub-THz Applications

Compared to traditional microwave engineering, the sub terahertz field is comparatively new. It is now described as the sector that includes techniques, manufacturing methods and devices that operate in the 100 GHz to 1 THz frequency band. This band includes wavelengths of submillimeter varying from 3 mm to 300 μm [3]. Owing to countless molecular spectral lines in this band carrying vital information for multi-gigabyte communications, astronomy as well as atmospheric research, the sub-THz frequencies take more and more attention. Unquestionably, astronomy applications played a significant role as a driver for the growth of THz technology and components. But soon, atmospheric window research in these bands as well as fast advances in the processing capacities of semiconductors, opened up a range of new THz applications for communication, defect detection, safety and biomedical imaging. Although there are several works in the biomedical at the microwave range, they suffer from limited resolution [114-117]. Thus the fast development seen over the last 20 years in THz frequencies made THz a field in communications systems [22, 118-124]. Therefore, a short review of distinct THz antennas is provided in this section. In this section, we focus on the antenna designs for sub-THz applications introduced in the literature. The number of antenna designs in the range from 0.1 THz to 1 THz is limited because the fabrication process in this range is still in progress. A key factor for THz applications is a technology platform for better performance of this band. Different technology platforms can be used for better performance of this band, such as CMOS, flip-chip, and hybrid techniques [3, 22, 121, 125-128].

## 2.5.1 THz Applications

### 2.5.1.1 Astronomy and Atmospheric

Certainly, astronomy and atmospheric sciences have been the first to explore the potential of THz technology for universe exploration as well as the earth features. The powerful water vapour absorption lines at 183 GHz and 557 GHz (as shown in Figure 1.1) have been used in universe water exploration and potentially life. When measuring the spectral lines of these molecules gave information about the surrounding temperature, pressure, gas velocities as well as magnetic fields within the observational region, the absorption of THz radiation by water, oxygen and other gasses led to many space missions.

### 2.5.1.2 Communications

The frequency band of 100 GHz to 1000 GHz, which has not yet been assigned for particular uses, is of particular concern for future wireless devices with information rates of more than 100 Gb/s. Despite the presence of different types of terahertz antenna, Koch [129] suggested waveguide horn and planar antenna to be used for next communication systems. Where, the horn antenna provides high efficiency and low loss. The horn-based imaging and communication systems were discussed in detail in [120, 123] at terahertz frequency. However, there is a higher potential for the planar antenna structure that has integration compatibility with planar systems. In [130] a 4 x 4 antenna array is printed on polypropylene substrate with $\varepsilon_r = 2.35\ and\ tan\delta = 0.0005$ at 300 GHz is introduced to achieve peak gain of 18.1 dBi. In [131] three series patches are introduced to operate at 0.1 THz and achieved 12 dBi of gain for short communication systems. This antenna is printed on thin Rogers substrate with height 0.127 mm and dielectric constant 2.2. The same research group improved this antenna by using polymer substrate with thickness 0.025 mm instead of Rogers substrate [132]. This antenna achieved 16 dBi of gain but with five series patches instead of three patches in the previous antenna. All antennas are shown in Figure 2.23.

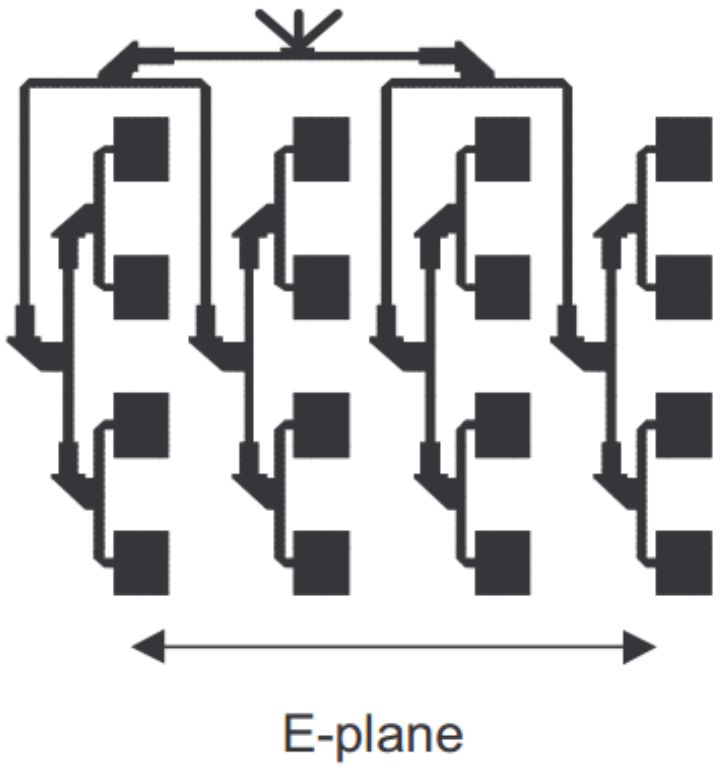

(a)Planar antenna array [130]

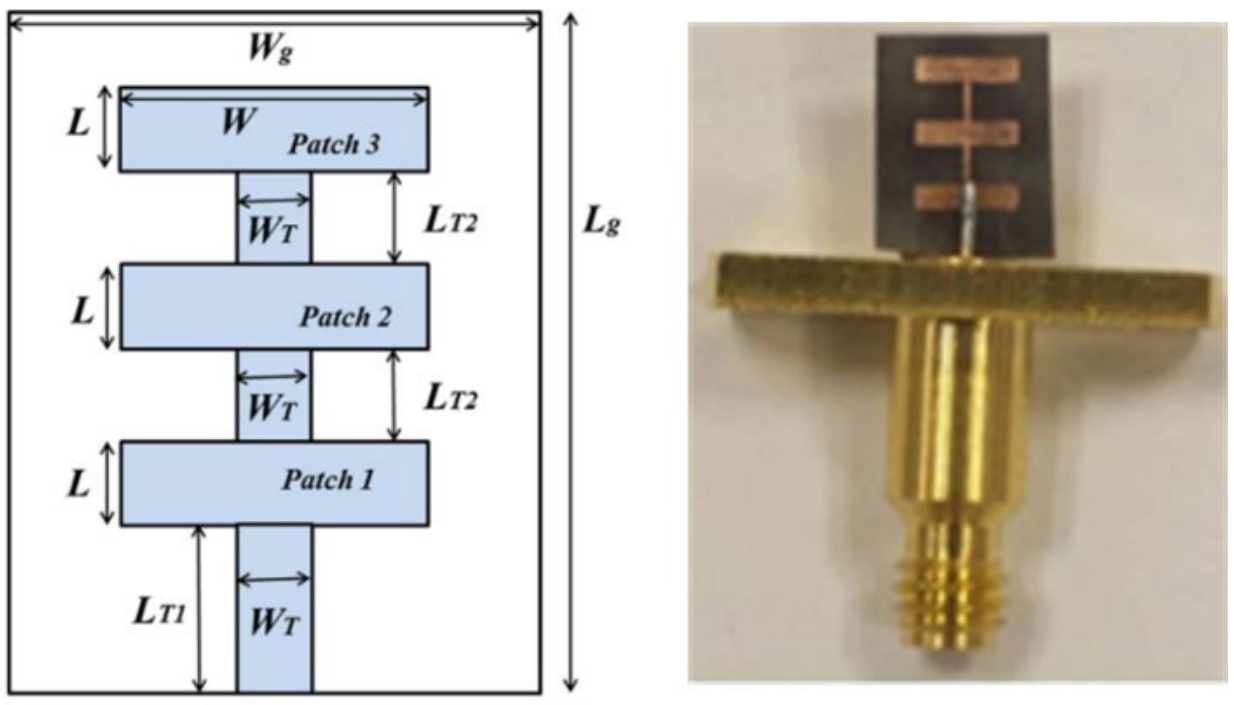

(b)Three series patches antenna array [131]

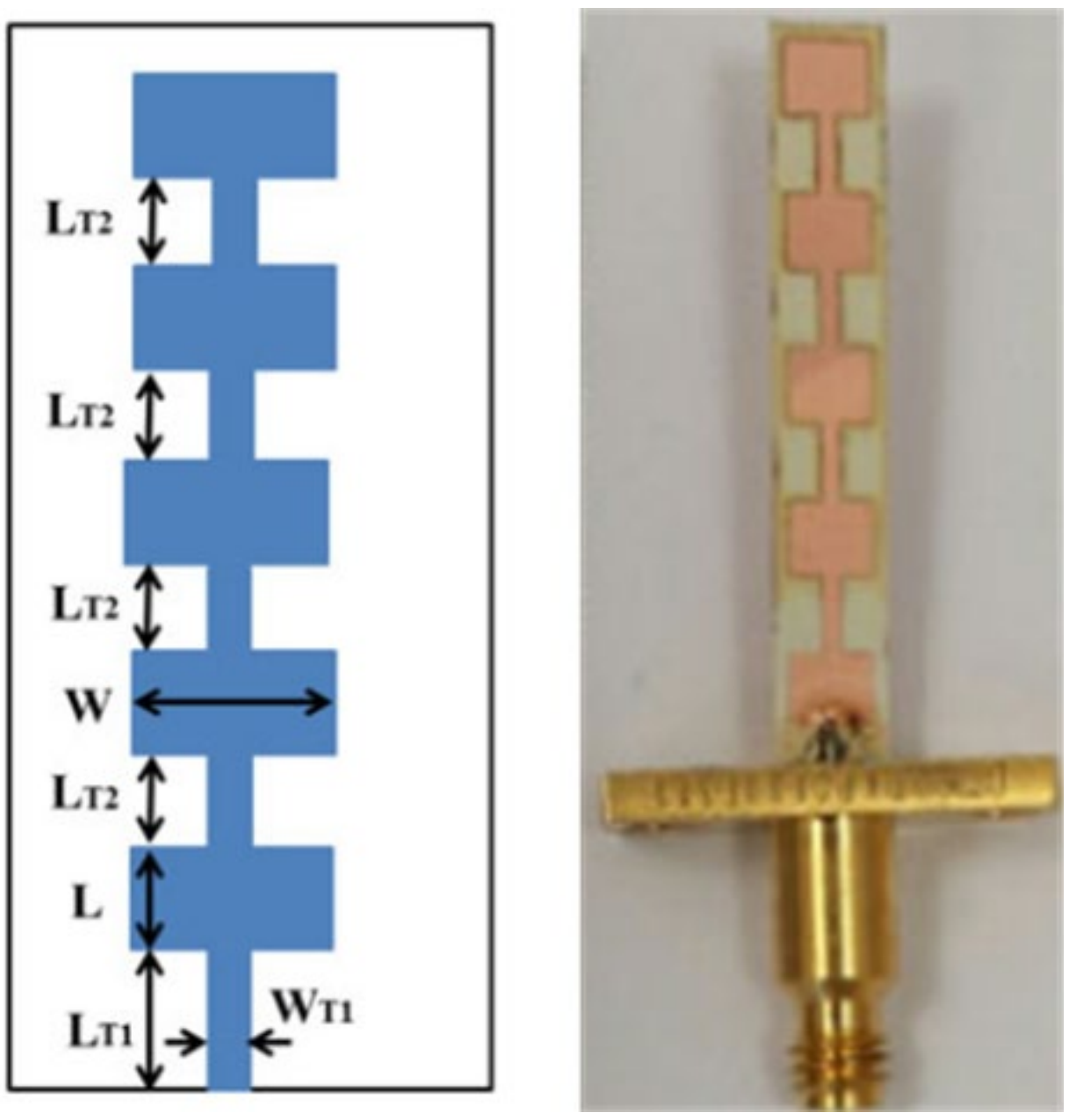

(c)Five series patches [132]

Figure 2.23 Different configurations of antennas for THz short communications.

In order to increase the antenna gain and directivity, different designs are introduced in [22, 133, 134]. In [134] an array of glass lens antennas arranged on a silicon (Si) substrate is introduced based on planar metallic rectangular waveguide structure. In [133], the authors presented a two tapered dielectric antenna that is designed and implemented in the suspended SOG waveguide platform. [135].

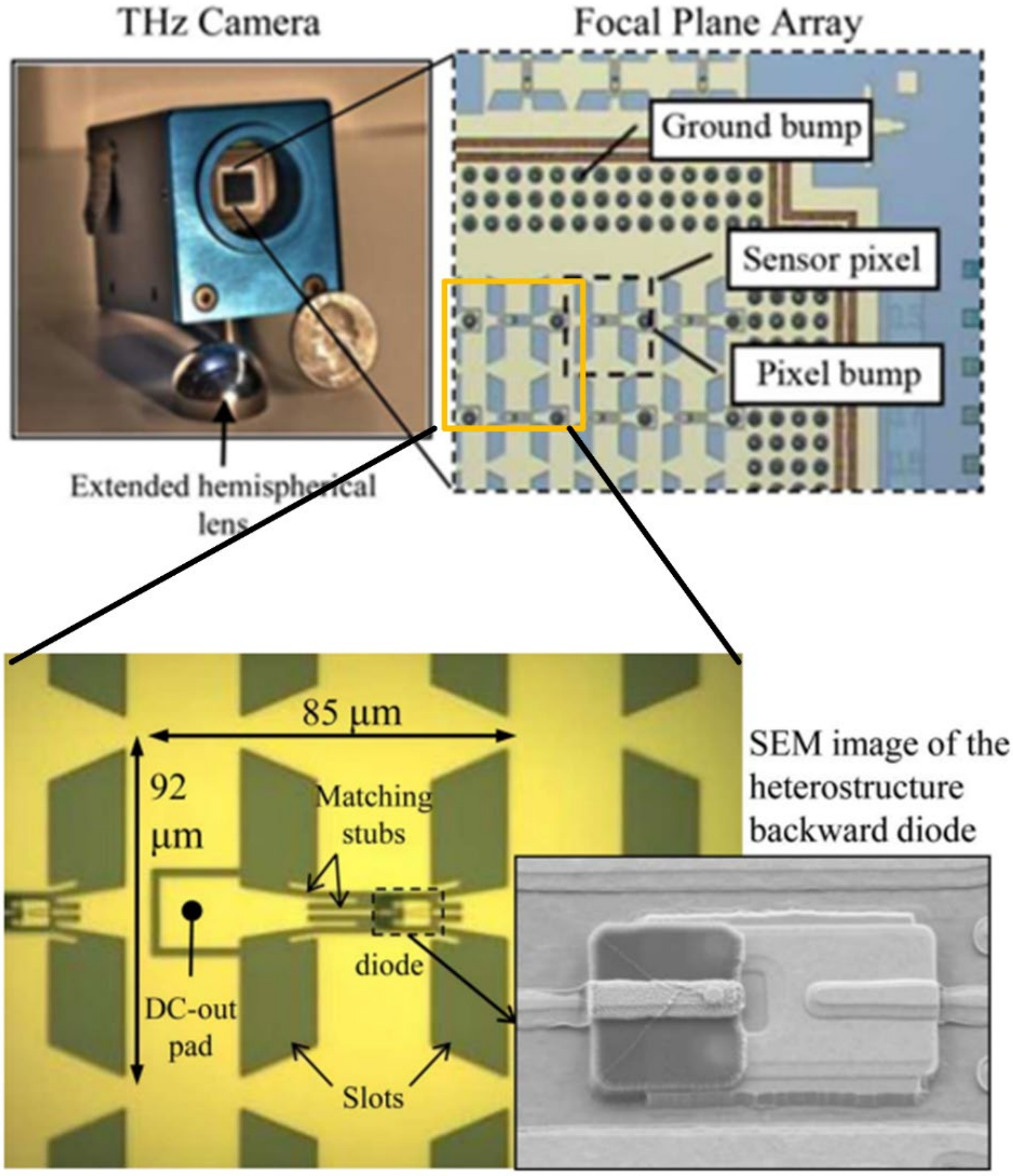

Figure 2.24 Real time focal plane antenna array camera [134].

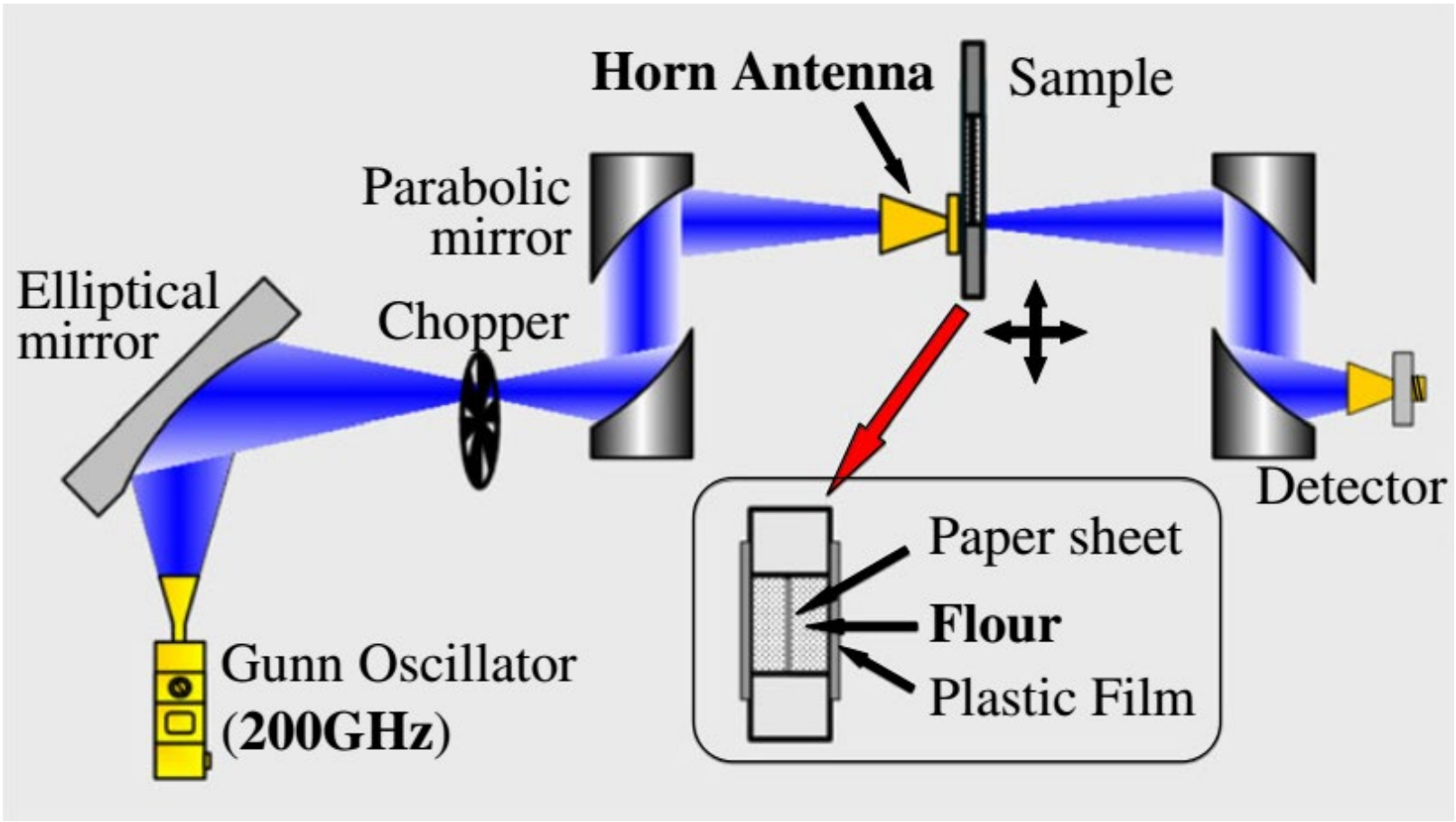

Figure 2.25 Imaging system based on horn antenna at 200 GHz [123].

#### 2.5.1.3 THz Imaging

The THz imaging is estimated as an attractive technique for reducing, or possibly eliminating, the effect of atmospheric circumstances with low visibility. In THz imaging, there are several contributing factors to attract the researchers to it such as:

- Low penetration depths of imaging in optical and IR bands are not attractive for non-destructive applications. Although, the THz image allows elevated capacity for penetration compared to the optical and IR.
- The THz band enhance far-field spatial resolution compared to the millimeter waves and reduced Rayleigh scattering compared to the infrared (IR). The THz imaging is considered as a one of non-ionizing radiation system that has benefits over ionized radiation systems such as positron emission tomography (PET), magnetic resonance imaging, planar X-rays, and X-ray CT scan.
- It has a low scattering loss compared to the infrared.

For optimum detection, the imaging systems should be wide-band and highly efficient. Therefore, Trichopoulos et. al [136] introduced a real-time THz imaging camera for broadband focal plane array, as Figure 2.24. The THz camera includes a hemispherical lens and objective lens. The hemispherical lens consists of a planar slot antenna array that redirects the receiving THz signals to high permittivity lens. On the other hand, Kim et.al [123] introduced a THz imaging system as shown in Figure 2.25. In this system the horn antenna is used to improve the imaging system's resolution. The waveguide antenna is used in this system instead of aperture antenna to reduce the power loss with small size (1.3x0.648 $mm^2$) to have a cutoff frequency of 115 GHz.

## 2.6 Conclusion

This chapter focused on the main applications in mm-wave and S-THz ranges. The literature review for automotive radar, 5G mobile, short-range communications and S-THz antennas are investigated. The bands, commercial sensors and the different antenna configurations that used for automotive radar are introduced to present a guide lines in the next chapter. Furthermore, the different techniques, antenna configurations and candidate bands of 5G applications are introduced as a one of the master applications in mm-wave ranges. The third part in this chapter presents the on-chip technology and its applications for short-range communications. Finally, the antenna designs in S-THz ranges and their applications such as astronomy, communications, and imaging are introduced.

# Chapter Three: ANTENNA DESIGN FOR AUTOMOTIVE RADARS

## 3.1 Introduction

This chapter introduces a novel antenna array for an automotive radar system based on the concept of a virtual antenna array (VAA). The proposed VAA is introduced to serve medium-range radar (MRR) and long-range radar (LRR) by the same antenna at the same time that has a flat shoulder shape (FSS) radiation pattern. Furthermore, an unequal power divider is introduced to feed the VAA and its excitation coefficient based on the method of moment and the genetic algorithm. The proposed VAA consists of two linear antenna arrays with a total number of elements equal to 16 patches and an overall size of $\mathbf{30} \times \mathbf{48}$ mm$^2$ to cover the band from 23.55 GHz to 24.7 GHz. The VAA is fabricated and measured; it presents a good agreement between all the simulated, synthesized and measured results is found. The second part of this chapter introduces an antenna array for LRR that operates at 77 GHz. This antenna depends on the hybrid radiator to provide wide bandwidth and high gain. Furthermore, the AMC technique reduces the surface wave and enhances the isolation between elements. The antenna is simulated by CST version 2018 and HFSS version 16 to verify its results.

## 3.2 MIMO/Phased Array

The phased array is considered one of the main important antennas widely used for radar in civilian and military applications. The phased array consists of multiple TX antennas and multiple Rx antennas that are often co-located. Another type of multi-antenna is the Single-Input-Multiple-Output (SIMO) radar. The radar has one TX antenna and multiple Rx antennas. The number of Rx antennas directly affects the angular resolution, where the angular resolution enhances with increasing the number of Rx antennas. Whilst, this type of radar limits the increasing number of Rx antennas because each antenna requires an individual receiver [137, 138].

Also, in Multiple-Input-Multiple-Output (MIMO) radar, the radar consists of multiple TX antennas and multiple RX antennas. But compared to the phased-array radars, the MIMO radars have more degrees of freedom that enhance the angular resolution, the characteristic identifiable, and give more flexibility for transmitting beam-pattern design. Furthermore, the MIMO monostatic radars can synthesize a more extensive virtual array to provide more enhancement of angular resolution and the number of targets that can be detected. In summarizing, the collected MIMO can divide into two categories: the bi-static MIMO radar, the radar that does not share any antenna between transmitter and receiver, and the monostatic radar, it's the radar that shares the same antenna array for transmitter and receiver. As an example, the SIMO radars that have one TX antenna and $M_{TX} \times N_{RX}$ RX antennas are equivalent to the MIMO radars that have $M_{TX}$ TX antennas and $N_{RX}$ RX antennas. So, the MIMO radar achieves a low cost and high angular resolution compared to the other system [139].

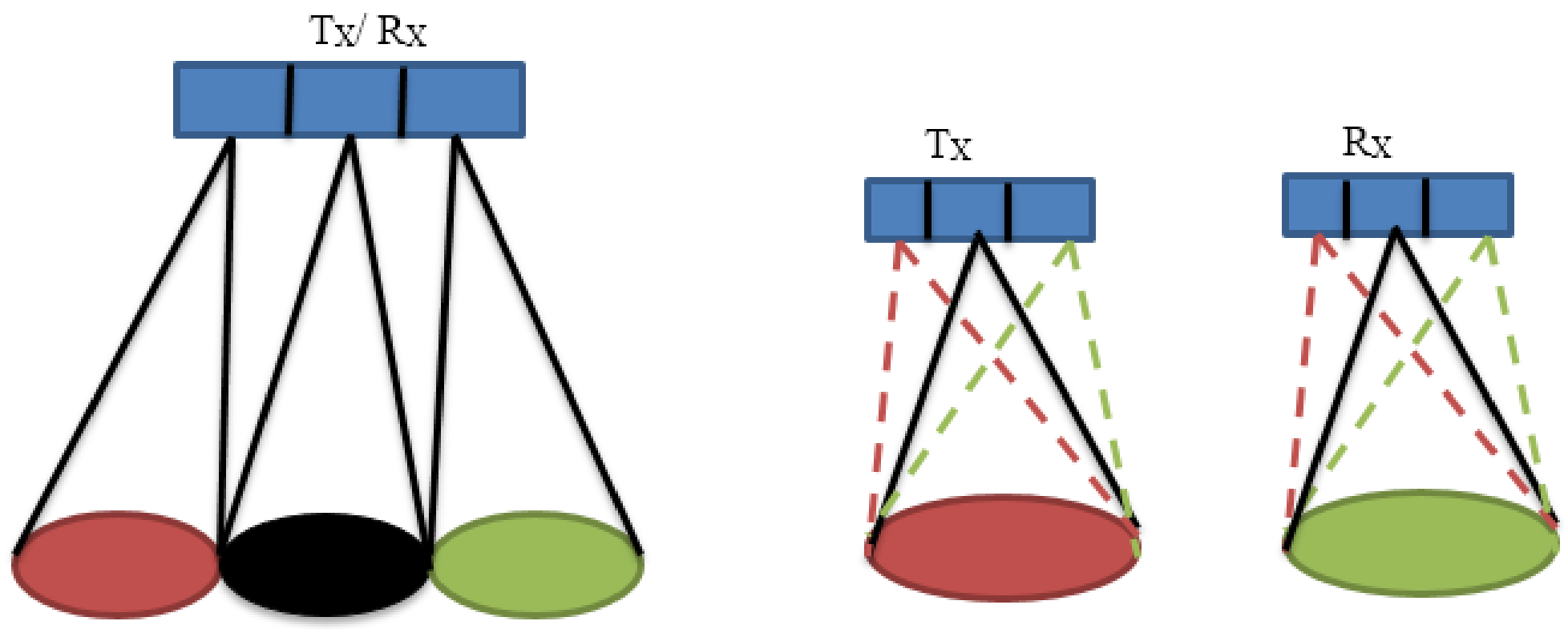

(a)SPCM (b)MPCM

Figure 3.1 Modes of MIMO radar antenna.

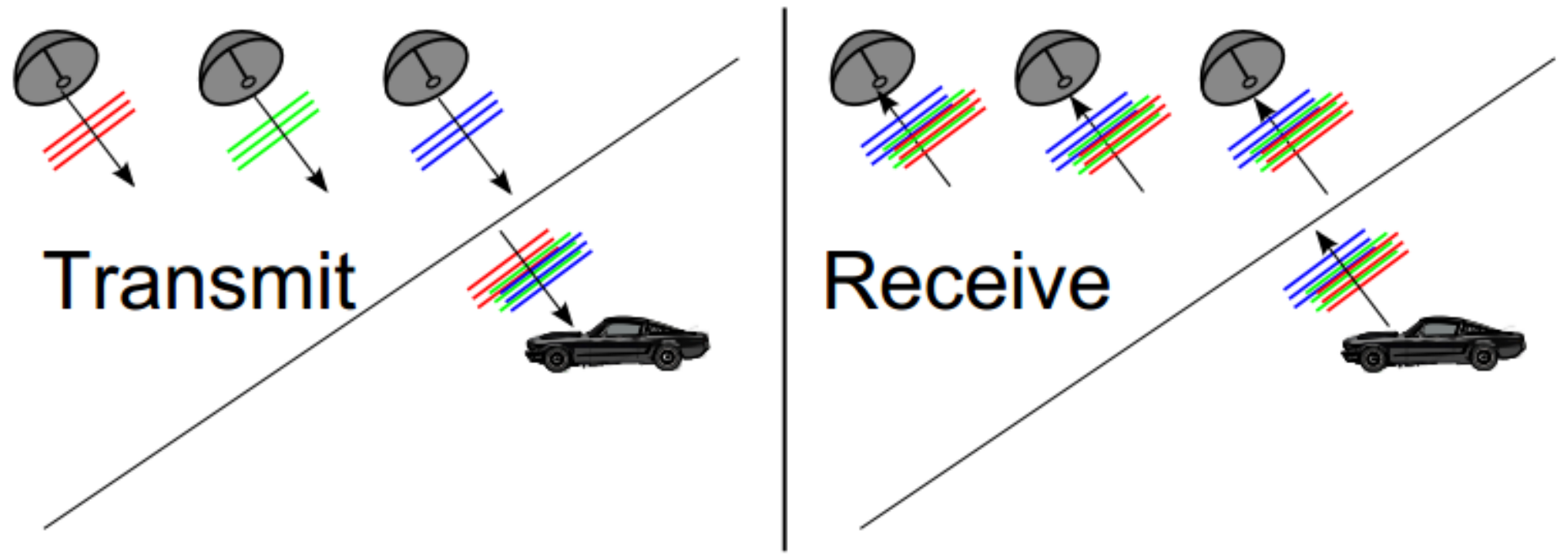

Figure 3.2 MIMO radar model [139].

On the other hand, the MIMO radar can be divided into two kinds: the first one is the radar with single-phase-centre-multibeam (SPCM) and the other one is the multiple-phase-centre-multibeam (MPCM) as shown in Figure 3.1. In the SPCM, the data are split/divided according to the angular position in the azimuth direction. This technique gives freedom to the sampling rate of each channel. In contrast, in the MPCM, the radar transmits broad multi-beam. This technique applies in the case of the requirement of broad beams.

Figure 3.2 shows the general case of the MIMO. In this case, we consider a MIMO radar that has M transmit antenna to transmit M orthogonal waveforms. The echoes signals will be received by N receive antenna. The antenna that used on the receive may or may not be the same antenna that used on transmit [138, 139].

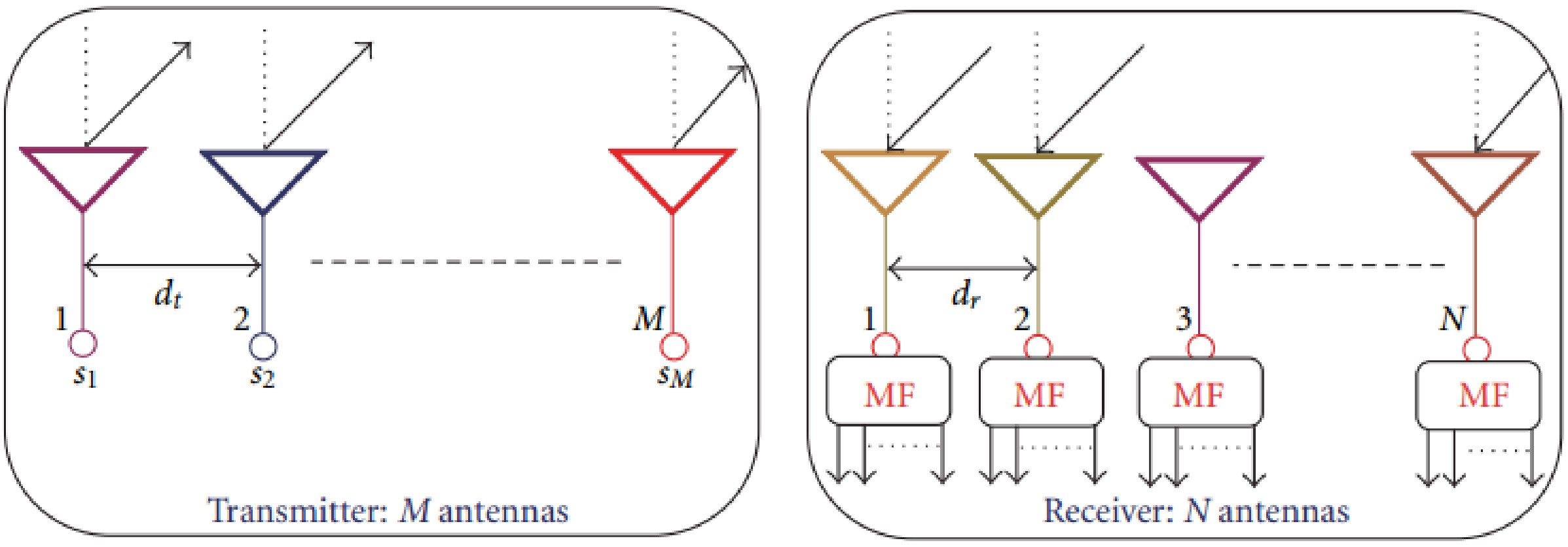

Figure 3.3 Example of MIMO system [138]

From Figure 3.3, the MIMO received signal at each receiving antenna is the weighted summation of all the transmitted waveforms.

$$S_{rn}(t) = \sum_{m=1}^{M} h_{n,m} S_{tm}(t) \quad for\ m = (1,2, \ldots \ldots, M)\ \ and\ n = (1, 2, \ldots \ldots . N) \tag{3.1}$$

Where$S_{rn}(t)$ is the received signal at antenna number (n), $h_{n,m}$ is the channel response/coefficient for channel number (n,m), and $S_{tm}(t)$ is the transmitted signal of antenna number (m). When the transmitted signals are orthogonal, the following relation should be achieved

$$\int S_{tm}(t) S_{tm'}(t)^* \, dt = \begin{cases} \delta(t), & m = m' \\ 0 & m \neq m' \end{cases} \tag{3.2}$$

The transmitted signals are designed to be orthogonal signals and these signals are extracted by M matched filter as shown in Figure 3.4 at each Rx antenna. So, the total number of extracted signals equal MN. The channel response is assumed to be unity (ideal channel) in all analysis in this chapter.

# 3.3 Linear Virtual Antenna Array

Consider the M -TX antennas and N- RX antennas that are parallel and collocated for each other, where the distance between the antennas is $d_{tm}, d_{rn}$ for the transmitter and the receiver, respectively. Where the mth TX antenna is located at $XT, m = d_{tm}$. Also, the nth RX antenna is located at $XR, n = d_{rn}$. Suggested that the proposed target in the far-field point, so we can represent the steering vectors of the transmitted signal $a(\theta_s)$ and the received signal $b(\theta_s)$ by:

$$a(\theta_s) = \left[e^{jkd_{t1}\sin\theta_s}, e^{jkd_{t2}\sin\theta_s}, \ldots \ldots \ldots, e^{jkd_{tM}\sin\theta_s}\right]^T \tag{3.3}$$

$$b(\theta_s) = \left[e^{jd_{r1}\sin\theta_s}, e^{jd_{r2}\sin\theta_s}, \dots\dots\dots, e^{jkd_{rN}\sin\theta_s}\right]^T \quad (3.4)$$

The MIMO received signal can be expressed as

$$S_r(t) = h\, S_t(t) a(\theta_s) b^T(\theta_s) \quad (3.5)$$

$$v(\theta_s) = a(\theta_s) b^T(\theta_s) \quad (3.6)$$

$$v(\theta_s) = \begin{bmatrix} e^{jkd_{t1}\sin\theta_s} \\ e^{jkd_{t2}\sin\theta_s} \\ \vdots \\ \vdots \\ e^{jkd_{tM}\sin\theta_s} \end{bmatrix}_{M\times 1} \left[e^{jkd_{r1}\sin\theta_s}, e^{jkd_{r2}\sin\theta_s}, \dots\dots\dots, e^{jkd_{rN}\sin\theta_s}\right]_{1\times N} \quad (3.7)$$

$$v(\theta_s) = \begin{bmatrix} e^{jk(d_{t1}+d_{r1})\sin\theta_s} & e^{jk(d_{t1}+d_{r2})\sin\theta_s} & \dots & e^{jk(d_{t1}+d_{rN})\sin\theta_s} \\ e^{jk(d_{t2}+d_{r1})\sin\theta_s} & e^{jk(d_{r2}+d_{t2})\sin\theta_s} & \dots & e^{jk(d_{t2}+d_{rN})\sin\theta_s} \\ \vdots & \vdots & \dots & \vdots \\ e^{jk(d_{tM}+d_{r1})\sin\theta_s} & e^{jk(d_{tM}+d_{r2})\sin\theta_s} & \dots & e^{jk(d_{tM}+d_{N})\sin\theta_s} \end{bmatrix}_{M\times N} \quad (3.8)$$

In other words, the steering vector can be expressed as the Kronecker product between the two steering vectors $v(\theta_s) = a(\theta_s)\otimes b(\theta_s)$, where $\otimes$ denotes the Kronecker Product [139].
In this example, we introduce an example of Kronecker Product:
Assume two matrices A, B with any dimensions

$$\mathrm{A} = \begin{bmatrix} a_{11} & \cdots & a_{1n} \\ \vdots & \ddots & \vdots \\ a_{n1} & \cdots & a_{nn} \end{bmatrix}, B = \begin{bmatrix} b_{11} & \cdots & b_{1n} \\ \vdots & \ddots & \vdots \\ b_{n1} & \cdots & b_{nn} \end{bmatrix}$$

$$A\otimes B = \begin{bmatrix} a_{11}B & \cdots & a_{1n}B \\ \vdots & \ddots & \vdots \\ a_{n1}B & \cdots & a_{nn}B \end{bmatrix} \quad (3.9)$$

To study different cases of virtual MIMO, we assume that $d_t = \beta d_r$

### 3.3.1 Equal Distance

- **Case 1: Uniform Array**

We assume that the antenna is linear uniform with equal distances and an equal number of elements between TX antennas and Rx antennas. So, in this case, $\beta = 1$, $M = N = L$, $d_t = d_r = d$. We assume that the first element is the reference element at origin for TX and RX to solve this case. The steering vector of the TX is the same steering vector of the RX and can be expressed as:

$$a(\theta_s) = b(\theta_s) = \left[1, e^{jkd\sin\theta_s}, e^{j2kd\sin\theta_s}, \cdots, e^{j(L-1)kd\sin\theta_s}\right]^T \quad (3.10)$$

$$v(\theta_s) = \begin{bmatrix} 1 & e^{jkd\sin\theta_s} & \cdots & e^{j(L-1)kd\sin\theta_s} \\ e^{jkd\sin\theta_s} & e^{j2kd\sin\theta_s} & \cdots & e^{jLkd\sin\theta_s} \\ \vdots & \vdots & \cdots & \vdots \\ e^{j(L-1)kd\sin\theta_s} & e^{j(L)kd\sin\theta_s} & \cdots & e^{j(2L-2)kd\sin\theta_s} \end{bmatrix}_{L\times L} \quad (3.11)$$

From equation (3.11), we noted that the number of effective virtual phase centers (EVFC) =2L-1. The following expression in (3.12) introduces the length/number of EVPC.

$$\begin{array}{ccccccccc|c} \mathbf{1} & \mathbf{2} & \ldots & \mathbf{L-1} & \mathbf{L} & \mathbf{L-1} & \ldots & \mathbf{2} & \mathbf{1} & \boldsymbol{L=M} \\ & \mathbf{1} & & \ldots & \vdots & \ldots & & \mathbf{1} & & \boldsymbol{L=M-1} \\ & & \ddots & \ldots & \vdots & \ldots & ⋰ & & & \vdots \\ & & & \ddots & \vdots & ⋰ & & & & \vdots \\ & & \mathbf{1} & \mathbf{2} & \mathbf{3} & \mathbf{2} & \mathbf{1} & & & \boldsymbol{L=3} \\ & & & \mathbf{1} & \mathbf{2} & \mathbf{1} & & & & \boldsymbol{L=2} \\ & & & & \mathbf{1} & & & & & \boldsymbol{L=1} \end{array} \quad (3.12)$$

For more investigation, we introduce this example with the assumption that $M = N = 4$. We assume that the positions of the transmitter, receiver, virtual antenna array are $X_t$, $X_r$, $X_v$, respectively. $X_t = [1111]$, $X_r = [1111]$ are the position of Tx and Rx elements, can be considered as a function of distance as $X_t = X_r = [0, d, 2d, 3d]$. So,

$$v(\theta_s) = \begin{bmatrix} 1 & e^{jkd\sin\theta_s} & e^{j2kd\sin\theta_s} & e^{j3kd\sin\theta_s} \\ e^{jkd\sin\theta_s} & e^{j2kd\sin\theta_s} & e^{j3kd\sin\theta_s} & e^{j4kd\sin\theta_s} \\ e^{j2kd\sin\theta_s} & e^{j3kd\sin\theta_s} & e^{j4kd\sin\theta_s} & e^{j5kd\sin\theta_s} \\ e^{j3kd\sin\theta_s} & e^{j4kd\sin\theta_s} & e^{j5kd\sin\theta_s} & e^{j6kd\sin\theta_s} \end{bmatrix}_{4\times 4} \quad (3.13)$$

$$= [1, 2e^{jkd\sin\theta_s}, 3e^{j2kd\sin\theta_s}, 4e^{jkd\sin\theta_s}, 3e^{jkd\sin\theta_s}, 2e^{jkd\sin\theta_s}, e^{jkd\sin\theta_s}]$$

So, the virtual vector can be expressed as $X_v = [1, 2, 3, 4, 3, 2, 1]$. Figure 3.5 depicts the corresponding EVPC of this example. We note that the number of virtual elements=7.

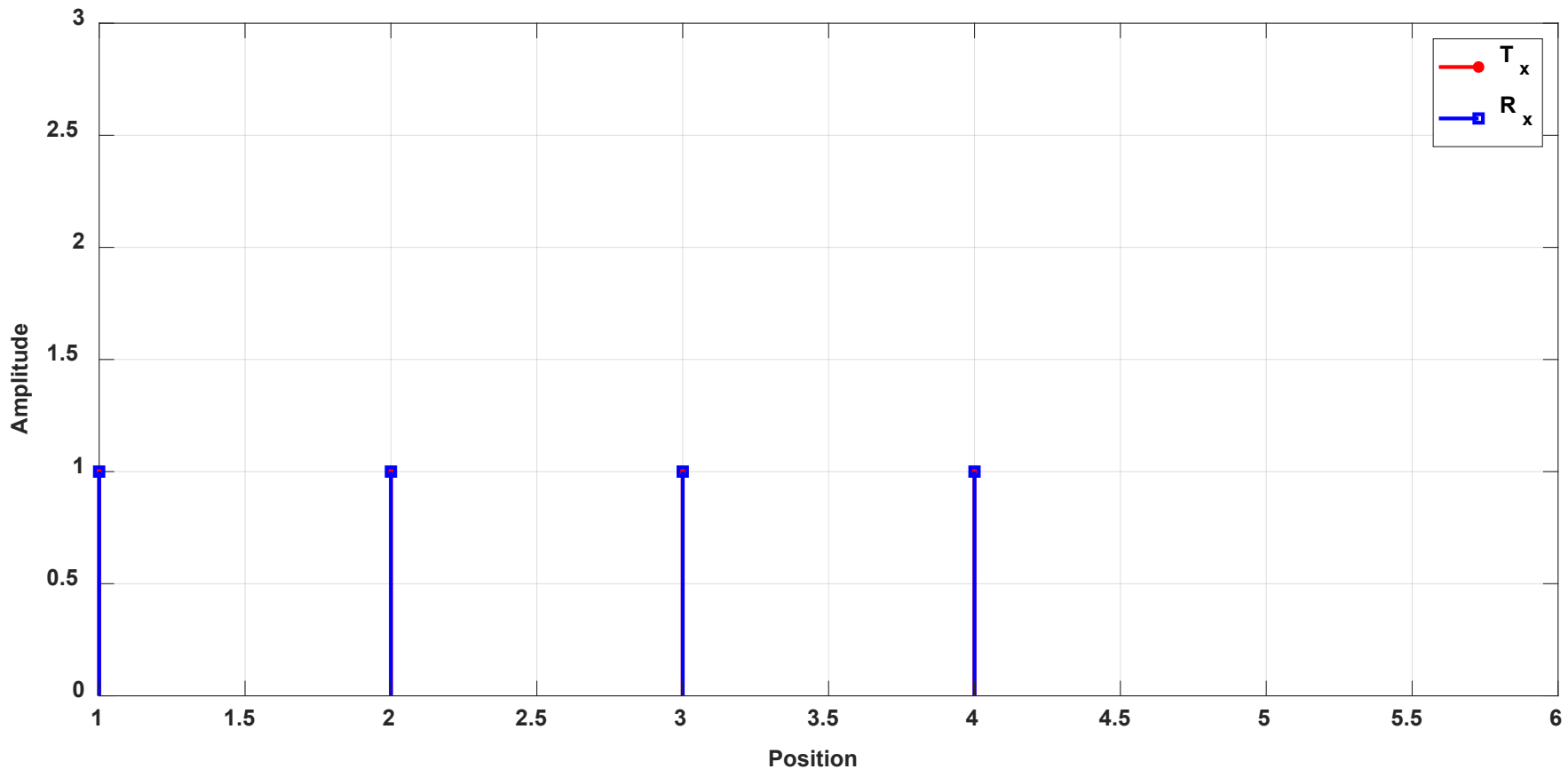

(a)Positions of TX and RX elements

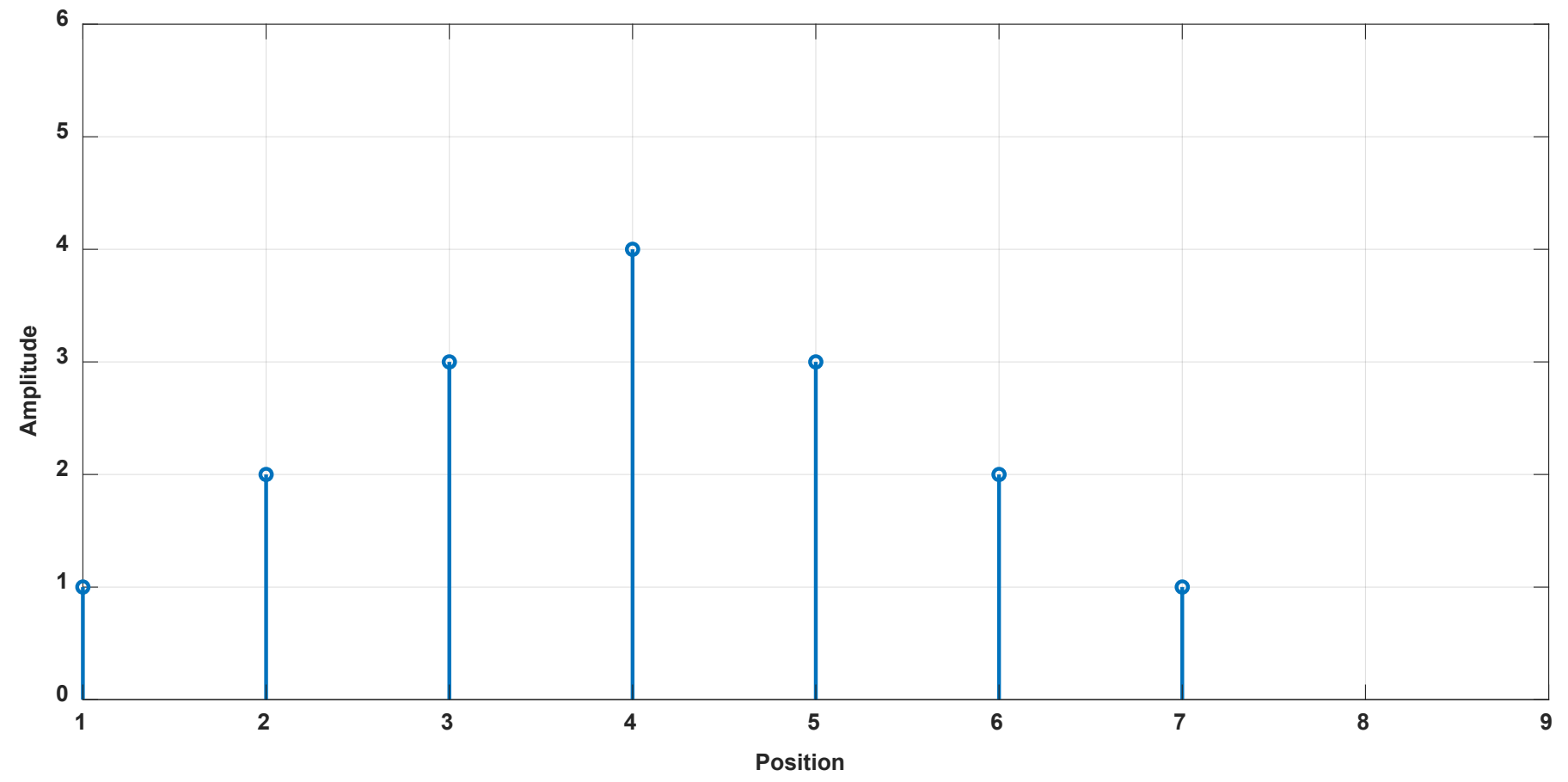

(b)EVPC

Figure 3.4 Virtual phase center of uniform array (M=4, N=4)

- **Case 2: Non-uniform array**

We assume that the antenna is linear uniform with an equal number of elements between Tx antennas and Rx antennas. So, in this case, $\beta = 1$, $M = N = L$, $d_{tm} = d_{rn} = d_m$. The steering vector of the TX is the same steering vector of the RX and can be expressed as:

$$a(\theta_s) = b(\theta_s) = \left[e^{jkd_1 \sin\theta_s}, e^{jkd_2 \sin\theta_s}, \cdots, e^{jkd_M \sin\theta_s}\right]^T \tag{3.14}$$

$$v(\theta_s) = \begin{bmatrix} e^{jk(2d_1)\sin\theta_s} & e^{jk(d_1+d_2)\sin\theta_s} & \cdots & e^{jk(d_1+d_M)\sin\theta_s} \\ e^{jk(d_2+d_1)\sin\theta_s} & e^{jk(d_2+d_2)\sin\theta_s} & \cdots & e^{jk(d_2+d_M)\sin\theta_s} \\ \vdots & \vdots & \cdots & \vdots \\ e^{jk(d_M+d_1)\sin\theta_s} & e^{jk(d_M+d_2)\sin\theta_s} & \cdots & e^{jk(2d_M)\sin\theta_s} \end{bmatrix} \tag{3.15}$$

It can be detected that the maximum number of virtual phase that can be achieved in the case of non-uniform array $L_v = L(L+1)/2$. As example suppose that M=4, N=4. In this case we assume that the position of transmitter and receiver are$X_t$, $X_r$, respectively. $X_t$=[0 1 0 1 0 0 1 0 0 0 1], $X_r$=[0 1 0 1 0 0 1 0 0 0 1], the position of Tx and Rx elements can be considered as $X_t = X_r = [d, 3d, 6d, 10d]$

$$v(\theta_s) = \begin{bmatrix} e^{jk(2d)\sin\theta_s} & e^{jk(4d)\sin\theta_s} & e^{jk(7d)\sin\theta_s} & e^{jk(11d)\sin\theta_s} \\ e^{jk(4d)\sin\theta_s} & e^{jk(6d)\sin\theta_s} & e^{jk(9d)\sin\theta_s} & e^{jk(13d)\sin\theta_s} \\ e^{jk(7d)\sin\theta_s} & e^{jk(9d)\sin\theta_s} & e^{jk(12d)\sin\theta_s} & e^{jk(16d)\sin\theta_s} \\ e^{jk(11d)\sin\theta_s} & e^{jk(13d)\sin\theta_s} & e^{jk(16d)\sin\theta_s} & e^{jk(20d)\sin\theta_s} \end{bmatrix} \tag{3.16}$$

$$= \left[\begin{matrix} e^{jk(2d)\sin\theta_s}, 2e^{jk(4d)\sin\theta_s}, 1e^{jk(6d)\sin\theta_s}, 2e^{jk(7d)\sin\theta_s}, 2e^{jk(9d)\sin\theta_s}, 2e^{jk(11d)\sin\theta_s}, \\ e^{jk(12d)\sin\theta_s}, e^{jk(13d)\sin\theta_s}, e^{jk(14d)\sin\theta_s}, 2e^{jk(16d)\sin\theta_s}, e^{jk(20d)\sin\theta_s} \end{matrix}\right]$$

The position of virtual elements can be expressed as $X_v$ = [0 0 1 0 2 0 1 2 0 2 0 2 1 1 0 0 2 0 0 0 1], So, $L_v = 10$. Figure 3.5 (a) shows the positions of TX and RX antennas

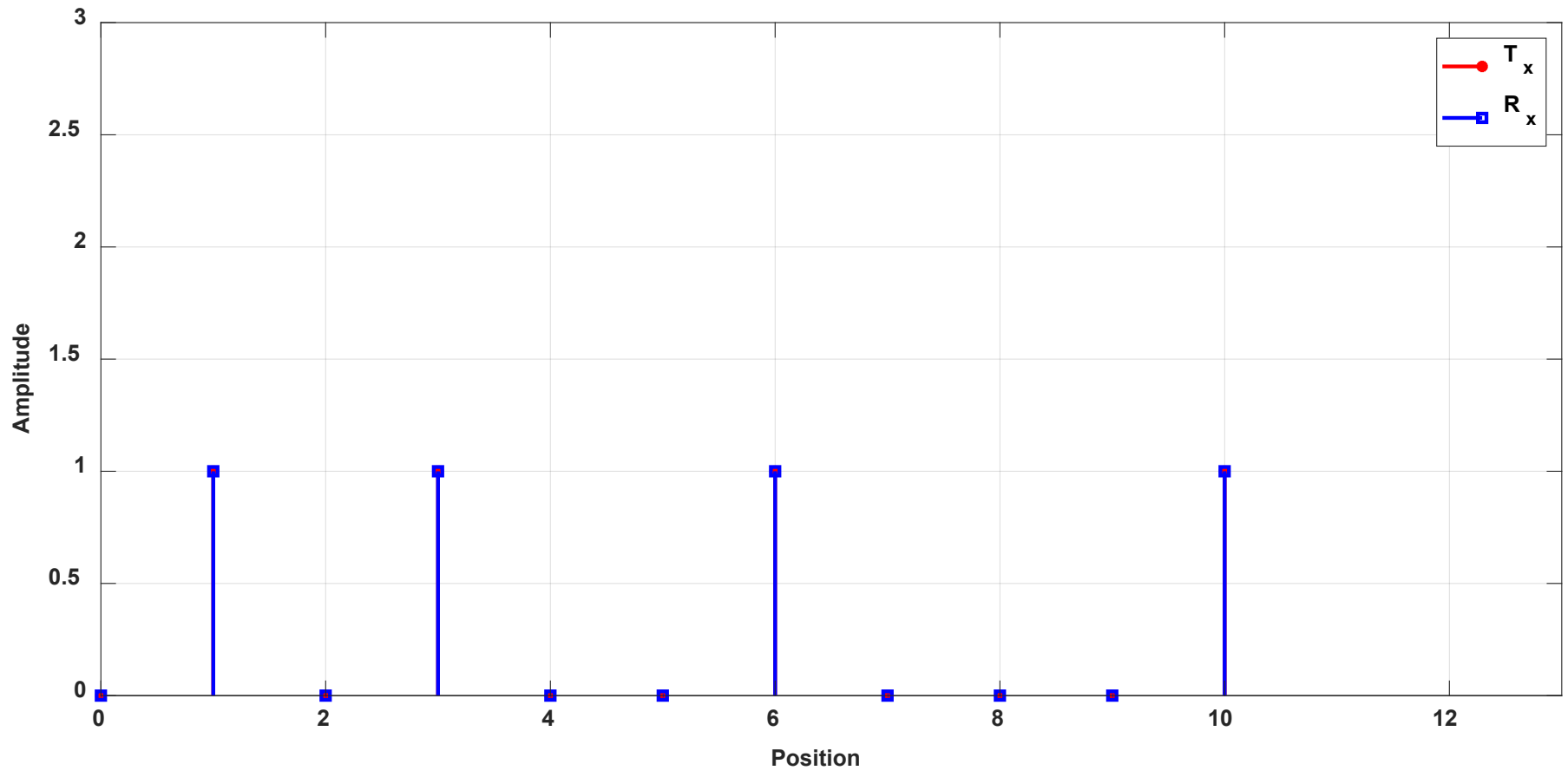

(a)positions of TX and RX antennas

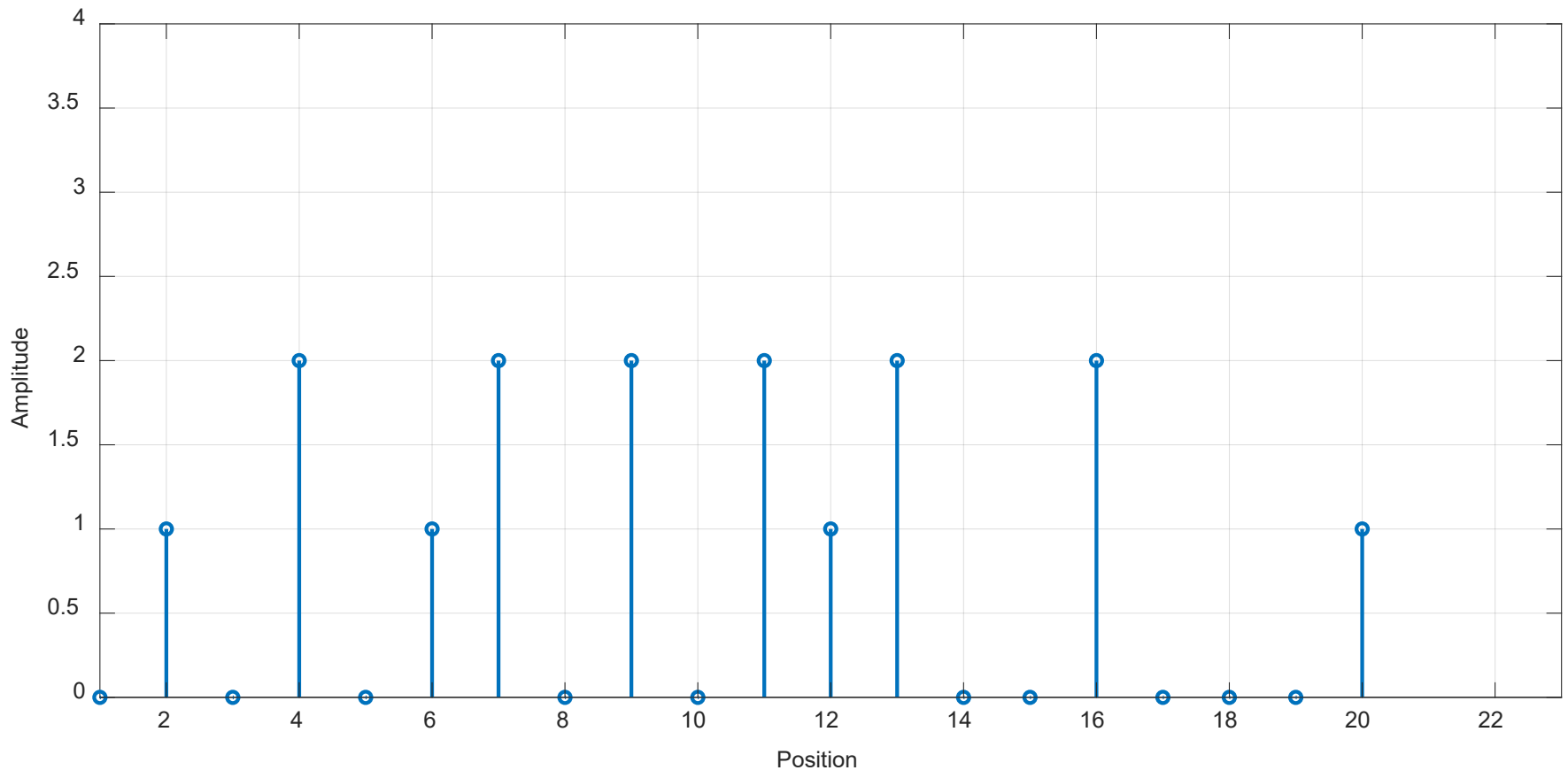

(b) EVPC

Figure 3.5 virtual phase center of non-uniform elements (M=N)

## 3.3.2 Unequal Distance

In this section we focus on case of $\beta = N$;

We assume that, the transmitter and receiver are overlapped and start at the same point. $\beta = N$, number of the transmitter antennas equal $M$, and the number of receiver antennas equal N. The distance between receiver elements is denoted by d and distance between transmitter elements denotes by Nd.

$$a(\theta_s) = \left[1, e^{jNkd\sin\theta_s}, \dots \dots \dots, e^{j(M-1)Nkd\sin\theta_s}\right]^T \tag{3.17}$$

$$b(\theta_s) = \left[1, e^{j2kd\sin\theta_s}, \dots \dots \dots, e^{j(N-1)kd\sin\theta_s}\right]^T \tag{3.18}$$

$$v(\theta_s) = \begin{bmatrix} 1 & e^{jdk\sin\theta_s} & \cdots & e^{j(N-1)kd\sin\theta_s} \\ e^{jNkd\sin\theta_s} & e^{j(N+1)kd\sin\theta_s} & \cdots & e^{j(2N-1)kd\sin\theta_s} \\ \vdots & \vdots & \vdots & \vdots \\ e^{j((M-1)N)kd\sin\theta_s} & e^{j((M-1)N+1)kd\sin\theta_s} & \cdots & e^{j(NM-1)kd\sin\theta_s} \end{bmatrix}_{M\times N} \tag{3.19}$$

As an example, we assume that M=3, N=4, the positions of transmitter and receiver are $X_t$, $X_r$ respectively. $X_t$=[1 0 0 0 1 0 0 0 1 0 0 0 1], $X_r$=[1 1 1 1], the positions of Tx and Rx elements can be consider as $X_t = [0, 4d, 8d]$, $X_r = [0, d, d, 3d]$ with the first element as reference in Tx and Rx.

$$v(\theta_s) = \begin{bmatrix} 1 & e^{jkd\sin\theta_s} & e^{j2kd\sin\theta_s} & e^{j3kd\sin\theta_s} \\ e^{j4kd\sin\theta_s} & e^{j5kd\sin\theta_s} & e^{j6kd\sin\theta_s} & e^{j7kd\sin\theta_s} \\ e^{j8kd\sin\theta_s} & e^{j9kd\sin\theta_s} & e^{j10kd\sin\theta_s} & e^{j11kd\sin\theta_s} \end{bmatrix} \tag{3.20}$$

So, the positon's vector of the virtual elements can be expressed as $X_v = [1\ 1\ 1\ 1\ 1\ 1\ 1\ 1\ 1\ 1\ 1\ 1]$

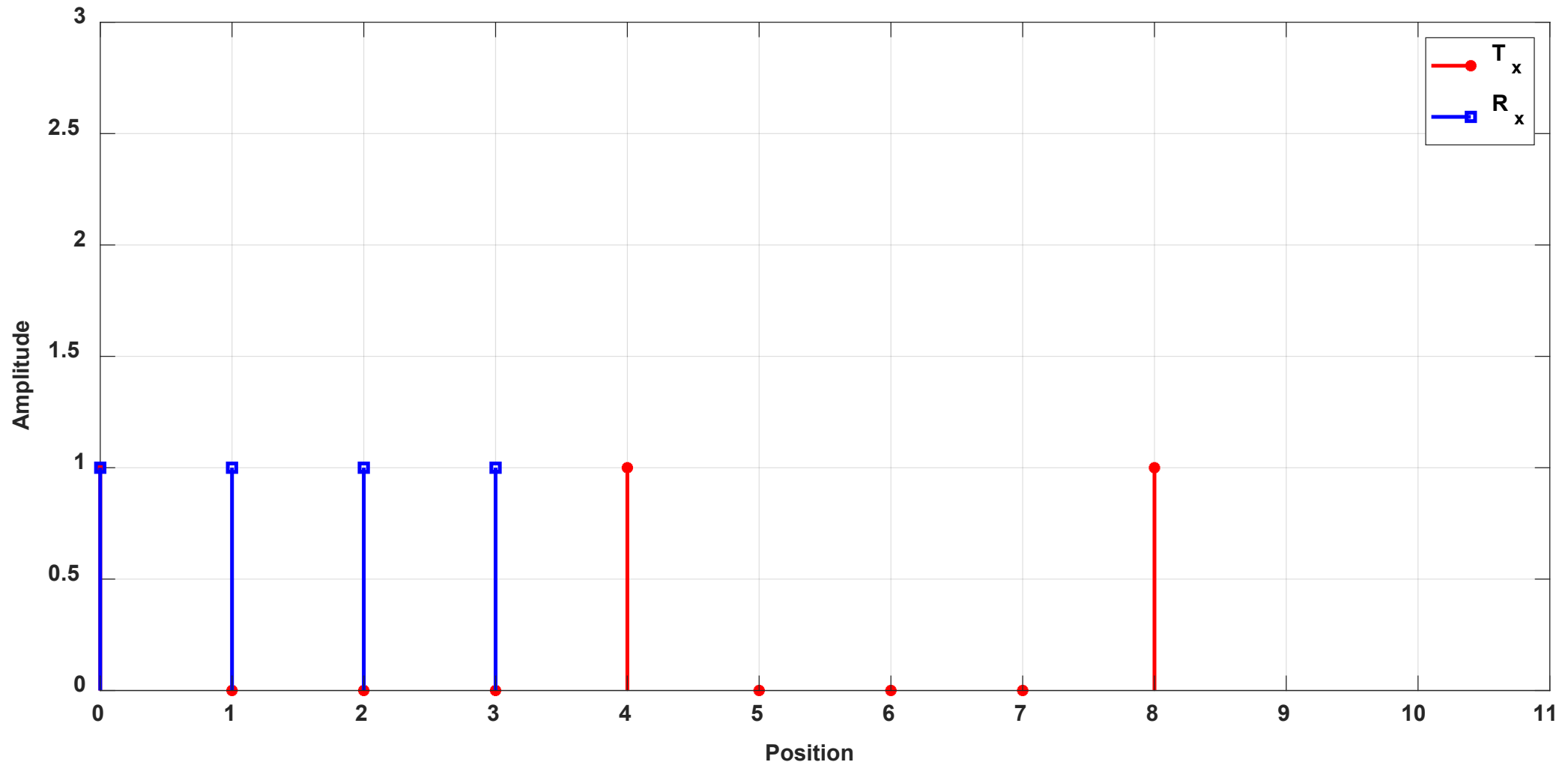

(a)

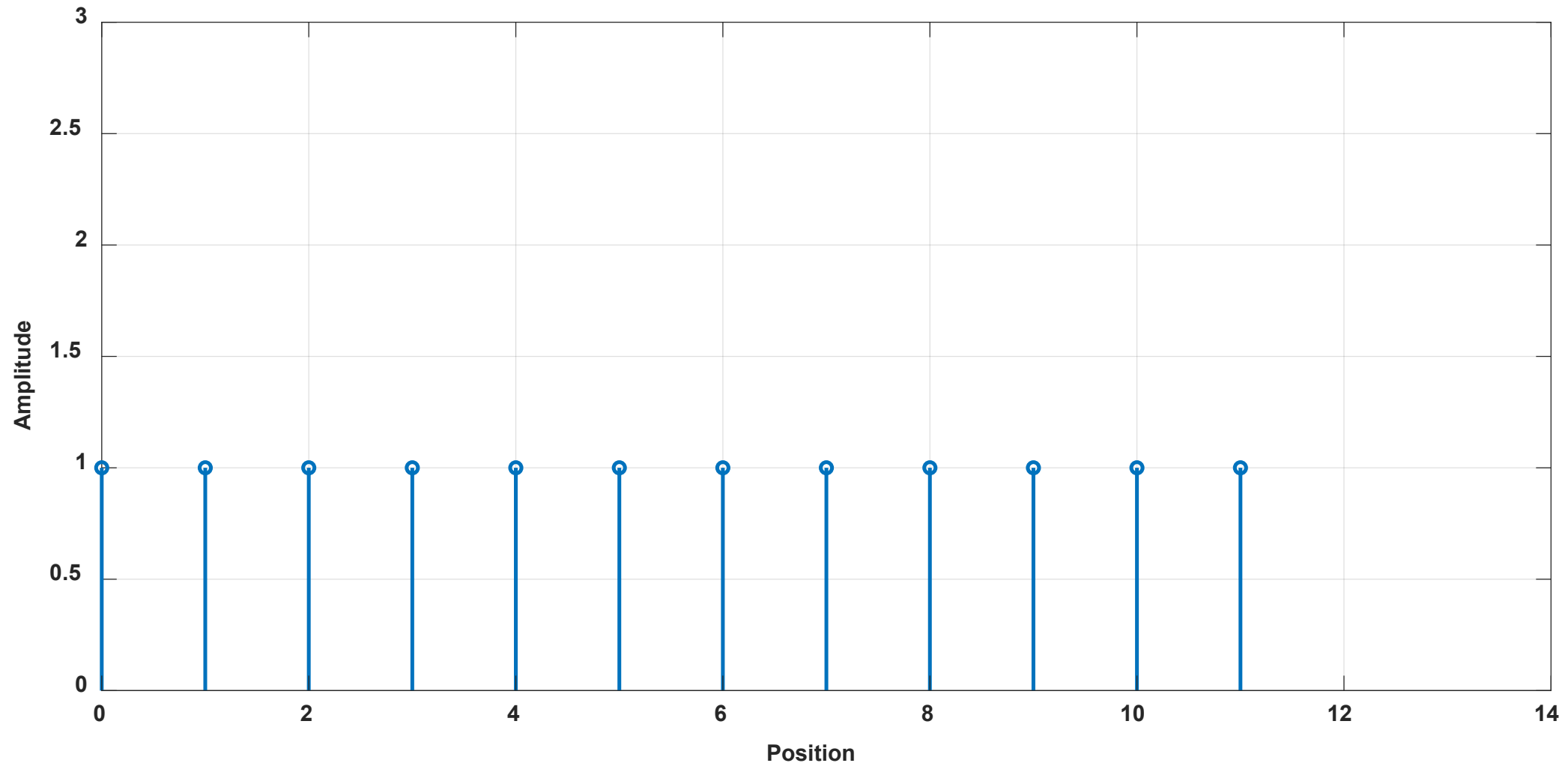

(b)

Figure 3.6 EVPC of M=3, N=4, $\boldsymbol{\beta} = \mathbf{4}$ overlapped

The same number of EVPC can be achieved with no-overlapped as shown in Figure 3.7.

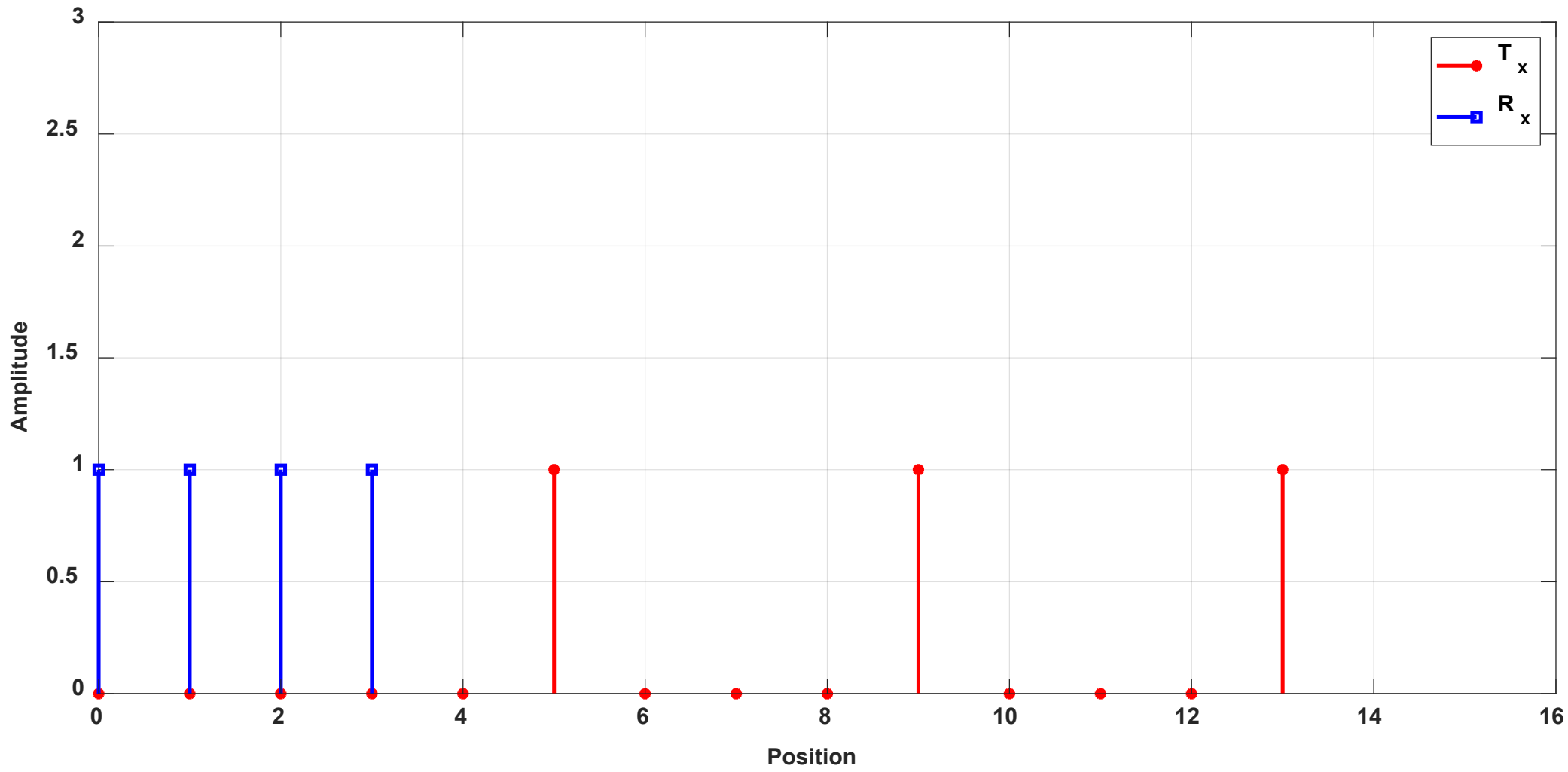

(a)

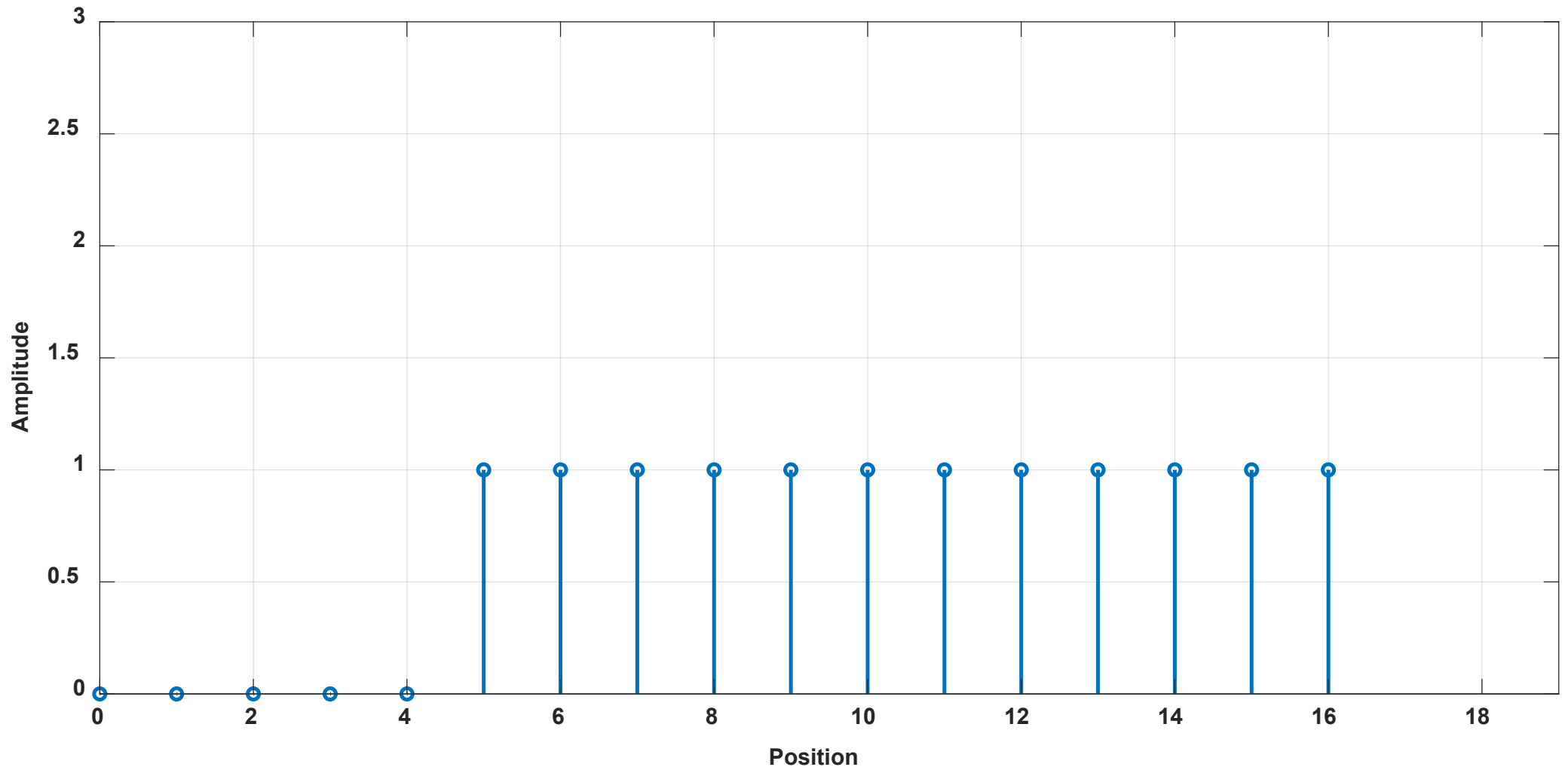

(b)

Figure 3.7 EVPC of M=3, N=4, $\boldsymbol{\beta} = \mathbf{4}$ No-overlapped

# 3.3.3 ULA Vs. VAA

In this case study, the comparison between conventional antenna array and Virtual array is introduced. We assume that we have a uniform linear array (ULA). The antenna array consists of 8 isotropic elements on Z-axis. The array factor can be considered as:

$$AF_C = 1 + e^{-jkdcos(\theta)} + e^{-j2kdcos(\theta)} + e^{-j3kdcos(\theta)} + e^{-j4kdcos(\theta)} + e^{-j5kdcos(\theta)} + e^{-j6kdcos(\theta)} + e^{-j7kdcos(\theta)} \tag{3.21}$$

The Virtual MIMO (VMIMO) provides the same effect of 8 elements with only 6 elements with the following specification $L_v = MN$, $d_t = Nd_r$, where $L_v$ is the total number of elements in case of VMIMO. The virtual array factor can be expressed as

$$AF_V = \left(1 + e^{-j4kdcos(\theta)}\right)\left(1 + e^{-jkdcos(\theta)} + e^{-j2kdcos(\theta)} + e^{-j3kdcos(\theta)}\right) \qquad (3.22)$$

$AF_C = AF_V$ If and only if $d_t = Nd_r$

Figure 3.8 shows the comparison between the array factor of uniform linear array (ULA) and VMIMO. We noted that the AF of the ULA and VMIMO are the same under the previous conditions.

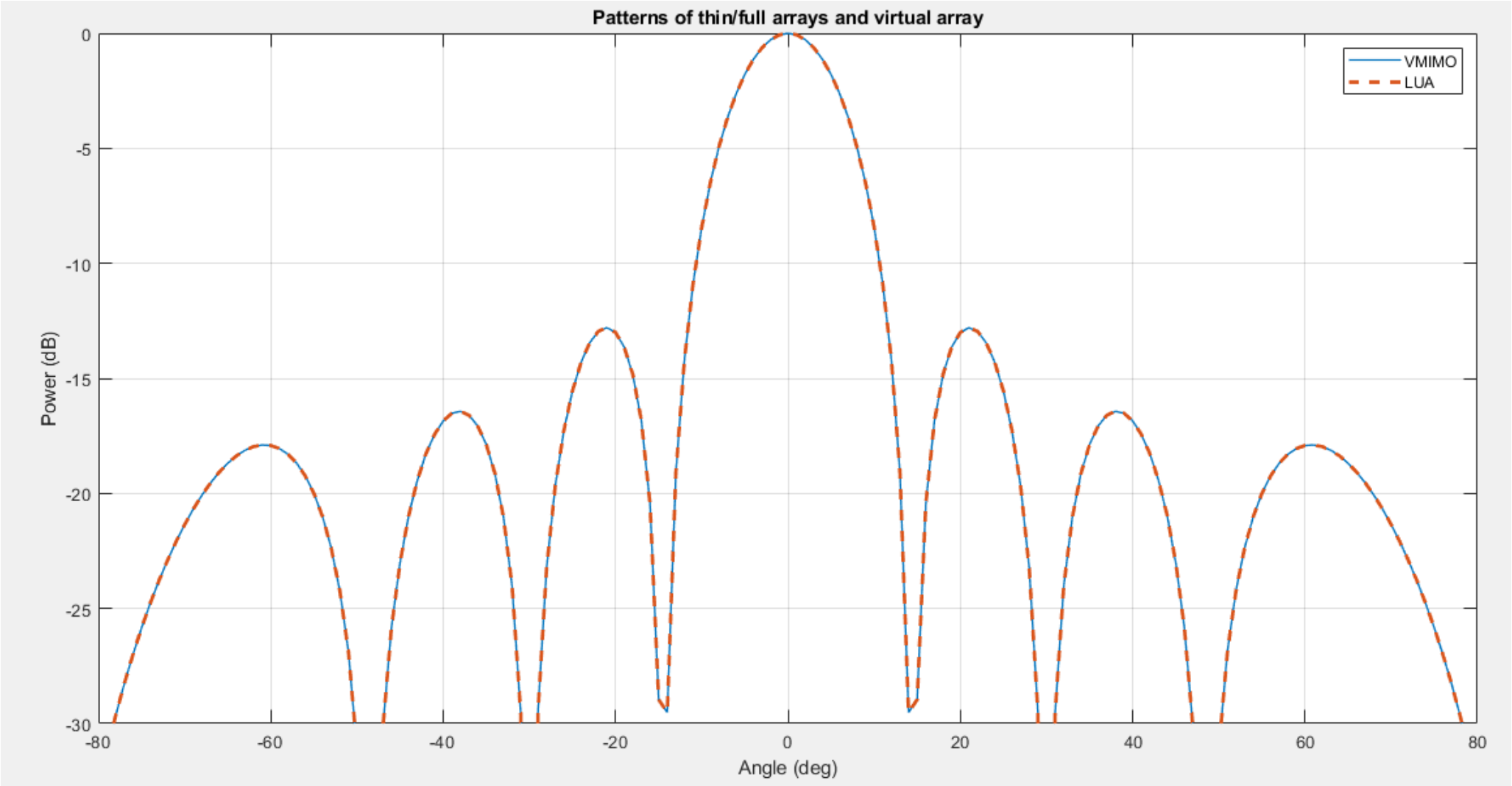

Figure 3.8 Array factor of linear uniform array and VMIMO

The relation between the number of antennas of LAA and VMIMO can be described as following:

$N * M|_{ULA} = N + M|_{VMIMO}$ If and only If $d_t = Nd_r$ in VMIMO case (3.23)

# 3.4 Virtual Antenna Array Design

The basic concept of VAA is introduced in [137-141]. Suppose that an M element transmitting linear array distributed along the x-direction and an N element receiving linear antenna array distributed along the y-direction. This distribution constitutes the basic building elements of the virtual antenna array. The concept of a virtual antenna array is illustrated in Figure 3.9. It appears clearly in radar applications where the received signal depends mainly on the multiplication of the radiation patterns for the transmitting and receiving antennas. As will be illustrated, the radiation pattern of the planar antenna shown in the right-hand side is equivalent to the multiplication of the radiation patterns of the two linear antenna arrays in the left-hand side of Figure 3.9.

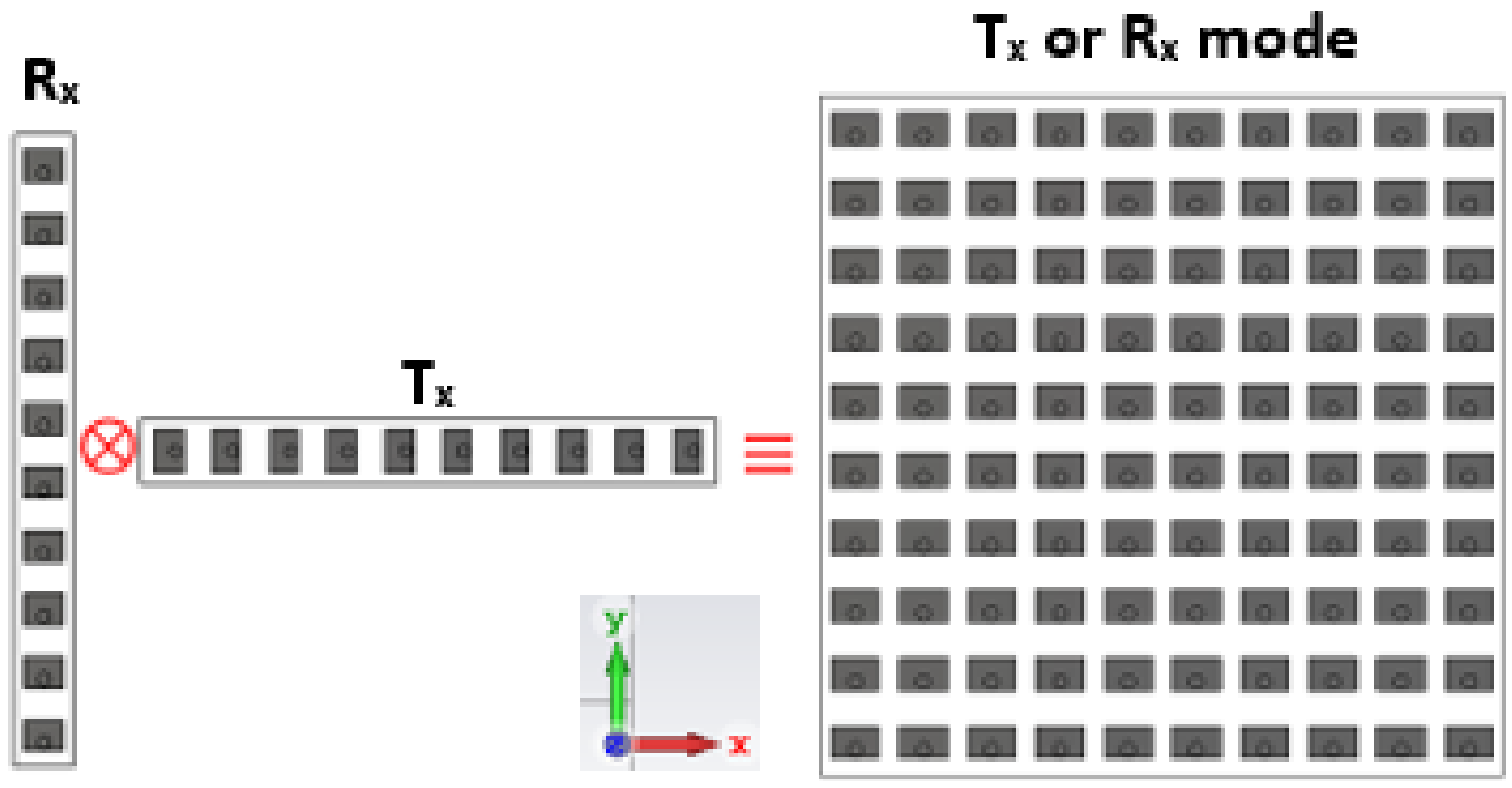

Figure 3.9 Virtual Antenna array concept

The array factor of the transmitter linear antenna array is calculated according to equation (3.24):

$$AF_T(\theta,\emptyset) = \sum_{m=1}^{M} a_m e^{jkd_m \sin\theta \cos\emptyset} \tag{3.24}$$

Where $a_m$ is the m[th] complex excitation coefficient. The gain of the transmitting antenna is proportional to the square of the array factor such that

$$G_T(\theta,\emptyset) = \alpha_T \left|\widehat{AF}_T(\theta,\emptyset)\right|^2 \tag{3.25}$$

Where $\widehat{AF}_T(\theta,\emptyset)$ is the normalized transmitter array factor and similarly, the receiver array factor is written as:

$$AF_R(\theta,\emptyset) = \sum_{n=1}^{N} a_n e^{jkd_n \sin\theta \sin\emptyset} \tag{3.26}$$

And the receiver gain is also written as

$$G_R(\theta,\emptyset) = \alpha_R \left|\widehat{AF}_R(\theta,\emptyset)\right|^2 \tag{3.27}$$

$\widehat{AF}_R(\theta,\emptyset)$ is the normalized receiving array factor.

Suppose that an object of radar cross section $\sigma$ is positioned in front of the transceiver system then the receiving power is calculated according to [49] so that:

$$P_{rV} = \frac{P_t G_R(\theta,\emptyset) G_T(\theta,\emptyset) \sigma \lambda^2}{(4\pi)^3 R^4} h(\theta,\emptyset) \tag{3.28}$$

Where $P_r$: Received Power in watts, $P_t$:Peak transmitted power in watts, $G_T$: Transmitter Gain, $G_R$: Receiver Gain, $\lambda$: Wavelength (m), $\sigma$: RCS of the target (m$^2$), R: Range between radar and target (m) and $h(\theta,\emptyset)$ is the channel response/coefficient for the wave impinging from the transmitter and reflected from the scattered under investigation and then back to the receiving point.

In case of normal radar operation where one antenna is utilized for both transmission and reception, the used antenna may be a planar array of MxN elements of array factor equals:

$$AF_P(\theta,\emptyset) = \sum_{m=1}^{M} a_m e^{jkd_m \sin\theta \cos\emptyset} \otimes \sum_{n=1}^{N} a_n e^{jkd_n \sin\theta \sin\emptyset} \tag{3.29}$$

Where ⊗ denotes to the Kronecker Product [142]. From equations (3.24), (3.26) and (3.29), it is noted that the array factor of the planar array in either receiving or transmitting modes equals the multiplication of the array factor of the transmitter and the array factor of the receiver of the virtual array. So that, the planar array factor is written as,

$$AF_P(\theta,\emptyset) = AF_T(\theta,\emptyset) \otimes AF_R(\theta,\emptyset) \tag{3.30}$$

The gain of the array factor equals

$$G_P(\theta,\emptyset) = \alpha_P\left|\widehat{AF}_P(\theta,\emptyset)\right|^2 \tag{3.31}$$

Where $\widehat{AF}_P$ is the normalized array factor of the planar array.

By noticing equation (3.25), (3.27), (3.29) and (3.31), the gain of the planar antenna array could be written in terms of the gain of the virtual array transceiver as follows

$$G_P(\theta,\emptyset) = G_T(\theta,\emptyset) \otimes G_R(\theta,\emptyset) \tag{3.32}$$

In case of the planar array, the received power is calculated as follows,

$$P_{rP} = \frac{P_t G_P^2(\theta,\emptyset)\sigma\lambda^2}{(4\pi)^3 R^4} h(\theta,\emptyset) \tag{3.33}$$

Comparing equation (3.28) with equation (3.33) taking into consideration equation (3.32) the following relation is held;

$$P_{rP} = G_P(\theta,\emptyset) \otimes P_{rv} \tag{3.34}$$

This means that the received power using the planar antenna array is greater than that of the virtual array by a factor of $G_P(\theta,\emptyset)$. Then what is the benefit of the virtual array? The reply to this question is introduced in the following section.

The second step, is to apply the concept of VAA to a two-dimensional antenna array that is created by placing two orthogonal linear antenna arrays (LAA) to each other. Each LAA consists of 10 rectangular patch microstrip antennas as shown in Figure 3.9. The number of elements in two LAA are the same to achieve the same angular resolution in x and y planes. The distance between the elements in X, and Y directions is half air wavelength from operating frequency to avoid the grating lobe (d=6 mm). At the same time we consider the PAA that consists of 100 elements. The patch is designed to resonate at 24 GHz on Rogers RO4003 substrate with dielectric constant 3.38 and thickness 0.2 mm. The patch width W=4.3 mm and length L=3.3 mm. The patch is designed to cover the band from 23.55 GHz to 24.7 GHz for short range automotive radar as shown in Figure 3.10.

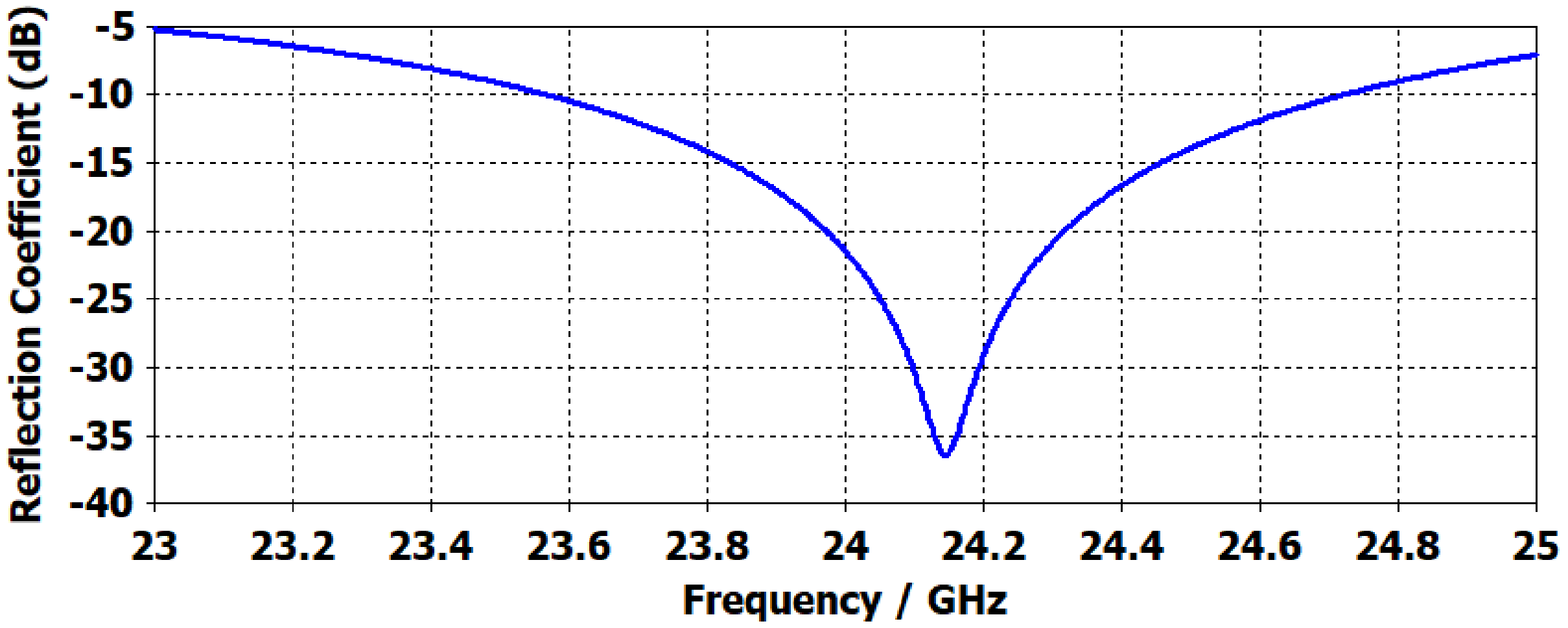

Figure 3.10 Reflection coefficient of the rectangular patch antenna, $\varepsilon_r = 3.38, h = 0.2\ mm$.

We introduce three different cases to compare between the PAA and VAA as following:

- **Case I:**

We consider the PAA that consists of 100 elements in TX mode as compared with the VAA that consist of M=10, N=10 (only 20 elements) orthogonal as shown in Figure 3.9. We notice that the radiation patterns are close together for the main lobe and the sidelobe and the difference in the back lobe is due to the mutual coupling effect in case of PAA that isn't present in the VAA, as shown in Figure 3.11. The comparisons are introduced at different angles ($\varphi = 0^0, 30^0, 45^0, 60^0, and\ 90^0$). So, we can summarized that the VAA with only 20 elements introduce the same radiation pattern of PAA with 100 elements and provide enhancement in the back lobe.

| **Phi** | **Case I** |
|---|---|
| $0^0$ | 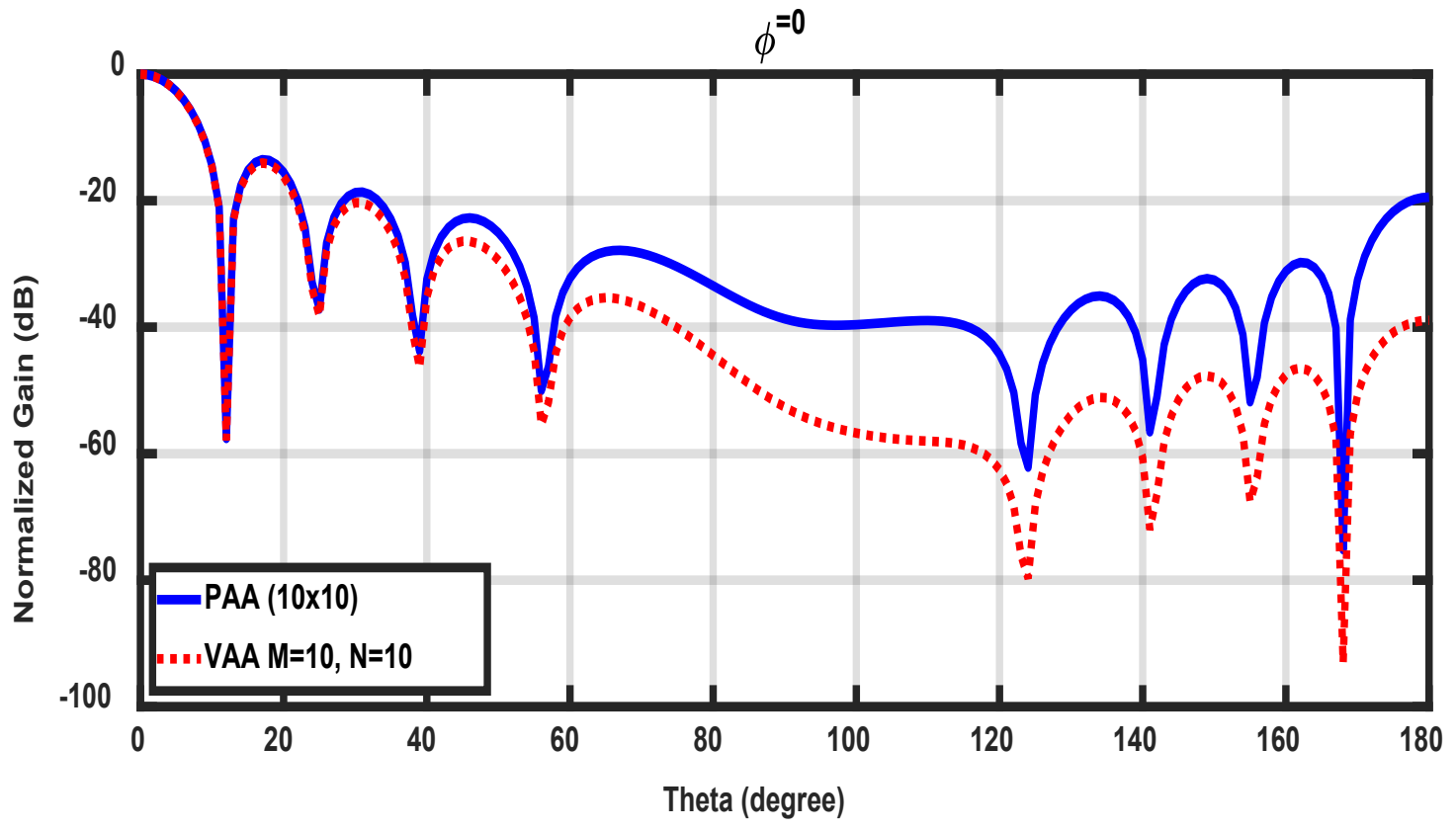 |

**$30^0$**

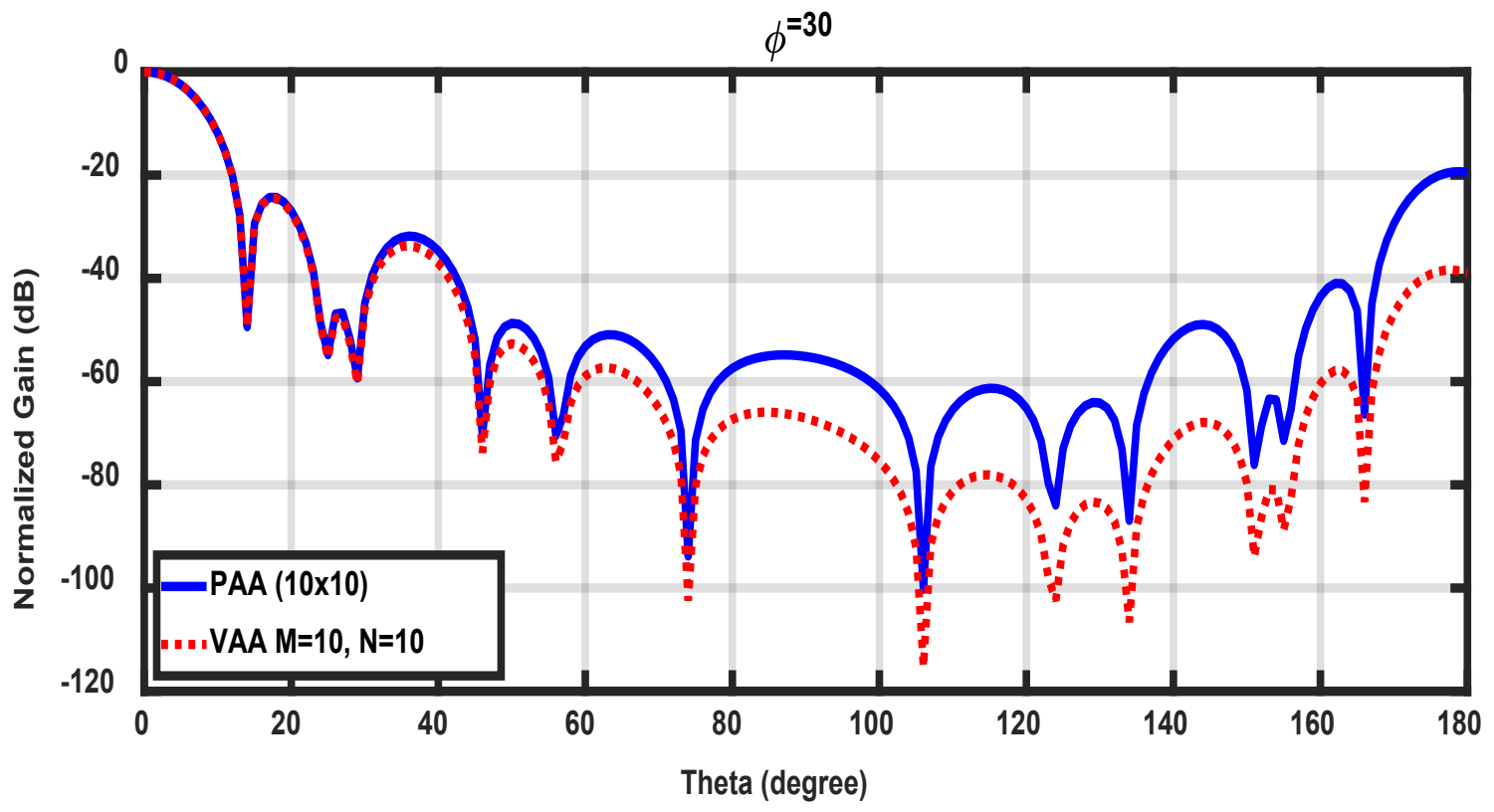

**$45^0$**

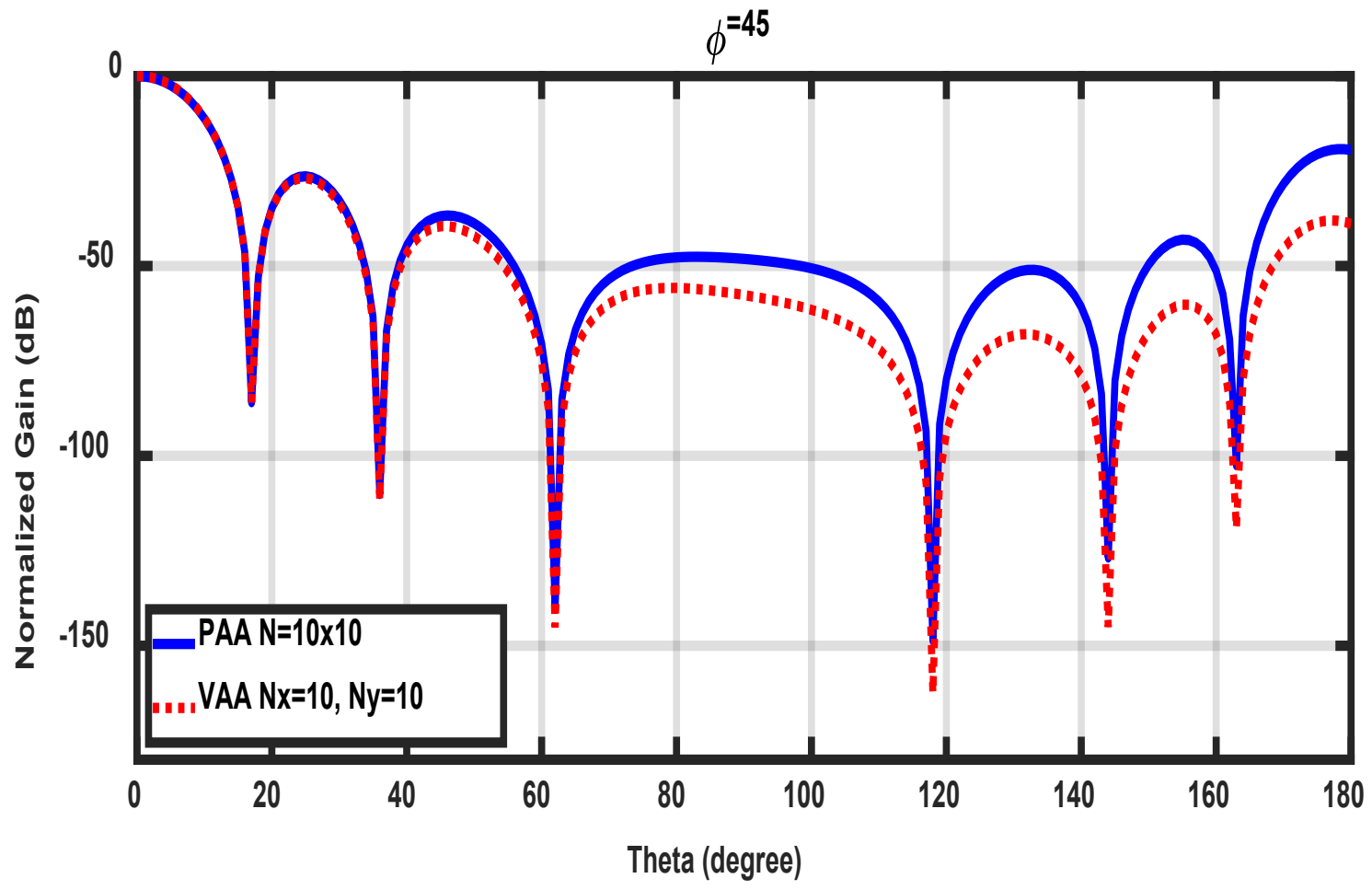

**$60^0$**

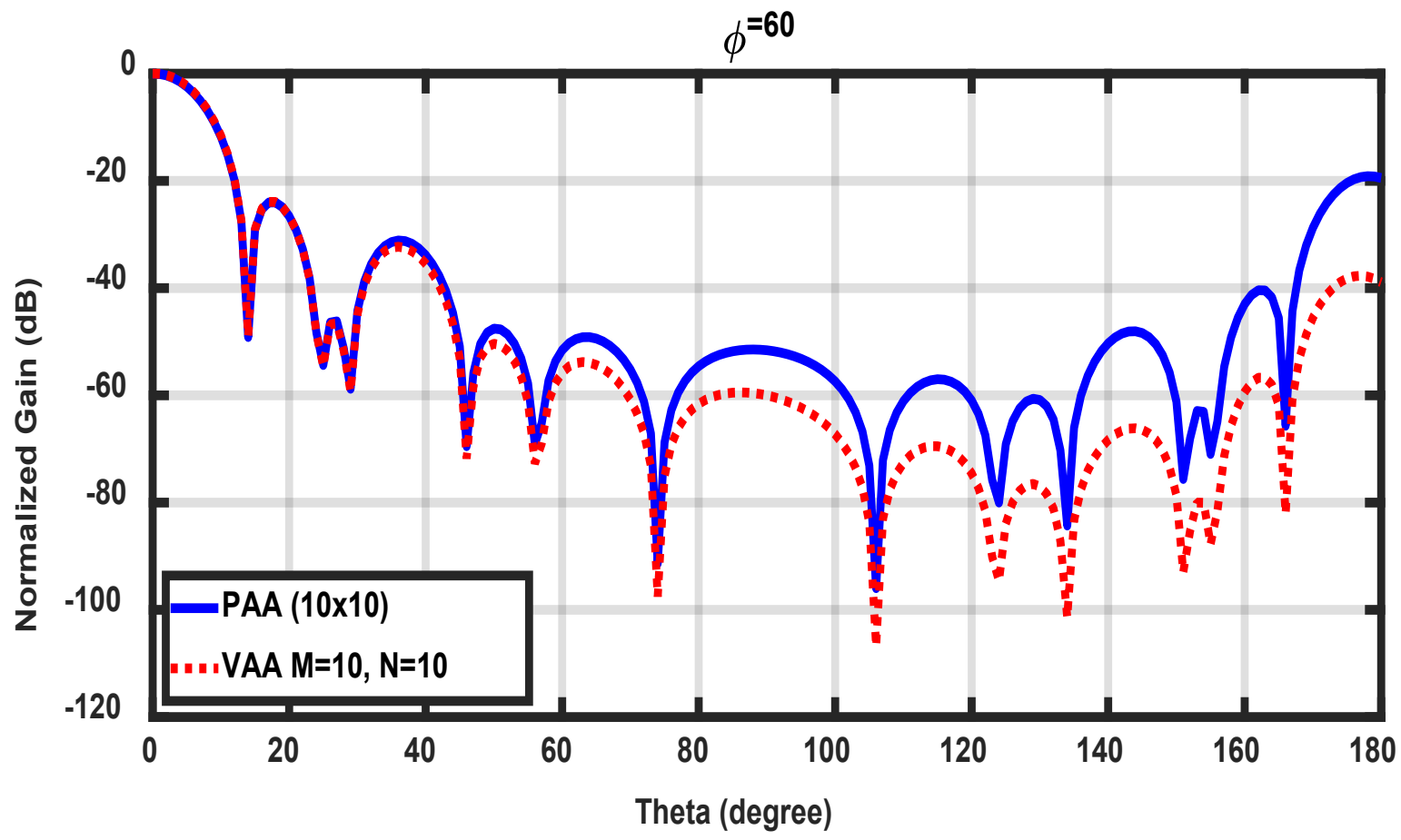

**$90^0$**

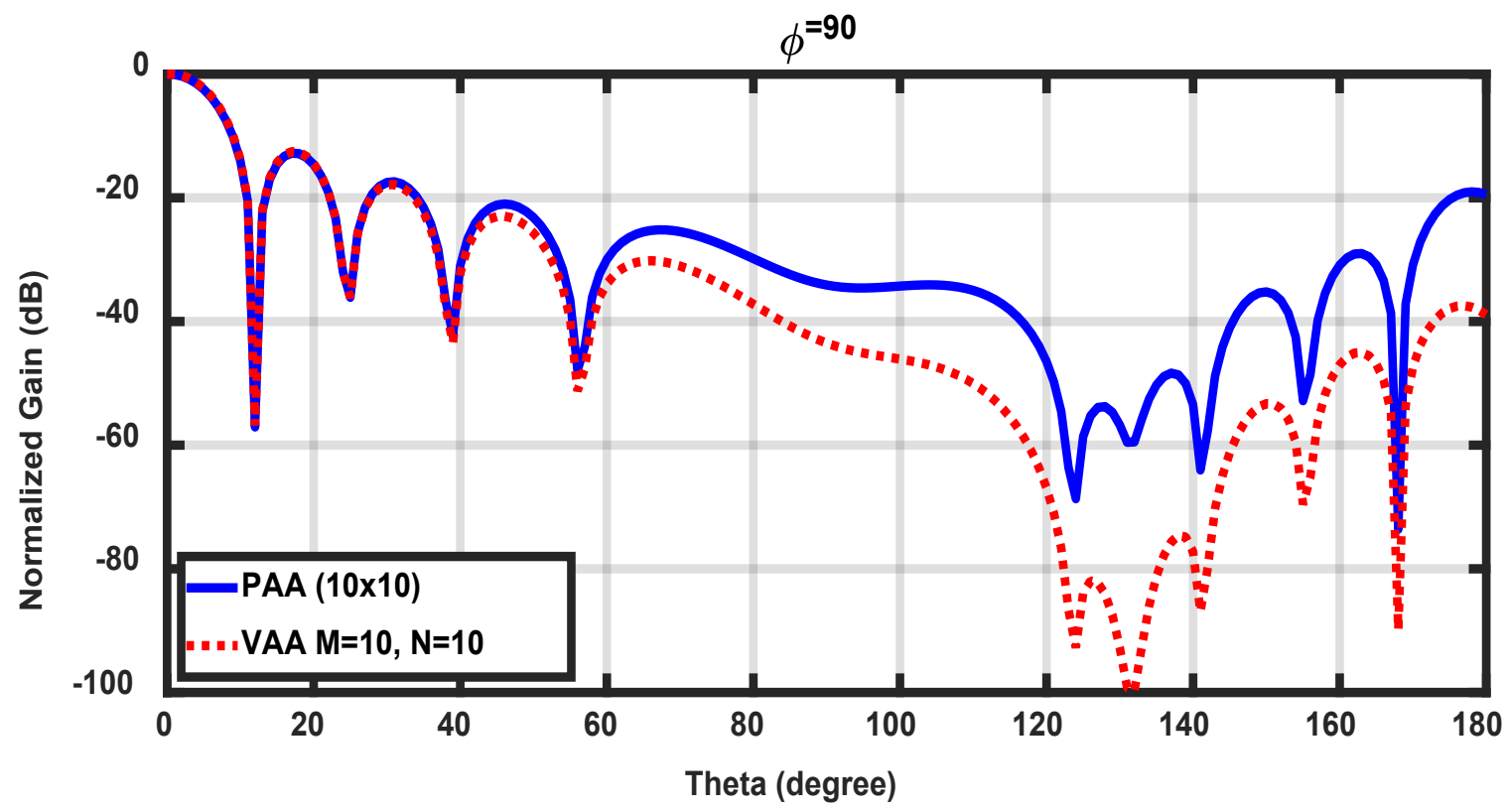

Figure 3.11 Comparison between VAA and PAA, case I.

- **Case II**

In the RX side, the equation of the reflected power from the target can be expressed as:

$$P_{ref=}\frac{P_t G_t(4\pi\sigma)}{(4\pi R)^2} \tag{3.35}$$

The reflected power back to the radar receiver is expressed as:

$$P_r = \frac{P_t G_t(4\pi\sigma)}{(4\pi R)^2}\frac{G_r\lambda^2}{(4\pi R)^2} = \frac{P_t G_t G_r \sigma \lambda^2}{(4\pi)^3 R^4} \tag{3.36}$$

Where $P_r$: Received Power in watts, $P_t$:Peak transmitted power in watts, $G_t$: Transmitter Gain, $G_r$: Receiver Gain, $\lambda$: Wavelength (m), $\sigma$: RCS of the target (m$^2$), R: Range between radar and target (m). In equation (3.36), we notice that the received power consists of gain of TX and RX antennas($G_{tr} = G_t G_r$). Therefore in case II we compare between $G_{tr}$ of the PAA and the $G_r$ of the VAA as shown inFigure 3.12. One can notice that the half power beam width (HPBW) of VAA is wider than that of PAA. Also, the side lobes and back lobes of VAA have larger values than that of PAA. So, we present case III that introduces the optimum number of M and N that gives the same HPBW of PAA

| Phi | Case II |
|---|---|
| $0^0$ | 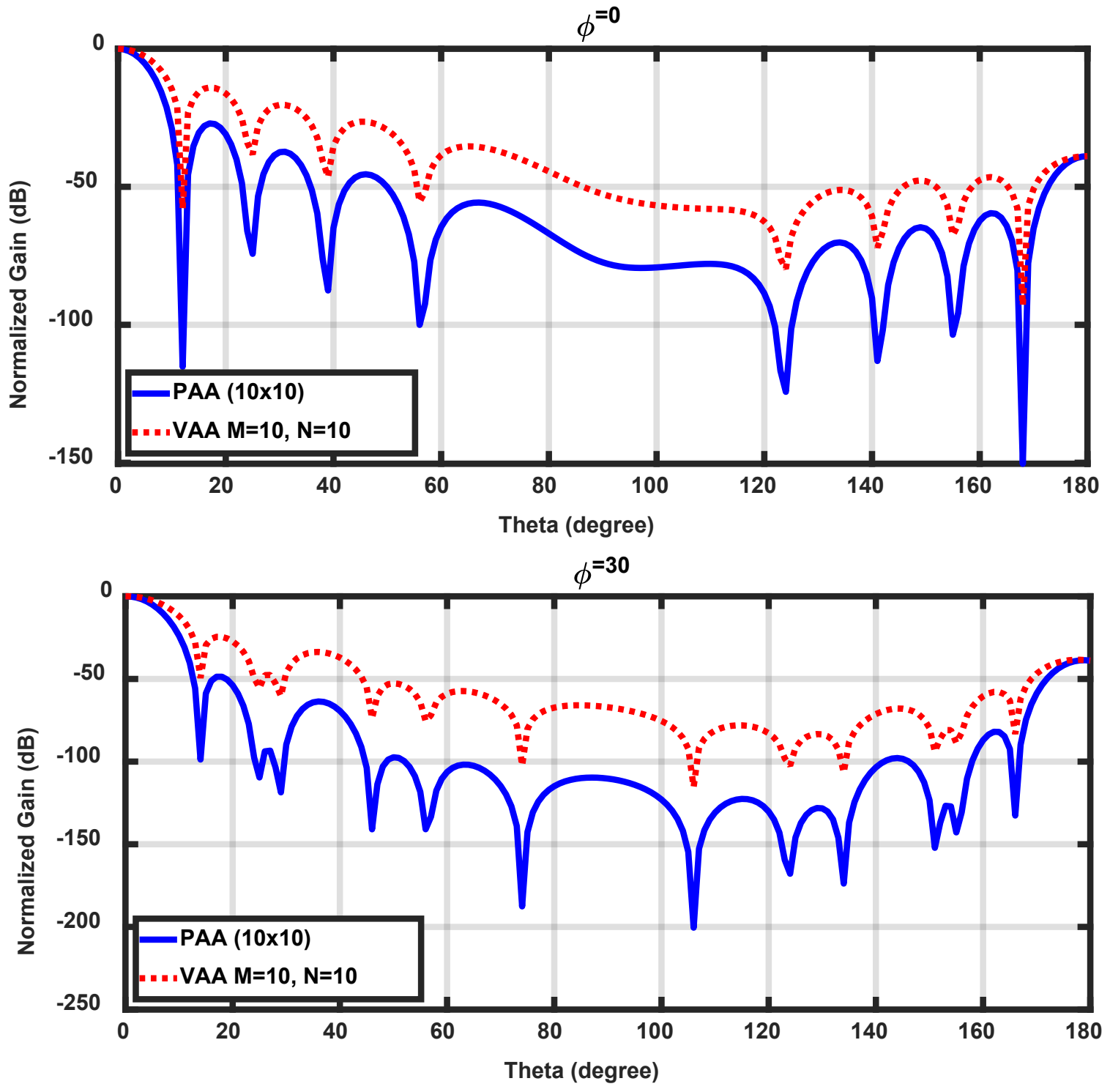 |
| $30^0$ | |

**$45^0$**

**$60^0$**

**$90^0$**

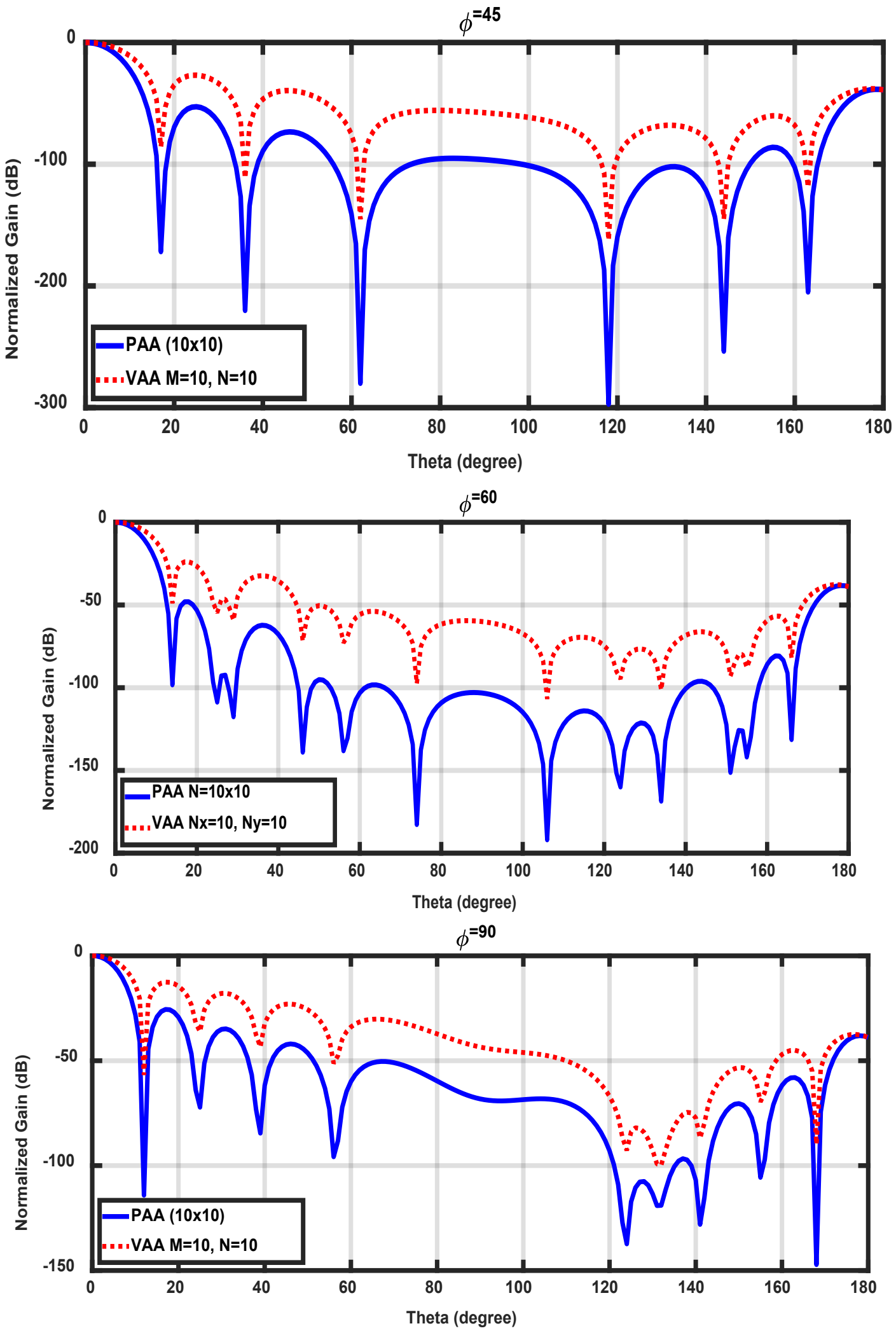

Figure 3.12 Comparison between VAA and PAA, case II.

- **Case III**

In this case, we need to introduce the VAA that is equivalent to the PAA in the receiving mode. So, we optimize the number of elements in VAA by using genetic algorithm (GA) (CST optimization tools). The comparison between VAA and PAA in the receiving mode with the same HPBW is depicted in Figure 3.13 . However, the VAA achieves the same HPBW of PAA but its side lobe and back lobe still have high values than that of PAA. Therefore, we need to use non-uniform excitation for VAA to give the same levels of side lobes and back lobes of PAA.

Phi **Case III**

$0^0$

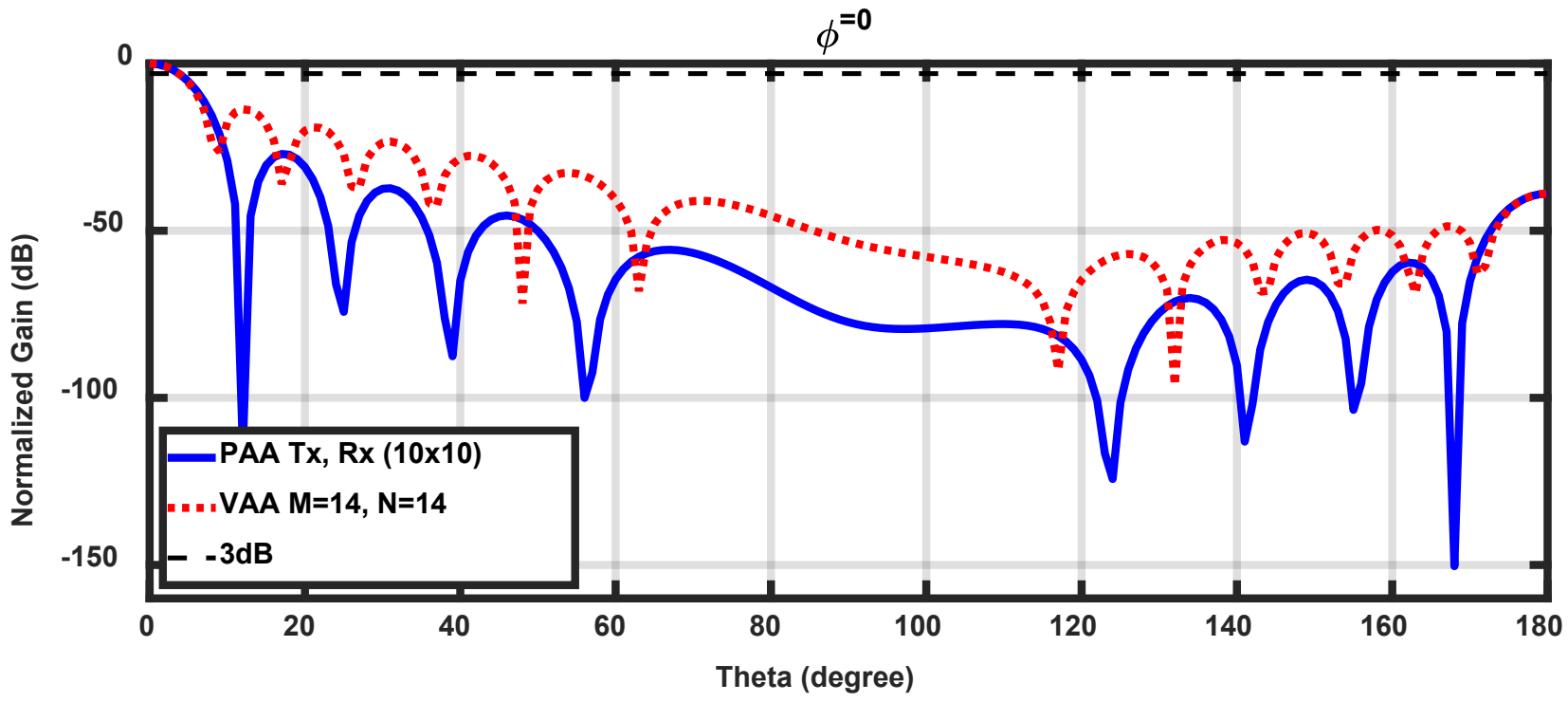

$30^0$

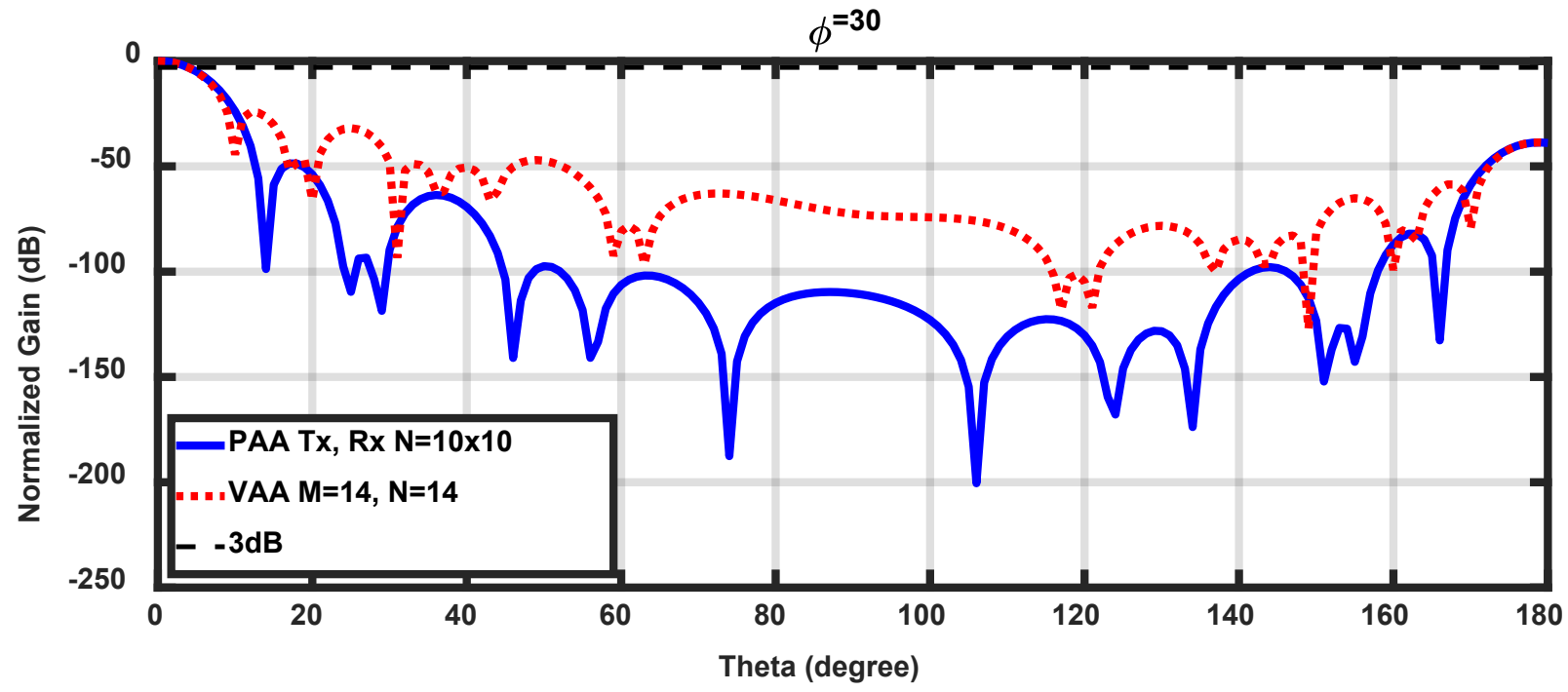

$45^0$

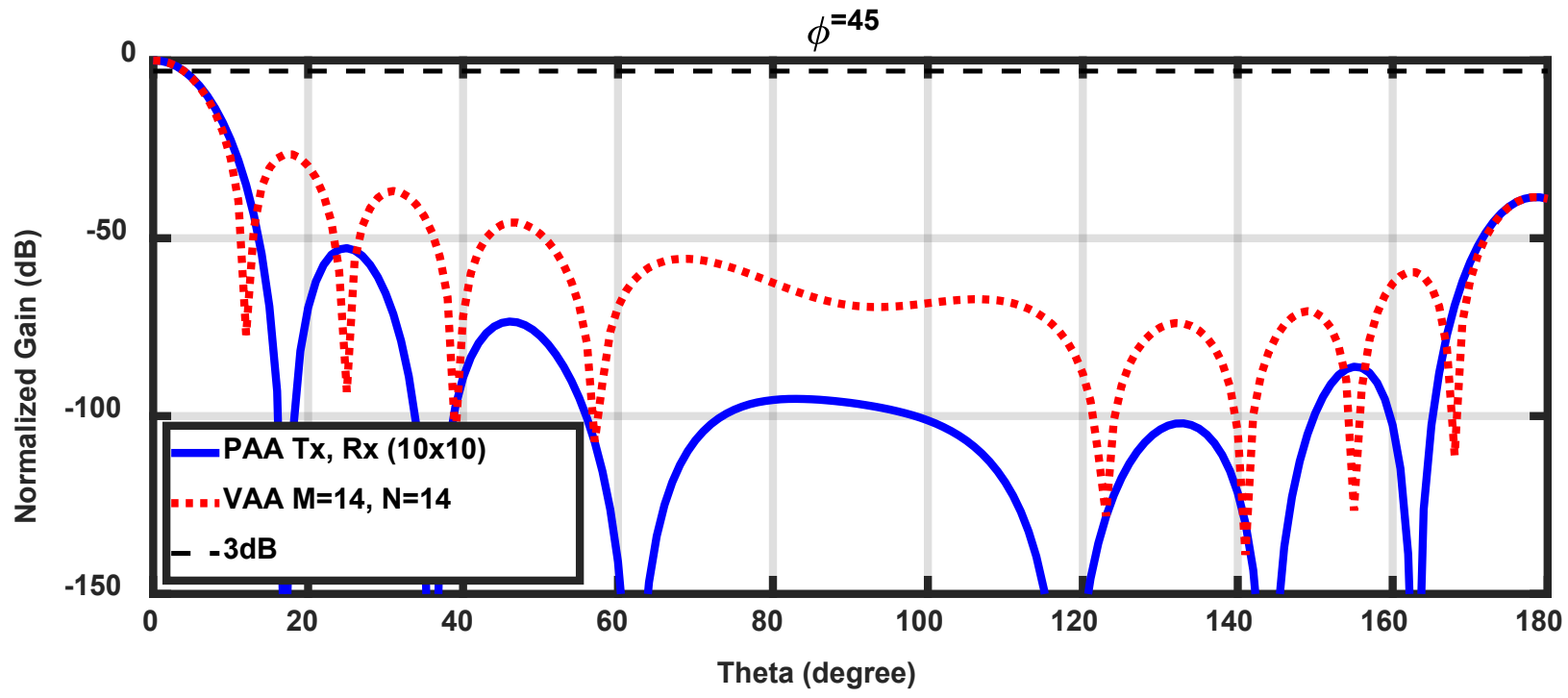

$60^0$

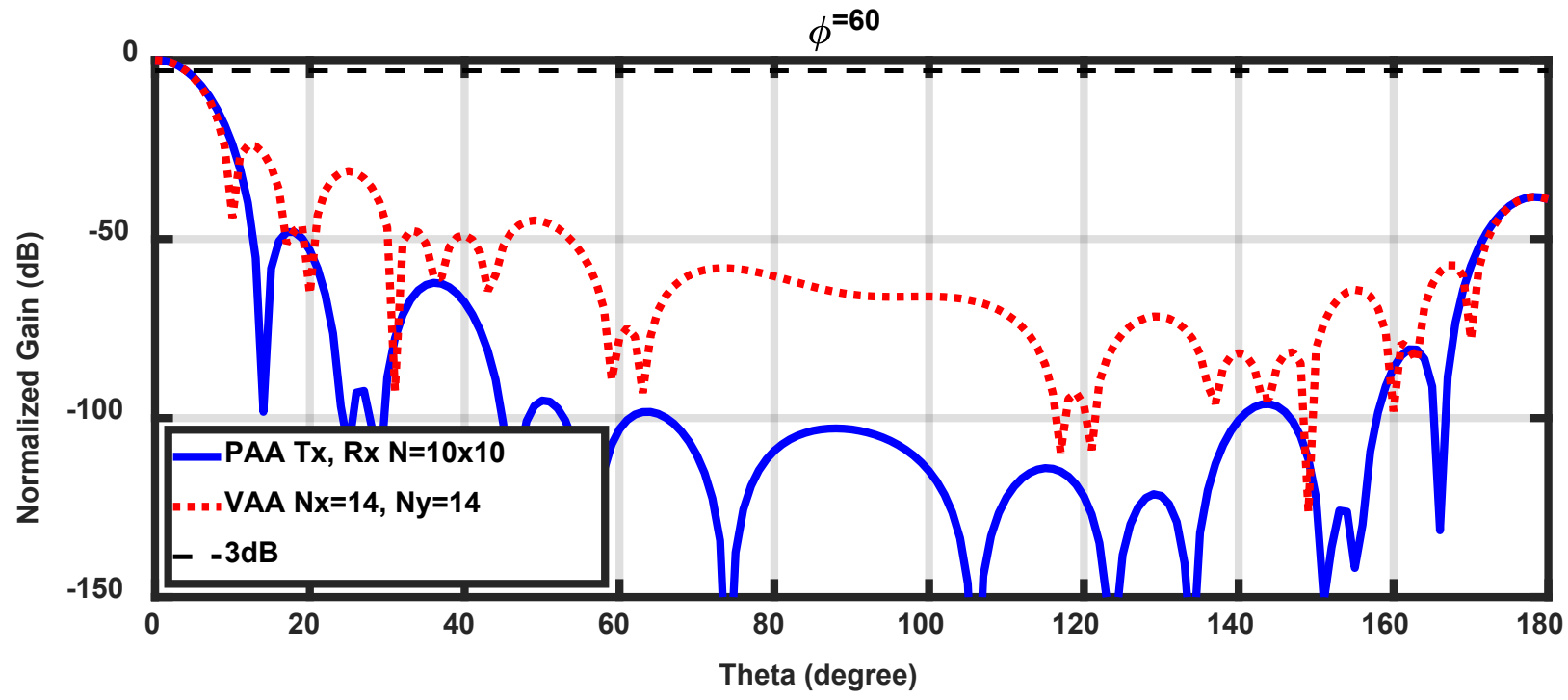

**90⁰**

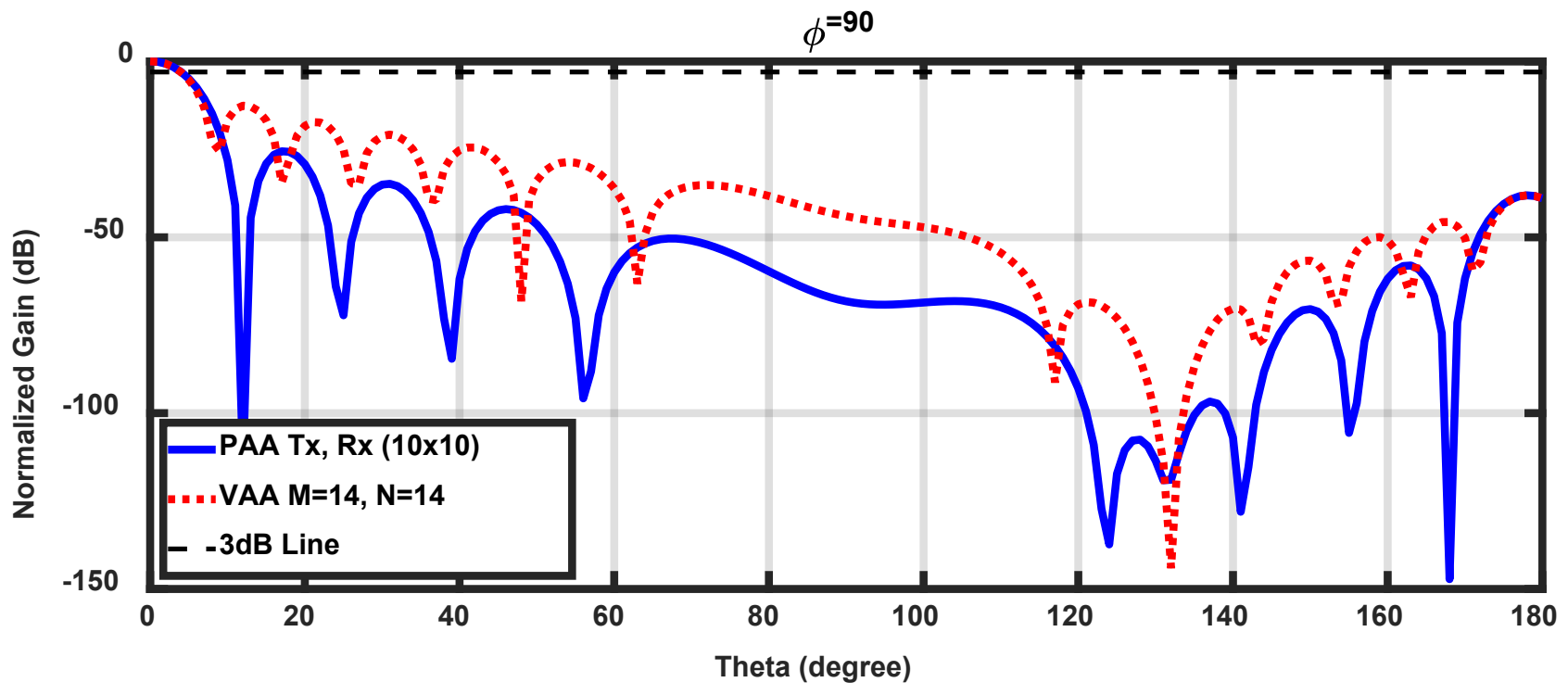

Figure 3.13 Comparison between VAA and PAA, case III.

In this work, the MoM/GA, a semi-analytical technique, is utilized to synthesize the two columns of the virtual array to mimic the planar array. Consider the configuration of isotropic sources antenna array along the Z-axis with equal distance between them [143]. The MoM/GA summarizes the synthesis problem to the following equation:

$$[Z]_{M\times M}[I]_{M\times 1} = [V]_{M\times 1} \tag{3.37}$$

Where the elements of the matrix $[Z]_{M\times M}$ are given by,

$$z_{mn} = \int_0^{\pi} e^{j(d_n - d_m)k\cos(\Theta)}\, d\Theta \tag{3.38}$$

and the elements of the vector $[V]_{M\times 1}$ are given by,

$$V_m = \int_0^{\pi} AF_d(\Theta)\, e^{-jkd_m\cos(\Theta)} d\Theta \tag{3.39}$$

The excitation coefficients $a_n$ are determined by solving the linear system of equation (3.37). Where $a_n$ are the elements of the matrix $[I]_{M\times 1}$, where $[I]_{M\times 1} = [a_1, a_2, a_3, \dots \dots, a_M]^T$

In order to get the same received power for both the planar and the virtual array, the number of elements for the transmitter and the receiver array should be adjusted so that

$$G_P(\theta, \emptyset) = \sqrt{G_{Tnew}(\theta, \emptyset) \otimes G_{Rnew}(\theta, \emptyset)} \tag{3.40}$$

From equation (3.40) and applying the MOM/GA to the VAA with only 18 elements we can have the same equivalent radiation pattern of the PAA with 100 elements as shown in Figure 3.14. The excitation coefficients of the first 9 elements of proposed VAA in this case according to the MOM/GA are [0.15, 0.257, 0.363, 0.462, 0.571, 0.667, 0.781, 0.868, and 1.0].

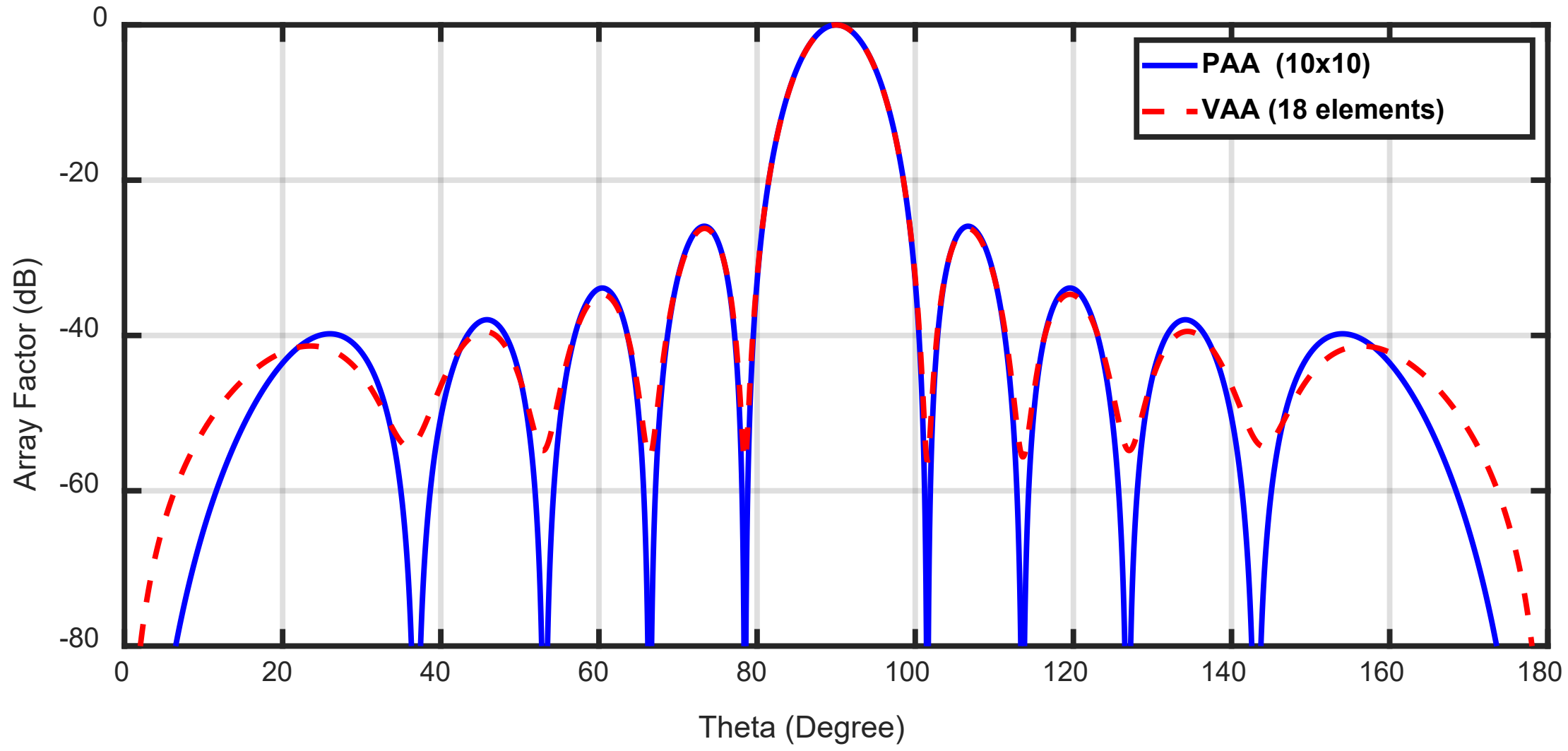

Figure 3.14 Comparision between radiation pattern of PAA and VAA

# 3.5 Design Antenna Array for LRR and MRR

The automotive radars are classified according to the operating range into; long- range radar (LLR) (10- 250 or 300 m), medium-range radar (MRR) (1-100 m), and short range radar (SRR) (0.15-30 m). Where both LRR and MRR are required to detect the forward obstacles. So, most of the radars use two transmitters with two unique radiation patterns to give the LRR mode and MRR mode. The performance of this system isn't efficient for the automotive radar due to the switching performance (time response and losses) between the two modes. Recently, Xu et.al [16, 144], developed the idea of the shaped beam antenna that was used in other applications such as satellites and radars to introduce a single antenna that meets the requirements of LRR and MRR in automotive radar applications. The authors introduce two papers in this direction using substrate integrated waveguide (SIW) power divider to feed the planar patches array and to feed the slot SIW array.

## 3.5.1 Proposed Radiation Pattern of LRR and MRR

In this study, we use the same transceiver for the LRR and MRR modes with the same antenna but we need to determine the antenna gain difference between two modes (LRR and MRR). We assume that the minimum received power at the radar is constant in the two modes $P_{rm}$, because the sensitivity of the receiver is constant in the two modes. $R_L$, and $R_m$ are considered the radar ranges for LRR and MRR, respectively. Also, we assume that $R_L = 300m$ and $R_m = 100m$, so, $R_L = KR_m$. Moreover, $G_T^L, G_T^M$ are the transmitting antenna gain for LRR and MRR, respectively. The received power is expressed as

$$P_{rm} = \frac{P_t G_t G_r \sigma \lambda^2}{(4\pi)^3 R^4} = \frac{P_t G_t^M G_r \sigma \lambda^2}{(4\pi)^3 R_m^4} = \frac{P_t G_t^L G_r \sigma \lambda^2}{(4\pi)^3 (K R_m)^4}. \quad (3.41)$$

We can determine the gain difference $G_d$ between the two operating mode from equation (3.41) as

$$[G_d] = [G_t^L] - [G_t^m] = 40 \log_{10} K \quad (3.42)$$

In case $K \geq 3$, $G_d \geq 19\, dB$, $G_d$ is define as the difference gain between the two scenario modes as show in Figure 3.15 that depicts the expected radiation pattern of proposed system. This shape is called flat-shoulder shape and shows the suggested ideal radiation pattern of the proposed antenna to support the two expected scenarios mode, where $\theta_L = 15^0$ at -3dB as half power beam width in case of LRR and $\theta_m = \pm 40^0$ at $G_d$ level as beam width in case of MRR.

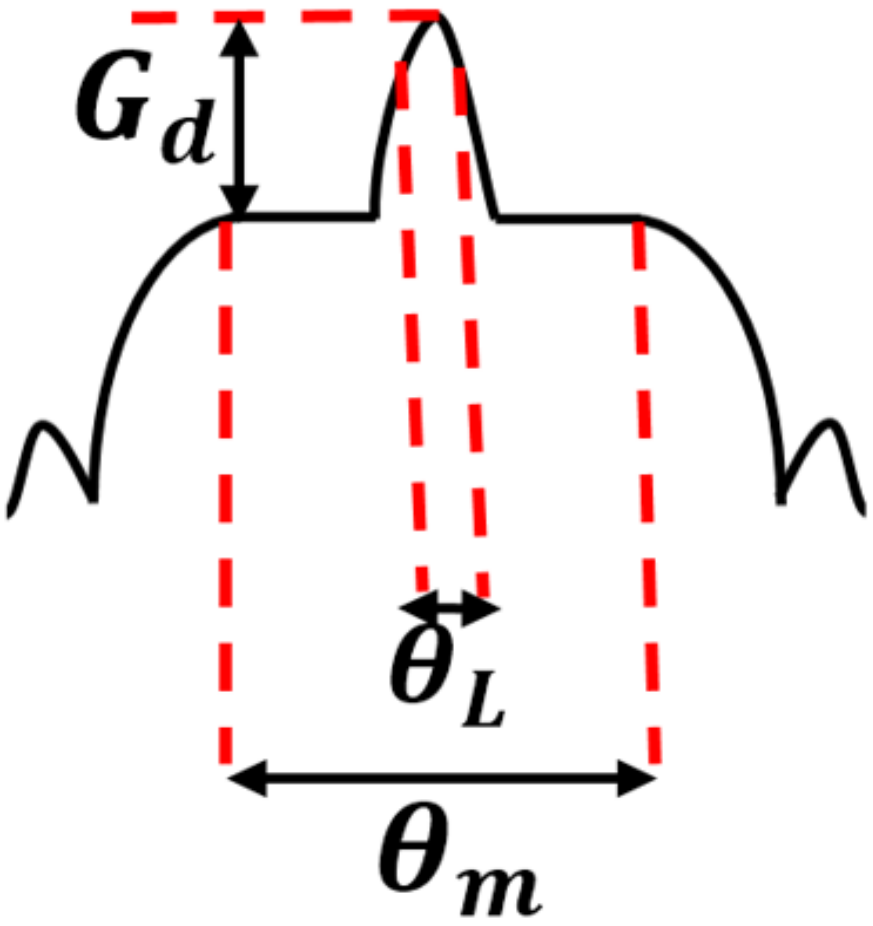

Figure 3.15 The suggested radiation pattern of the antenna to support MRR and LRR (LMRR)

## 3.5.2 Excitation Coefficients

To achieve the flat shoulder shape (FSS) pattern, we propose radiation pattern by combination of two Dolph-Chebyshev antenna arrays [145]. The radiation pattern (FSS) in 3. 15 is applied to the MoM/GA algorithm as a desired radiation pattern. The synthesis process using equations (3.37 to 3.39) results in the excitation coefficients that guarantee the occurrence of the FSS pattern when exciting an 8-element antenna array whose elements are half-wavelength spaced from each other.

As the FSS pattern is considered a broadside radiation pattern, the phase shift between elements should vanish. After applying the MoM/GA algorithm, it is noticed that the excitation coefficients pattern is symmetric around the center of the array. The left-handed and right-handed four elements are excited by the proposed excitation coefficient as shown in Table 3. 1 Then according to these coefficients, we need to design unequal $8 \times 1$ power divider.

### 3.5.3 Power Divider Design

We designed unequal Willikson 8x1 power divider to feed the proposed antenna as shown in Figure 3.16 according to the excitation coefficients from MOM/GA. The Willikson power divider is designed on Rogers RO4003 substrate with dielectric constant 3.38 and thickness 0.2 mm. The dimensions of the proposed power divider is based on the equations presented in [146]. We use the quarter wave length transformer and stepped impedance matching to achieve the required excitation coefficient at frequency 24 GHz. Here, we use 24 GHz as an example of automotive radar frequency instead of 77 GHz because of the limitation of our measurements and fabrications facilities in our laboratory. The proposed power divider consists of 3 stages to introduce 8 excitation coefficients. PD1 is an equal power divider, PD2, PD3 and PD4 are unequal power dividers. All the simulated S-parameters magnitude and phase of the proposed power divider are depicted in Figure 3.17. We notice that all ports achieve the required magnitude and phase according to the output from the MOM/GA technique. Furthermore, the comparison between the required values of magnitude and phase from MOM/GA and CST are shown in Table 3. 1 with equal phase between the power divider ports.

Table 3. 1 Amplitude of power divider outputs

| | **Synthesis (Ideal)** | | **Simulated (CST)** | |
|---|---|---|---|---|
| **Power** | **Amplitude (dB)** | **Phase (Degree)** | **Amplitude (dB)** | **Phase (Degree)** |
| **$P_1$** | -3 | (0,0) | -3.21 | (-96.15, -96.15) |
| **$P_2$** | -10.838 | (0,0) | -10.62 | (-58.75, -58.75) |
| **$P_3$** | -3.788 | (0,0) | -4.765 | (-57.939, -57.939) |
| **$A_1$** | -17.122 | (0,0) | -17.72 | (101.544,101.544) |
| **$A_2$** | -11.04 | (0,0) | -11.6 | (103.324,103.324) |
| **$A_3$** | -7.642 | (0,0) | -8.19 | (102.73,102.73) |
| **$A_4$** | -6.38 | (0,0) | -6.88 | (105.253,105.253) |

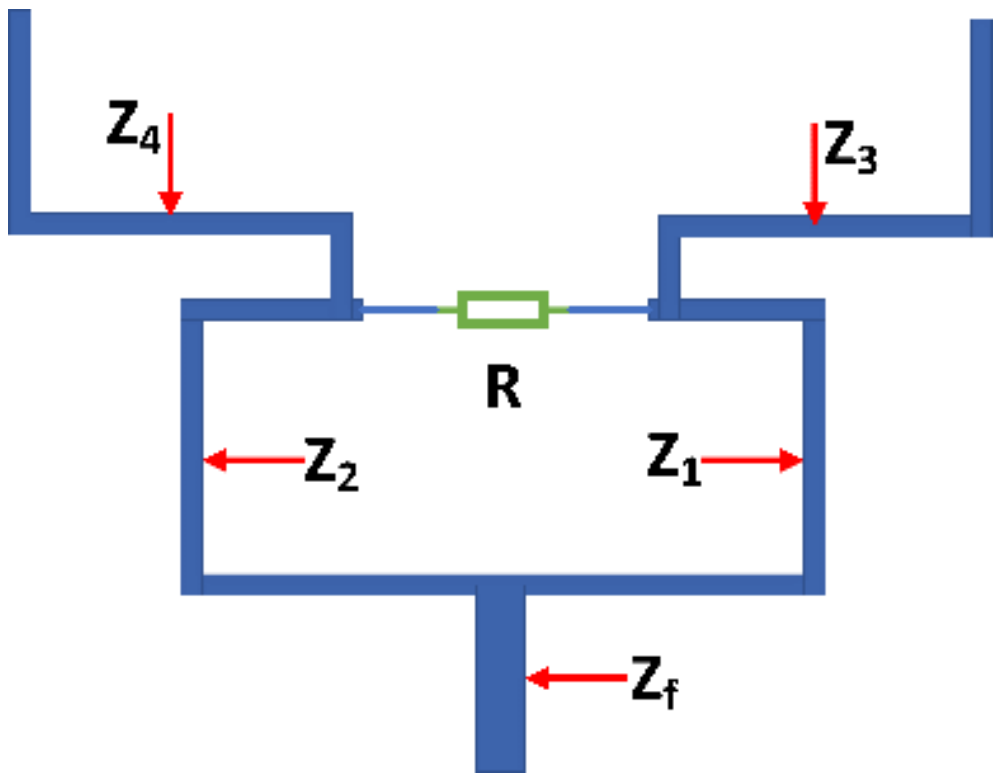

(a)Reference power divider

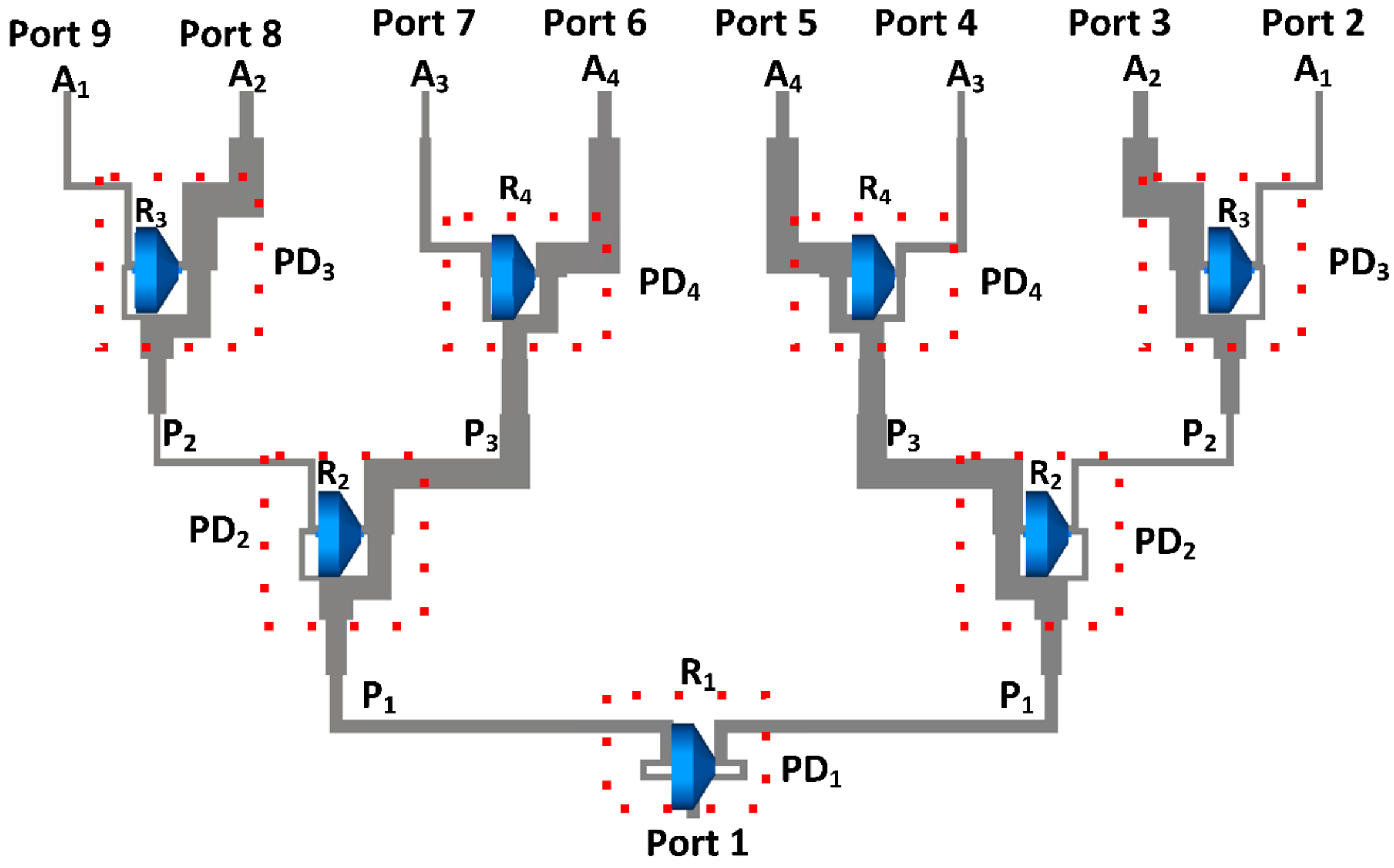

(b)The Proposed power divider

Figure 3.16 The configuration of Wilkinson power divider with impedance distributions

Table 3. 2 Impedance of power dividers (optimization values), (all values in Ω)

| PD | $Z_f$ | $Z_1$ | $Z_2$ | $Z_3$ | $Z_4$ | R |
|---|---|---|---|---|---|---|
| **1** | 50 | 70.7 | 70.7 | 50 | 50 | 100 |
| **2** | 26.2 | 76.1 | 33.46 | 68.6 | 28.5 | 135 |
| **3** | 26.79 | 76.217 | 33.311 | 66.494 | 25.77 | 68 |
| **4** | 33.31 | 65.1 | 39.94 | 55.94 | 27.5 | 135 |

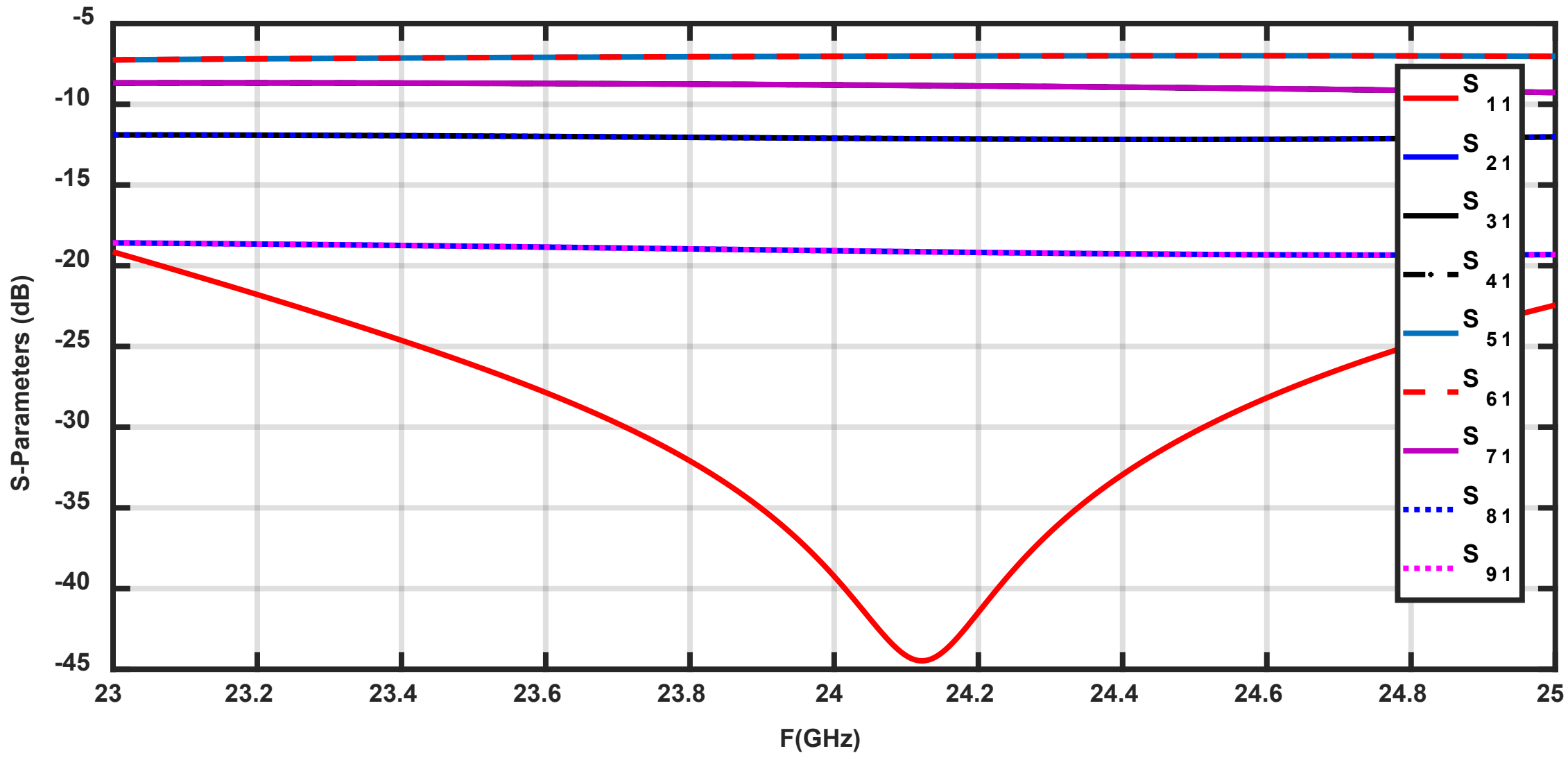

(a)Magnitude (dB)

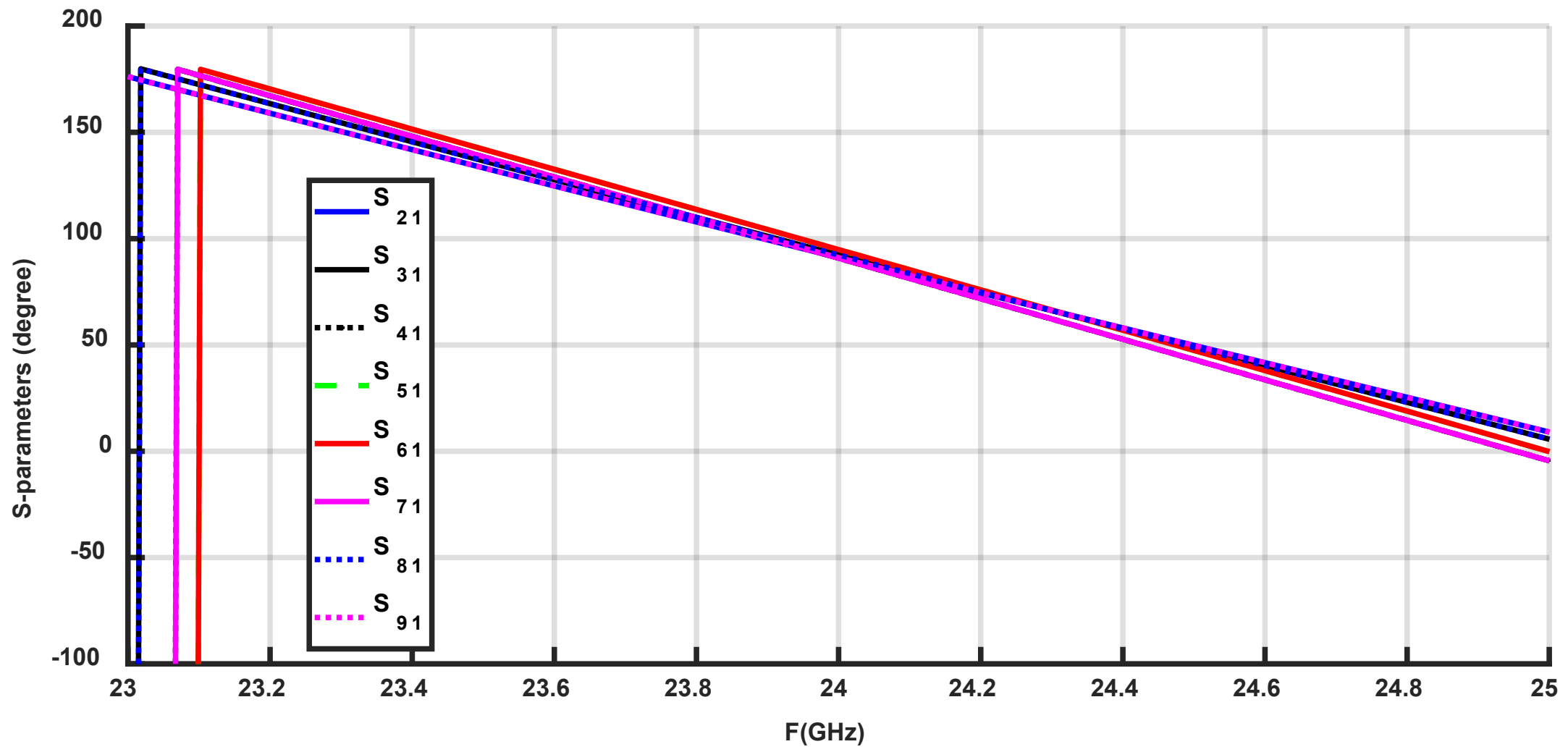

(b)Phase (degree)

Figure 3.17 S-parameters of power divider

### 3.5.4 Antenna Array Design

Figure 3.18 (a) shows geometry of the rectangular patch linear antenna array with half wave length as a distance between the elements. The FSS radiation pattern is achieved as shown in Figure 3.18 (b) and there are good agreements between the simulated FSS pattern from CST and required FSS pattern from MOM/GA. The concept of VAA is applied to the proposed antenna as shown in Figure 3.19 (a). The VAA is fabricated on Rogers RO4003C substrate with dielectric constant 3.38 and thickness 0.2 mm as shown in Figure 3.19 (b). We notice that the VAA operates at 24.25 GHz with 400 MHz bandwidth and it has a very good isolation coefficient between the transmitter and receiver antennas (more than 43 dB) as shown in Figure 3.20. The achieved isolation is one of the main factors that support our proposed VAA. Therefore, we don't need to integrate circulator to the system because we use two antennas with very good isolation. The second advantage in our design is that the VAA needs simple feeding network comparable with the same structure in case of the PAA. Also, Figure 3.20 shows the comparison between the simulated and measured reflection coefficient of VAA. The VAA antenna with its feeding network that is based on a cascaded network of Wilkinson power dividers printed on the same substrate of overall dimensions of 30x48x0.2 $mm^3$.

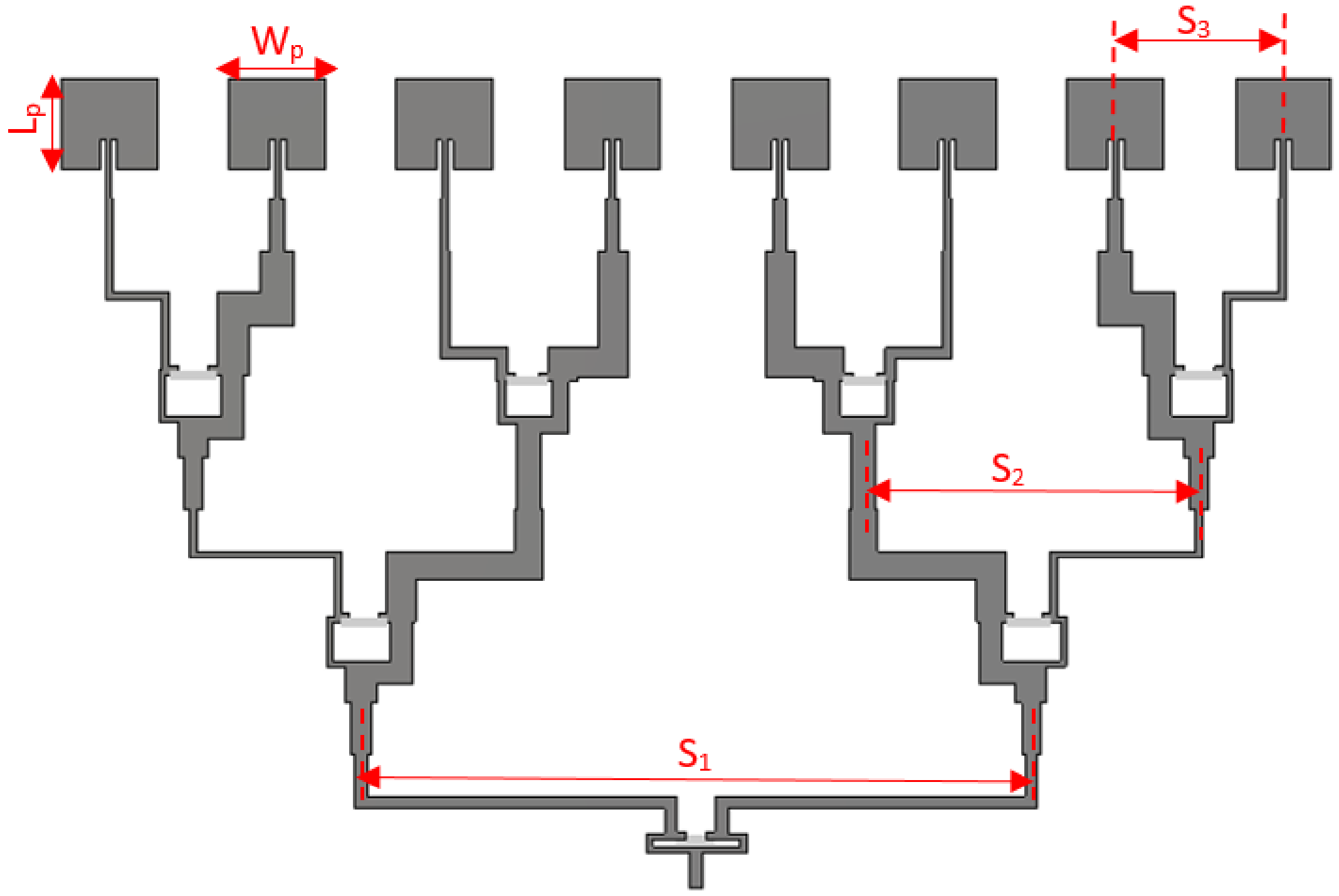

(a)Geometry of LAA

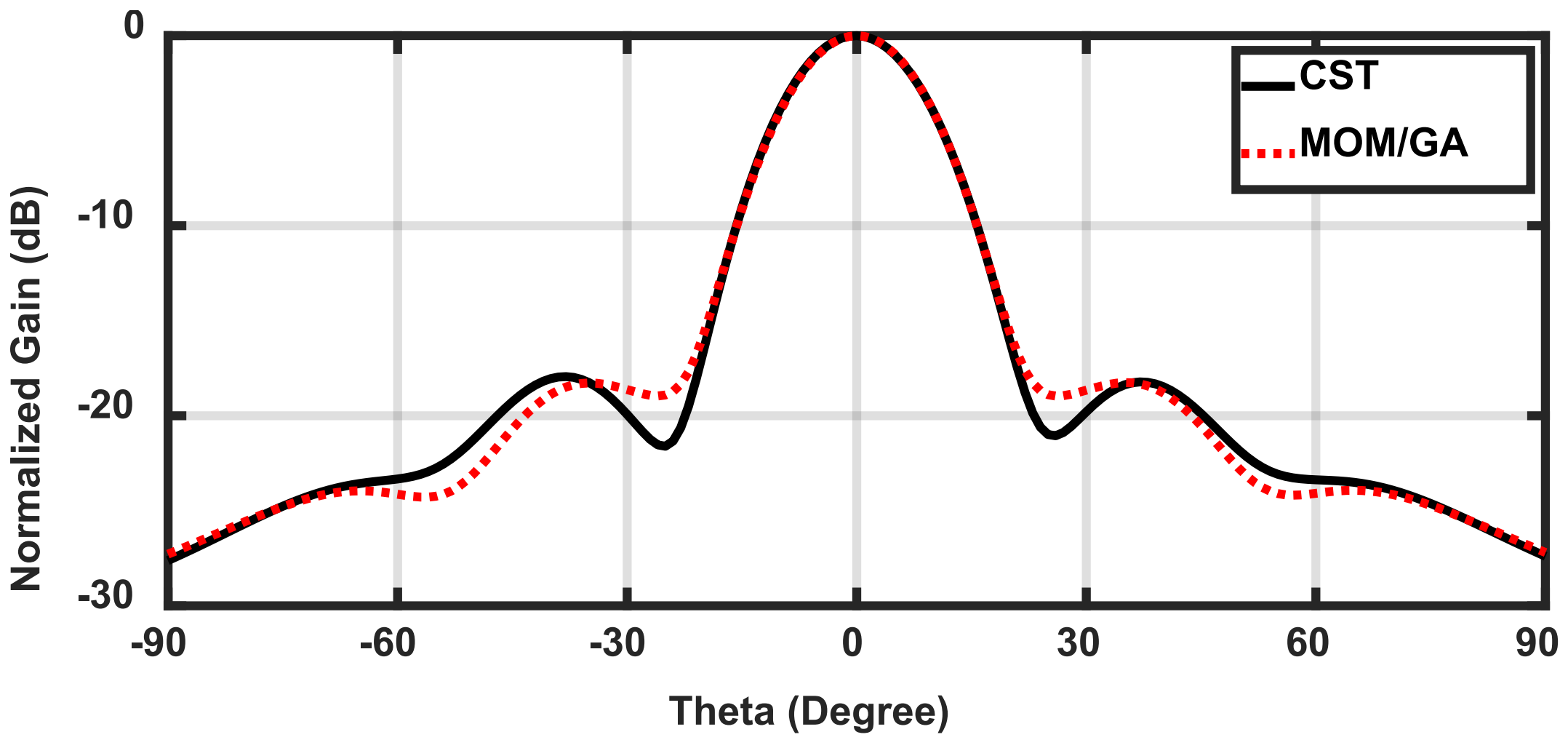

(b) Normalized XZ plane gain pattern

Figure 3.18 Linear antenna array ($L_p$=3.2mm, $W_p$=3.45 mm, $S_1$=24 mm, $S_2$= 12mm, and $S_3$=6 mm)

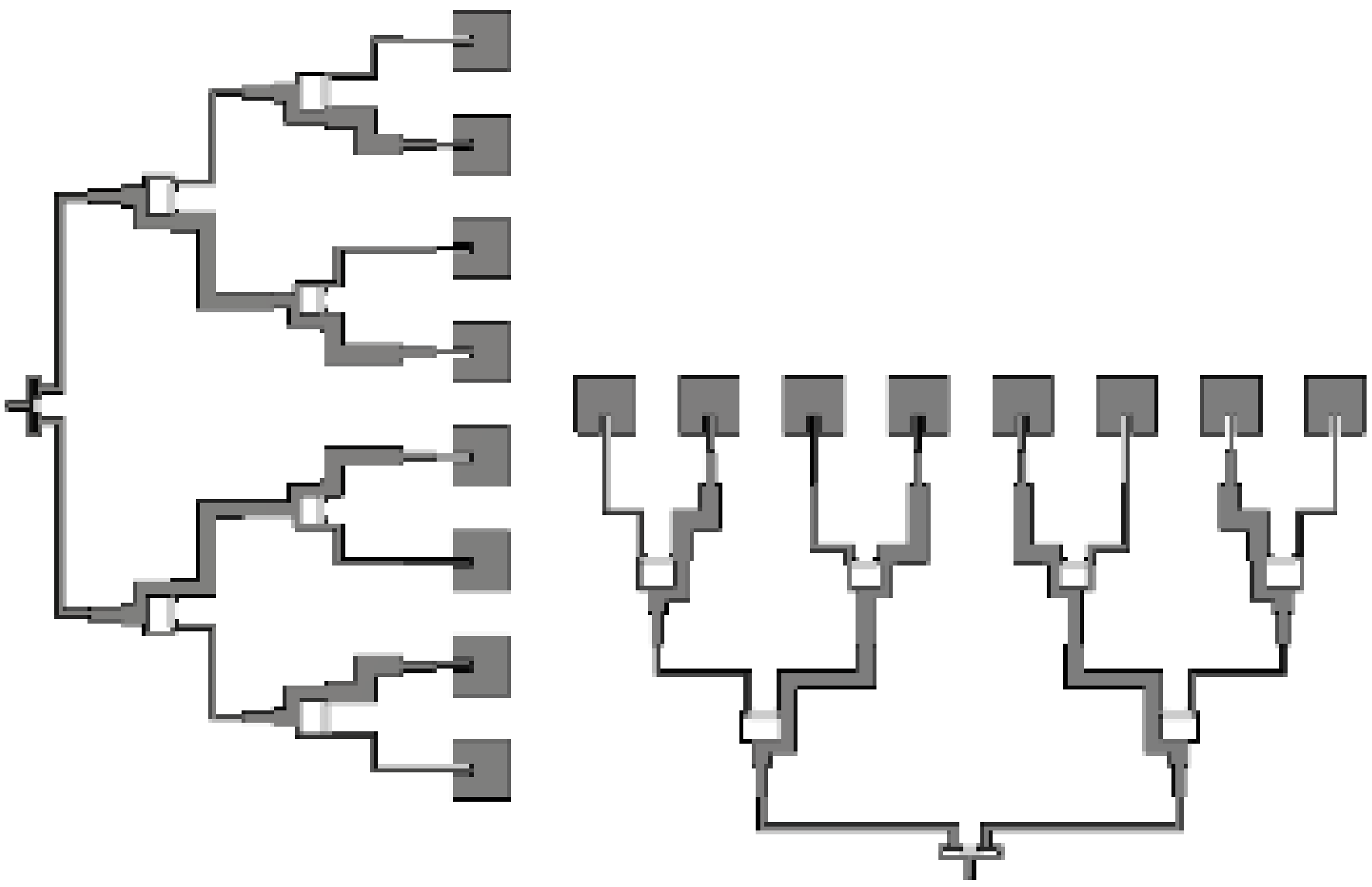

(a)VAA configuration

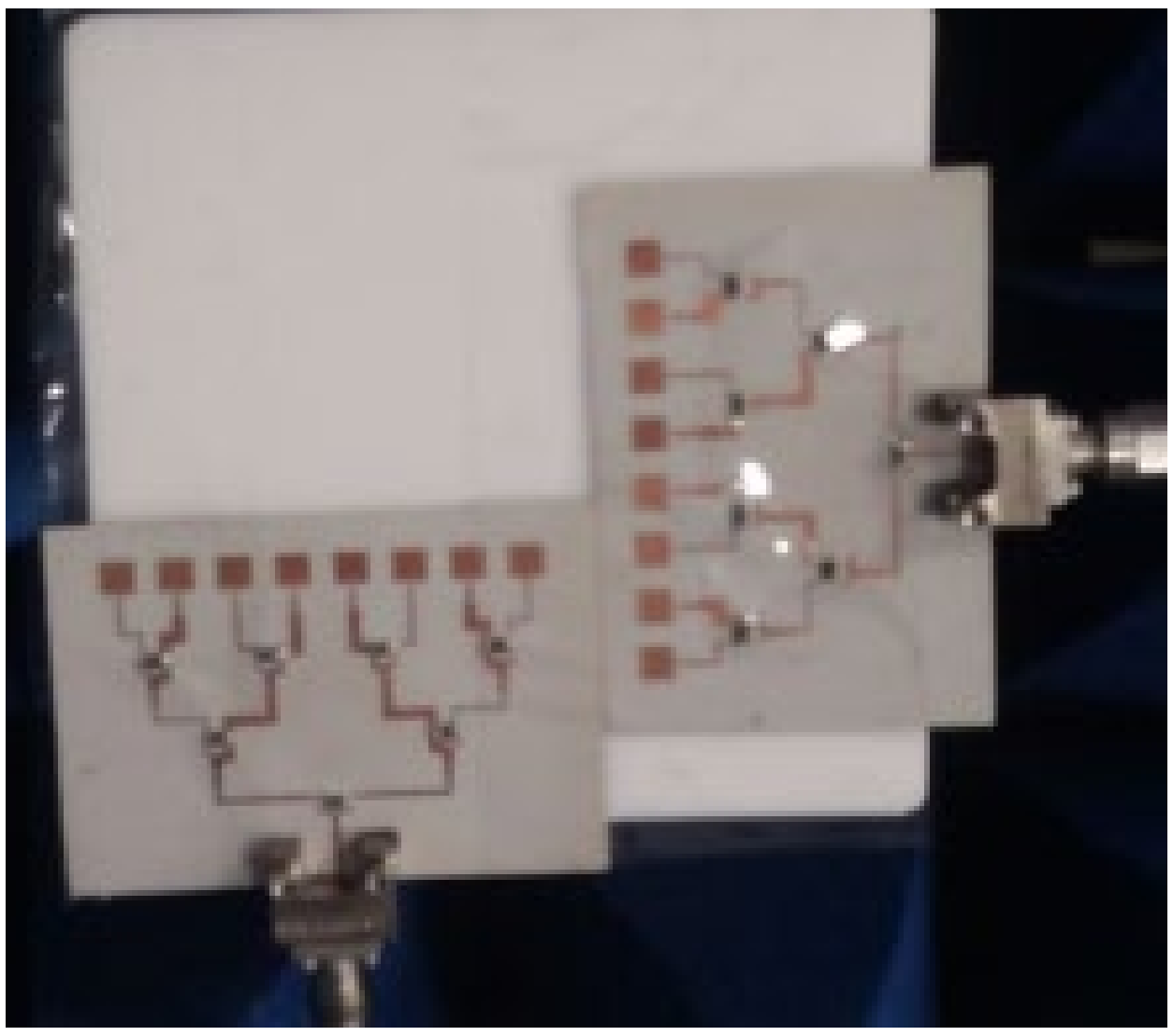

(b) Photo of VAA

Figure 3.19 Geometry and photo of fabricated VAA

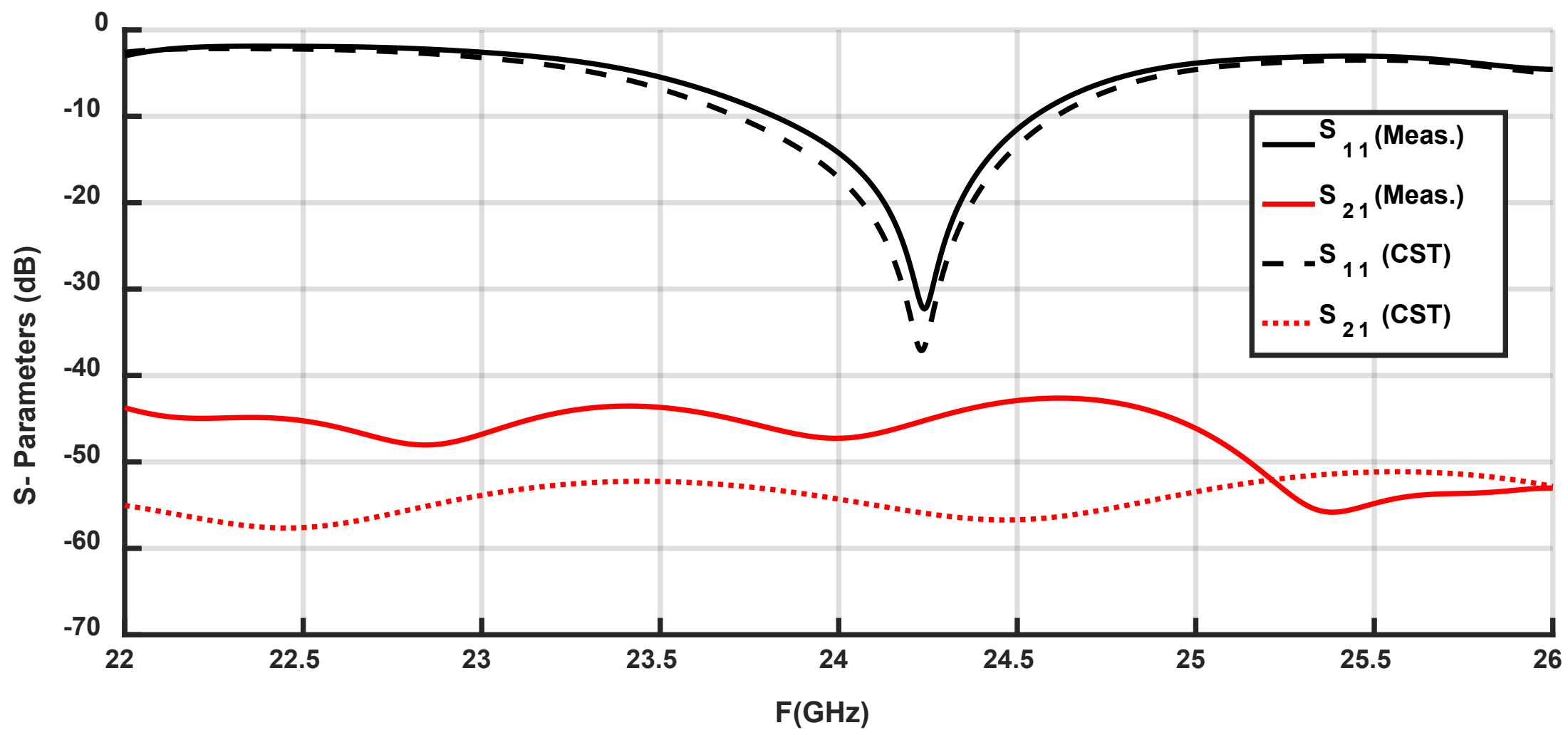

Figure 3.20 Simulated and measured S-parameters of VAA

The radiation pattern measurements have two stages; the first stage, we measure the TX antenna's radiation pattern as common radiation pattern measurements in the anechoic chamber. The second stage is to measure the receiving mode radiation pattern of VAA. Figure 3.21 (a) shows the setup structure of our system to measure the radiation pattern of VAA; the measurement system is homemade. The VAA antenna is positioned on the plate. Two DC motors rotate this plate in two planes (azimuth and elevation plane). The DC motors are controlled by a microcontroller kit (Arduino kit), and the VAA is connected to the VNA through two channels to send and receive the power. We use the 2-D reflector as a target. Figure 3.21(c) shows a good agreement between the simulated and measured radiation pattern for E-Plane and H-Plane.

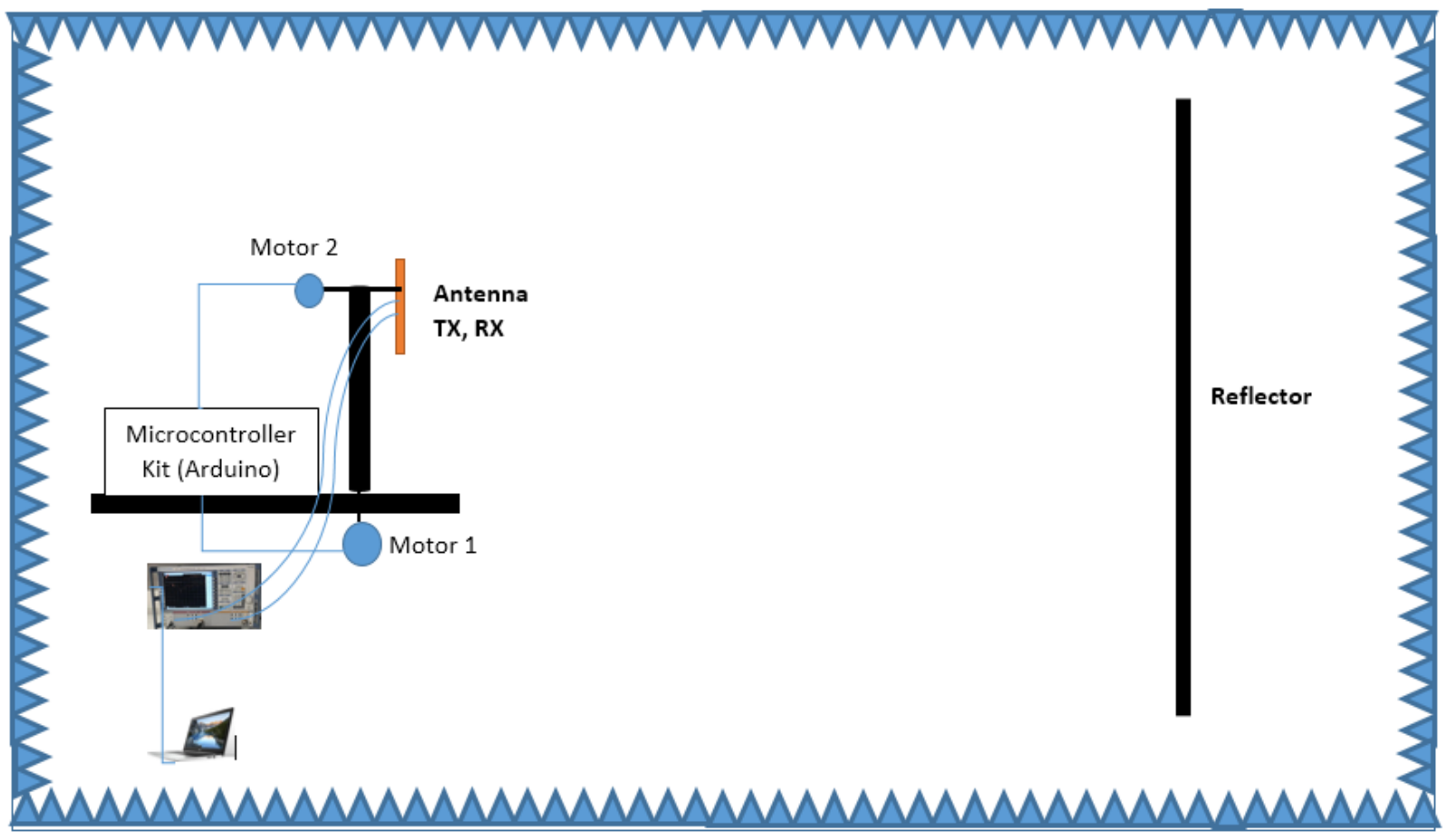

(a)Proposed measurements system

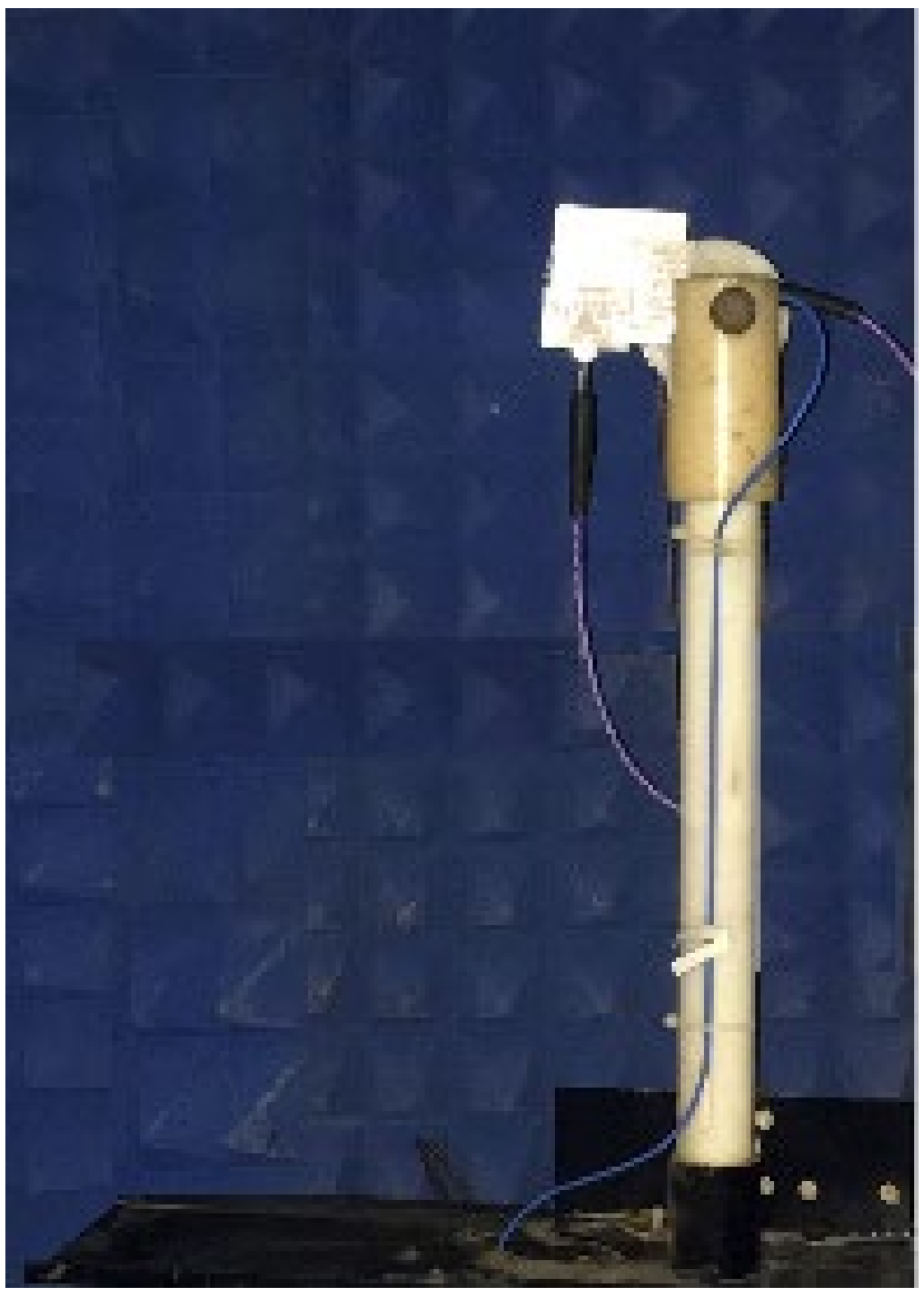

(b)Photo inside anechoic chamber

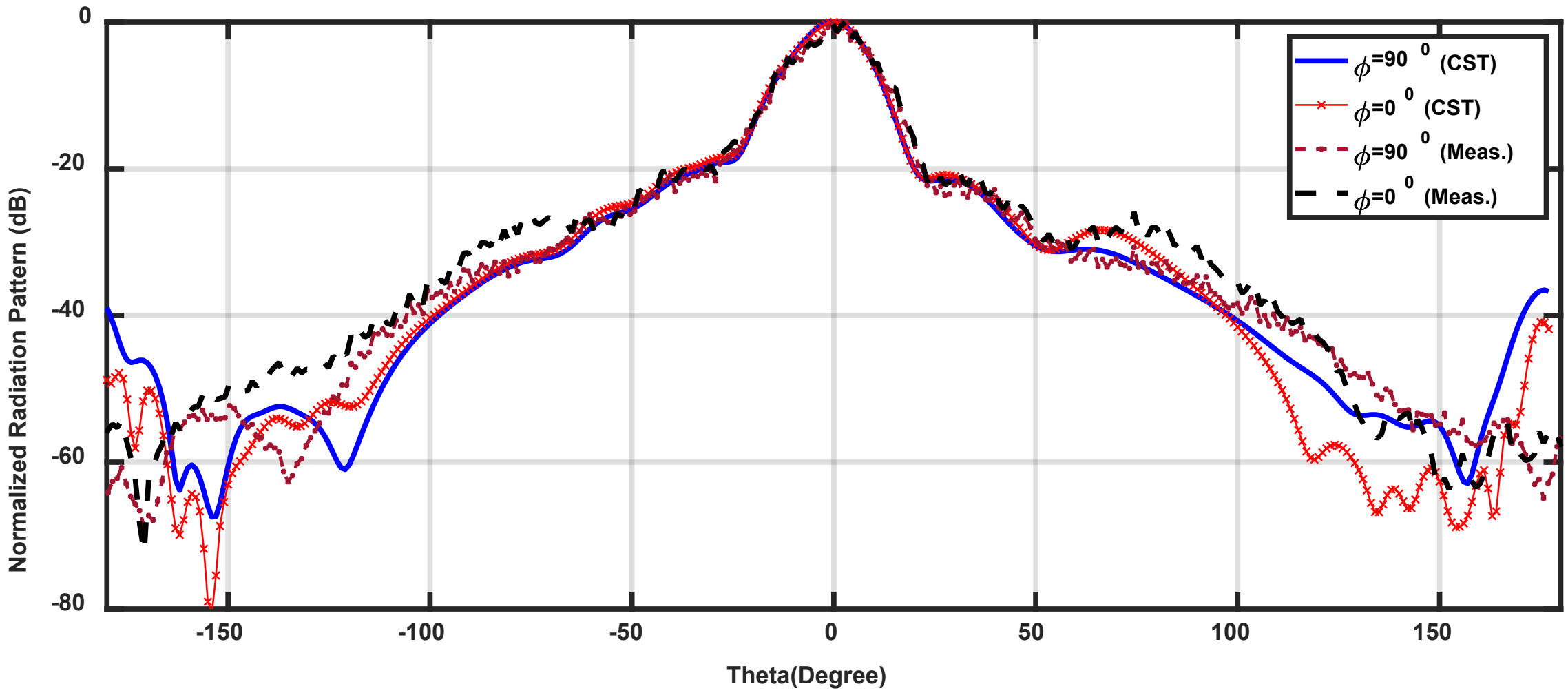

(c) Radiation patterns

Figure 3.21 Setup system for measurements and radiation patterns.

Table 3. 3 shows the comparison between our proposed VAA for automotive radars and the antennas that are presented in the literature. As far as we know, only two papers in the literature combine between the LRR mode and MRR mode [16, 144], but these papers have a complicated feeding SIW network with antenna size $13.2\lambda_0 \times 5.1\lambda_0$ and $3.9\,\lambda_0 \times 7.9\lambda_0$ excluding the feeding network size for antenna in [16] and antenna in [144], respectively. Our design has small size compared to the aforementioned antennas and achieves the same radiation properties. Furthermore, the proposed design used VAA concept that reduces the number of antenna elements to eight patches for Tx and eight patches for Rx compare with the 60 patches presented in [16] and 96 slot antennas presented in [144]. Furthermore, the proposed design achieved the requirements of LRR and MRR beam angles in E-plane and H-plane. Otherwise, the automotive radar antennas that are introduced in literatures have a large size, low gain and multiple number of elements.

Table 3. 3 Comparison between proposed antenna and antennas in literature

| Paper/Year | F (GHz) | Radar | Antenna description | HPBW | | Gain (dBi) | BW (GHz) | Size | Substrate $\varepsilon_r$ |
|---|---|---|---|---|---|---|---|---|---|
| | | | | E Plane | H plane | | | | |
| [147]/2010 | 77 | LRR | RP (8× 8) | $\pm 9^0$ | - | 20 | 1 | $5.9\lambda_0 \times 6.4\lambda_0$ | 3.38 |
| [148]/2012 | 77 | LRR | RP(5× 14) | $10^0$ | - | 20.5 | 1.5 | $8.1\lambda_0 \times 2.95\lambda_0$ | Silicon 11.8 |
| [149]/2013 | 77 | LRR | VP (8× 18) | $4.8^0$ | $18.3^0$ | 20.8 | 1 | $17.96\lambda_0 \times 7.7\lambda_0$ | 2.2 |
| [150]/2014 | 80 | LRR | VP (16× 16) RX, (2× 16) TX | $\pm 7^0$ | - | 25.6 | 1.5 | (WFN) $5.9\lambda_0 \times 6.4\lambda_0$ | 3.38 |
| [151]/2015 | 77 | LRR | Varying patch (2× 1) | $\pm 15^0$ | - | 18.5 | 2.4 | $10.3\lambda_0 \times 10.3\lambda_0$ | 3.38 |
| [152]/2016 | 77 | LRR | MC (32× 32) | $10^0$ | -- | 24.7 | 0.7 | $10.8\lambda_0 \times 19.2\lambda_0$ | 3.38 |
| [58]/2018 | 77 | LRR | 10-element series, AMC | $20^0$ | -- | 4 | - | (WOFN) $9.7\lambda_0 \times 3.6\lambda_0$ | 3 |
| [144]/2018 | 77 | LRR/ MRR | SIW (6× 16) | $\pm 7.5^0$, $\pm 40^0$ | -- | 21.7 | 2.9 | (WOFN) $3.9\ \lambda_0 \times 7.9\lambda_0$ | 2.2 |
| [16]/2017 | 77 | LRR/ MRR | Microstrip +SIW (6*10) | $\pm 7.5^0$, $\pm 40^0$ | -- | 20 | 9.5 | (WOFN) $13.2\lambda_0 \times 5.1\lambda_0$ | 2.2 |
| [153]/2015 | 24 | SRR | Grid 33 elements | 7 | $90^0$ | 13.8 | 6 | $1.44\lambda_0 \times 11.7\lambda_0$ | 3 |
| [15]/2019 | 23.7 | SRR | Patches with mushroom | 10 | $150^0$ | 11 | 1 | $2.9\lambda_0 \times 5.3\lambda_0$ | 3.38 |
| **This Work** | 24.1 | LRR/ MRR | 16 RP VAA | $\pm 7^0$ $\pm 38^0$ | $\pm 7^0$ $\pm 38^0$ | 17 | 1.15 | (WFN) $3.6\lambda_0 \times 2\lambda_0$ | 3.38 |

*RP: rectangular patch, VP: varying patch, MC: microstrip combline, WOFN: without feeding network, WF: with feeding network, VAA: virtual antenna array.

# 3.6 Hybrid Antenna for 77 GHz Automotive Radar

This section introduces a linear antenna array for LRR automotive radar to operate at 77 GHz. The proposed antenna consists of a hybrid radiator and dielectric resonator. The hybrid radiator is a circular patch that feed by aperture method and the dielectric resonator is ring that is fed by the circular patch to operate at 77 GHz. The EBG structure is implemented on the top layer to widen the proposed band, gives high gain and to reduce surface waves. The compact size of one element 3mm×3mm×1.19mm to operate from 75.3 GHz to 80 GHz with gain of 12.3 dBi.

## 3.6.1 Antenna Design

### 1. Configuration

Figure 3.22 shows the geometry of the proposed antenna. All the labeled dimensions are tabulated in Table 3. 4 Antenna parameters (all dimensions in mm). The antenna consists of three layers. The 50-ohm microstrip feed line is designed on a first layer, this layer is made from silicon material with $\varepsilon_r$ = 11.7 and $H_1$=0.278 mm. The second layer is Rogger RT5880 material with $\varepsilon_r$ = 2.2, and $H_2$=0.278 mm then the third layer is a ring dielectric resonator from silicon material with $\varepsilon_r$ = 11.7 and $H_3$=0.635mm.

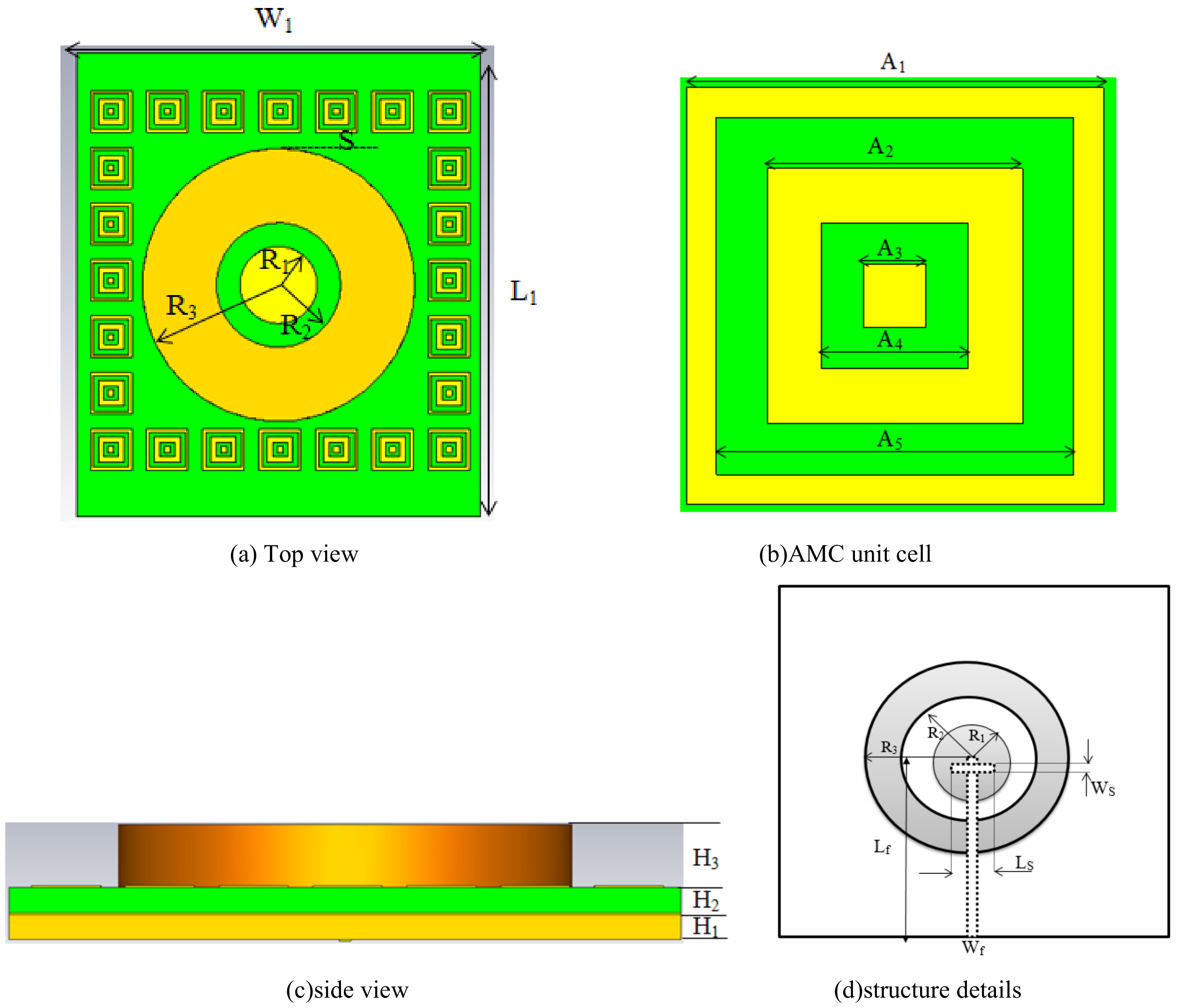

Figure 3.22 Antenna geometry

Table 3. 4 Antenna parameters (all dimensions in mm)

| **$L_1$** | **$W_1$** | **$R_1$** | **$R_2$** | **$R_3$** | **$A_1$** | **$A_2$** | **$A_3$** | **$A_4$** |
|---|---|---|---|---|---|---|---|---|
| 3 | 3 | 0.4 | 0.7 | 0.9 | 0.35 | 0.3 | 0.22 | 0.12 |
| **$A_5$** | **$L_f$** | **$L_s$** | **$W_s$** | **$W_f$** | **$H_1$** | **$H_2$** | **$H_3$** | |
| 0.1 | 1.8 | 0.85 | 0.08 | 0.12 | 0.278 | 0.278 | 0.635 | |

## 2. Analysis of Dielectric Ring

To design dielectric ring resonator, the hybrid mode (hybrid electromagnetic ($HEM_{mn\delta}$)) should be calculated as a first step in our design. The hybrid mode is calculated in first time by Cohn [154] in 1968 and after this developed in [155-157]. The ring resonator parameter can be calculated from the following equations:

$$f_{mn\delta} = \frac{c}{2\pi\sqrt{\varepsilon_{eff}}}\sqrt{\left(\frac{X_{mn}}{r}\right)^2 + \left(\frac{\delta\pi}{2h_{eff}}\right)^2} \quad (3.43)$$

$$h_{eff} = h_2 + h_3 \quad (3.44)$$

$$\varepsilon_{eff} = \frac{{r_{out}}^2 h_{eff}}{\frac{{r_{in}}^2 h_3}{\varepsilon_{rair}} + \frac{({r_{out}}^2 - {r_{in}}^2)\, h_3}{\varepsilon_{r_ring}} + \frac{{r_{out}}^2 h_2}{\varepsilon_{r_sub}}} \quad (3.45)$$

$$\delta = \beta \frac{2h_{eff}}{\pi} \quad (3.46)$$

$$\beta = \sqrt{K_0^2 \varepsilon_{eff} - \left(\frac{X_{mn}}{r}\right)^2} \quad (3.47)$$

$$X_{mn} = \begin{cases} J'_m(x) = 0 & \text{n odd} \\ J_m(x) = 0 & n\ even \end{cases} \quad (3.48)$$

Where $f_{mn\delta}$ resonant frequency, m,n are order of resonant mode, $\delta$ value between 0 and 1 to identify the number of half wavelength changing in z direction, c speed in free space, $X_{mn}$ zero of derivative Bessel function and Bessel function for n odd and n even, respectively.

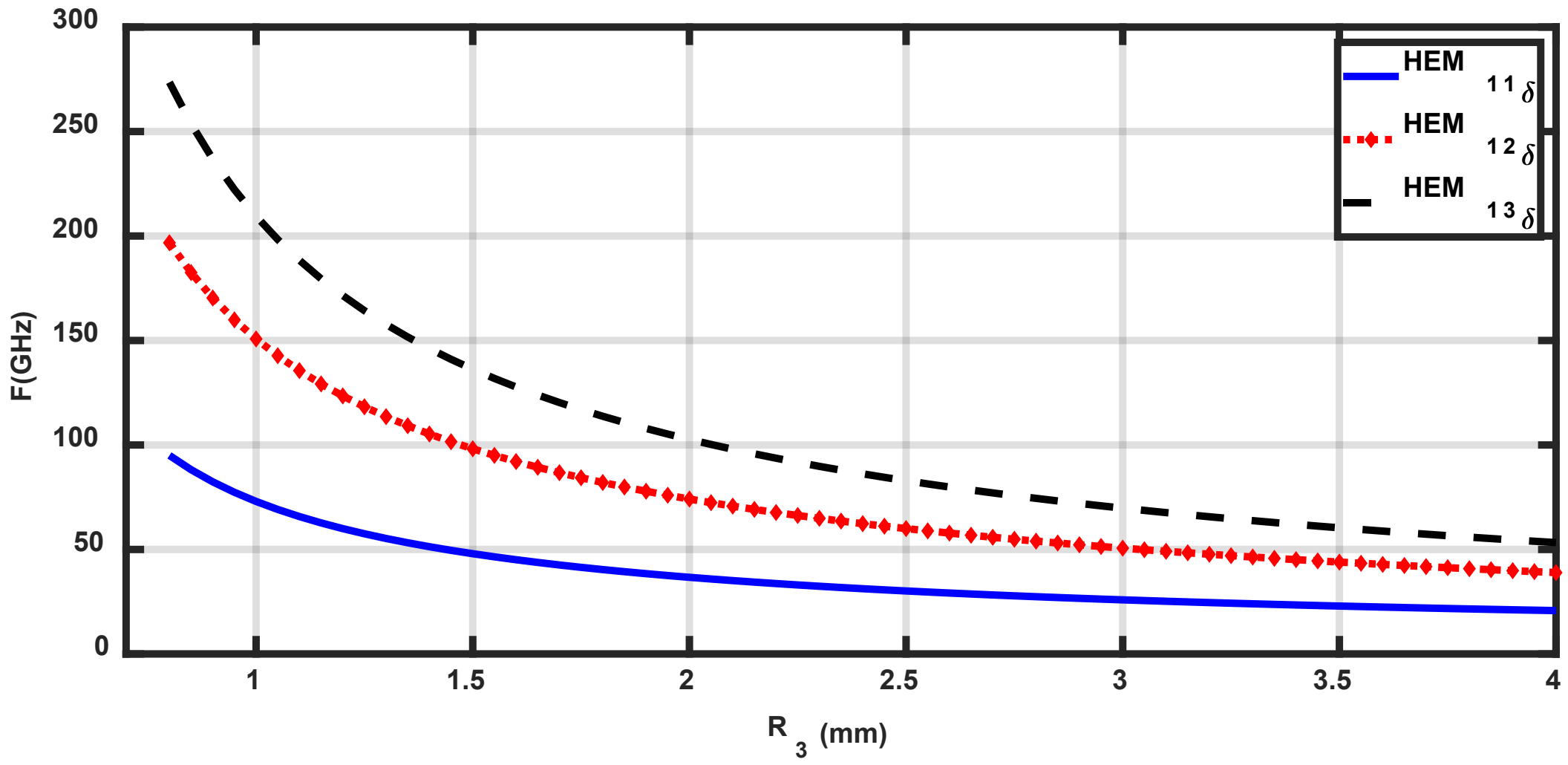

Figure 3.23 Outer radius of dielectric resenator versus frequency for the first three modes
$(\delta = 0.6, R_2 = R_3 - 0.6, H_3 = 0.635\ mm, H_2 = 0.278$ mm).

The antenna is designed to operate with the first hybrid mode $HEM_{11\delta}$. Figure 3.23 shows the relation between the different values of DR outer radius versus the resonant frequency. We notice that at $R_3$=0.9 mm the resonant frequency of first mode is around 77 GHz. The antenna consists of two radiators; the first radiator is circular patch to operate at 77 GHz, and the second radiator is the ring dielectric antenna that is optimized to operate at 77 GHz to give wide bandwidth, and high gain. The patch is designed on top of second layer and it is fed by the aperture slot on the ground plane between first and second layer. The design process; in the first, the circular patch is coupled to the feeder through aperture without ring resonator as conventional techniques and adjust its resonant frequency at 77 GHz. The resonant frequency can be controlled by the slot length which is equal to the half-guided wavelength. A hybrid technique is used and adjusts the dimensions of ring resonator to operate at 77 GHz.

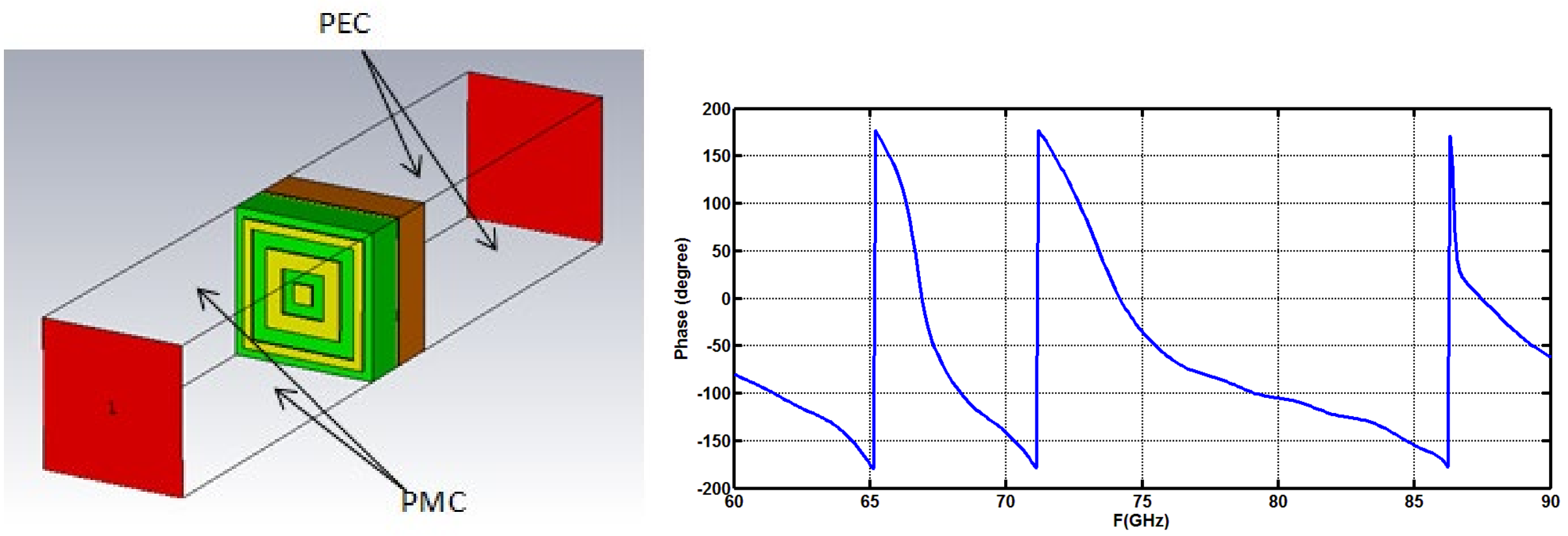

(a)Simulation setup of AMC (b) phase of AMC

Figure 3.24 Artificial Magnetic conductor result

## 3.6.2 Results and Discussion for Single Element

A structure of unit cell of EBG with periodic boundary is simulated to model an infinite periodic surface. The wave port is positioned a half wavelength above periodic surface of the structure, and normal plane waves are launched to illuminate it. The periodic surface is chosen as the phase reference plane. Figure 3.24(a) shows the simulation setup for the planar artificial magnetic conductor (AMC). With this setup, the observation plane and periodic surface are in different locations, so the reflection phase has to be translated to the periodic surface.

The bandwidth of AMC performance is generally defined in the range from $90^0$ to $-90^0$ in this range the BW = 10 GHz with center frequency at 77 GHz as shown in Figure 3.24(b). The EBG structures are added to act as an artificial magnetic conductor, AMC, within its stop bands. The proposed EBG configuration reveals stop bands at 77 GHz for automotive radar applications. This means that it has high surface impedance within this band, where the tangential magnetic field is small, even with a large electric field along the surface. The EBG structure is positioned perpendicular to the antenna.

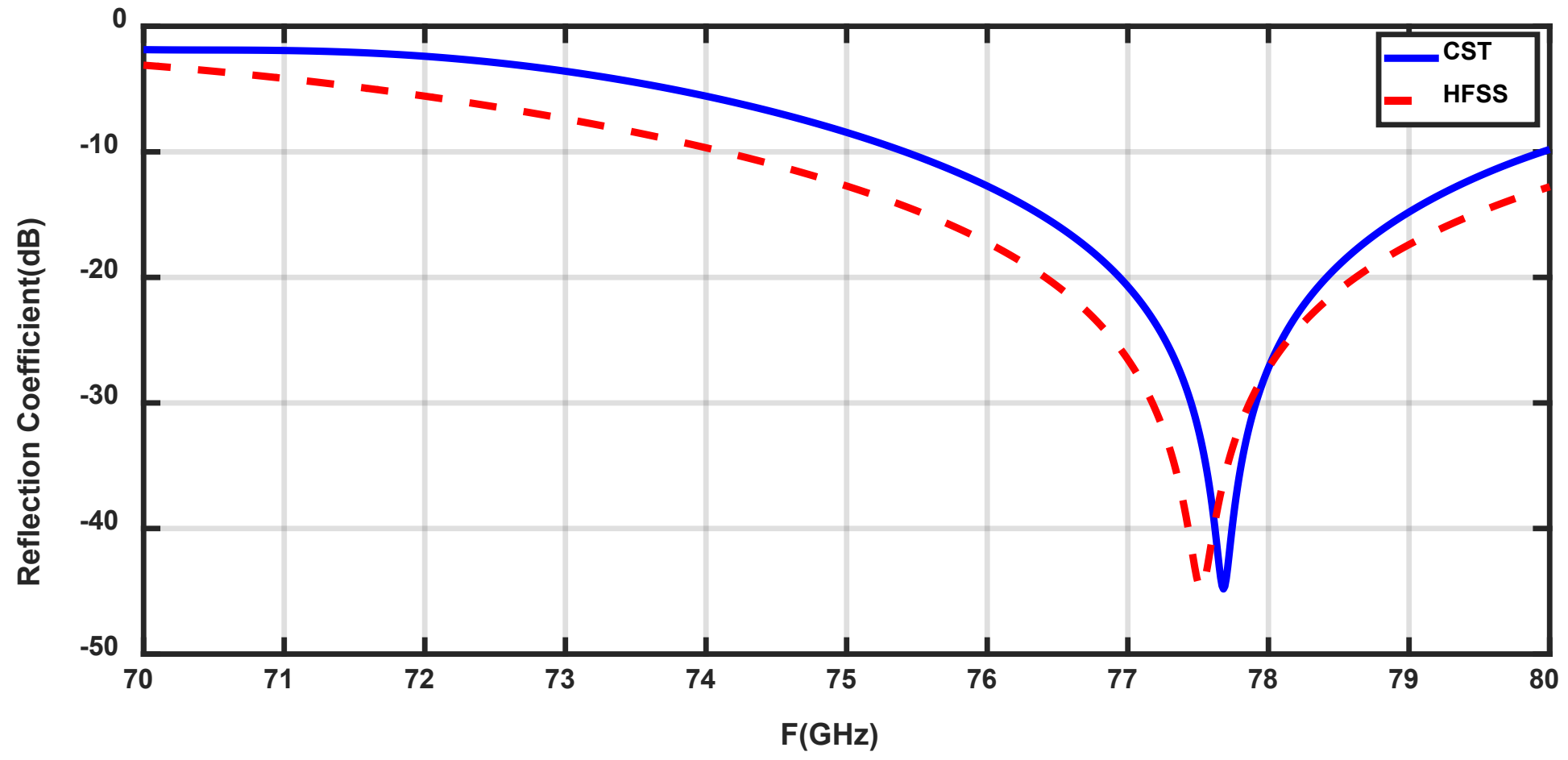

Figure 3.25 Simulated reflection coefficient of the proposed antenna

After applying EBG, the size of antenna is reduced with the final dimensions shown in Table 3. 4. Figure 3.25 shows the return loss of the proposed antenna. Taking the -10 dB return loss as a reference, the antenna operates from 75.3 GHz to 80 GHz. Figure 3.26 shows the radiation pattern of the antenna with high gain 12.3 dBi at 77 GHz. The radiation pattern is directive in the perpendicular plane of the antenna with HPBW $=35^0$ in XZ plane and YZ plane. The gain of the proposed antenna and the radiation efficiency are shown in Figure 3.27.

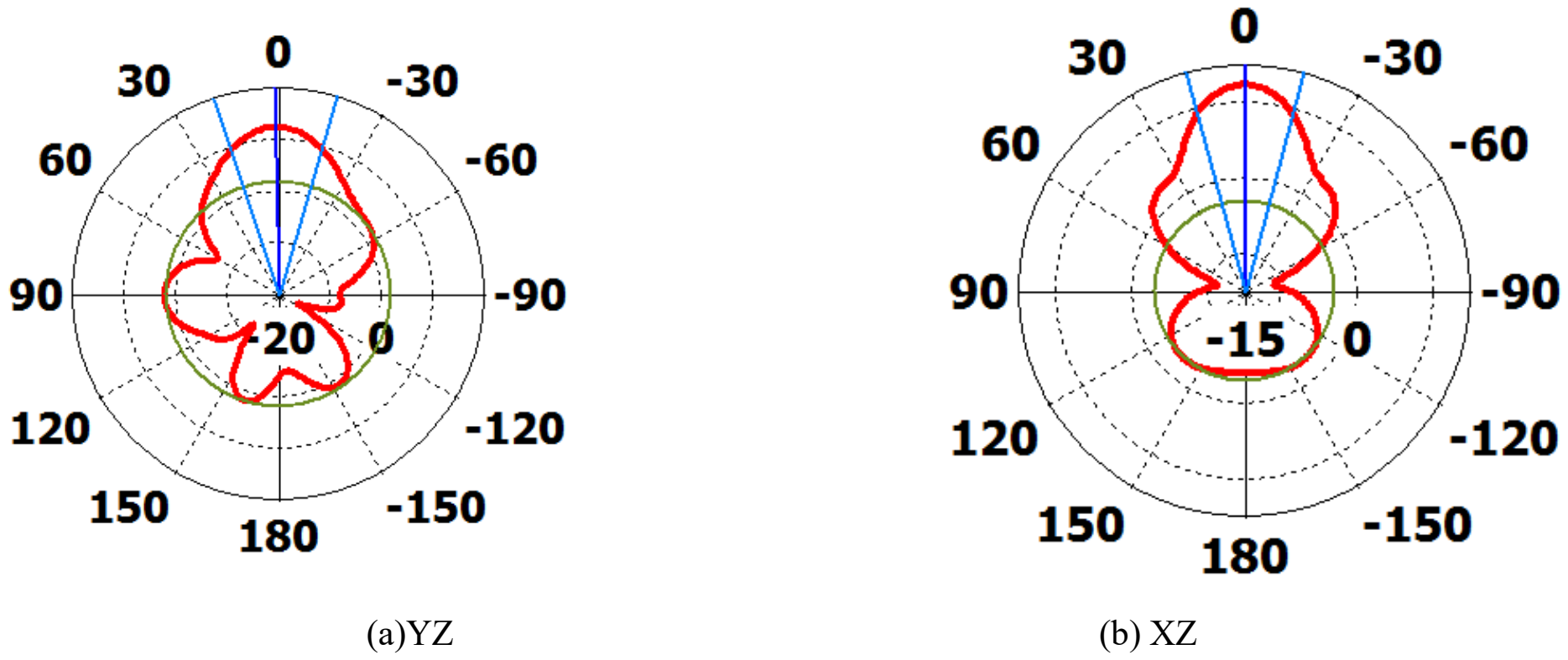

Figure 3.26 Radiation pattern at 77 GHz

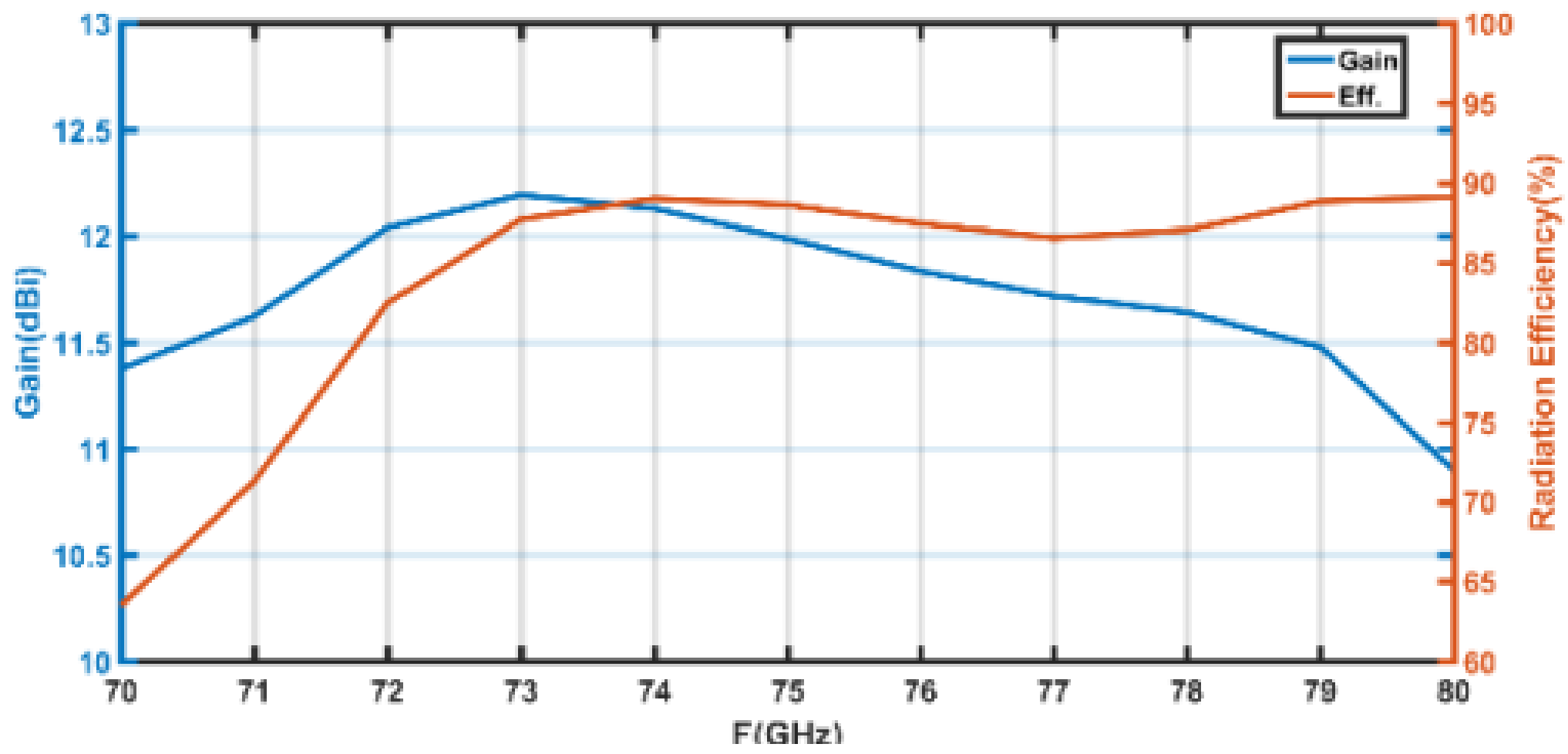

Figure 3.27 Gain and radiation efficiency of the proposed antenna

## 3.6.3 Antenna Array

The recent automotive radar needs to detect the objective at longer distance than 250m. So, the system needs an antenna with high gain, narrow beamwidth and sometimes beam scanning. Therefore, in this section, the antenna array with 8 and 16 elements in series and in parallel are introduced to achieve the requirements of LRR as shown in Figure 3.28 and Figure 3.29, respectively. The antenna arrays have a total size 20.5mm×3mm and 40.5× 3mm for 8 elements and 16 elements, respectively. The thickness of two arrays are the same thickness as one element 1.19 mm with $0.4\lambda_0$ between elements at 77 GHz. The EBG is used to reduce the mutual coupling which has the advantages of reducing the distance between elements to be less than the half wavelength. In series antenna array, the optimized values of slot length and width are ($L_s$=0.9 mm and $W_s$= 0.1 mm) but in the parallel antenna arrays are the same of the aforementioned single element. The EBG structure is used to give high gain, wide bandwidth and isolation between elements to reduce mutual coupling.

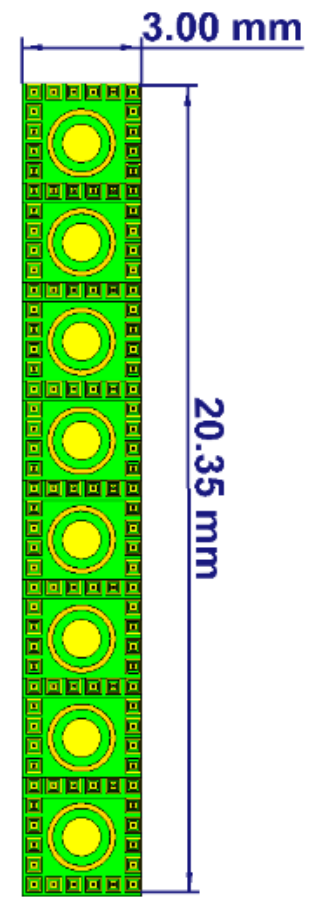

(a)Front of series antenna array

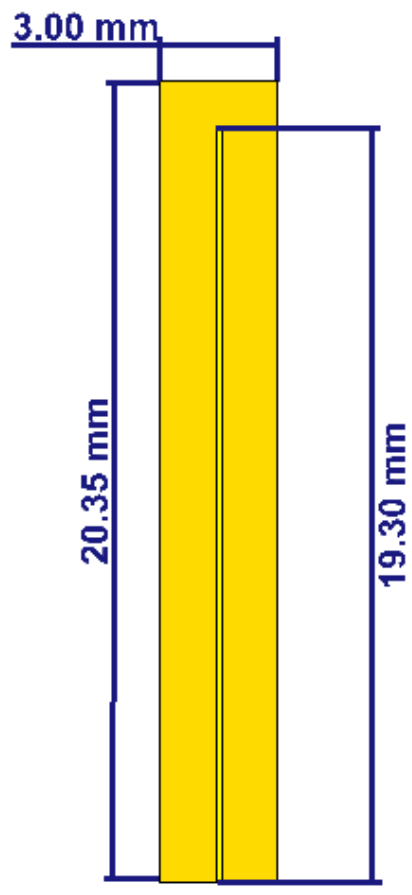

(b)Back of serise antenna array

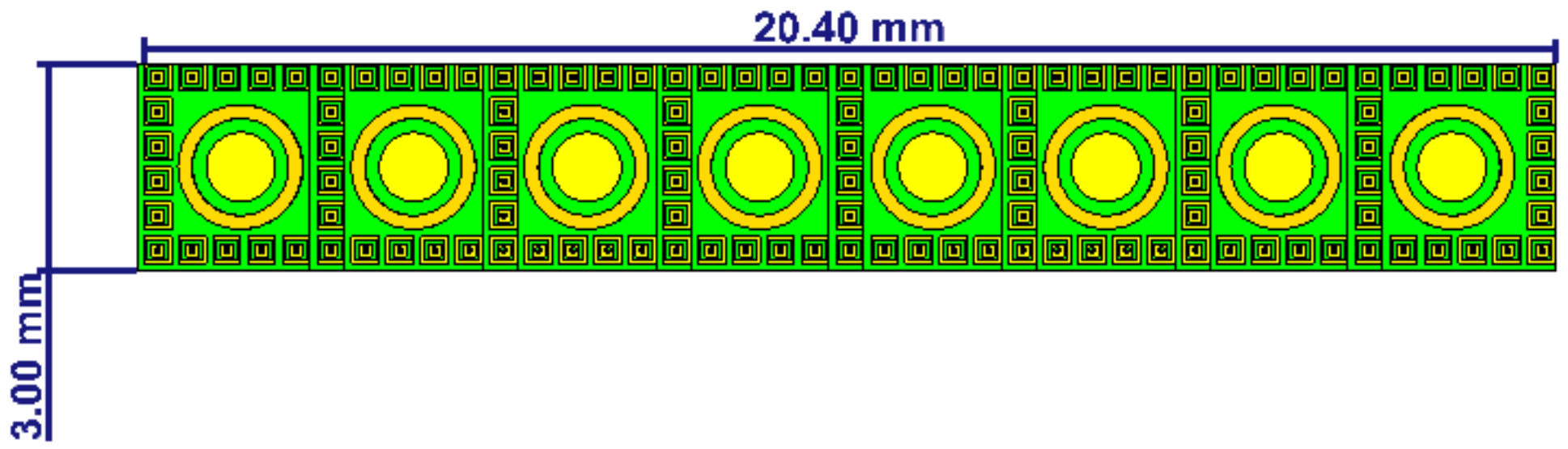

(c)Front of parallel antenna array

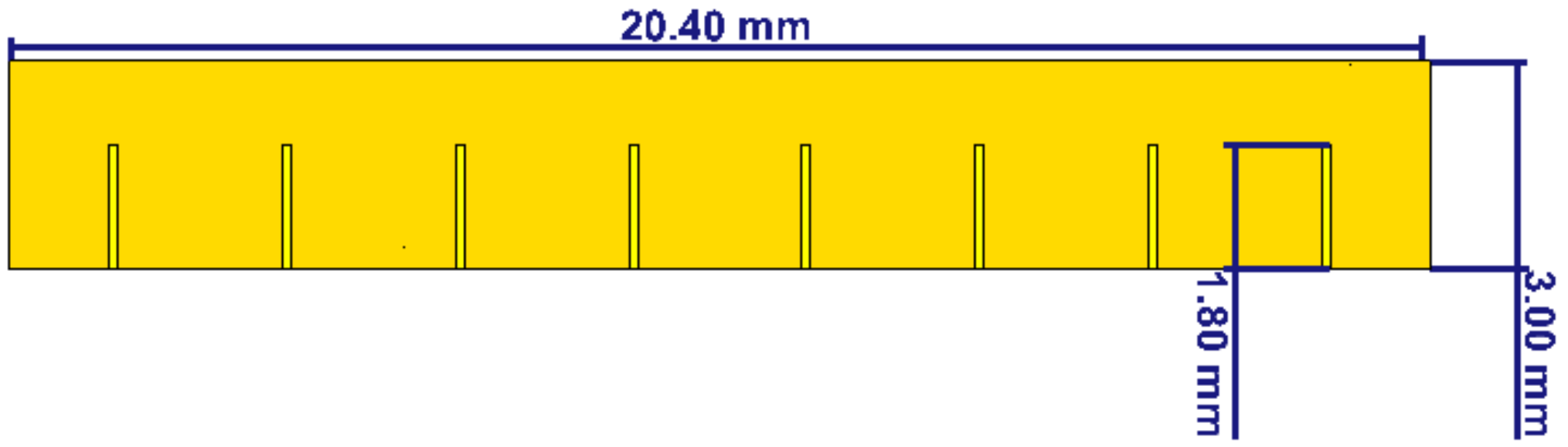

(d)Back of parallel antenna array

Figure 3.28 Geometry of the proposed antenna arrays with 8 elements

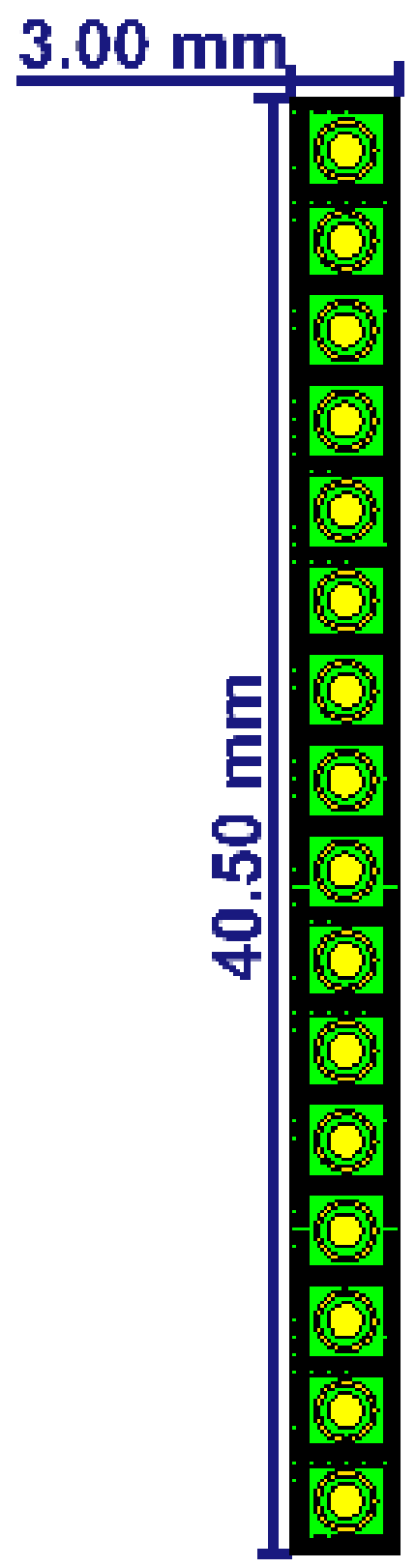

(a)Front of series antenna array

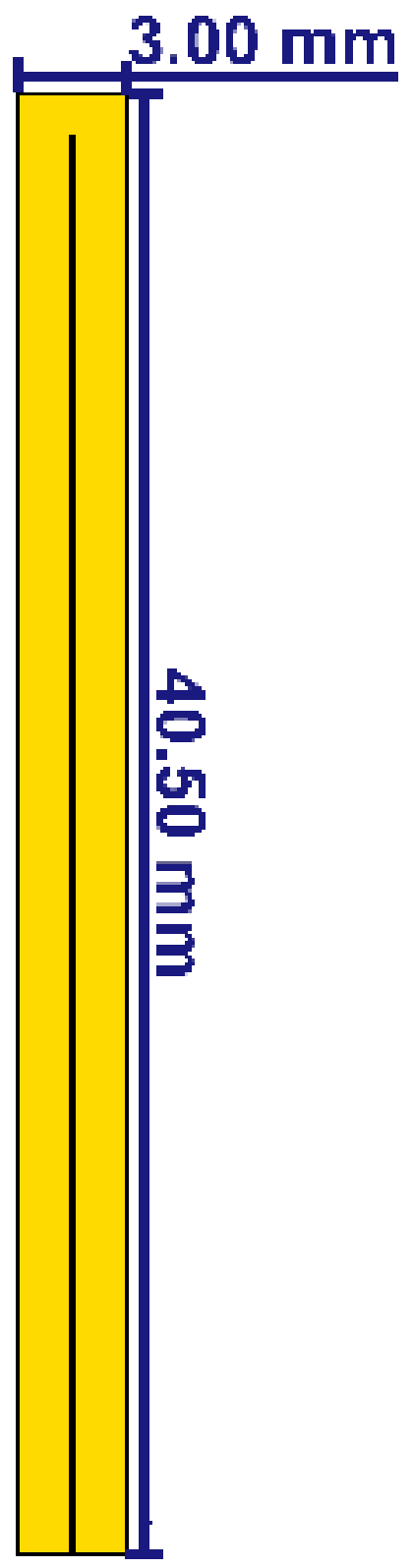

(b)Back of serise antenna array

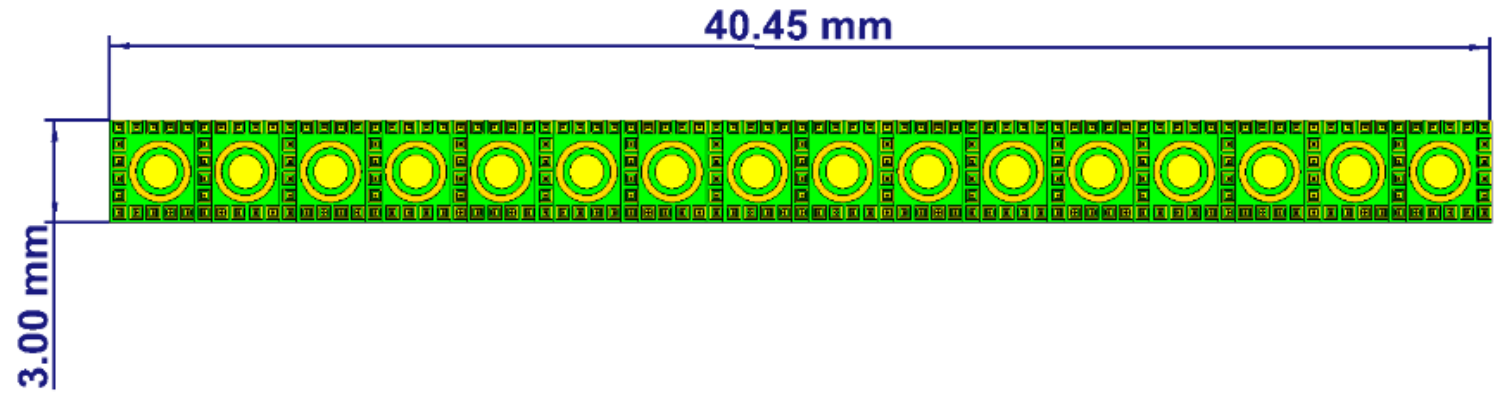

(c)Front of parallel antenna array

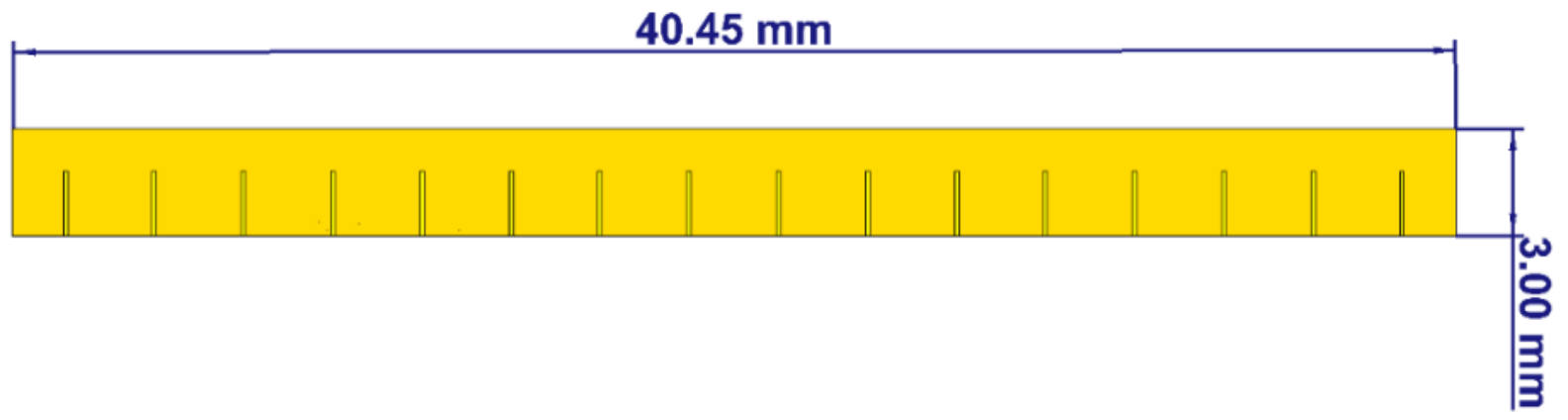

(d)Back of parallel antenna array

Figure 3.29 Geometry of the proposed antenna arrays with 16 elements

The S-parameters of the proposed antenna arrays are shown in Figure 3.30, we notice that the antenna arrays still operate in the proposed band for LRR and the isolation coefficients in case of parallel antenna arrays are below 30 dB. The advantage of series antenna arrays is that they don't need feeding network, with the same number of elements and the same size.

Table 3. 5 presents the comparison between the performances of the proposed antenna arrays. We notice that the two antenna arrays have similar bandwidth but the series antenna arrays have small gain, and small HPBW than that of parallel antenna arrays because the power coupling for the last two or three elements in series antenna array is less than its value for the start elements. The HPBW in parallel less than in series, therefore, longer range is achieved in parallel than in series. The antenna array gain is larger in parallel than in series. 16 elements has narrower BW in parallel than in series.

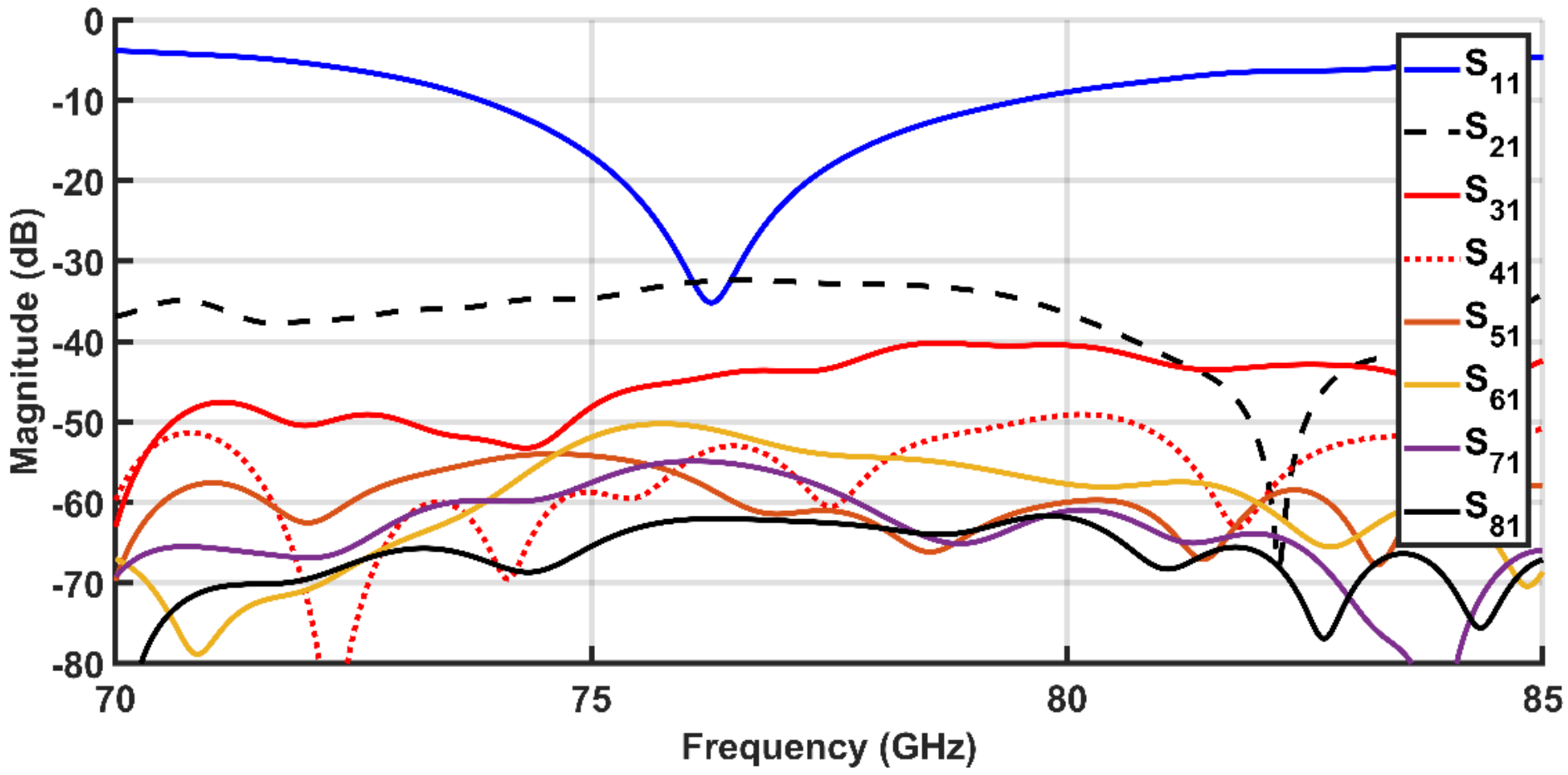

(a)parallel N=8

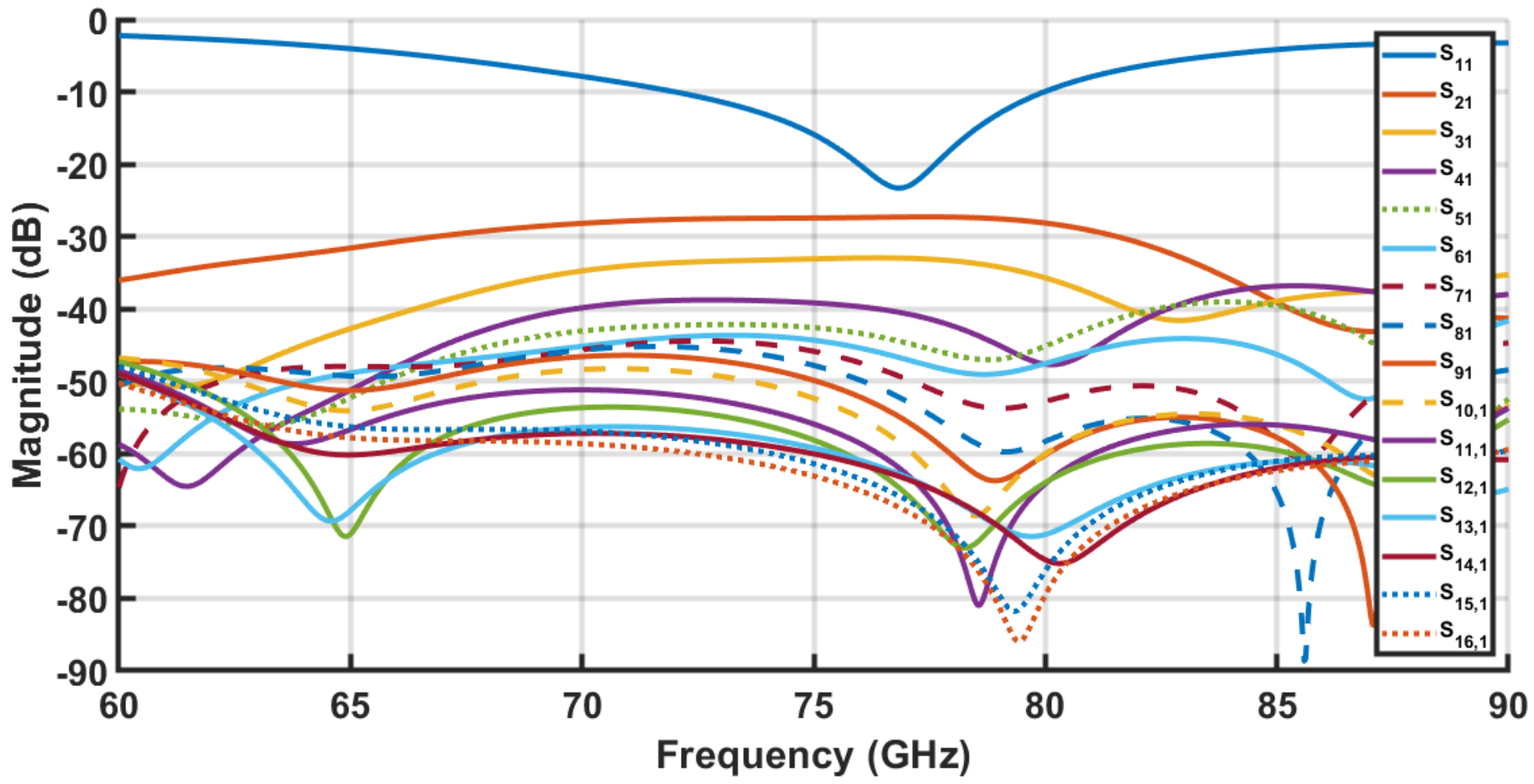

(b)parallel N=16

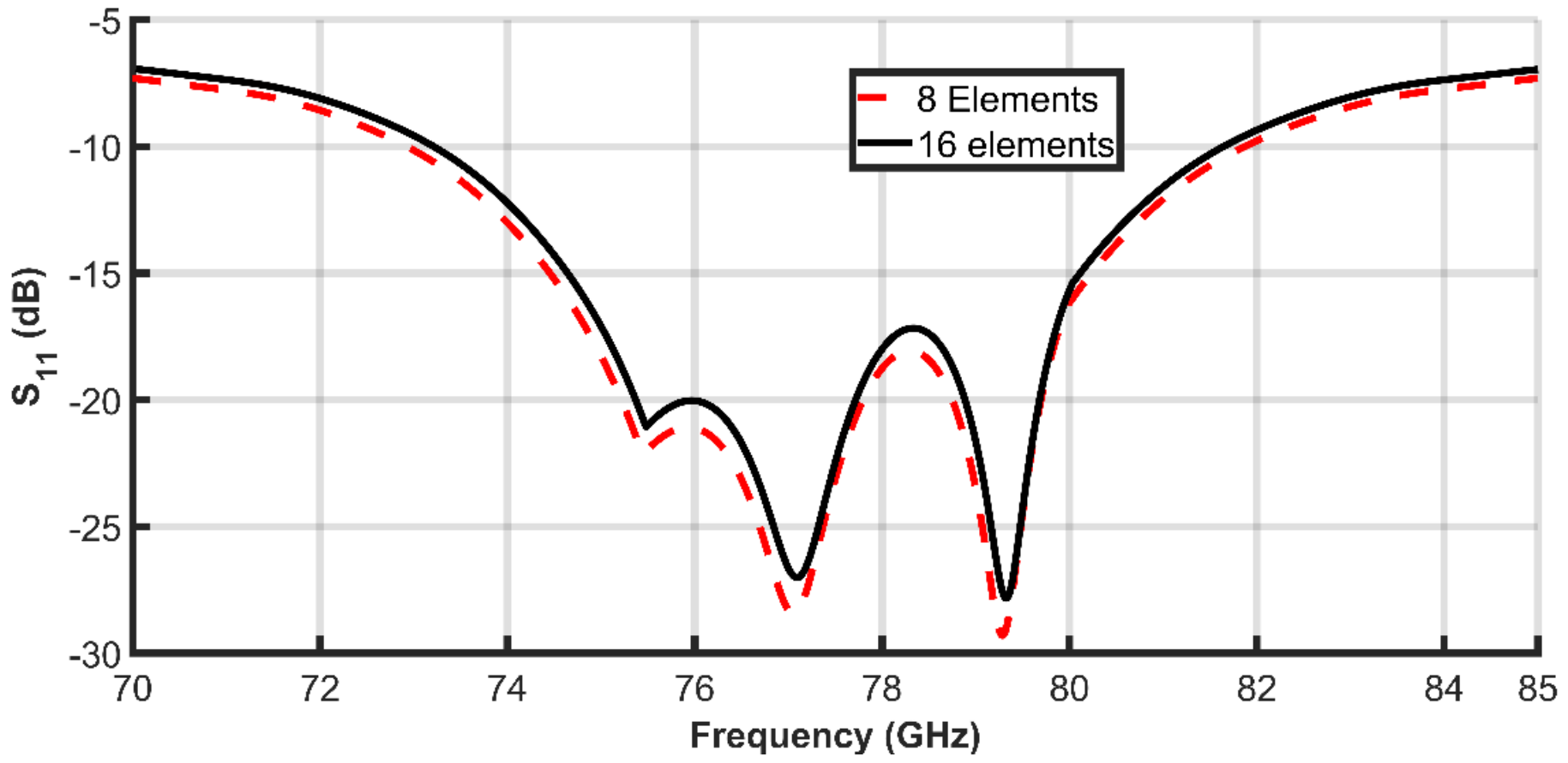

(c)series

Figure 3.30 S-Parameters of proposed antenna array configurations.

The 3-D radiation patterns of the proposed different configurations of antenna array are shown in Figure 3.31. Noted that the parallel configuration has high gain than that of series configuration by about 2.4dBi. Figure 3.32 depicts the comparison between 2-D radiation patterns of antenna arrays in XZ plane and YZ plane. The two configurations have little difference in gain and HPBW. The two antenna arrays have average gain and average efficiency (17.5 dBi, 84.5%) and (19.5 dBi, 85%) for 8 elements series and parallel configurations, respectively. On the other hand, the gain and efficiency of antenna arrays of 16 elements are (20.2 dBi, 82%) and (21.5 dBi, 84.5%) for series and parallel configurations, respectively. The comparison between the parameters of the two configurations are summarized in Table 3. 5. We notice that the HPBW of parallel configurations are better than that of series configuration by about $0.7^0$. Therefore, the parallel configuration can give long range in the radar applications than the series configuration but it still needs a feeding network that will reduce its efficiency.

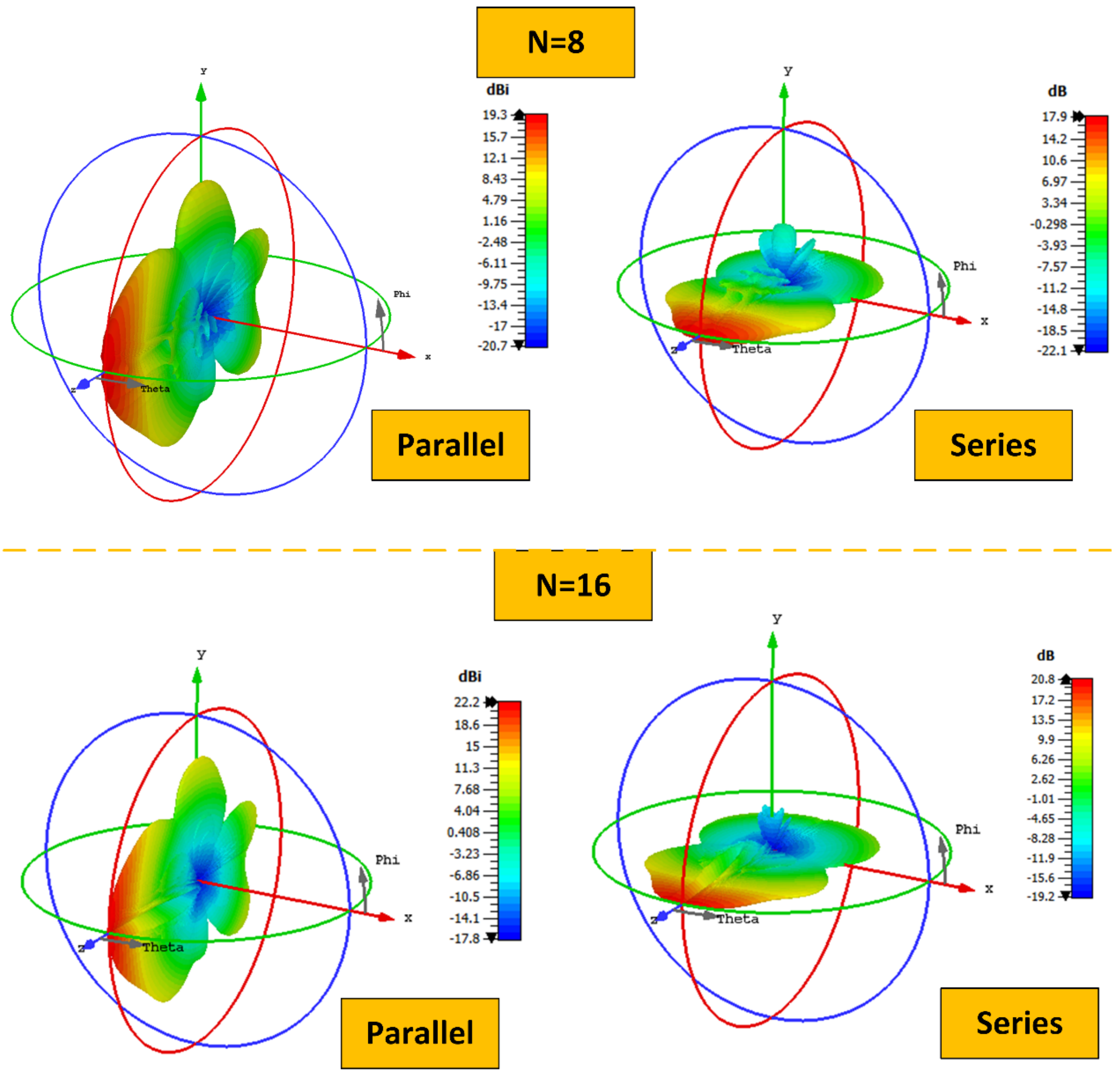

Figure 3.31 3-D radiation pattern of proposed antenna array configurations.

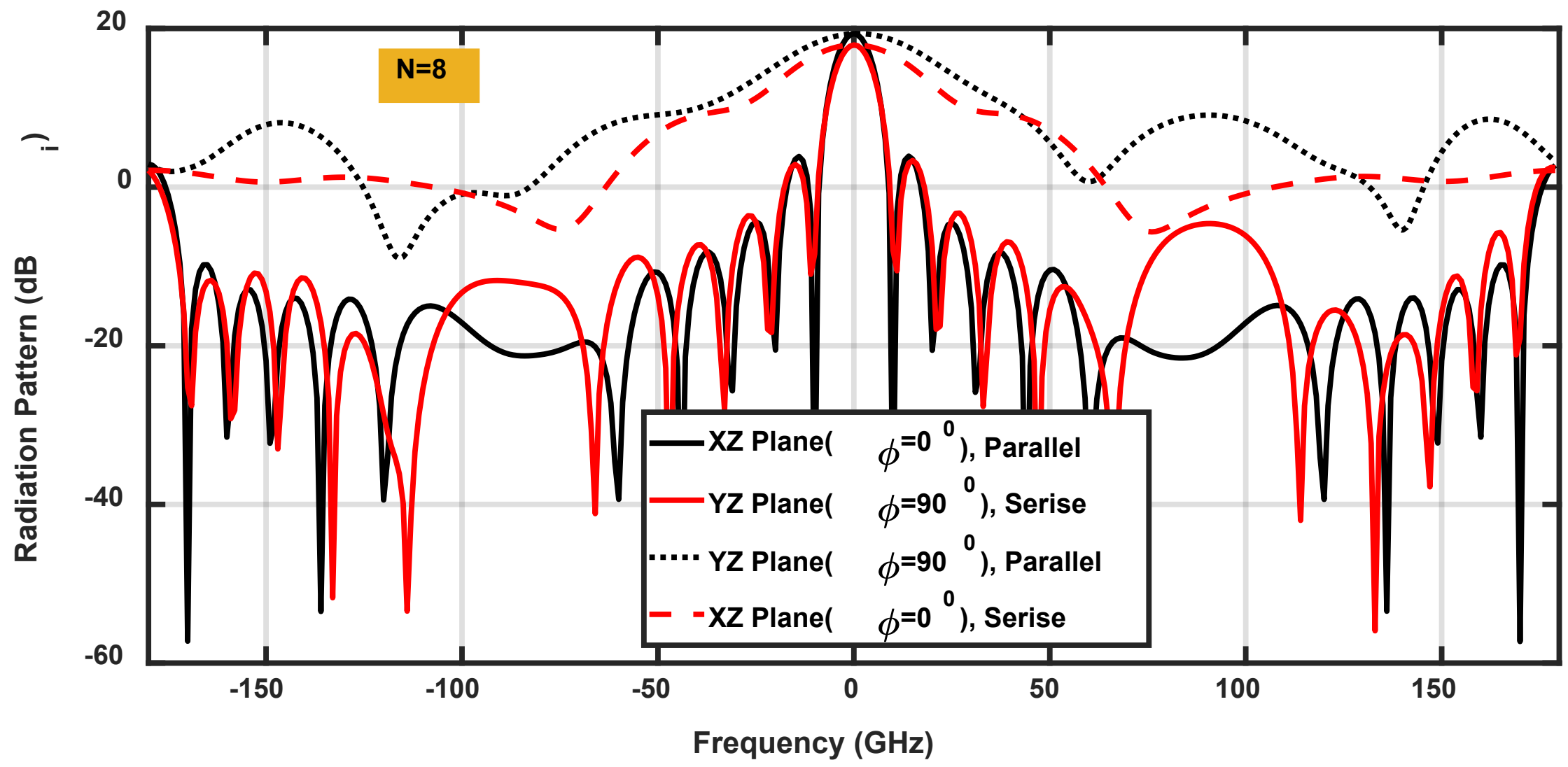

(a)N=8

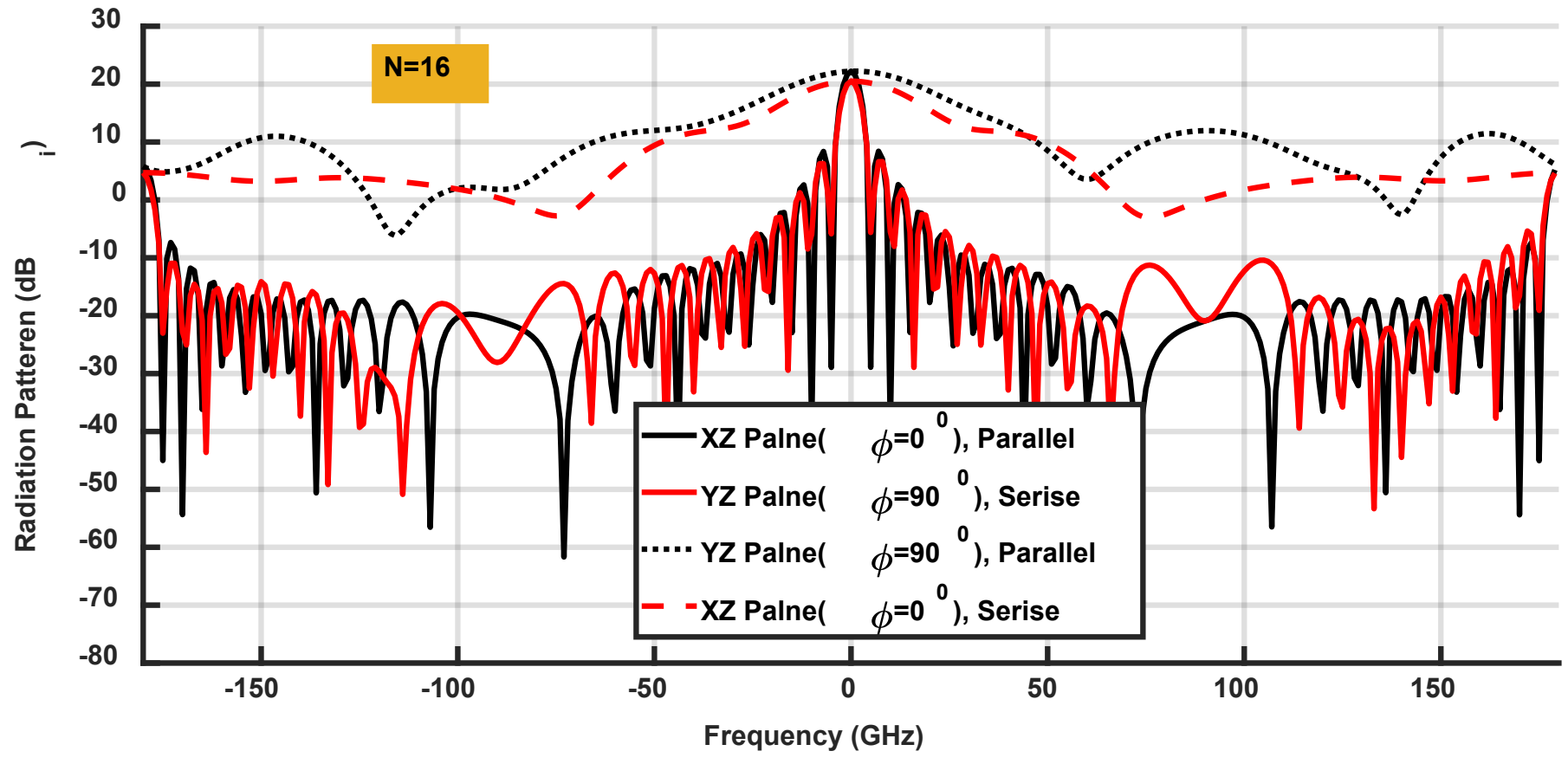

(b)N=16

Figure 3.32 2-D radiation pattern of proposed antenna array configurations.

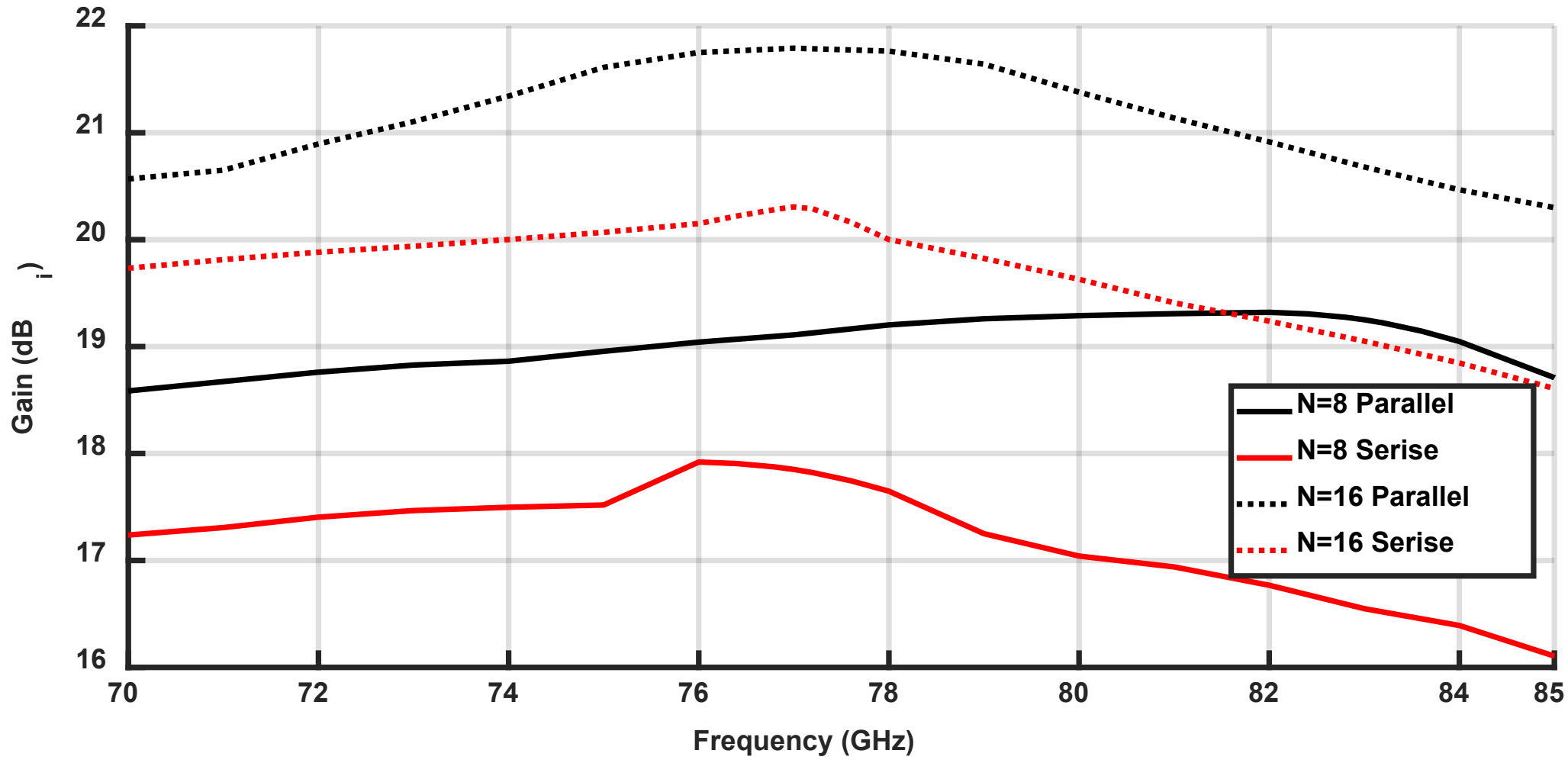

(a)Gain

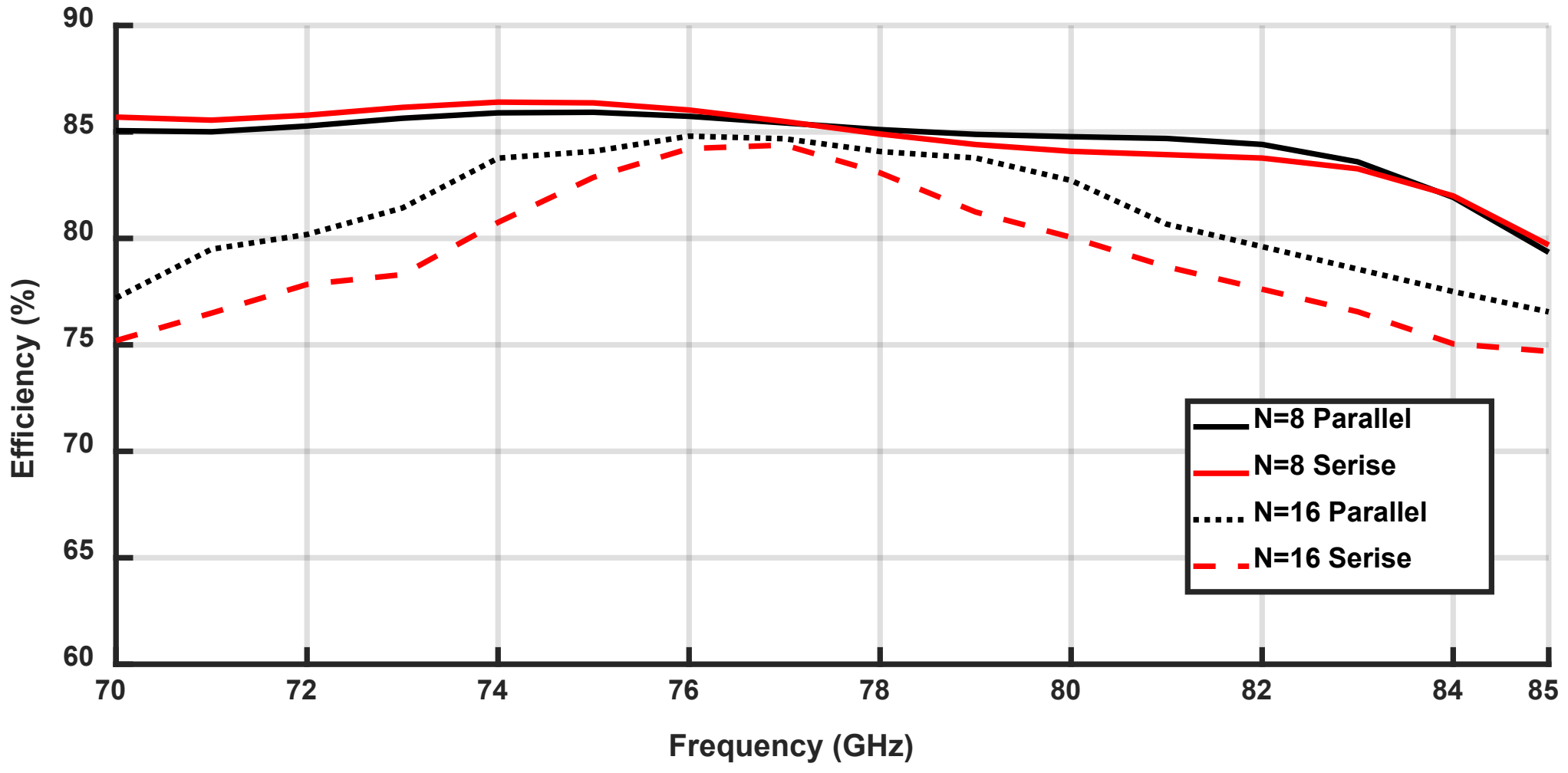

(b)Efficiency

Figure 3.33 Gain and efficiency of proposed different antenna array configurations.

Table 3. 5 Comparison between antenna arrays performance

| Number of Elements | N=8 | | N=16 | |
|---|---|---|---|---|
| Geometry | Series | Parallel | Series | Parallel |
| Size ($mm^2$) | 3×20.5 | 20.5×3 | 3×40.5 | 40.5×3 |
| BW (GHz) | 73-82 | 74-80 | 73-82 | 72-80 |
| Gain (dBi) | 17.5 | 22.2 | 19.7 | 24.7 |
| Efficiency (%) | 86 | 87 | 83 | 84 |
| HPBW(XZ) | $32.5^0$ | $4.5^0$ | $30.5^0$ | $2.8^0$ |
| HPBW(YZ) | $5.3^0$ | $32^0$ | $3.4^0$ | $28.7^0$ |
| Feeding network | No need | Need | No need | Need |

# 3.7 Conclusion

In this chapter, a novel concept in the design of an automotive radar antenna system is introduced. The VAA concept is utilized to have a simple, highly isolated and a high efficient antenna array that competes for the PAA in most of its characteristics. The proposed concept is analyzed and verified using an analytical solution and used the MoM/GA algorithm to get the optimum characteristics of the VAA compared to the PAA antenna. The work ended by a design of the VAA antenna with its feeding network that is based on a cascaded network of Wilkinson power dividers on the same substrate of overall dimensions of 30x48x0.2 $mm^3$. The antenna has an FSS radiation pattern that supports both the MRR and the LRR application simultaneously. The experimental results agree well with simulated results. Furthermore, two configurations of antenna arrays (series and parallel configurations) for LRR at 77 GHz are introduced. These antenna arrays are used to achieve high resolution by introducing small HPBW. Each antenna element consists of two hybrid resonators and AMC surrounding it. The two resonators are used to achieve wide bandwidth in addition to high gain and AMC pattern to achieve high isolation between elements in addition to high gain and wide bandwidth.

# Chapter Four:
# Antenna Design for 5G

## 4.1 Introduction

The researchers still introduce different studies to obtain the optimum specifications of the antenna that can be used for 5G applications, especially for the range from 28 GHz to 38 GHz. Therefore, the antenna design for 5G applications introduced in this chapter has a dual-polarized property to avoid the high attenuation in the millimeter range and give high data rate communications. Furthermore, the MIMO antenna configuration is presented with a comprehensive study for all parameters. The characteristic mode analysis is used here to analyse the antenna performance in addition to using metasurface to enhance the proposed antenna. Section 4.2 introduces the characteristic mode theory, while section 4.3 presents a dual-polarized antenna for a 5G application. Section 4.4 discusses the metasurface, while the antenna using these metasurfaces is given in section 4.5. The conclusion is given in section 4.7.

## 4.2 Characteristics Mode Theory

Over the previous few years, significant progress has been introduced in the field of antennas engineering. To overcome various of difficulties faced in antennas designs, many different methods and design ideas have been introduced. One of the exciting breakthroughs in antenna engineering is the characteristic mode theory (CMT) or theory of characteristic mode (TCM) or characteristic mode analysis (CMA), which focuses on its comprehensive implementation in many critical antenna models. Its promising potential has always attracted antenna engineers ' attention [158-167].

The characteristic modes (CMs) are set of current modes for arbitrarily shaped conducting bodies that are presented numerically. They are independent of any excitation and depend only on the conductive object's size and shape. For PEC, the CM theory is widely defined in [168, 169]. After this, it has been spread for composite metal-dielectric objects, where the results of the characteristic modes are achieved using a surface integral equation-based MoM.

A set of currents (i.e. Current distributions) is detected when TCM is applied to an object; each current is unique in its distribution. The other significant benefits of TCM in comparison to other computational electromagnetic methods are better described. While there are benefits for all computational electromagnetic solvers, TCM is unique as unlike any other. This is partly due to the fact that TCM does not require excitation sources to be placed when determining the electromagnetic characteristics of an analyzed object. These solvers usually involve a physical excitation or feeding component to excite an antenna/object. Once solutions are provided from these kinds of computational solvers, only single current distribution will be achieved, which generates a specific pattern of radiation.

The CMT has a history that is brief but complex. In 1965 [168], Garbacz introduced the idea that the electromagnetic characteristics of an object can be defined by a linear combination of modal field patterns, which are determined only by the shape and material properties of the object. This initial definition used a scattering matrix that describes an object's interaction with an exciting wave as a mathematical representation. This definition offered theoretical proof of the possibility of decomposing any excitable current on an object into an infinite set of radiating currents, with the most significant having the lowest magnitude of its eigenvalue.

Each eigenvalue is described in terms of the radiation resistance of the respective current. However, this theoretical introduction to CMs needed a previous understanding of the distinctive patterns of far-field radiation and their related values. Garbacz [170, 171] has provided two possible solutions to find the unknown modes; both methods are difficult to implement and cannot be used on all geometries of objects. There has been no published study on this subject for another six years with this important drawback.

The researches and studies of theoretical or experimental applications of CMT have been developed and grown significantly over the last 10 years. Most of these applications are summarized as a synthesis of the antennas, small/compact antennas, MIMO systems, pattern synthesis, and scattering problems [158-167, 172-176].

## 4.2.1 Analysis of CMs

As described in [169], the characteristic modes or characteristics currents are obtained by solving a particular eigenvalue equation that is derived from the Method of Moments (MoM) impedance matrix,

$$ZJ = E_{tan}^{i} \text{ where } Z = R + jX \tag{4.1}$$

From (4.1), the characteristic modes (CMs) are introduced based on the coefficient matrix's generalized values by Harrington [177]:

$$Z(\overrightarrow{J_n}) = \nu_n R(\overrightarrow{J_n}) \tag{4.2}$$

Where $\nu_n$: eigenvalues, R is a real part of the impedance matrix/operator for the MoM, $\overrightarrow{J_n}$ is an eigencurrent (eigenfunction) and from equation (4.1) and (4.2)

$R(\overrightarrow{J_n}) + jX(\overrightarrow{J_n}) = v_n R(\overrightarrow{J_n})$ (4.3)

$X(\overrightarrow{J_n}) = \lambda_n R(\overrightarrow{J_n})$ (4.4)

$\lambda_n = \frac{v_n - 1}{j}, \quad v_n = 1 + j\lambda_n$ (4.5)

Where $\lambda_n$ corresponding characteristic values to eigenvalues, sometimes called eigenvalues because they have a direct relation with $v_n$. R gives an orthogonal radiation pattern of CMs. To solve the MoM equation, the CMs are used as a basis function to expand the unknown total current, J, on the surface of the metal as

$J = \sum_n \frac{V_n^i J_n}{1 + j\lambda_n}$ , The excitation coefficient can be expressed as $V_n^i = \sum_n E^i J_n$ (4.6)

Where $V_n^i$ the excitation coefficient on the conductor is surface and $E^i$ is the impressed E-field. . The eigenvalues have a range from $-\infty < \lambda_n < +\infty$ that can be divided into three regions:

- $-\infty < \lambda_n < 0$ , CM has capacitive and store electric energy.
- $\lambda_n = 0,$ CM has a resonate frequency and radiates efficiently.
- $0 < \lambda_n < +\infty$, CM has inductive and store magnetic energy.

The first step in the analysis of the CMs, is to get the eigenvalues because they introduce the data on how the related modes $(J_n)$ radiate and how they relate to the resonance. In CMA, many modes can be determined dependent or equal to the number of unknown numbers in the equation of MoM at every frequency. The eigencurrents, unaffected by the type of source or excitation method, depend only on the shape and dimensions of the structure and its operating frequency. Also, the total current on an antenna's surface can be calculated as a summation of eigencurrents of the antenna. Therefore, the eigenvalues are used as an indicator to know the resonant frequency for each characteristic mode.

The multilayer solver and integral equation solver from CST microwave studio is chosen to implement the CMT with very high quality [158, 159, 166]. These solvers are used to calculate the characteristic modes and those related parameters.

The second parameter for CMA is a characteristic angle $(\alpha_n)$ which is used to describe the antenna operation and performance. The Characteristic angle calculates the difference in phase between the characteristic currents $(J_n)$ and its related characteristic field. The characteristics angle can be calculated from the following relation

$\alpha_n = 180^0 - tan^{-1}(J_n)$ (4.7)

The characteristic angles values are within this range $0^0 \leq \alpha_n \leq 360^0$ that can be divided into three regions:

- $\lambda_n > 0, \ \alpha_n < 180^0$ (CM has inductive and store magnetic energy)

- $\lambda_n < 0, \alpha_n > 180^0$ (CM has capacitive and store electric energy)
- $\lambda_n = 0$ (CM is in resonance)

The third parameter for CMT is a model significance (MS). Equation (4.5) demonstrates the main parameters that affect the significance of each CM to a radiated field and from equation (4.6) the term $\left|\frac{1}{1+j\lambda_n}\right|$ seems more compatible to express the variation of the $J_n$ rather than the variation of $\lambda_n$. This term represents the inherent normalized amplitude for each current mode $J_n$ and it is named the modal significance. If its value close to 1, the mode meets the resonance condition.

$$MS_n = \left|\frac{1}{1+j\lambda_n}\right| \tag{4.8}$$

From MS equation (4.8), we are able to calculate the resonance of the CM in addition to the operating bandwidth of CM. The CM that has a resonance must be at $\lambda_n = 0$ and $MS_n = 1$ and the CM that doesn't contribute to the radiated field is at $MS_n = 0$. Furthermore, the half-power radiating bandwidth can be calculated from the following approximation formula [178]:

$$BW_n \approx \frac{F_h(MS_n = 0.707) - F_l(MS_n = 0.707)}{F_{res}(MS_n = 1)} \tag{4.9}$$

$F_h$ and $F_l$ are the edges of the high and low frequency bands of any maximum where the MS is equal to or higher than 0.707. $F_{res}$ is the resonant frequency where $MS_n = 1$ or the place of the highest MS. Modal bandwidth is often a significant parameter in many CMT antennas, as it helps to determine the radiation characteristics of the CM. It is important to understand, however, that the modal bandwidth corresponds to the half-power bandwidth of the radiated pattern (for single-mode excitation) and not an exciting structure's impedance bandwidth.

## 4.3 Dual Polarized Antenna for 5G

Due to the mastering of the dual-polarized antennas to introduce a solution in enhancing the isolation and channel capacity, makes these antennas a good candidate for MIMO smartphone designs [69, 87-90, 179, 180]. In [87] Yang Li et al., introduce a hybrid eight ports orthogonal dual-polarized antenna for 5G smartphones; this antenna consists of 4 L-shaped monopole slot elements and C-shaped coupled fed elements. The 4 L-shaped are printed at the corners and the 4 C-shaped are printed in the middle on a thick 1mm FR-4 substrate. This design achieves 12.5 dB, 15 dB for the isolation and the cross-polarization, respectively. Over the past months, Zaho et al. [88] presented a 5G/WLAN dual-polarized antenna based on the integration between inverted cone monopole antenna and cross bow-tie antenna for VP and HP, respectively, where a 90◦ phase difference feeding network feeds the cross bow-tie antenna, so, the separated power divider and phase shifter are introduced to be used as a feeding network. In [89] Huang et al. introduce a dual-polarized antenna that consists of a main radiator, an annulus, and a reflector. The main radiator consists of two pairs of differentially-driven feedlines to transmit the energy to the coplanar patch. This structure achieves 26 dB and 35dB for the isolation and the cross-polarization, respectively. Eight-ports dual-polarized antenna array is reported in [90], the proposed antenna array is composed of four square loops and each loop is excited by two orthogonal fed coupled feeding strip.

## 4.3.1 Antenna Design

To achieve a dual-polarized antenna, two slot antennas are introduced and etched on the same substrate as shown in Figure 4.1. The proposed antenna consists of two orthogonal slots to achieve pure dual-polarization and gives two pattern diversity.

The proposed antenna is implemented on a Rogers 4003C substrate with a dielectric constant 3.38, tangential loss of 0.0027 and thickness h=0.2 mm. The thin substrate is selected to reduce the losses at the millimeter band and to be compatible with the end launch connector (1.85 mm) with a very thin pin. All the simulation, optimization and CMA were conducted with the computer simulation technology (CST): time-domain solver and multilayer solver.

The slot antenna is one of the popular antennas used for smartphones in the last few years [65, 181-188] due to its simple structure and its multiple operating modes. Figure 4.1(a and b) shows the proposed slot antenna (Ant. I), where a slot is etched on the top ground plane and is fed by the 50-ohm microstrip line. The feed line is printed on the opposite side of the ground plane to feed the slot antenna. The second antenna (Ant. II) is a slot antenna fed by the CPW line, as depicted in Figure 4.1 (c and d). As a result of the thin layer substrate used, the width of the CPW-feeding line is wider than the inner pin of the end-launch connector (The width of the CPW feed line without bottom ground is $W_{f2}$=2.5 mm, while the diameter of the end-lunch pin is 0.18 mm). Therefore, we used a CPW with a ground plane to decrease the width of the CPW-feeding line and achieve 50-ohm input impedance and be compatible with the end-launch connector. After this, the CPW-feeding line is tapered to match the aperture/slot impedance.

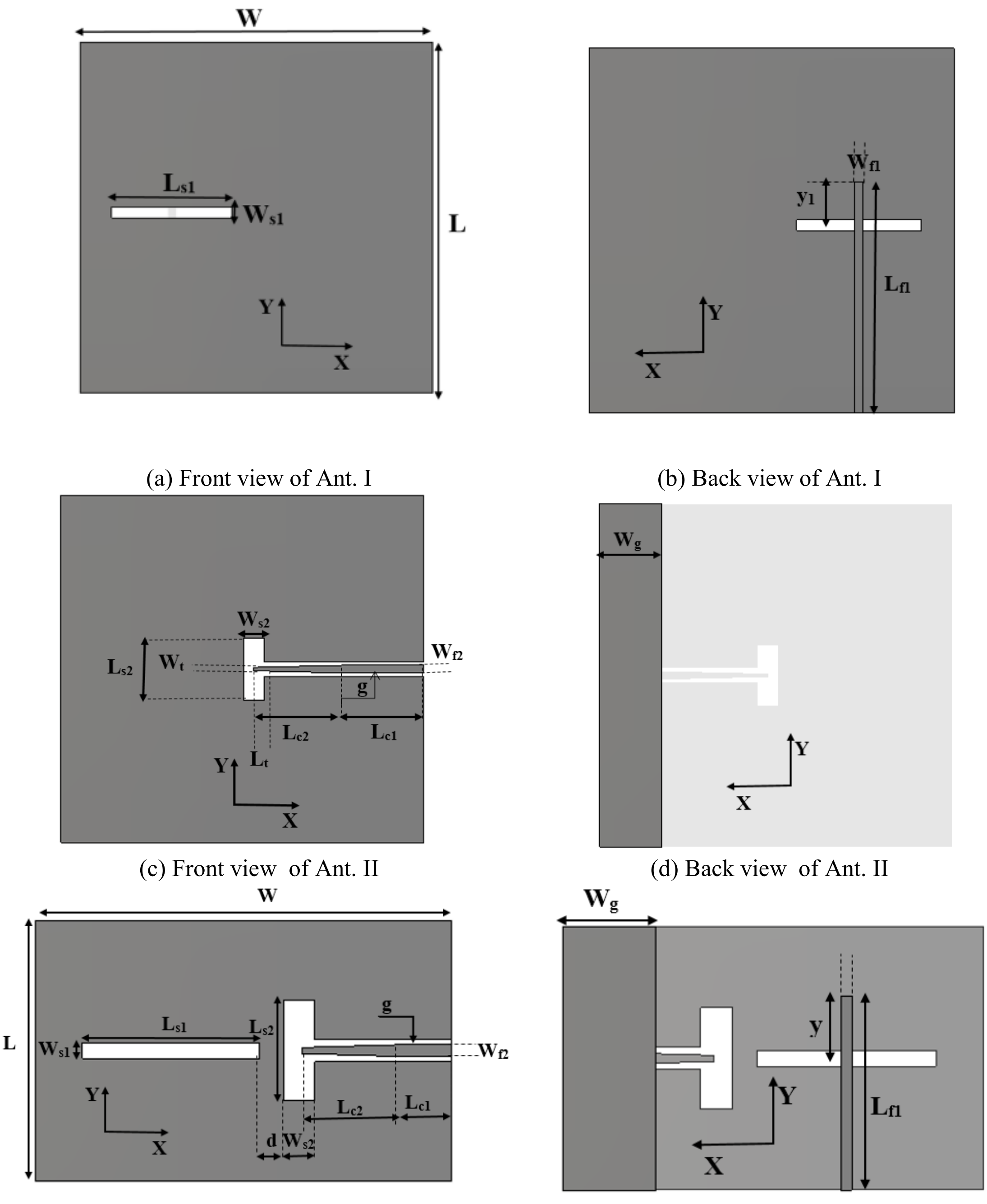

(a) Front view of Ant. I
(b) Back view of Ant. I
(c) Front view of Ant. II
(d) Back view of Ant. II
(e) Front view of Ant. III
(f) Back view of Ant. III

Figure 4.1 Geometry of proposed antenna

The third antenna (Ant. III: proposed antenna) consists of the integration between Ant. I and Ant. II configurations to achieve dual-polarization from the two antennas. The feed lines of antenna I and antenna II are printed on a different face of the substrate as shown in Figure 4.1(e and f), furthermore, the feed lines are orthogonal together. Moreover, the two slots are used to prevent the coupling between the two feed lines. Table 4. 1 gives the geometric dimensions of the proposed antenna (all dimensions in mm).

Table 4. 1 Geometric parameters of the proposed antennas (mm)

| Parameters | An. I | Ant. II | Ant. III |
|---|---|---|---|
| L | 20 | 20 | 10 |
| W | 20 | 20 | 20 |
| $L_{s1}$ | 5 | --- | 5 |
| $W_{s1}$ | 0.6 | --- | 0.6 |
| $Y_1$ | 1.6 | --- | 1.6 |
| $W_{f1}$ | 0.428 | --- | 0.428 |
| $L_{f1}$ | 11.6 | --- | 6.6 |
| $L_{s2}$ | --- | 3.8 | 3.7 |
| $W_{s2}$ | --- | 1.2 | 1.2 |
| $W_t$ | --- | 0.225 | 0.225 |
| $L_t$ | --- | 0.4 | 0.4 |
| $L_{c1}$ | --- | 3.5 | 3.5 |
| $L_{c2}$ | --- | 4 | 4 |
| G | --- | 0.2 | 0.2 |
| $W_c$ | --- | 0.45 | 0.45 |
| $W_g$ | --- | 3.5 | 3.5 |
| D | --- | --- | 0.9 |

## 4.3.2 CM Analysis of Antennas

The CMA is introduced as a first step to provide a method to discover the physical essence of the proposed antennas. In addition to enhancing the bandwidth of the proposed antennas by creating multiple resonant modes using CMA. Figure 4.2 shows the characteristic mode analysis of Ant. I, we notice that the slot antenna achieves multiple operating modes around 28 GHz. Any mode can be considered as a resonant mode when achieve MS=1 (MS>0.707) in addition to achieve $\lambda_n = 0$ and $\alpha_n = 180^0$. However, modes number 4-8 don't have eigenvalue equal to zero but they are very closed to the zero and their modal significant values more than 0.98, therefore they don't consider as a pure resonant mode. Therefore, these modes consist of a small capacitive loaded and a small amount from electric energy. Figure 4.3 depicts the characteristic mode analysis of Ant. II, one can notice that the CPW slot antenna gives a wide bandwidth for resonant modes than that of Ant. I. Figure 4.4 shows the characteristic mode analysis for Ant. III, we can notice that their modes are a combination between modes of Ant. I and Ant. II. Moreover, this antenna achieves wide bandwidth according to their MS values. Table 4. 2 shows a comparison between the resonant frequencies for each characteristic mode of the aforementioned antennas. We notice that Ant. III has a large number of resonant modes within the simulated range.

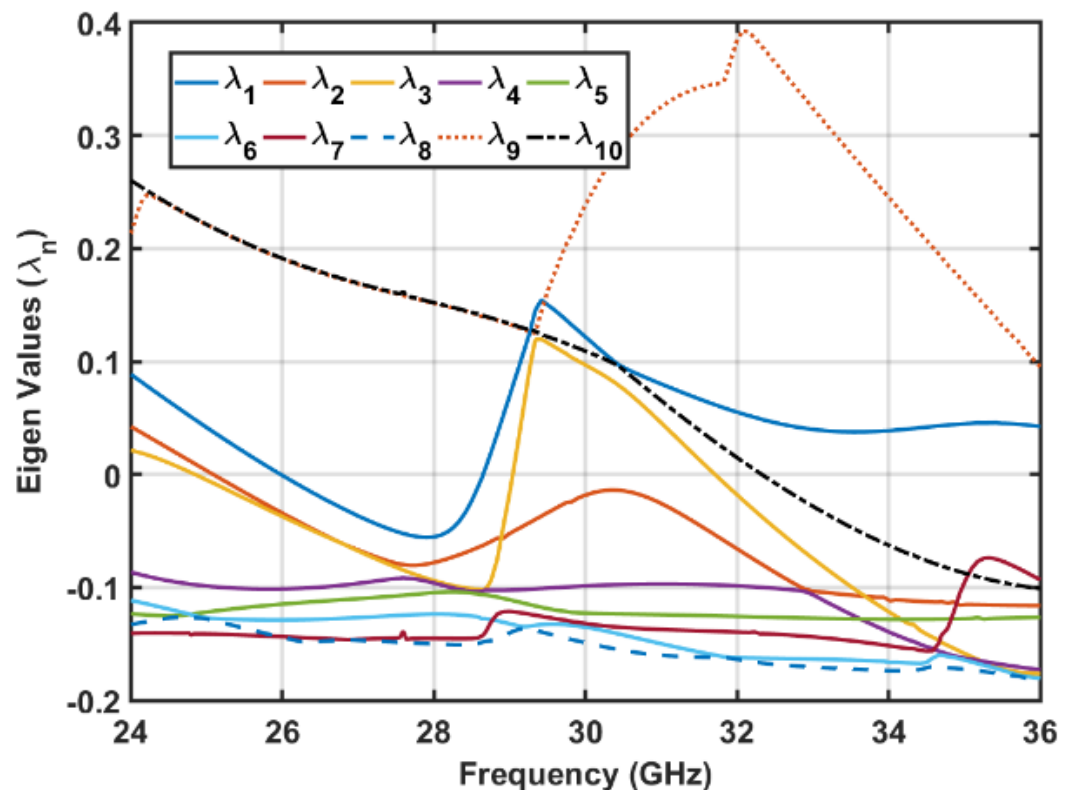

(a) Eigen values

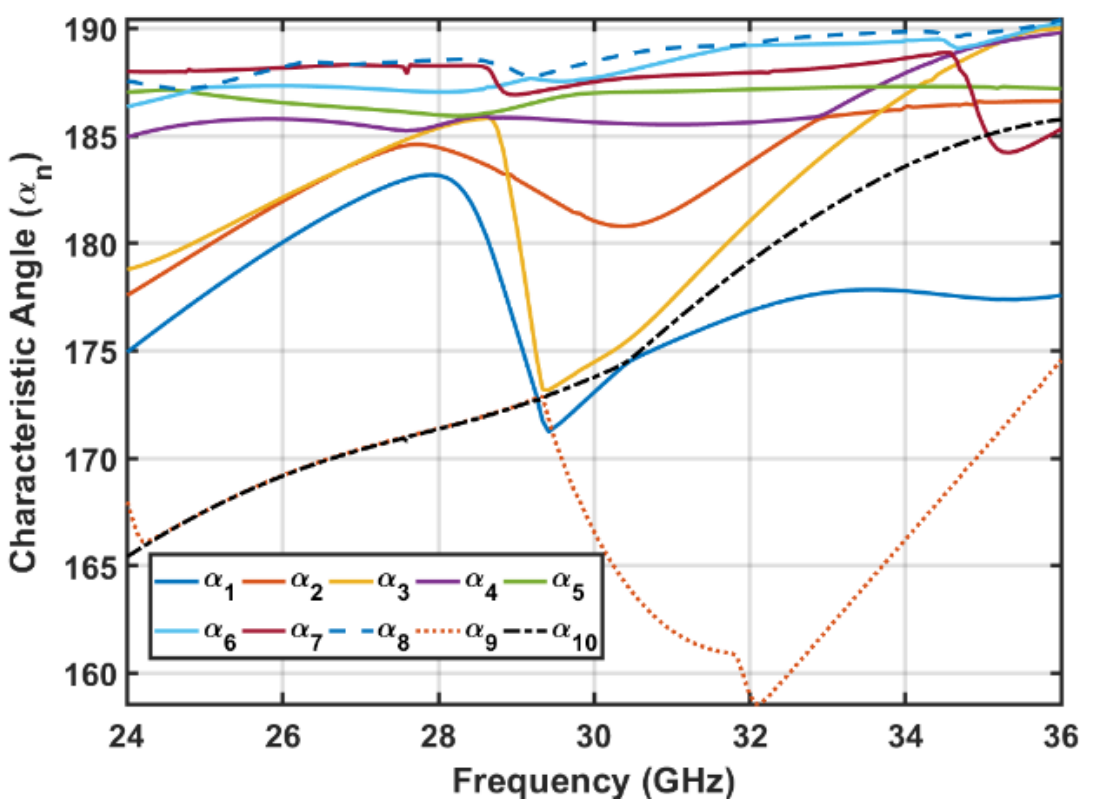

(b) Characteristic angles

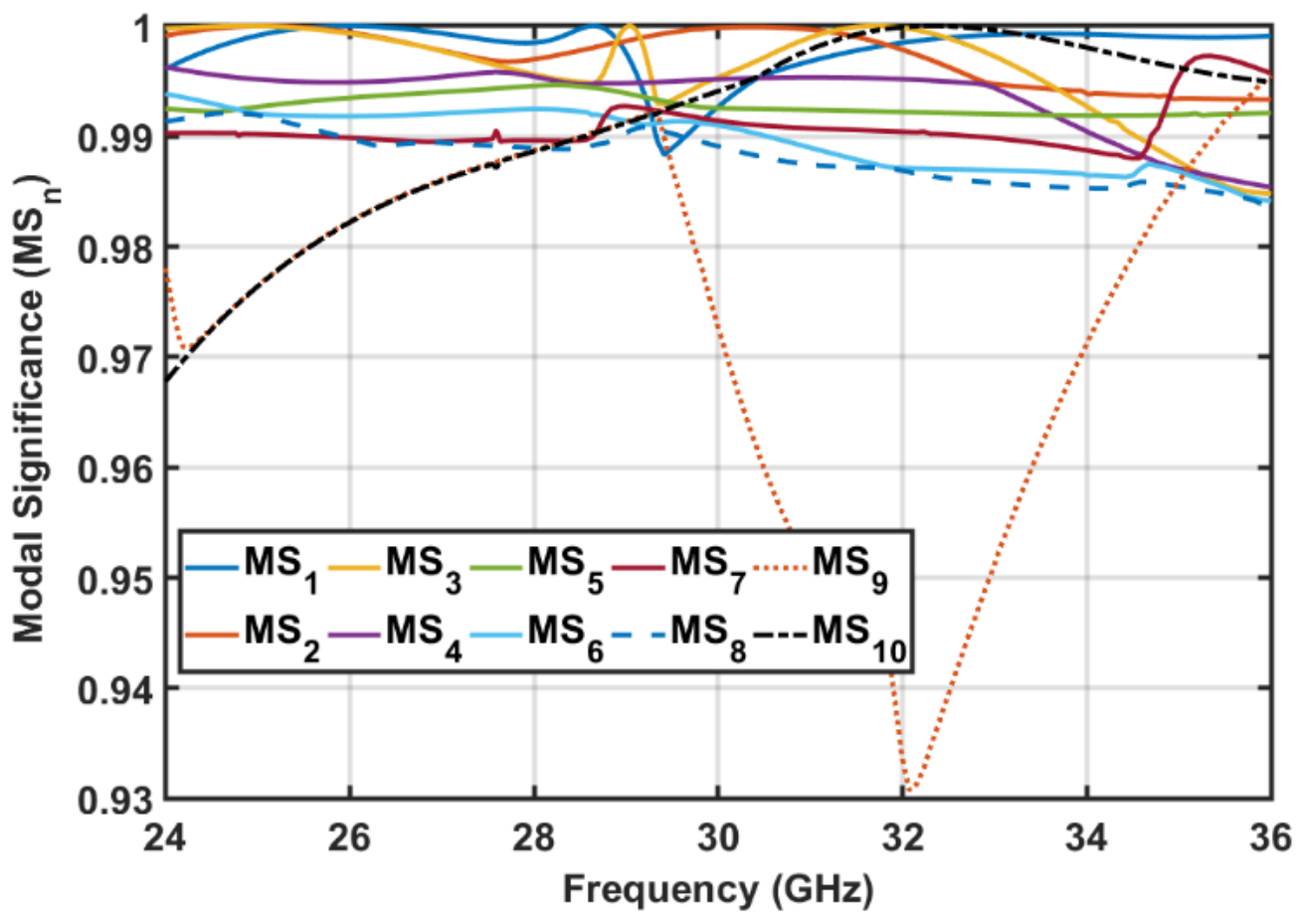

(c) Modal Significance

Figure 4.2 Characteristic mode analysis of Ant. I

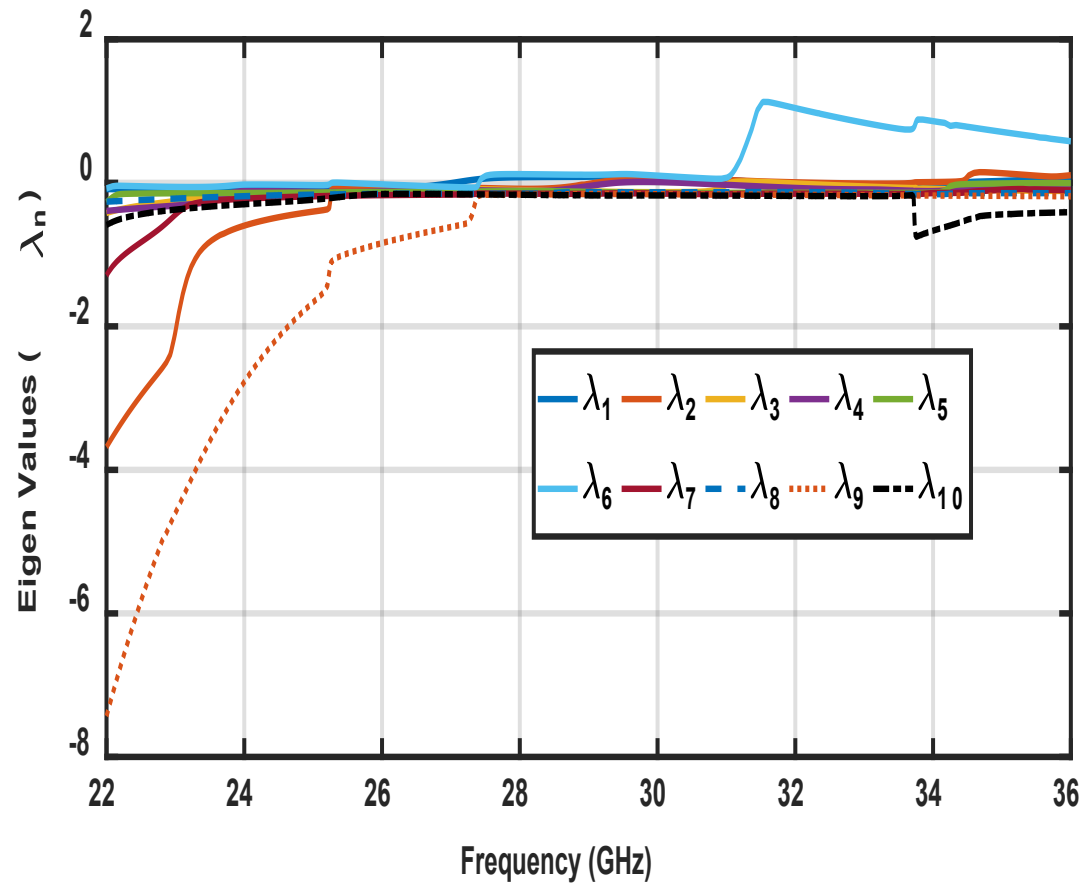

(a) Eigen values

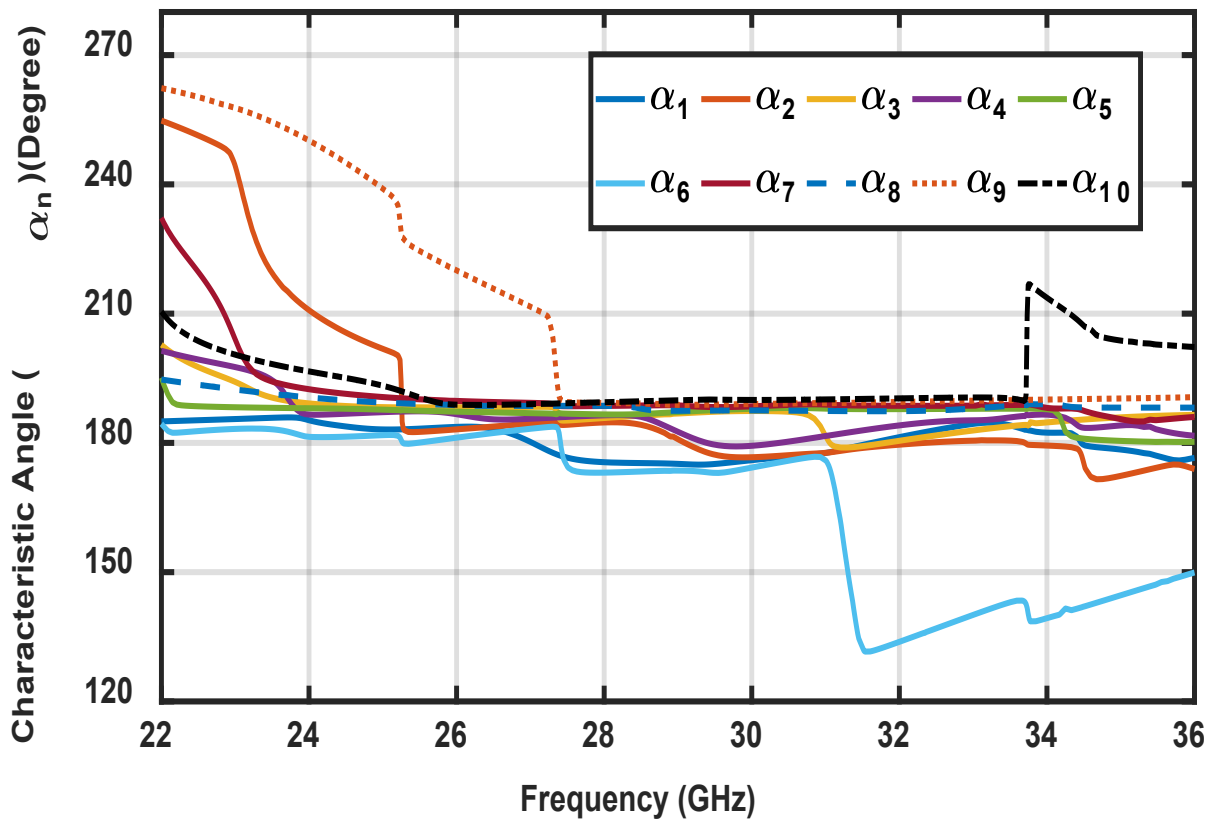

(b) Characteristic angles

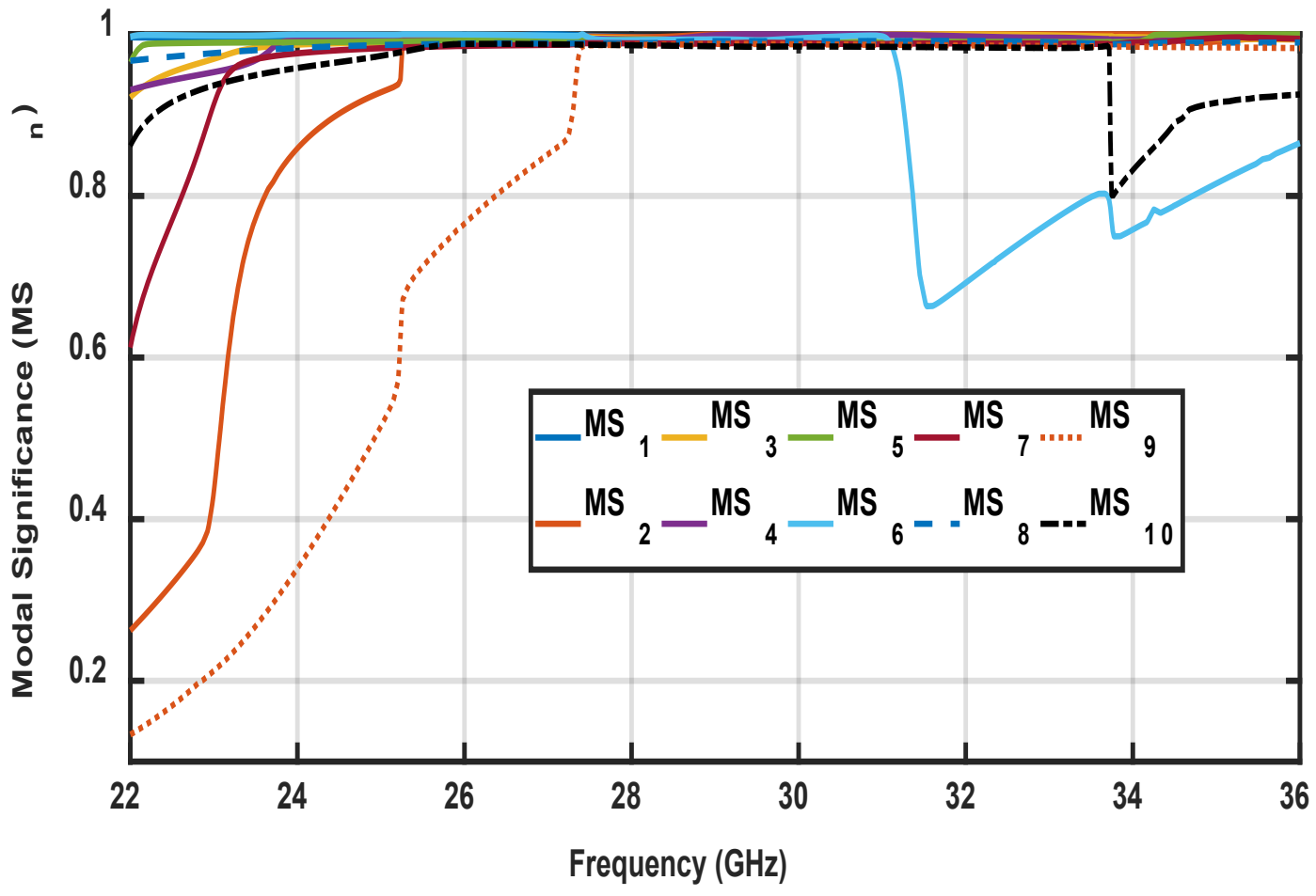

(c) Modal Significance

Figure 4.3 Characteristic mode analysis of Ant. II

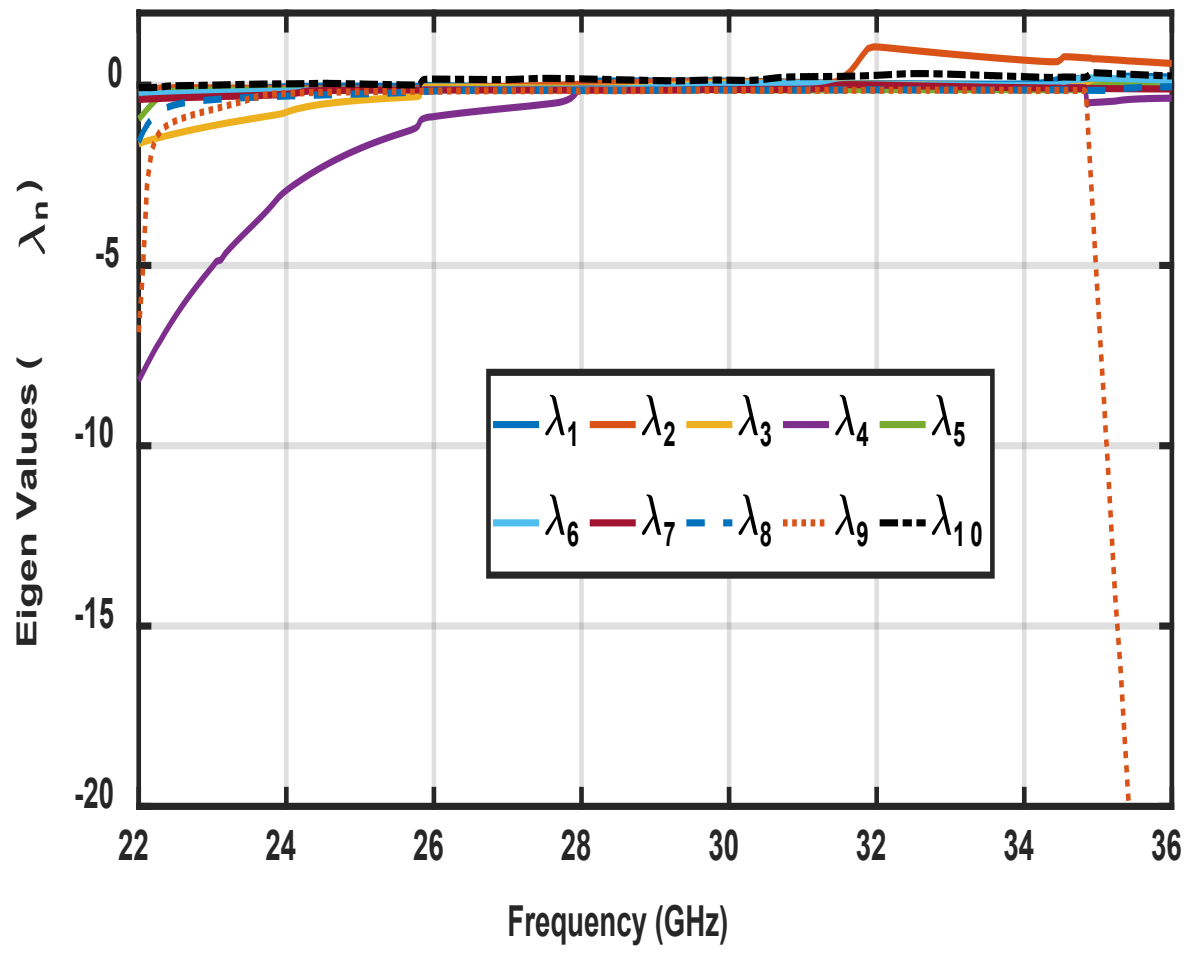

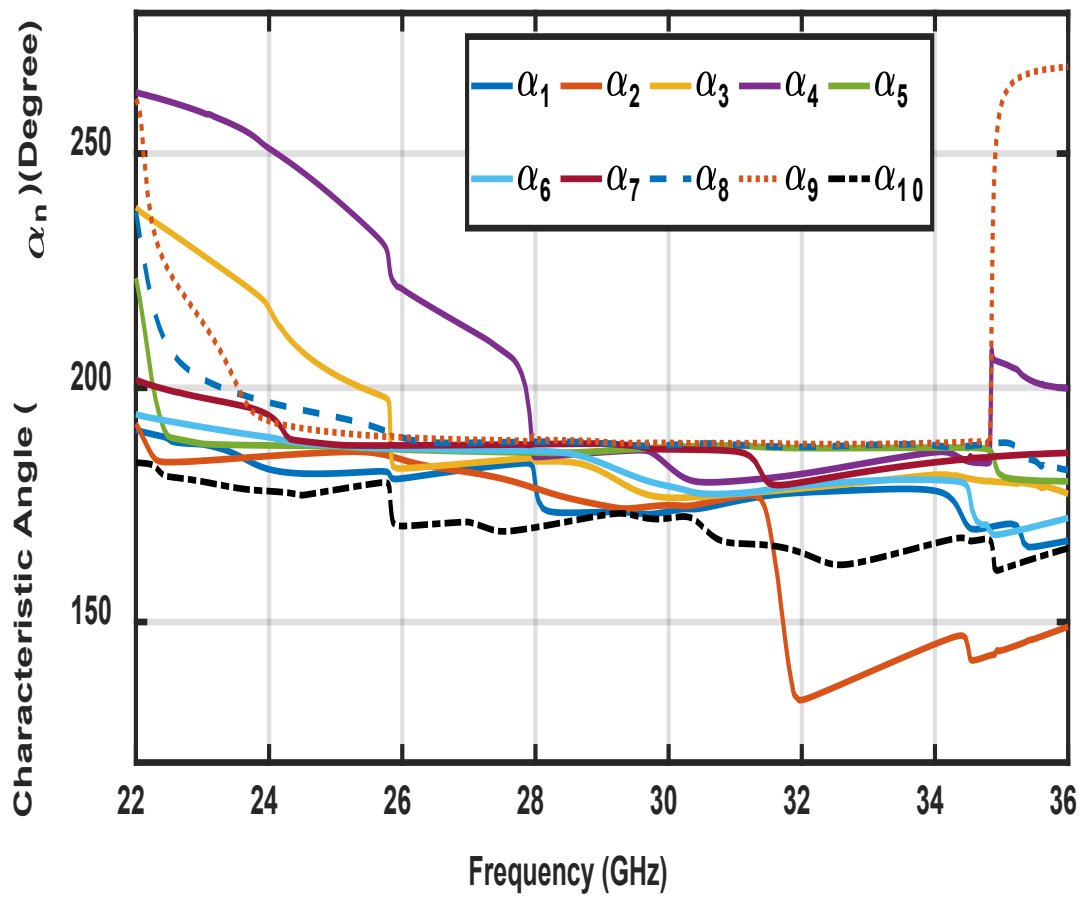

| (a) Eigen values | (b) Characteristic angles |
|---|---|

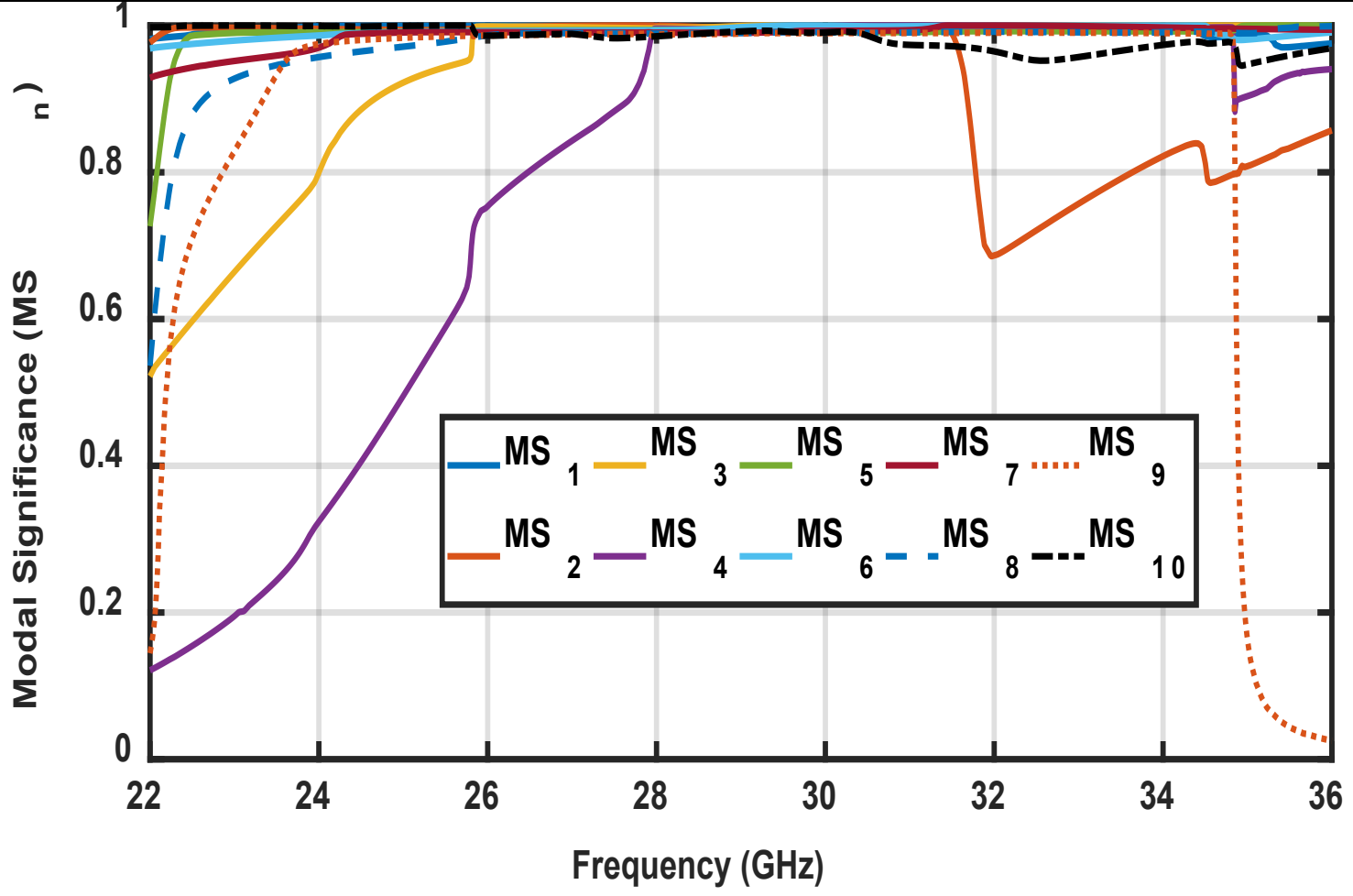

(c) Modal Significance

Figure 4.4 Characteristic mode analysis of Ant. III

Table 4. 2 Comparison between resonant frequencies for CM of three antennas

| Modes | Ant. I | Ant. II | Ant. III |
|---|---|---|---|
| 1 | 28.6 | 27.03 | 28 |
| 2 | 25.04 | 29.1 | 27.6 |
| 3 | 29 | 31.07 | 29.28 |
| 4 | 32.96 | 29.45 | 30.6 |
| 5 | X | 36 | 36 |
| 6 | X | 25.25 | 29.69 |
| 7 | X | X | 31.4 |
| 8 | X | X | X |
| 9 | X | X | X |
| 10 | 32.3 | X | 23 |

*X: mean that this mode doesn't have resonant frequency through the simulated bandwidth.

# 4.3.3 Slot Antenna Designs

Figure 4.5 shows the current and the electric field distributions of the slot antennas I and II. In the case of Ant. I, the magnitude of surface current on the ground plane are the strongest above the microstrip line thus the slot executes the highest disruptive impact on this current. The current is completely impeded near the center of the slot and induces a charge build-up on the long sides of the slot and this like the capacitor. Moreover, the current near to the parts of the slot bends around its ends to continue to pass and this like the inductor, Therefore, the slot antenna is equivalents to two shunt transmission lines shunted by a parallel integration of an inductor and a capacitor. Also, a radiation resistor can be added to the equivalent circuit of slot antenna for long slots.

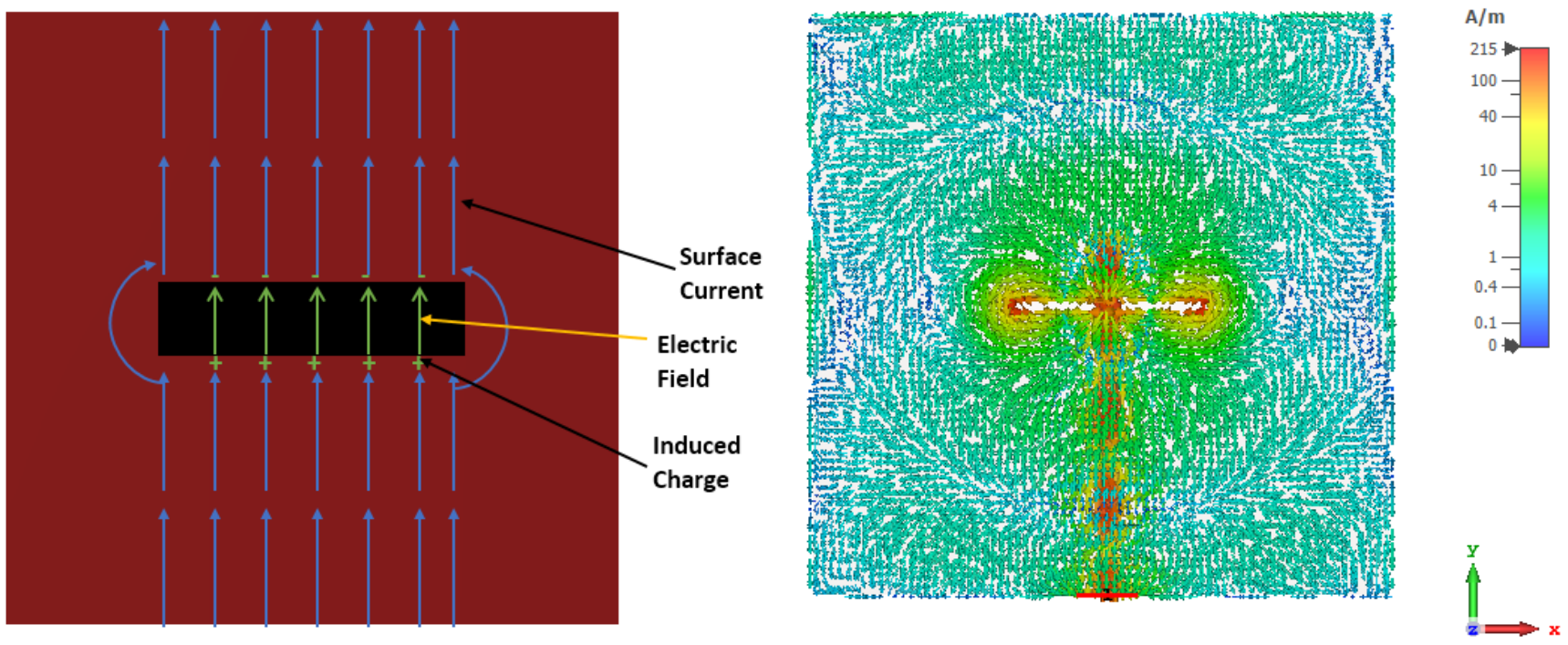

(a)Current distribution and electric field for Ant. I

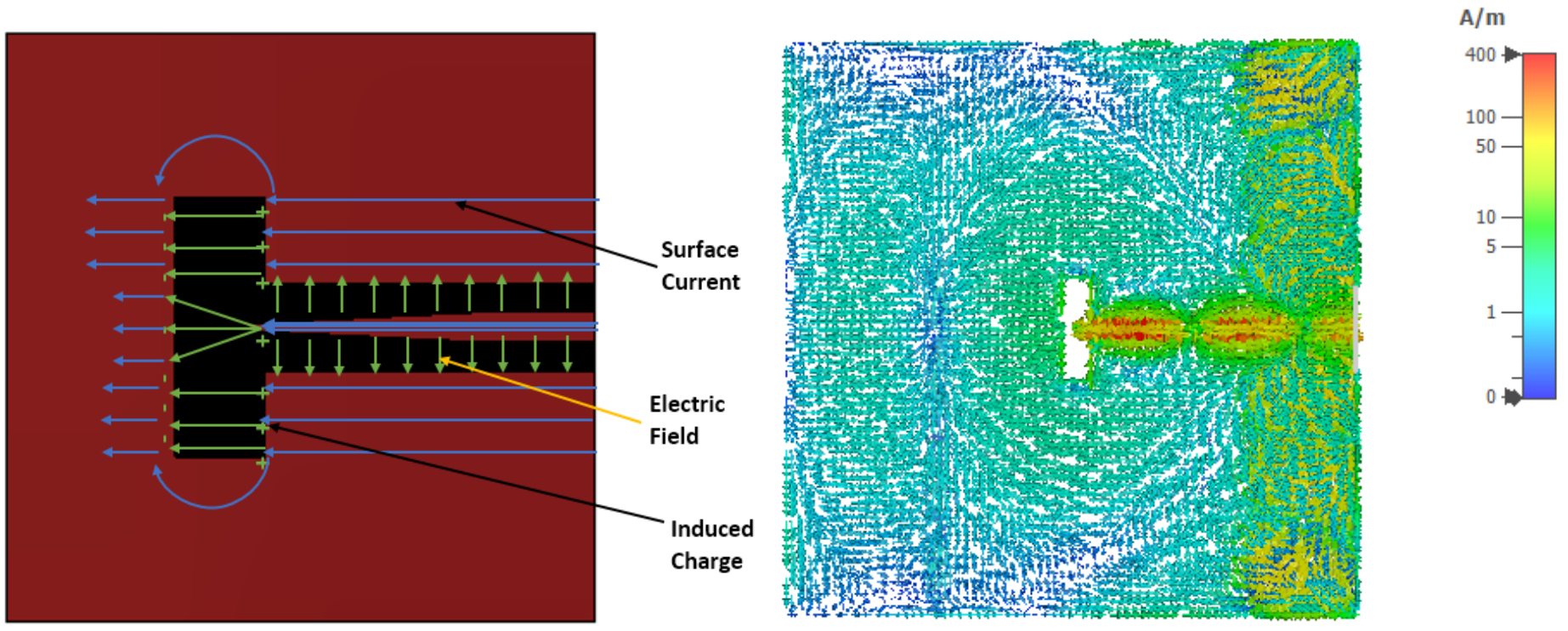

(b) Current distribution and electric field for Ant. II

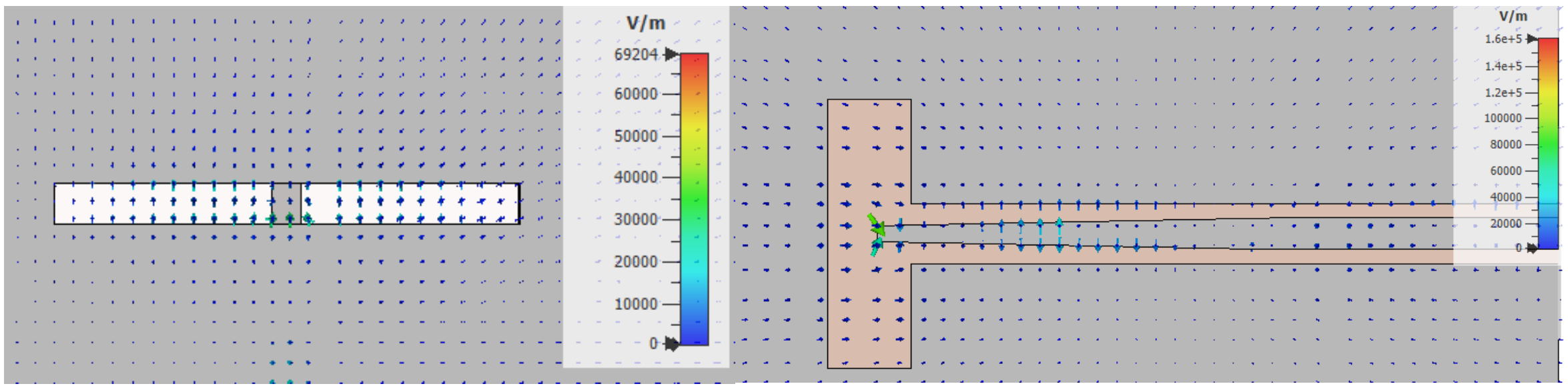

(c)Simulated electric field Ant. I (d) Simulated electric field Ant. II

Figure 4.5 Expected and simulated electric field and current distributions for Ant. I and Ant. II.

The proposed slot antenna operates with a length equivalent to $\left(\lambda_g/2\right)$ from resonant frequency at the working:

$$L_s = \frac{c}{2f_r\sqrt{\epsilon_{eff}}} \tag{4.10}$$

where $L_s$ is slot length, c is the velocity of light free space, $f_r$ is the resonant frequency, and $\epsilon_{eff}$ is the relative effective permittivity of proposed antenna. Figure 4.6 shows the reflection coefficient of Ant. I at different values of length and width of the slot. For Ant. II, the wide bandwidth is achieved due to the multiple resonance mode are excited by the combination of the CPW and aperture antenna. The resonant frequency and BW are tuned by the length and width of aperture antenna ($L_{s2}$, $W_{s2}$). Figure 4.7 shows the effect of length and width of the aperture antenna on the operating bandwidth.

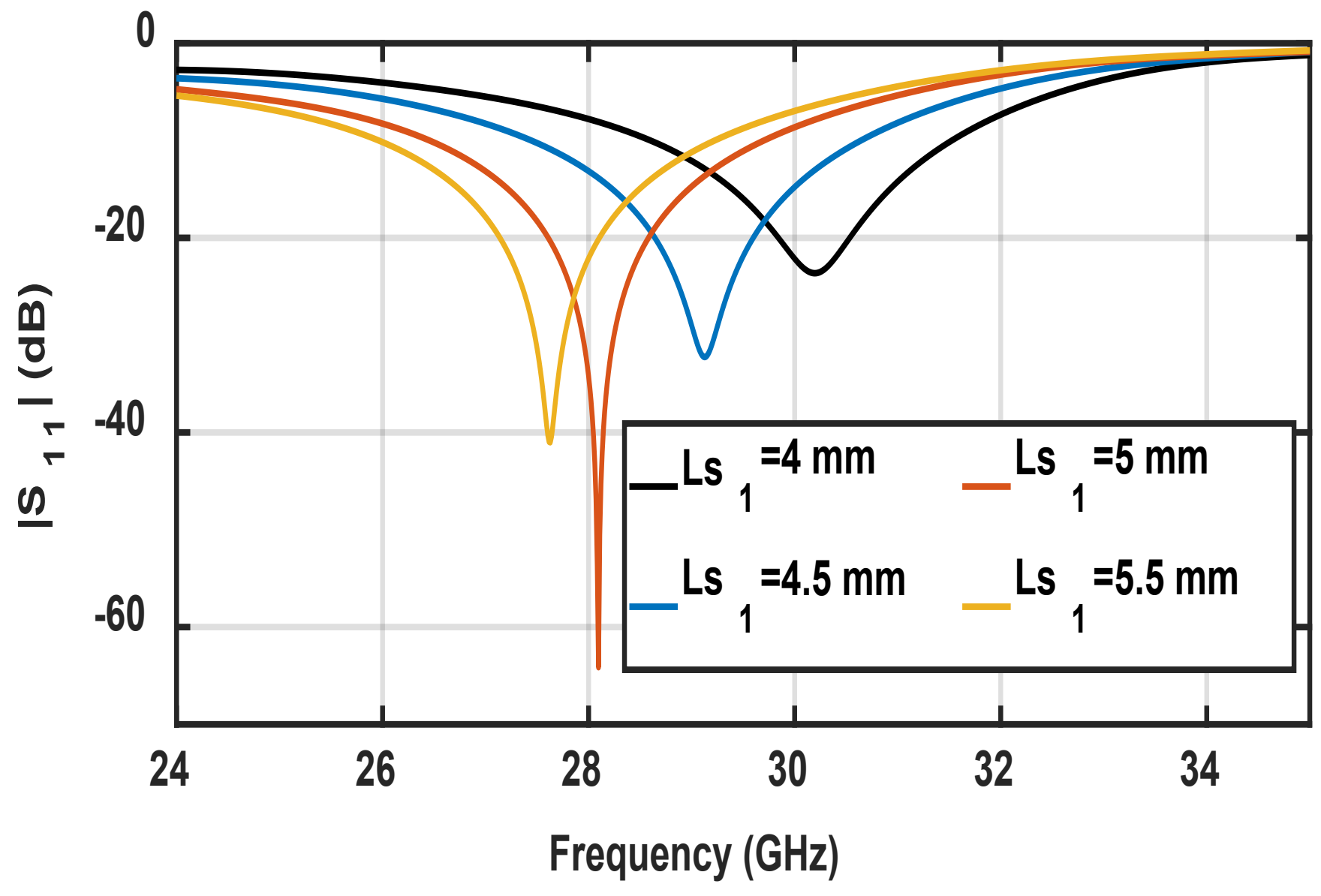

(a)Reflection coefficient at different values of slot length

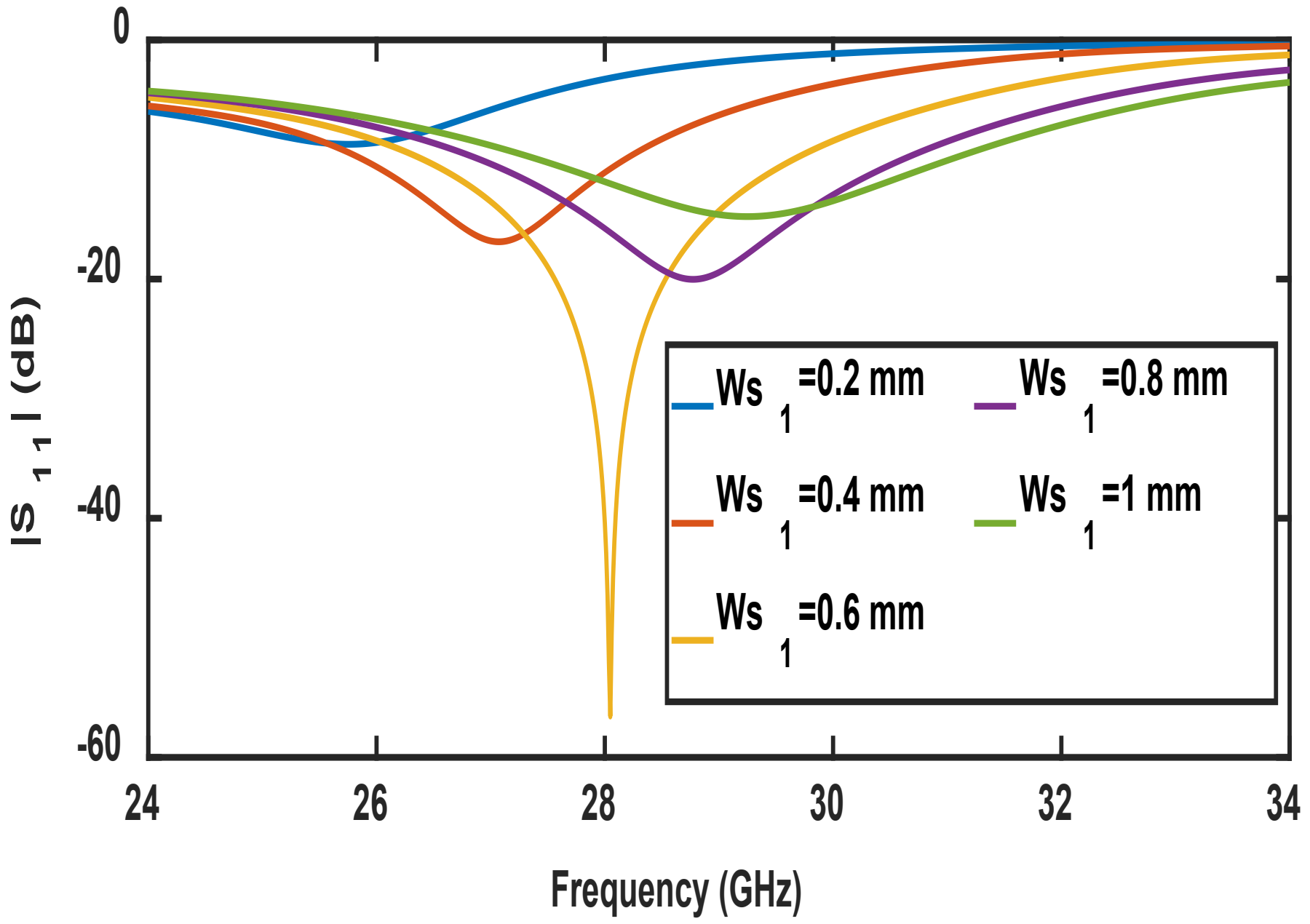

(b) Reflection coefficient at different values of slot width

Figure 4.6 Reflection coefficient of Ant. I at different values of slot width and slot length.

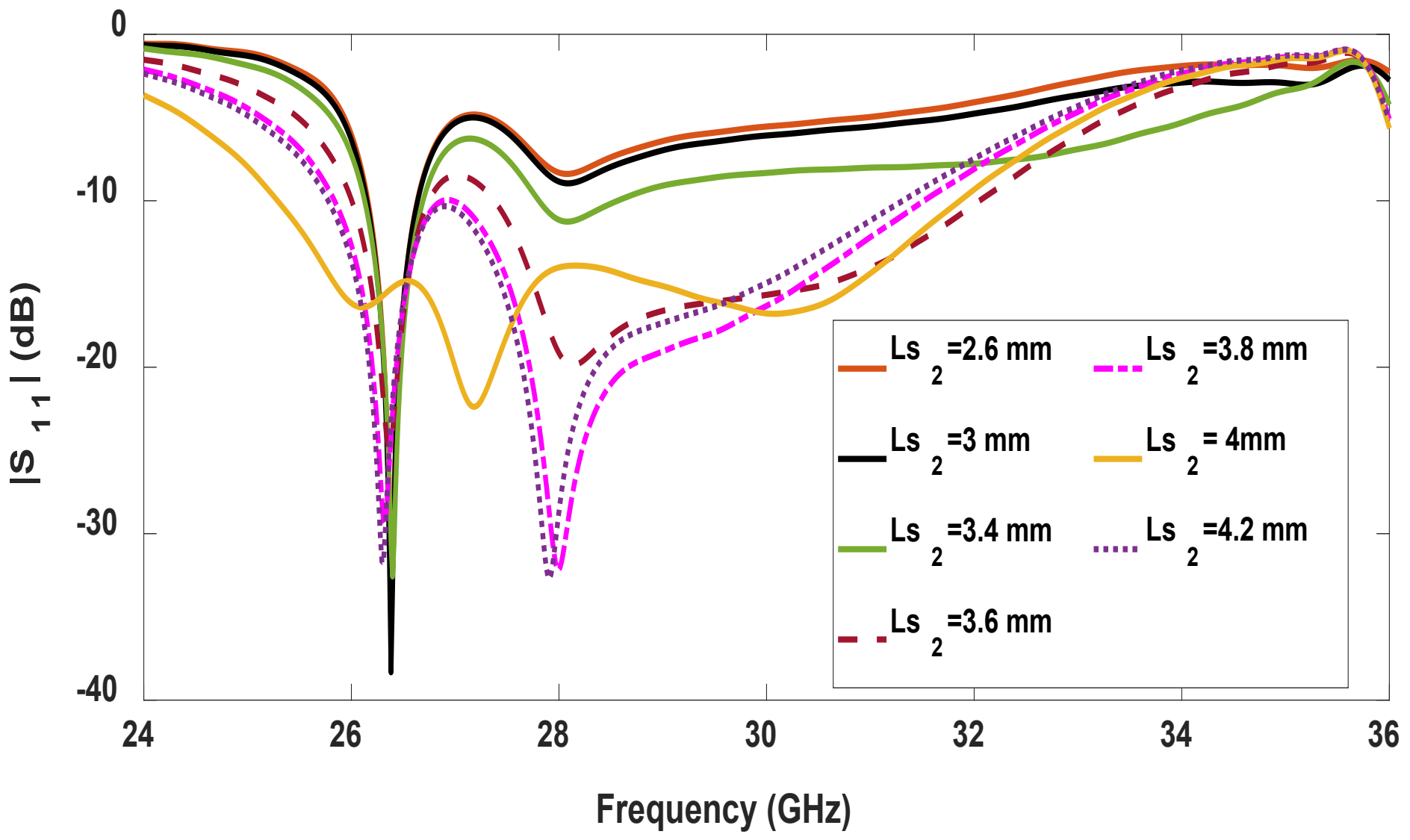

(a)Reflection coefficient at different values of slot length

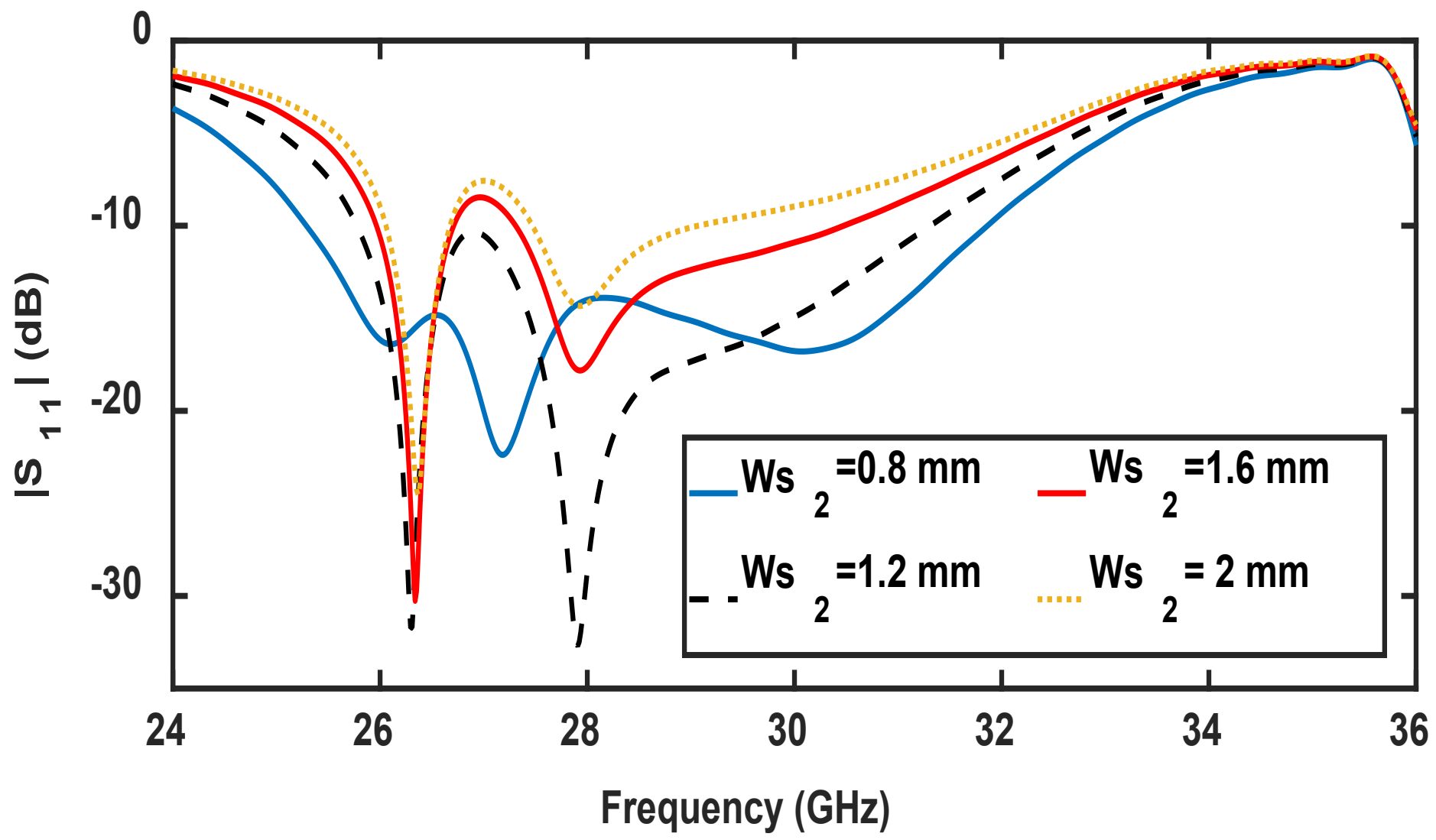

(a)Reflection coefficient at different values of slot width

Figure 4.7 Reflection coefficient of Ant. II at different values of slot width and slot length.

Ant. I has vertical polarization and Antenna II has horizontal polarization. Ws1, Ws2 are the dimensions of slot widths to control the matching of vertical and horizontal modes, respectively. Furthermore, $y_1$ is a tuning parameter for matching port 1, and $W_t$, $L_t$ are parameters to optimize the matching at port 2. To consider the practical case, we consider the end launch connector in our designs as shown in Figure 4.8. Therefore, the feed lines are increased by 5 mm in x and y direction to avoid the interconnection between the two connectors.

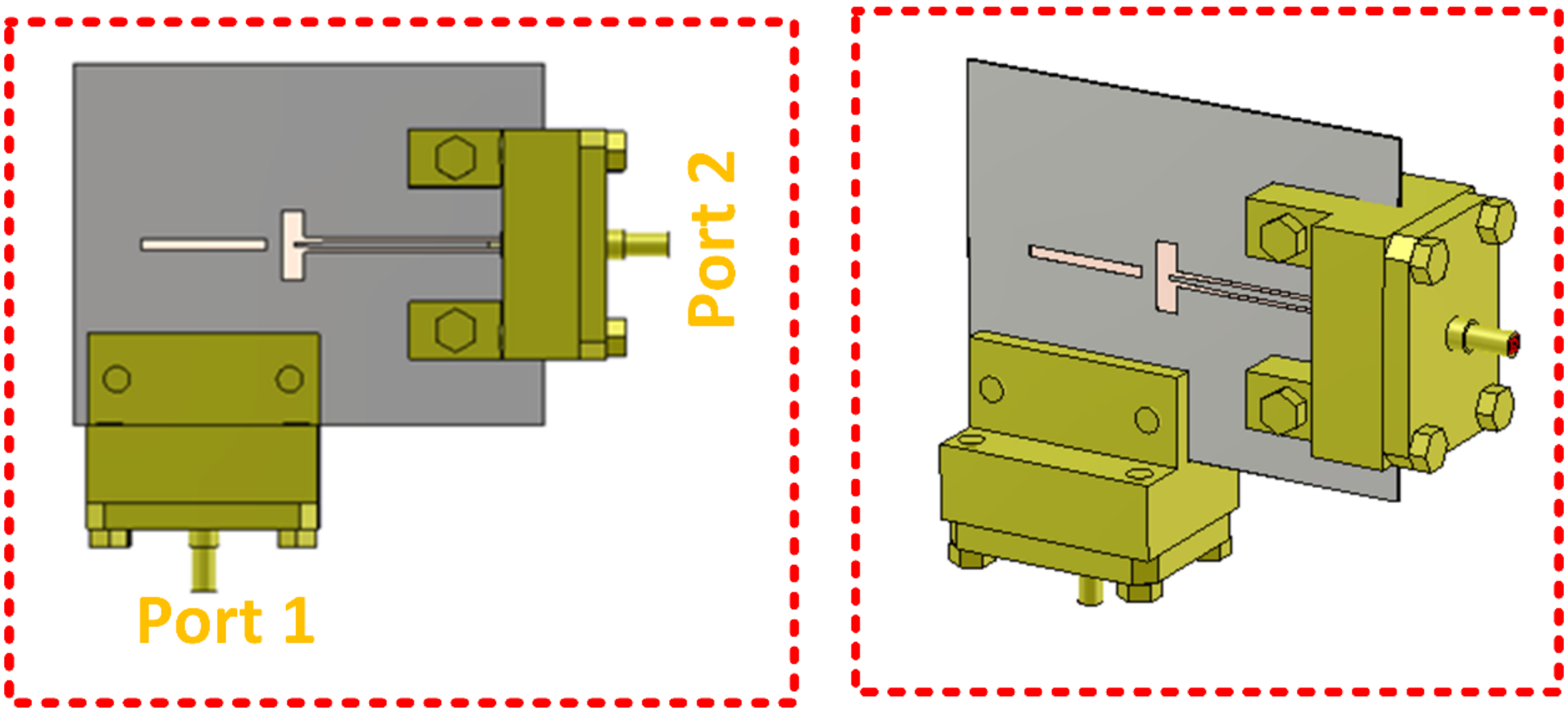

(a)Front view (b)3-D view

Figure 4.8 Proposed antenna design with connectors.

A high separation between the two ports can be achieved due to the orthogonality characteristics and symmetric/antisymmetric characteristics of the three modes of CPW (with the ground).

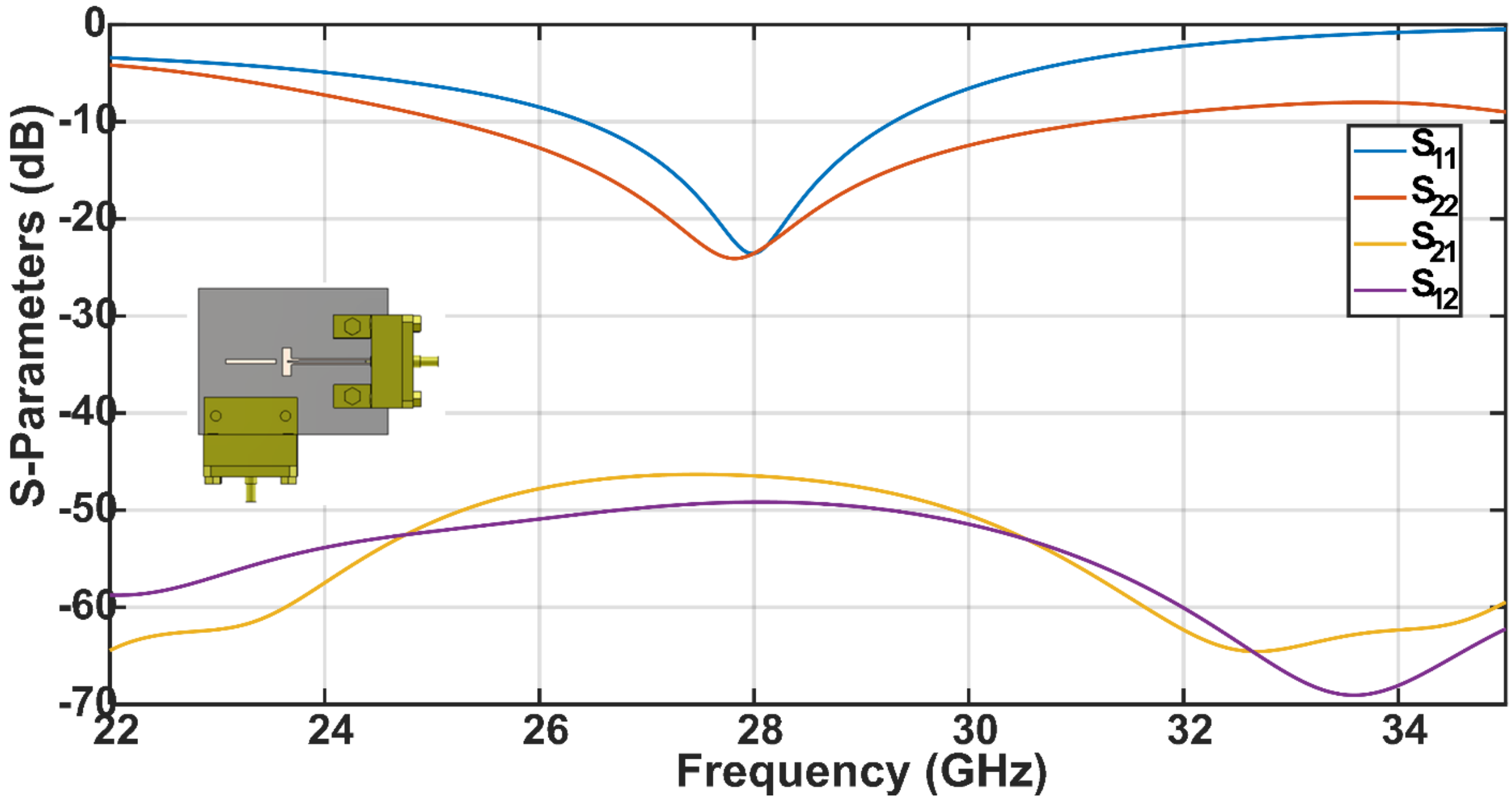

Figure 4.9 Simulated S-parameters of two ports antenna.

The reflection coefficients and isolation coefficients of the proposed two ports antenna are shown in Figure 4.9. One can notice that the isolation coefficients between the two ports have a high value through the operating bandwidth (more than 45 dB) and the antenna has good matching for ($S_{11}$ and $S_{22}$). The antenna achieves 2.2 GHz as a wide bandwidth from 27 GHz to 29.2 GHz when port 1 is excited and 5 GHz from 25.6 GHz to 31.6 GHz for port 2. The proposed antenna achieved common bandwidth (2.2 GHz) to cover the 5G applications at 28 GHz.

The Ant. III is redesigned without ground in the bottom (the width of CPW line without ground is recalculated) to make the antenna with only one common ground in the top. The S-parameters of this design is shown in Figure 4.10. The results still ensure that the antenna has good matching and high isolation between its ports.

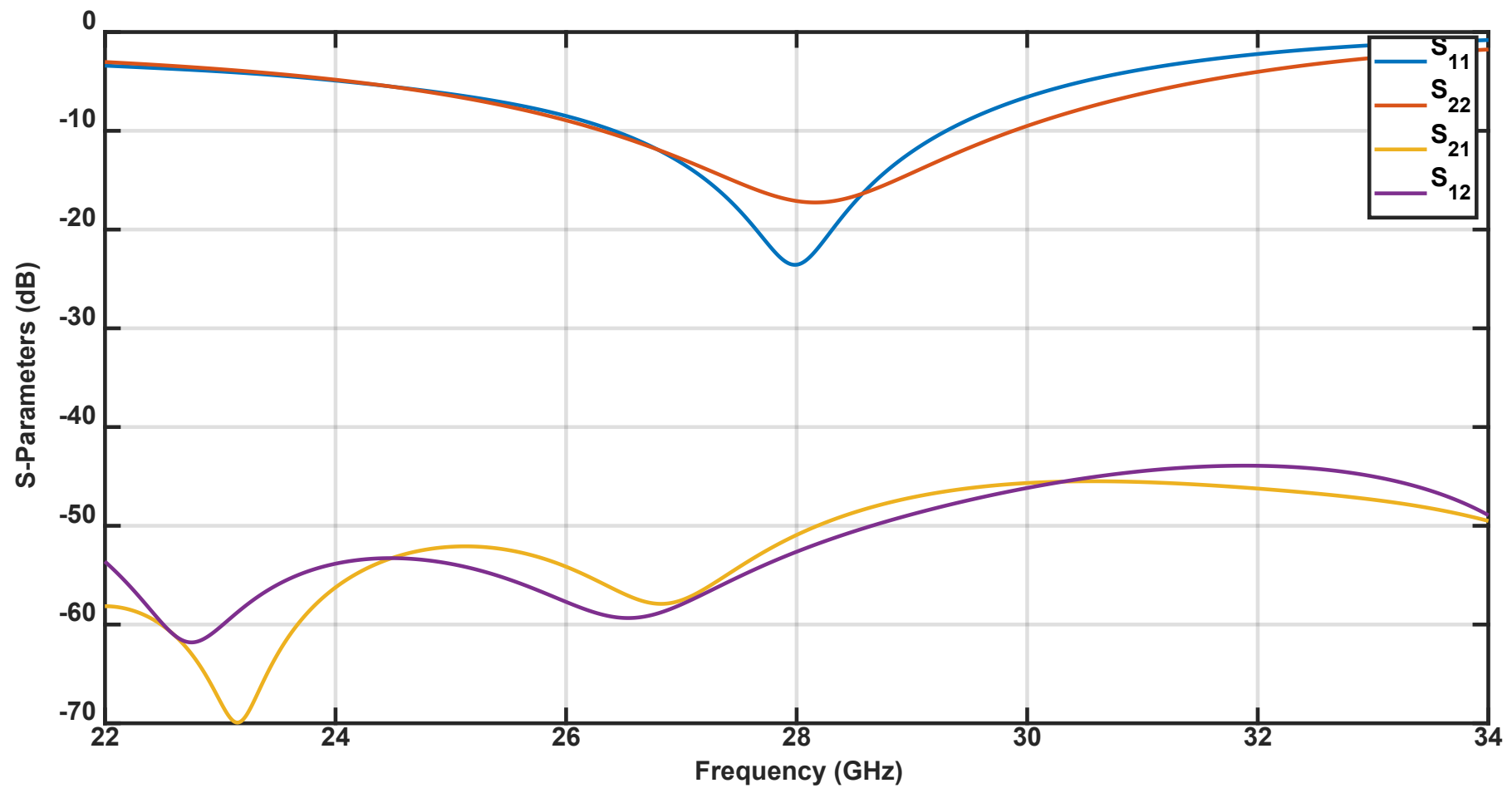

Figure 4.10 Simulated S-parameters of the two ports antenna (CPW without ground in the bottom).

Figure 4.11 and Figure 4.12 show the surface current and electric field distributions for two ports. The surface currents and electric field of the proposed antenna at 28 GHz for two ports are presented to ensure that the proposed antenna achieves dual-polarization between their ports. It is clear to note that the surface current and electric field flow along the y-axis when port 1 is fed. While they flow along the *x*-axis when port 2 is excited. Therefore, dual-polarization is achieved.

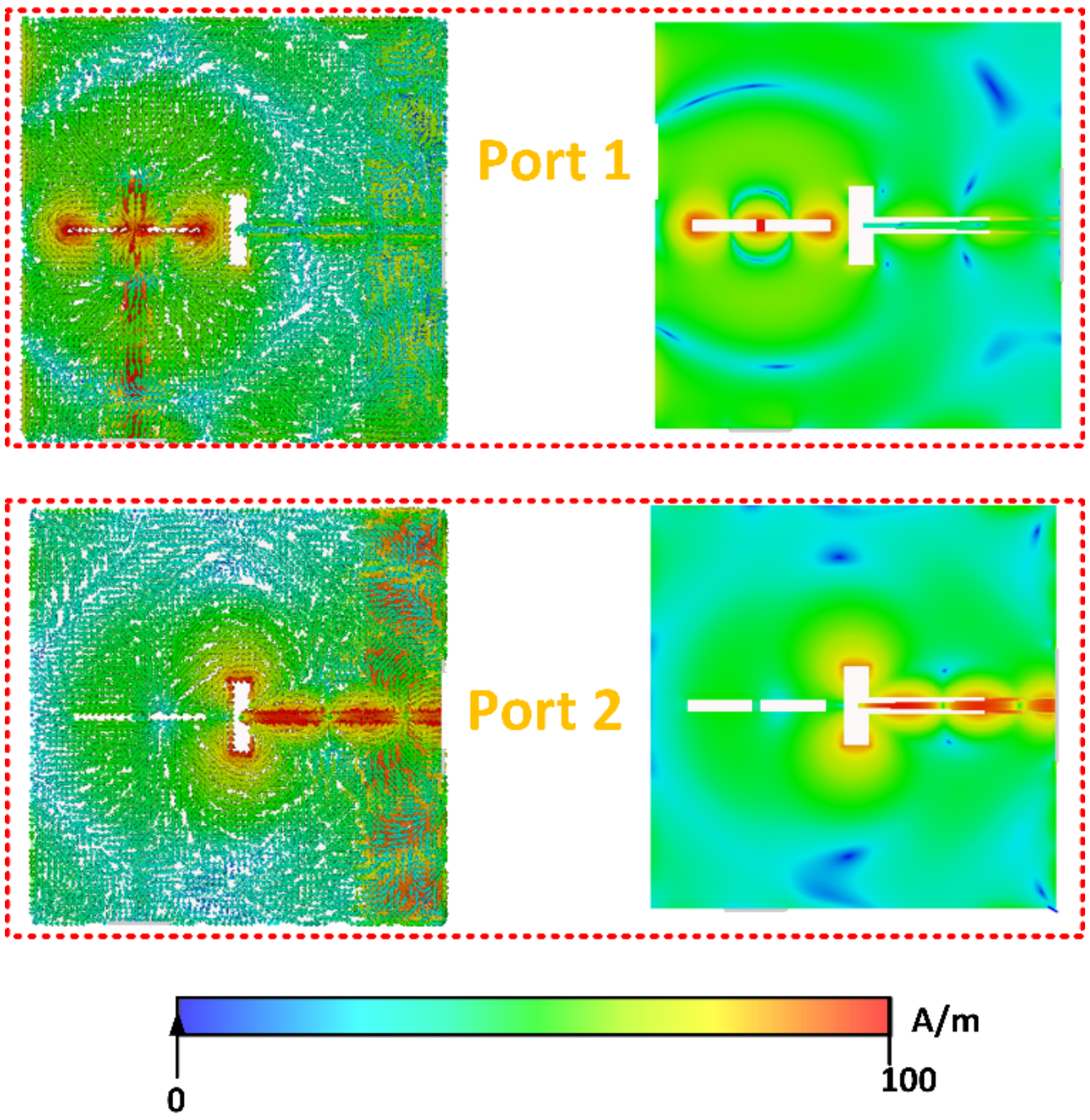

Figure 4.11 Current distribution of proposed antenna at 28 GHz from two ports

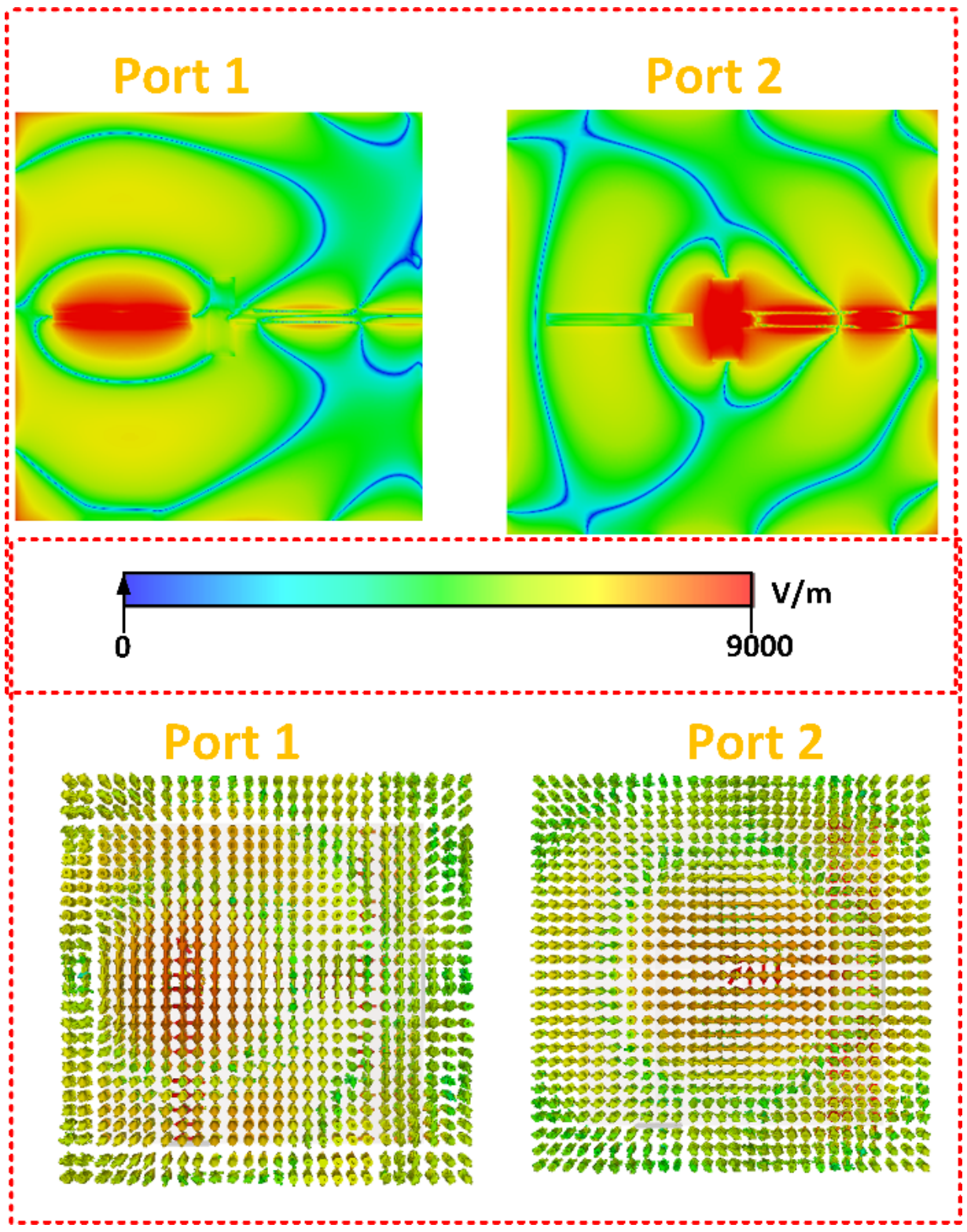

Figure 4.12 Electric field distributions from two ports.

The radiation patterns of the proposed antenna from port 1 and port 2 in both XZ and YZ planes at 28 GHz are shown in Figure 4.13. We can observe that the cross-polarization levels in both planes are less than 40 dB as compared with the co-polarizations. The antenna achieved gain is 6.23 dBi and 6.85 dBi as shown in Figure 4.14.

| Port | XZ Plane (Phi=0) | YZ Plane (phi=90) |
|---|---|---|
| Port 1 | 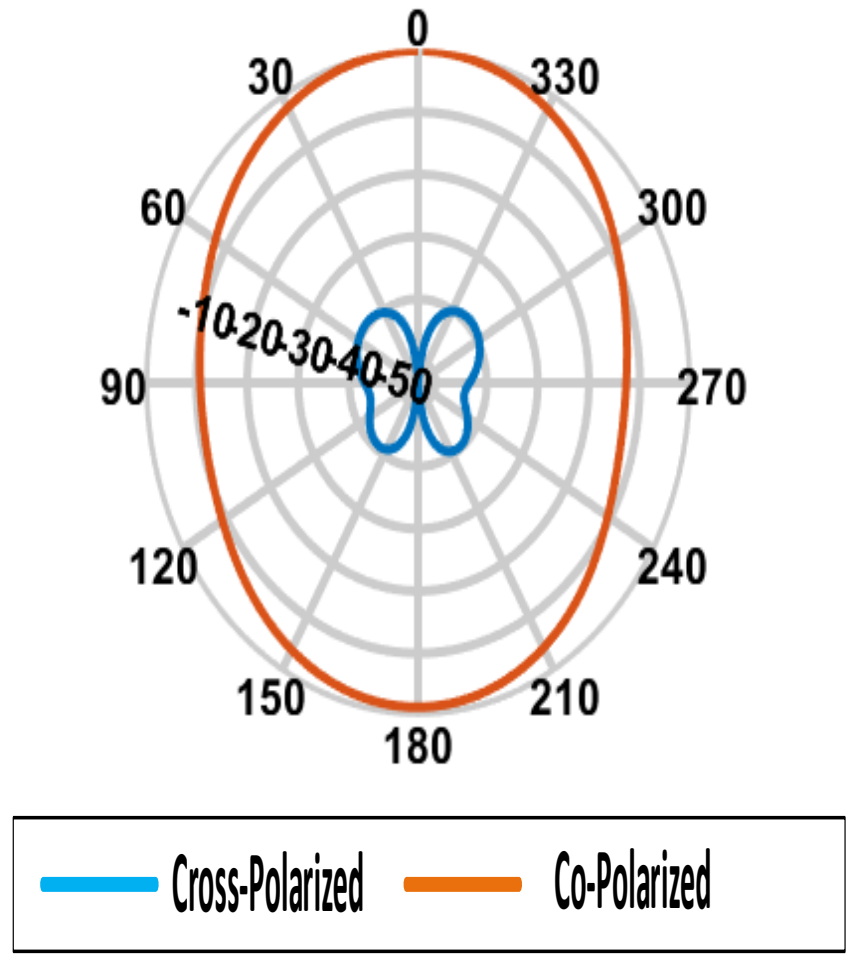 (a) | 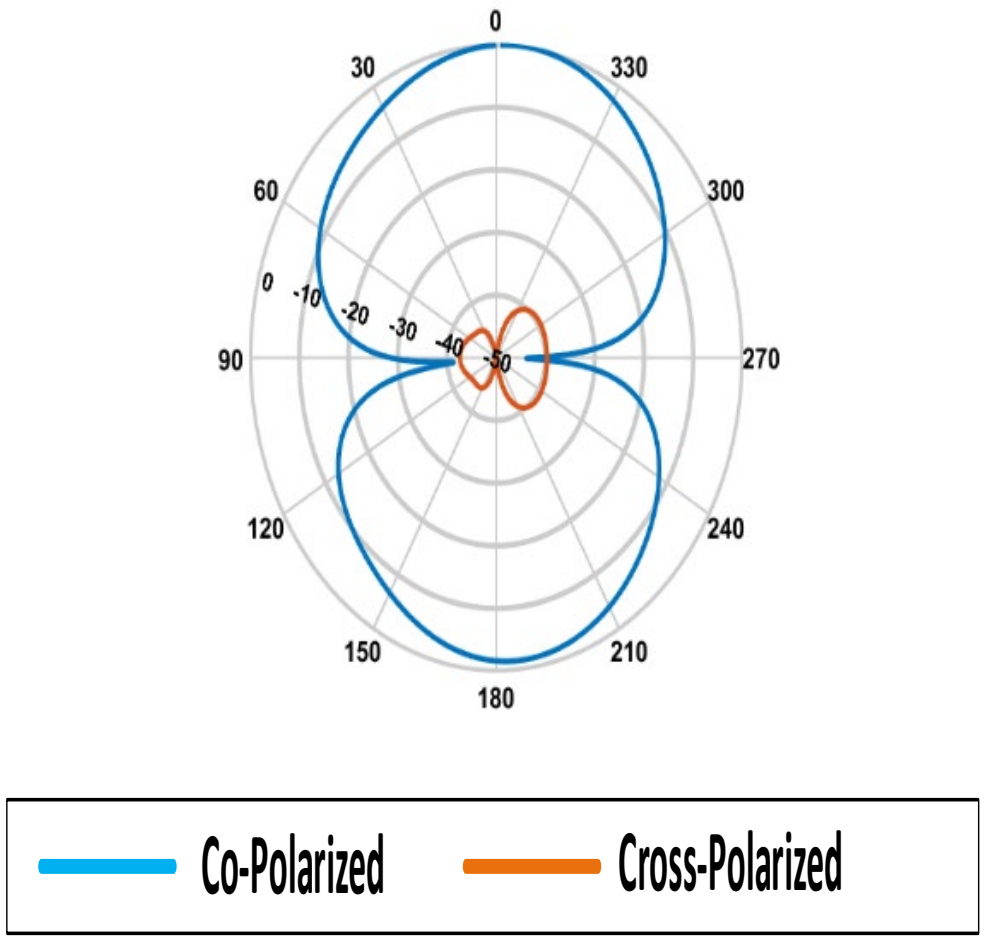 (b) |
| Port 2 | 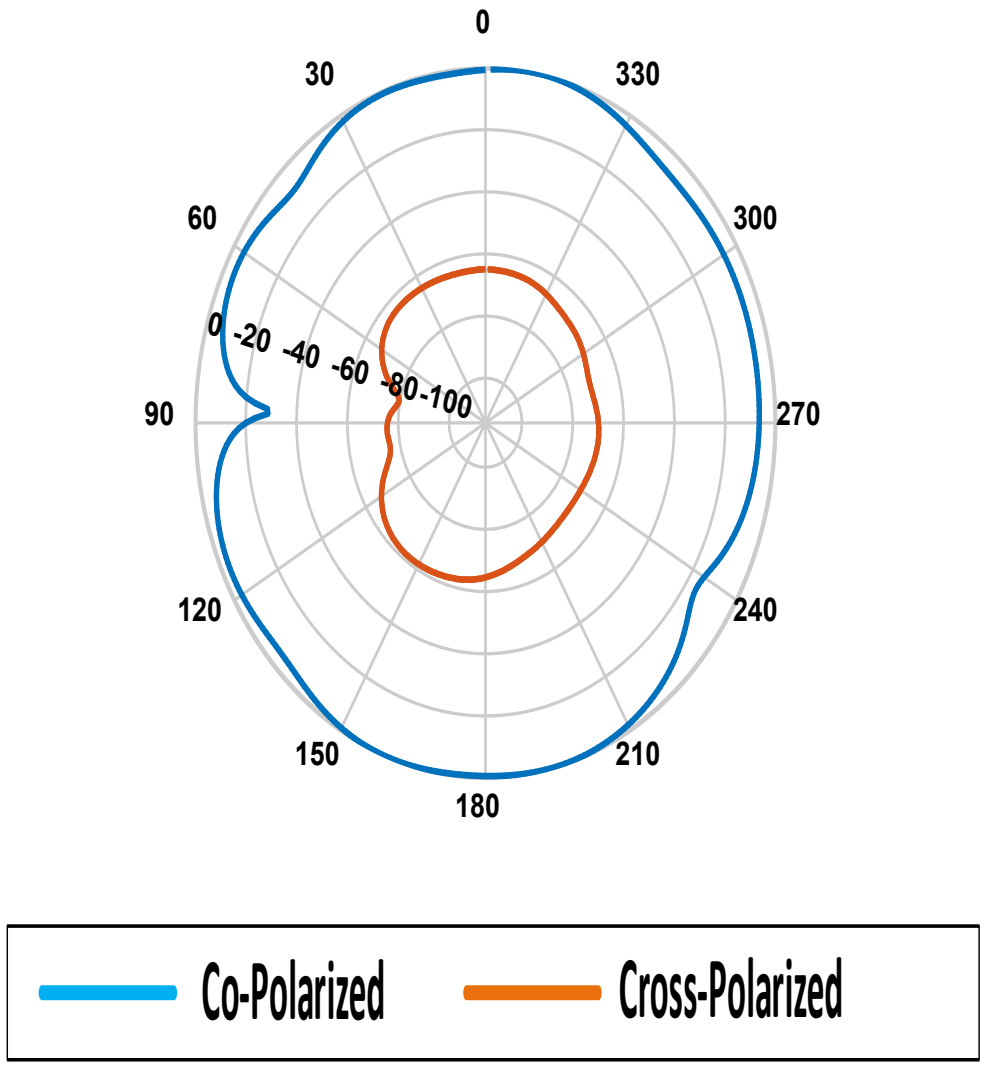 (c) | 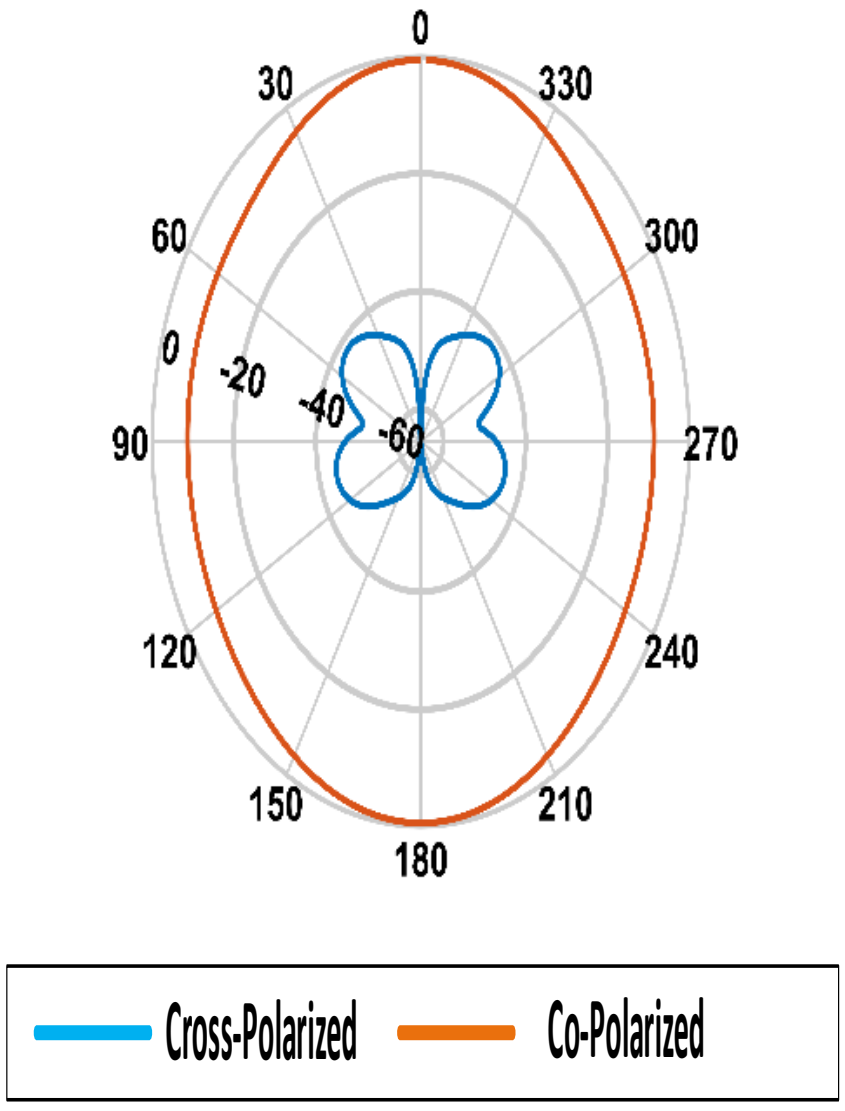 (d) |

Figure 4.13 Co-Polarized and Cross Polarized for port1 and port 2.

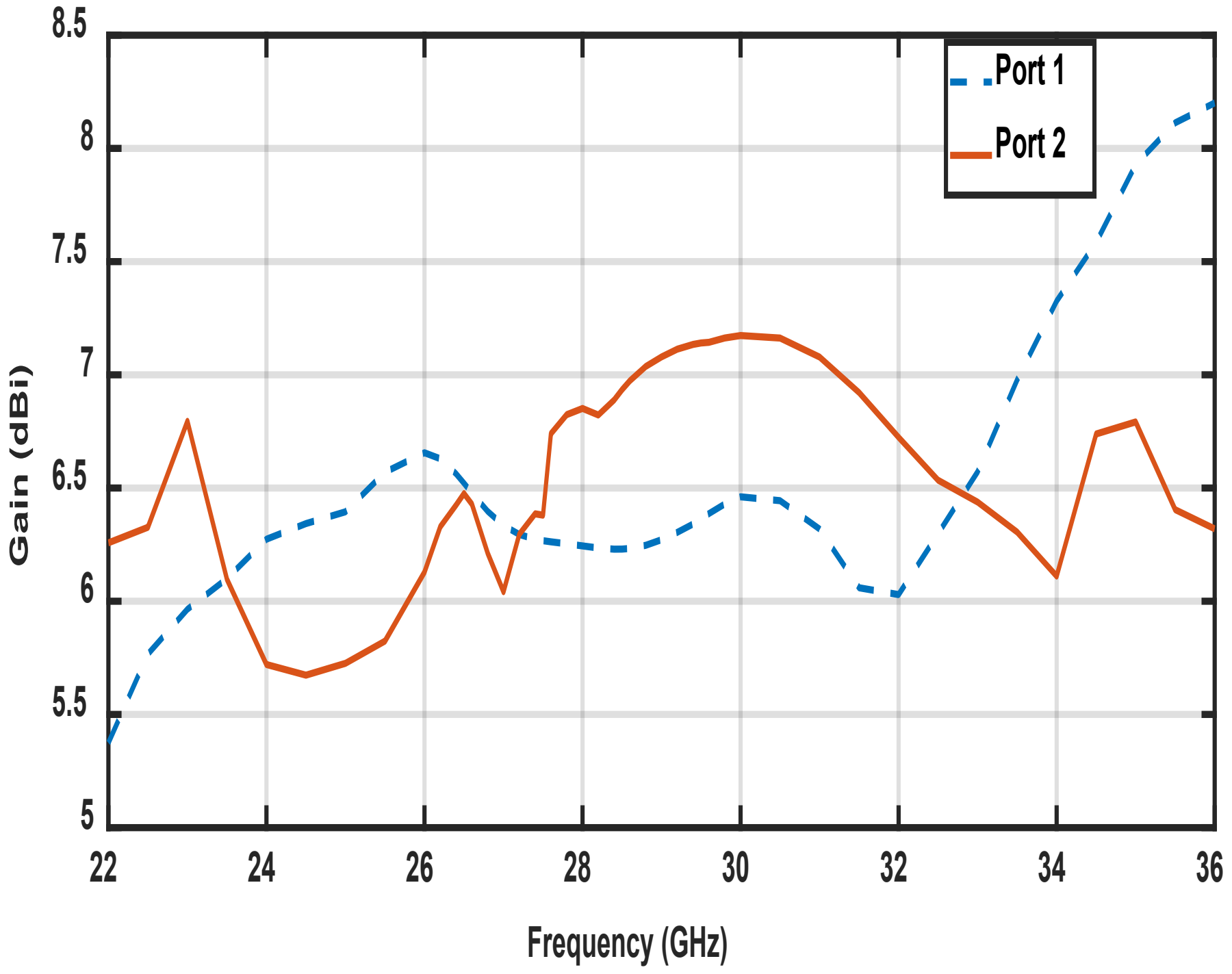

Figure 4.14 Gain of proposed antenna.

## 4.4 Metasurface Design

Recently, the metasurface (MTS), which is a 2-D structure of metamaterial, has been used to improve the bandwidth and give a low profile of microstrip antennas by placing the metasurface unit cells in manipulating electromagnetic waves direction. The MTS is considered one of radiating surfaces that can radiate electromagnetic waves by exciting it through a nearby external antenna. Extensive works have been introduced in order to analyze the characteristics of MTS such as the equivalent circuit, dispersion diagram, periodic structures, surface impedance tensor, transmission and scattering coefficients, theory of effective medium, and reflection phase diagram [189-191]. The CMA has been used in the past few years to predict the characteristics and behaviour of MTS in the accurate form [192-195].

### 4.4.1 Analyze Metasurface using CMA

This section introduces the analysis of MTS and the comparison between the antenna and MTS unit cells. The metasurface of 7x7 unit cells (square patches) is proposed at 28 GHz, as shown in Figure 4.15. The square patches are printed on Rogers RO 4003C substrate with dielectric constant 3.38 and a thickness of 0.2 mm. the CMA solver from CST microwave studio is used with free-space boundary conditions to analyze this structure. To examine the metasurface modal behaviors, the first ten modes are calculated (Eigenvalues, characteristic angles, and modal significances) in the range from 18 GHz to 38 GHz.  Over the band, we notice that only the modes from 1 to 9 have resonant frequencies. In this design, we have two groups from degenerate modes (J1/J2 and J7/J8) resonant around 28 GHz and 31 GHz, respectively. In this design, vertical and horizontal polarizations are desired. Therefore, we need two similar modes (one is VH and the other is HP) over the proposed band. In this design, we have two degenerate modes (J1/J2 and J7/J8) that can be used, but the other modes are not considered (6). On the other hand, J7/J8 are at the end of the operating band (31 GHz). Figure 4.16 (a)-(c) shows that the modes J1/J2 are the only two modes with pure resonant at 28 GHz. Also, J1\J2 have a characteristic angle equal to 1800 at 28GHz in addition to zero Eigenvalue at 28GHz.

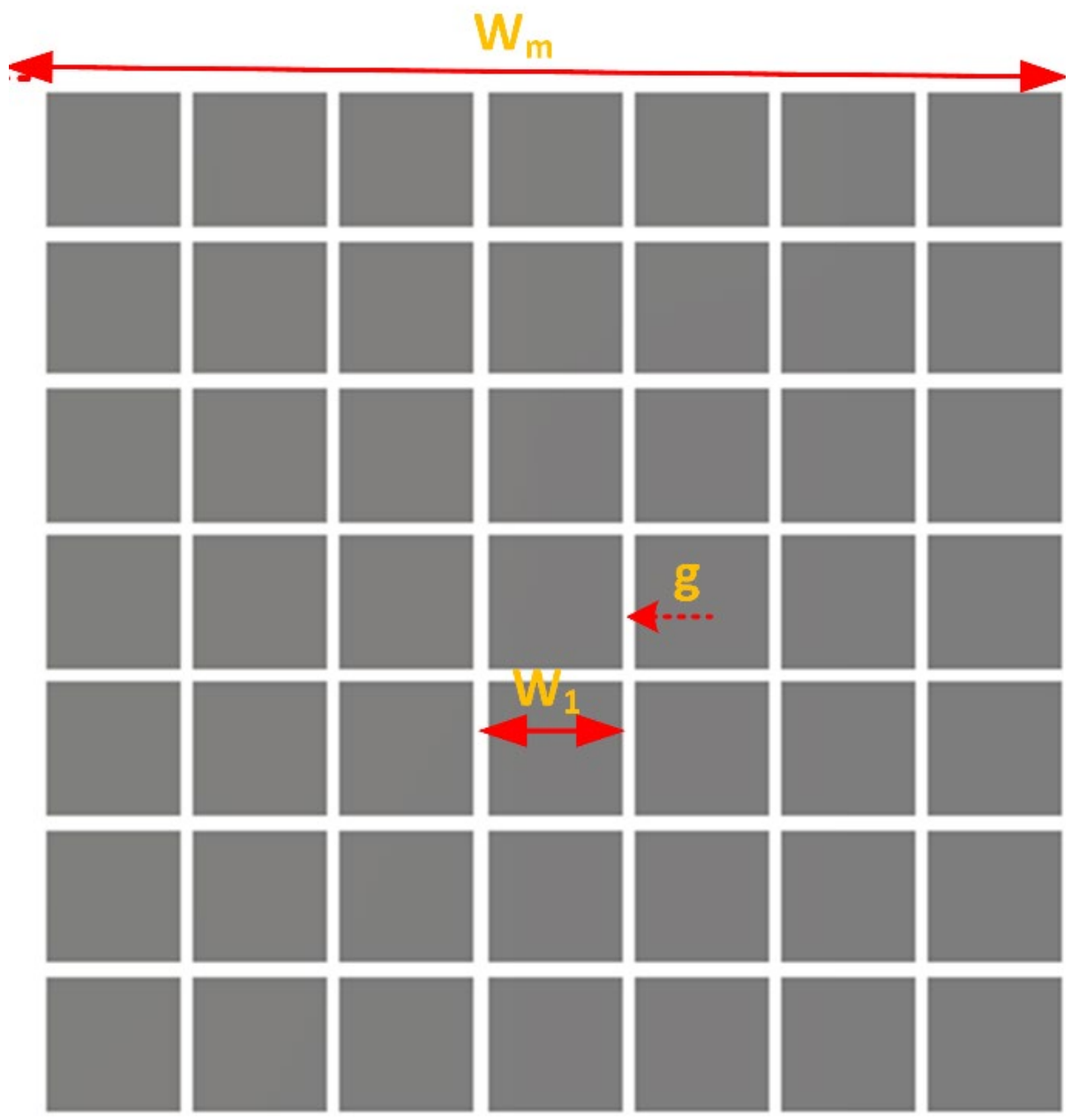

Figure 4.15 Metasurface structure ($W_1$=1.7 mm, g=0.2 mm, Wm=13.3 mm).

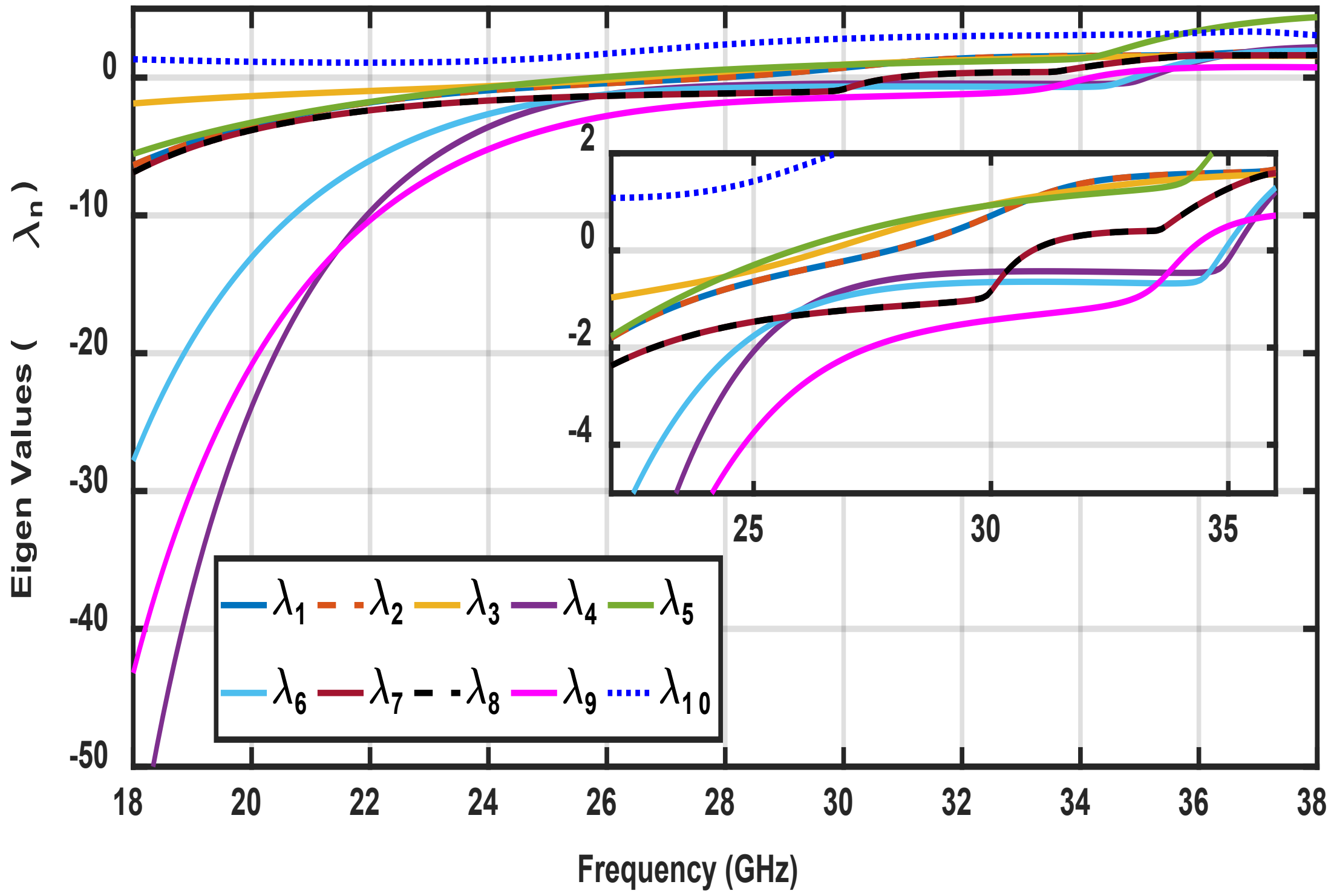

(a)Eigen values

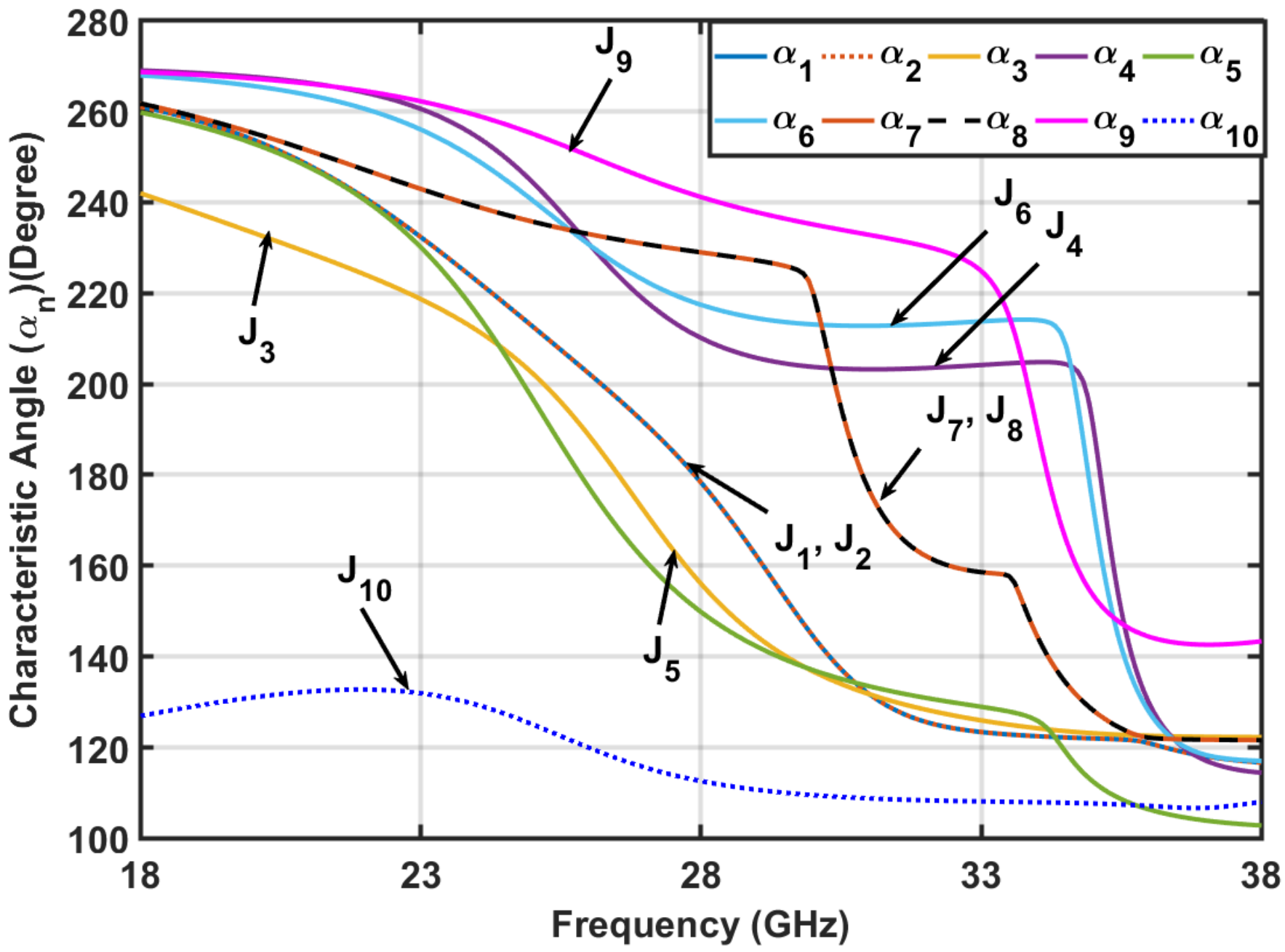

(b)Characteristic angles

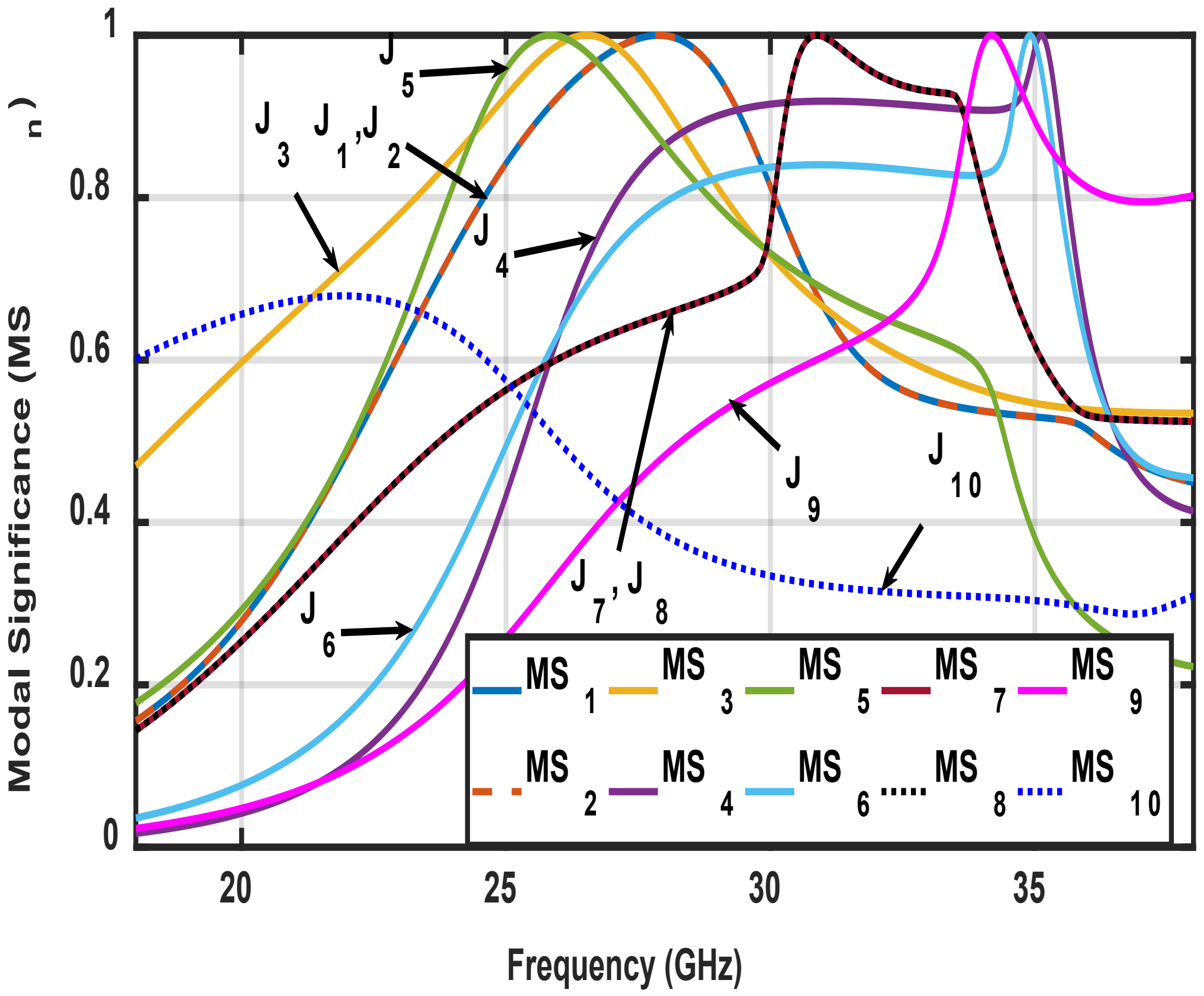

(c)Modal Significance

Figure 4.16 CMA parameters of MTS.

The modal electrical surface currents are shown in Figure 4.17 and the current directions of each mode are indicated by black arrows. All the field and current in this section are calculated at 28 GHz. As can be noticed, first modal current ($J_1$) is in phase through the MTS and its polarization in the y-direction. Also, the second modal current ($J_2$) is typical as ($J_1$) but with $90^0$ out of phase. In other words, $J_2$ directs in the x-direction through the MTS. Therefore, $J_1/J_2$ are a pair of orthogonal modes. As a result of all currents of first and second modes are in phase, they have pure broadside radiation as shown in Figure 4.18. $J_3$ and $J_5$ have symmetrical distribution about the y-axis, and x-axis, respectively, with null current at the center of MTS and null along the z-axis in the radiation pattern as shown in Figure 4.18. The current of the third mode ($J_3$) flows as a closed loop with null at the center and thus it is like the behavior of inductive, which can be verified from its characteristic angle about 28 GHz which is below $180^0$. The currents of mode $J_4$ and mode $J_6$ are self-symmetrical about y and x, respectively. $J_7$ and $J_8$ are $90^0$ out of phase and symmetrical around $45^0$ from the x-axis and y-axis, respectively. The last two modes have quasi-quadrature symmetric about the x-axis and y-axis at the same time. Clearly, the only modes $J_1$ and $J_2$, have good main lobes, whereas the other modes have split main lobes. Therefore, these are unacceptable modes according to (6) and the theory of mode expansion.

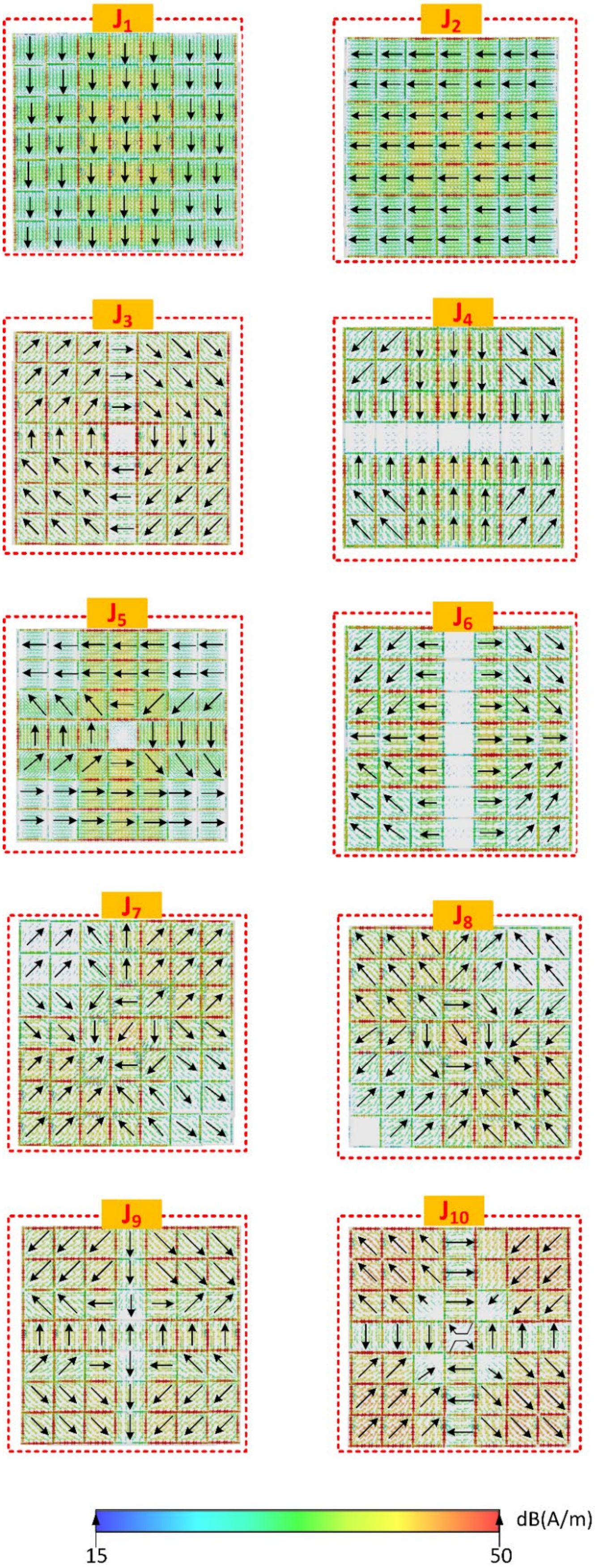

Figure 4.17 Modal surface current of MTS.

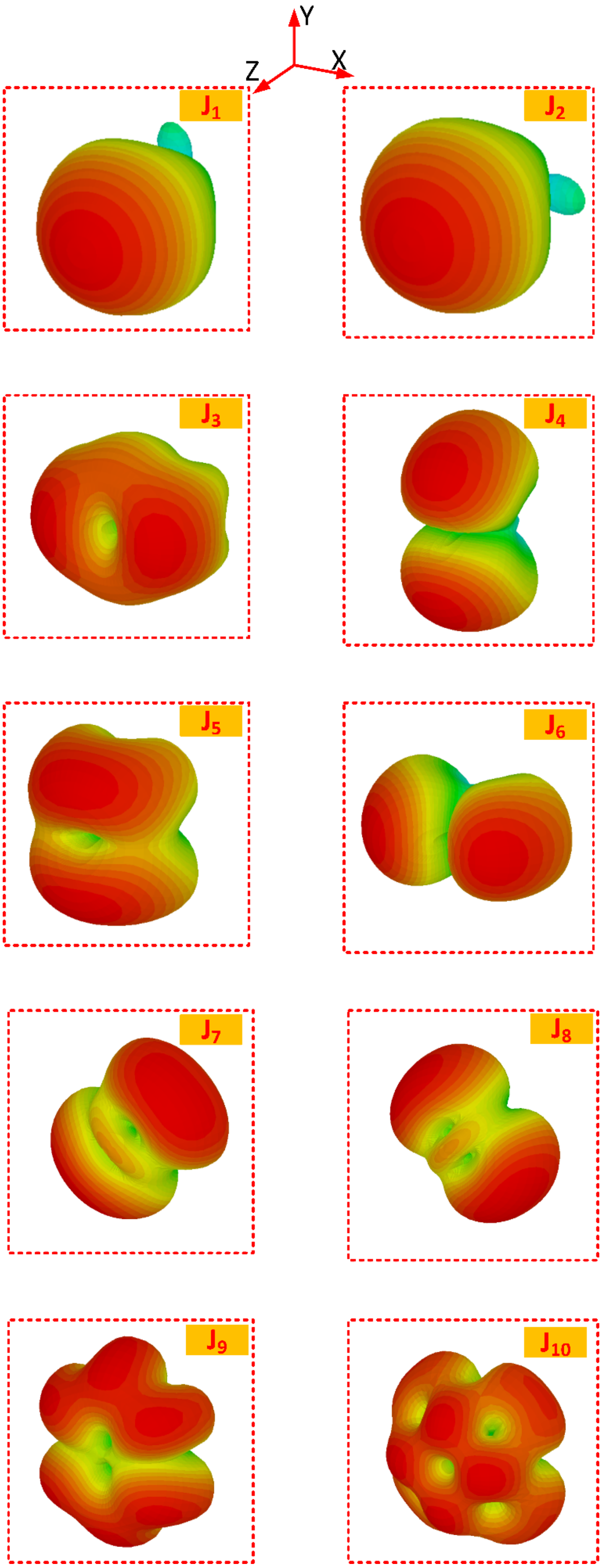

Figure 4.18 Modal radiation patterns of MTS.

## 4.5 Proposed Antenna with MTS

In this section, the MTS is used to provide high gain, wide bandwidth, reduce the antenna size and reduce the back radiation from slot antenna. The integration between the proposed antenna and MTS is as following:

- The dual polarized antenna is based on two slots are introduced to operate at 28 GHz.
- Then we optimize the dimensions of metasurface unit cell to have the pair of orthogonal modes ($J_1$-$J_2$) with broadside radiation. Moreover, the others modes of MTS are out of the focused band and they have spilt the main lobe.
- This section introduces the integration between the dual slots antenna and the MTS. The slots that are adjusted to operate at 28 GHz are used to excite the modes $J_1$ and $J_2$ of the MTS that have pure resonant at 28 GHz.

The MTS is fed by the two slot antennas; therefore, the two small slots are etched from the MTS and aligned to the slots of the antenna to increase the coupling between the antenna and MTS. Figure 4.19 delicates the configuration of proposed antenna with MTS layer. The optimized dimension of the antenna after integrated with MTS are shown in Table 4. 3. The overall dimensions of antenna is extended by 5 mm in x and y to be compatible with the end launch connector (1.85 mm). The proposed antenna printed on Rogers 4003C with dielectric constant 3.38 and thickness 0.2 mm. The prototype of the proposed antenna is shown in Figure 4.19 (c).

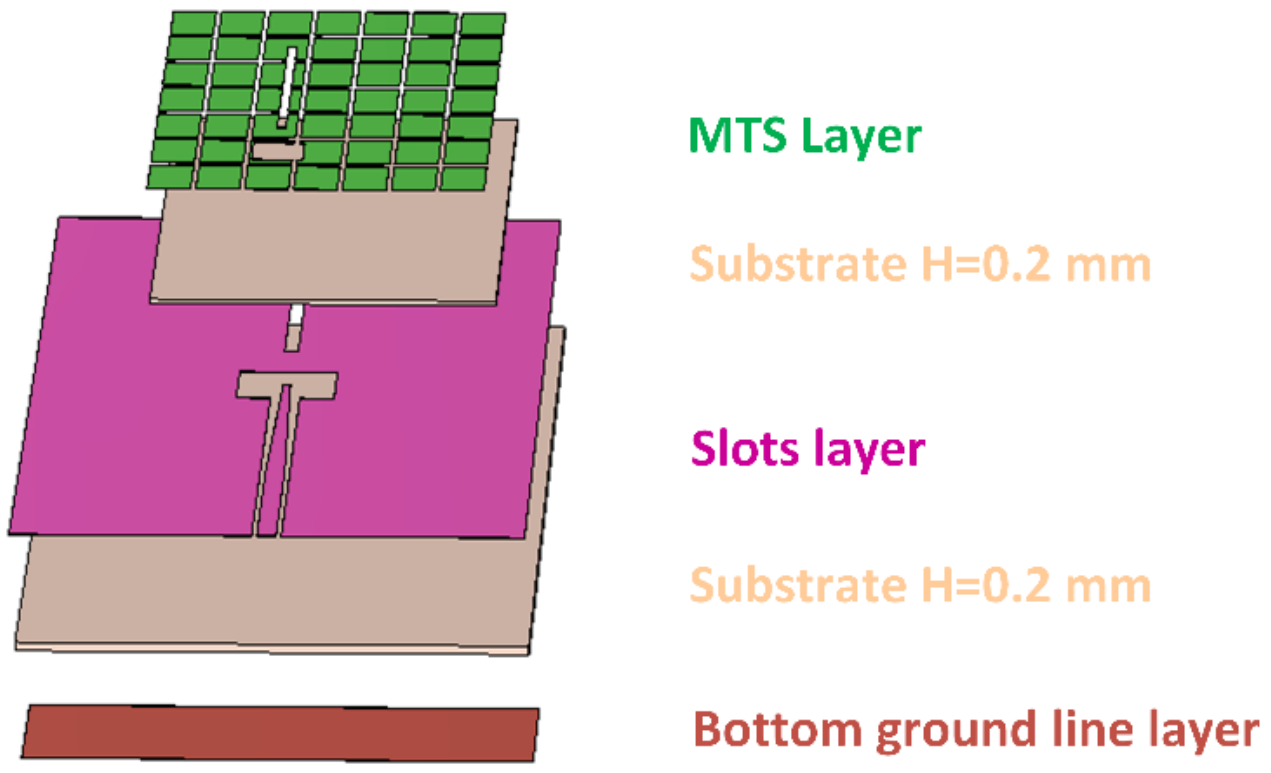

(a)3-D structure

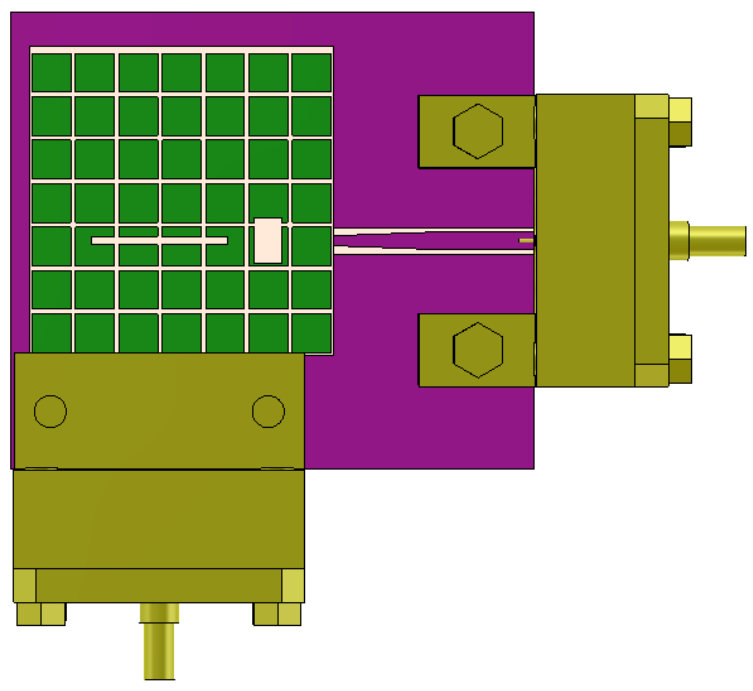

(b)Top view

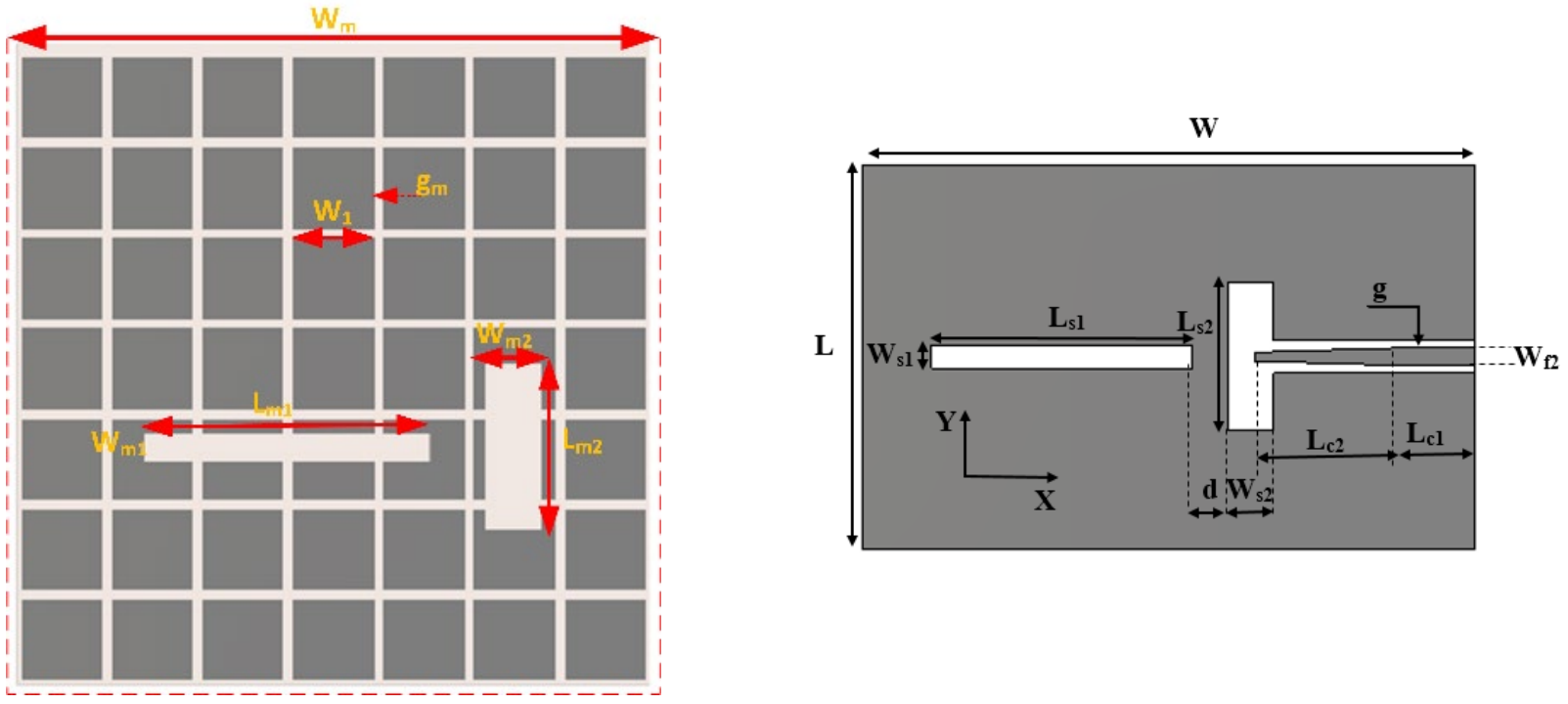

(c)MTS layer (d)Slots layer (e)Bottom layer

Figure 4.19 Configuration of proposed antenna with MTS.

Table 4. 3 Dimensions of proposed MTS antenna (mm)

| | | | | | | | | | | | |
|---|---|---|---|---|---|---|---|---|---|---|---|
| L | W | $L_{s1}$ | $W_{s1}$ | $Y_1$ | $W_{f1}$ | $L_{f1}$ | $L_{s2}$ | $W_{s2}$ | $W_t$ | $L_t$ | $L_{c1}$ |
| 20 | 20 | 4.8 | 0.6 | 1.6 | 0.428 | 6 | 3.2 | 1.1 | 0.21 | 0.4 | 3 |
| $L_{c2}$ | G | $W_c$ | $W_g$ | d | $W_m$ | $W_1$ | $g_m$ | $W_{m1}$ | $W_{m2}$ | $L_{m1}$ | $L_{m2}$ |
| 3.5 | 0.2 | 0.45 | 3.5 | 0.9 | 13.3 | 1.7 | 0.2 | 0.4 | 1 | 4.8 | |

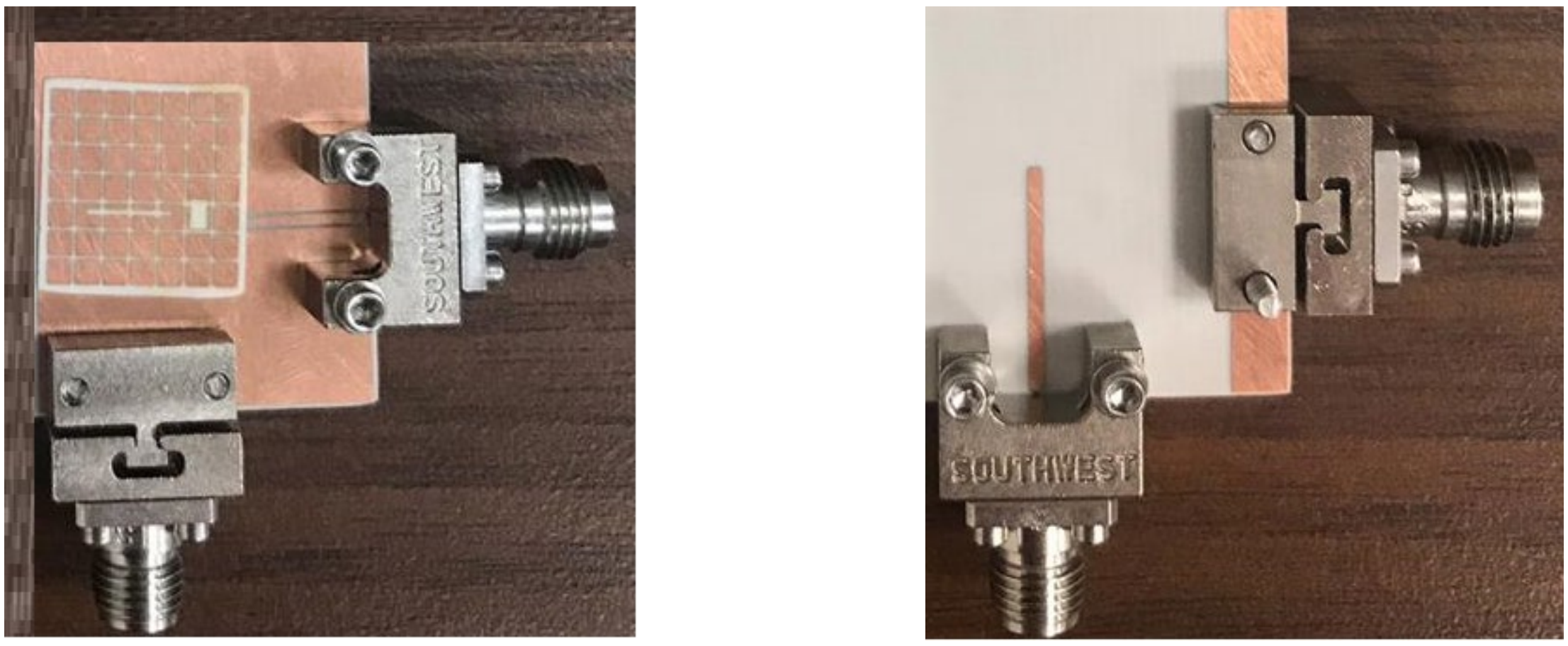

(a)Front view (b)Back view

Figure 4.20 Prototype of the proposed antenna with MTS.

Figure 4.21 shows the measured and simulated reflection coefficients and isolation coefficients of the ports for the proposed antenna. The results confirmed that the proposed antenna achieves wide bandwidth from two ports (26.5-29.5 GHz for port 1 and 25.5 – 30 GHz for port 2) that satisfy the requirements of 5G in term of bandwidth. The measured operating frequency of the proposed antenna is at 28 GHz, which is in good agreement with the simulated result. The proposed antenna achieves good isolation between its ports (more than 40 dB).

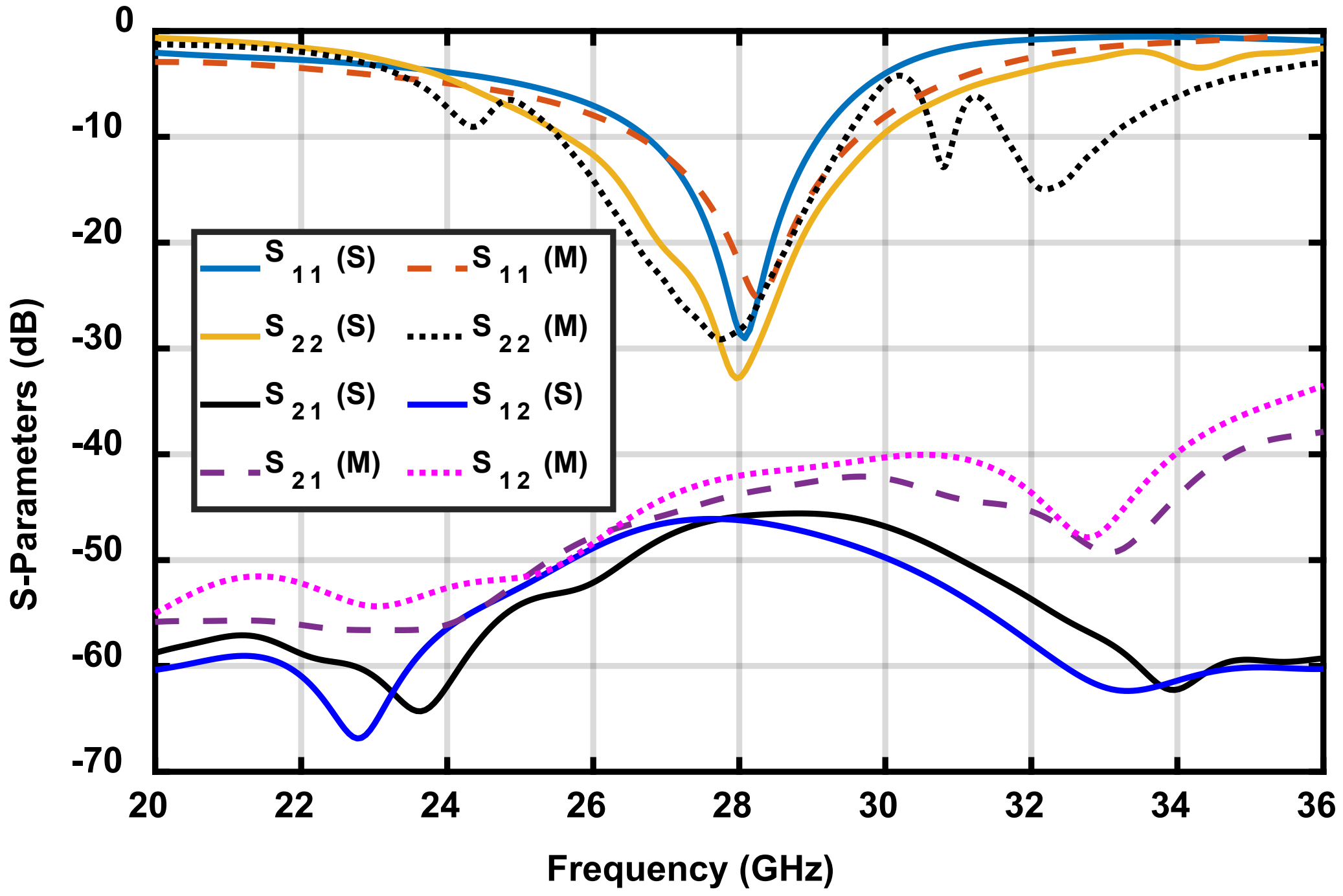

Figure 4.21 Simulated and measured S-parameters of single antenna (Ports 1 and 2).

The normalized radiation patterns of MTS antenna are shown in Figure 4.22 for port 1 and port 2 in the x-z plane and y-z plane at 28GHz. The co and cross components are introduced with more than 40 dB as a difference between them. Furthermore, the MTS achieved low back radiation at the x-z and y-z planes. The radiation efficiency of the two ports is illustrated in Figure 4.23. The efficiency of the proposed MTS antenna is around 95% within the whole band. The gain of the proposed antenna is shown in Figure 4.24; one can notice that the MTS is used to increase the gain of the proposed antenna by about 4 dBi.

| Port | XZ Plane (Phi=0) | YZ Plane (phi=90) |
|---|---|---|
| Port 1 | 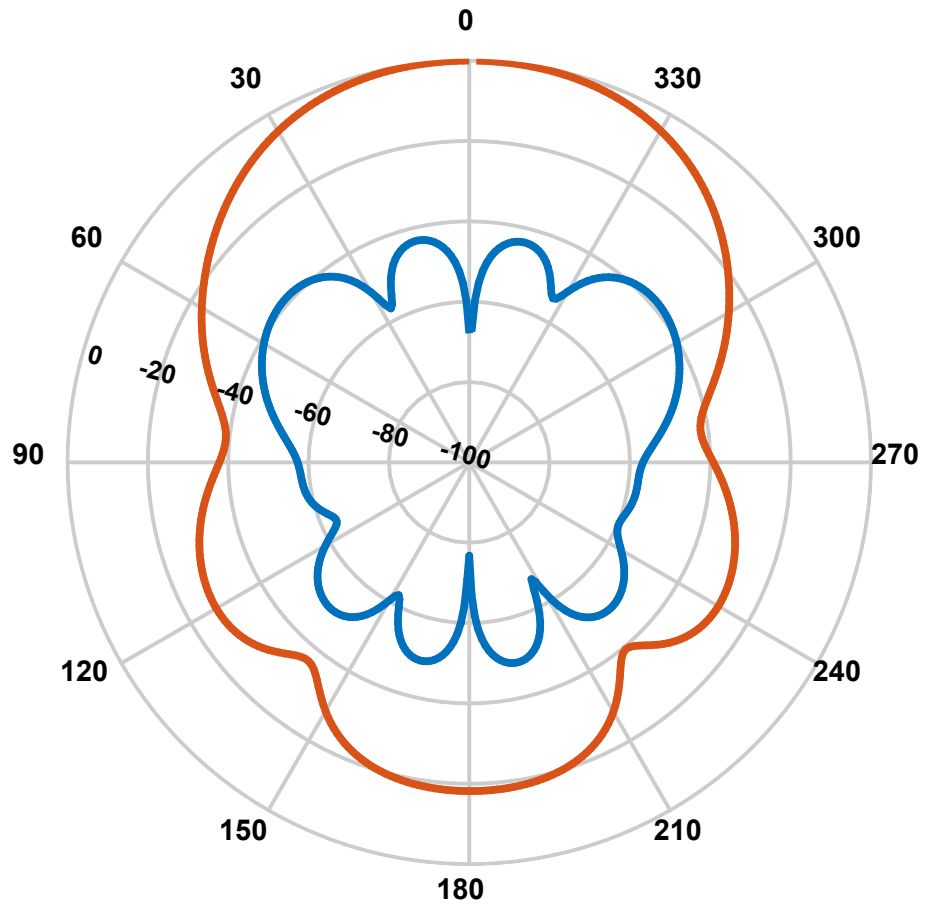 | 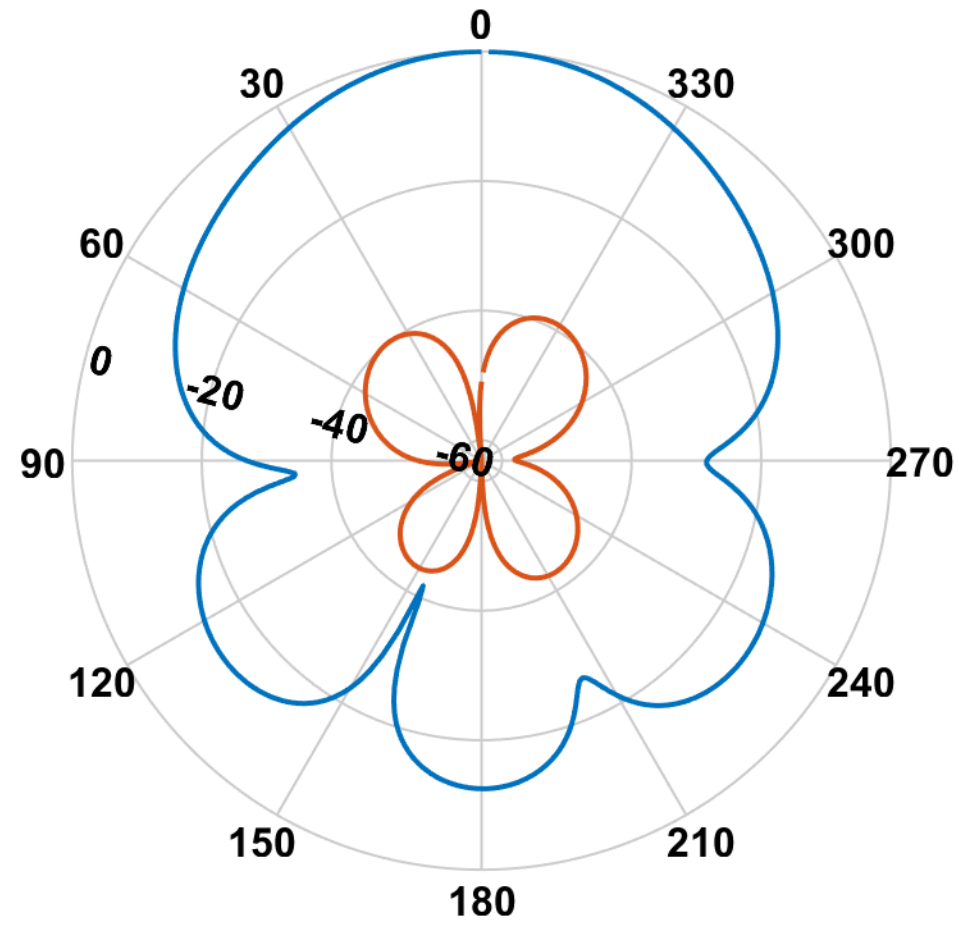 |
| | Cross-Polarized Co-Polarized | Co-Polarized Cross-Polarized |
| | (a) | (b) |
| Port 2 | 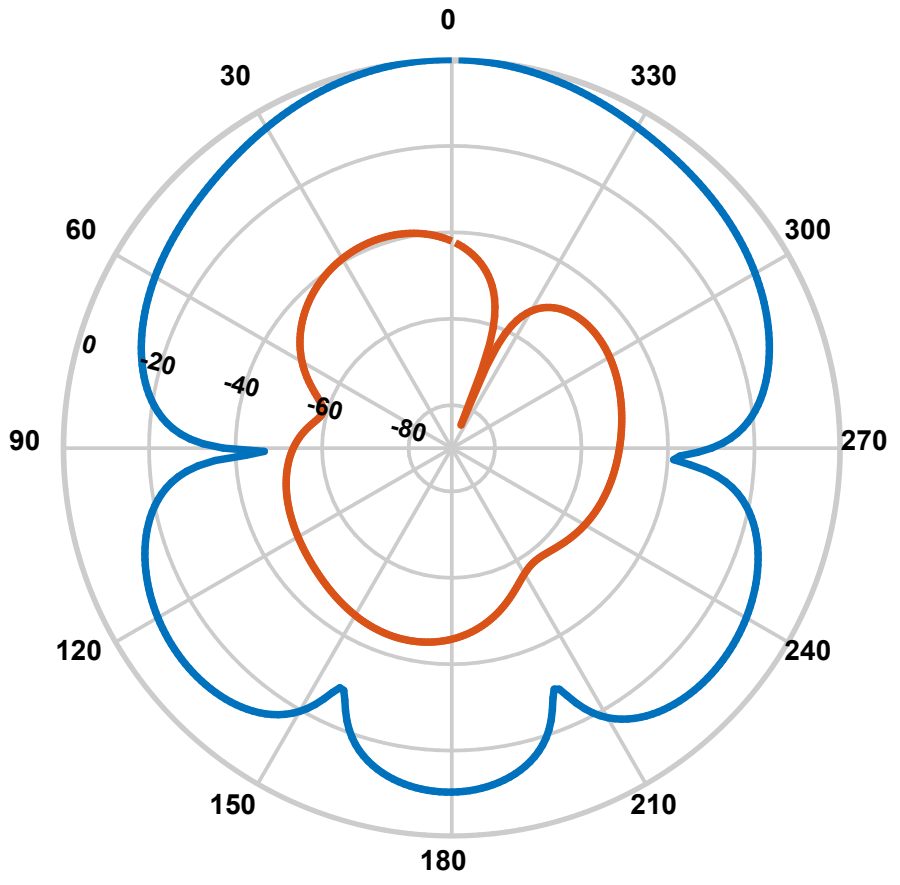 | 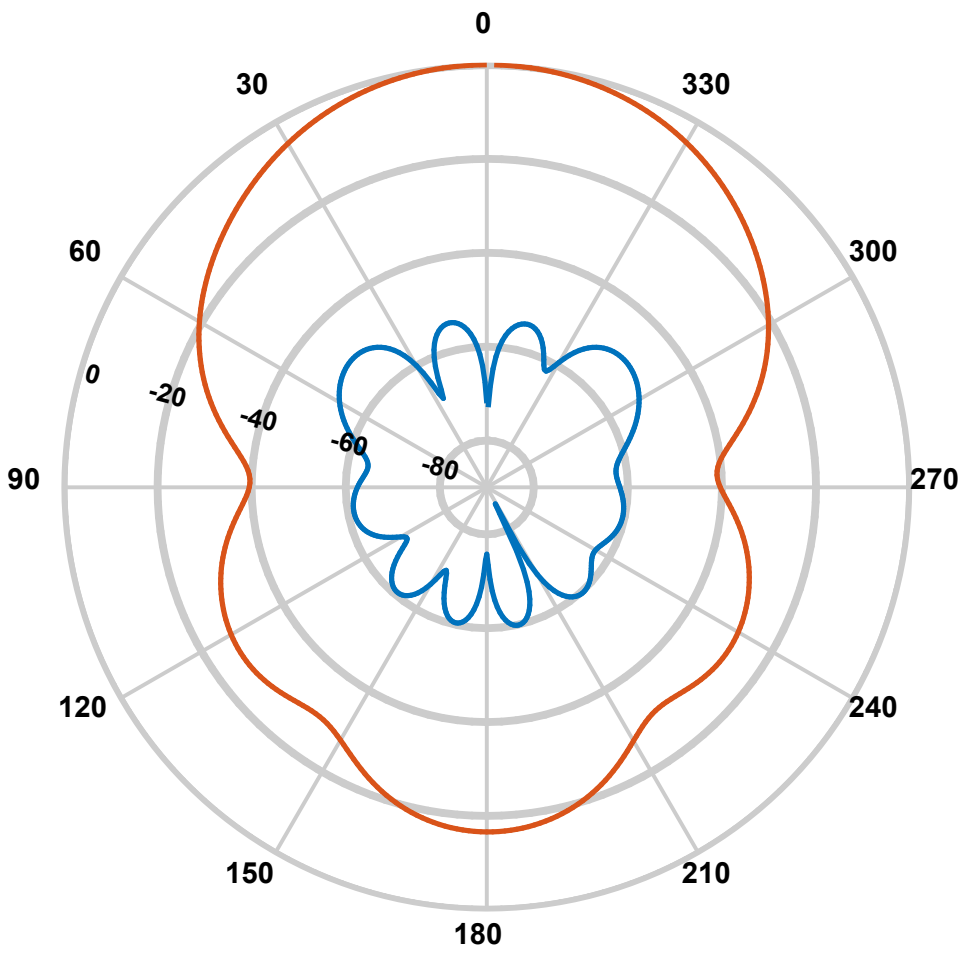 |
| | Co-Polarized Cross-Polarized | Cross-Polarized Co-Polarized |
| | (c) | (d) |

Figure 4.22 Co-Polarized and Cross-Polarized for port1 and port 2 of MTS antenna at 28GHz.

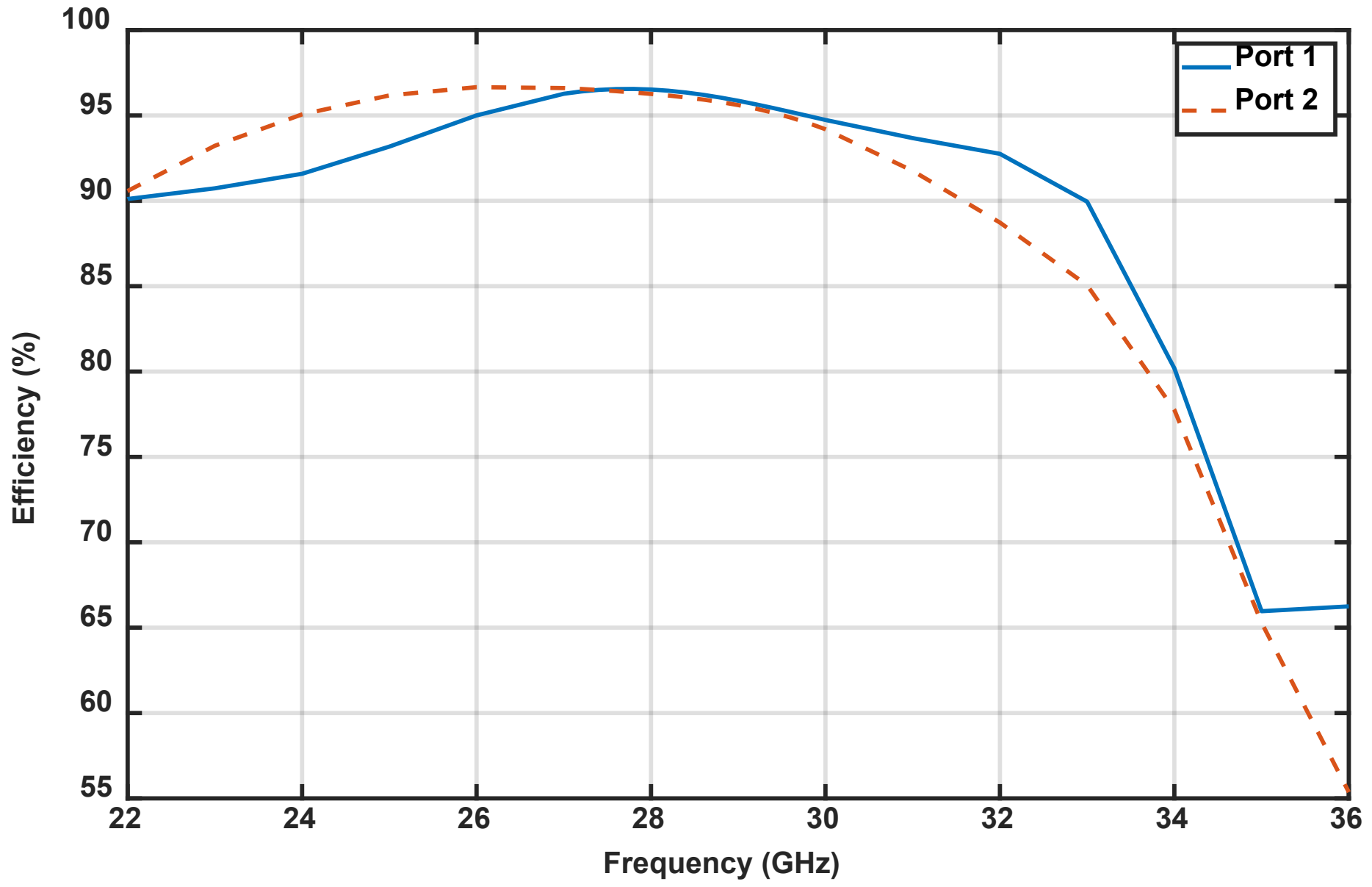

Figure 4.23 Efficiency of proposed antenna.

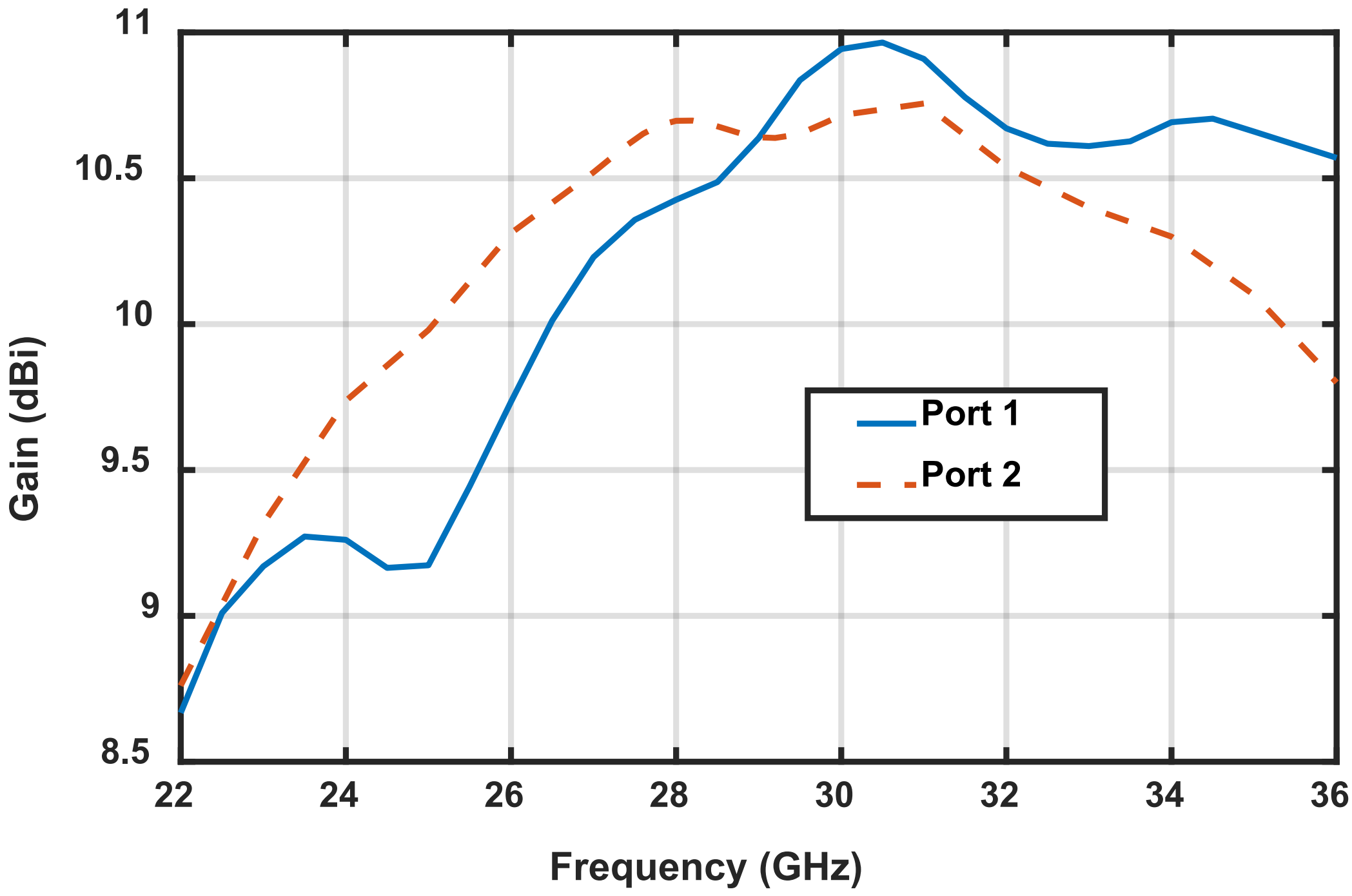

Figure 4.24 Gain of proposed MTS antenna.

# 4.6 MIMO Antenna Design

## 4.6.1 Fabrication and Measurements

The multiple-input-multiple-output (MIMO) system is preferred for 5G smartphone applications to meet the high demand to maximize throughput and quality of service. In other words, the MIMO antenna technology is one of the most significant components of future wireless communication schemes as it improves throughput without raising input power and bandwidth. However, the incorporation of the MIMO antenna scheme into the same board for handheld devices that have a small size is challenging owing to the high mutual coupling between the adjacent antenna components, particularly when they are spaced less than a half-wavelength apart. In our proposed MIMO antenna, the antenna elements are positioned at the corner of the handset board with a total dimension $100 \times 60\ mm^2$ as shown in Figure 4.25 for design configurations and prototypes of MIMO antenna with MTS.

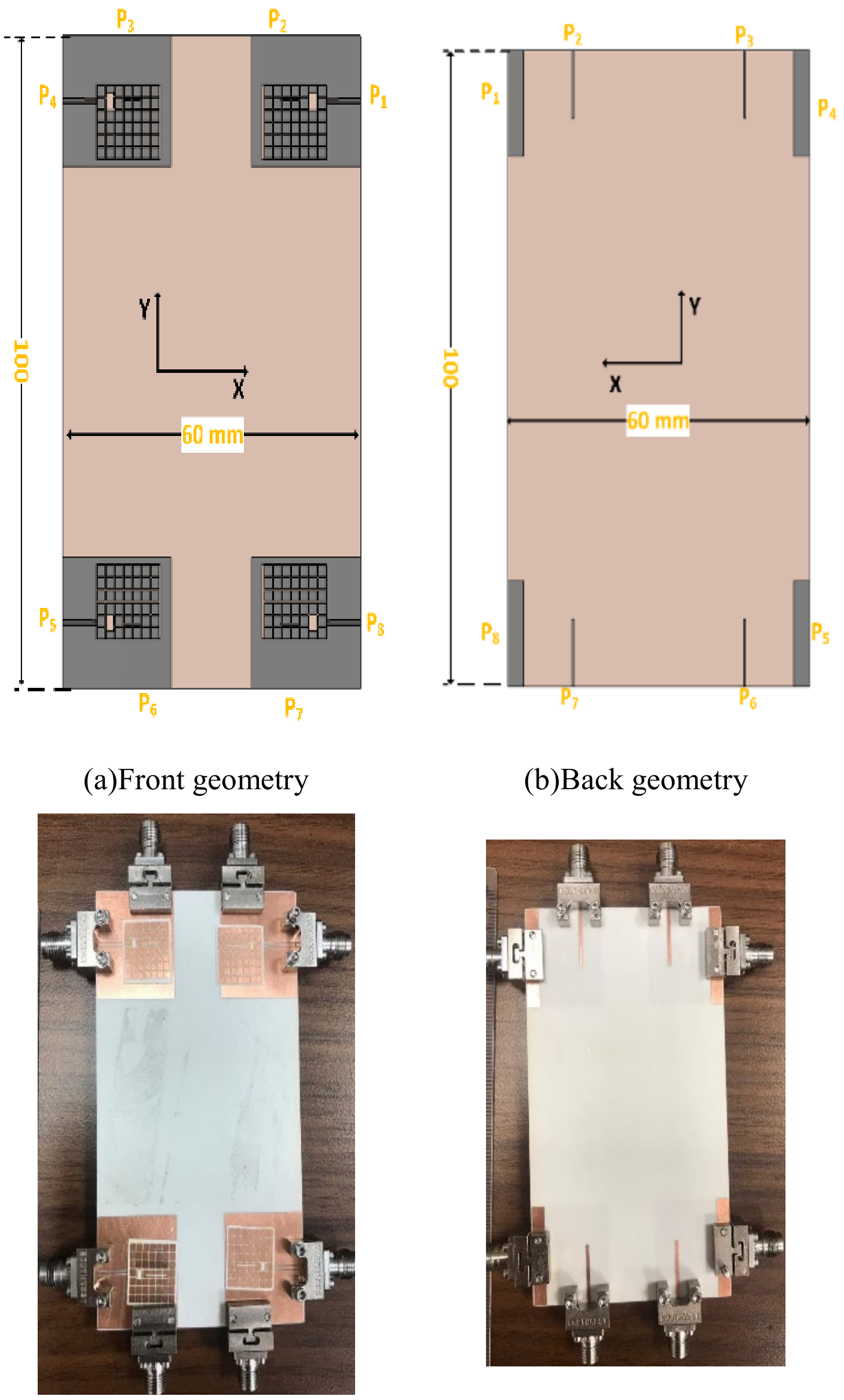

(a)Front geometry (b)Back geometry

(c)Front photo (d)Back photo

Figure 4.25 MIMO configuration and prototype of proposed antenna with MTS.

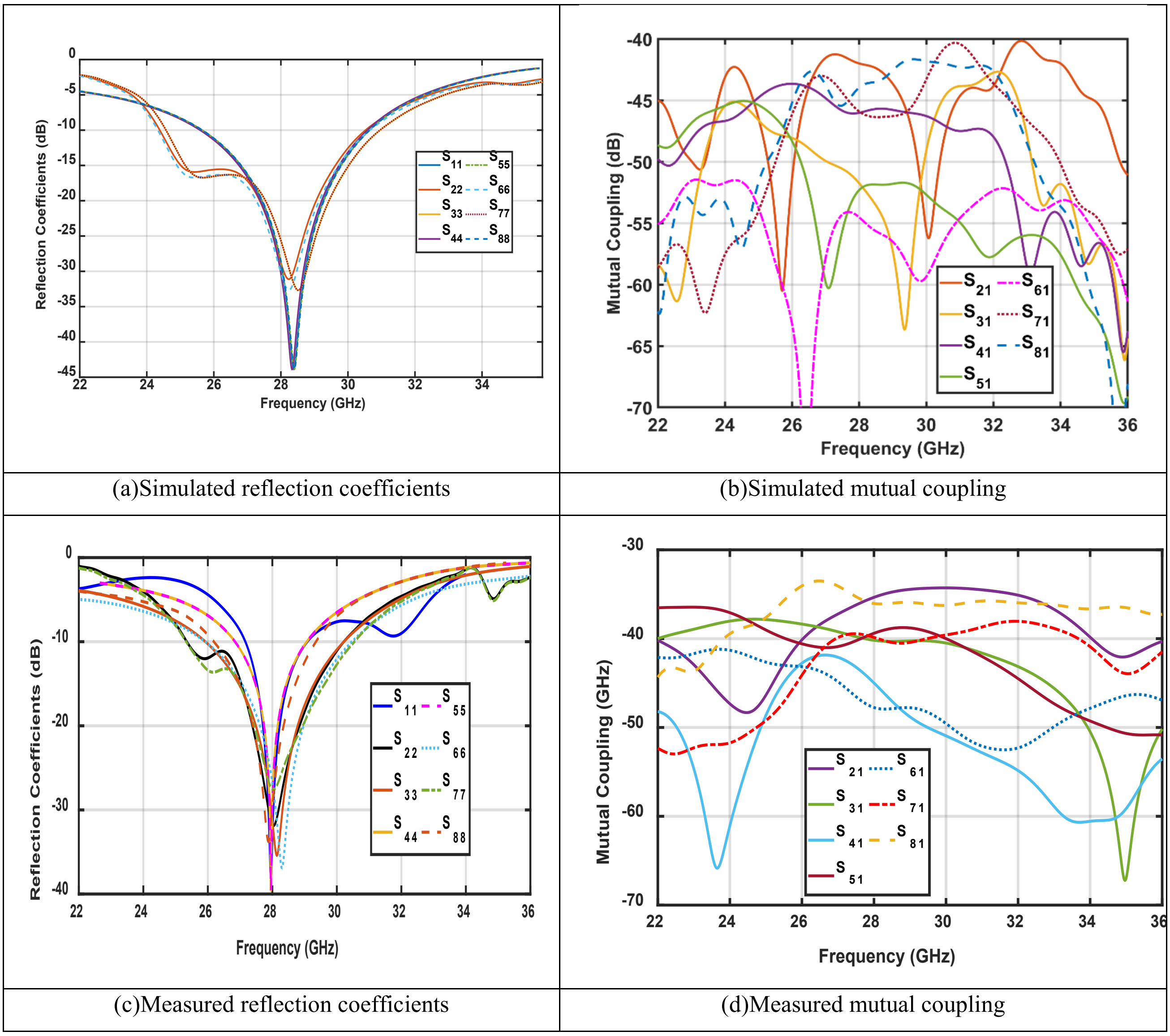

(a)Simulated reflection coefficients

(b)Simulated mutual coupling

(c)Measured reflection coefficients

(d)Measured mutual coupling

Figure 4.26 S-parameters of proposed antenna with MTS.

Some of the antennas in smartphones require a common ground plane between its MIMO elements. Therefore, the MIMO antenna with printed common ground plane on the top layer is presented in Figure 4.27. The common ground plane does not have any significant changes on the reflection coefficients of the MIMO elements, in contrast, it reduces the isolation between ports by small significant amount as shown in Figure 4.28. One can notice that the worst isolation coefficient is higher than 37 dB between port 1 and port 2. Furthermore, all ports have good matching and achieve the required BW for 5G applications.

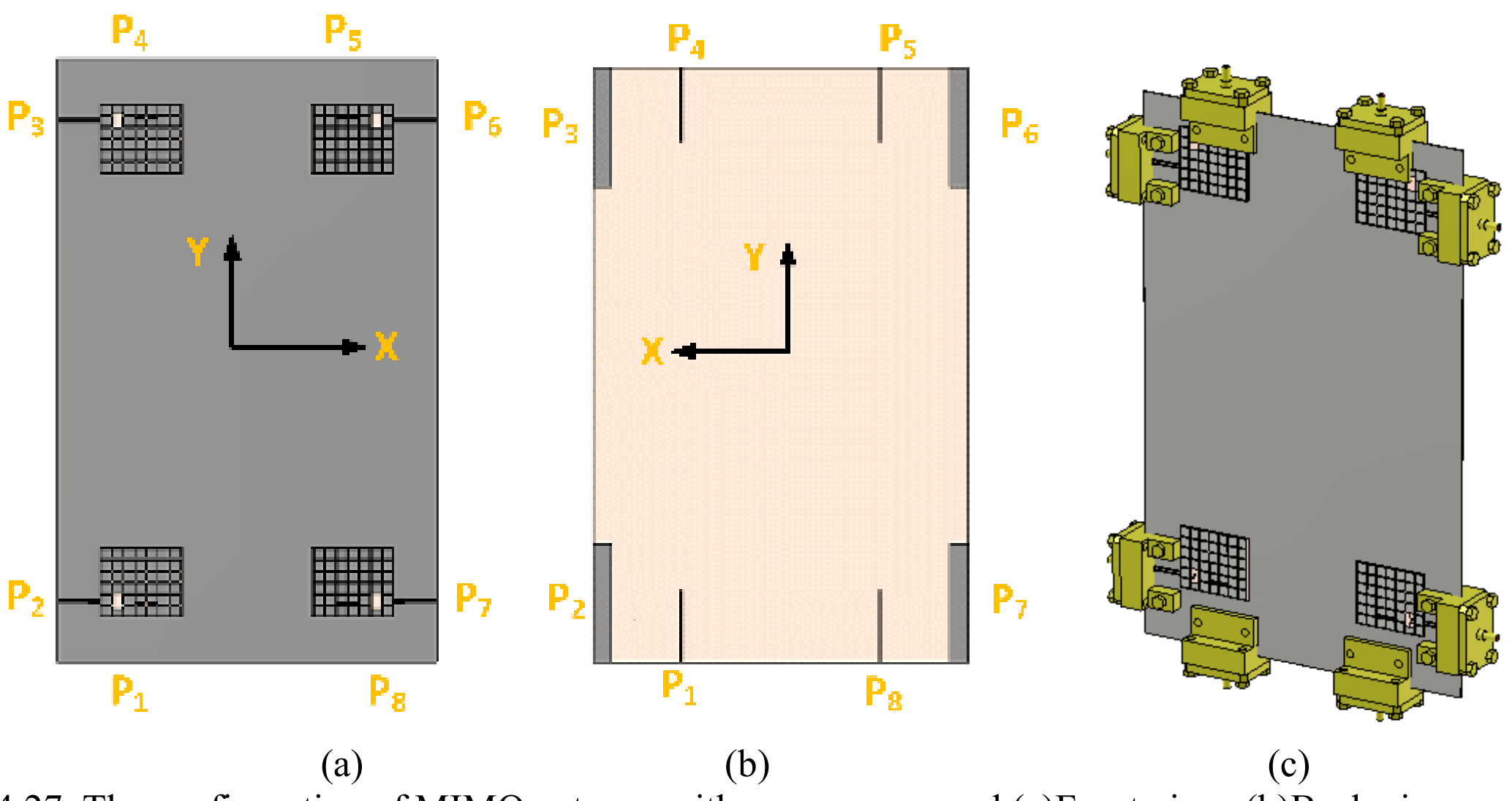

Figure 4.27 The configuration of MIMO antenna with common ground (a)Front view, (b)Back view and (c)3-D view.

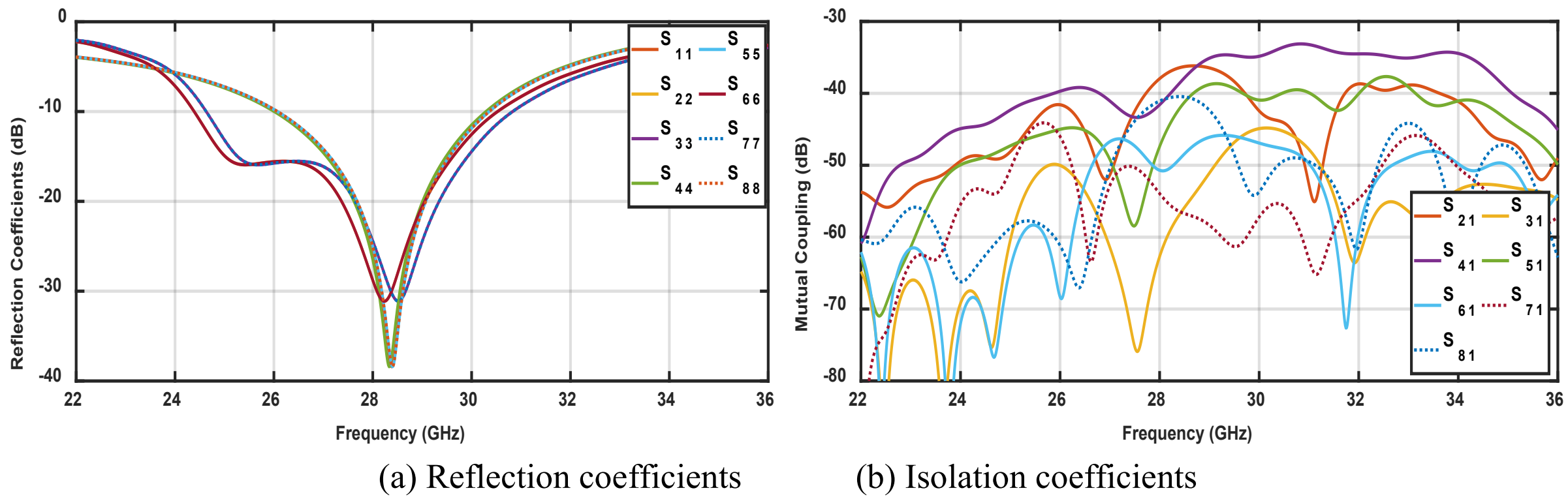

Figure 4.28 S-parameters of proposed MIMO antenna with common ground plane.

## 4.6.2 Envelope Correlation Coefficient

The envelope correlation coefficient (ECC) is one of the main parameters to evaluate the MIMO performance. Where the ECC is used to calculate the similarity between the antenna performances and evaluate the diversity between the elements of MIMO. The acceptable value of ECC should be less than 0.5 [196-199]. Whereas the lower values of ECC mean that the two antennas are good isolated. The ECC can be calculated as [200]:

$$\rho_{nm} = \frac{|S_{nn}^{*}S_{nm} + S_{mn}^{*}S_{mm}|^2}{\left(1-(|S_{nn}|^2+|S_{mn}|^2)\right)\left(1-(|S_{mm}|^2+|S_{nm}|^2)\right)} \tag{4.11}$$

Where $\rho$: ECC, S:S-parameter, S*: complex conjugate of S-parameters, m, and n are number of antenna m,n =1,2,….,8.

Figure 4.29 shows the ECC between MIMO elements from simulated and measured data. It is obvious from the figure that the ECC is less than 0.001 within the operating band.

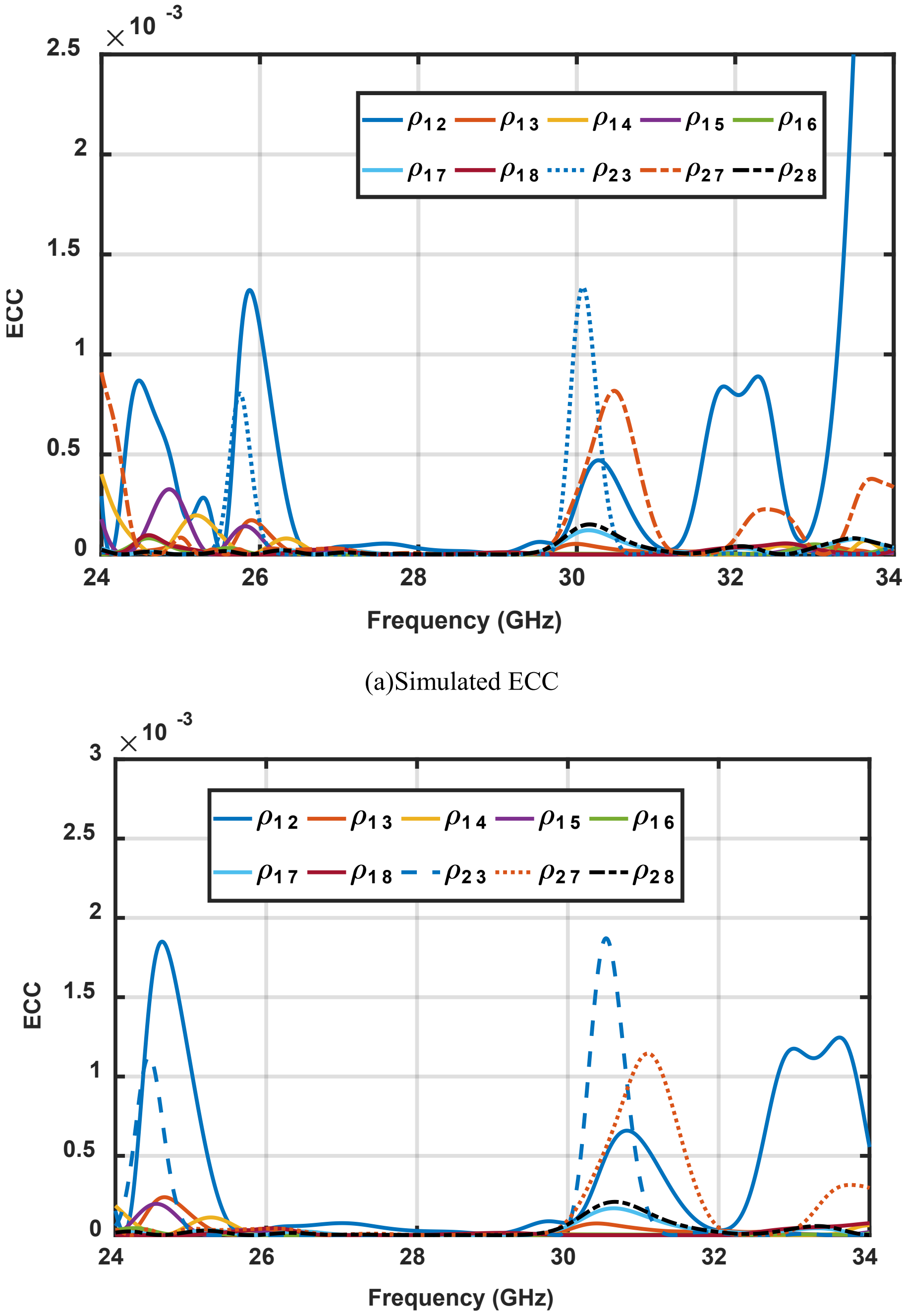

(a)Simulated ECC

(b)Measured ECC

Figure 4.29 ECC parameter of proposed antenna with MTS.

The ECC can be calculated based on the radiation pattern as:

$$\rho_{mn} = \frac{\left|\iint_0^{4\pi}\left[\overrightarrow{F_{\mathrm{m}}}(\theta,\emptyset) \times \overrightarrow{F_{\mathrm{n}}}(\theta,\emptyset)\right]d\Omega\right|^2}{\iint_0^{4\pi}\left|\overrightarrow{F_{\mathrm{m}}}(\theta,\emptyset)\right|^2 d\Omega \iint_0^{4\pi}\left|\overrightarrow{F_{\mathrm{n}}}(\theta,\emptyset)\right|^2 d\Omega} \tag{4.12}$$

Where $\rho$: ECC, $\mathrm{F}(\theta,\emptyset)$: radiation patterns of antenna #m or #n, m and n are number of the antenna m,n =1,2,….,8.

Figure 4.30 shows the different ECC values between the MIMO elements (two elements each time) based on the 3-D radiation pattern of each element. The values of ECC is very small due to the different polarization between the neighbour antennas. It is observed that the values of ECC is less than 0.02 and this means that the MIMO antenna has a good diversity performance.

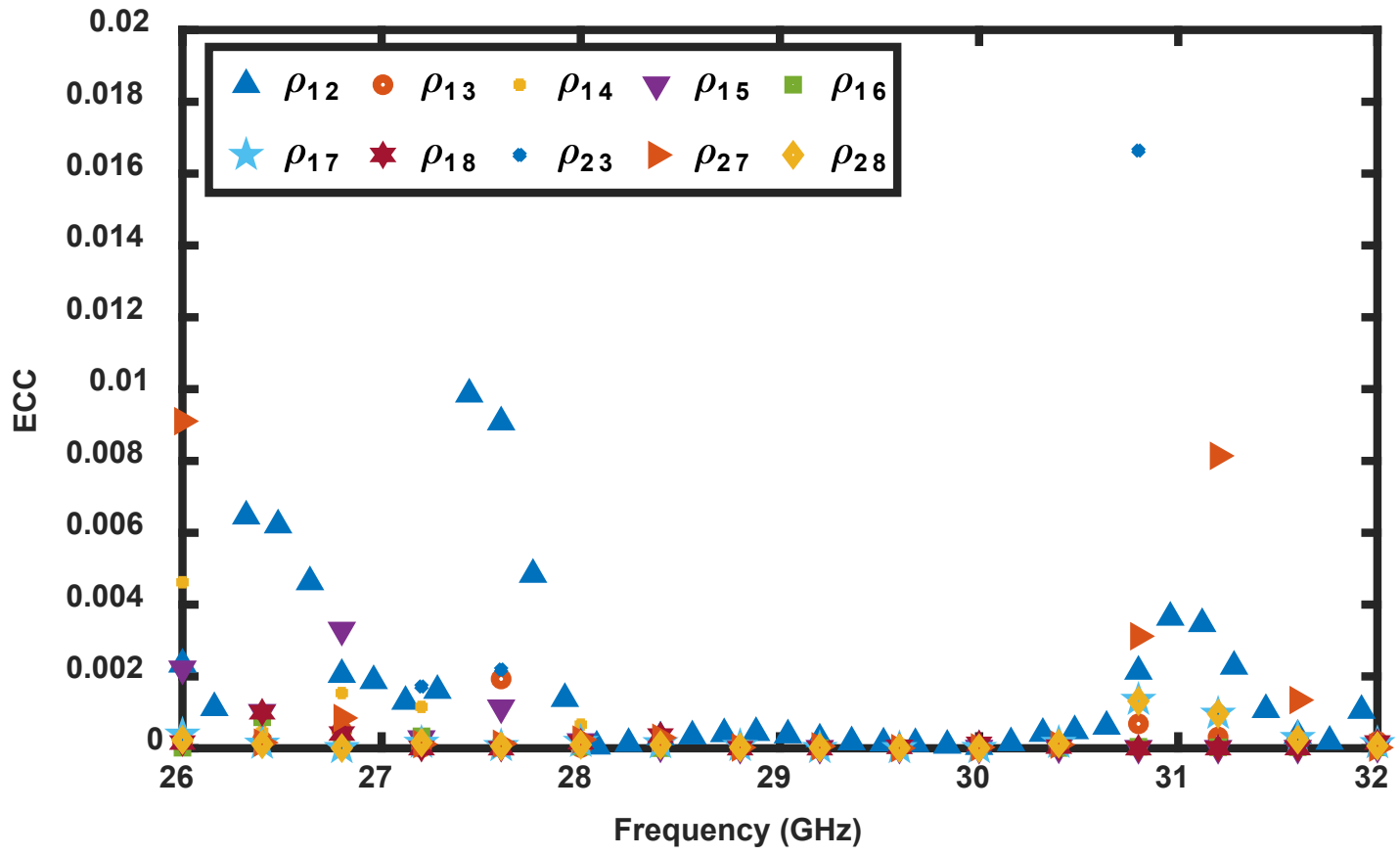

(a)Simulated ECC

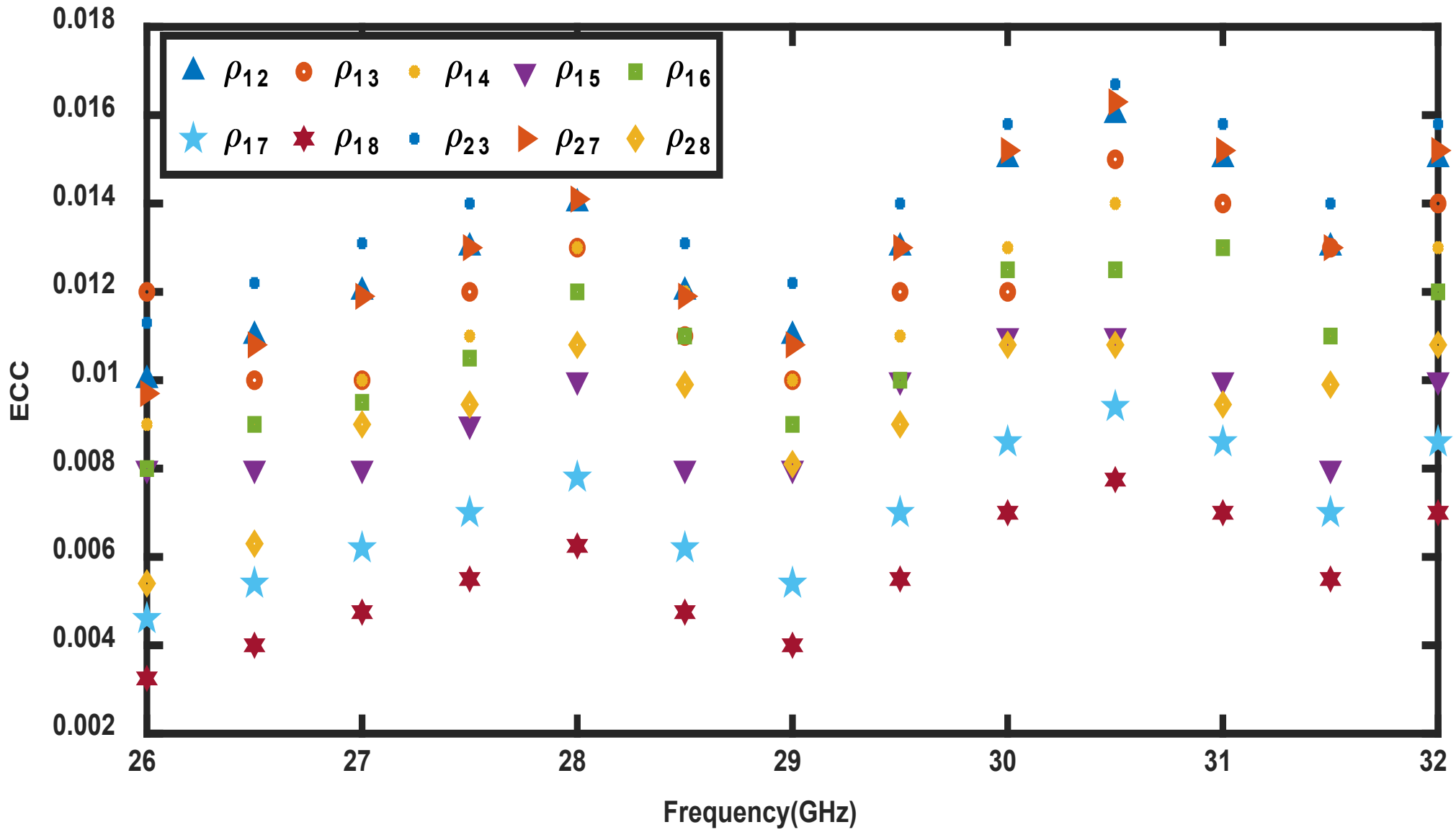

(b)Measured ECC

Figure 4.30 ECC parameter of the proposed antenna with MTS.

### 4.6.3 Smartphone Modeling

#### (A) *Impacts of Housing and Phone Components*

To simulate the real environment of the smartphones, the antenna is integrated with the housing and the components of mobile as shown in Figure 4.31. The screen, speaker, camera, battery, and other components are considered with the proposed MIMO antenna and all components are covered by plastic material. The materials of each part are tabled in Table 4. 4. The module of liquid crystal display (LCD) consists of two parts; the LCD panel and the LCD shield that have the same size of PCB. The battery cell is placed inside the battery shield as shown in Figure 4.31. There are top and bottom fillers to fix the board. Four plastic holders are used to fix the LCD panel and the LCD shield on the filler. The dimensions of all components are compatible with commercial smartphones. The MIMO antenna is tested inside the phone taking the housing and the components into considerations. The S-parameters of the proposed antenna are shown in Figure 4.32. One can notice that the reflection coefficients from all elements are affected due to the existence of the housing. The reflection coefficient of the ports are slightly shifted but are still have good matching and achieve the requirements for millimeter 5G. On the other hand, there is a high isolation between ports. The 3-D radiation patterns of MIMO elements are presented in Figure 4.33.We can notice that the radiation patterns are in different directions due to the diversity between the elements.

Table 4. 4 Materials of smartphone

| **Part** | **Material** | **Properties** |
|---|---|---|
| Housing | Plastic | $\varepsilon_r$=2.2, $tan\delta$=0.005 |
| Camera | Glass(Pyrex) | $\varepsilon_r$=4.82, $tan\delta$=0.0054 |
| LCD shield | Metal(Copper) | $\sigma$=5.8e7 (S/m) |
| LCD panel | Glass(Pyrex) | $\varepsilon_r$=4.82, $tan\delta$=0.0054 |
| Holder | Plastic | $\varepsilon_r$=2.2, $tan\delta$=0.005 |
| Connectors | Plastic | $\varepsilon_r$=2.2, $tan\delta$=0.005 |
| Battery cell | Metal(Copper) | $\sigma$=5.8e7 (S/m) |
| Battery shell | Plastic | $\varepsilon_r$=1.5 |
| Filler | Plastic_HDPE | $\varepsilon_r$=2.3 |
| Speaker | Plastic | $\varepsilon_r$=2.2, $tan\delta$=0.005 |
| Antenna PCB | Rogers 4003C | $\varepsilon_r$=3.38, $tan\delta$=0.0027 |

*HDPE: High Density Polyethylene

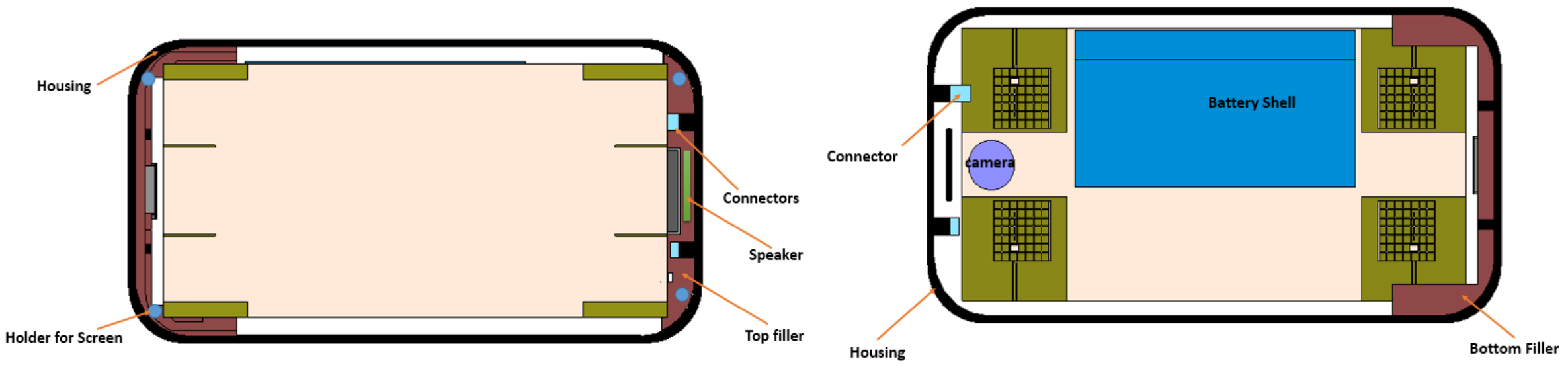

(a)Top view without cover and LCD (b)Bottom view without cover

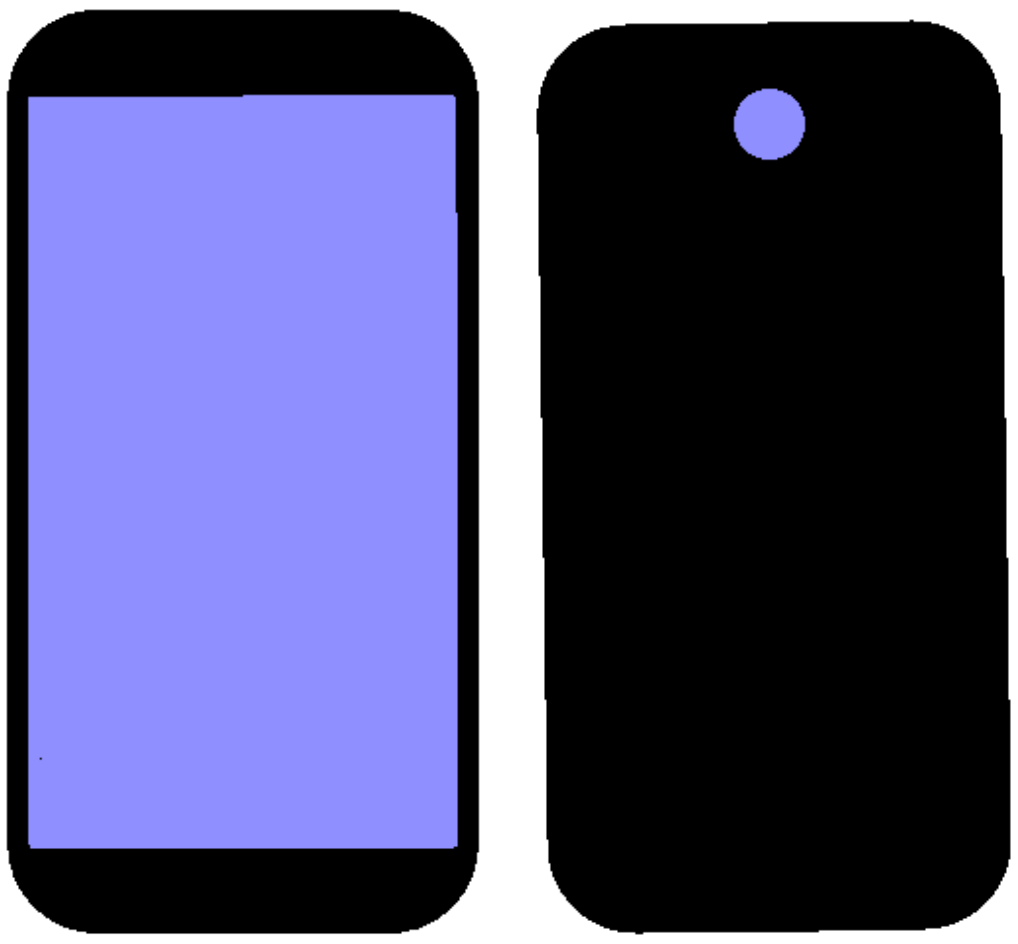

(c)Top view (d)Bottom view

Figure 4.31 Mobile modelling with the components.

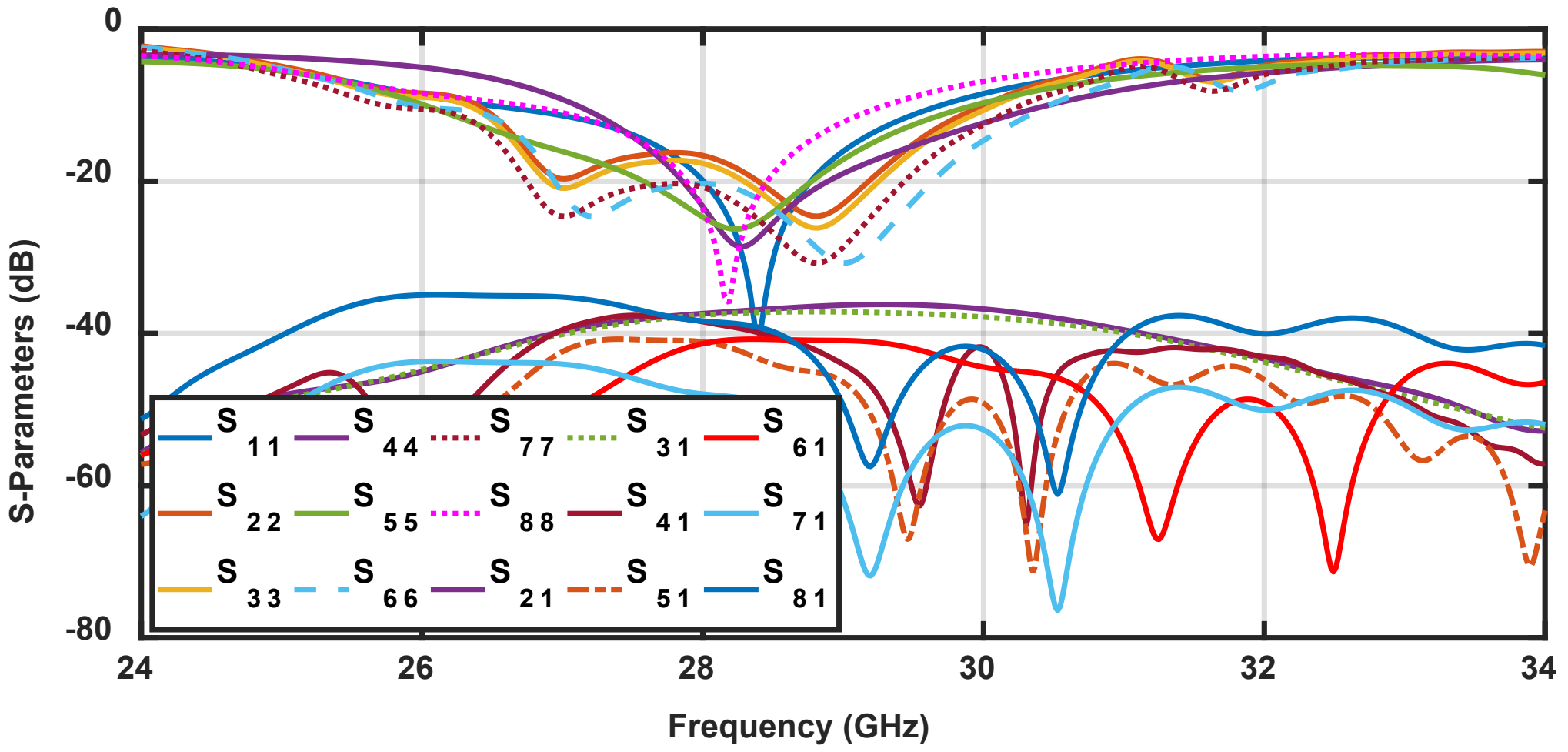

Figure 4.32 S-Parameters of the proposed MIMO antenna inside housing.

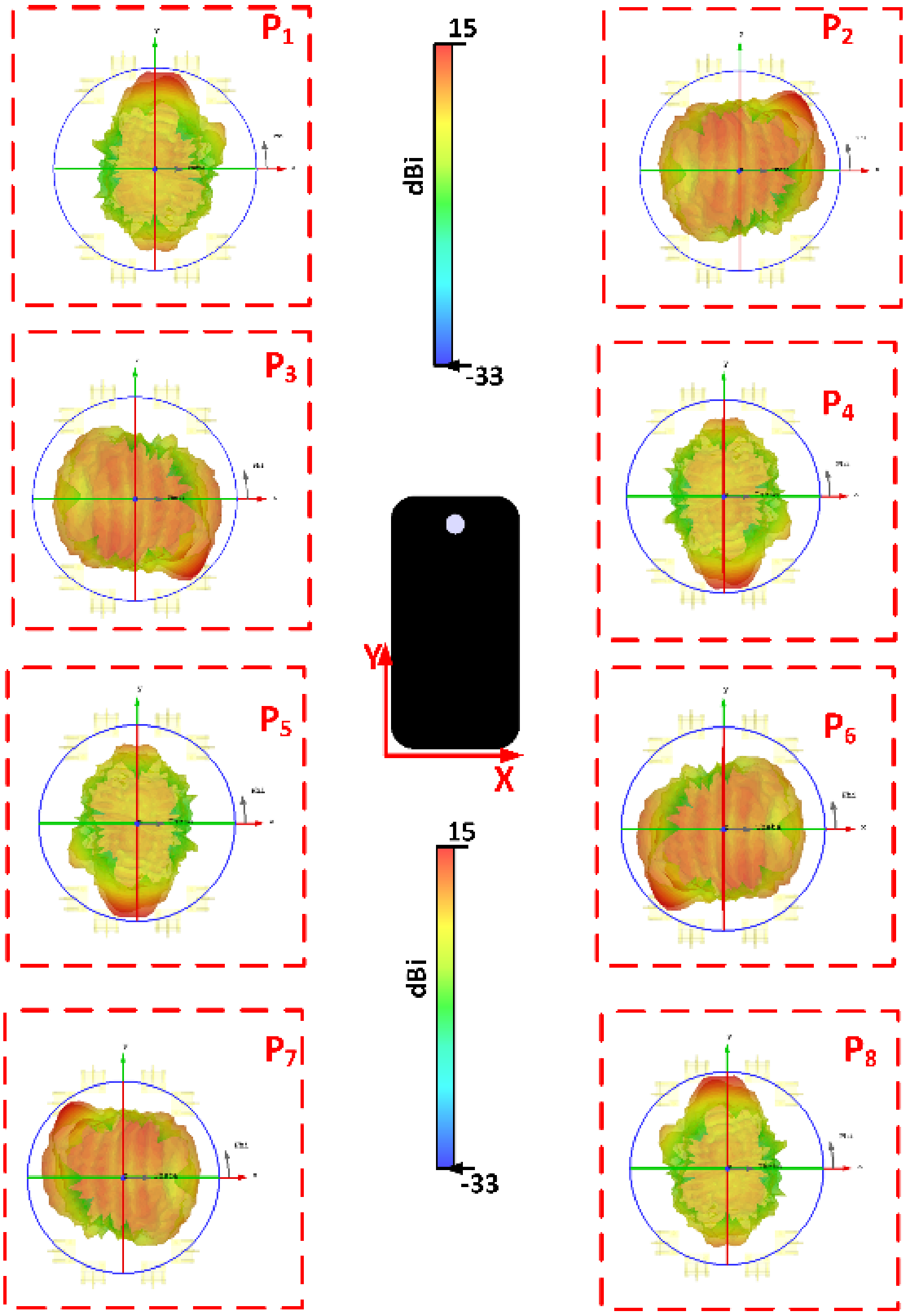

Figure 4.33 3-D radiation pattern of all ports at 28 GHz (with housing).

The Research on health risk from the electromagnetic waves produced from wireless terminals is introduced in the literature. The Specific absorption rate (SAR) is a figure of merit for evaluating the power absorbed by the human tissues. For the frequencies used by current mobile communications networks of second, third and fourth generation (2 G, 3 G and 4 G), basic constraints on RF-EMF exposure are defined in terms of the Specific Absorption Rate (SAR) to avoid, broad safety margins, adverse health effects associated with excessive localized tissue heating and heat stress of the whole body [201-211]. The SAR quantifies the absorbed energy per unit of tissue volume. The SAR values should follow one of two standards: the American standard (1.6 w/kg) for each 10 g and the European standard (2 w/kg) for each 1g [212-215].

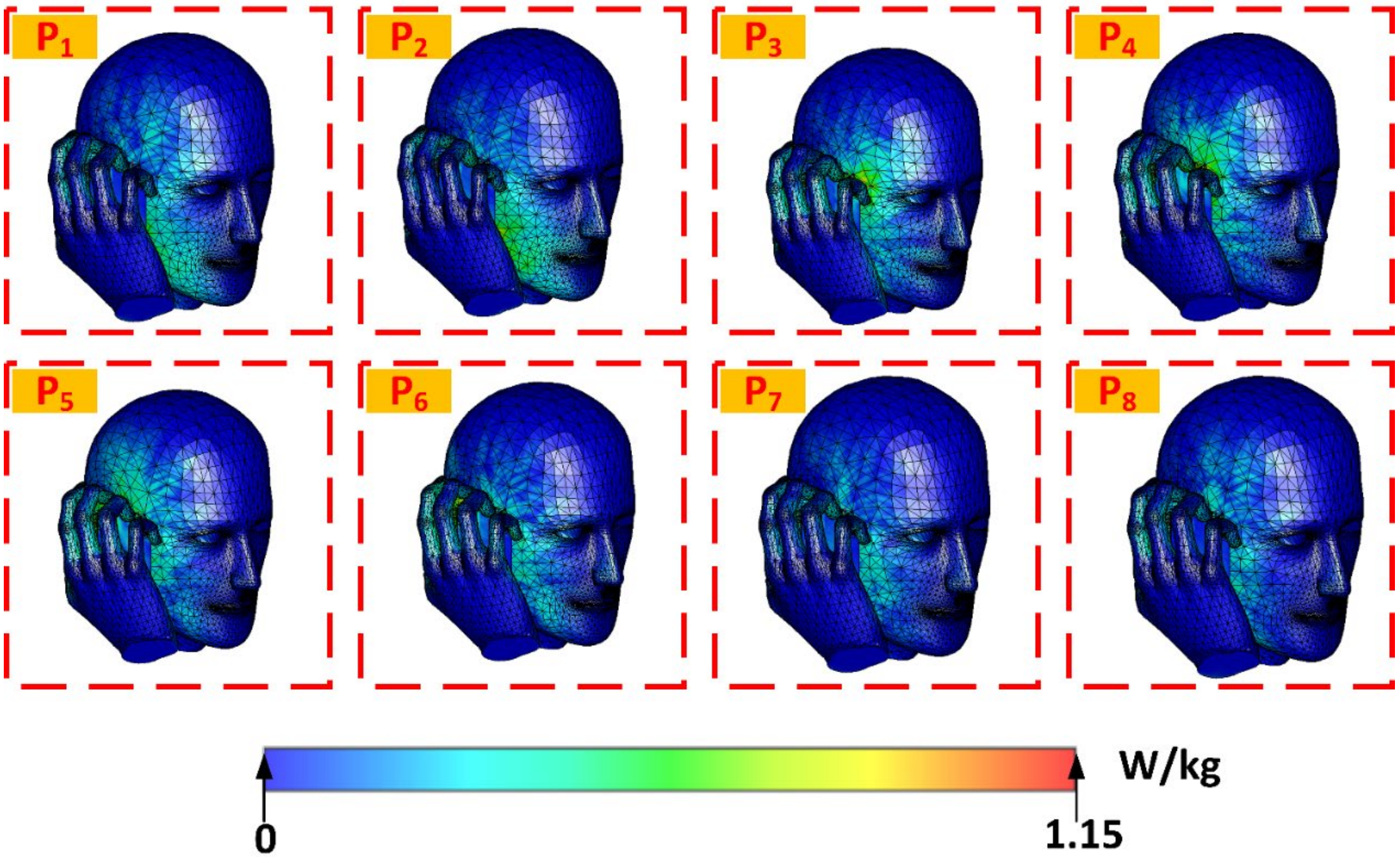

Figure 4.34 SAR distribution from MIMO elements.

Table 4. 5 SAR values (W/kg)

| Port | 1 | 2 | 3 | 4 | 5 | 6 | 7 | 8 |
|---|---|---|---|---|---|---|---|---|
| SAR (1g) | 0.69 | 0.71 | 0.64 | 0.73 | 0.75 | 0.78 | 0.67 | 0.67 |
| SAR (10g) | 0.48 | 0.49 | 0.501 | 0.52 | 0.54 | 0.496 | 0.57 | 0.45 |

For the millimeter wave range, there is two approaches to calculate the electromagnetic exposure to the human:

1. SAR: Some papers in the literature evaluate electromagnetic exposure by the same previous definition of SAR [9].
2. Power density: the term to calculate the electromagnetic exposure into the human body changed from SAR to the term of power density (Pd) because the absorption becomes more superficial due to the fact that penetration is very low at higher frequencies [216-219].

Figure 4.34 shows SAR distribution from 8 elements for 10g standard. The SAR values for the two standards are summarized in Table 4. 5. The antenna is proximity close to the human head model with 0.5 mm distance and inclined as take mode by ($60^0$). The reference power of the proposed antenna elements at 28 GHz is set to 24 dBm for each element.

Table 4. 6 Power density limits from different standards

| | ICNIRP [213] | FCC[212] | IEEE[214, 215] | |
|---|---|---|---|---|
| F(GHz) | 10-300 | 6-100 | 3-30 | 30-100 |
| Pd (W/m$^2$), A | 10, (20 cm$^2$)<br>200, (1 cm$^2$) | 10, (1 cm$^2$) | 10, (100 $\lambda^2$)<br>18.56$f^{0.699}$ | 10, (100 cm$^2$)<br>200, (1 cm$^2$) |

Table 4. 7 Power density values at 28 GHz from different ports according to different standards.

| | Port 1 | Port 2 | Port 3 | Port 4 | Port 5 | Port 6 | Port 7 | Port 8 |
|---|---|---|---|---|---|---|---|---|
| ICNIRP | 1.32 | 1.32 | 1.35 | 1.32 | 1.28 | 1.25 | 1.35 | 1.35 |
| FCC | 1.97 | 1.96 | 1.99 | 1.99 | 1.98 | 1.98 | 1.97 | 1.97 |
| IEEE | 1.71 | 1.69 | 1.78 | 1.77 | 1.68 | 1.69 | 1.71 | 1.71 |

In the second approach, IEEE, FCC and International Commission on Non- Ionizing Radiation Protection (ICNIRP) introduced frequency limits at which the definition of SAR calculation shifts to power density calculation as shown in Table 4. 6. The conversion frequency at which this shift in exposure metric is 3 GHz, 6 GHz and 10 GHz, for IEEE, FCC, and ICNIRP, respectively. In other words, at mm-Wave frequencies, PD is currently preferred due to the difficulty of determining a reasonable volume for SAR assessment when penetration depths are very low [212-215]. The power density exposure into the human model is calculated as shown in Table 4. 7 for all ports and compare its values from different standards. We noted that all the power densities satisfy the safety guidelines. Table 4. 8 lists two comparison sections; the first section makes a comparison between the proposed antenna and the referenced dual-polarized antennas, and the second section makes a comparison between the proposed antenna and the referenced MIMO antennas of smartphones. High isolation, low profile, low complexity, compact size, high efficiency, high gain, high cross-polarization are achieved in the proposed antenna.

Table 4. 8 Comparison between referenced antennas and the proposed antenna

| **Ref** | **Size ($\lambda_0^3$)** | **Isolation (dB)** | **Gain (dBi)** | **X-pol (dB)** | **Freq. (GHz)** | **Eff. (%)** | **Complicated** | **Remarks** |
|---|---|---|---|---|---|---|---|---|
| **[220]** | 1.37×1.37×0.222 | 39 | 13 | 42 | 8.16-11.15 | NA | High | • Six layers.<br>• Based on integrated cavity. |
| **[221]** | 3.2×3.2×0.1 | 20 | 3.8 | 25 | 30.1-30.9 | | High | • Multilayer organic buildup substrates. |
| **[86]** | 0.93×0.93×0.004 | 26 | 4.5 | 28 | 1.86-2.97 | NA | High | • H-shaped slot antenna.<br>• $90^0$ phase shift feeding network. |
| **[222]** | 1.75×1.75×0.02 | 35 | 8.6 | 20 | 2.4-4.12 | NA | Medium | • 2-substrate<br>• Dual-pol.<br>• Circular dipole<br>• Microstrip line Balun feed |
| **[223]** | 1.73×1.03×0.144 | 30 | 9.4 | 20 | 1.88-2.9 | NA | High | • Fed by parallel strip line balun.<br>• Bulk structure |
| **[224]** | 1.3×1.3×0.05 | 27 | 9.6 | 29 | 4.8-5 | | Medium | • |
| **[225]** | 2.7×2.54×0.10 | 18 | 7.48 | 10 | 27.5-29.5 | | High | • SIW horn antenna<br>• Works as a waveguide antenna.<br>• 3 layers |
| **Proposed antenna** | 0.83×0.83×0.03 | 40 | 11 | 40 | 25.5-30 | 92 | Low | • Low profile<br>• Two orthogonal slots.<br>• Dual feed. |

**MIMO Antenna for 5G Smartphones**

| **Ref.** | **Phone Board ($mm^2$)** | **$H_0$ (mm)** | **MIMO Order** | **Gain (dBi)** | **ECC** | **Isolation (dB)** | **Dual-Pol. (X.Pol.)** | **Eff. (%)** | **Remarks** |
|---|---|---|---|---|---|---|---|---|---|
| **[87]** | 136×68 | 5 | 8 | | 0.15 | 12.5 | yes (15) | 55 | • C-shaped coupled-fed and L-shaped monopole slot. |
| **[226]** | 150×75 | 6.2 | 8 | NA | 0.08 | 11 | No | 42 | • 3-D folded monopole.<br>• Dual band @3.5 GHz, and 5 GHz. |
| **[183]** | 150×80 | 0.8 | 8 | NA | 0.05 | 17.5 | No | 62 | • Open slot antenna |
| **[227]** | NA | 1.93 | 8 | 7 | NA | 20 | Yes (18.3) | 90 | • 3 layers.<br>• Yagi-uda<br>• Endfire radiation |
| **Proposed antenna** | 100×60 | 0.4 | 8 | 11 | 0.001 | 40 | Yes (40) | 90 | • Low profile<br>• Two orthogonal slots. Dual feed. |

# 4.7 Conclusion

This chapter introduces a dual-polarized MIMO antenna with eight elements for a 5G smartphone. The MIMO configuration is based on the diversity between elements. The dual-polarization antenna is introduced to overcome the high attenuation in 5G communication system and give high data rates. Furthermore, the orthogonal polarization between the antenna ports is used to achieve high isolation between antenna ports. The antenna achieved a good matching bandwidth of more than 2GHz at center frequency of 28GHz. The antenna is combined with MTS to increase its gain and bandwidth. CMT analyzes the MTS and all the parameters are investigated. The antenna is fabricated and measured. The electromagnetic exposures into the human model from the proposed antenna at 28 GHz are investigated and analyzed in terms of SAR and power density.

# Chapter Five:
# Antenna Design for Short Range Communications

## 5.1 Introduction

Nowadays, low-frequency bands are very crowded and with the rapid growth of communication technologies, high-speed short-range wireless communications require a wide band, higher data rate, and compact size. In order to achieve the desired requirements, the millimeter-wave (mmW) band at 60 GHz has more and more attention because it offers unlicensed bandwidth (from 57 GHz to 64 GHz) for several applications such as video streaming, wireless gaming, short-distance communication WPAN [228, 229]. The complementary metal-oxide-semiconductor (CMOS) technology is considered a good solution to cost and circuit integration issues at this frequency. However, the CMOS substrate is inherited losses due to its high permittivity ($\varepsilon r$=11.9) and low resistivity ($\sigma$=10 s/m). Additionally, CMOS antennas at 60 GHz require more enhancements of antenna efficiency and antenna gain [228, 230, 231].

The inherent losses in CMOS substrate is a key factor in RF CMOS designs. So, several studies have been performed to solve the problem of inherent losses in CMOS substrate due to its high permittivity and low resistivity causes performance degradation. Different methodologies are presented to improve the antenna on-chip performance, such as micromachining [232] and proton implantation [233]. Nevertheless, these techniques suffer from reeducation of system level integration and increase the overall cost. On the other hand, Barakat et al. [31], introduce an Artificial Magnetic Conductor (AMC) and High Impedance Surface (HIS) to improve radiation characteristics in the broadside antennas, also introduce a shield plane inserted between the AOC and the lossy CMOS substrate to minimize the loss [234].

This chapter focuses on end-fire antennas. Thus, the Yagi-Uda antenna and Slot Tapered antennas are common end-fire antennas reported in previous works [29, 112, 235]. The previous 60-GHz Yagi antenna designs using CMOS technology suffer low radiation efficiency and low gain communications to replace the metal interconnects between chips [29, 235]. Bao et-al. [235], presented 60-GHz differential Yagi antenna using 0.18-µm CMOS technology combine with AMC to improve the radiation performance with overall size 2.45mm ×1.8mm and feed by differential feeding G-S-G-S-D. However, the achievable gain is - 2.64 dB and the F/B ratio is 16.6 dB. Recently, El-saidy et-al [29], improved gain of Yagi-Uda antenna by changing antenna location from -1.7 dBi at center of CMOS to -0.7 dBi at 1700 μm far from the center. Moreover, in [112], the Vivaldi antenna is introduced in order to give a gain of -0.4 dBi and radiation efficiency of 32% with an overall size of 785μm×930μm.

This chapter presents a solution for the low gain of the on-chip antenna systems and the poor efficiency of this system with outdoor systems. The high gain antennas on the two side walls of the on-chip system (OCS) are introduced to communicate between the on-chip antenna (OCA) and the outdoor systems (ODS). Where the high transmission between the OCS and the ODSs is achieved by using high gain antennas to transmit between them. So, a Quasi Yagi Antenna (QYA) and a Tapered Slot Vivaldi Antenna (TSVA) are introduced to enhance the radiation properties of the end-fire radiator in millimetre wave range for OCS. The proposed two antennas are introduced to use for point to point communications. The antennas are designed using standard 0.18μm six metal-layer CMOS technology. The first antenna is a Quasi Yagi-Uda consisting of a T-shaped meandered line that operates as a driven dipole element connected to a Coplanar Wave Guide (CPW) through a Coplanar Slot (CPS) line transition. A meandered parasitic strip on front of the driver operates as a director and a planar arc is used as a reflector to reduce the back radiation. The overall size of the Antenna on Chip (AOC) is $0.72\times0.85mm^2$. The proposed antenna gives an end-fire radiation pattern with 0.3 dBi simulated average gain and 45% radiation efficiency. The second introduced, antenna is a Vivaldi antenna with three techniques to enhance the radiation properties. The first technique is the insertion of an elliptical patch parasitic radiating element in the Vivaldi aperture to enhance the coupling between arms and produce strong radiation in the end-fire direction. The second technique is the addition of corrugation at the antenna edge to improve the antenna characteristics. The third technique is the insertion of a planar reflector at the backend of the antenna, which greatly improves the front-to-back (F/B) ratio. The overall size of the antenna on-chip is $0.5\times0.87mm^2$. The proposed antenna reveals an end-fire radiation pattern with 0.8 dBi simulated average gain and 37% radiation efficiency.

We need to introduce a 3-D mm-wave system in our proposed system, as shown in Figure 5.1. The high gain antennas are required to be on the sides to communicate with another system at a long distance. It is designed to operate at the same proposed band (57-64 GHz). In this chapter, we focus on the design of the on-chip antennas to serve the connection between layers.

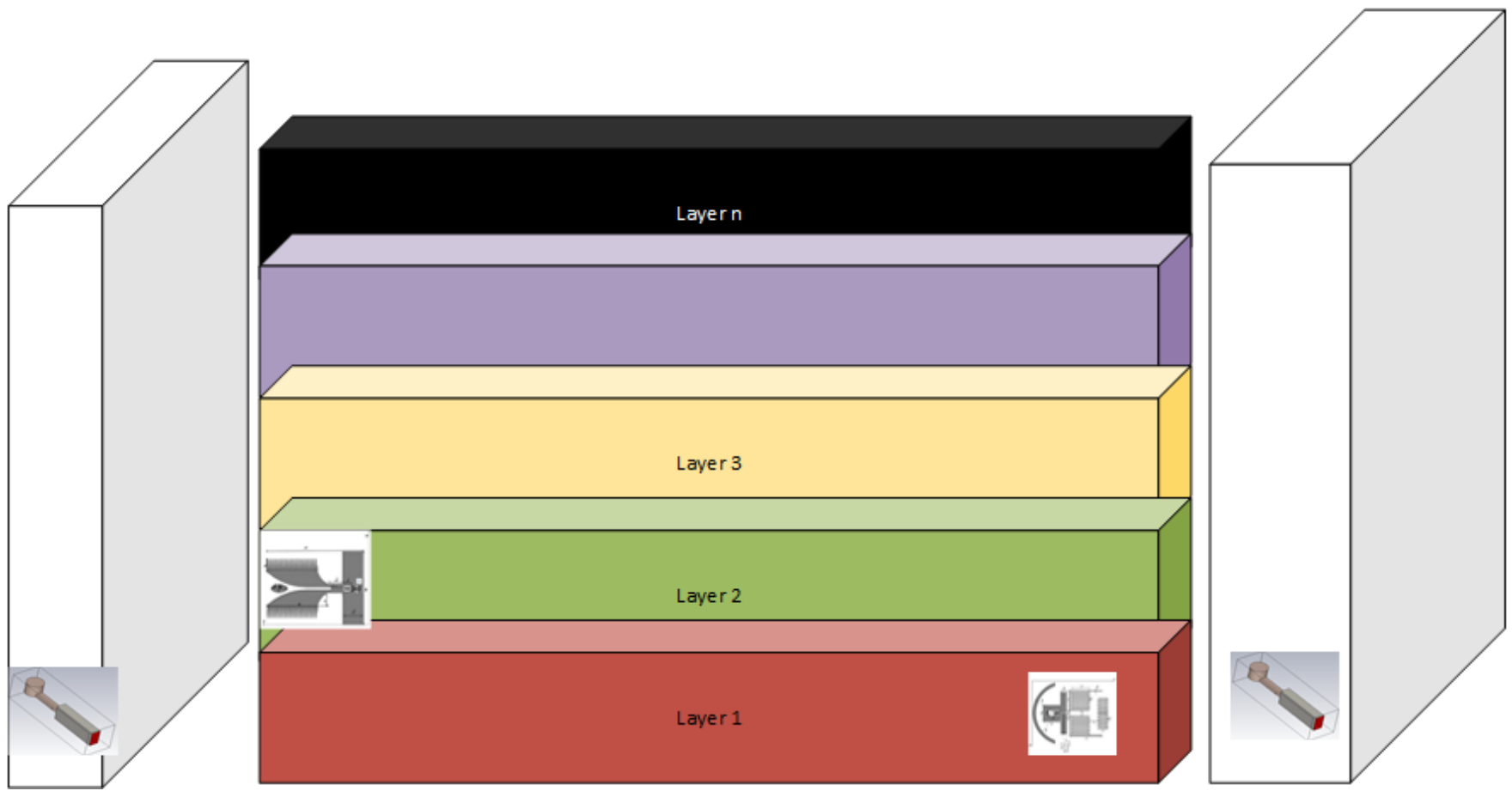

Figure 5.1 Build 3-D system Packaging

# 5.2 Yagi-Uda Antenna

The proposed antenna is designed using 0.18µm standard six metal layers CMOS (5×5 $mm^2$ standard dimensions). The Quasi-Yagi antenna is designed in metal six with overall size of 720µm×850µm as shown in Figure 5.2. The antenna dimensions are tabulated in Table 5.1. The proposed Yagi antenna is composed of a driven element, a parasitic director and a reflector. The driven element is a T-shaped dipole with meander line shape in order to increase the path over which the surface current flows. It is fed by a CPW transmission line so the need for a transition from CPW to CPS appears since the antenna is fed by a CPS line. In the millimetre wave circuit, the CPW is used to be suitable with the Ground-Signal-Ground G-S-G feeding standard. The analysis of this transition is introduced in [236]. Generally, in Yagi antenna design, a metallic strip is always used as a director to improve directivity by directing the wave into the end-fire direction and to enhance the impedance matching in the high-frequency band. The ground plane acts as a reflector in conjunction with a planar arc that acts as a second reflector to prevent any radiation in the back side and direct all the radiated power to the front direction.

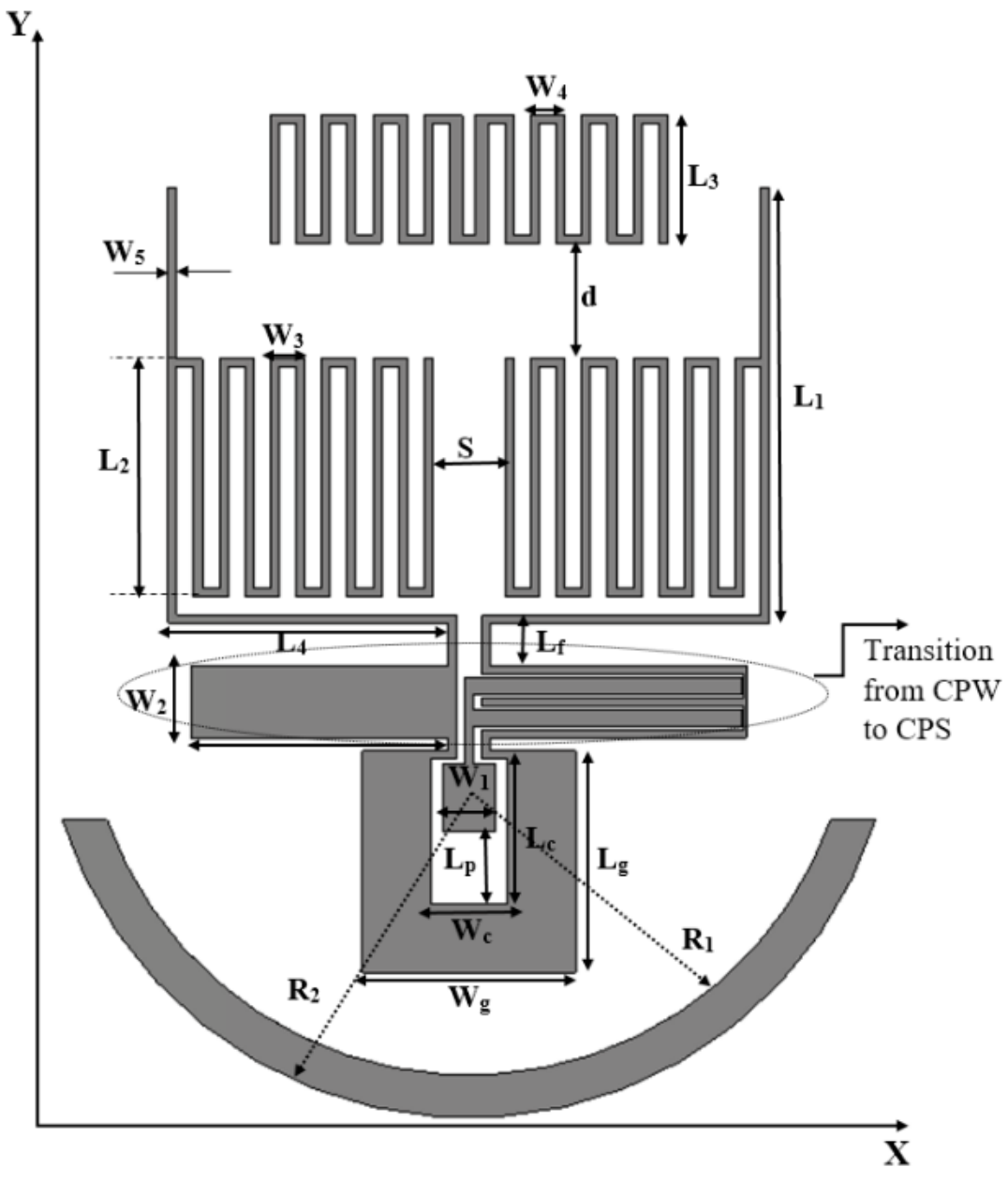

(a)Top view (CST)

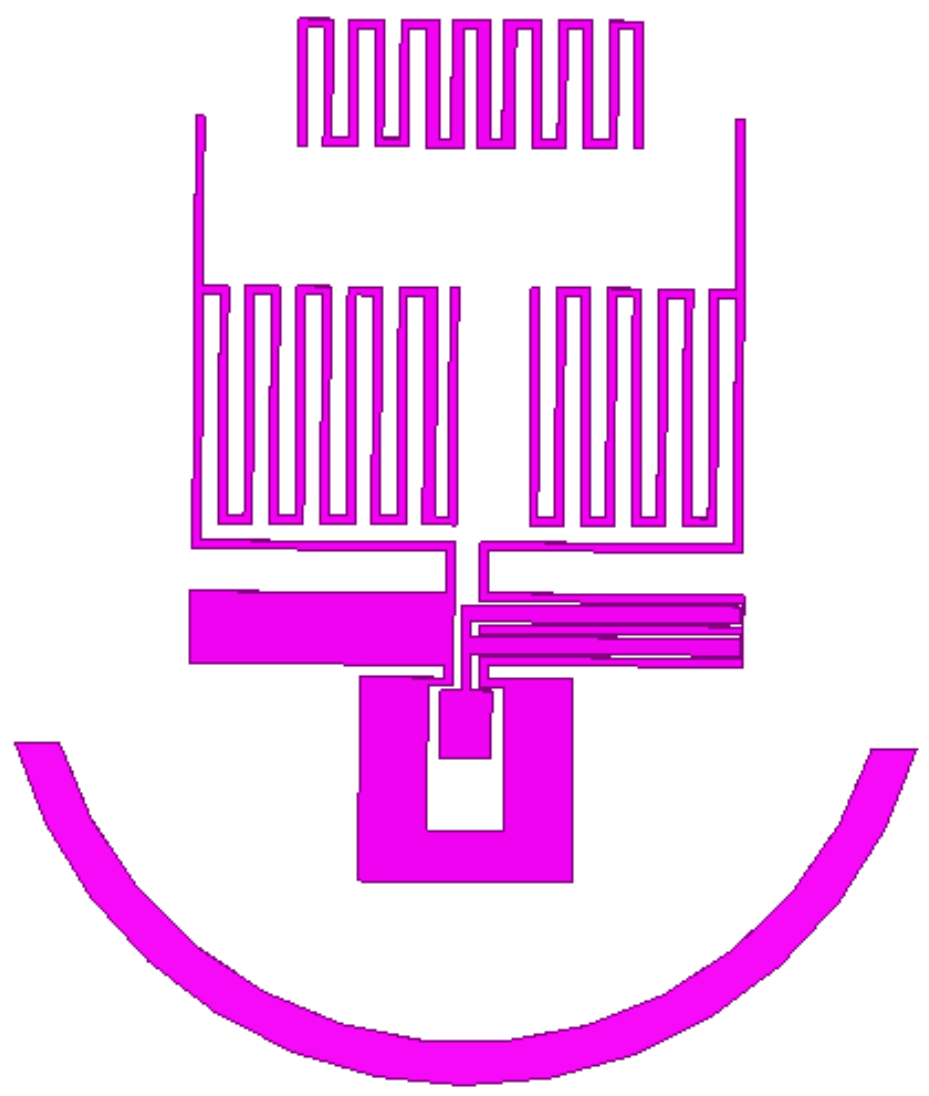

(b)Top view (HFSS)

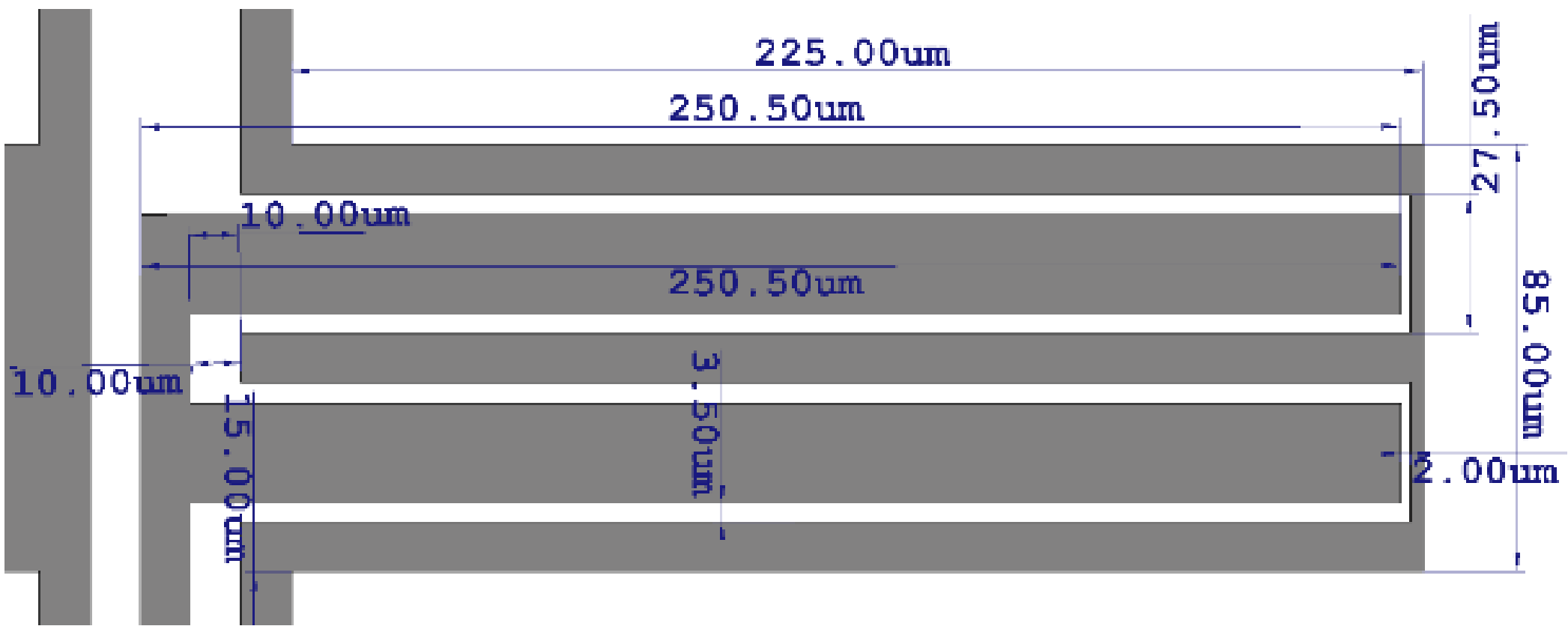

(c)Transition from CPW to CPS

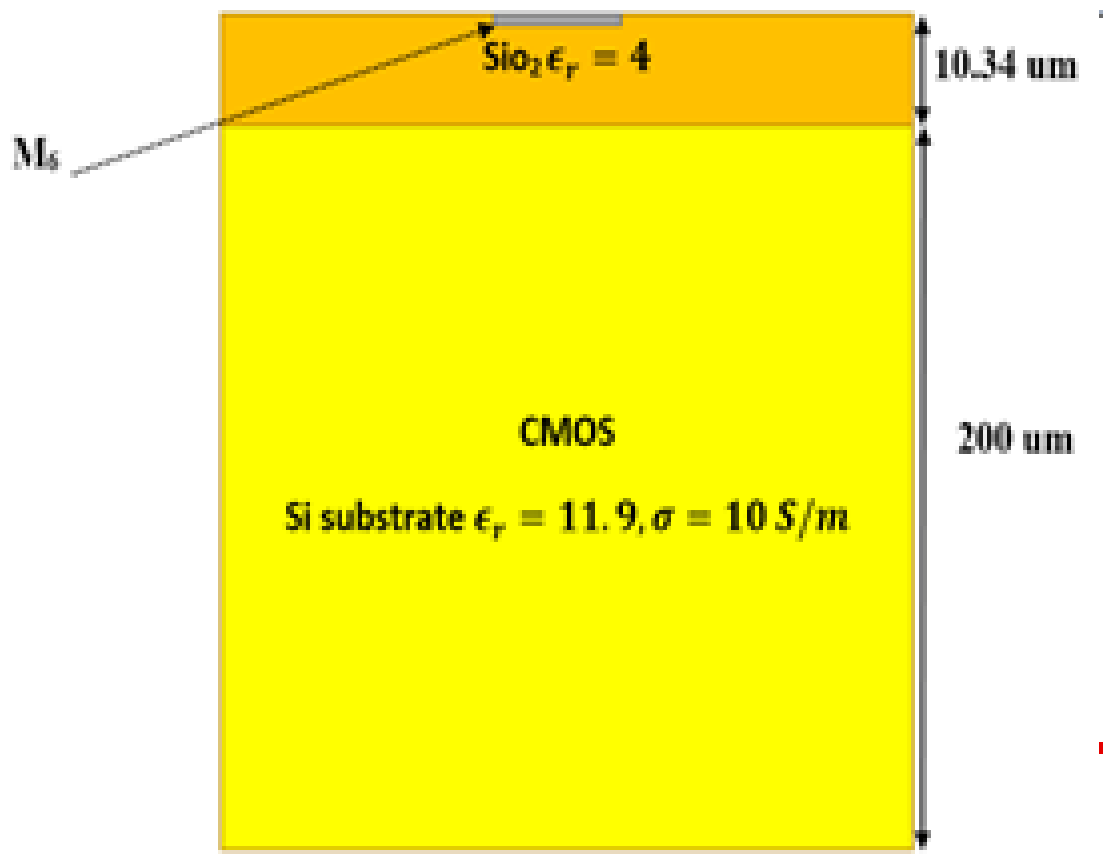

(d) Side view

Figure 5.2 Yagi antenna geometry

Table 5.1 Antenna dimensions (μm)

| $L_g$ | $L_p$ | $L_c$ | $L_f$ | $L_1$ | $L_2$ | $L_3$ | $L_4$ |
|---|---|---|---|---|---|---|---|
| 150 | 40 | 100 | 30 | 400 | 160 | 100 | 300 |
| $W_g$ | $W_c$ | $W_1$ | $W_2$ | $W_3$ | $W_4$ | $W_5$ | D |
| 180 | 60 | 30 | 85 | 30 | 30 | 10 | 80 |
| $R_1$ | $R_2$ | S | | | | | |
| 300 | 350 | 85 | | | | | |

The proposed antenna is simulated using commercial CST Microwave Studio 2017 and HFSS version 16. All the results are verified by the aforementioned softwares. The antenna performance is studied in two cases, with using planar arc as a main reflector and using ground as a main reflector without arc. Figure 5.3 shows the return loss of the antenna to cover band from 50 GHz to 80 GHz with good matching in two different cases; case 1, the ground is used as reflector with Wg=350 μm and in case 2, the planar arc used as reflector with Wg=180μm. The return losses from CST and HFSS are close together.

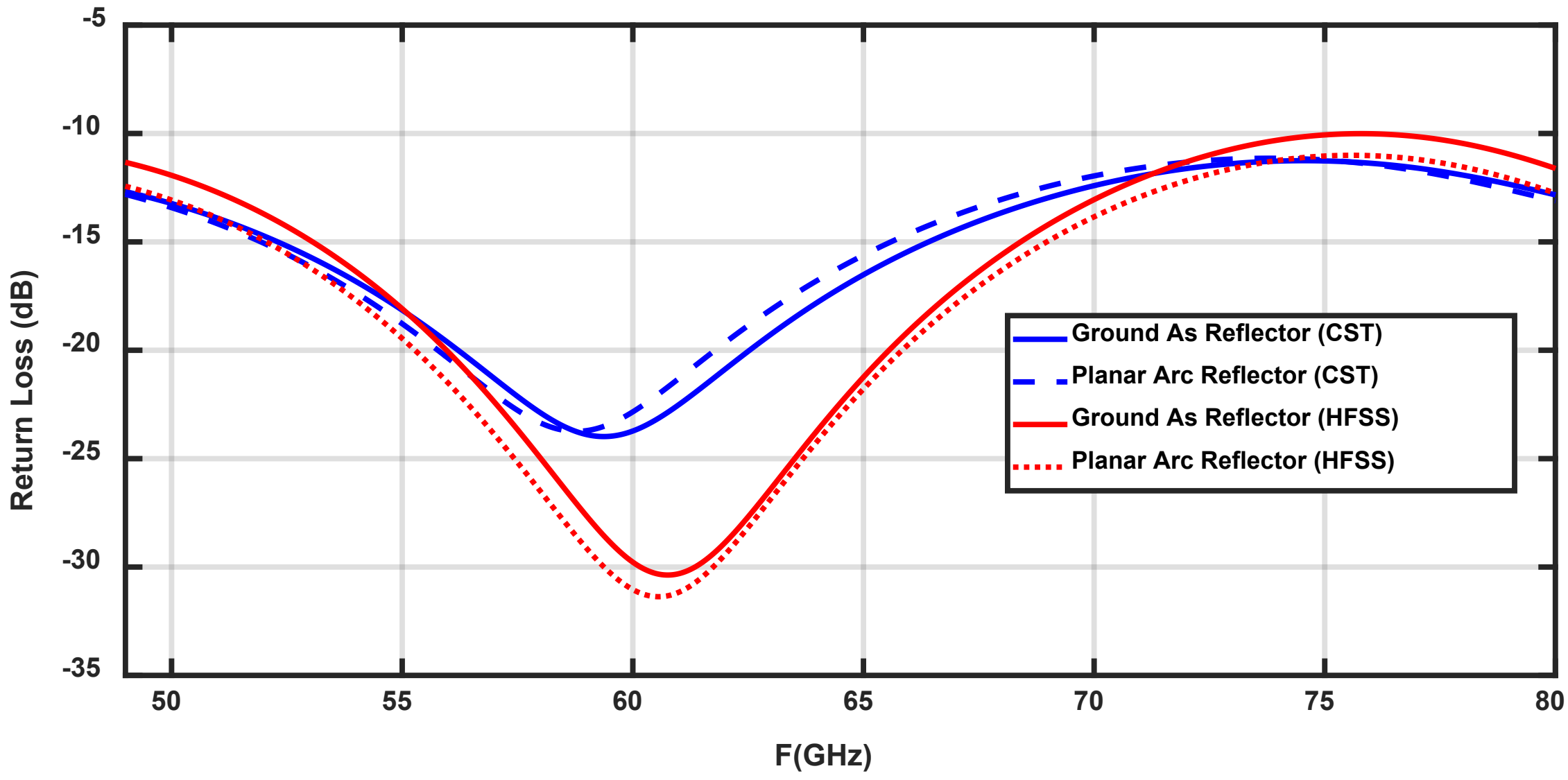

Figure 5.3 Return loss of the Yagi antenna.

Figure 5.4 and Figure 5.5show the gain and radiation efficiency of the proposed antenna with and without arc, respectively. We notice that the arc enhance the value of gain by 0.8 dBi because it reflects the back radiation from the Yagi antenna. Furthermore, the gain of the antenna is verified by using HFSS in the two cases and there are good agreement between the results. Furthermore, the antenna efficiency is enhanced by using the arc to be about 45%.

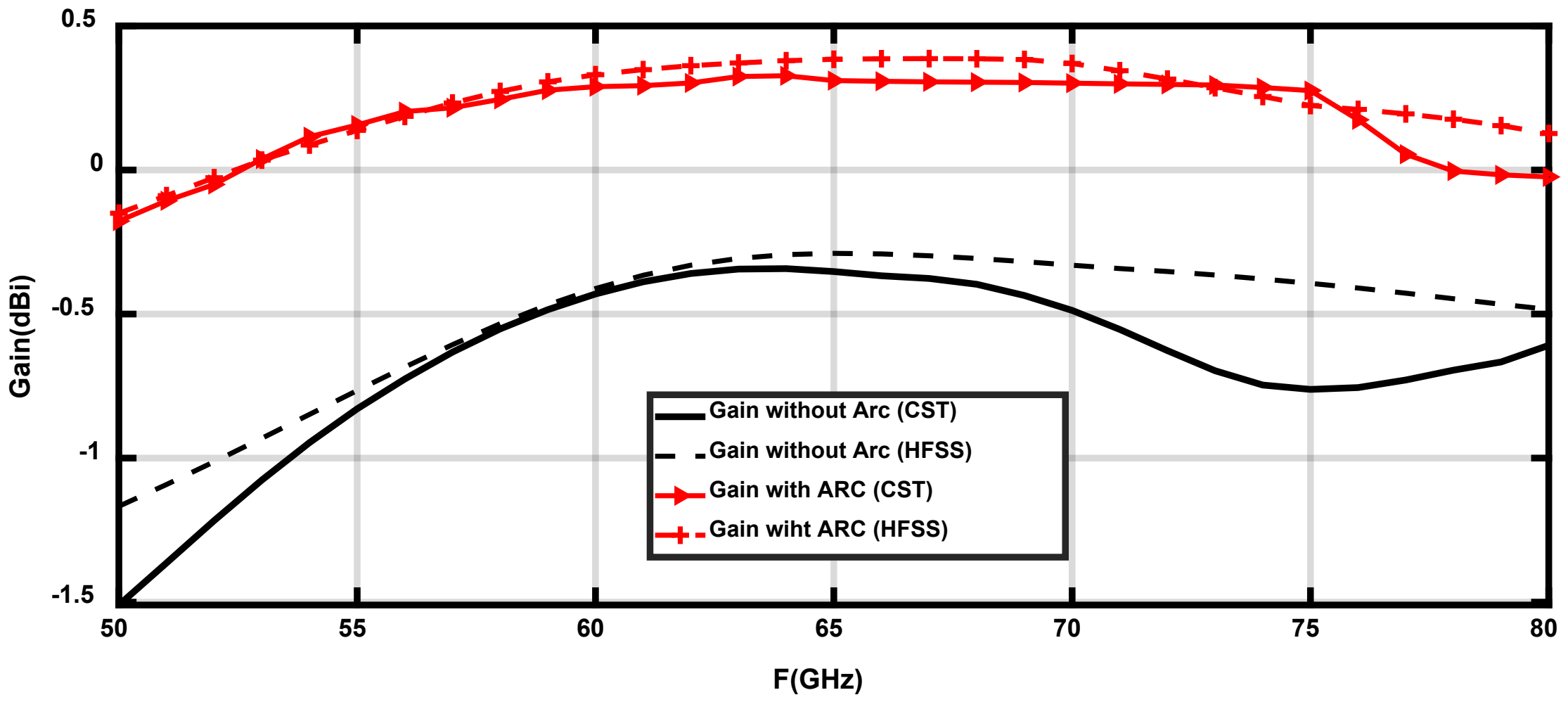

Figure 5.4 Gain of the Yagi-Uda antenna.

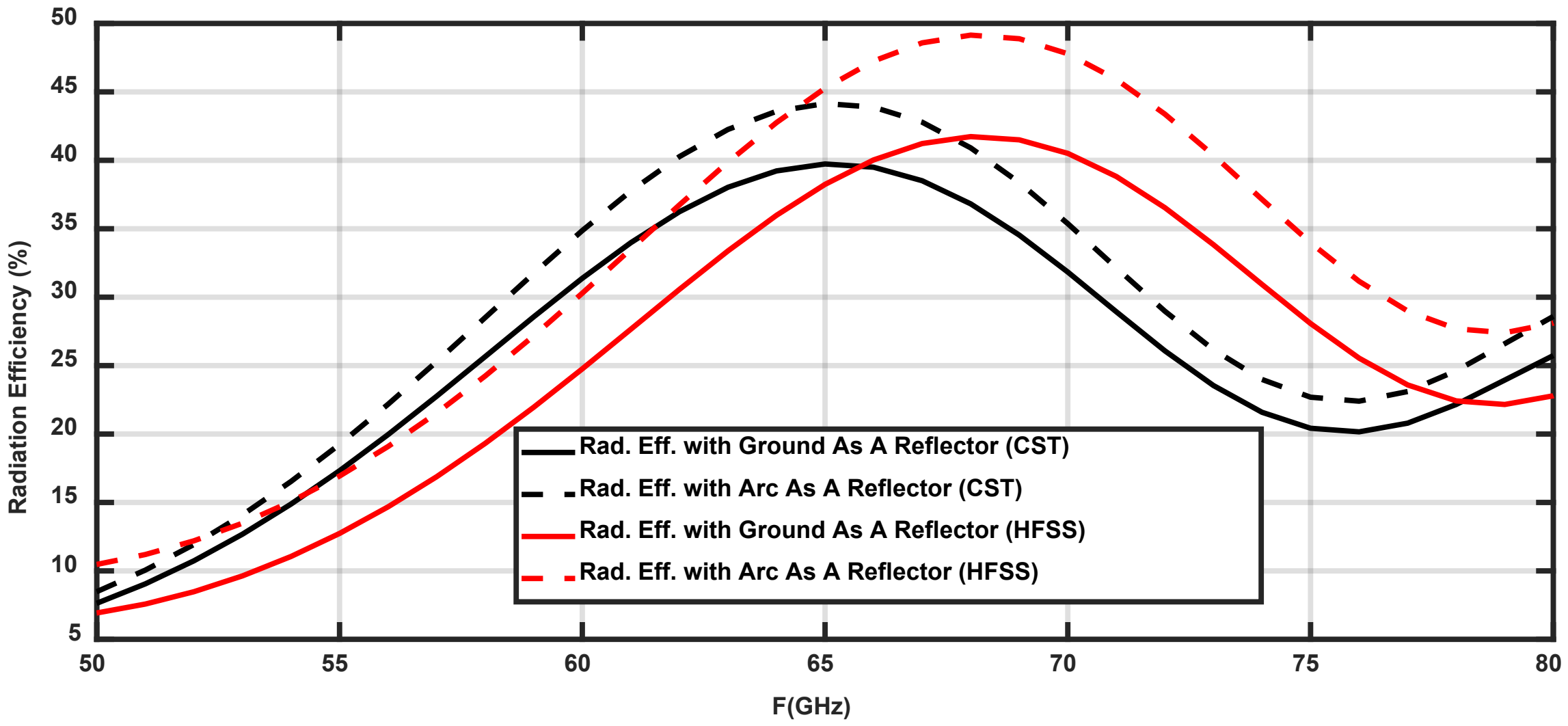

Figure 5.5 Radiation efficiency of the Yagi-Uda antenna.

Due to the inherent losses of the substrate, the antenna position on the substrate effects on the antenna performance. So, the performance of the Yagi antenna is presented at different three positions, as shown in Figure 5.6. The gain and the radiation efficiency are increased for P2 and P3, as shown in Table 5.2 due to decreasing the effect of substrate. Figure 5.7 shows the radiation pattern of the antenna in the XY plane and ZY plane at 60 GHz and 65 GHz. There is a good agreement between the simulated radiation pattern from CST and HFSS, as depicted in Figure 5.7. Also, we notice that the radiation pattern of the antenna is in the direction of the Y-axis to ensure that the antenna has end-fire radiation.

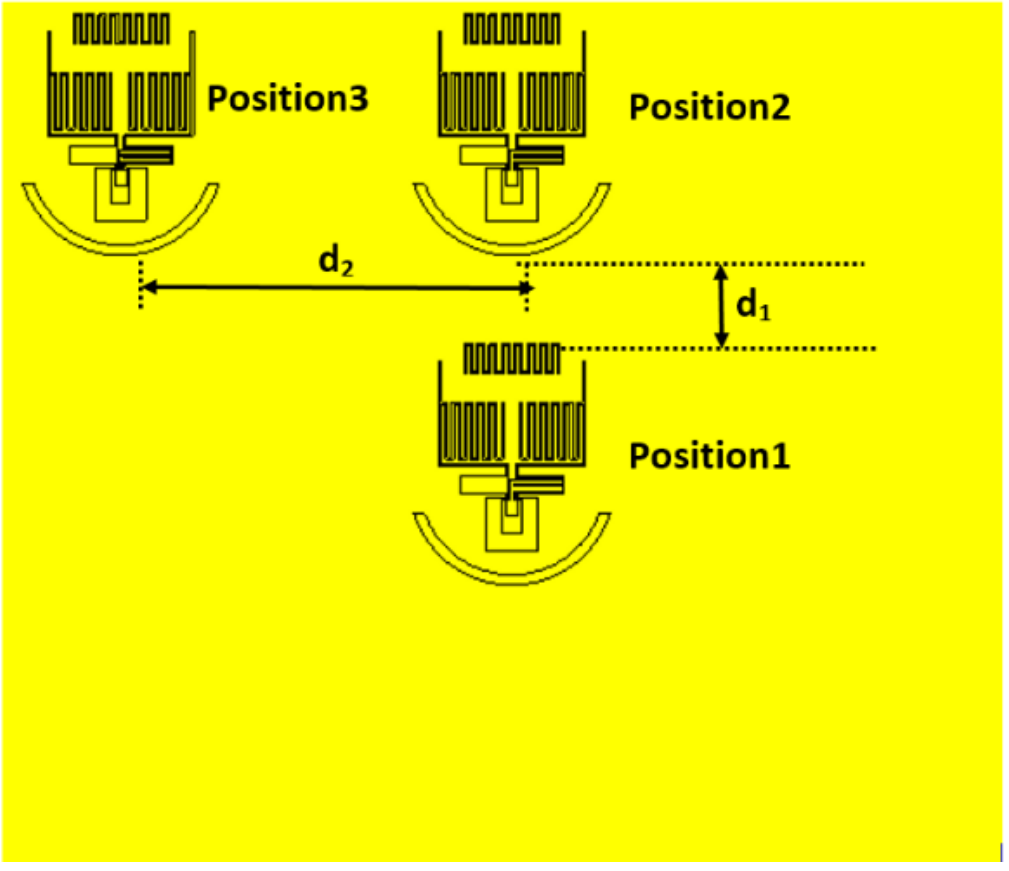

Figure 5.6 Yagi antenna positions.

Table 5.2 Comparison between different positions of the antenna

| Positions | Return Loss (dB) | Gain($dB_i$) | Rad. Eff. |
|---|---|---|---|
| P1 | -24 | -0.18 | 30 |
| P2 | -18 | 0.2 | 36 |
| P3 | -17 | 0.31 | 45 |

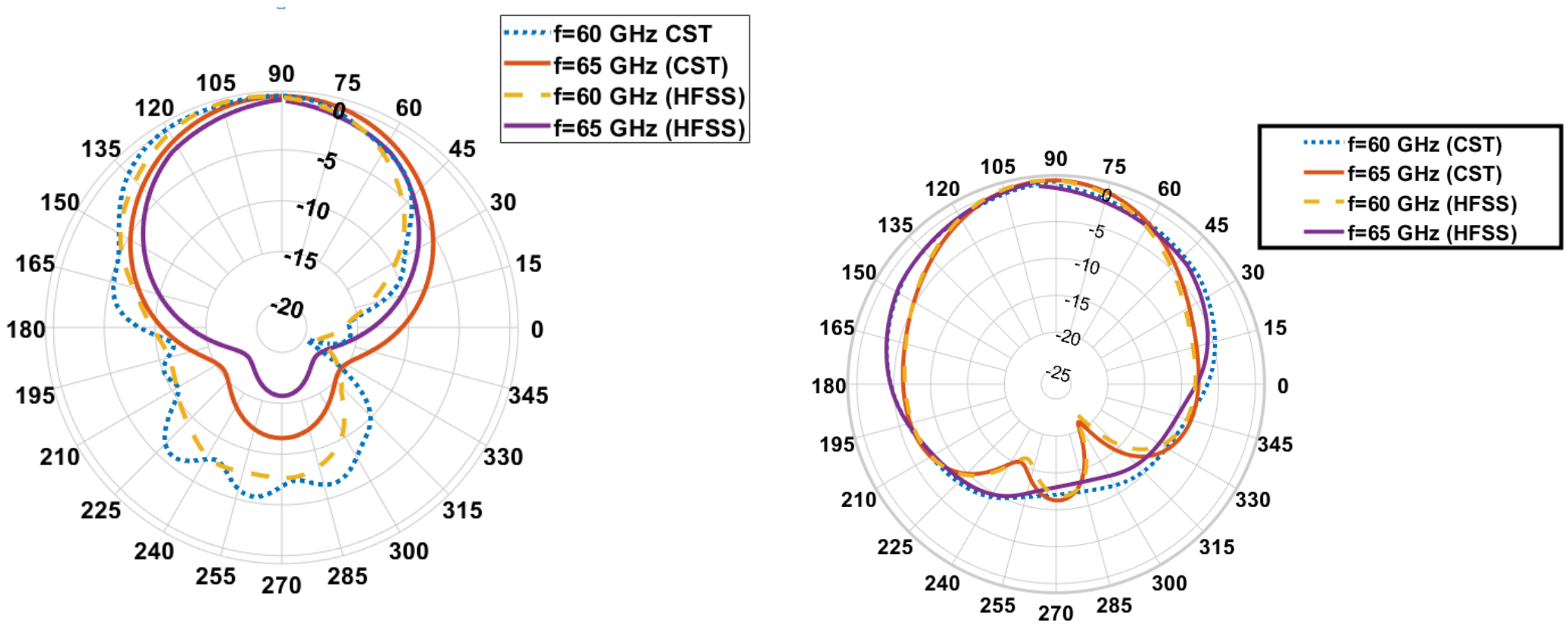

(a) (b)

Figure 5.7 The radiation pattern of the antenna in (a)XY plane and (b)YZ plane.

# 5.3 Tapered Slot Antenna

The Vivaldi antenna is designed using 0.18μm CMOS technology with the standard dimensions. The CMOS consists of the silicon substrate (5×5 mm$^2$ with height 200 μm, high permittivity ($\varepsilon_r$=11.9) and low resistivity (σ=10S/m)) and a thin $SiO_2$ layer (5×5 mm$^2$ with thickness 10.4 μm and dielectric constant $\varepsilon_r$=4) as shown in Figure 5.2 (d). The Vivaldi antenna is designed in a metallic six-layer with an overall size of 500μm×870μm, as shown in Figure 5.8(a). The antenna dimensions are tabulated in Table 5.3. The proposed Vivaldi antenna consists of two arms flared in opposite directions and symmetrically rotated around the antenna aperture axis. The antenna is fed by the transition from CPW to CPS according to the theory and the analysis introduced in [236] and the transition is optimized as shown in Figure 5.8(c). This transition is introduced to make the feeding method suitable with the millimeter circuit. Three different techniques are (as shown in Figure 5.9) introduced to enhance the radiation characteristics of the antenna. The first technique is the parasitic elliptical patch in the aperture area between the Vivaldi arms to increase the coupling between the two arms and to produce strong radiation in the end fire. The second technique is the Sin corrugation of the two arm outer edges. The corrugated edges are defined by $A_c sin(my)$, m is fraction of variable y-axis and $A_c$ amplitude. The corrugation enlarges the effective aperture size to improve the gain. The final technique is the addition of a planar reflector to prevent back radiation and to improve the front to back ratio.

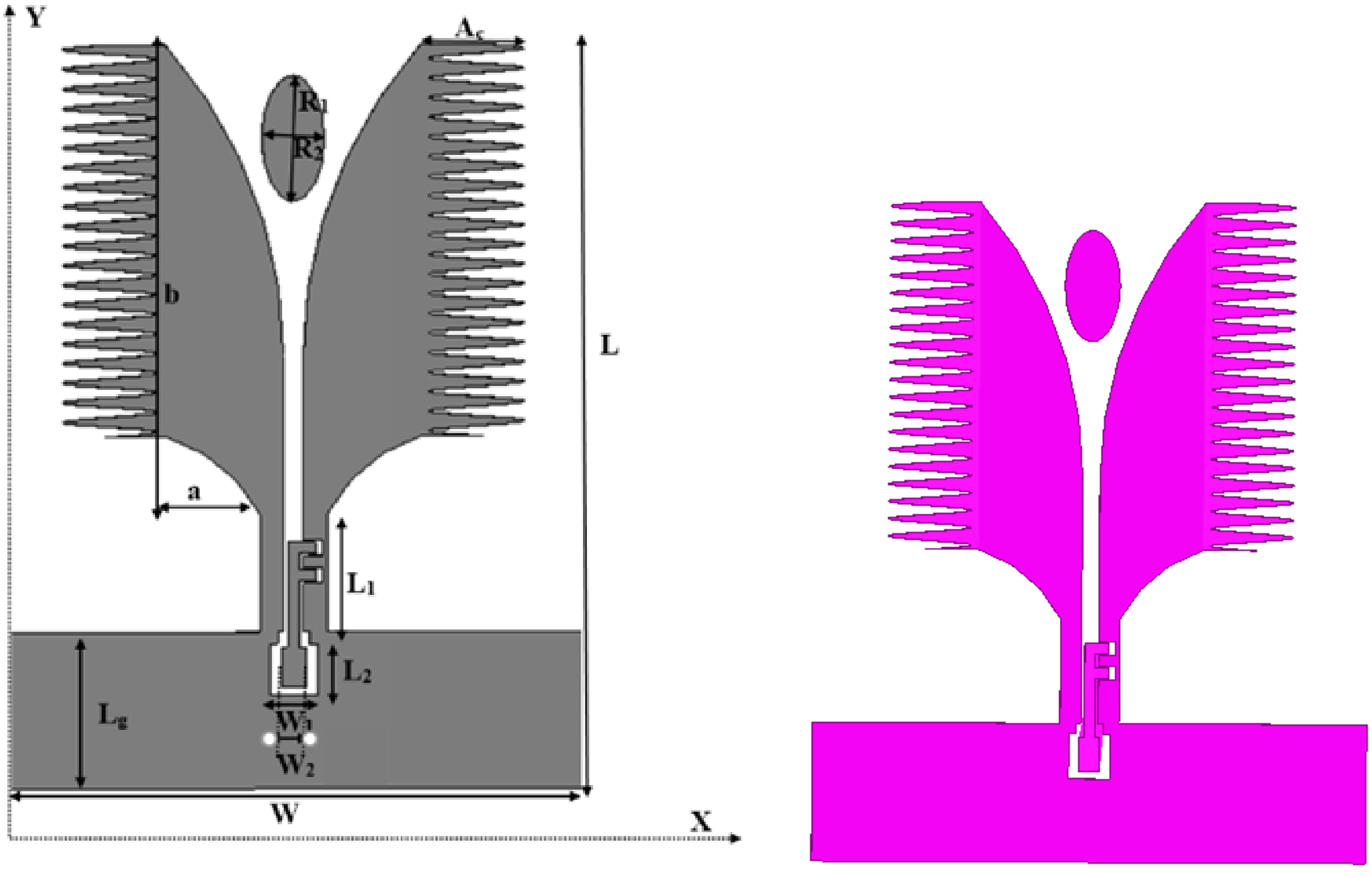

(a)Top view (CST)

(b)Top view (HFSS)

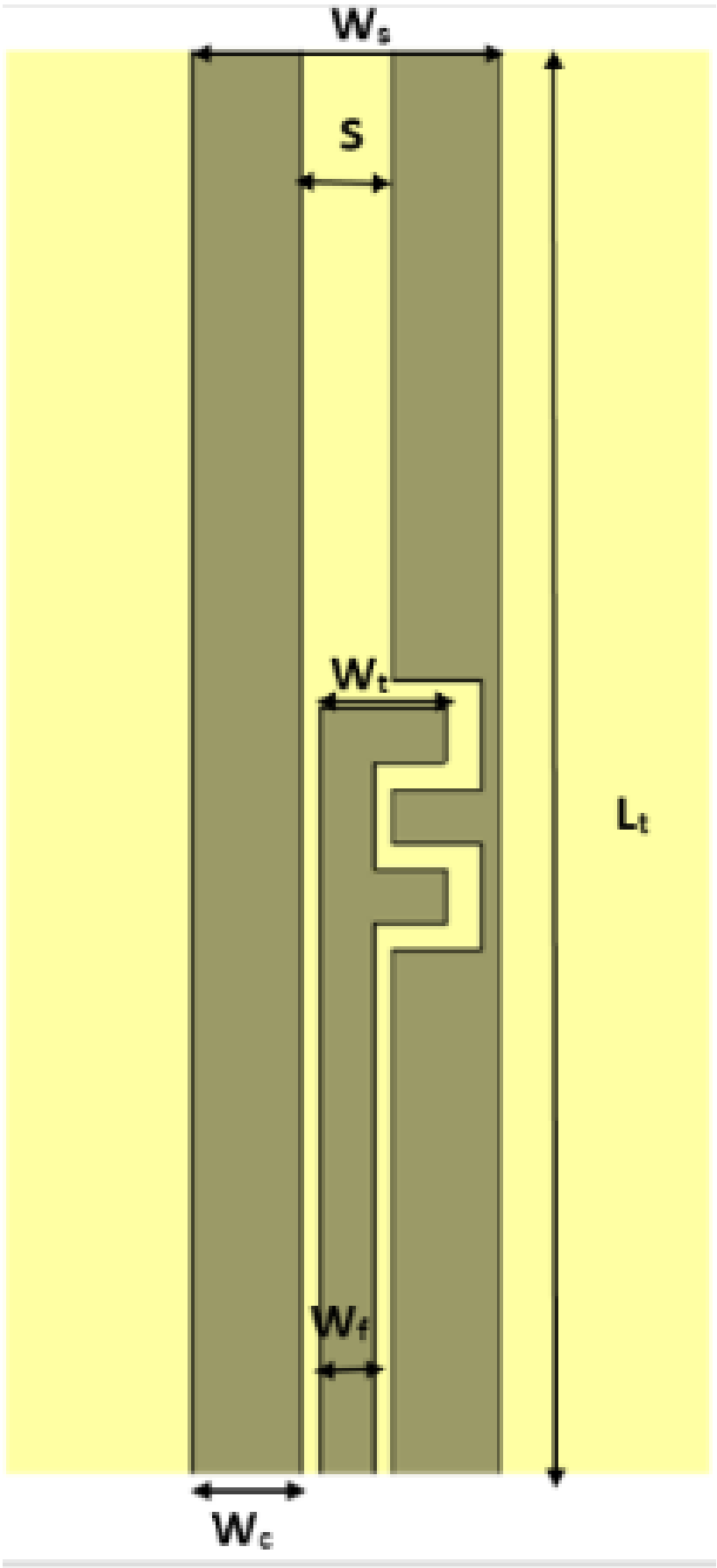

(c) Transition from CPW to CPS

Figure 5.8 Vivaldi antenna geometry

Table 5.3 Antenna dimensions (µm)

| L | W | $L_g$ | $L_1$ | $L_2$ | $W_1$ | $W_2$ | A | B |
|---|---|---|---|---|---|---|---|---|
| 870 | 500 | 150 | 120 | 50 | 50 | 30 | 100 | 500 |
| $A_c$ | $W_c$ | $W_s$ | $W_t$ | $W_f$ | S | $L_t$ | $R_1$ | $R_1$ |
| 50 | 25 | 55 | 22 | 12 | 20 | 110 | 80 | 40 |

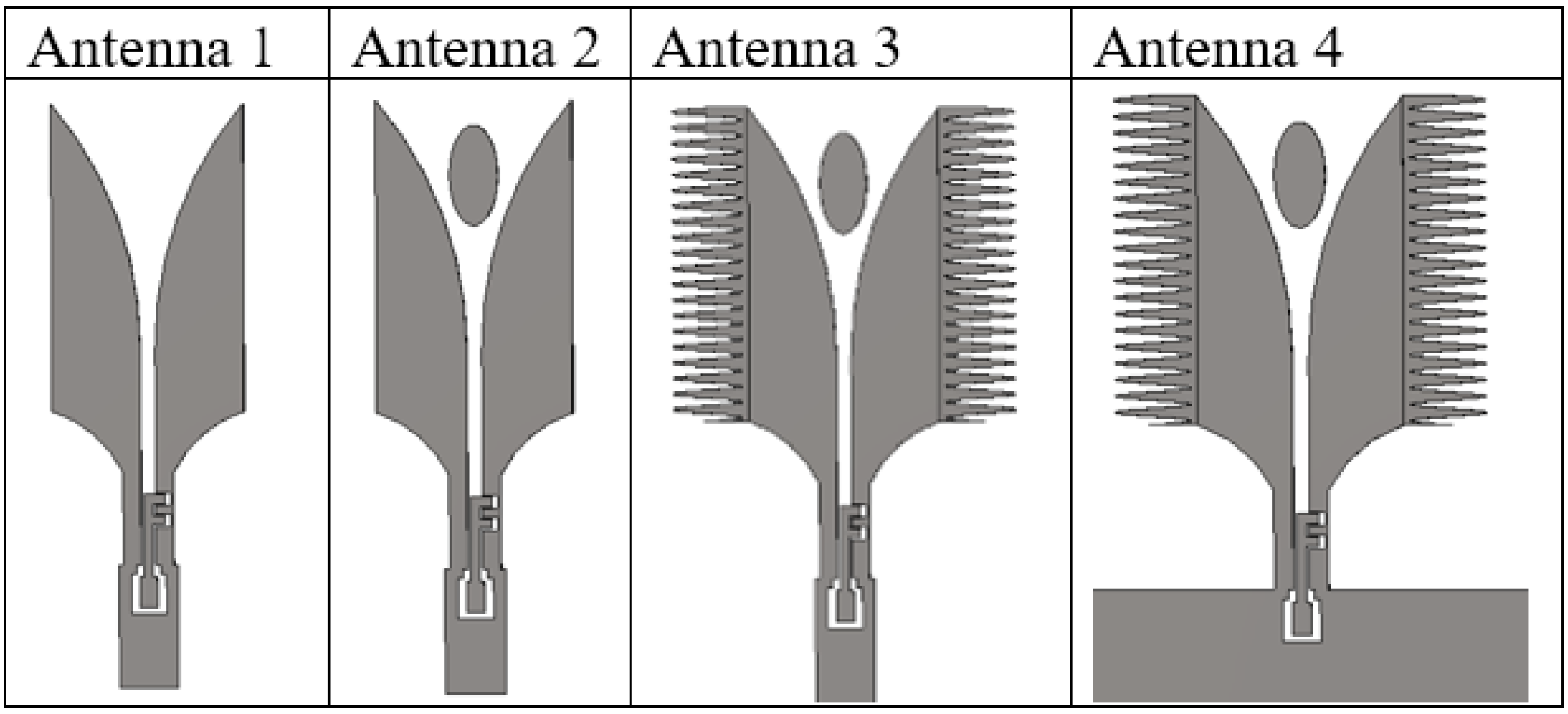

Figure 5.9 Design steps of Vivaldi antenna

The transition from CPW to CPS is designed as the first step to show the amount of transmission losses. The analysis of the transition from CPW to CPS is introduced in [236]. Figure 5.10 shows the S-parameters of transition from CPW to CPS on CMOS technology to ensure that the transmission coefficient is acceptable. The simulated curves of return loss are shown in Figure 5.11 to ensure that the antennas operate from 50 GHz to 70 GHz with good matching. The gain of four antennas are introduced in Figure 5.12 to see the effect of ellipse shape is about 2dBi. Moreover, the sin corrugation and the ground reflector increase the gain by 0.4 dBi and 0.9 dBi, respectively. The radiation efficiency for the four types of antenna are presented in Figure 5.13.  Figure 5.14 shows the radiation pattern of the antenna in the XY plane and ZY plane at 60 GHz to ensure that it is end-fire.

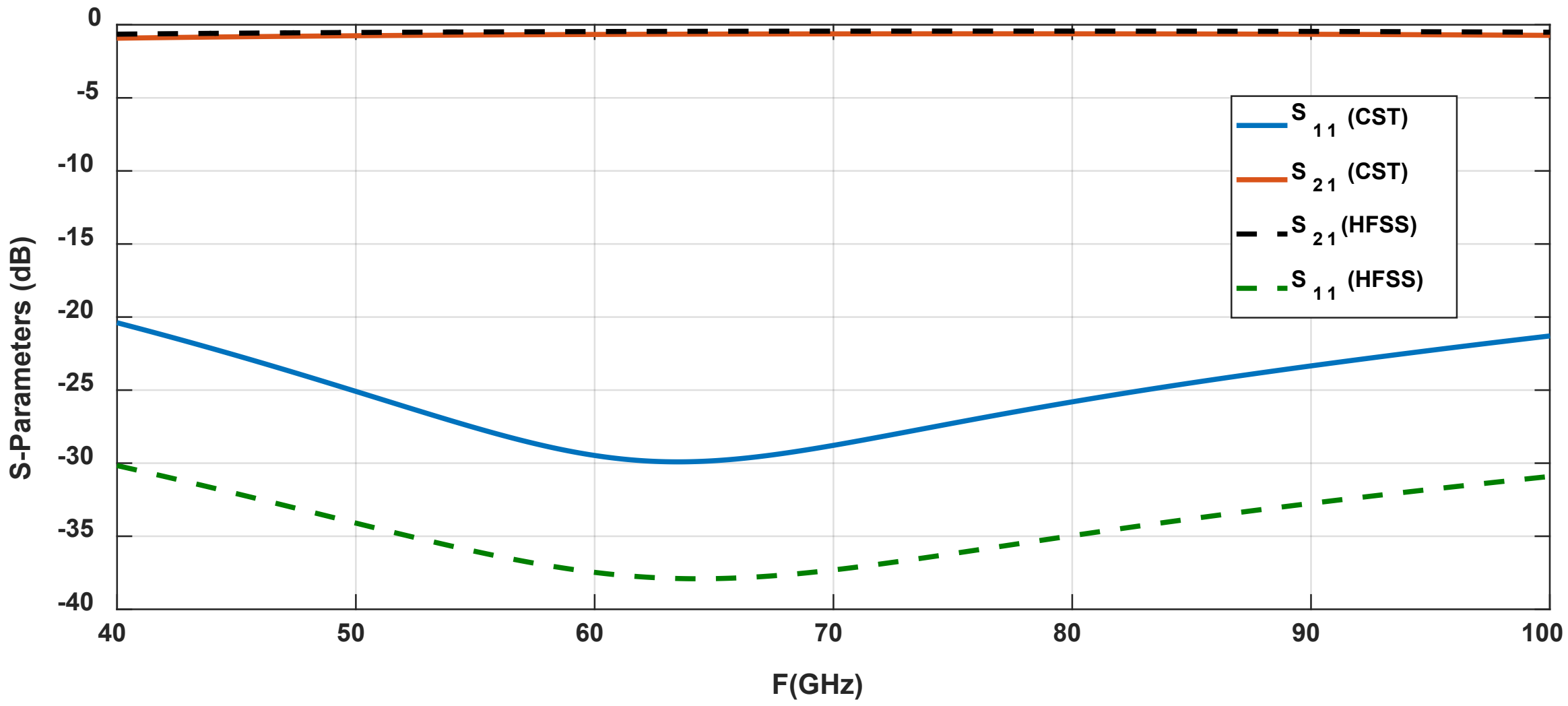

Figure 5.10 S-Parameters of transition from CPW to CPS

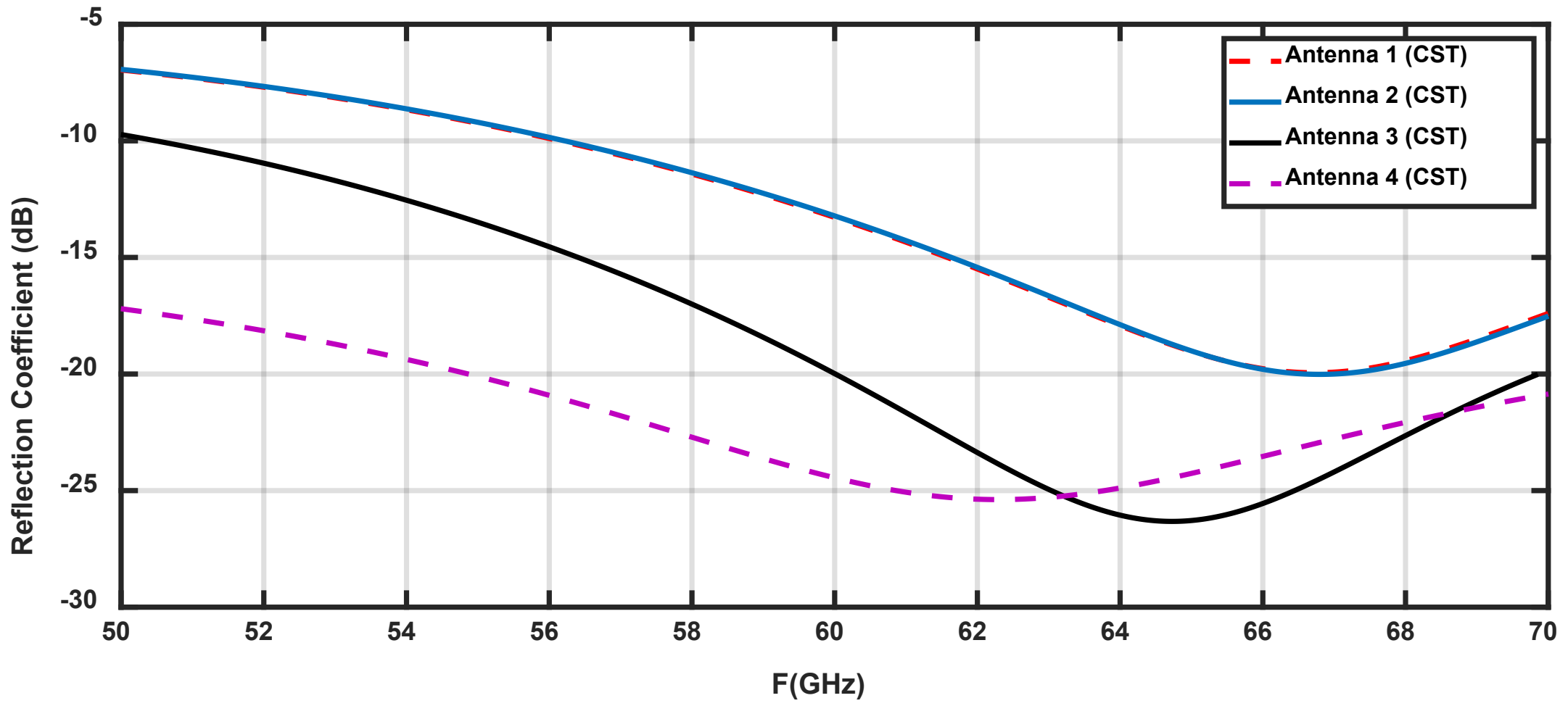

(a)$S_{11}$ using CST

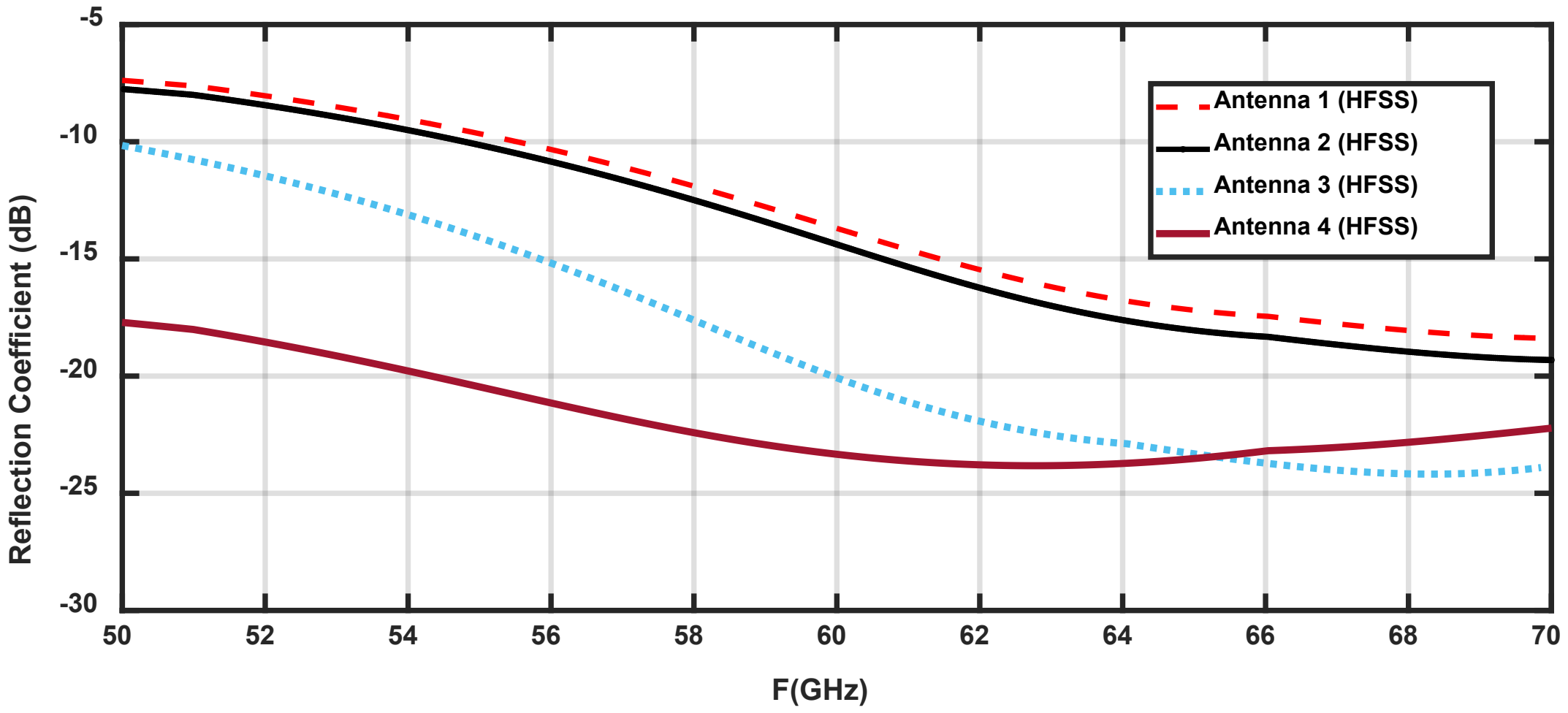

(b)$S_{11}$ using HFSS

Figure 5.11 Reflection coefficient of the proposed Vivaldi antenna for four cases using CST and HFSS

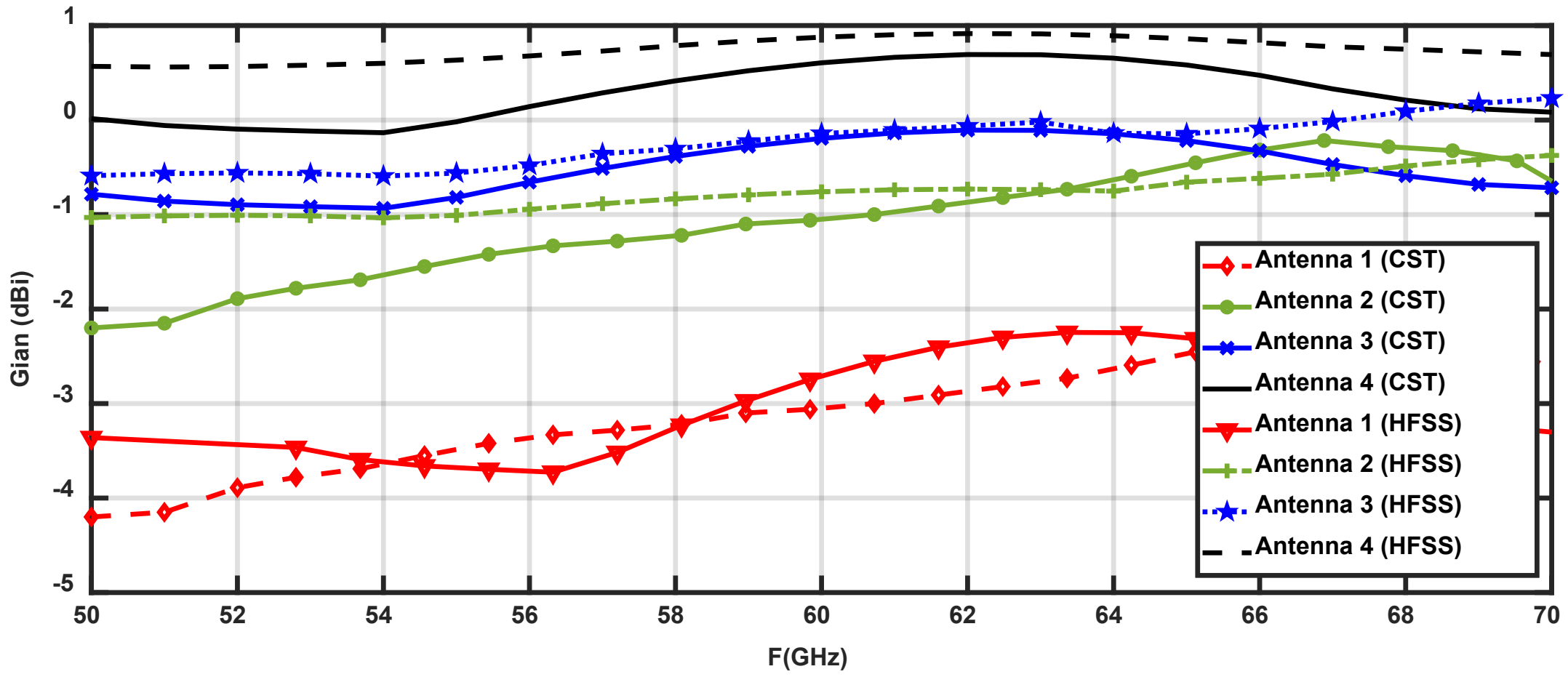

Figure 5.12 Gain of Vivaldi antennas

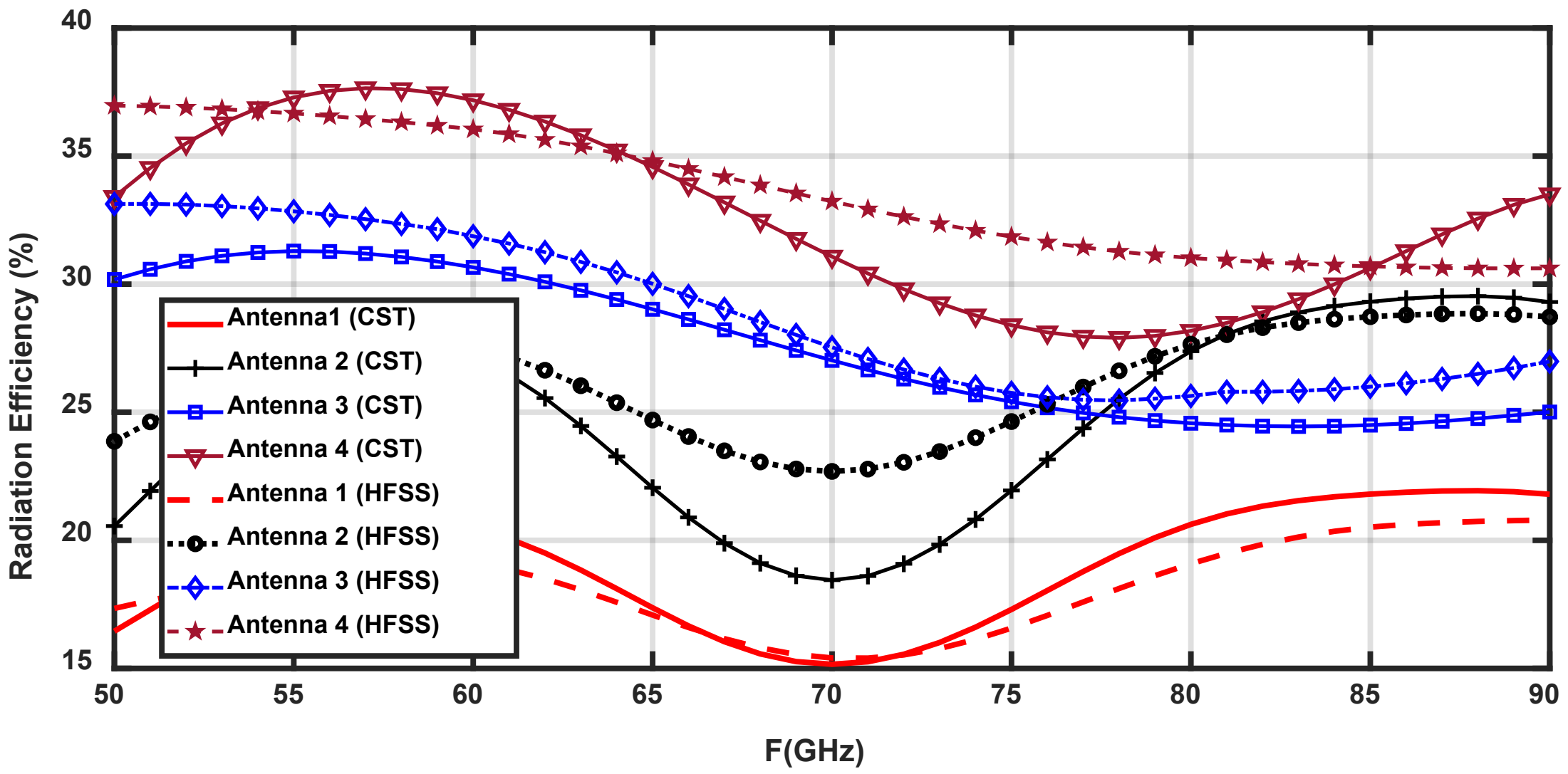

Figure 5.13 Radiation efficiency of Vivaldi antennas

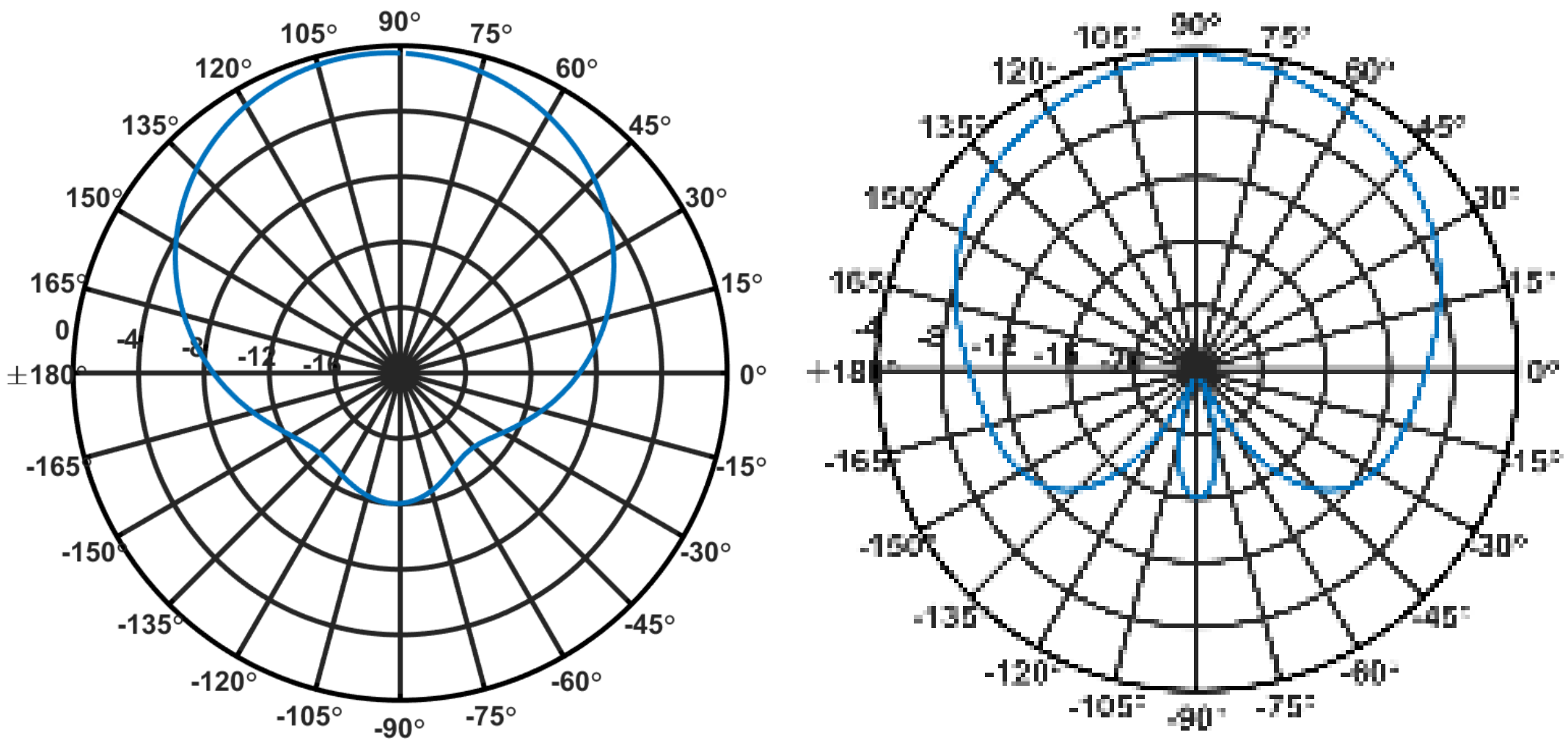

(a) (b)

Figure 5.14 The radiation pattern of the Vivaldi antenna in (a)XY plane and (b)YZ plane

The characteristics of proposed antennas are compared with other published papers, as shown in Table 5.4. We notice that the proposed two antennas introduce high gain and efficiency compared with the antenna in literature due to using different techniques for each antenna to enhance its performance. Furthermore, the gain of the Vivaldi antenna is best than that of the Yagi-Uda antenna; in contrast, the efficiency of Yagi-Uda is better than that of Vivaldi because the metallic loss in the Yagi-Uda is small compared with the Vivaldi antenna.

Table 5.4 Comparison with previous CMOS antennas

| Reference | CMOS (μm) | F(GHz) | Size ($mm^2$) | Gain (dBi) | Eff. (%) | F/B ratio (dB) |
|---|---|---|---|---|---|---|
| [234] | 0.18 | 60 | 2.45×1.8 | -2.64 | 16.8 | 16.66 |
| [235, 237] | 0.18 | 60 | 1×0.87 | -0.746 | 21.4 | 11.01 |
| [29] | 0.18 | 60 | 0.785×0.94 | -0.4 | 32 | NA |
| [96] | 0.18 | 60 | 1.1× 0.95 | -8 | 10 | 9 |
| [238] | 0.13 | 60 | 2.2× 1.3 | -0.2 | 19.6 | NA |
| [239] | CMOS | 60 | 0.5×0.15 | -20 | NA | NA |
| [240] | 0.18 | 9.45 | 2×2.1 | -29 | 21.1 | NA |
| [91] | CMOS | 60 | 1.86×1.86 | -1 | NA | NA |
| [241] | 28 | 33 | 0.66×0.85 | -1.8 | 30 | NA |
| [242] | 65 | 24 | 2.5× 2.5 | -1 | 41 | 10 |
| [243] | Bi-CMOS | 165 | 1.26×0.92 | 0.3 | 32 | NA |
| Proposed Antenna (QYA) | 0.18 | 60 | 0.631× 0.46 | 0.35 | 45 | 16.3 |
| Proposed Antenna (TSVA) | 0.18 | 60 | 0.5× 0.87 | 0.8 | 37 | 18 |

# 5.4 Design of MIMO on-chip Antenna

In recent years, Multiple-Input-Multiple-Output (MIMO) have been developed to be essential in most of wireless communications systems to make best use of the multipath scattering phenomena either to increase the system capacity, to enhance the overall channel gain or to mitigate the fading effect according to the type of channel. Furthermore, the MIMO technology is used to improve reliability, excessive increase of data rates and enhancement of capacity without exhaustion of transmitted power and bandwidth [244]. The antenna designers confront two main problems in any MIMO design; the first one is the implementation of multiple antennas in the close size of the portable devices whereas the second one is the isolation between elements to reduce mutual coupling in the proposed band. Therefore, the performance of the antenna is deteriorated with increasing the mutual coupling between the elements.

## 5.4.1 Design of Two Elements

This part introduces three different configurations of the MIMO On-chip (MOC) for two elements. Configuration I (Conf. I) introduces two elements side by side configuration with gap=120$\mu m$, Conf. II introduces two elements side by back configuration and Conf. III introduces two elements back by back configuration. The three configurations of the MIMO antenna structure have two elements that provide better isolation between elements without using any additional technique or decoupling circuit. The optimized geometry of the proposed three different configurations based on achieving maximum isolation is shown in Figure 5.15. Here, the ground plane and the arc play a significant role in the isolation performance of the proposed antenna. Furthermore, the diversity of the antenna positions helps to reduce coupling and achieve better isolation among them. The CST microwave studio and HFSS are used together to verify the simulated results for the S-parameters as illustrated in Figure 5.16. In this figure, only $S_{11}$, and $S_{21}$ are simulated because of the symmetrical arrangement of antenna elements in the structure. A good agreement between the simulated results of CST and HFSS is obtained. Table 5.5 shows a comparison between the three aforementioned configurations. One can notice that all three configurations offer good matching and high isolation because of using arc as reflector between the elements. The isolation between elements is more than 40 dB that provides an enhancement in the radiation properties of antennas. We notice that Conf. II has low isolation because the distance between its ports (1 and 2) is small compared to the other configurations but it still has isolation of 40 dB. On the other hand, Conf. II, and Conf. III achieve diversity in the radiation pattern and this diversity is the main factor in the MIMO designs.

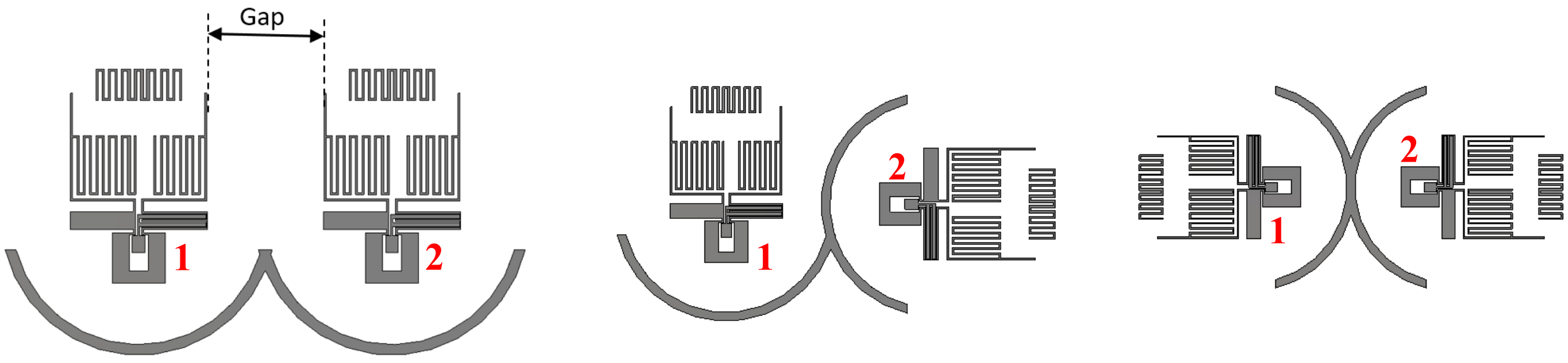

(a)Conf. I (b)Conf. II (c)Conf. III

Figure 5.15 Different configurations of two elements MIMO Yagi-Uda antenna

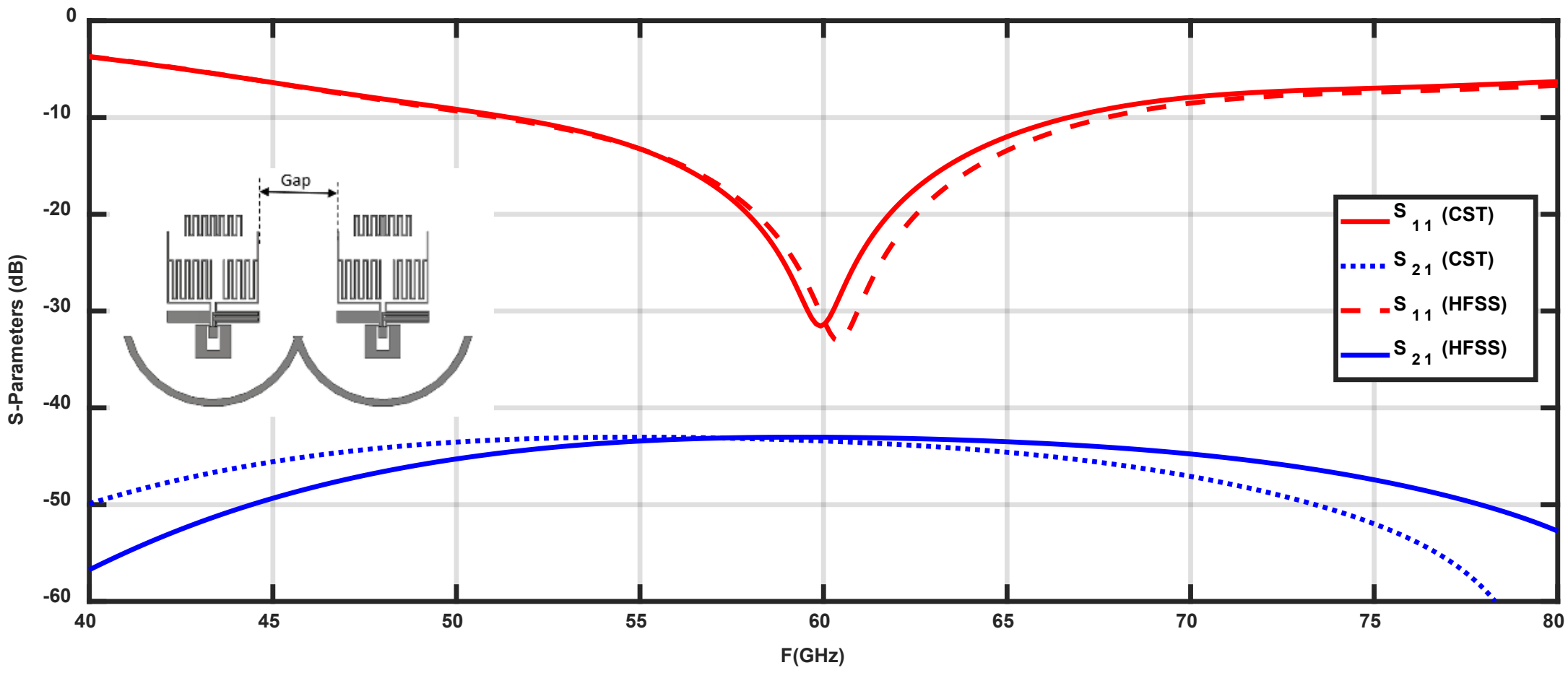

(a)Conf. I

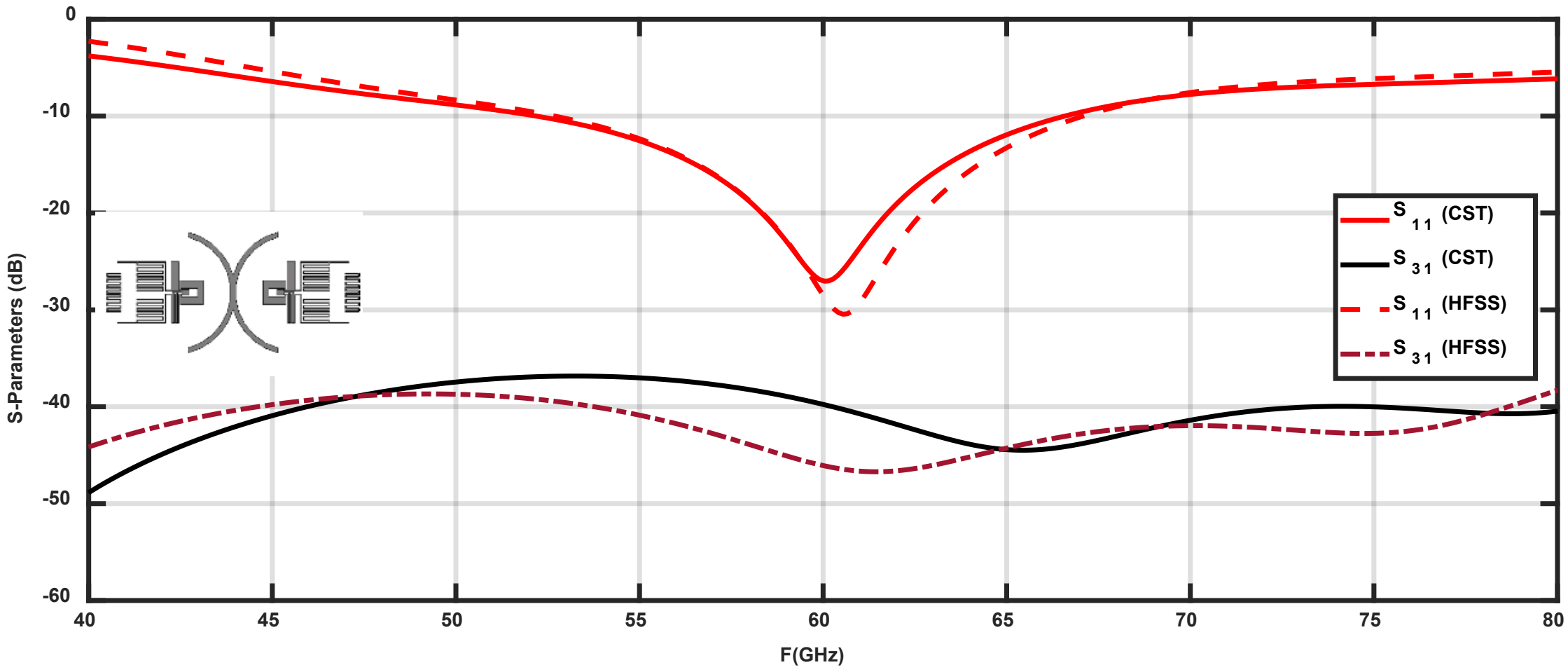

(b)Conf. II

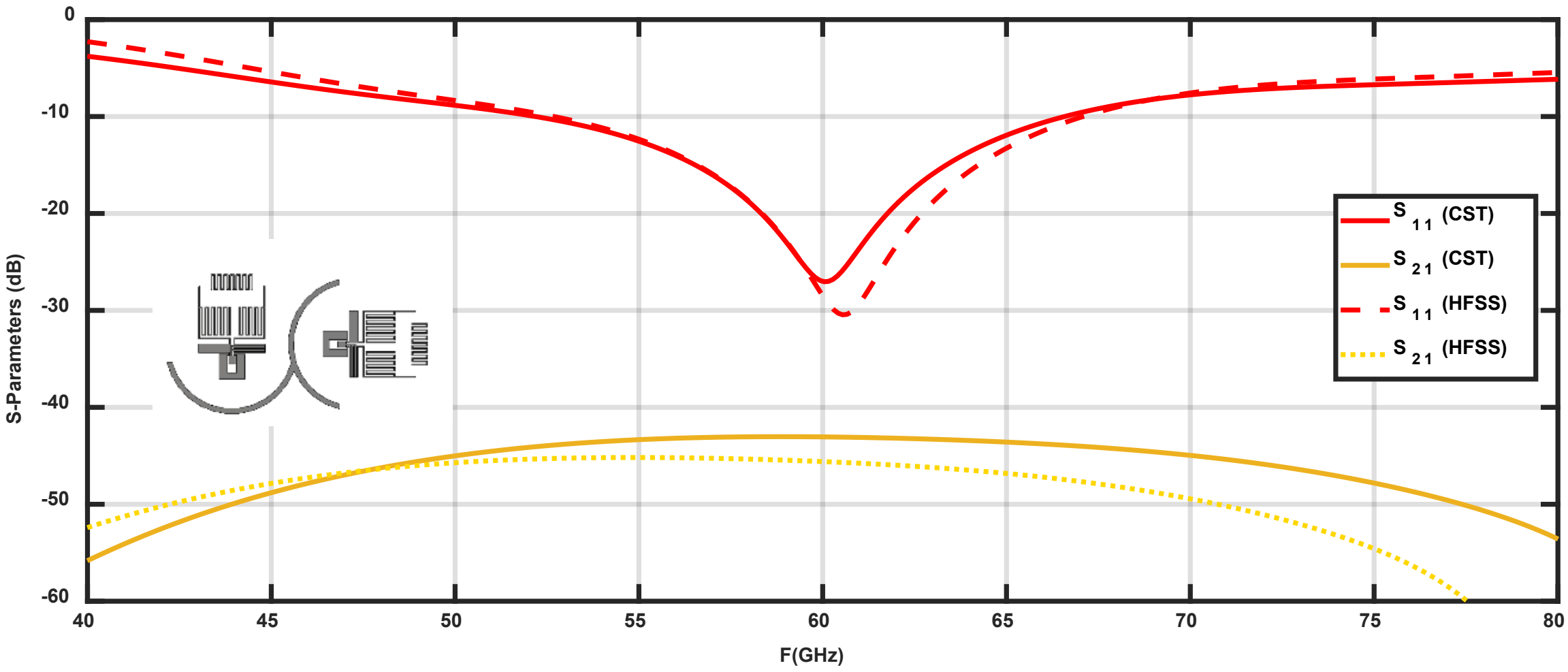

(c)Conf. III

Figure 5.16 S-parameters of different configuration of two elements MIMO Yagi-uda antenna.

Table 5.5 Comparison between three configurations of MIMO antennas

| Parameters | Conf. I | Conf. II | Conf. III |
|---|---|---|---|
| $\lvert S_{11}\rvert$ (dB) (at 60 GHz) | 32 | 27 | 27 |
| $\lvert S_{21}\rvert$ (dB) (at 60 GHz) | 43 | 40 | 43 |
| Diversity | No | Yes | Yes |
| BW (GHz) | 51-67 | 51.5-67 | 51.5-67 |

## 5.4.2 Four Elements MIMO

This section introduces four elements MIMO antenna as shown in Figure 5.17. The provided MIMO consists of two back by back and two side by back antennas. This configuration is introduced to provide high isolation between all elements. Moreover, the antennas in the proposed configuration are orthogonal together, giving diversity in the radiation patterns. These properties of the proposed MIMO indicate that antenna can be used for spatial multiplexing or pattern diversity. Figure 5.18 illustrates the S-parameters of the proposed MIMO antenna to ensure that the antenna covers the band from 51 GHz to 67 GHz with good matching and high isolation. All the results are verified by CST in addition to HFSS. From the 3-D radiation pattern of four elements that is introduced in Figure 5.19, we notice that the radiation pattern of 4 elements is in different directions because of the diversity between elements.

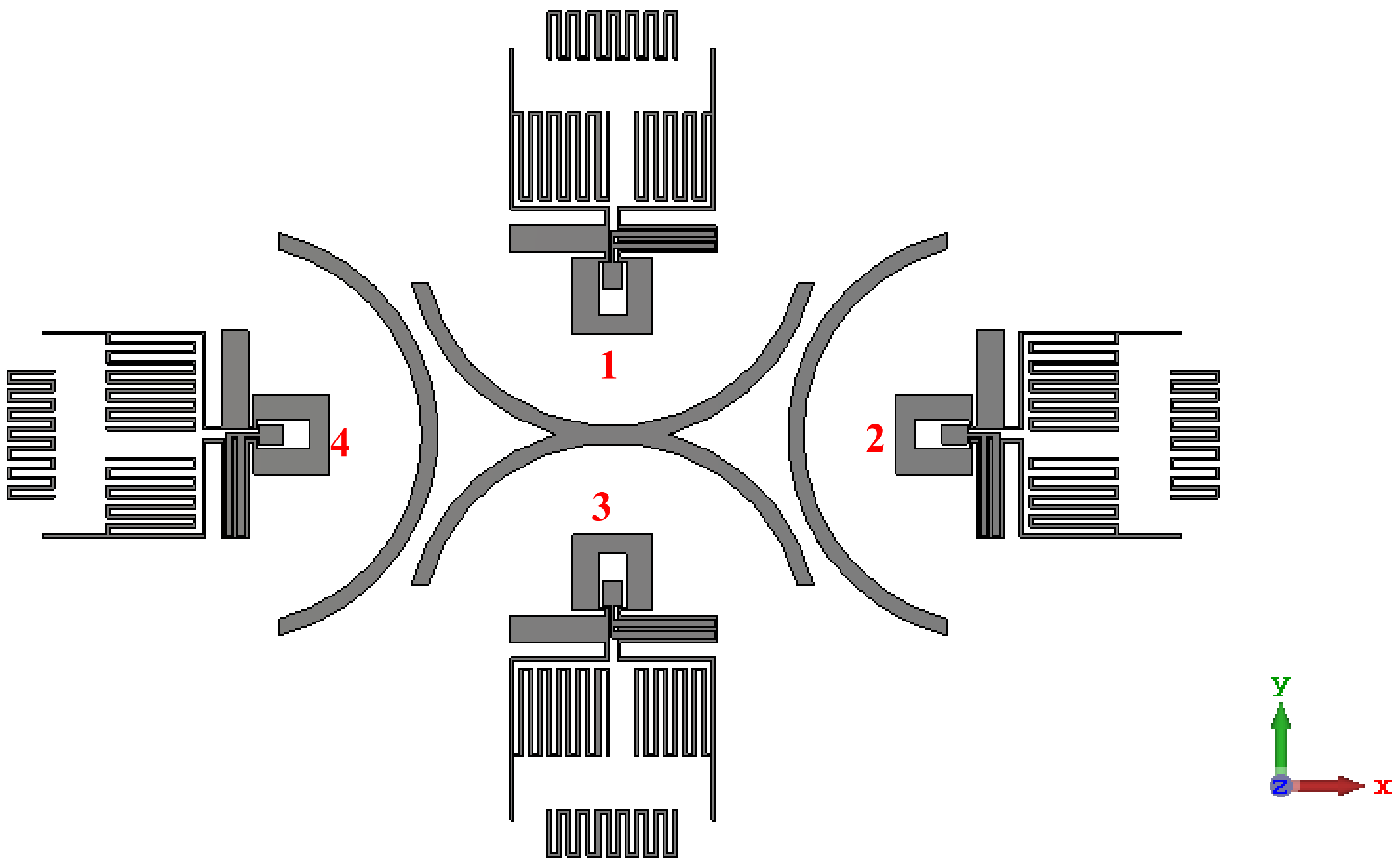

Figure 5.17 Configuration of four elements MIMO antenna.

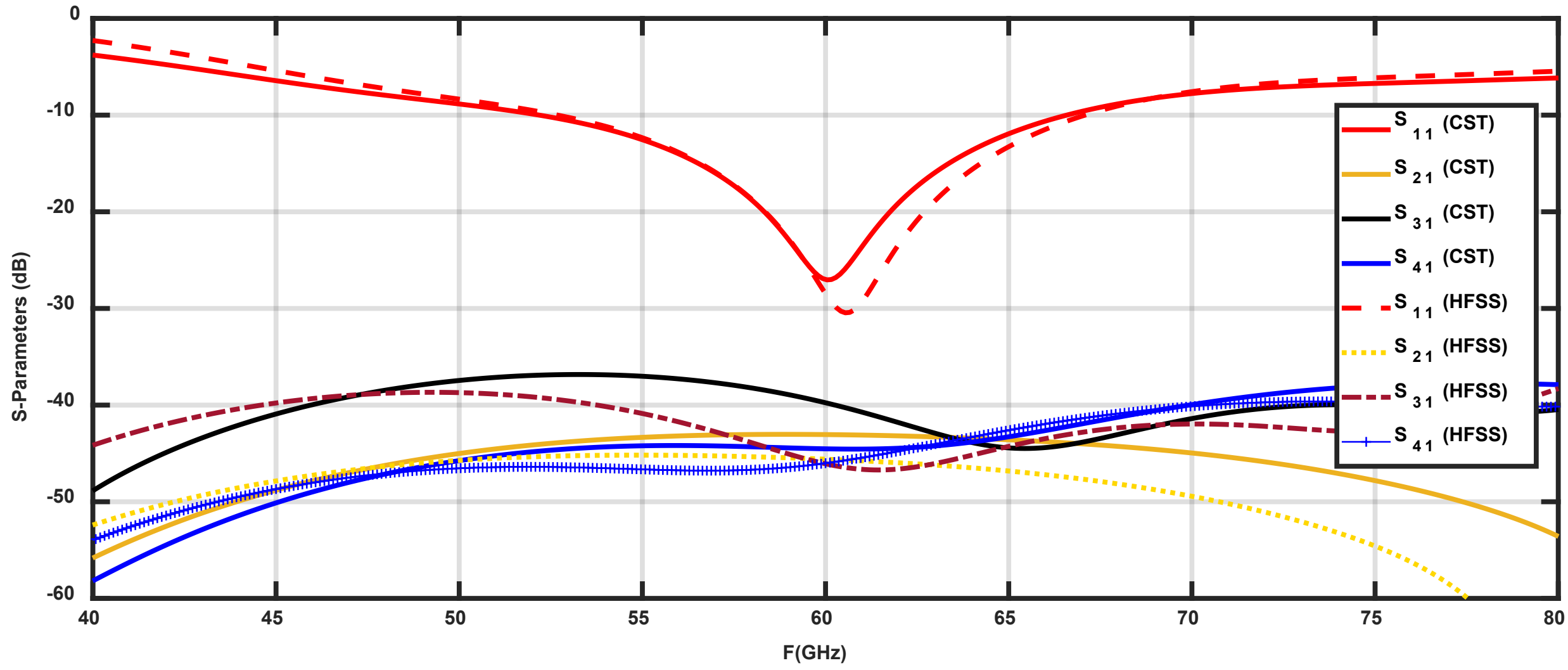

Figure 5.18 S-parameters of proposed four elements MIMO antenna

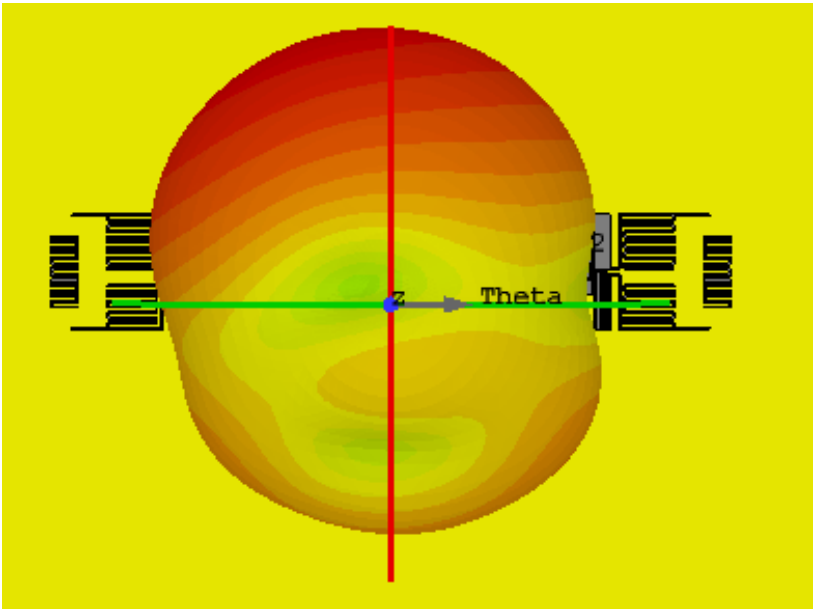

(a)Port 1

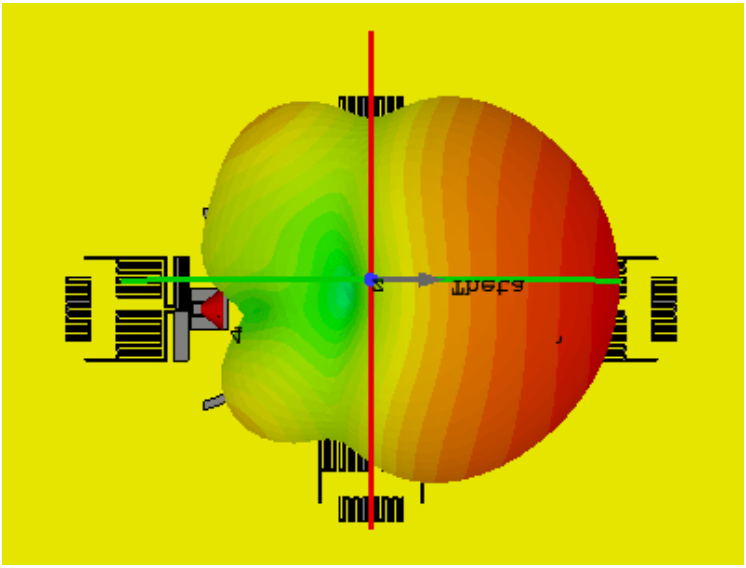

(b)Port 2

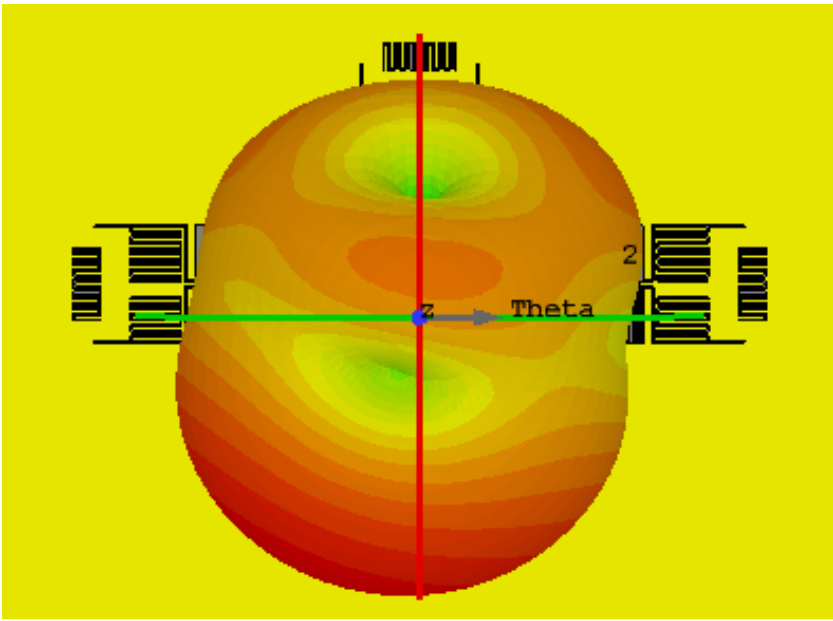

(c)Port 3

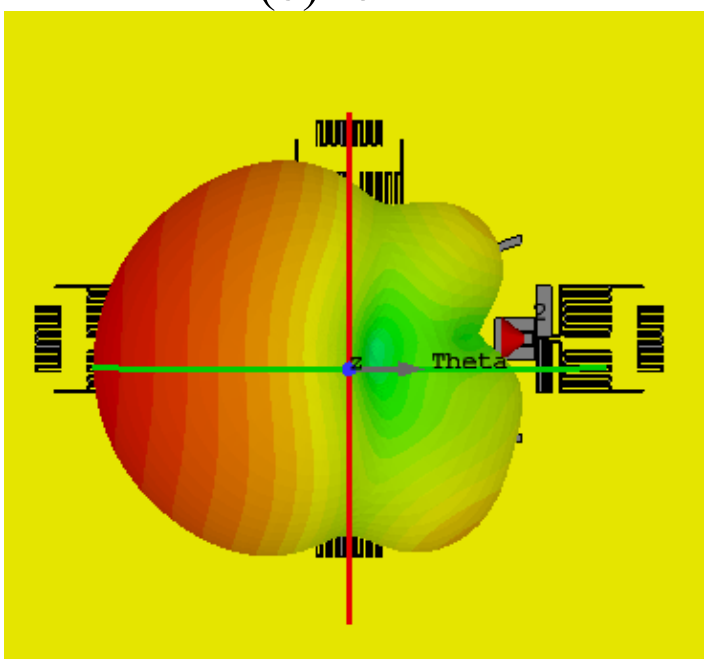

(d)Port4

Figure 5.19 Radiation pattern of proposed MIMO antenna at different ports

## 5.4.3 MIMO Parameters

The MIMO have four different parameters that should be tested to ensure that the MIMO gives a good performance:

- **Envelope Correlation Coefficient (ECC)**

The ECC is one of the main parameters used to characterize the performance of MIMO antenna. Where it measures the similarity between the antennas performance especially its radiation patterns. The ECC can be calculated from the following formula [200]:

$$\rho_{nm} = \frac{|S_{nn}^{*}S_{nm} + S_{mn}^{*}S_{mm}|^{2}}{(1-(|S_{nn}|^{2}+|S_{mn}|^{2}))\,(1-(|S_{mm}|^{2}+|S_{nm}|^{2}))} \quad (5.1)$$

Where $\rho$: ECC, S:S-parameter, S*: complex conjugate of S-parameters, m, and n are number of antenna m,n =1,2,3,4.

The value of ECC should be less than 0.5 over the operating band according to the published standards [196-199]. Whereas the lower values of ECC mean that the two antennas are good isolated. Figure 5.20 shows the ECC between MIMO elements. It is obvious from the figure that the ECC is less than 0.0003 within the operating band.

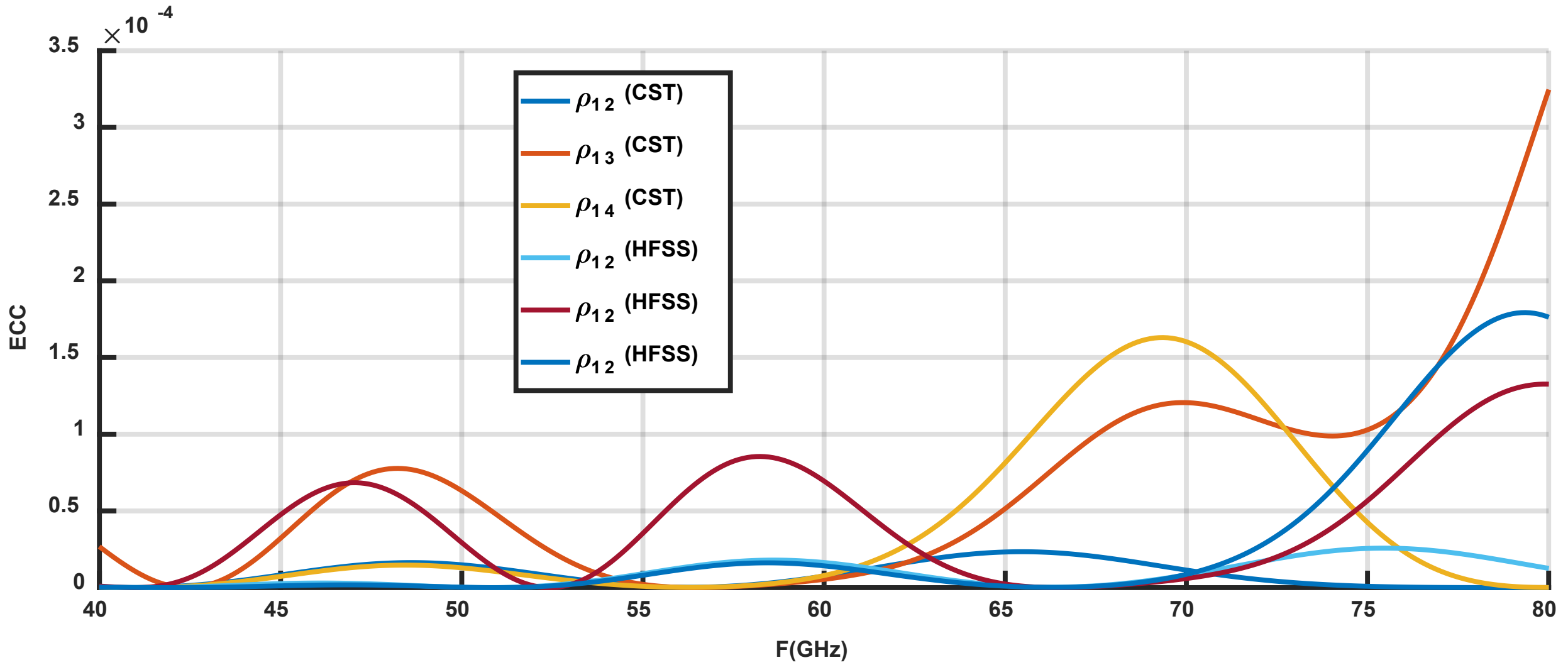

Figure 5.20 ECC of four elements proposed MIMO antenna

- **Diversity Gain (DG)**

The second parameter is a diversity gain, where the diversity gain (DG) can be expressed as

$$DG = 10\sqrt{1 - ECC^{2}} \quad (5.2)$$

As shown in Figure 5.21, the DG has 10 dB as a high value during the operating band due to the fact that ECC is extremely low ($ECC \cong 0$).

- **Total Active Reflection Coefficient**

The total active reflection coefficient (TARC) is the third parameter that indicates the coupling between ports. Its minimum value is 0 which means that all incident power is radiated whereas the maximum value is 1 which means that all the incident power is reflected. The TARC affects greatly the operating bandwidth of the MIMO antenna system [200]. The TARC is calculated according to the following equations [200]:

$$\Gamma_a{}^t = \frac{\sqrt{\sum_{i=1}^{N}|b_r|^2}}{\sqrt{\sum_{i=1}^{N}|a_i|^2}} \quad (5.3)$$

$$[b] = [S][a] \quad (5.4)$$

Where $a_i$, $b_r$ are the incident signals and reflected signals, respectively. [S], [a] and [b] represent scattering matrix, excitation vector, and scattered vector of the antenna, respectively.

Figure 5.22 shows the TARC curves upon exciting port one at $1e^{j0}$ when others ports have the same amplitude but with different excitation phases. Some of possible cases are introduced as shown in the Figure. The values of the TARC is referred to the effective BW of MIMO system. We can observe that the operating BW of the proposed antenna is not affected by different excitation phase of the other ports.

- **Channel Capacity Loss**

The final parameter is the channel capacity loss (CCL). The standard of CLL is CCL<0.4 b/s/Hz [197]. The capacity of MIMO system grow up with the increase of antenna numbers

$$CLL = -log_2 \det(\psi^R) \quad (5.5)$$

$$\psi^R = \begin{bmatrix} \rho_{11} & \rho_{12} & \rho_{13} & \rho_{14} \\ \rho_{21} & \rho_{22} & \rho_{23} & \rho_{24} \\ \rho_{31} & \rho_{32} & \rho_{33} & \rho_{34} \\ \rho_{41} & \rho_{42} & \rho_{43} & \rho_{44} \end{bmatrix}, \rho_{mm} = 1 - \left|\sum_{n=1}^{4} S_{mn}^* S_{nm}\right|, \rho_{mp} = -\left|\sum_{n=1}^{4} S_{mn}^* S_{np}\right|, \text{ for m,p=1,2.3 or 4}$$

Figure 5.23 shows the simulation and the measurement of the CCL for MIMO antenna. One can notice that the CCL values within the proposed band is less than 0.3 b/s/Hz.

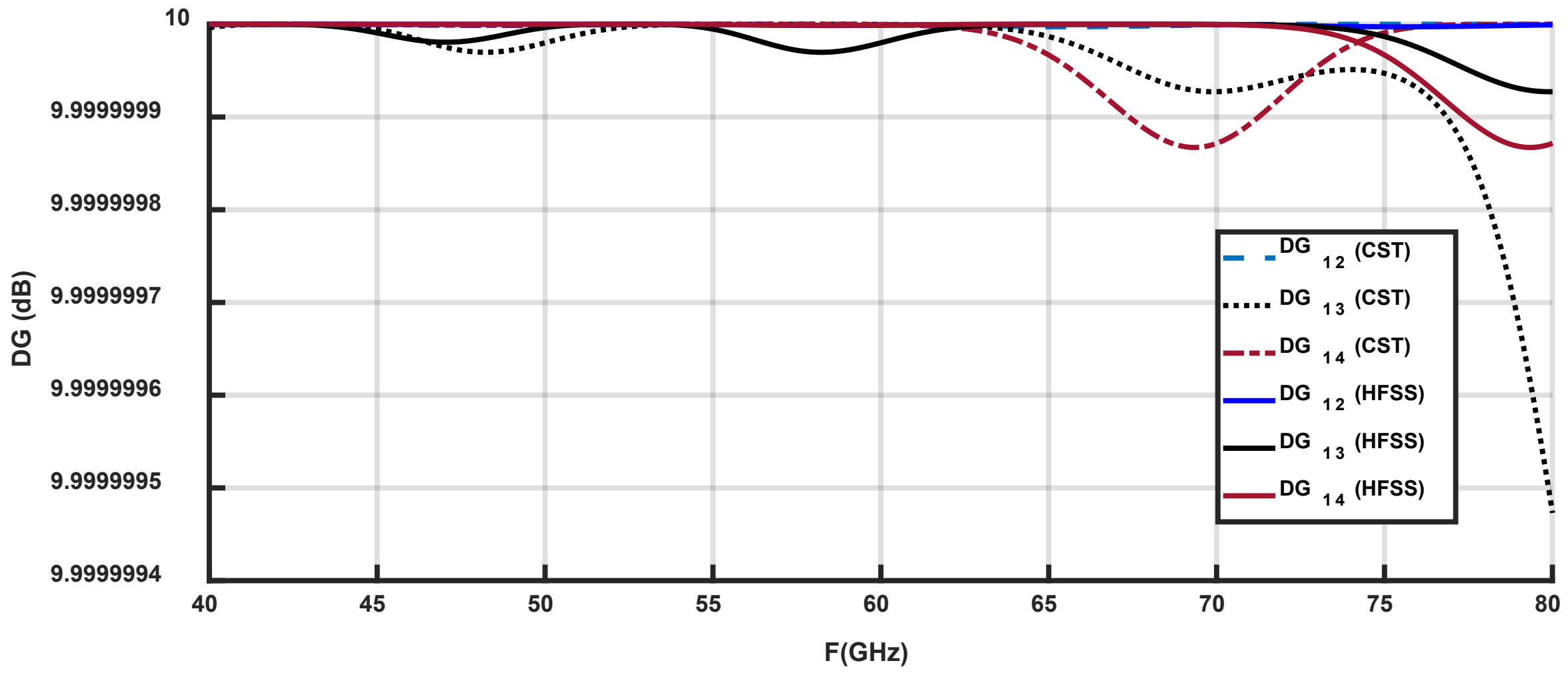

Figure 5.21 DG of four elements proposed MIMO antenna

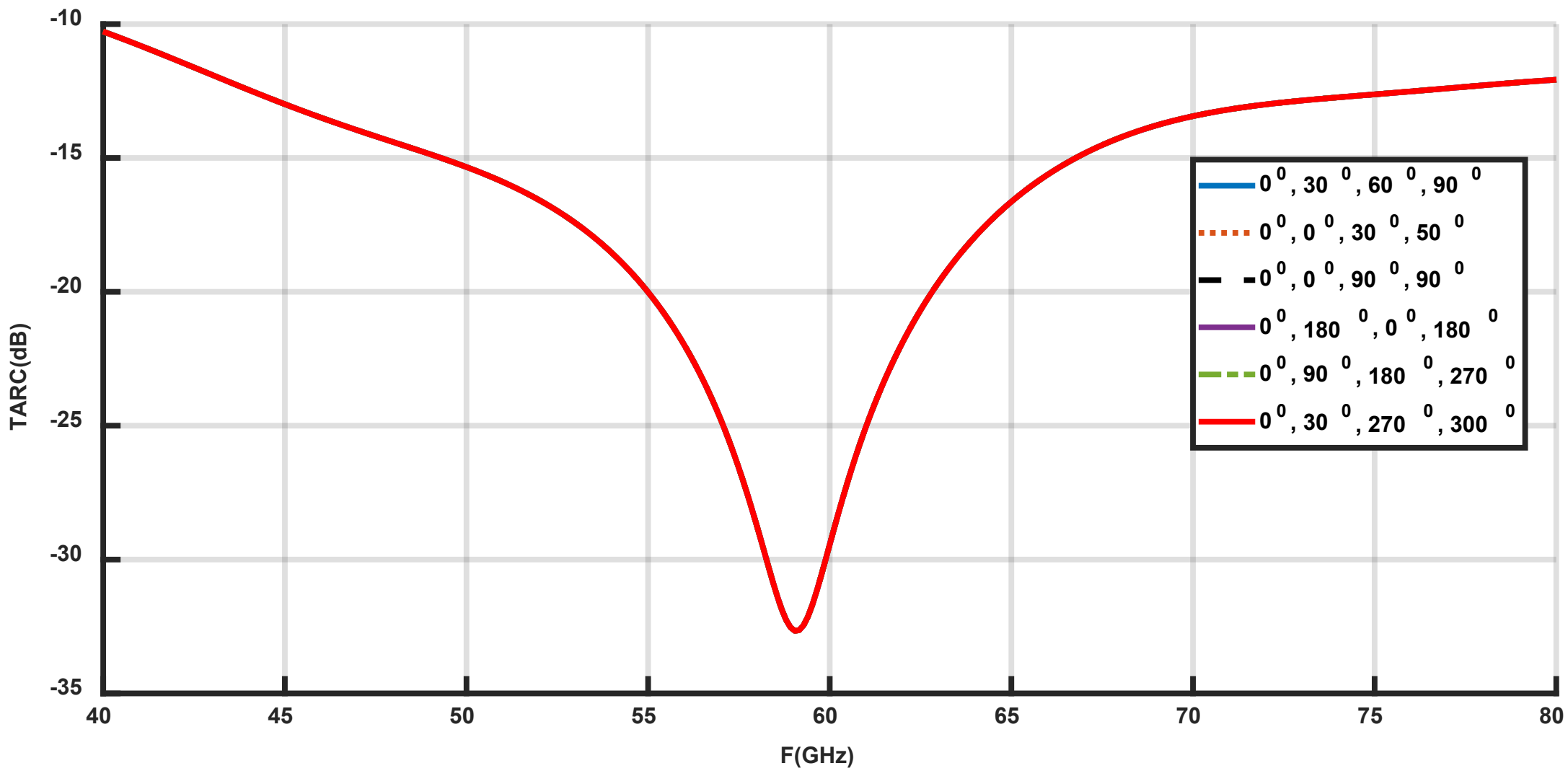

Figure 5.22 TARC of proposed four elements MIMO antenna

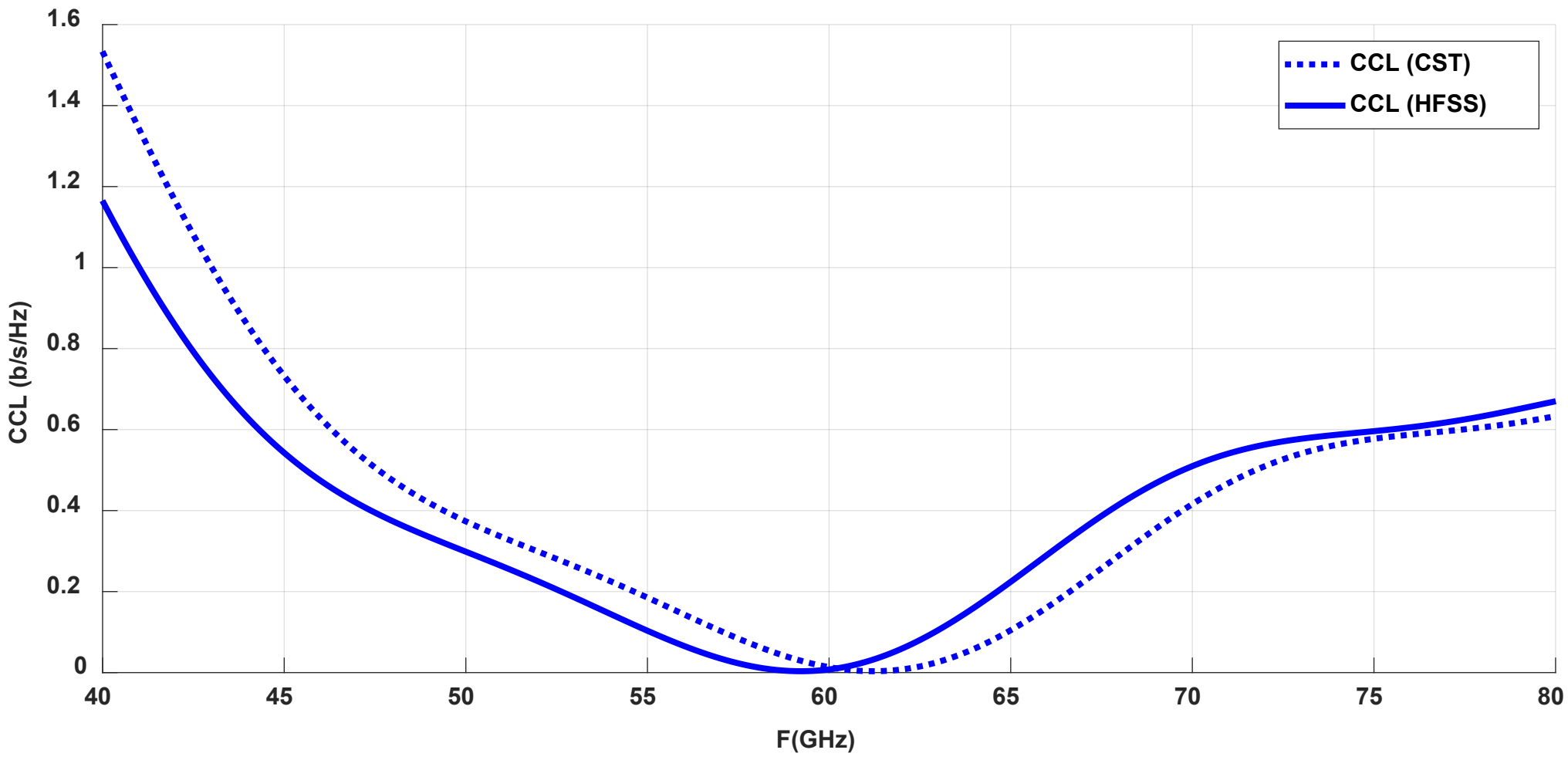

Figure 5.23 CCL of four elements proposed MIMO antenna

# 5.5 Conclusion

A Yagi-Uda and Vivaldi antennas are presented to support impedance bandwidth from 51 GHz to more than 67 GHz. The proposed antennas are designed using 0.18 µm CMOS technology with substrate size 5×5 mm$^2$ and overall antennas size 0.72×0.85 mm$^2$ and 0.5×0.87 mm$^2$ for Yagi-Uda and Vivaldi, respectively. These antennas are introduced to support point to point communications. Different techniques are used to enhance the radiation pattern properties of the Yagi Uda and the Vivaldi antennas. The antennas have a maximum gain of 0.4 dBi, 0.7 dBi and maximum radiation efficiency of 38%, 37% for the QYA and TSVA, respectively. The simulated antennas radiation pattern shows that the antennas have end-fire radiation characteristics and F/B ratio of 16 dB for Yagi-uda and 18 dB for Vivaldi. Furthermore, three configurations of two elements MIMO antenna and four elements MIMO antenna are introduced in this chapter to provide a high-performance antenna that can be used for short-range communications or indoor networks. The ECC, CCL, DG and TARC are introduced to investigate that the proposed antenna has good performance over the proposed band.

# Chapter Six:
# Antenna Designs for THz Applications

## 6.1 Introduction

This chapter introduces a disc resonator antenna array with compact size and wide bandwidth for THz applications. The disc antenna is design based on modified silicon on glass (SOG) technology platform from high resistivity Si. A Dielectric Waveguide (DWG) is matched with the disc dielectric antenna using CPW feed. The CPW feed is designed on the Pyrex side of the Si wafer bonded to the Pyrex. The proposed antenna is designed to operate from 325 GHz to 500 GHz with good return loss. The end-fire and broadside antennas are introduced with a high gain of about 17 dBi. The antenna has high efficiency and low cost. Also, the antenna array is introduced with compact size 1 mm× $\mathbf{0.72\ mm}$ with H=0.11 mm for endfire and H=0.31 mm for the broadside.

## 6.2 THz

The terahertz waves offer bands from 0.3 THz to 10 THz. The terahertz frequency range offers new specifications over other spectrum for a large number of applications, such as high-resolution imagers ultra-high-speed short-distance communication systems, bio-medical, pharmaceutical, security, sensing, and spectroscopy. This indicates that wireless devices are required to support different technologies and operate in different frequency bands [3, 125].

In order to increase the antenna gain and directivity, several methods are introduced in [134, 245, 246]; In [134], an array of glass lens antennas arranged on a silicon (Si) substrate is introduced based on planar metallic rectangular waveguide structure. In [245], the authors present a two tapered dielectric antenna designed and implemented in the suspended SOG waveguide platform.

## 6.3 The transition from MWG to DRW

According to the dimensions of the metallic waveguide (MWG) in the x and y-axis, it is excited by $TE_{10}$ mode or $TE_{01}$ mode. In this case, we assume that a>b and the proposed MWG is excited by $TE_{10}$ mode with the main component of electric field Ey, as shown in Figure 6.1. A part of the dielectric rod waveguide (DRW) with length Li is inserted into the waveguide without any modification for impedance matching between the dielectric rod and the metallic waveguide, as shown in Figure 6.1 (b). Part of excited power is transformed into the surface wave power at the end of feed, which is $TE^{y}_{11}$ mode and the analysis of this coupling method and converting the mode from MWG to DRW are introduced in [247-252].

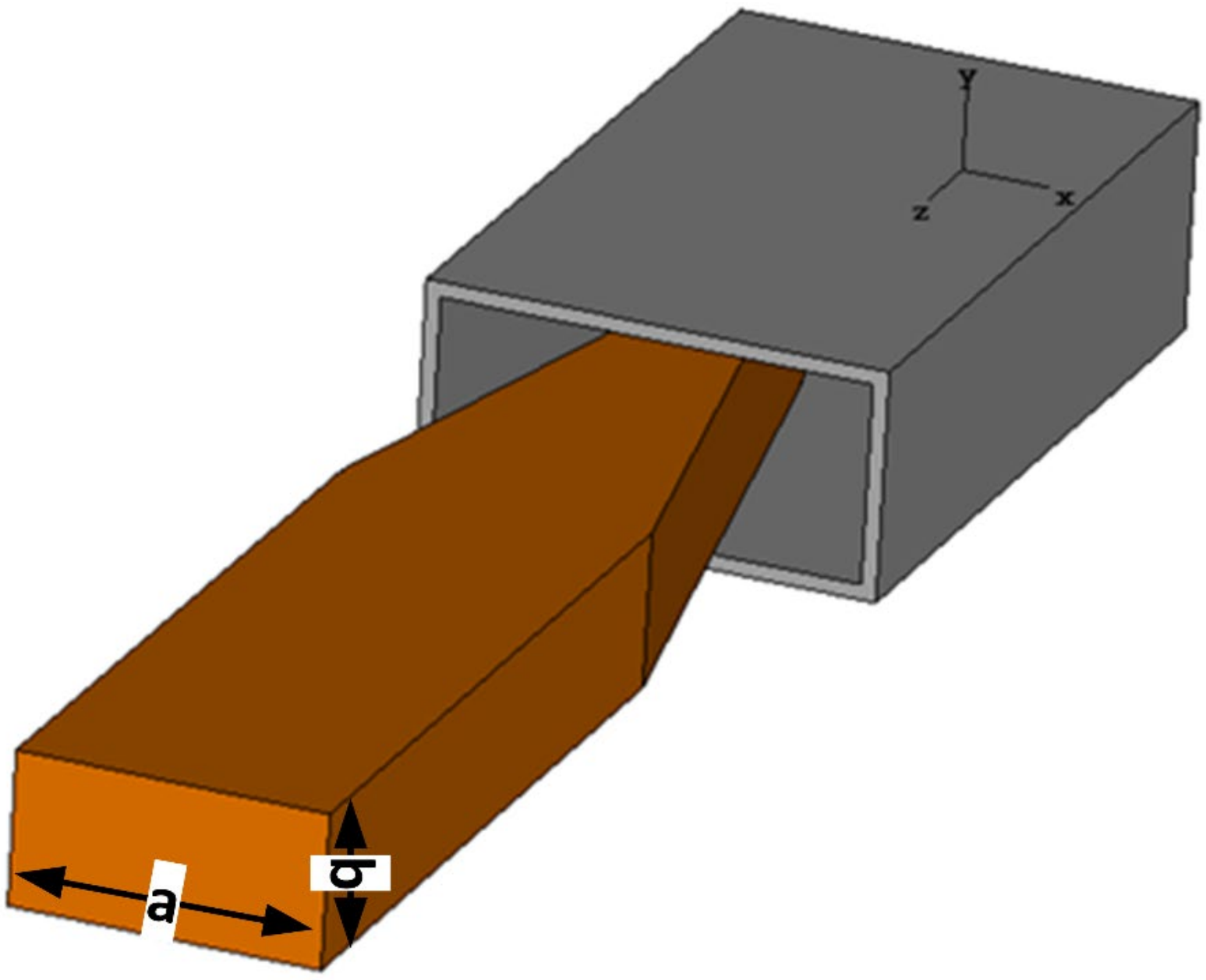

(a)3-D

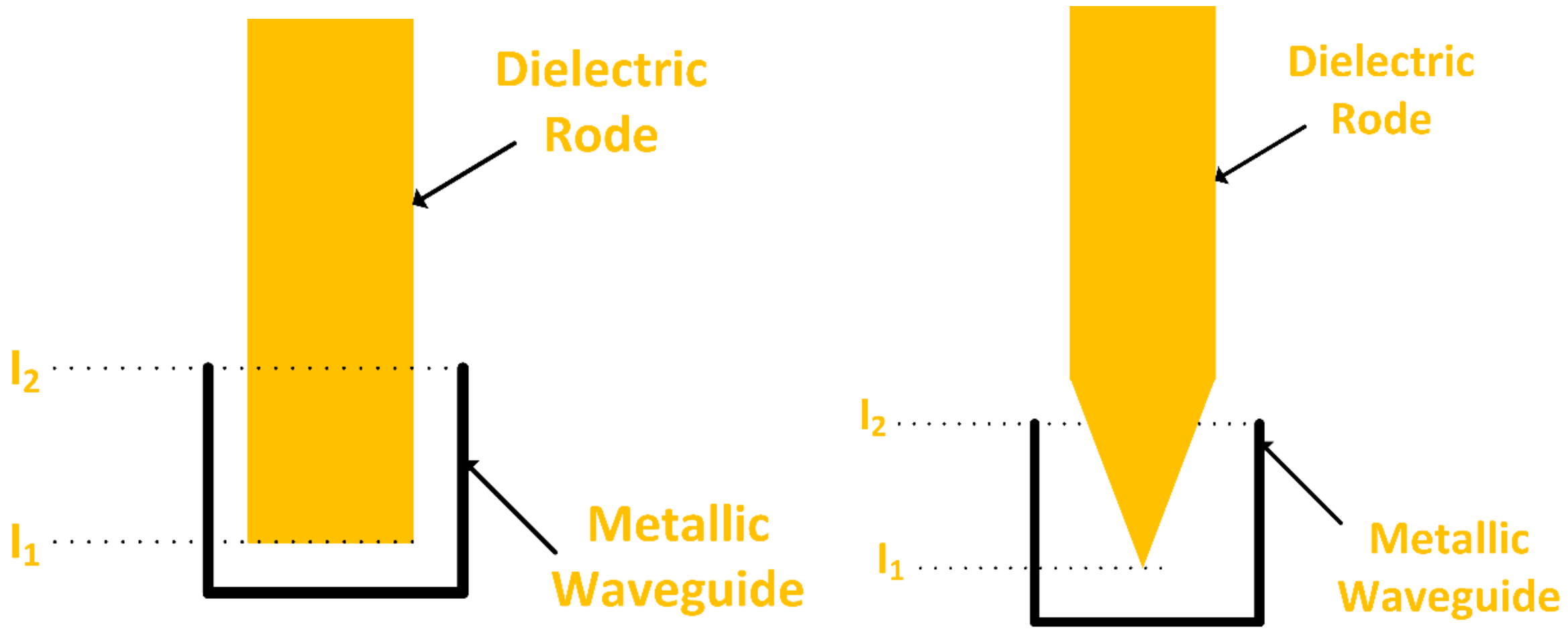

(b)2-D without tapered section (c)2-D with tapered section

Figure 6.1 Geometry of the DRW fed by MWG

In the MWG, two interfaces I1 and I2, partly reflect the excited $TE_{10}$ mode, as shown in Figure 6.1 (b and c). The first interface, $I_1$, is the interface inside the MWG between the air-filled and dielectric rod filled, and the second interface, $I_2$, is the interface at the edge of MWG between the MWG and the dielectric rod. The minimum reflection coefficient from the conversion between the MWG and DRG can be calculated as follows:

$$\Gamma_{min} = \frac{\Gamma_1 - \Gamma_2}{1 - \Gamma_1 \Gamma_2} \tag{6.1}$$

Where $\Gamma_1$ and $\Gamma_2$ are reflection coefficients at interface $I_1$ and interface $I_2$, respectively.

$$\Gamma_1 = \frac{Z_a - Z_d}{Z_a + Z_d} \tag{6.2}$$

To calculate $\Gamma_2$, we need to calculate wave impedance of $TE^{y}_{11}$

$$Z_{TE} = \frac{k\eta}{\beta} \tag{6.3}$$

Where $Z_{TE}$ mode impedance, $\beta$ propagation constant, $\eta$ intrinsic impedance, and k wave number.

$$\Gamma_2 = \frac{Z_d - Z_{TE}}{Z_d + Z_{TE}} \tag{6.4}$$

To minimize the total reflection coefficient at interfaces, the taper section is used to increase the excitation of DRW mode as shown in Figure 6.1 (c).

## 6.4 Design of DRW

In this part, the transition between metallic waveguide (MWG) and the dielectric rod waveguide (DRW) is studied; for this transition, there are two main points that should be considered:

- The TE (transverse electric) field distribution of the high dielectric rod has contrast in the core relative to the outside of the rod in contrast to the MWG that have equal electric field distribution.
- The continuity of field from the MWG to the DRW should be considered to be a smoothly transition between them.

Firstly, the DWG is designed to show the amount of losses. The taper section of the DRW can be matched between the DWG and the metal rectangular waveguide used for excitation without any additional structures. The fundamental mode $TE_{10}$ of the metal waveguide (WR2.2) with standard dimensions excites the $TE_{11}$ mode in the high-permittivity DWG, providing good matching and transmission. The proposed DRW is designed using high-resistivity Si with a relative permittivity of $\varepsilon_r = 11.7$ and a conductivity of $\sigma = 0.01$ S/m. Figure 6.2 shows the design of DRW and excited with WR-2.2 for band from 325 GHz to 500 GHz and WR-1.5 for the band from 500 to 700 GHz. Figure 6.3 shows the transmission coefficient of the DRW with small losses in the proposed band from 325 GHz to 700 GHz. CST and HFSS verify all the results; results are closed together. The losses at this frequency band are very small compared with the microstrip line. Figure 6.4 shows the 2-D simulation using COMSOL and 3-D simulation using CST software in order to show the single-mode operation of DRW; it is easy to note that the field confines in the DRW and there is no leakage outside the DRW cross-section. Therefore, this dielectric rod is used as a transmission line in our proposed design. The width (a) and height (b) of DRW is selected to ensure that there is one dominant mode at the proposed frequency ${}^{a}/_{b} \approx 0.5$.

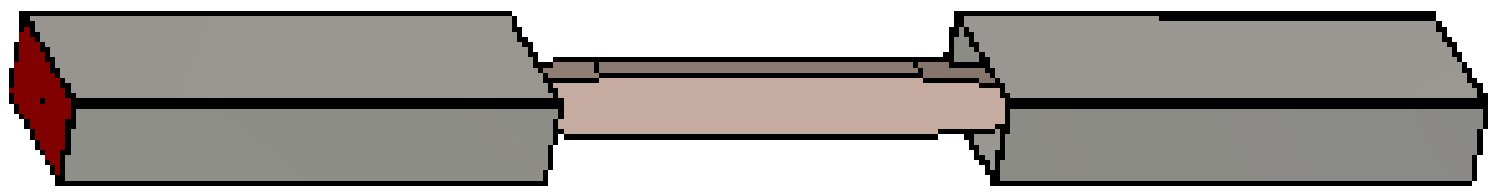

(a) DRW with metallic waveguide

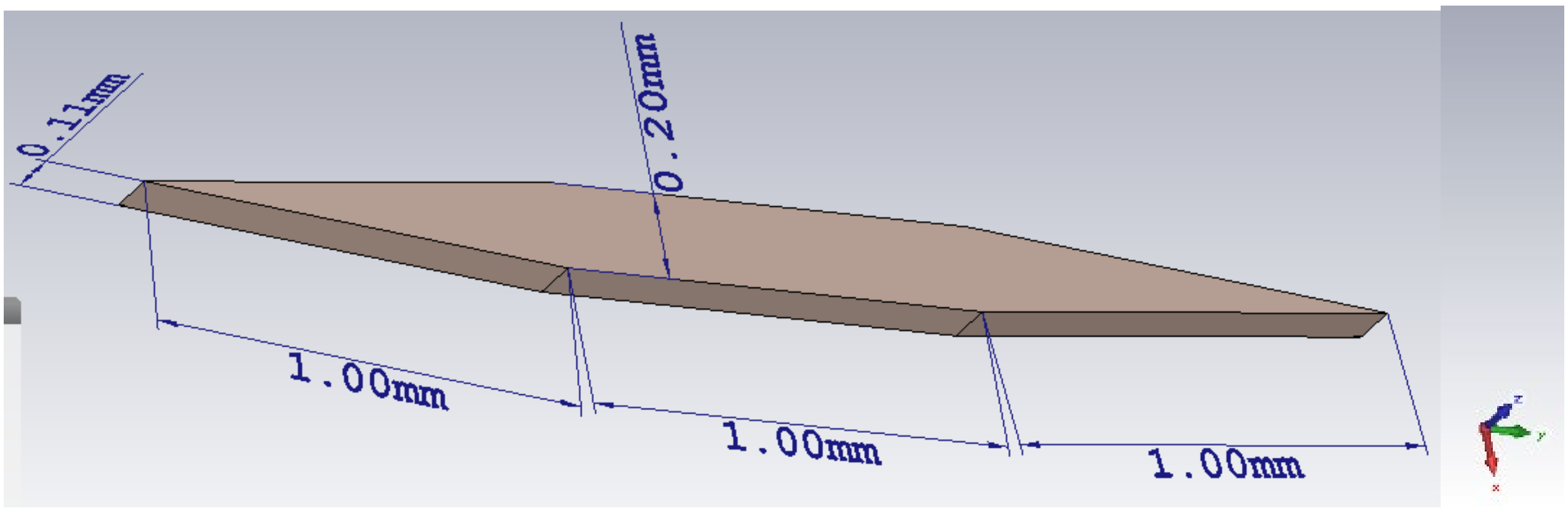

(b)DRW

Figure 6.2 Geometry of the proposed DRW

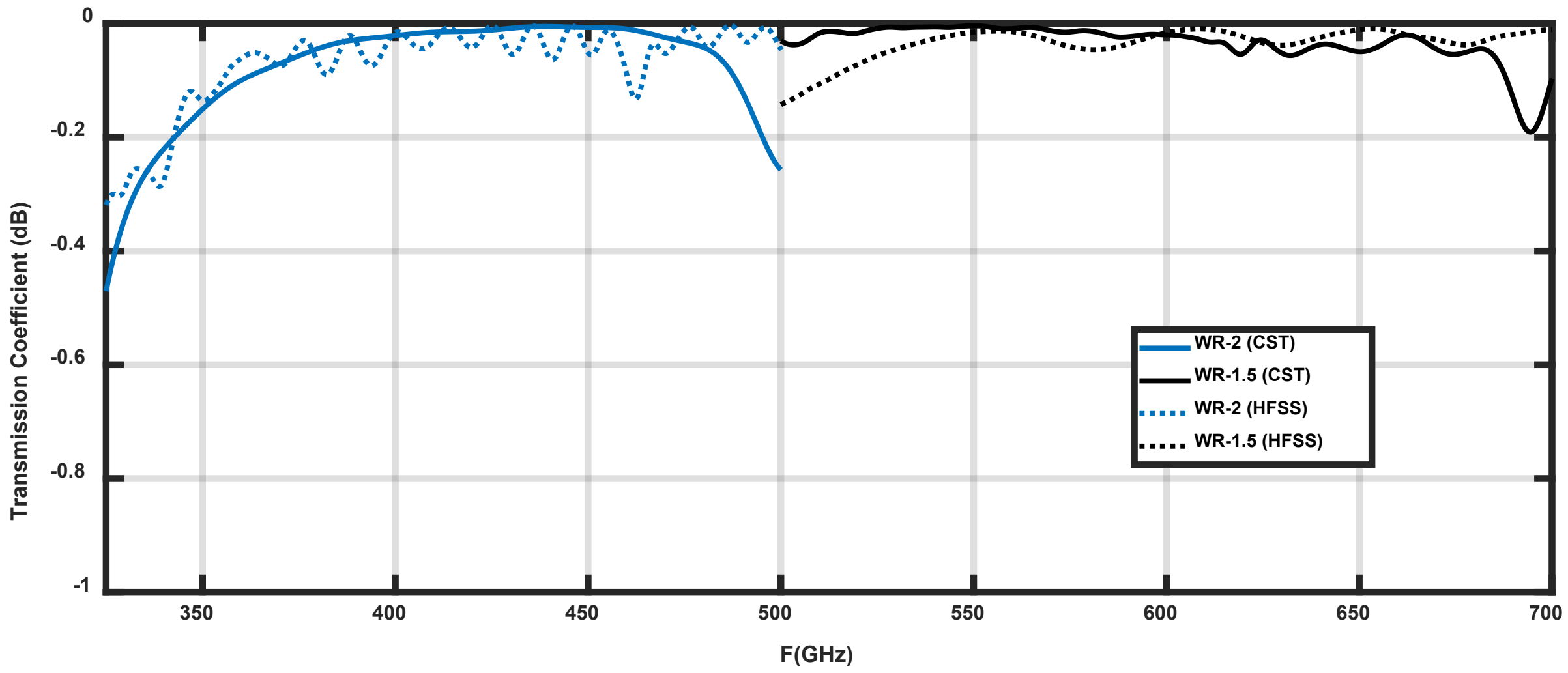

Figure 6.3 Transmission coefficient of the DRW

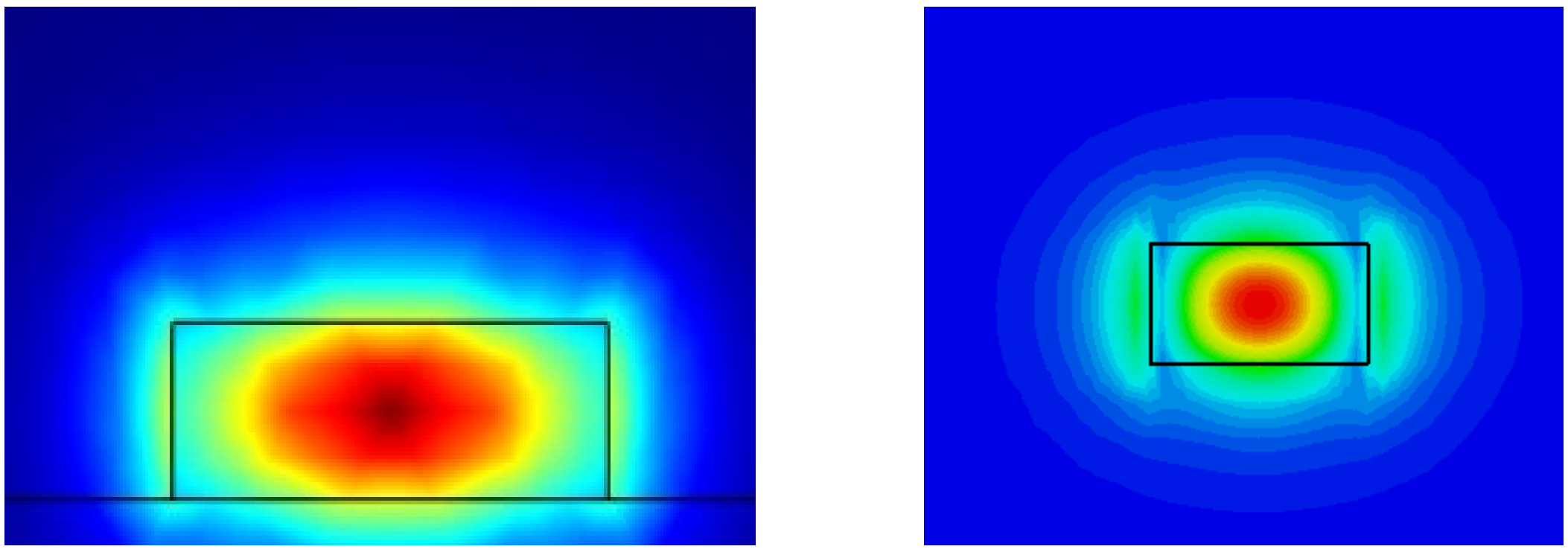

(a) 2-D using COMSOL (b) 3-D using CST

Figure 6.4 First mode in DRW

# 6.5 The Transition from CPW to DRW

In this section, we study the transition from CPW to DRW. The geometry of the CPW transition is shown in Figure 6.5. The CPW transmission line is the best candidate line to realize the transition from CMOS and the components to the dielectric waveguide structures. The dielectric waveguide is tapered at one end to transition from DRW to WR-2.2 rectangular metallic waveguide. At the other end of the DRW, the DRW to CPW transition is patterned over the DRW. The transition from DRW to CPW is designed to couple the fundamental mode of the dielectric waveguide to the CPW line. The two ground planes are extended gradually to the sides of the DRW. The two extended ground planes, forces the electric field to align horizontally along the CPW line. All the dimensions of this transition are shown in Figure 6.5. Figure 6.6 shows the electric field distribution of the CPW transition, most of the electric field is coupled to the CPW line where a small portion of the energy is still traveling through the dielectric waveguide. The simulated S-parameters is shown in Figure 6.7 from the s-parameters one can note that the insertion loss of the transition is between 0.2-1.3 dB. The return loss is better than 18 dB over the frequency range. The HFSS and CST simulated results are introduced to verify the structure.

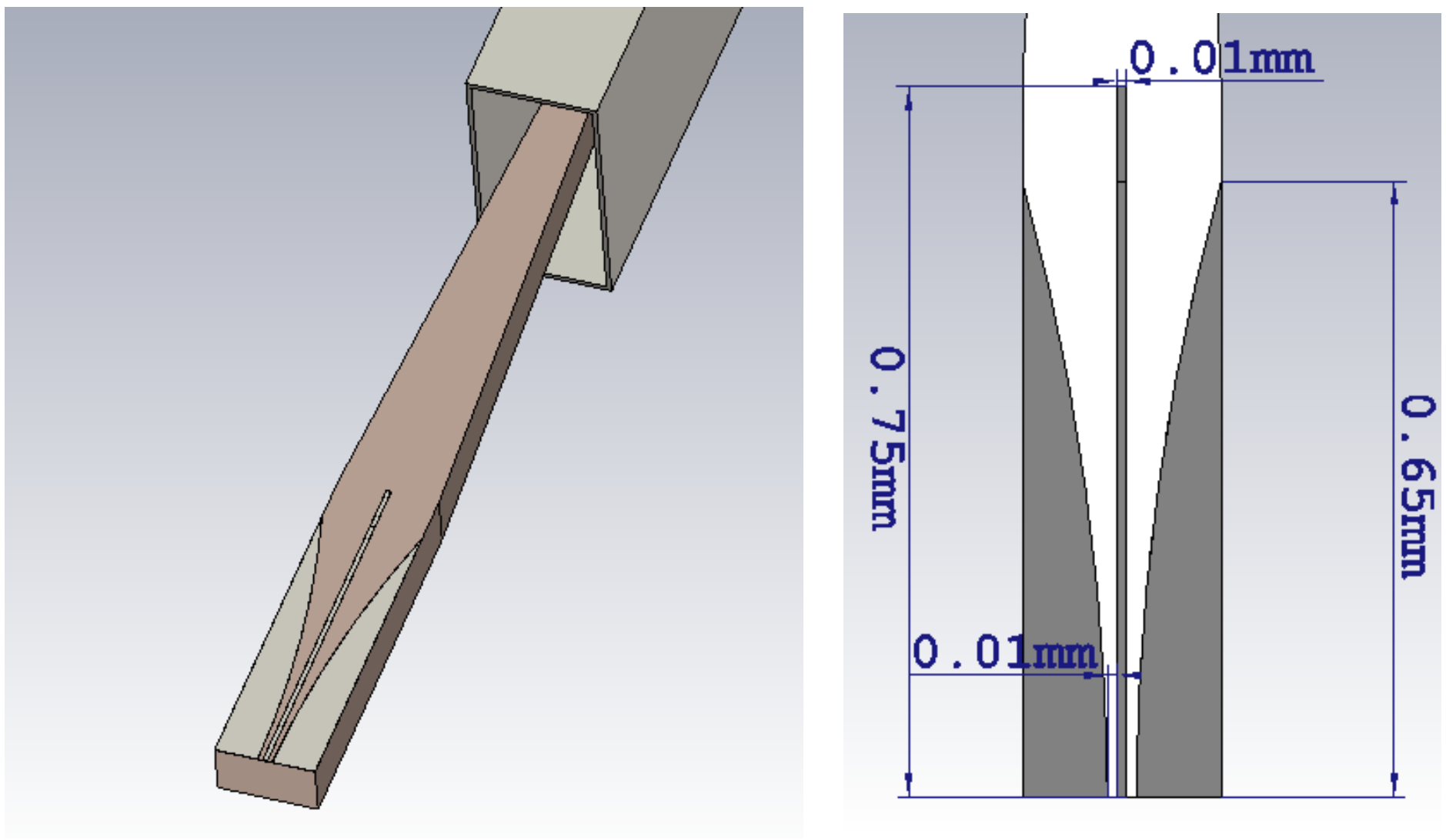

Figure 6.5 Transition from CPW to DRW

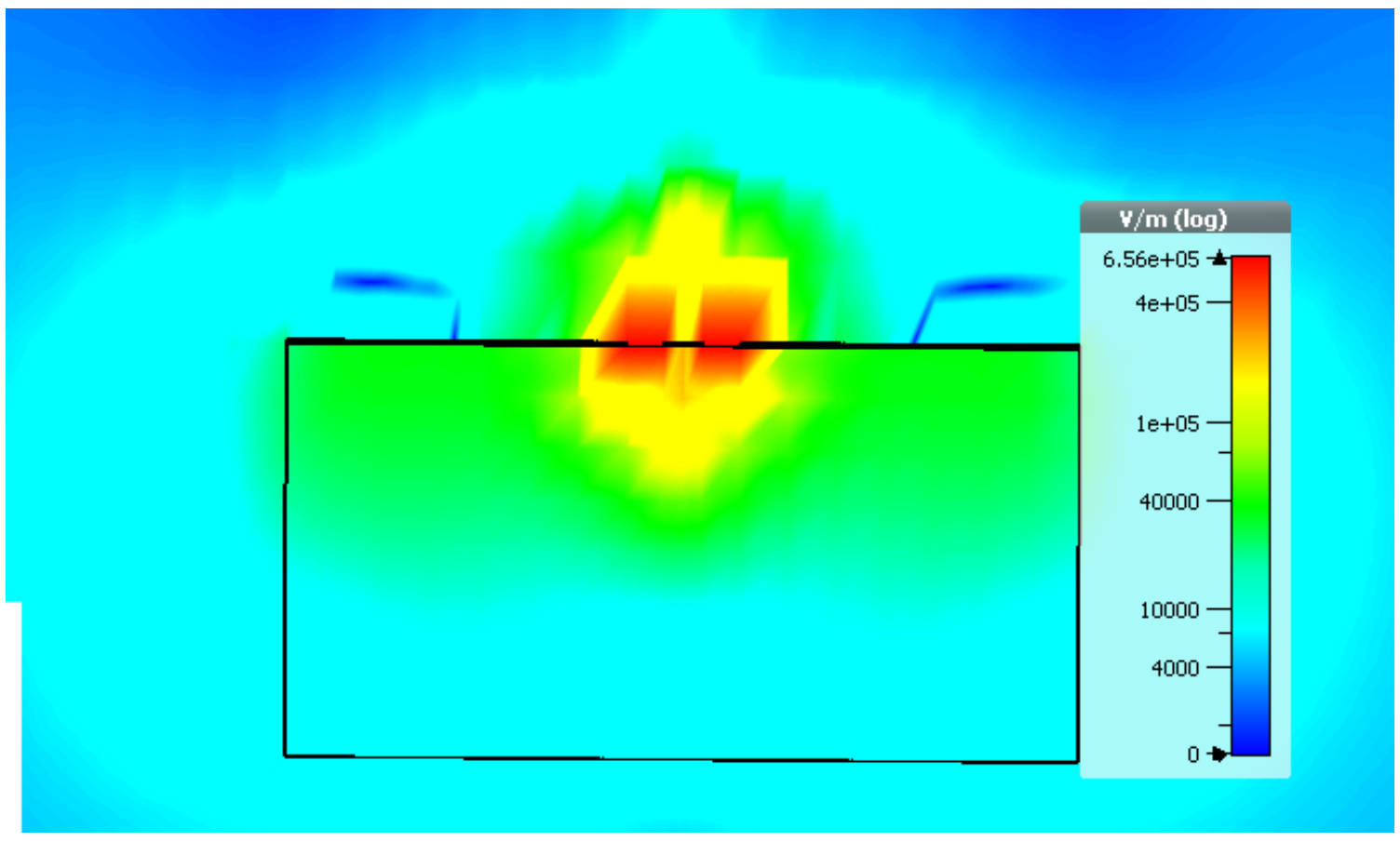

(a)Electric field contrast

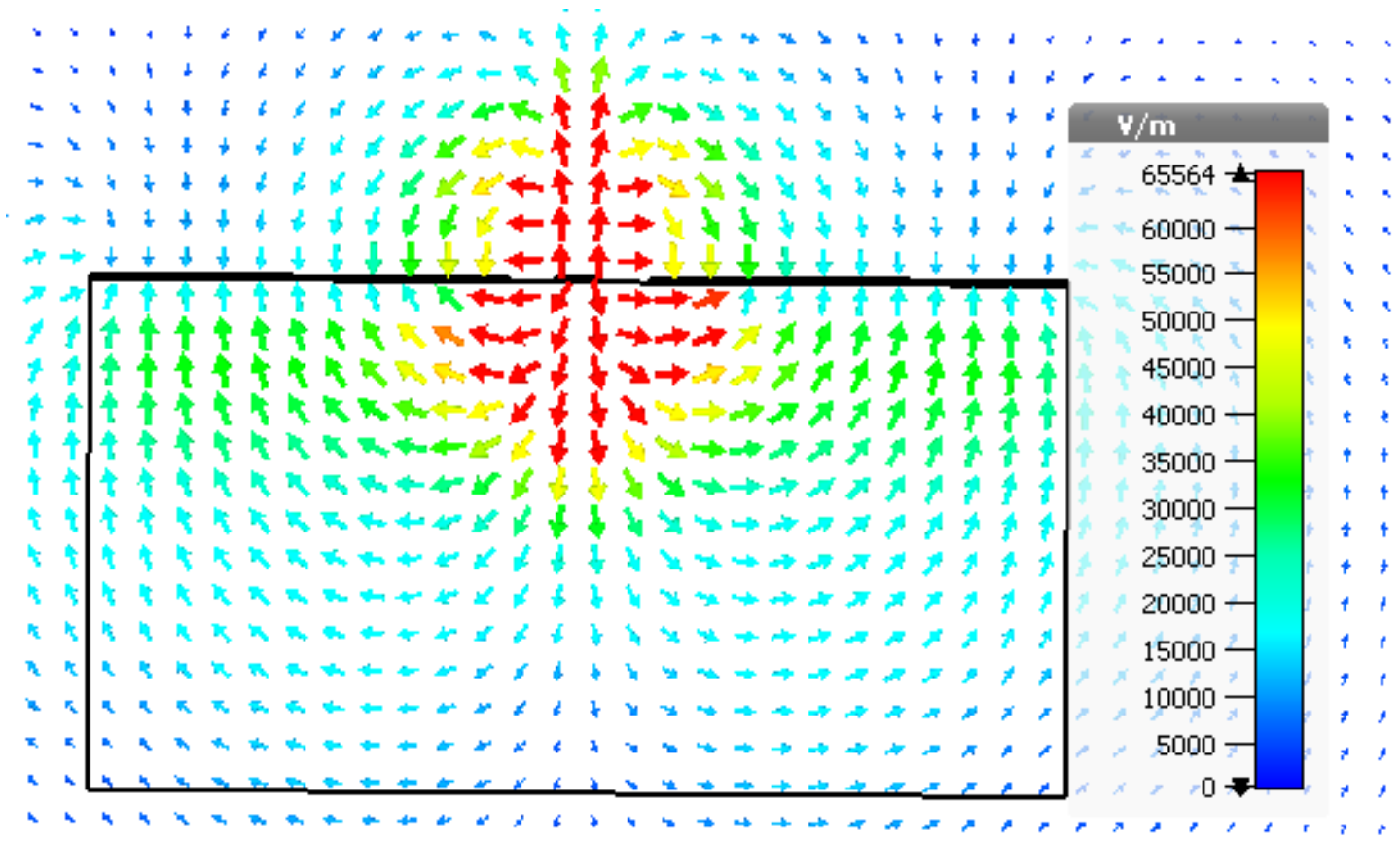

(b)Electric field lines

Figure 6.6 Electric field distribution in transition from CPW to DRW

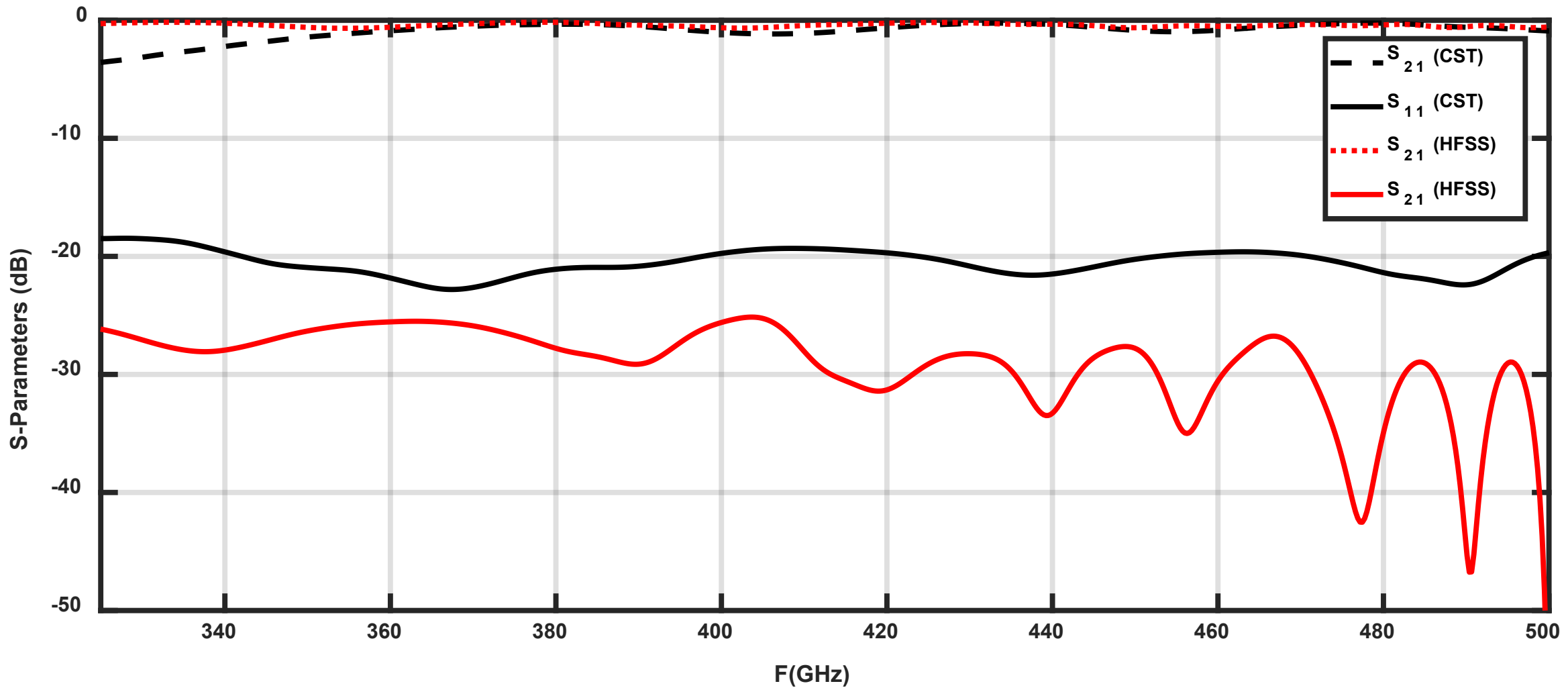

Figure 6.7 S-parameters of transition from CPW to DRW

## 6.6 Design of One Element

The proposed antenna consists of a silicon straight section waveguide segment connected in series with a disc resonator acting as a radiating element. The CPW is used to couple the power from the DWG to the disc resonator. The CPW is tapered at the input of the antenna for providing a smooth transition to the disc antenna, as shown in Figure 6.8.

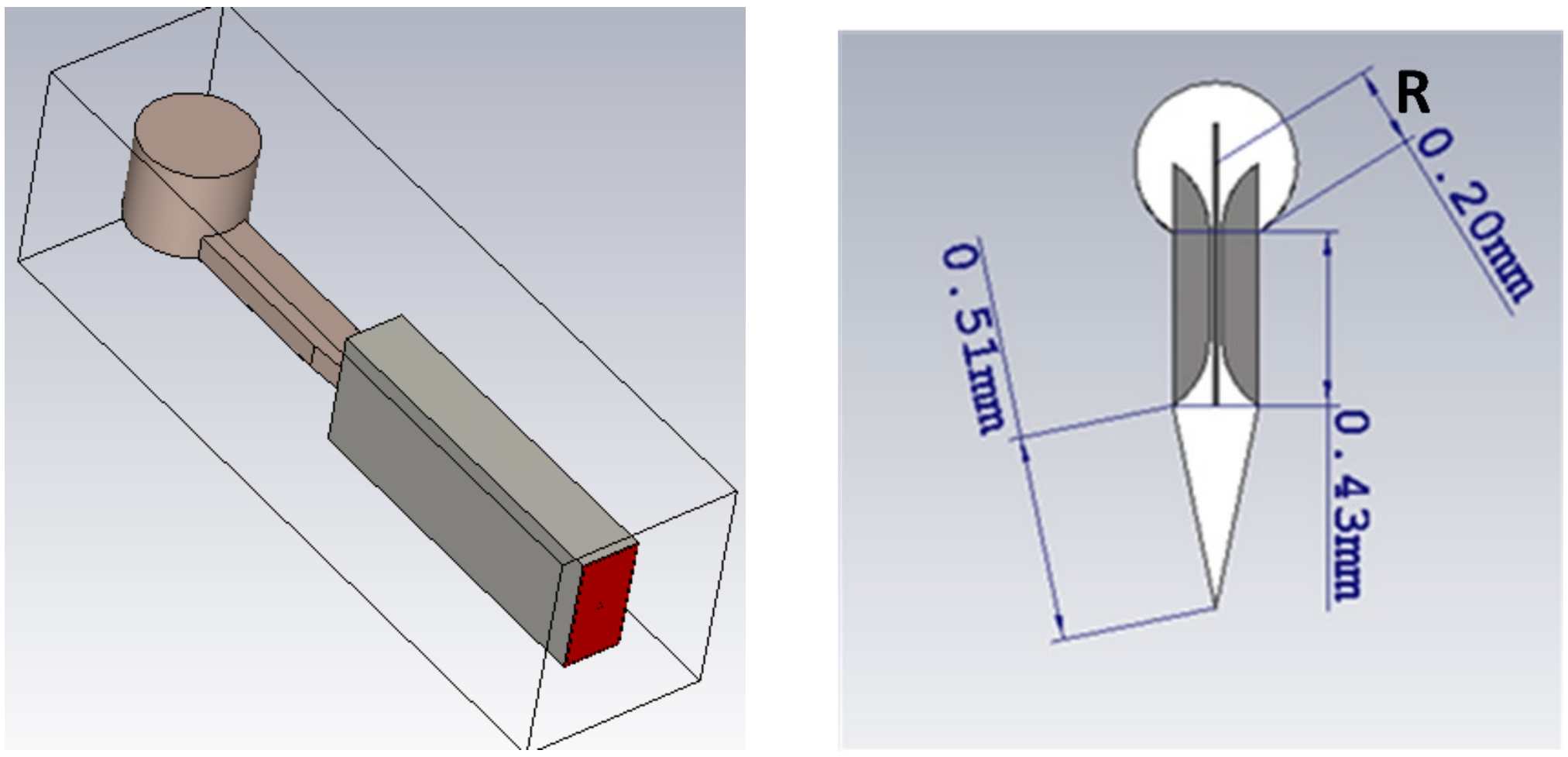

(a)3-D view (b)Back-side view

Figure 6.8 Antenna Geometry

The proposed disc dimensions are calculated according to the equations that introduced in chapter 3 (from eq. (3.43) to eq. (3.48)). Furthermore, in 1976 Yih [253] introduces an approximate relation to calculate the endfire tapered dielectric rode antenna.

$$R = \frac{\lambda_0}{\sqrt{\frac{\pi}{2}(\epsilon_r - 1)}} \tag{6.5}$$

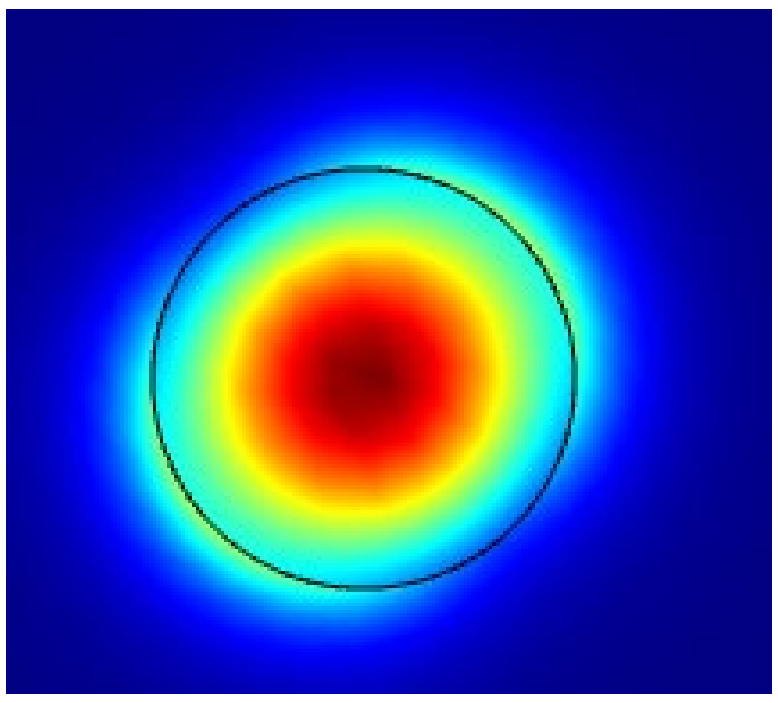

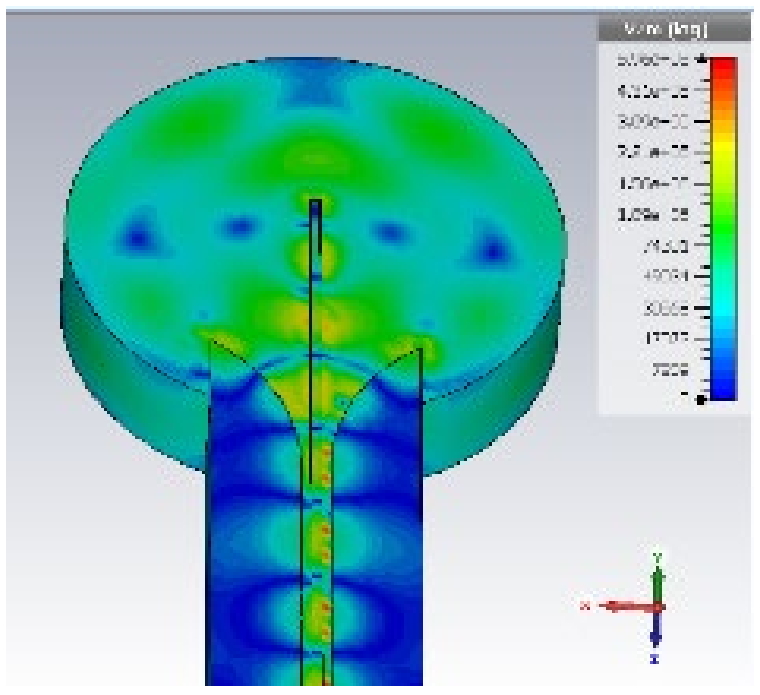

(a)First mode of Disc resenator (b)Electric field in the CPW feed

Figure 6.9 Electric field distribution.

Figure 6.9 (a) shows the 2-D simulation using COMSOL software in order to show the single mode operation of DRA and Figure 6.9 (b) shows the 3-D simulation using CST microwave studio to show the electric field distribution on CPW and DRA. The antenna operates from 400 GHz to 500 GHz with height H=0.11mm, and with H=0.31 mm the antenna operates from 325 GHz to 500 GHz with the same height of DWG (0.11 mm). The antenna is simulated by CST and HFSS simulators and the return loss of antenna with two different height of disc is shown in Figure 6.10. The most common type of dielectric antenna is the tapered antenna, which is inherently long and radiates in the end-fire direction [121, 128, 133, 134, 245, 246, 254, 255]. The radiation pattern of the proposed antenna with H=0.11 mm is end-fire radiation, while with H=0.31 mm is broadside with high gain as shown in Figure 6.11. The radiation patterns at 490 GHz is selected as an example because this frequency has the best matching for the two antennas. At the higher frequency the dielectric resonators have hybrid modes $HE_{mn}$ or $EH_{mn}$ which are described by a combination of two linear modes TE mode and TM mode. The HE or EH is described by the dominant linear mode TE or TM, respectively. The variation in the radiation pattern between two heights is due to the variation of propagation hybrid mode in each case as shown in Figure 6.12. Figure 6.13 shows the gain of the two antennas over the operating band, we notice that the antennas have average gain more than 11 dBi over the operating bands.

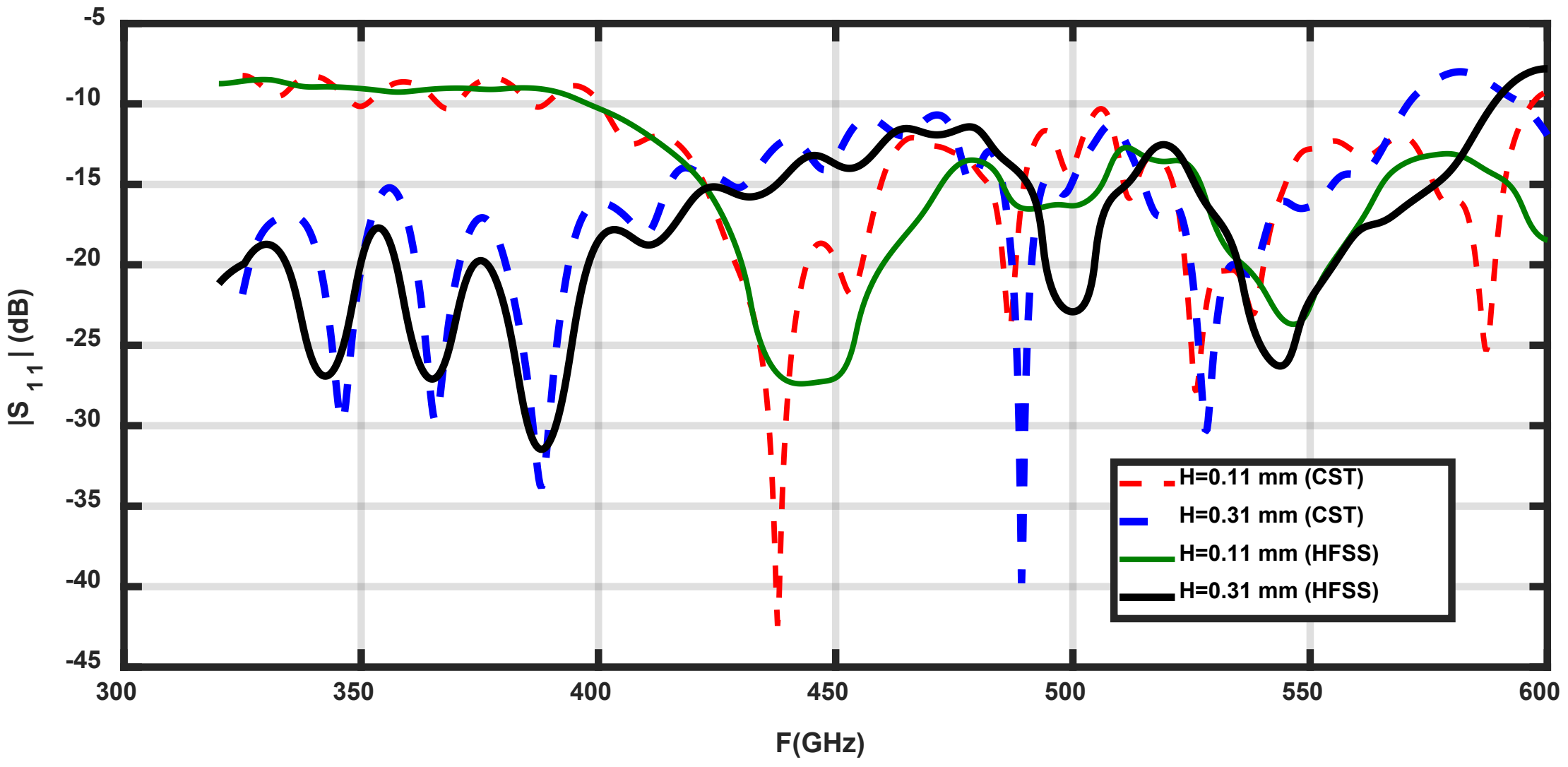

Figure 6.10 Return losses of the one element antenna

`

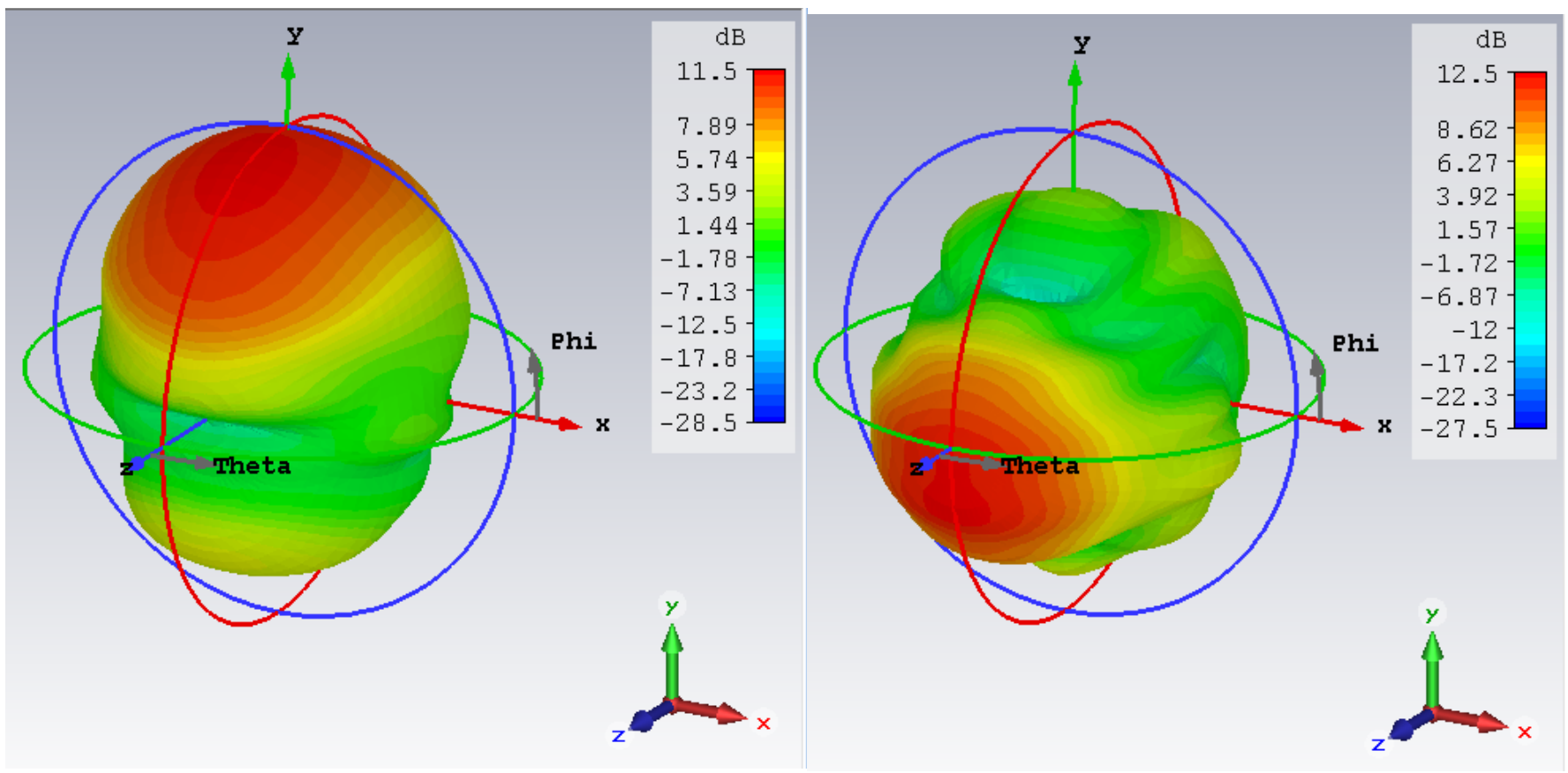

(a) H = 0.11 mm (b) H=0.31 mm

Figure 6.11 3-D radiation pattern of the antenna at 490 GHz.

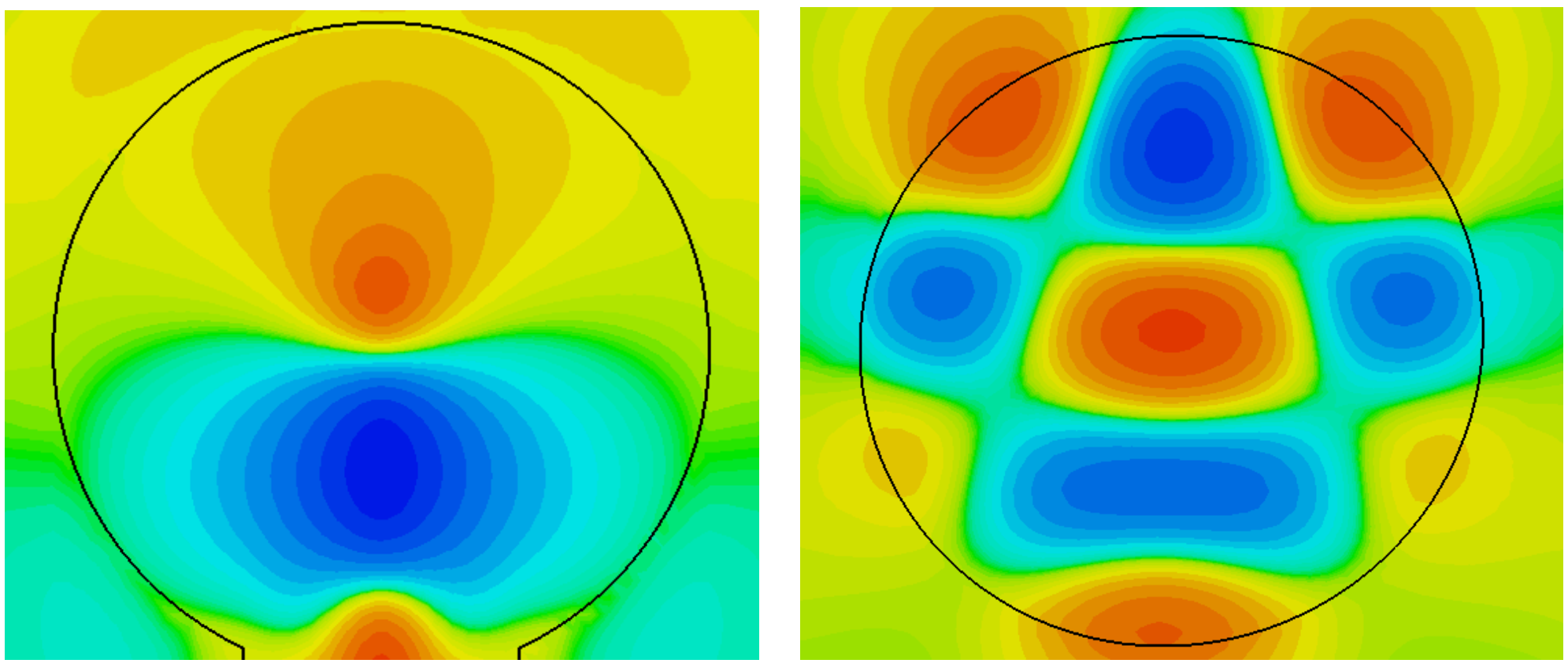

(a) $EH_{12}$ at H=0.11 mm (b)$EH_{25}$ at H=0.31 mm

Figure 6.12 Hybrid mode distribution for two antennas at 490 GHz

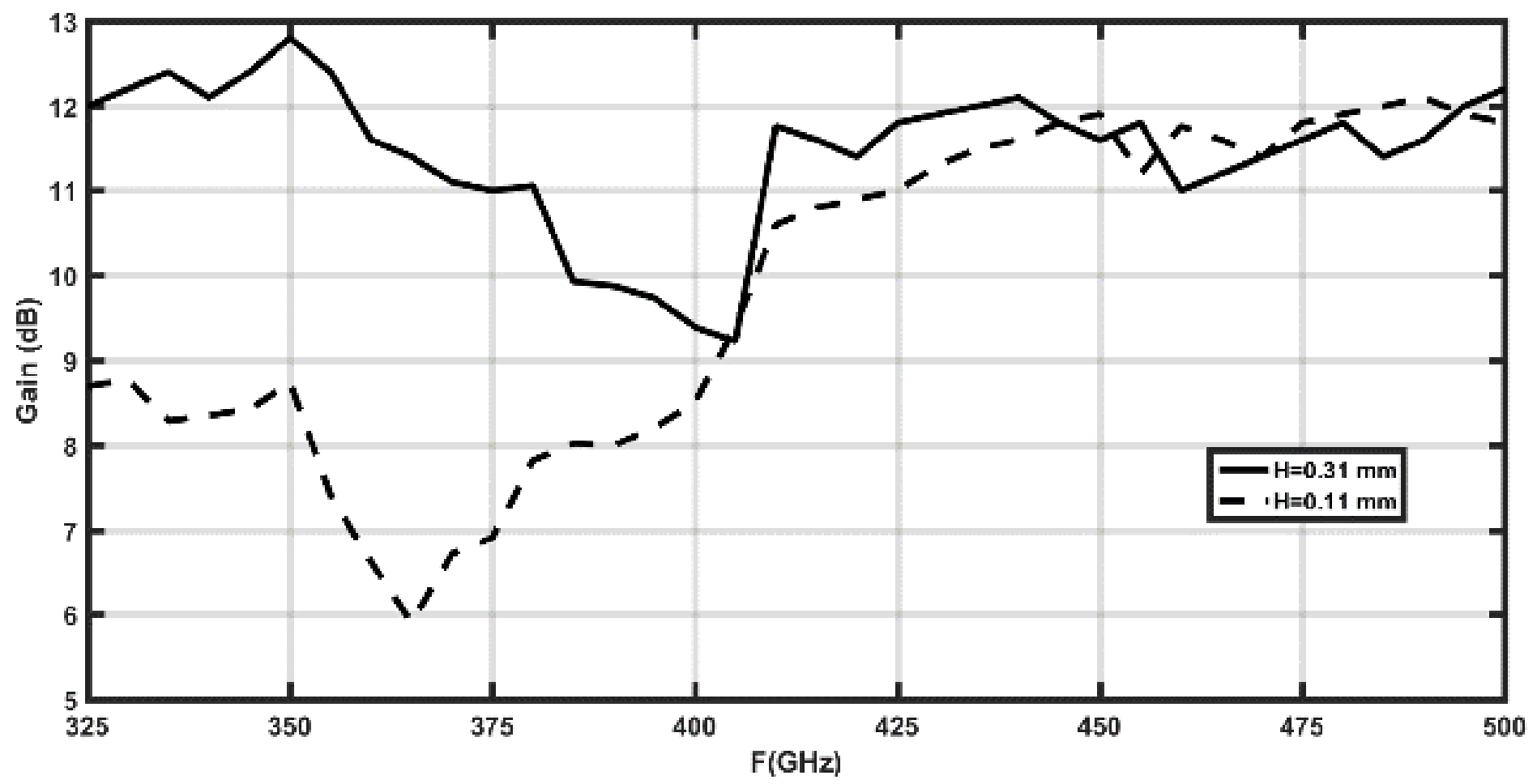

Figure 6.13 Gain of the proposed antenna

# 6.7 Design of Two Elements

The antenna array with two elements is introduced with CPW power divider with compact size as shown in Figure 6.14. The parameters of this power divider is optimized using built particle swarm optimization (PSO) tools in CST microwave studio. The proposed antenna consists of a silicon straight section waveguide segment connected in series with a disc resonator, which acts as a radiating element. The antenna array operates from 400 GHz to 500 GHz with height H=0.11mm, and with H=0.31 mm the antenna operates from 325 GHz to 500 GHz with the same height of DWG (0.11 mm) as shown in Figure 6.15. On the other hand, the CPW with curvature at the end of CPW operates as the Vivaldi antenna at a higher frequency, so at H=0.11 mm, the radiation pattern is end-fire. Still, with H=0.31 mm, the broadside radiation pattern is achieved, as shown in Figure 6.16. The gain of the antenna array is increased, as shown in Figure 6.17.

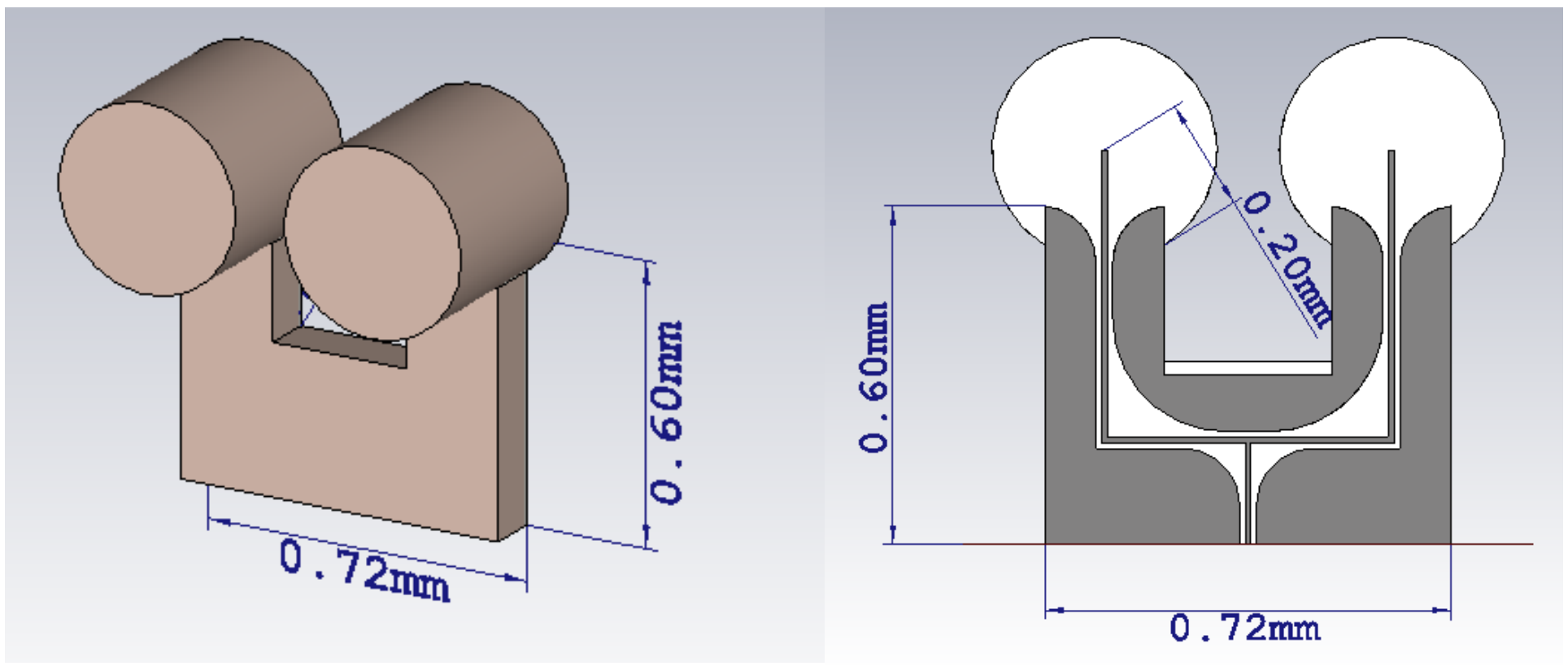

(a)3-D structure (b)Back view

Figure 6.14 Geometry of the antenna array

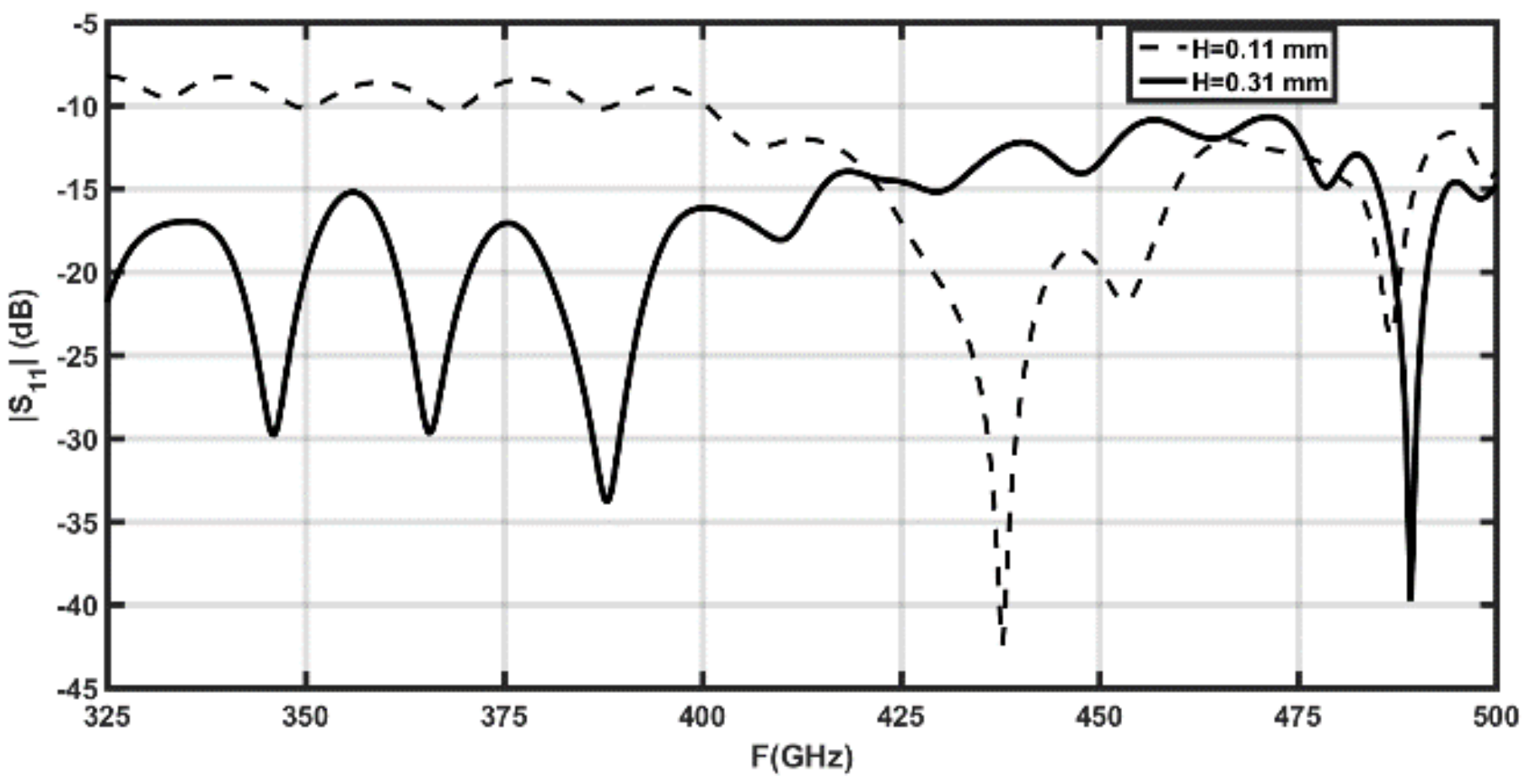

Figure 6.15 Return loss of the antenna

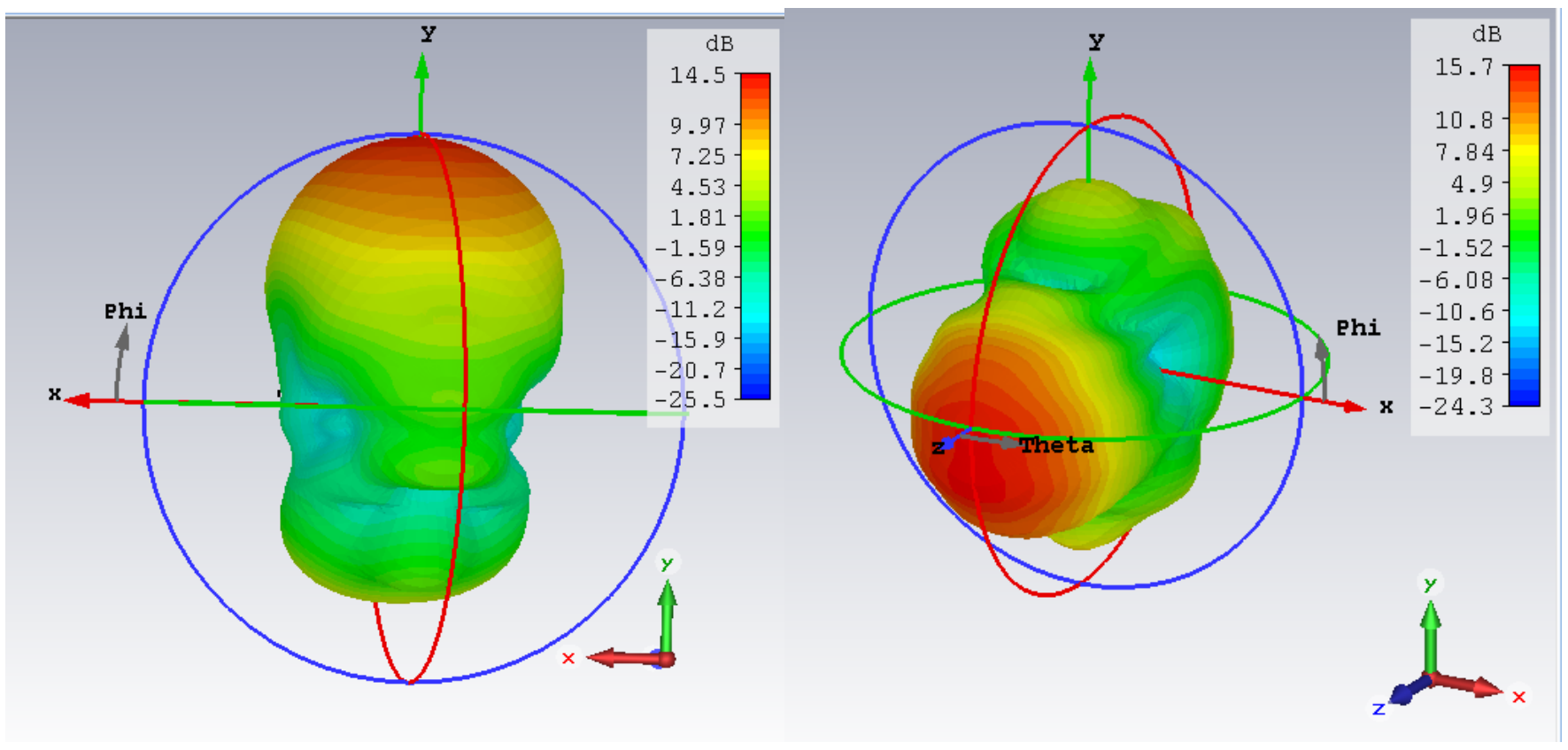

(a)H=0.11 mm (b)H=0.31 mm

Figure 6.16 3-D radiation pattern of antenna array at 490 GHz

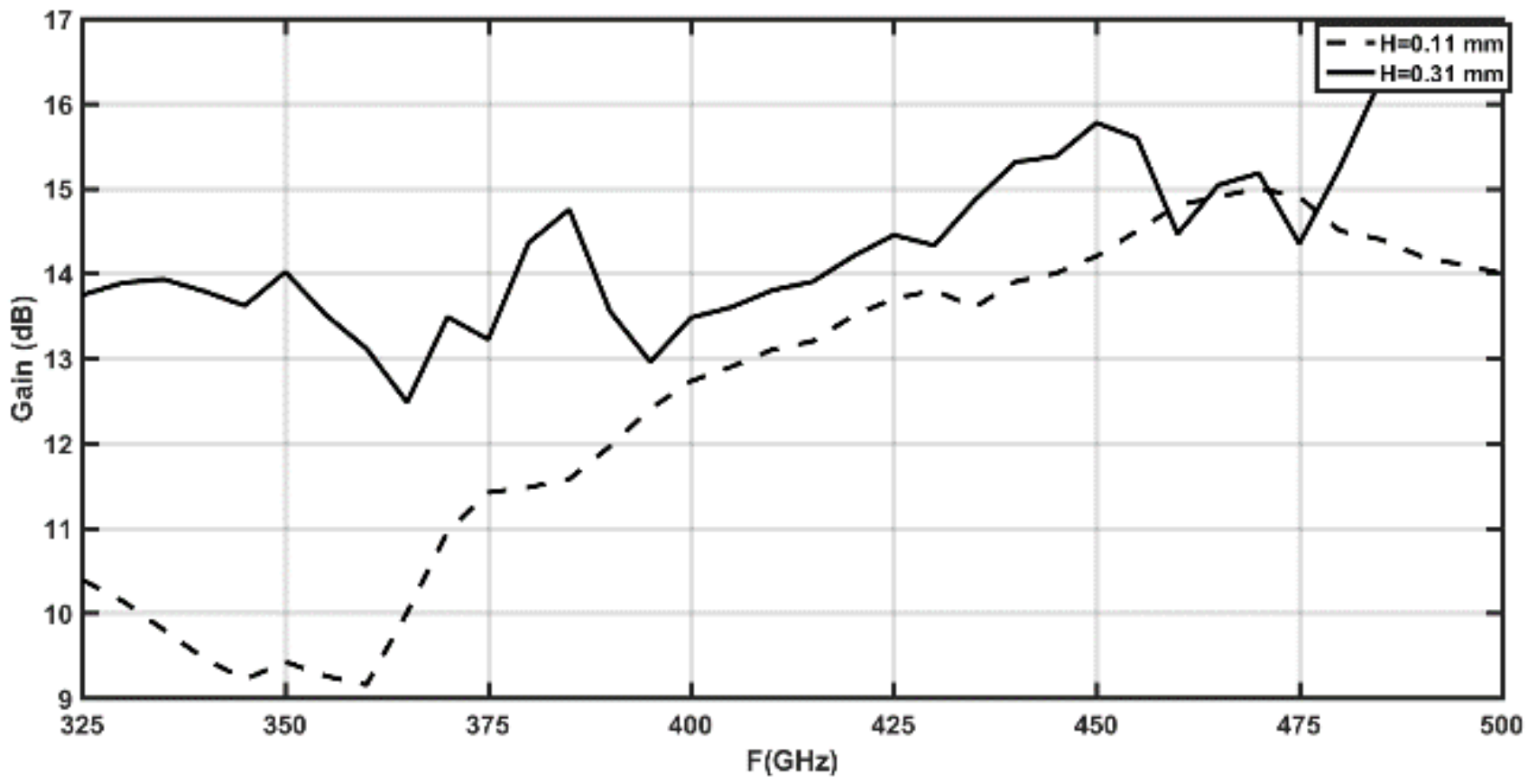

(a)Gain

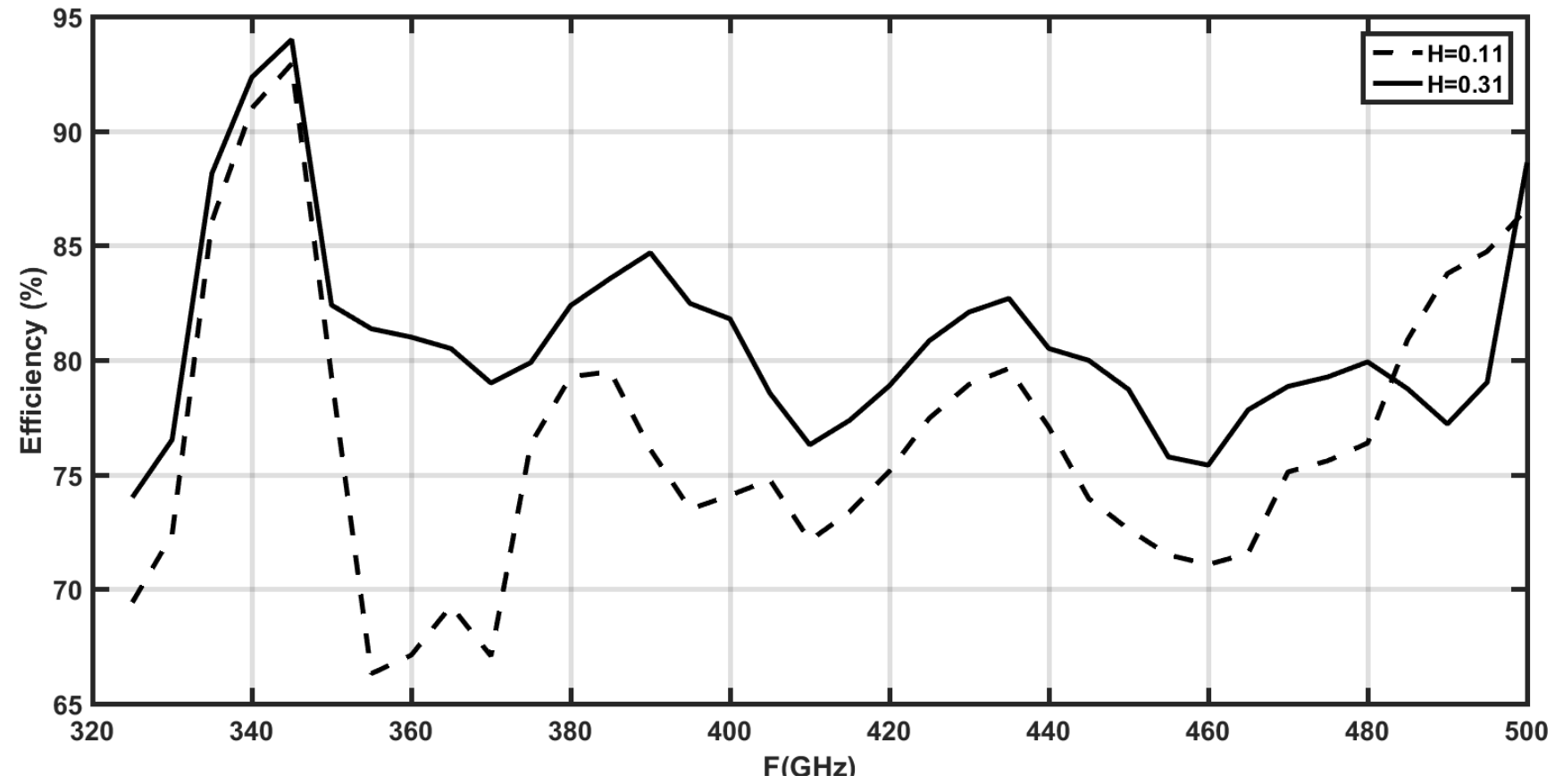

(b)Rradiation Efficiency

Figure 6.17 Radiation properties of antenna array.

# 6.8 Conclusion

A new dielectric antennas for one element disc and two elements are presented to support wide band for THz wireless communication applications. The proposed antenna consists of DRW section and disc resonator antenna feeding with CPW. The antenna array introduces more directivity and narrow beam width. The antenna array has gain of 16 dBi with efficiency 80%. The endfire and broadside radiation patterns are achieved with different disc height.

# Chapter Seven:
# Conclusion and Future work

## 7.1 Thesis Conclusions

The study in this thesis is provided to design, develop and implement different antennas for the millimeter and Sub-THz applications. All research work carried out aims at fulfilling the objectives of this thesis as outlined in section 1.2 and overcoming the limitations of existing designs for the millimeter and sub-THz applications as well as the challenges faced by each application. This chapter summarizes the contributions introduced in this thesis and indicates some future directions for further studies. The work in this thesis is divided to cover the most of millimeter and S-THz applications and to provide most of the technologies that can be used in these frequency ranges.

Antennas are one of the most important main parts of communications, radars, and imaging systems because the system's achievement depends on their performance. Therefore, the research work carried out in this thesis aims to introduce antennas for automotive radars, 5G mobile handsets, short communications, and multi gigs communications.

After the comprehensive studies that are introduced in chapter 2 for the main applications in the millimeter and sub-THz ranges, the thesis provides a complete analysis, verification and design for the concept of VAA. The VAA is introduced to solve the problems of large size, low isolations, low efficiency, low resolution and small range for the antenna arrays of automotive radar. The performance of the VAA is compared with the performance of the PAA. The VAA is designed and implemented with an overall size $30 \times 48 \times 0.2$ mm$^3$. The antenna achieved a high gain of 17 dBi, FSS radiation pattern shape to support LRR and MRR with beam width $\pm 7^0$ and $\pm 37^0$, respectively. The experimental results agree well with the simulated results. Furthermore, the second section of chapter 3 introduced a hybrid antenna at 77 GHz, consisting of a disc patch fed by an aperture coupled and ring dielectric resonator to give wide bandwidth from 75 GHz to 80 GHz. Furthermore, the AMC is used to provide a high gain of the antenna and isolate between the elements in case of the antenna array. Two configurations of antenna arrays (series and parallel configurations) for LRR at 77 GHz are introduced. These antenna arrays are used to achieve high resolution by introducing small HPBW. The second contribution in this thesis is introducing the antenna for the future mobile communications (5G). We introduce a concept of characteristic mode analysis to give complete analysis for the proposed antenna. The proposed 5G antenna combined between two slot antennas, one slot antenna is fed by micrstrip line from the opposite direction and the other slot antenna is fed in the same layer by CPW. The proposed antenna for 5G achieves dual polarization to overcome the problems of high attenuation and propagation losses at 28 GHz. Furthermore, the proposed antenna is combined with metasurface to enhance the bandwidth and the gain. Finally, a MIMO antenna with four elements is introduced. The interaction between the antenna and the human body for this antenna is taken in our consideration. The antenna is fabricated and measured. Good agreement is found between simulated and measured results.

The on-chip technology especially CMOS is used in this thesis as one of the popular technology in this band of frequency (especially for unlicensed band 57-64 GHz). Therefore, two different antenna configurations are introduced using CMOS technology to solve the problems of low efficiency and low gain associated to this technology. The first antenna is a Quazi-Yagi-uda antenna that composed of a driven element, a parasitic director and a reflector. The second antenna is a tapper slot Vivaldi antenna with different technique to enhance its radiation characteristics. The two antennas are designed using 0.18 μm CMOS technology with substrate size 5×5 $mm^2$ and the antennas size are 0.72×0.85 $mm^2$ and 0.5×0.87 $mm^2$ for Yagi-uda and Vivaldi, respectively. The two antennas achieved a gain of 0.4 dBi, 0.7 dBi and maximum radiation efficiency of 38%, 37% for the QYA and TSVA, respectively. The simulated antennas radiation pattern shows that the antennas have end-fire radiation characteristics and F/B ratio of 16 dB for Yagi-uda and 18 dB for Vivaldi. Finally, to solve the inherent propagation losses and increase the data rate for this application, a four elements MIMO antenna is introduced that can be used for short-range communications or indoor networks. The ECC, CCL, DG and TARC are introduced to prove that the proposed antenna has good performance over the proposed band.

The antenna for the S-THz applications is also presented with the technology of dielectric waveguide that can be suitable for this range of frequencies. The antenna array with two elements is introduced. The antenna consists of a dielectric waveguide (DWG) and disc resonator. The CPW power divider is used to convert the feeding method from a metallic waveguide to traditional feed method that is compatible with PCB designs. The antenna array achieved a gain of 16 dBi with high efficiency of 80% in addition to the fact that the antenna achieved end-fire and broadside radiation characteristics depending on the height of the disc resonator.

## 7.2 Suggestions for Future Works

Although the research studies carried out in this dissertation have effectively solved a number of significant problems in introducing effective antennas for the aforementioned applications, some proposed works may be recommended to further enhance for these applications:

1. Apply the antennas with the radar system and introducing an antenna for LRR, MRR and SRR is a challenge and can be considered a future work of this part. Furthermore, the VAA can be applied for different applications such as radars, satellites, and communication systems.
2. The work in the 5G applications is still under investigation to select the best mechanisms for it. Therefore, an intensive effort still needs to exert in this direction. The antenna with Omni-direction can be introduced. The work for the 5G base stations is needed for more studies.
3. The antenna arrays using CMOS with metamaterial can be used to enhance the indoor applications at 60 GHz.
4. The dielectric wave guide technology can be used to introduce compact antennas and sensors that can be used for imaging and security applications by introducing imaging algorithms to analyze the reflected signals from them.

# References

[1] S. Salous *et al.*, "Millimeter-wave propagation: characterization and modeling toward fifth-generation systems," *IEEE Antennas and Propagation Magazine,* vol. 58, no. 6, pp. 115-127, 2016.

[2] G. Liu *et al.*, "3-D-MIMO with massive antennas paves the way to 5G enhanced mobile broadband: from system design to field trials," *IEEE Journal on Selected Areas in Communications,* vol. 35, no. 6, pp. 1222-1233, 2017, doi: 10.1109/jsac.2017.2687998.

[3] P. H. Siegel, "Terahertz technology," *IEEE Transactions on Microwave Theory and Techniques,* vol. 50, no. 3, pp. 910-928, 2002.

[4] Y. S. Zhang and W. Hong, "A millimeter-wave gain enhanced multi-beam antenna based on a coplanar cylindrical dielectric lens," *IEEE Transactions on Antennas and Propagation,* vol. 60, no. 7, pp. 3485-3488, 2012.

[5] J. Hasch, E. Topak, R. Schnabel, T. Zwick, R. Weigel, and C. Waldschmidt, "Millimeter-wave technology for automotive Radar sensors in the 77 GHz frequency band," *IEEE Transactions on Microwave Theory and Techniques,* vol. 60, no. 3, pp. 845-860, 2012.

[6] J. Zhang, X. Ge, Q. Li, M. Guizani, and Y. Zhang, "5G millimeter-wave antenna array: design and challenges," *IEEE Wireless Communications,* vol. 24, no. 2, pp. 106-112, 2017.

[7] W. Hong, K. Baek, and S. Ko, "Millimeter-wave 5G antennas for smartphones: overview and experimental demonstration," *IEEE Transactions on Antennas and Propagation,* vol. 65, no. 12, pp. 6250-6261, 2017.

[8] T. S. Rappaport, Y. Xing, G. R. MacCartney, A. F. Molisch, E. Mellios, and J. Zhang, "Overview of millimeter wave communications for fifth-generation (5G) wireless networks—with a focus on propagation models," *IEEE Transactions on Antennas and Propagation,* vol. 65, no. 12, pp. 6213-6230, 2017.

[9] J. Bang and J. Choi, "A SAR reduced mm-wave beam-steerable array antenna with dual-mode operation for fully metal-covered 5G cellular handsets," *IEEE Antennas and Wireless Propagation Letters,* vol. 17, no. 6, pp. 1118-1122, 2018.

[10] B. Yu, K. Yang, C.-Y.-D. Sim, and G. Yang, "A novel 28 GHz beam steering array for 5G mobile device with metallic casing application," *IEEE Transactions on Antennas and Propagation,* vol. 66, no. 1, pp. 462-466, 2018.

[11] S. Zhang, I. Syrytsin, and G. F. Pedersen, "Compact beam-steerable antenna array with two passive parasitic elements for 5G mobile terminals at 28 GHz," *IEEE Transactions on Antennas and Propagation,* vol. 66, no. 10, pp. 5193-5203, 2018.

[12] P. Gupta, L. Malviya, and S. V. Charhate, "5G multi-element/port antenna design for wireless applications:a review," *International Journal of Microwave and Wireless Technologies,* vol. 11, no. 9, pp. 918-938, 2019.

[13] J. Wenger, "Automotive radar - status and perspectives," in *IEEE Compound Semiconductor Integrated Circuit Symposium, CSIC '05.*, Palm Springs, CA, USA,, 30 Oct.-2 Nov. 2005 2005, pp. 21-24.

[14] Y. He, K. Ma, N. Yan, Y. Wang, and H. Zhang, "A cavity-backed endfire dipole antenna array using substrate-integrated suspended line technology for 24 GHz band applications," *IEEE Transactions on Antennas and Propagation,* vol. 66, no. 9, pp. 4678-4686, 2018.

[15] C.-A. Yu *et al.*, "24 GHz horizontally polarized automotive antenna arrays with wide fan beam and high gain," *IEEE Transactions on Antennas and Propagation,* vol. 67, no. 2, pp. 892-904, 2019.

[16] J. Xu, W. Hong, H. Zhang, G. Wang, Y. Yu, and Z. H. Jiang, "An array antenna for both long- and medium-range 77 GHz automotive radar applications," *IEEE Trans. Antennas Propag.,* vol. 65, no. 12, pp. 7207-7216, Dec. 2017.

[17] C. Cui, S.-K. Kim, R. Song, J.-H. Song, S. Nam, and B.-S. Kim, "A 77-GHz FMCW radar system using on-chip waveguide feeders in 65-nm CMOS," *IEEE Transactions on Microwave Theory and Techniques,* vol. 63, no. 11, pp. 3736-3746, 2015.

[18] *Recommendation ITU-R P.676-10: Attenuation by atmospheric gases*, I. T. U. (ITU), 2013.

[19] J. S. Seybold, *Introduction to RF propagation*. Canda: John Wiley & Sons, Inc., 2005, p. 330.

[20] J. Wells, "Multigigabit wireless technology at 70 GHz, 80 GHz and 90 GHz," *Tx-Rx Technology,* pp. 50-58, May 2006.

[21] S. Shishanov *et al.*, "Height-finding for automotive THz radars," *IEEE Transactions on Intelligent Transportation Systems,* vol. 20, no. 3, pp. 1170-1180, 2019, doi: 10.1109/tits.2018.2845542.

[22] N. Ranjkesh, H. Amarloo, S. Gigoyan, N. Ghafarian, M. A. Basha, and S. Safavi-Naeini, "1.1 THz U-silicon-on-Ggass (U-SOG) waveguide: a low-loss platform for THz high-density integrated circuits," *IEEE Transactions on Terahertz Science and Technology,* vol. 8, no. 6, pp. 702-709, 2018.

[23] K. R. Jha and G. Singh, "Terahertz planar antennas for future wireless communication: A technical review," *Infrared Physics and Technology,* vol. 60, p. 71, 2013.

[24] W. Menzel, "Antennas in automobile radar," in *Handbook of Antenna Technologies*, 2016, ch. Chapter 96, pp. 2475-2500.

[25] ITU. "5G - fifth generation of mobile technologies." https://www.itu.int/en/mediacentre/backgrounders/Pages/5G-fifth-generation-of-mobile-technologies.aspx (accessed.

[26] A. Abdalrazik, A. S. A. El-Hameed, and A. B. Abdel-Rahman, "A three-port MIMO dielectric resonator antenna using decoupled modes," *IEEE Antennas and Wireless Propagation Letters,* vol. 16, pp. 3104-3107, 2017.

[27] B. Ahn, H.-W. Jo, J.-S. Yoo, J.-W. Yu, and H. L. Lee, "Pattern reconfigurable high gain spherical dielectric resonator antenna operating on higher order mode," *IEEE Antennas and Wireless Propagation Letters,* vol. 18, no. 1, pp. 128-132, 2019.

[28] H. Chu, Y. X. Guo, F. Lin, and X. Q. Shi, "Wideband 60GHz on-chip antenna with an artificial magnetic conductor," in *2009 IEEE International Symposium on Radio-Frequency Integration Technology (RFIT)*, Singapore, Singapore, 9 Jan.-11 Dec. 2009 2009, pp. 307-310.

[29] E. Elsaidy, A. Barakat, A. B. Abdel-Rahman, A. Allam, and R. K. Pokharel, "Radiation performance enhancement of a 60 GHz CMOS Quasi-Yagi antenna," in *2016 IEEE 17th Annual Wireless and Microwave Technology Conference (WAMICON)*, Clearwater, FL, USA, 11-13 April 2016 2016, pp. 1-4.

[30] S. Abdelhamied, S. H. Zainud-Deen, and H. A. Malhat, "Comparative study on the transmission gain between on-chip cylindrical dielectric resonators antenna and on-chip circular microstrip patch antenna for 60-GHz communications," in *2017 34th National Radio Science Conference (NRSC)*, Alexandria, Egypt, 13-16 March 2017 2017, pp. 22-29, doi: 10.1109/NRSC.2017.7893472.

[31] A. Barakat, A. Allam, H. Elsadek, A. B. Abdel-Rahman, S. M. Hanif, and R. K. Pokharel, "Miniaturized 60 GHz triangular CMOS Antenna-on-Chip using asymmetric artificial magnetic conductor," in *2015 IEEE 15th Topical Meeting on Silicon Monolithic Integrated Circuits in RF Systems*, San Diego, CA, USA, 26-28 Jan. 2015 2015, pp. 92-94, doi: 10.1109/SIRF.2015.7119885.

[32] A. Bronner, F. Schwarze, and F. Ellinger, "60 GHz On-Chip BiCMOS Bow-Tie Antenna," in *2018 IEEE-APS Topical Conference on Antennas and Propagation in Wireless Communications (APWC)*, Cartagena des Indias, 10-14 Sept. 2018 2018, pp. 769-772, doi: 10.1109/APWC.2018.8503749.

[33] K. Sultan, H. Abdullah, E. Abdallah, and H. El-Hennawy, "MOM/GA-Based Virtual Array for Radar Systems," *Sensors (Basel, Switzerland),* vol. 20, no. 3, 2020, doi: 10.3390/s20030713.

[34] K. S. Sultan, H. H. Abdullah, E. A. Abdallah, M. A. Basha, and H. H. El-Hennawy, "Dielectric resonator antenna with AMC for long range automotive radar applications at 77 GHz," in *2018 IEEE International Symposium on Antennas and Propagation & USNC/URSI National Radio Science Meeting*, 8-13 July 2018 2018, pp. 1617-1618, doi: 10.1109/APUSNCURSINRSM.2018.8608494.

[35] K. S. Sultan, H. H. Abdullah, E. A. Abdallah, and H. S. El-Hennawy, "Metasurface-Based Dual Polarized MIMO Antenna for 5G Smartphones Using CMA," *IEEE Access,* vol. 8, pp. 37250-37264, 2020, doi: 10.1109/ACCESS.2020.2975271.

[36] K. S. Sultan, E. A. Abdallah, and H. El Hennawy, "A multiple-input-multiple-output on-chip Quasi-Yagi-Uda antenna for multigigabit communications: Preliminary study," *Engineering Reports,* p. e12133, 2020.

[37] K. S. Sultan, M. A. Basha, H. H. Abdullah, E. A. Abdallah, and H. El-Hennawy, "A 60-GHz CMOS Quasi-Yagi antenna with enhanced radiation properties," in *12th European Conference on Antennas and Propagation (EuCAP 2018)*, 9-13 April 2018 2018, pp. 1-3, doi: 10.1049/cp.2018.1113. [Online]. Available: https://ieeexplore.ieee.org/document/8568791

[38] K. S. Sultan, H. H. Abdullah, E. A. Abdallah, M. A. Basha, and H. H. El-Hennawy, "A 60-GHz Gain Enhanced Vivaldi Antenna On-Chip," in *2018 IEEE International Symposium on Antennas and Propagation & USNC/URSI National Radio Science Meeting*, 8-13 July 2018 2018, pp. 1821-1822, doi: 10.1109/APUSNCURSINRSM.2018.8608914.

[39] K. S. Sultan, M. A. Basha, and S. Safavi-Naeini, "High gain disc resonator antenna array with CPW coupled for THz applications," in *2018 IEEE International Symposium on Antennas and Propagation & USNC/URSI National Radio Science Meeting*, 8-13 July 2018 2018, pp. 603-604, doi: 10.1109/APUSNCURSINRSM.2018.8608891.

[40] K. S. Sultan and M. A. Basha, "High gain CPW coupled disc resonator antenna for THz applications," in *2016 IEEE International Symposium on Antennas and Propagation (APSURSI)*, 26 June-1 July 2016 2016, pp. 263-264, doi: 10.1109/APS.2016.7695840.

[41] F. C. C. (FCC), "Operation within the bands 46.7-46.9 GHz and 76.0-77.0GHz," in "47 - Telecommunication," 61 FR 14503, 61 FR 41018, 63 FR 42279, Oct. 2010.

[42] F. C. C. (FCC), "Operation of radar services in the 76-81 GHz band," in "Engineering & Technology," Feb. 2015, vol. FCC-15-16.

[43] H. Winner, "Automotive radar," in *Handbook of Driver Assistance Systems*, 2015, ch. Chapter 17-1, pp. 1-63.

[44] D. Freundt and B. Lucas, "Long range radar sensor for high-volume driver assistance systems market," in *SAE World Congress & Exhibition*, Detroit, U.S.A, 2008, pp. 117-124.

[45] W. Menzel, "Millimeter-wave radar for civil applications," in *in Proc. Eur. Radar Conf*, Paris, France, Oct. 2010, pp. 89-92.

[46] H. Winner, "Automotive radar," in *Handbook of Driver Assistance Systems*, 2016, ch. Chapter 17, pp. 325-403.

[47] H. H. Meinel, "Evolving automotive radar – from the very beginnings into the future," in *8th European Conf. Antennas and Propagation*, The Hague, Netherlands, 2014, pp. 3107-3114.

[48] W. Menzel and A. Moebius, "Antenna concepts for millimeter-wave automotive radar sensors," *Proceedings of the IEEE,* vol. 100, no. 7, pp. 2372-2379, July 2012.

[49] V. Rabinovich and N. Alexandrov, *Antenna arrays and automotive applications*. New York: Springer Science Business Media, 2013.

[50] H. Rohling, "Milestones in radar and the success story of automotive radar systems," in *11th International Radar Sym.*, Lithuania, 2010, pp. 1-6.

[51] M. K. Saleem, H. Vettikaladi, M. A. S. Alkanhal, and M. Himdi, "Lens Antenna for Wide Angle Beam Scanning at 79 GHz for Automotive Short Range Radar Applications," *IEEE Transactions on Antenna and Propagation,* vol. 65, no. 4, pp. 2041 - 2046, 2017.

[52] G. C. V. Colome, G. Dassano, and M. Orefice, "Optimization of a lens-patch antenna for radar sensor applications," in *in ICECom*, Turin, 2005, pp. 1-4.

[53] P. Wenig, R. Weigel, and M. Schneider, "Dielectric lens antenna for digital beamforming and superresolution DOA estimation in 77 GHz automotive radars," in *Proceedings International ITG Workshop on Smart Antennas*, Vienna, Austria, 2008, pp. 184-189.

[54] T. Binzer, M. Klar, and V. Grob, "Development of 77 GHz radar lens antennas for automotive applications based on given requirements," in *2nd International ITG Conference on Antennas*, Munich, 2007, pp. 205-209.

[55] T. Metzler, "Microstrip series array," *IEEE Trans. Antennas Propag.,* vol. 29, no. 1, pp. 174–178, Jan. 1981.

[56] I. Hideo, K. Sakakibara, T. Watanabe, K. Sato, and K. Nishikawa, "Millimeter-wave microstrip array antenna with high efficiency for automotive radar systems," *R&D Review of Toyota CRDL,* vol. 37, no. 2, pp. 7-12, Apr. 2002.

[57] H. Lizuka, K. Sakakibara, T. Watanabe, K. Sato, and K. Nishikawa, "Millimeter wave microstrip array antenna with high efficiency for automative radar systems," *R&D Review of CRDL,* vol. 37, no. 2, pp. 7-12, 2002.

[58] C. Vasanelli, F. Bogelsack, and C. Waldschmidt, "Reducing the radar cross section of microstrip arrays using AMC structures for the vehicle integration of automotive radars," *IEEE, Transaction on Antennas and Propagation,* vol. 66, no. 3, pp. 1456-1464, March 2018.

[59] C. Buey, "Design and measurement of multi-antenna systems toward future 5G technologies," PhD, Université Côte d'Azur, 2018AZUR4023, 2018.

[60] T. Deckmyn, M. Cauwe, D. Vande Ginste, H. Rogier, and S. Agneessens, "Dual-band (28,38) GHz coupled quarter-mode substrate-integrated waveguide antenna array for next-generation wireless systems," *IEEE Transactions on Antennas and Propagation,* vol. 67, no. 4, pp. 2405-2412, 2019.

[61] M. Ikram, R. Hussain, and M. S. Sharawi, "4G/5G antenna system with dual function planar connected array," *IET Microwaves, Antennas & Propagation,* vol. 11, no. 12, pp. 1760-1764, 2017.

[62] S. Krishna, "Design and development of 5G spectrum massive MIMO array antennas for base station and access point applications," S. K. Sharma, Ed., ed: ProQuest Dissertations Publishing, 2018.

[63] H. G. D. Løvaas, "Multiband UWB antenna design for WiFi, LTE and 5G," E. Eide, Ed., ed: NTNU, 2017.

[64] N. Ojaroudiparchin, M. Shen, S. Zhang, and G. F. Pedersen, "A Switchable 3-D-coverage-phased array antenna package for 5G mobile terminals," *IEEE Antennas and Wireless Propagation Letters,* vol. 15, pp. 1747-1750, 2016.

[65] N. O. Parchin *et al.*, "Eight-element dual-polarized MIMO slot antenna system for 5G smartphone applications," *IEEE Access,* vol. 7, pp. 15612-15622, 2019, doi: 10.1109/access.2019.2893112.

[66] A. Zhao and Z. Ren, "Size reduction of self-isolated MIMO antenna system for 5G mobile phone applications," *IEEE Antennas and Wireless Propagation Letters,* vol. 18, no. 1, pp. 152-156, 2019.

[67] M. Ikram, K. Sultan, M. F. Lateef, and A. S. M. Alqadami, "A Road towards 6G Communication—A Review of 5G Antennas, Arrays, and Wearable Devices," *Electronics,* vol. 11, no. 1, p. 169, 2022. [Online]. Available: https://www.mdpi.com/2079-9292/11/1/169.

[68] D. Jackson, "Phased Array Antenna Handbook (Third Edition) [Book Review]," *IEEE Antennas and Propagation Magazine,* vol. 60, no. 6, pp. 124-128, 2018.

[69] A. Li, K. Luk, and Y. Li, "A dual linearly polarized end-fire antenna array for the 5G applications," *IEEE Access,* vol. 6, pp. 78276-78285, 2018.

[70] R. A. Alhalabi and G. M. Rebeiz, "High-efficiency angled-dipole antennas for millimeter-wave phased array applications," *IEEE Transactions on Antennas and Propagation,* vol. 56, no. 10, pp. 3136-3142, 2008.

[71] R. Hussain, A. T. Alreshaid, S. K. Podilchak, and M. S. Sharawi, "Compact 4G MIMO antenna integrated with a 5G array for current and future mobile handsets," *IET Microwaves, Antennas & Propagation,* vol. 11, no. 2, pp. 271-279, 2017.

[72] S. X. Ta, H. Choo, and I. Park, "Broadband printed-dipole antenna and its arrays for 5G applications," *IEEE Antennas and Wireless Propagation Letters,* vol. 16, pp. 2183-2186, 2017.

[73] J. Ala-Laurinaho *et al.*, "2-D beam-steerable integrated lens antenna system for 5G E-band access and backhaul," *IEEE Transactions on Microwave Theory and Techniques,* vol. 64, no. 7, pp. 2244-2255, 2016.

[74] M. M. S. Taheri, A. Abdipour, S. Zhang, and G. F. Pedersen, "Integrated millimeter-wave wideband end-fire 5G beam steerable array and low-frequency 4G LTE antenna in mobile terminals," *IEEE Transactions on Vehicular Technology,* vol. 68, no. 4, pp. 4042-4046, 2019.

[75] H. Tanaka and T. Ohira, "Beam-steerable planar array antennas using varactor diodes for 60-GHz-band applications," in *33rd European Microwave Conference Proceedings (IEEE Cat. No.03EX723C)*, Munich, Germany,, 7-7 Oct. 2003 2003, vol. 3, pp. 1067-1070 Vol.3, doi: 10.1109/EUMC.2003.177667.

[76] H. Tanaka and T. Ohira, "A single-planar integrated self-heterodyne receiver with a built-in beam-steerable array antenna for 60-GHz-band video transmission systems," in *IEEE MTT-S*

*International Microwave Symposium Digest (IEEE Cat. No.04CH37535)*, Fort Worth, TX, USA,, 6-11 June 2004, vol. 2, pp. 735-738 Vol.2.

[77] P. Wu and S. Chen, "Design of beam-steerable dual-beam reflectarray," in *2017 IEEE International Symposium on Antennas and Propagation & USNC/URSI National Radio Science Meeting*, San Diego, CA, USA, 9-14 July 2017 2017, pp. 2081-2082.

[78] Y. Yazid and G. Xun, "Beam-steerable patch antenna array using parasitic coupling and reactive loading," in *2007 IEEE Antennas and Propagation Society International Symposium*, Honolulu, HI, USA, 9-15 June 2007 2007, pp. 4693-4696.

[79] K. S. Sultan, M. Ikram, and N. Nguyen-Trong, "A Multi-band Multi-beam Antenna for Sub-6 GHz and Mm-Wave 5G Applications," *IEEE Antennas and Wireless Propagation Letters,* pp. 1-1, 2022, doi: 10.1109/lawp.2022.3164627.

[80] M. Ikram, K. Sultan, M. F. Lateef, and A. S. M. Alqadami, "A Road towards 6G Communication—A Review of 5G Antennas, Arrays, and Wearable Devices," *Electronics,* vol. 11, no. 1, p. 169, 2022, doi: 10.3390/electronics11010169.

[81] Q. Chen and H. Zhang, "Dual-patch polarization conversion metasurface-based wideband circular polarization slot antenna," *IEEE Access,* vol. 6, pp. 74772-74777, 2018.

[82] H.-L. Chu, "Investigations and design of wideband dual linear polarized massive MIMO panel array antenna for 5G communication applications," S. K. Sharma, Ed., ed: ProQuest Dissertations Publishing, 2018.

[83] X. Hu, S. Yan, and G. A. E. Vandenbosch, "Compact circularly polarized wearable button antenna with broadside pattern for U-NII worldwide band applications," *IEEE Transactions on Antennas and Propagation,* vol. 67, no. 2, pp. 1341-1345, 2019.

[84] A. Li and K. Luk, "Millimeter-wave dual linearly polarized end-fire antenna fed by 180-degree hybrid coupler," *IEEE Antennas and Wireless Propagation Letters,* pp. 1390 - 1394, 2019.

[85] H. Li, L. Kang, F. Wei, Y.-M. Cai, and Y.-Z. Yin, "A Low-profile dual-polarized microstrip antenna array for dual-mode OAM applications," *IEEE Antennas and Wireless Propagation Letters,* vol. 16, pp. 3022-3025, 2017.

[86] C. Wang, Y. Chen, and S. Yang, "Bandwidth enhancement of a dual-polarized slot antenna using characteristic modes," *IEEE Antennas and Wireless Propagation Letters,* vol. 17, no. 6, pp. 988-992, 2018.

[87] Y. Li *et al.*, "Eight-port orthogonally dual-polarized antenna array for 5G smartphone applications," *IEEE Transactions on Antennas and Propagation,* vol. 64, no. 9, pp. 3820-3830, Sep. 2016.

[88] L. Zhao, Z.-M. chen, and W. Jun, "A wideband dual-polarized omnidirectional antenna for 5G/WLAN," *IEEE Access,* vol. 7, pp. 14266-14272, Feb. 2019.

[89] H. Huang, X. Li , and Y. Liu, "A low profile, dual-polarized patch antenna for 5G MIMO application," *IEEE Transactions on Antennas and Propagation,* vol. 67, no. 2, pp. 1275 - 1279, Feb. 2019.

[90] M.-Y. Li, Z.-Q. Xu , Y.-L. Ban, C.-Y. D. Sim , and Z.-F. Yu, "Eight-port orthogonally dual-polarised MIMO antennas using loop structures for 5G smartphone," *IET Microwaves, Antennas & Propagation,* vol. 11, no. 12, pp. 1810 - 1816, 2017.

[91] P. Baniya, A. Bisognin, K. L. Melde, and C. Luxey, "Chip-to-chip switched beam 60 GHz circular patch planar antenna array and pattern considerations," *IEEE Transactions on Antennas and Propagation,* vol. 66, no. 4, pp. 1776-1787, 2018.

[92] X. Bao, Y. Guo, and Y. Xiong, "60-GHz AMC-based circularly polarized on-chip antenna using standard 0.18 μm CMOS technology," *IEEE Transactions on Antennas and Propagation,* vol. 60, no. 5, pp. 2234-2241, 2012.

[93] T. Chi, J. S. Park, S. Li, and H. Wang, "A millimeter-wave polarization-division-duplex transceiver front-end with an on-chip multifeed self-interference-canceling antenna and an all-passive reconfigurable canceller," *IEEE Journal of Solid-State Circuits,* vol. 53, no. 12, pp. 3628-3639, 2018.

[94] H. Chuang, L. Yeh, P. Kuo, K. Tsai, and H. Yue, "A 60-GHz millimeter-wave CMOS integrated on-chip antenna and bandpass filter," *IEEE Transactions on Electron Devices,* vol. 58, no. 7, pp. 1837-1845, 2011.

[95] F. Gutierrez, S. Agarwal, K. Parrish, and T. S. Rappaport, "On-chip integrated antenna structures in CMOS for 60 GHz WPAN systems," *IEEE Journal on Selected Areas in Communications,* vol. 27, no. 8, pp. 1367-1378, 2009.

[96] S. Hsu, K. Wei, C. Hsu, and H. Ru-Chuang, "A 60-GHz millimeter-wave CPW fed yagi antenna fabricated by using 0.18- μm CMOS technology," *IEEE Electron Device Letters,* vol. 29, no. 6, pp. 625-627, 2008.

[97] F. Huang, C. Lee, C. Kuo, and C. Luo, "MMW antenna in IPD process for 60-GHz WPAN applications," *IEEE Antennas and Wireless Propagation Letters,* vol. 10, pp. 565-568, 2011.

[98] H. Kuo, H. Yue, Y. Ou, C. Lin, and H. Chuang, "A 60-GHz CMOS sub-harmonic rf receiver with integrated on-chip artificial-magnetic-conductor Yagi antenna and balun bandpass filter for very-short-range gigabit communications," *IEEE Transactions on Microwave Theory and Techniques,* vol. 61, no. 4, pp. 1681-1691, 2013.

[99] Y. P. Zhang, M. Sun, and L. H. Guo, "On-chip antennas for 60-GHz radios in silicon technology," *IEEE Transactions on Electron Devices,* vol. 52, no. 7, pp. 1664-1668, 2005.

[100] C. Chan, C. Chou, and H. Chuang, "Integrated packaging design of low-cost bondwire interconnection for 60-GHz CMOS vital-signs radar sensor chip with millimeter-wave planar antenna," *IEEE Transactions on Components, Packaging and Manufacturing Technology,* vol. 8, no. 2, pp. 177-185, 2018.

[101] T. Mitomo *et al.*, "A 2-Gb/s throughput CMOS Transceiver Chipset With In-Package Antenna for 60-GHz Short-Range Wireless Communication," *IEEE Journal of Solid-State Circuits,* vol. 47, no. 12, pp. 3160-3171, 2012.

[102] Y. Tsutsumi, T. Ito, K. Hashimoto, S. Obayashi, H. Shoki, and H. Kasami, "Bonding wire loop antenna in standard ball grid array package for 60-GHz short-range wireless communication," *IEEE Transactions on Antennas and Propagation,* vol. 61, no. 4, pp. 1557-1563, 2013.

[103] T. Hirano, N. Li, T. Inoue, H. Yagi, K. Okada, and A. Matsuzawa, "Gain measurement of 60 GHz CMOS on-chip dipole antenna by proton irradiation," in *2017 International Symposium on Antennas and Propagation (ISAP)*, Phuket, Thailand, 30 Oct.-2 Nov. 2017 2017, pp. 1-2.

[104] T. Hirano *et al.*, "Design of 60 GHz CMOS on-chip dipole antenna with 50 % radiation efficiency by helium-3 ion irradiation," in *2015 IEEE Conference on Antenna Measurements & Applications (CAMA)*, Chiang Mai, Thailand, 30 Nov.-2 Dec. 2015 2015, pp. 1-2.

[105] T. Hirano, T. Yamaguchi, N. Li, K. Okada, J. Hirokawa, and M. Ando, "60 GHz on-chip dipole antenna with differential feed," in *2012 Asia Pacific Microwave Conference Proceedings*, Kaohsiung, Taiwan, 4-7 Dec. 2012 2012, pp. 304-306.

[106] S. Upadhyay and S. Srivastava, "A 60-GHz on-chip monopole antenna using silicon technology," in *2013 IEEE Applied Electromagnetics Conference (AEMC)*, Bhubaneswar, India, 18-20 Dec. 2013 2013, pp. 1-2.

[107] Y. Wu, C. Chou, W. Ruan, C. Yu, S. Huang, and H. Chuang, "60-GHz CMOS 2 × 2 artificial-magnetic-conductor monopole on-chip antenna array for phased-array RF receiving system," in *IEEE Antennas and Propagation Society International Symposium (APSURSI)*, Memphis, TN, USA, 6-11 July 2014 2014, pp. 436-437.

[108] A. Barakat, A. Allam, R. K. Pokharel, H. Elsadek, M. El-Sayed, and K. Yoshida, "60 GHz triangular monopole Antenna-on-Chip over an Artificial Magnetic Conductor," in *6th European Conference on Antennas and Propagation (EUCAP)*, Prague, Czech Republic, 26-30 March 2012 2012, pp. 972-976.

[109] D. Gang, H. Ming-Yang, and Y. Yin-Tang, "Wideband 60-GHz on-chip triangular monopole antenna in CMOS technology," in *Proceedings of 2014 3rd Asia-Pacific Conference on Antennas and Propagation*, Harbin, China, 26-29 July 2014 2014, pp. 623-626, doi: 10.1109/APCAP.2014.6992572.

[110] C. Lin, S. Hsu, C. Hsu, and H. Chuang, "A 60-GHz millimeter-wave CMOS RFIC-on-chip triangular monopole antenna for WPAN applications," in *2007 IEEE Antennas and Propagation Society International Symposium*, Honolulu, HI, USA, 9-15 June 2007 2007, pp. 2522-2525.

[111] A. S. A. El-Hameed, A. Barakat, A. B. Abdel-Rahman, A. Allam, and R. K. Pokharel, "A60-GHz double-Y balun-fed on-chip Vivaldi antenna with improved gain," in *27th International Conference on Microelectronics (ICM)*, Casablanca, Morocco, 20-23 Dec. 2015 2015, pp. 307-310.

[112] A. S. A. El-Hameed, N. Mahmoud, A. Barakat, A. B. Abdel-Rahman, A. Allam, and R. K. Pokharel, "A 60-GHz on-chip tapered slot Vivaldi antenna with improved radiation characteristics," in *10th European Conference on Antennas and Propagation (EuCAP)*, 10-15 April 2016, pp. 1-5, doi: 10.1109/EuCAP.2016.7481426.

[113] T. Lin, T. Chiu, Y. Chang, C. Hsieh, and D. Chang, "High-gain 60-GHz on-chip PIFA using IPD technology," in *IEEE International Symposium on Radio-Frequency Integration Technology (RFIT)*, Taipei, Taiwan, 24-26 Aug. 2016, pp. 1-3.

[114] K. Sultan and A. M. Abbosh, "Wearable Dual Polarized Electromagnetic Knee Imaging System," *IEEE Trans Biomed Circuits Syst,* vol. PP, pp. 1-1, Apr 5 2022, doi: 10.1109/TBCAS.2022.3164871.

[115] K. S. Sultan, B. Mohammed, M. Manoufali, A. Mahmoud, P. C. Mills, and A. Abbosh, "Feasibility of Electromagnetic Knee Imaging Verified on Ex-Vivo Pig Knees," *IEEE Trans Biomed Eng,* vol. 69, no. 5, pp. 1651-1662, May 2022, doi: 10.1109/TBME.2021.3126714.

[116] K. S. Sultan, B. Mohammed, M. Manoufali, and A. M. Abbosh, "Portable Electromagnetic Knee Imaging System," (in English), *Ieee Transactions on Antennas and Propagation,* vol. 69, no. 10, pp. 6824-6837, Oct 2021, doi: 10.1109/Tap.2021.3070015.

[117] K. Sultan, A. Mahmoud, and A. Abbosh, "Textile Electromagnetic Brace for Knee Imaging," *IEEE Trans Biomed Circuits Syst,* vol. 15, no. 3, pp. 522-536, Jun 2021, doi: 10.1109/TBCAS.2021.3085351.

[118] M. Koch, "Terahertz indoor communications: fundamental considerations and recent developments," in *39th International Conference on Infrared, Millimeter, and Terahertz waves (IRMMW-THz)*, Tucson, AZ, USA, 14-19 Sept. 2014, pp. 1-1.

[119] A. Abdellatif, S. Safavi-Naeini, A. Taeb, S. Gigoyan, and N. Ranjkesh, "W-band piezoelectric transducer-controlled low insertion loss variable phase shifter," *Electronics Letters,* vol. 50, no. 21, pp. 1537-1538, 2014.

[120] C. Jastrow, K. Münter, R. Piesiewicz, T. Kürner, M. Koch, and T. Kleine-Ostmann, "300 GHz transmission system," *Electronics Letters,* vol. 44, no. 3, pp. 1-2, 2008.

[121] W. M. Abdel-Wahab, M. Abdallah, J. Anderson, Y. Wang, H. Al-Saedi, and S. Safavi-Naeini, "SIW-integrated parasitic DRA array: analysis, design, and measurement," *IEEE Antennas and Wireless Propagation Letters,* vol. 18, no. 1, pp. 69-73, 2019.

[122] J. Federici and L. Moeller, "Review of terahertz and subterahertz wireless communications," *Journal of Applied Physics,* vol. 107, no. 11, 2010.

[123] G.-J. Kim, J.-I. Kim, S.-G. Jeon, J. Kim, K.-K. Park, and C.-H. Oh, "Enhanced continuous-wave terahertz imaging with a horn antenna for food inspection," *Journal of Infrared, Millimeter, and Terahertz Waves,* vol. 33, no. 6, pp. 657-664, 2012.

[124] T. Kleine-Ostmann and T. Nagatsuma, "A review on terahertz communications research," *Journal of Infrared, Millimeter, and Terahertz Waves,* vol. 32, no. 2, pp. 143-171, 2011/02/01 2011.

[125] A. A. Generalov, D. V. Lioubtchenko, and A. V. Räisänen, "Dielectric rod waveguide antenna at 75 – 1100 GHz," in *2013 7th European Conference on Antennas and Propagation (EuCAP)*, Gothenburg, Sweden, 8-12 April 2013 2013, pp. 541-544.

[126] W. Heinrich, "The flip-chip approach for millimeter wave packaging," *IEEE Microwave Magazine,* vol. 6, no. 3, pp. 36-45, 2005.

[127] N. Llombart, G. Chattopadhyay, A. Skalare, and I. Mehdi, "Novel terahertz antenna based on a silicon lens fed by a leaky wave enhanced waveguide," *IEEE Transactions on Antennas and Propagation,* vol. 59, no. 6, pp. 2160-2168, 2011.

[128] A. Patrovsky and K. Wu, "Active 60 GHz front-end with integrated dielectric antenna," *Electronics Letters,* vol. 45, no. 15, pp. 765-766, 2009.

[129] M. Koch, "Terahertz communications: A 2020 vision," ed, 2007, pp. 325-338.

[130] R. Piesiewicz, M. N. Islam, M. Koch, and T. Kurner, "Towards short-rangeterahertz communication systems: basic considerations," in *18th International Conference on Applied Electromagnetics and Communications*, Dubrovnik, Croatia, 12-14 Oct. 2005, pp. 1-5.

[131] M. S. Rabbani and H. Ghafouri-Shiraz, "Size improvement of rectangular microstrip patch antenna at MM-wave and terahertz frequencies," *Microwave and Optical Technology Letters,* vol. 57, no. 11, pp. 2585-2589, 2015, doi: 10.1002/mop.29400.

[132] M. S. Rabbani and H. Ghafouri-Shiraz, "Liquid crystalline polymer substrate-based THz microstrip antenna arrays for medical applications," *IEEE Antennas and Wireless Propagation Letters,* vol. 16, pp. 1533-1536, 2017.

[133] N. Ranjkesh, A. Taeb, N. Ghafarian, S. Gigoyan, M. A. Basha, and S. Safavi-Naeini, "Millimeter-wave suspended silicon-on-glass tapered antenna with dual-mode operation," *IEEE Transactions on Antennas and Propagation,* vol. 63, no. 12, pp. 5363-5371, 2015.

[134] N. Ranjkesh, A. Taeb, S. Gigoyan, M. Basha, and S. Safavi-Naeini, "Millimeter-wave silicon-on-glass Integrated tapered antenna," *IEEE Antennas and Wireless Propagation Letters,* vol. 13, pp. 1425-1428, 2014.

[135] R. Wu *et al.*, "A 60-GHz efficiency-enhanced on-chip dipole antenna using helium-3 ion implantation process," in *44th European Microwave Conference*, Rome, Italy, 6-9 Oct. 2014, pp. 108-111.

[136] G. C. Trichopoulos, H. L. Mosbacker, D. Burdette, and K. Sertel, "A broadband focal plane array camera for real-time THz imaging applications," *IEEE Transactions on Antennas and Propagation,* vol. 61, no. 4, pp. 1733-1740, 2013.

[137] Y. Qu, G. S. Liao, S. Q. Zhu, X. Y. Liu, and H. Jiao, "Performance comparisons of MIMO and phased-array radar," in *17th International Conference on Microwaves, Radar and Wireless Communications*, Wroclaw, Poland, May 2008, pp. 1-4.

[138] W.-Q. Wang, "Virtual antenna array analysis for MIMO synthetic aperture radars," *International Journal of Antennas and Propagation,* pp. 1-10, 2012.

[139] W.-Q. Wang, "Range-angle dependent transmit beampattern synthesis for linear frequency diverse arrays," *IEEE Transactions on Antennas and Propagation,* vol. 61, no. 8, pp. 4073-4081, 2013.

[140] J. Li, *MIMO radar signal processing*. Hoboken, N.J.: J. Wiley, 2009.

[141] C.-Y. Chen, "Signal processing algorithms for MIMO radar," P. P. Vaidyanathan, Ed., ed: ProQuest Dissertations Publishing, 2009.

[142] W.-Q. Wang, "Range-angle dependent transmit beampattern synthesis for linear frequency diverse arrays," *IEEE Transactions on Antennas and Propagation,* vol. 61, no. 8, pp. 4073 - 4081, Aug. 2013.

[143] A. H. Hussein, H. H. Abdullah, A. M. Salem, S. Khamis, and M. Nasr, "Optimum design of linear antenna arrays using a hybrid MoM/GA algorithm," *IEEE Antennas and Wireless Propagation Letters,* vol. 10, pp. 1232-1235, 2011.

[144] Y. Yu, W. Hong, H. Zhang, J. Xu, and Z. H. Jiang, "Optimization and Implementation of SIW Slot Array for Both Medium- and Long-Range 77 GHz Automotive Radar Application," *IEEE TRANSACTIONS ON ANTENNAS AND PROPAGATION,* vol. 66, no. 7, pp. 3769-3774, 2018.

[145] C. A. Balanis *Antenna theory: analysis and design*. John Wiley & Sons, Inc, 2016.

[146] D. M. Pozar, *Microwave engineering*. John Wiley & Sons, Inc, 2012.

[147] S. Honma and N. Uehara, "A fully-integrated 77 GHz FMCW radar transceiver in 65-nm CMOS technology," *IEEE Journal of Solid State Circuits,* vol. 45, no. 12, pp. 2746 - 2756, Dec. 2010.

[148] I. Hamieh, "A 77 GHz reconfigurable micromachined microstrip antenna array," Electronic Theses and Dissertations. University of Windsor, 2012.

[149] D. h. Shin, K.-b. Kim, J.-g. Kim, and S.-o. Park, "Design of low side lobe level milimeter-wave microstrip array antenna for automotive radar," in *Proceedings of the International Symposium on Antennas and Propagation*, Nanjing, China, 2013, pp. 1-4.

[150] B.-H. Ku *et al.*, "A 77–81-GHz 16-element phased-array receiver with +50 beam scanning for advanced automotive radars," *IEEE, Transactions on Microwave Theory and Techniques,* vol. 62, no. 11, pp. 2823 - 2832, Nov. 2014.

[151] S. B. Yeap, X. Qing, and Z. N. Chen, "77-GHz dual-layer transmit-array for automotive radar applications," *IEEE, Transactions on Antennas and Propagation,* vol. 63, no. 6, pp. 2833 - 2837, June 2015.

[152] S. Yasini and K. M. Aghdam, "Design and simulation of a comb-line fed microstrip antenna array with low side lobe level at 77GHz for automotive collision avoidance radar," in *2016 Fourth International Conference on Millimeter-Wave and Terahertz Technologies*, Tehran, Iran, 2016, pp. 87-90.

[153] M. G. N. Alsath, L. Lawrance, and M. Kanagasabai, "Bandwidth enhanced grid array antenna for UWB automotive radar sensors," *IEEE Transactions on Antennas and Propagation,* vol. 63, no. 11, pp. 5215 - 5219, 2015.

[154] S. B. Cohn, "Microwave bandpass filters containing high-Q dielectric resonators," *IEEE Transactions on Microwave Theory and Techniques,* vol. 16, no. 4, pp. 218-227, 1968.

[155] A. Perron, T. A. Denidni, and A. R. Sebak, "High gain dielectric resonator/microstrip hybrid antenna for millimeter-wave applications," in *2008 IEEE Antennas and Propagation Society International Symposium*, 5-11 July 2008 2008, pp. 1-4, doi: 10.1109/APS.2008.4618955.

[156] A. Perron, T. A. Denidni, and A. R. Sebak, "A low-cost and high-gain dual-polarized wideband millimeter-wave antenna," in *2009 3rd European Conference on Antennas and Propagation*, Berlin, Germany, 23-27 March 2009 2009, pp. 3558-3561.

[157] A. Perron, T. A. Denidni, and A. R. Sebak, "High-gain circularly polarized millimeter-wave antenna," in *IEEE Antennas and Propagation Society International Symposium*, Charleston, SC, USA, 1-5 June 2009, pp. 1-4.

[158] Z. Zhang, Y. Zhao, N. Liu, L. Ji, S. Zuo, and G. Fu, "Design of a dual-beam dual-polarized offset parabolic reflector antenna," *IEEE Transactions on Antennas and Propagation,* vol. 67, no. 2, pp. 712-718, 2019, doi: 10.1109/TAP.2018.2882593.

[159] S. Yan, Z. K. Meng, W. Y. Wei, W. Zheng, and L. Li, "Characteristic mode cancellation method and its application for antenna RCS reduction," *IEEE Antennas and Wireless Propagation Letters,* pp. 1784-1788, 2019, doi: 10.1109/LAWP.2019.2929834.

[160] Y. Su, X. Q. Lin, and Y. Fan, "Dual-band coaperture antenna based on a single-layer mode composite transmission line," *IEEE Transactions on Antennas and Propagation,* vol. 67, no. 7, pp. 4825-4829, 2019.

[161] W. Su, Q. Zhang, S. Alkaraki, Y. Zhang, X. Zhang, and Y. Gao, "Radiation energy and mutual coupling evaluation for multimode MIMO antenna based on the theory of characteristic mode," *IEEE Transactions on Antennas and Propagation,* vol. 67, no. 1, pp. 74-84, 2019, doi: 10.1109/TAP.2018.2878078.

[162] N. Peitzmeier and D. Manteuffel, "Upperbounds and design guidelines for realizing uncorrelated ports on multimode antennas based on symmetry analysis of characteristic modes," *IEEE Transactions on Antennas and Propagation,* vol. 67, no. 6, pp. 3902-3914, 2019, doi: 10.1109/TAP.2019.2905718.

[163] C. Guo, X. Zhao, C. Zhu, P. Xu, and Y. Zhang, "An OAM patch antenna design and its array for higher order OAM mode generation," *IEEE Antennas and Wireless Propagation Letters,* vol. 18, no. 5, pp. 816-820, 2019.

[164] L. Guan, Z. He, D. Ding, and R. Chen, "Efficient characteristic mode analysis for radiation problems of antenna arrays," *IEEE Transactions on Antennas and Propagation,* vol. 67, no. 1, pp. 199-206, 2019, doi: 10.1109/TAP.2018.2876705.

[165] X. Bi, G. Huang, X. Zhang, and T. Yuan, "Design of wideband and high-gain slotline antenna using multi-mode radiator," *IEEE Access,* vol. 7, pp. 54252-54260, 2019.

[166] L. Akrou and H. J. A. d. Silva, "Enhanced modal tracking for characteristic modes," *IEEE Transactions on Antennas and Propagation,* vol. 67, no. 1, pp. 356-360, 2019.

[167] C. Zhao and C.-F. Wang, "Characteristic mode design of wide band circularly polarized patch antenna consisting of h-shaped unit cells," *IEEE Access,* vol. 6, pp. 25292-25299, 2018, doi: 10.1109/access.2018.2828878.

[168] R. J. Garbacz, "Modal expansions for resonance scattering phenomena," *Proceedings of the IEEE,* vol. 53, no. 8, pp. 856-864, 1965, doi: 10.1109/PROC.1965.4064.

[169] R. Garbacz and R. Turpin, "A generalized expansion for radiated and scattered fields," *IEEE Transactions on Antennas and Propagation,* vol. 19, no. 3, pp. 348-358, 1971, doi: 10.1109/TAP.1971.1139935.

[170] R. Garbacz and E. Newman, "Characteristic modes of a symmetric wire cross," *IEEE Transactions on Antennas and Propagation,* vol. 28, no. 5, pp. 712-715, 1980, doi: 10.1109/TAP.1980.1142388.

[171] R. Garbacz and D. Pozar, "Antenna shape synthesis using characteristic modes," *IEEE Transactions on Antennas and Propagation,* vol. 30, no. 3, pp. 340-350, 1982, doi: 10.1109/TAP.1982.1142820.

[172] X. Yang, Y. Liu, and S.-X. Gong, "Design of a wideband omnidirectional antenna with characteristic mode analysis," *IEEE Antennas and Wireless Propagation Letters,* vol. 17, no. 6, pp. 993-997, 2018, doi: 10.1109/lawp.2018.2828883.

[173] D. Su, Z. Yang, and Q. Wu, "Characteristic mode assisted placement of antennas for the isolation enhancement," *IEEE Antennas and Wireless Propagation Letters,* vol. 17, no. 2, pp. 251-254, 2018, doi: 10.1109/LAWP.2017.2783328.

[174] Z. Liang, J. Ouyang, and F. Yang, "Design and characteristic mode analysis of a low-profile wideband patch antenna using metasurface," *Journal of Electromagnetic Waves and Applications,* vol. 32, no. 17, pp. 2304-2313, 2018, doi: 10.1080/09205071.2018.1507843.

[175] M. Bouezzeddine and W. L. Schroeder, "Design of a wideband, tunable four-port MIMO antenna system with high isolation based on the theory of characteristic modes.(multiple input multiple output)(Technical report)," vol. 64, ed: Institute of Electrical and Electronics Engineers, Inc., 2016, p. 2679.

[176] K. S. Sultan and B. Mohammed, "Compressed higher order modes slot loaded trapezoidal antenna for electromagnetic imaging," in *2020 4th Australian Microwave Symposium (AMS)*, 13-14 Feb. 2020 2020, pp. 1-2.

[177] R. Harrington and J. Mautz, "Theory of characteristic modes for conducting bodies," *IEEE Transactions on Antennas and Propagation,* vol. 19, no. 5, pp. 622-628, 1971, doi: 10.1109/TAP.1971.1139999.

[178] Y. Chen and C.-F. Wang, *Characteristics modes : theory and applications in antenna engineering*. Hoboken, New Jersey: John Wiley and Sons, Inc., 2015.

[179] M. Abdullah, S. H. Kiani, and A. Iqbal, "Eight element multiple-input multiple-output (MIMO) antenna for 5G mobile applications," *IEEE Access,* vol. 7, pp. 134488-134495, 2019, doi: 10.1109/ACCESS.2019.2941908.

[180] B. Feng, C. Zhu, J. Cheng, C. Sim, and X. Wen, "A dual-wideband dual-polarized magneto-electric dipole antenna with dual wide beamwidths for 5G MIMO microcell applications," *IEEE Access,* vol. 7, pp. 43346-43355, 2019, doi: 10.1109/ACCESS.2019.2906882.

[181] Q. Chen *et al.*, "Single Ring Slot-Based Antennas for Metal-Rimmed 4G/5G Smartphones," *IEEE Transactions on Antennas and Propagation,* vol. 67, no. 3, pp. 1476-1487, 2019, doi: 10.1109/TAP.2018.2883686.

[182] X. Zhang, Y. Li, W. Wang, and W. Shen, "Ultra-Wideband 8-Port MIMO Antenna Array for 5G Metal-Frame Smartphones," *IEEE Access,* vol. 7, pp. 72273-72282, 2019, doi: 10.1109/ACCESS.2019.2919622.

[183] Y. Li, C. Sim, Y. Luo, and G. Yang, "High-isolation 3.5 GHz eight-antenna MIMO array using balanced open-slot antenna element for 5G smartphones," *IEEE Transactions on Antennas and Propagation,* vol. 67, no. 6, pp. 3820-3830, 2019, doi: 10.1109/TAP.2019.2902751.

[184] M. S. Sharawi, M. Ikram, and A. Shamim, "A Two Concentric Slot Loop Based Connected Array MIMO Antenna System for 4G/5G Terminals," *IEEE Transactions on Antennas and Propagation,* vol. 65, no. 12, pp. 6679-6686, 2017, doi: 10.1109/TAP.2017.2671028.

[185] M. Ikram, N. Nguyen-Trong, and A. Abbosh, "Multiband MIMO Microwave and Millimeter Antenna System Employing Dual-Function Tapered Slot Structure," *IEEE Transactions on Antennas and Propagation,* vol. 67, no. 8, pp. 5705-5710, 2019, doi: 10.1109/TAP.2019.2922547.

[186] L. Liu, C. Liu, Z. Li, X. Yin, and Z. N. Chen, "Slit-Slot Line and Its Application to Low Cross-Polarization Slot Antenna and Mutual-Coupling Suppressed Tripolarized MIMO Antenna," *IEEE Transactions on Antennas and Propagation,* vol. 67, no. 1, pp. 4-15, 2019, doi: 10.1109/tap.2018.2876166.

[187] M. M. Mansour, K. S. Sultan, and H. Kanaya, "High-Gain Simple Printed Dipole-Loop Antenna for RF-Energy Harvesting Applications," in *2020 IEEE International Symposium on Antennas and Propagation & USNC/URSI National Radio Science Meeting*, 2020.

[188] M. Mansour, K. Sultan, and H. Kanaya, "Compact Dual-Band Tapered Open-Ended Slot-Loop Antenna For Energy Harvesting Systems," *Electronics (Basel),* vol. 9, no. 9, p. 1394, 2020, doi: 10.3390/electronics9091394.

[189] H. L. Zhu, S. W. Cheung, K. L. Chung, and T. I. Yuk, "Linear-to-circular polarization conversion using metasurface," *IEEE Transactions on Antennas and Propagation,* vol. 61, no. 9, pp. 4615-4623, 2013, doi: 10.1109/tap.2013.2267712.

[190] H. L. Zhu, S. W. Cheung, X. H. Liu, and T. I. Yuk, "Design of polarization reconfigurable antenna using metasurface," *IEEE Transactions on Antennas and Propagation,* vol. 62, no. 6, pp. 2891-2898, 2014, doi: 10.1109/TAP.2014.2310209.

[191] K. Konstantinidis, A. P. Feresidis, and P. S. Hall, "Broadband sub-wavelength profile high-gain antennas based on multi-layer metasurfaces," *IEEE Transactions on Antennas and Propagation,* vol. 63, no. 1, pp. 423-427, 2015, doi: 10.1109/TAP.2014.2365825.

[192] "Recent advances in metamaterials and metasurfaces," *IEEE Antennas and Propagation Magazine,* vol. 60, no. 6, pp. 129-129, 2018, doi: 10.1109/map.2018.2875252.

[193] T. Li and Z. N. Chen, "A dual-band metasurface antenna using characteristic mode analysis," *IEEE Transactions on Antennas and Propagation,* vol. 66, no. 10, pp. 5620-5624, 2018, doi: 10.1109/tap.2018.2860121.

[194] T. Li and Z. N. Chen, "Metasurface-based shared-aperture 5G S/K-band antenna using characteristic mode analysis," *IEEE Transactions on Antennas and Propagation,* vol. 66, no. 12, pp. 6742-6750, 2018, doi: 10.1109/TAP.2018.2869220.

[195] F. H. Lin and Z. N. Chen, "Low-profile wideband metasurface antennas using characteristic mode analysis," *IEEE Transactions on Antennas and Propagation,* vol. 65, no. 4, pp. 1706-1713, 2017, doi: 10.1109/tap.2017.2671036.

[196] P. Gao, S. He, X. Wei, Z. Xu, N. Wang, and Y. Zheng, "Compact printed UWB diversity slot antenna with 5.5-GHz band-notched characteristics," *IEEE Antennas and Wireless Propagation Letters,* vol. 13, pp. 376-379, 2014.

[197] S. Tripathi, A. Mohan, and S. Yadav, "A compact koch fractal UWB MIMO antenna with WLAN band-rejection," *IEEE Antennas and Wireless Propagation Letters,* vol. 14, pp. 1565-1568, 2015.

[198] Y. Li, C.-Y.-D. Sim, Y. Luo, and G. Yang, "12-Port 5G massive MIMO antenna array in sub-6GHz mobile handset for LTE bands 42/43/46 applications," *IEEE Access,* vol. 6, pp. 344-354, 2018.

[199] K. R. Jha and S. K. Sharma, "Combination of MIMO antennas for handheld devices," *IEEE Antennas and Propagation Magazine,* vol. 60, no. 1, pp. 118-131, 2018.

[200] R. Chandel, A. K. Gautam, and K. Rambabu, "Tapered fed compact UWB MIMO-diversity antenna with dual band-notched characteristics," *IEEE Transactions on Antennas and Propagation,* vol. 66, no. 4, pp. 1677-1684, 2018.

[201] H. A. Mohamed and K. Sultan, "Quad band monopole antenna for IoT applications," in *2018 IEEE International Symposium on Antennas and Propagation & USNC/URSI National Radio Science Meeting*, 2018: IEEE, pp. 1015-1016.

[202] K. S. Sultan, H. H. Abdullah, and E. A. Abdallah, "Low-SAR Miniaturized Handset Antenna Using EBG," in *Trends in Research on Microstrip Antennas*: Intech, 2017, pp. 127-147.

[203] K. S. Sultan, O. M. A. Dardeer, and H. A. Mohamed, "Low SAR, compact printed meander antenna for mobile and wireless applications," *International Journal of Microwave and Optical Technology,* vol. 12, no. 6, pp. 419-423, 2017.

[204] K. S. Sultan, H. H. Abdullah, and E. A. Abdallah, "Comprehensive study of printed antenna with the handset modeling," *Microwave and Optical Technology Letters,* vol. 58, no. 4, pp. 974-980, 2016.

[205] K. Sultan and H. Mohamed, "Low SAR, Novel compact Textile Wearable Antenna for Body Communications," in *PIET Conference*, 2015.

[206] K. S. Sultan, H. H. Abdullah, and E. A. Abdallah, "Low SAR, simple printed compact multiband antenna for mobile and wireless communication applications," *International Journal of Antennas and Propagation,* vol. 2014, no. 7, 2014, doi: 10.1155/2014/946781.

[207] K. Sultan, H. Abdullah, E. Abdallah, and E. Hashish, "Low SAR, Planar Monopole Antenna with Three Branch Lines for DVB, Mobile, and WLAN," *International Journal of Engineering & Technology IJET-IJENS,* vol. 14, no. No. 1, pp. 70-74, 2014.

[208] H. H. Abdullah and K. S. Sultan, "Multiband compact low SAR mobile hand held Antenna," *Progress in Electromagnetics Research Letters,* vol. 49, pp. 65-71, 2014, doi: 10.2528/PIERL14061605.

[209] K. S. Sultan, H. H. Abdullah, E. A. Abdallah, and E. A. Hashish, "Low SAR, Compact and Multiband Antenna," in *Progress in Electromagnetics Research Symposium Proceedings*, Taipei, Taiwan, 2013, pp. 748-751, doi: 10.13140/2.1.2145.2166.

[210] K. S. Sultan, H. H. Abdullah, E. A. Abdallah, and E. A. Hashish, "Low-SAR, miniaturized printed antenna for mobile, ISM, and WLAN services," *IEEE Antennas and Wireless Propagation Letters,* vol. 12, pp. 1106-1109, 2013, doi: 10.1109/LAWP.2013.2280955.

[211] K. S. Sultan, H. H. Abdullah, E. A. Abdallah, and E. A. Hashish, "Low SAR, compact and multiband antenna for mobile and wireless communication," in *The 2nd Middle East Conference on Antennas and Propagation*, 29-31 Dec. 2012 2012, pp. 1-5, doi: 10.1109/MECAP.2012.6618206.

[212] FCC, "Code of federal regulations CFR title 47, part 1.1310, radiofrequency radiation exposure limits," Federal Commun. Commission, Washington, DC, USA, 1997.

[213] ICNIRP, "Guidelines for limiting exposure to time-varying electric, magnetic, and electromagnetic fields (up to 300 GHz)," *Health Phys.,* vol. 74, no. 4, pp. 494–522, 1998.

[214] *IEEE standard for safety levels with respect to human exposure to radio frequency electromagnetic fields, 3 kHz to 300 GHz*, IEEE, 2005.

[215] *IEEE standard for safety levels with respect to human exposure to radio frequency electromagnetic fields, 3 kHz to 300 GHz. amendment 1: specifies ceiling limits for induced and contact current, clarifies distinctions between localized exposure and spatial peak power density*, IEEE, 2010.

[216] B. Thors, D. Colombi, Z. Ying, T. Bolin, and C. Törnevik, "Exposure to RF EMF from array antennas in 5G mobile communication equipment," *IEEE Access,* vol. 4, pp. 7469-7478, 2016, doi: 10.1109/ACCESS.2016.2601145.

[217] C. Leduc and M. Zhadobov, "Impact of antenna topology and feeding technique on coupling with human body: application to 60-GHz antenna arrays," *IEEE Transactions on Antennas and Propagation,* vol. 65, no. 12, pp. 6779-6787, 2017. [Online]. Available: https://ieeexplore.ieee.org/ielx7/8/8124133/07918591.pdf?tp=&arnumber=7918591&isnumber=8124133&ref=.

[218] B. Xu *et al.*, "Power density measurements at 15 GHz for RF EMF compliance assessments of 5G user equipment," *IEEE Transactions on Antennas and Propagation,* vol. 65, no. 12, pp. 6584-6595, 2017.

[219] B. Xu, M. Gustafsson, S. Shi, K. Zhao, Z. Ying, and S. He, "Radio frequency exposure compliance of multiple antennas for cellular equipment based on semidefinite relaxation," *IEEE Transactions on Electromagnetic Compatibility,* vol. 61, no. 2, pp. 327-336, 2019, doi: 10.1109/TEMC.2018.2832445.

[220] J. Wang, W. Wang, A. Liu, M. Guo, and Z. Wei, "Cross polarization suppression of a dual-polarized microstrip antenna using enclosed substrate integrated cavities," *IEEE Antennas and Wireless Propagation Letters,* pp. 1-1, 2019, doi: 10.1109/LAWP.2019.2953076.

[221] D. Liu, X. Gu, C. W. Baks, and A. Valdes-Garcia, "Antenna-in-package design considerations for Ka-band 5G communication applications," *IEEE Transactions on Antennas and Propagation,* vol. 65, no. 12, pp. 6372-6379, 2017, doi: 10.1109/tap.2017.2722873.

[222] C. Wu, C. Lu, and W. Cao, "Wideband dual-polarization slot antenna with high isolation by using microstrip line balun feed," *IEEE Antennas and Wireless Propagation Letters,* vol. 16, pp. 1759-1762, 2017, doi: 10.1109/LAWP.2017.2672538.

[223] J. Zhang, X. Q. Lin, L. Y. Nie, J. W. Yu, and Y. Fan, "Wideband dual-polarization patch antenna array with parallel strip line balun feeding," *IEEE Antennas and Wireless Propagation Letters,* vol. 15, pp. 1499-1501, 2016, doi: 10.1109/LAWP.2016.2514538.

[224] Y. Luo, Z. N. Chen, and K. Ma, "A single-layer dual-polarized differentially-fed patch antenna with enhanced gain and bandwidth operating at dual compressed high-order modes using characteristic mode analysis," *IEEE Transactions on Antennas and Propagation,* pp. 1-1, 2019, doi: 10.1109/TAP.2019.2951536.

[225] J. Zhang, K. Zhao, L. Wang, S. Zhang, and G. F. Pedersen, "Dual-polarized phased array with endfire radiation for 5G handset applications," *IEEE Transactions on Antennas and Propagation,* pp. 1-1, 2019, doi: 10.1109/TAP.2019.2937584.

[226] J. Guo, L. Cui, C. Li, and B. Sun, "Side-edge frame printed eight-port dual-band antenna array for 5G smartphone applications," *IEEE Transactions on Antennas and Propagation,* vol. 66, no. 12, pp. 7412-7417, 2018, doi: 10.1109/TAP.2018.2872130.

[227] Y.-W. Hsu, T.-C. Huang, H.-S. Lin, and Y.-C. Lin, "Dual-polarized quasi yagi–uda antennas with endfire radiation for millimeter-wave MIMO terminals," *IEEE Transactions on Antennas and Propagation,* vol. 65, no. 12, pp. 6282-6289, 2017, doi: 10.1109/tap.2017.2734238.

[228] P. Smulders, "Exploiting the 60 GHz band for local wireless multimedia access: prospects and future directions," *IEEE Communications Magazine,* vol. 40, no. 1, pp. 140-147, 2002.

[229] K. S. Sultan, T. A. Ali, N. A. Fahmy, and A. El-Shibiny, "Using millimeter-waves for rapid detection of pathogenic bacteria in food based on bacteriophage," *Engineering Reports,* vol. 1, no. 1, 2019, doi: 10.1002/eng2.12026.

[230] Y. P. Zhang and D. Liu, "Antenna-on-Chip and Antenna-in-Package Solutions to Highly Integrated Millimeter-Wave Devices for Wireless Communications," *IEEE Transactions on Antennas and Propagation,* vol. 57, no. 10, pp. 2830-2841, 2009.

[231] S. Beer, H. Gulan, C. Rusch, and T. Zwick, "Coplanar 122-GHz antenna array with air cavity reflector for integration in plastic packages," *IEEE Antennas and Wireless Propagation Letters,* vol. 11, pp. 160-163, 2012.

[232] K. Jeong-Geun, L. Hyung Suk, L. Ho-Seon, Y. Jun-Bo, and S. Hong, "60-GHz CPW-fed post-supported patch antenna using micromachining technology," *IEEE Microwave and Wireless Components Letters,* vol. 15, no. 10, pp. 635-637, 2005.

[233] K. T. Chan, A. Chin, Y. B. Chen, Y. Lin, T. S. Duh, and W. J. Lin, "Integrated antennas on Si, proton-implanted Si and Si-on-quartz," in *International Electron Devices Meeting. Technical Digest (Cat. No.01CH37224)*, Washington, DC, USA, 2-5 Dec. 2001, pp. 40.6.1-40.6.4.

[234] A. Barakat, A. Allam, H. Elsadek, H. Kanaya, and R. K. Pokharel, "Small size 60 GHz CMOS Antenna-on-Chip: Gain and efficiency enhancement using asymmetric Artificial Magnetic Conductor," in *2014 44th European Microwave Conference*, Rome, Italy, 6-9 Oct. 2014, pp. 104-107.

[235] X. Bao, Y. Guo, and S. Hu, "A 60-GHz differential on-chip Yagi antenna using 0.18-µm CMOS technology," in *2012 IEEE Asia-Pacific Conference on Antennas and Propagation*, Singapore, Singapore, 27-29 Aug. 2012, pp. 277-278.

[236] K. Ma, Y. Qian, and T. Itoh, "Analysis and applications of a new CPW-slotline transition," *IEEE Transactions on Microwave Theory and Techniques,* vol. 47, no. 4, pp. 426-432, 1999.

[237] X. Bao, Y. Guo, and S. Hu, "A 60-GHz Differential on-chip Yagi antenna using 0.18-µm CMOS technology," in *2012 IEEE Asia-Pacific Conference on Antennas and Propagation*, 27-29 Aug. 2012 2012, pp. 277-278, doi: 10.1109/APCAP.2012.6333253.

[238] Y. Huo, X. Dong, and J. Bornemann, "A wideband Artificial Magnetic Conductor Yagi antenna for 60-GHz standard 0.13-µm CMOS applications," in *2014 12th IEEE International*

*Conference on Solid-State and Integrated Circuit Technology (ICSICT)*, 28-31 Oct. 2014 2014, pp. 1-3, doi: 10.1109/ICSICT.2014.7021678. [Online]. Available: https://ieeexplore.ieee.org/ielx7/7001798/7021153/07021678.pdf?tp=&arnumber=7021678&isnumber=7021153&ref=

[239] M. S. Shamim, N. Mansoor, R. S. Narde, V. Kothandapani, A. Ganguly, and J. Venkataraman, "A Wireless Interconnection Framework for Seamless Inter and Intra-Chip Communication in Multichip Systems," *IEEE Transactions on Computers,* vol. 66, no. 3, pp. 389-402, 2017, doi: 10.1109/TC.2016.2605093.

[240] H. Singh, S. Mandal, S. K. Mandal, and A. Karmakar, "Design of miniaturised meandered loop on-chip antenna with enhanced gain using shorted partially shield layer for communication at 9.45 GHz," *IET Microwaves, Antennas & Propagation,* vol. 13, no. 7, pp. 1009-1016, 2019, doi: 10.1049/iet-map.2018.5974.

[241] M. K. Hedayati *et al.*, "Challenges in on-chip antenna design and integration with RF receiver front-end circuitry in nanoscale CMOS for 5G communication systems," *IEEE Access,* vol. 7, pp. 43190-43204, 2019, doi: 10.1109/ACCESS.2019.2905861.

[242] P. Burasa, T. Djerafi, N. G. Constantin, and K. Wu, "On-chip dual-band rectangular slot antenna for single-chip millimeter-wave identification tag in standard CMOS technology," *IEEE Transactions on Antennas and Propagation,* vol. 65, no. 8, pp. 3858-3868, 2017, doi: 10.1109/TAP.2017.2710215.

[243] W. A. Ahmad, M. Kucharski, A. D. Serio, H. J. Ng, C. Waldschmidt, and D. Kissinger, "Planar highly efficient high-gain 165 GHz on-chip antennas for integrated radar sensors," *IEEE Antennas and Wireless Propagation Letters,* vol. 18, no. 11, pp. 2429-2433, 2019, doi: 10.1109/LAWP.2019.2940110.

[244] T. Kaiser, F. Zheng, and E. Dimitrov, "An overview of ultra-wide-band systems With MIMO," *Proceedings of the IEEE,* vol. 97, no. 2, pp. 285-312, 2009.

[245] N. Ranjkesh, M. Basha, A. Taeb, and S. Safavi-Naeini, "Silicon-on-glass dielectric waveguide—part II: for THz applications," *IEEE Transactions on Terahertz Science and Technology,* vol. 5, no. 2, pp. 280-287, 2015.

[246] N. Ranjkesh, M. Basha, A. Taeb, A. Zandieh, S. Gigoyan, and S. Safavi-Naeini, "Silicon-on-glass dielectric waveguide—Part I: for millimeter-wave integrated circuits," *IEEE Transactions on Terahertz Science and Technology,* vol. 5, no. 2, pp. 268-279, 2015.

[247] A. A. Generalov, J. A. Haimakainen, D. V. Lioubtchenko, and A. V. Räisänen, "Wide band mm- and sub-mm-wave dielectric rod waveguide antenna," *IEEE Transactions on Terahertz Science and Technology,* vol. 4, no. 5, pp. 568-574, 2014.

[248] N. Ghassemi and K. Wu, "Planar Dielectric Rod Antenna for Gigabyte Chip-to-Chip Communication," *IEEE Transactions on Antennas and Propagation,* vol. 60, no. 10, pp. 4924-4928, 2012, doi: 10.1109/TAP.2012.2207359.

[249] S. M. Hanham and T. S. Bird, "High efficiency excitation of dielectric rods using a magnetic ring current," *IEEE Transactions on Antennas and Propagation,* vol. 56, no. 6, pp. 1805-1808, 2008, doi: 10.1109/TAP.2008.923335.

[250] D. V. Lioubtchenko, S. N. Dudorov, J. A. Mallat, and A. V. Raisanen, "Dielectric rod waveguide antenna for W band with good input match," *IEEE Microwave and Wireless Components Letters,* vol. 15, no. 1, pp. 4-6, 2005.

[251] S. A. Yahaya, M. Yamamoto, K. Itoh, and T. Nojima, "Dielectric rod antenna based on image NRD guide coupled to rectangular waveguide," *Electronics Letters,* vol. 39, no. 15, pp. 1099-1101, 2003.

[252] R. S. Yaduvanshi and H. Parthasarathy, *Rectangular dielectric resonator antennas: theory and design*, 1st ed. 2016 ed. New Delhi: Springer India, 2016.

[253] S. Yih, "Dielectric rod antennas for millimeter-wave integrated circuits," *IEEE Transactions on Microwave Theory and Techniques,* vol. 24, no. 11, pp. 869-872, 1976.

[254] L. K. Yeh, C. Y. Chen, and H. R. Chuang, "A millimeter-wave CPW CMOS on-chip bandpass filter using conductor-backed resonators," *IEEE Electron Device Letters,* vol. 31, no. 5, pp. 399-401, 2010.

[255] M. S. Abdallah, Y. Wang, W. M. Abdel-Wahab, and S. Safavi-Naeini, "Design and optimization of SIW center-fed series rectangular dielectric resonator antenna array with 45° linear polarization," *IEEE Transactions on Antennas and Propagation,* vol. 66, no. 1, pp. 23-31, 2018.